\documentclass[epj]{svjour}
\pdfoutput=1

\usepackage[utf8]{inputenc}
        \usepackage[T1]{fontenc}
\usepackage{lmodern}
\usepackage{textcomp}
\usepackage{microtype}

\usepackage[english]{babel}
\usepackage[autostyle, english = american]{csquotes}
\MakeOuterQuote{"}

\usepackage{amsmath}
\usepackage{amssymb}
\usepackage{mathtools}
\usepackage{upgreek}
\usepackage{units}
\usepackage{slashed}
\usepackage{physics}
\usepackage{authblk}

\usepackage{booktabs}
\usepackage{multirow}
\usepackage{graphicx}
\usepackage{graphics}
\usepackage{adjustbox}
\usepackage{xcolor}
\usepackage{wrapfig}
\usepackage{xspace}

\usepackage{tcolorbox}

\usepackage{rotating}
\usepackage{lineno}
\begin{document}


\pagenumbering{roman}



\title{  Feebly-Interacting Particles: FIPs 2020 Workshop Report}


\author{P.~Agrawal\inst{1} \and
M.~Bauer\inst{2} \and
J.~Beacham\inst{3} \and 
A.~Berlin\inst{4} \and
A.~Boyarsky\inst{5} \and
S.~Cebrian\inst{6} \and
X.~Cid-Vidal\inst{7} \and
D.~d'Enterria\inst{8} \and
A.~De~Roeck\inst{8} \and
M.~Drewes\inst{9} \and
B.~Echenard\inst{10} \and
M.~Giannotti\inst{11} \and
G.~F.~Giudice\inst{8} \and
S.~Gninenko\inst{12} \and
S.~Gori\inst{13} \and
E.~Goudzovski\inst{14} \and
J.~Heeck\inst{15} \and
P.~Hernandez\inst{16} \and
M.~Hostert\inst{17,18} \and
I.~G.~Irastorza\inst{6} \and
A.~Izmaylov\inst{12} \and
J. Jaeckel\inst{19} \and 
F.~Kahlhoefer\inst{20} \and
S.~Knapen\inst{8} \and  
G.~Krnjaic\inst{21} \and
G.~Lanfranchi\inst{22} \and
J.~Monroe\inst{23} \and
V.~I.~Martinez~Outschoorn\inst{24} \and
J.~Lopez-Pavon\inst{16} \and
S.~Pascoli\inst{2,25} \and
M.~Pospelov\inst{17} \and
D.~Redigolo\inst{8,26} \and
A.~Ringwald\inst{27} \and
O.~Ruchayskiy\inst{28} \and
J.~Ruderman\inst{4,27} \and
H.~Russell\inst{8} \and
J.~Salfeld-Nebgen\inst{29} \and
P.~Schuster\inst{30} \and
M.~Shaposhnikov\inst{31} \and
L.~Shchutska\inst{31} \and
J.~Shelton\inst{32} \and 
Y.~Soreq\inst{33} \and
Y.~Stadnik\inst{34} \and
J.~Swallow\inst{14} \and
K.~Tobioka\inst{35,36} \and
Y.-D.~Tsai\inst{24,37}}


\institute{ 
Rudolf Peierls Centre for Theoretical Physics, University of Oxford, Oxford, UK 
\and
Institute for Particle Physics Phenomenology, Department of Physics Durham University, Durham,  UK  
\and
Department of Physics, Duke University, Durham NC, USA 
\and
Center for Cosmology and Particle Physics, Department of Physics, New York University, New York, US 
\and
Instituut-Lorentz for Theoretical Physics, UniversiteitLeiden, Leiden, The Netherlands 
\and
Centro de Astropart\'iculas y F\'isica de Altas Energ\'ias (CAPA), Universidad de Zaragoza, Zaragoza, Spain 
\and
Instituto Galego de F\'isica de Altas Enerx\'ias, Universidade de Santiago de Compostela, Santiago, Spain  
\and
European Organization for Nuclear Research (CERN), Geneva, Switzerland 
\and
Centre for Cosmology, Particle Physics and Phenomenology, Universit\'e catholique de Louvain,  Louvain-la-Neuve, Belgium 
\and
Division of Physics, Mathematics and Astronomy, California Institute of Technology, Pasadena, US 
\and
Physical Sciences, Barry University,  Miami Shores, US  
\and
Institute for Nuclear Research of the Russian Academy of Sciences,  Moscow, Russia 
\and
Santa Cruz Institute for Particle Physics, University of California, Santa Cruz,  US 
\and
School of Physics and Astronomy, University of Birmingham, B15 2TT, United Kingdom 
\and
Department of Physics, University of Virginia,
Charlottesville, Virginia, US 
\and
Institut de F\'isica  Corpuscular - CSIC/Universitat de Val\`encia, Val\`encia, Spain 
\and
School of Physics and Astronomy and William I. Fine Theoretical Physics Institute, University of Minnesota, Minneapolis, US 
\and
Perimeter Institute for Theoretical Physics, Waterloo, Canada 
\and
Institute for Theoretical Physics, Heidelberg University, Heidelberg, Germany 
\and
Institute for Theoretical Particle Physics and Cosmology, Aachen University, Aachen, Germany 
\and
University of Chicago, Department of Astronomy and Astrophysics
and Kavli Institute for Cosmological Physics , Chicago, US 
\and
Laboratori Nazionali di Frascati, INFN,  Frascati (Rome), Italy 
\and
Royal Holloway, University of London, Egham Hill, United Kingdom
\and
University of Massachusetts, Amherst MA, US 
\and
Dipartimento di Fisica e Astronomia, Universita` di Bologna, Bologna, Italy 
\and
INFN, Sezione di Firenze and University of Florence, Sesto Fiorentino, Italy 
\and
DESY, Deutsches Elektronen-Synchrotron (DESY) Hamburg, Germany 
\and
Niels Bohr Institute, Copenhagen University, Copenhagen, Denmark 
\and
Department of Physics, Princeton University, Princeton, US 
\and
SLAC National Accelerator Laboratory Menlo Park,  US 
\and
Institute of Physics,  \'Ecole Polytechnique F\'ed\'erale de Lausanne (EPFL), Lausanne, Switzerland 
\and
University of Illinois, Urbana,  US 
\and
Physics Department, Technion, Institute of Technology, Haifa 3200003, Israel 
\and
Kavli Institute for the Physics and Mathematics of the Universe (KIPMU), University of Tokyo, Japan 
\and
Department of Physics, Florida State University, Tallahassee, US 
\and
High Energy Accelerator Research Organization (KEK), Tsukuba, Japan 
\and
Kavli Institute for Cosmological Physics, University of Chicago, Chicago, US 
}

\abstract{With the establishment and maturation of the experimental programs searching for new physics with sizeable couplings at the LHC, there is an increasing interest in the broader particle and astrophysics community for exploring the physics of light and feebly-interacting particles as a paradigm complementary to a New Physics sector at the TeV scale and beyond. 
FIPs 2020 has been the first workshop fully dedicated to the physics of feebly-interacting particles and was held virtually from 31 August to 4 September 2020. The workshop has gathered together experts from collider, beam dump, fixed target experiments, as well as from astrophysics, axions/ALPs searches, current/future neutrino experiments, and dark matter direct detection communities to discuss progress in experimental searches and underlying theory models for FIPs physics, and to enhance the cross-fertilisation across different fields. FIPs 2020 has been complemented by the topical workshop "Physics Beyond Colliders meets theory", held at CERN from 7 June to 9  June 2020. 
This document presents the summary of the talks presented at the workshops and the outcome of the subsequent discussions held immediately after. It aims to provide a clear picture of this blooming field and proposes a few recommendations for the next round of experimental results.}



\maketitle

\pagestyle{empty}  
\tableofcontents

\cleardoublepage
\pagestyle{plain} 
\setcounter{page}{1}
\pagenumbering{arabic}


\clearpage
\noindent{\large \bf Executive Summary}

\vskip 2mm
\noindent

Today particle physics faces the challenge of explaining the mystery of dark matter, the origin of matter over anti-matter in the Universe, the apparent fine-tuning of the electroweak scale, the strong CP problem, and many other aspects of fundamental physics. Perhaps the most striking frontier to emerge in the search for answers involves new physics at mass scales comparable to familiar matter ($\sim$ GeV scale) or below but with very feeble interaction strength with Standard Model (SM) particles.

\vskip 2mm
Feebly interacting particles (FIPs) have received a growing interest over the last decade, as demonstrated by several community planning documents such as the Cosmic~\cite{Feng:2014uja} and Intensity ~\cite{Hewett:2014qja} Frontier Reports of the 2013 Snowmass process, the Dark Sector community report~\cite{Alexander:2016aln}, the Cosmic Vision report~\cite{Battaglieri:2017aum}, the LHC Long-Lived Particle community white paper~\cite{Alimena:2019zri}, the Physics Beyond Colliders BSM report~\cite{Beacham:2019nyx}, the White Paper on new opportunities for next generation neutrino experiments~\cite{Arguelles:2019xgp}, and the Briefing Book of the European Strategy for Particle Physics~\cite{Strategy:2019vxc}.

\vskip 2mm
FIPs are defined by the presence of extremely suppressed interactions between new particles and the SM bosons and/or fermions. The smallness of the coupling can be generated by an approximate symmetry almost unbroken, and/or a large mass hierarchy between particles. While searches for new particles have generally been focused on weak scale masses,
mostly driven by the most popular theoretical ideas to extend the Standard Model, such as supersymmetry, 
and various low-energy limits of string theory,
the attention has recently shifted towards scales lighter than the electroweak one, after realizing that sub-GeV FIPs present many attractive features (see, e.g.~\cite{Alexander:2016aln,Battaglieri:2017aum,Beacham:2019nyx,Lanfranchi:2020crw}). 
As a matter of fact, light FIPs provide a well-motivated dark matter (DM) candidate, as well as an explanation for the origin of neutrino masses, the CP symmetry in strong interactions, or the baryon asymmetry in the universe. Quite surprisingly, these possibilities are only weakly constrained by current experimental observations.

\vskip 2mm
Indirect interactions of FIPs with SM particles are possible through low-dimensional operators.
These interactions are referred to as {\it portals} (see e.g. \cite{Beacham:2019nyx,Batell:2009di} for reviews). The lowest dimensional portals includes: the {\it vector portal}, mediated by a dark gauge boson dubbed dark photon; {\it the scalar portal}, mediated by a new scalar mixing to the SM Higgs boson; {\it the fermion portal}, mediated by a heavy neutral lepton (HNL) interacting with one of the left-handed SM doublets and the Higgs boson; and {\it the pseudoscalar portal}, mediated by an axion (or axion-like particle) coupling to gauge and fermion fields. The first three cases are renormalizable and unsuppressed by any (potentially very large) physics scale, while the axion portal has dimension five and is suppressed by the axion decay constant. Extensions of these portals can be constructed by gauging accidental symmetries of the SM or individual flavor numbers~\cite{Ilten:2018crw}. 

\vskip 2mm
Viable DM scenarios can be constructed for each portal, with the role of DM played by either the mediator or a new dark sector particle. The vector portal offers perhaps one of the most appealing possibilities: light freeze-out WIMP-like DM~\cite{Pospelov:2007mp,ArkaniHamed:2008qn}. A MeV-GeV dark gauge boson could enable the annihilation of a pair of DM particles into SM fermions, depleting the DM density in the early universe. The thermal relic abundance defines specific values of the parameters of the theory as a function of the DM mass and spin~\cite{Izaguirre:2013uxa}.

\vskip 2mm
Freeze-in DM, in which DM interactions with the SM are too weak to allow full thermalization in the early Universe, could also be realized with many FIPs, including light gauge bosons or HNLs~\cite{Hall:2009bx}. The axion and axion-like particles are  prime candidates for bosonic DM as well~\cite{Preskill:1982cy,Abbott:1982af,Dine:1982ah,Arias:2012az}, and the QCD axion is often invoked as a solution to the strong CP problem~\cite{Peccei:1977hh}. 
Right-handed neutrino fields, involved in neutrino mass generation mechanisms~\cite{Minkowski:1977sc}, are another prominent example of FIPs
belonging to the fermion portal. In all cases, a substantial fraction of the parameter space remains to be explored.

The use of the {\it portal framework} is mostly driven by the aim of embracing a large fraction of the phenomenology relevant to FIP searches with the attitude that more complex models that address broader SM puzzles would inevitably be covered by this approach. 
As such, the portal framework is expected to be superseded by more complete models that will provide clues to help prioritize regions of parameter space deserving of special experimental  attention. 
On the other hand, this systematic approach is allowing the community to combine/compare different experimental results in order to identify promising and still uncovered regions of parameter space to guide future proposals/efforts, and also to crucially test possible new physics explanations in the case of positive signals.

\vskip 2 mm
In order to explore the broad range in mass and couplings that FIP physics suggests, a {healthy diversity of small to medium scale experiments} operating at several facilities and employing a range of techniques is required. Deriving meaningful scientific conclusions from these varied studies will require extensive collaboration and cross-fertilization across different communities. The FIPs 2020 workshop has been an important step in this direction.

\vskip 2mm
This document follows closely the structure of the workshop where each Session aimed at approaching the same topic from multiple different viewpoints.
In the {\it Introductory  section} (Section~\ref{sec:introduction}), the search for FIPs is put into context: after a brief introduction of the phenomenological framework (Section~\ref{ssec:knapen}),
the strict connection of FIPs physics with cosmology (Section~\ref{ssec:pospelov}) and astrophysics (Section~\ref{ssec:giannotti}) are presented along with the most common techniques used to explore a large spectrum of possible FIP candidates, ranging from ultra-light bosonic DM (Section~\ref{ssec:stadnik}) to FIP searches at accelerator-based experiments (Sections~\ref{ssec:echenard}-\ref{ssec:russell}).
Section~\ref{sec:vector} is fully dedicated to {\it models and experimental searches for DM with thermal origin in the MeV-few~GeV range} using direct detection techniques (Section~\ref{ssec:cebrian}) and experiments at extracted beams (Section~\ref{ssec:gninenko}) and colliders (Section~\ref{ssec:salfeld}).
{\it Axions and axion-like particles (ALPs)} are presented in Section~\ref{sec:pseudoscalar}:  After a theoretical overview (Section~\ref{ssec:ringwald}), the most used experimental techniques are presented covering a vast range of couplings and mass ranging from $10^{-12}$~eV to the TeV scale (Sections~\ref{ssec:irastorza}--\ref{ssec:denterria}).
{\it The scalar portal}, and its deep connection to Higgs physics, is outlined in Section~\ref{sec:scalar}. 
Finally, {\it heavy neutral leptons or HNL} as responsible  of the origin of the neutrino masses and oscillations and leptogenesis~(Section~\ref{ssec:drewes}) are presented in Section~\ref{sec:fermion} along with multiple studies that can be performed to test their Majorana nature (or not). These tests include correlations with active neutrino physics parameters and neutrinoless double-beta decays (Section~\ref{ssec:lopezpavon}). The current status and prospects for direct searches for HNLs at extracted beams~ and colliders is presented in Section~\ref{ssec:izmaylov} and ~\ref{ssec:shschutska}, respectively.

At the end of each Chapter, a Section with the summary of current experimental bounds and future projections is presented along with a few recommendations for the next round of experimental results.
{\it Conclusions and outlook} are summarised in Section~\ref{sec:conclusions}.

\clearpage

\section{Introductory talks}
\label{sec:introduction}

\subsection{Models for feebly-interacting particles}
\label{ssec:knapen}
{\it Author: Simon Knapen, <simon.knapen@cern.ch>} 



\subsubsection{Theory priors for FIPS}
\label{intro}
The concept of feebly interacting particles (FIPs) has (re)generated a lot of interest in the recent years, both on the theoretical and experimental ends. 
My impression is that this is largely driven by pragmatic considerations, as
FIPs can be accessible to a very diverse set of probes, which includes dark matter direct detection experiments, astrophysical/cosmological observations and various low and high energy (future) accelerator facilities.  
%
That said, there is solid theory motivation for FIPs, though it is substantially more subtle and diverse than more well known theoretical paradigms such as weak scale supersymmetry or WIMPs. Perhaps a comprehensive review on models for FIPs is in order, given the interest of the experimental community and the very large body of theory literature that is out there.
%
This talk does not attempt to do that. I merely hope to supply some thoughts on how one may organize the available models for FIPs and provide a few concrete examples. The list of topics and especially references are therefore highly incomplete, and largely due to personal bias.

A set of crisp criteria of what commonly counts as ``feebly interacting particles'' are not so trivial to pin down: One may be tempted to define ``feeble'' as ``weaker than weak'', as compared to the SM weak force. I don't particularly like this definition.
This definition does not reflect the FIPs world, as, for example, millicharged particles can interact much more strongly with our detectors than the SM neutrinos\footnote{The SM neutrinos are themselves sometimes counted as FIPs, though in this write-up I will exclusively focus on extensions of the SM.}.
Necessary conditions do appear to be that the FIP is not charged under the SM strong force and has a mass $\lesssim 10$ GeV. The latter is largely for pragmatic reasons, though there are theoretical arguments for this mass range as well. 

\begin{figure*}\centering
\includegraphics[width=0.8\textwidth]{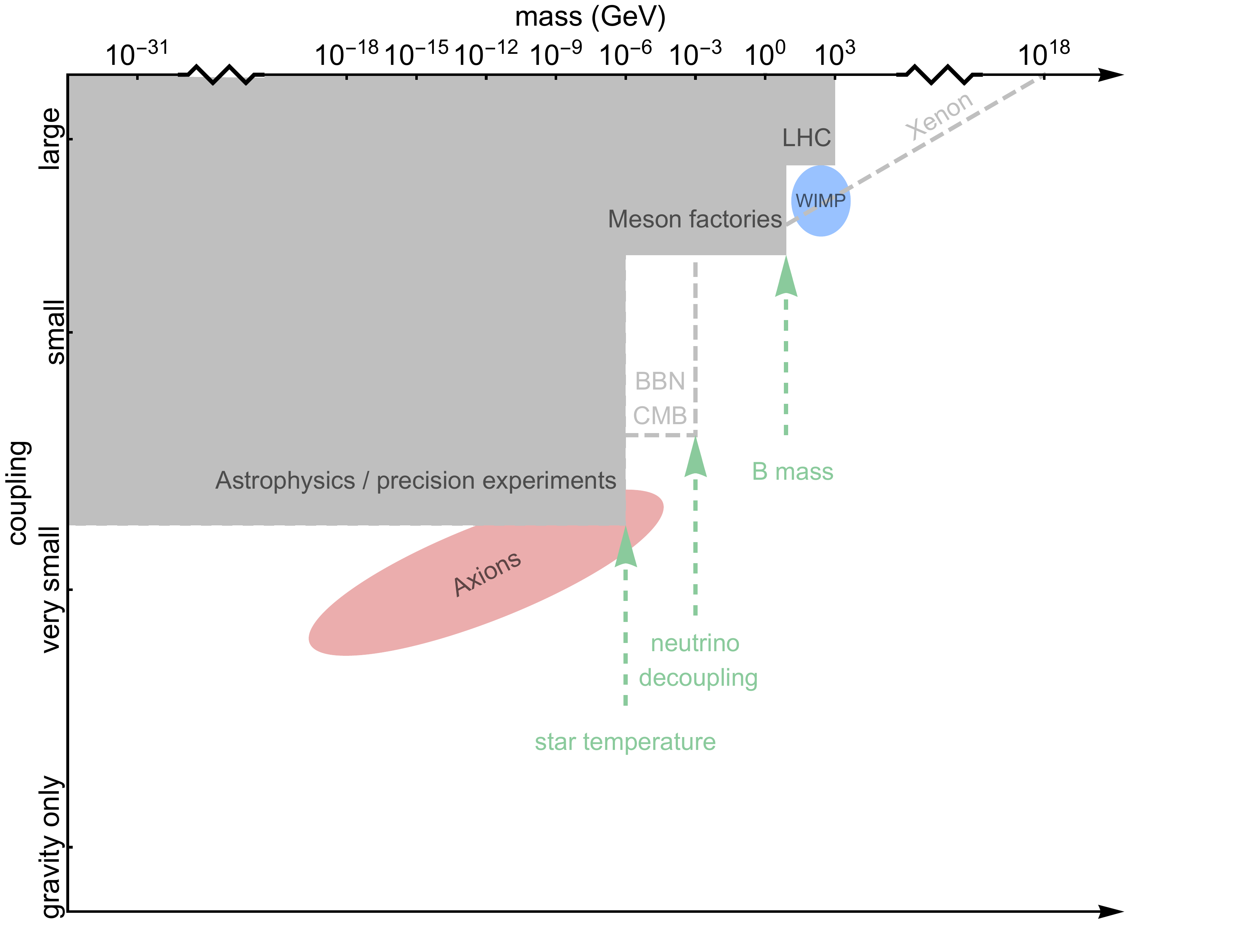}
\caption{Cartoon representation of the parameter space of FIPs, schematically indicating the existing constraints are the parameter space preferred by classic models such as WIMPs and the QCD axion. Dashed constraints depend to various degrees on the cosmological history.  \label{fig:cartoon}}
\end{figure*}

\vskip 2mm
Since FIPs are by definition feebly coupled to the SM, this begs the question of how we produce them in the first place. I know broadly speaking of 4 ways:
\begin{enumerate}
\item We do not produce them, as they are already around as a \textbf{cosmological relic}. This either be as dark radiation or (a component of) the dark matter. Critical energy scales for this class of models are the temperatures at which the neutrinos decouple ($T\approx 10$ MeV) and the temperature of the QCD phase transition ($T\approx 1$ GeV).
\item We can produce them in decay or oscillation of extremely \textbf{narrow SM resonances}, primarily in meson factories. This works well for FIPs with mass below the $b$ mass, though in some cases the SM Higgs can also be important.
\item We can detect or produce them through a \textbf{coherent enhancement}, such as the Primakoff effect or in $5^{\mathrm{th}}$ force experiments.
\item They can be produced in a \textbf{very large and hot oven}, such as various stars, or in some cases, nuclear reactors. Especially the stellar probes tend to be extremely powerful, as long as the FIP has a mass below the temperature of the star under consideration ($T\approx 100$ MeV for supernovas, $T\approx 1$ keV for most other stars).
\end{enumerate}
Most of the experimental probes of FIPs that I am aware of fall in one of these categories, though there are exceptions, such as invisibly decaying dark photons in experiments such as LDMX or NA64. I have attempted to qualitatively summarize the various scales and probes in the cartoon in Fig.~\ref{fig:cartoon}.

\vskip 2mm

At the present day, the vast majority of the FIPs-related activities are focused on \textit{millicharged particles, axion-like particles (ALP), heavy neutral leptons (HNL), dark photons and dark Higgs bosons}, to the extent that often this set is taken as the \emph{definition} of what constitutes a FIP. Perhaps driven by some degree of childhood nostalgia, I will refer to these 5 models as the \textbf{``the fabulous five''}, which is meant to reflect their enormous popularity in the FIPs community. They are popular by virtue of their theoretical simplicity, which in turn is dictated by the cardinal principle in effective field theory, stating that in perturbative models operators with higher operator dimensions should be parametrically suppressed. Indeed, 4 out of the fabulous five merely represent the leading interactions for the 4 lowest dimensional representations of the Lorentz group: the scalar (dark Higgs), pseudoscalar (ALP), vector (dark photon) and spinor (HNL). 


\begin{table}
\caption{The fabulous five. The ``milicharged particle'' can be charged under a nearly massless dark photon which mixes with the SM photon. Its phenomenology then resembles that of a particle carrying a small electric charge, as written in the table. }
\label{tab:portals}
\begin{center}
\begin{tabular}{rl}
Portal & Coupling \\
\hline
\smallskip
Dark Photon, $A'$ &  $-\frac{\varepsilon}{2 \cos\theta_W} F'_{\mu\nu} B^{\mu\nu}$ \\ \smallskip 
Axion-like particles, $a$ &  $\frac{a}{f_a} F_{\mu\nu}  \tilde{F}^{\mu\nu}$, $\frac{a}{f_a}  G_{i, \mu\nu}  \tilde{G}^{\mu\nu}_{i}$,
$\frac{\partial_{\mu} a}{f_a} \overline{\psi} \gamma^{\mu} \gamma^5 \psi$
\\
Dark Higgs, $S$  & $(\mu S + \lambda_{\rm HS} S^2) H^{\dagger} H$  \\ \smallskip Heavy Neutral Lepton, $N$ & $y_N L H N$ \\
milicharged particle, $\chi$ & $\epsilon A^\mu\bar \chi \gamma_\mu \chi$
\end{tabular}
\end{center}
\end{table}

\vskip 2mm
The simplicity of the fabulous five means that they can be fairly easily sliced into appealing 2D plots on which many experiments can be overlaid. This is both a blessing and a curse: While they are good tools to investigate the complementarity between various experimental approaches, there is a danger that this apparent simplicity can be interpreted as fabulous five, as defined in Table~\ref{tab:portals}, being very \emph{predictive} models. They are not. On their own, the fabulous five \emph{do not address any of the deficiencies of the SM}.


They therefore do not even predict their own existence, let alone a particular choice of parameters. On the other hand, over the years we have found that the fabulous five are very frequent \emph{guest characters in more complete models which do address important problems}, such as the hierarchy problem, the strong CP problem, the origin of dark matter, the neutrino masses and the baryon asymmetry of the Universe.  It are those connections which in my mind justify their status as ``fabulous'', and any of their predictions and limitations of applicability must thus be viewed in the context of this broader model space.  

\subsubsection{The fabulous five and where to find them}
Comprehensively reviewing how the fabulous five are featured in the modern model building literature is a gargantuan task, which at the present I feel unequipped to undertake. For these proceedings, I will therefore merely provide one example for each of the fabulous five. The examples are chosen among those models which I happen to be most familiar with, and such that the hierarchy, the strong CP, neutrino, dark matter and baryogenesis problems are all represented. Many different identifications are of course possible.

\textbf{The dark photon:}
The dark photon \cite{Holdom:1985ag} is perhaps the model builder's primary workhorse, and often the first ingredient one reaches for when constructing a dark sector with some coupling to the SM. I think its main appeal is perhaps due to the following salient features: \emph{(i)} all its parameters can be made technically natural, \emph{(ii)} one can easily couple it to the SM fermions without fearing gauge anomalies, anomalous flavor violation or Yukawa suppression factors for the lower generations and \emph{(iii)} its phenomenology is reassuringly familiar from our experience with the SM photon and $Z$ bosons. The dark photon is particularly useful in dark matter models, where it can act as the mediator in the freeze-out process. 
The connection of the dark photon with light DM models is presented in Section~\ref{ssec:berlin}.

\vskip 2mm
Here I chose to highlight another handy feature of the dark photon: It is very natural to have an $\mathcal{O}(1)$ coupling of the dark photon ($A'$) with dark sector matter while maintaining a very small coupling with the SM. This is achieved simply by setting its gauge coupling to be $\mathcal{O}(1)$ but its mixing with the photon to be tiny ($\epsilon \ll1$), which is technically natural. This is particularly convenient in models were one has a nasty, light relic floating around which is in danger of overclosing the universe. Let us call this relic ``$\pi$'' for now. By turning on the $\pi\pi\to A'A'$ annihilation or $\pi\to A'A'$ decay channels, one can easily deplete the relic abundance of $\pi$ by building up the abundance of $A'$. As long as $m_{A'}\gtrsim 10$ MeV, the $A'$ can then harmlessly decay back to the SM before to onset of BBN, without running afoul with other constraints on the dark photon itself. 

\vskip 2mm
This mechanism has been used in many occasions, and I must limit myself to a single recent example by Eleanor Hall~et.~al.~\cite{Hall:2019rld} which I happen to be familiar with: Hall.~et.~al.~constructed a dark sector similar in structure to the SM, but which harbors a first order phase transition. This phase transition can be used to generate the SM baryon asymmetry as well as an asymmetric dark matter candidate, which is the dark sector neutron in their case. Having a dark sector with baryons, their model naturally also contains dark pions which are lighter than the baryons, and whose relic density could create a phenomenological problem. This is resolved by adding the ever-so-convenient dark photon, as explained above. 

\textbf{Axion-like particles:}
%
In 1977 Roberto Peccei and Helen Quinn \cite{Peccei:1977hh,Peccei:1977ur} argued that the strong CP problem could be addressed by promoting the CP-violating $\theta$ angle in the QCD sector to a pseudoscalar field, the axion, whose potential sets its expectation value to zero. This clever trick however only works if all other contributions to the axion's potential are tiny compared to those from the SM QCD sector. In other words, if the potential gets tilted even slightly at the origin, the minimum will no longer occur at zero, which reintroduces the strong CP problem. Quantum gravity in particular is thought to induce such dangerous contributions, unless the axion's decay constant is chosen to be very low. In practice, this means that at face value, the Peccei-Quinn solution is already excluded by stellar cooling bounds and exotic Kaon transitions. This is the \emph{axion quality problem}. 

\vskip 2mm
The axion quality problem can be addressed by either suppressing the operators quantum gravity is expected to induce or by raising the axion mass rather than its decay constant. Either options requires additional model building in the ultraviolet, but deserves to be taken seriously. In the second case, we can effectively treat the axion mass and decay constants as independent parameters. We usually refer to such particles as ``axion-like particles'' (ALPs), to emphasize that the connection with the strong CP problem is more indirect. In other words, if an ALP with different couplings from those predicted by the standard QCD axion prediction is discovered, it may or may not play a role in solving the strong CP problem. This question can then only be answered in the context of a concrete UV completion, and measuring the various couplings of the ALP the best we can would therefore be essential to pin down its origin and role in our universe. 
For more details see Section~\ref{ssec:agrawal}.

\vskip 2mm
It is also important to bear in mind that even just as an effective theory, independent of their specific theoretical motivation, all ALP models still demand a UV completion. The nature of this UV completion predicts the relative strength of the couplings of the ALP to the SM fermions and gauge bosons, especially if the approximate unification of the SM gauge couplings is to be preserved. Generally UV completions fall in the class of models proposed by Dine, Fischler, Srednicki and Zhitnitsky (DFSZ) \cite{DINE1981199,Zhitnitsky:1980tq} or those following Kim-Shifman-Vainshtein-Zakharov (KSVZ) \cite{PhysRevLett.43.103,SHIFMAN1980493} scheme. In DFSZ models, the couplings to fermions tend to dominate, while in KSVZ models the ALP's couplings to the gauge bosons are typically larger.

\textbf{Heavy neutral leptons:}
The (type I) seesaw mechanism 
famously explains the smallness of neutrino masses in the SM in a very minimal and elegant manner. It relies simply on a set of neutral fermions $N_i$ which carry lepton number and a simple Yukawa interaction with the SM Higgs ($H$) and lepton doublets $(L)$
\begin{equation}\label{eq:HNLlag}
\mathcal{L}\supset y_{ij} H L_i N_j + \frac{1}{2} m_{ij} N_iN_j. 
\end{equation}
where the $m_{ij}$ are Majorana mass terms for the $N_i$. In the limit $m_{ij}\gg m_\nu$, \eqref{eq:HNLlag} reduces to the famous Weinberg's effective interaction $y_{ik}m^{-1}_{k\ell}y_{\ell j} HL_i HL_j$, which in turn reproduces the neutrino mass term upon substituting $H$ with its vacuum expectation value. Motivated by grand unification, one may obtain the correct range for the neutrino masses by setting $y\sim 1$ and $m\sim10^{16}\,$GeV. 

\vskip 2mm
On the other hand, the Majorana masses $m$ may also be within the kinematic range of our accelerators, as long as $y\ll1$. This scenario is typically parametrized in terms of a numerically small mixing matrix $U_{ij}$ between the active and the sterile heavy neutrinos.
In the minimal setup in \eqref{eq:HNLlag}, one finds parametrically $U \sim \sqrt{m_\nu/m_N}$, with $m_N\sim m$ the mass scale associated with the heavy neutral lepton's mass eigenstates. 
In practice, values of $U_{ij}$ which are experimentally accessible tend to over-predict the $m_\nu$.


\vskip 2mm
Of course there is also the possibility that HNLs are more indirectly related to the generation of the neutrino masses. This can be obtained by adding
a Dirac mass term in \eqref{eq:HNLlag} for the $N$
\begin{equation}\label{eq:HNLlagdirac}
\mathcal{L}\supset y_{ij} H L_i N_j + \frac{1}{2} m_{ij} N_iN_j+M_{ij}N_i\overline N_j.
\end{equation}
The presence of the Dirac mass term effectively decouples $U$ and $m_N$ from the masses of the active neutrinos. 
In this more general scenario, observable HNLs are possible.  
This is because $m_N$ and $U$ now depend on the lepton number preserving parameters $y$ and $M$, while the active neutrino masses are kept small by the small, lepton number violating Majorana masses in $m_{ij}$ \cite{Kersten:2007vk}. The price we pay for this slight generalization is that the direct connection with the SM neutrino masses is lost by construction, and model is less predictive than the more minimal setup in \eqref{eq:HNLlag}. 

In this very brief description, I have of course not done justice to the rich phenomenology of this subject. For instance, in certain parts of the parameter space, the model may generate the baryon and lepton asymmetry in the Universe \cite{Asaka:2005pn}. 
A rather complete overview of the HNLs phenomenology and model building activity is provided in Section~\ref{ssec:drewes}.

\textbf{The dark Higgs:}
\label{sec:intro_DarkHiggs} 
The dark Higgs model is a simple extension of the SM Higgs sector with a single, real scalar $S$
\begin{equation}
\mathcal{L}\supset-\frac{1}{2}{m}_S^2 S^2-\mu S H^\dagger H -\frac{1}{2}\lambda_{SH} S^2 H^\dagger H\label{eq:benchmark}
\end{equation}
which develops a small mixing with the SM Higgs. $S$ thus inherits its couplings to the SM fermions proportional to their mass. The dark Higgs is therefore often the most convenient way to couple a hidden sector to the SM if suppressed couplings to the lower generations are desired. Depending on the size of $\lambda_{SH}$, the exotic Higgs decay $h\to SS$ can also be relevant. This model often features in dark matter models or models with extended Higgs sectors, such as the next-to-minimal supersymmetric SM.

\vskip 2mm
There I touch on the example of the relaxion mechanism~\cite{Graham:2015cka}. In this setup, the cosmological evolution of the vev of $S$ is used to dynamically drive the vacuum expectation of the SM Higgs to a relatively small value, which provides a solution to the hierarchy problem. The available parameter space for the relaxion extends down to masses as low as $m_S\sim 10^{-20}$~eV and can be partially accessed through $5^{\mathrm{th}}$ force experiments, stellar cooling bounds and exotic meson decays. (See \cite{Banerjee:2020kww} for a recent overview.) 
 I refer the reader to Section~\ref{ssec:gori} for a discussion about other models.

\textbf{Milli-charged particles:}
The name ``milli-charged'' particles is a bit of a misnomer for two reasons: We use this name to refer to all particles with charge $\ll 1$, regardless on whether the charge is $\sim 10^{-3}$. Moreover, in most models one considers a particle coupled through a very light kinetically mixed dark photon. Such a particle behaves for all intends and purposes as a particle with a tiny electric charge, though it does not carry SM electroweak quantum numbers.\footnote{ This is particularly good news for those readers who are anxious about grand unification.} Among the fabulous five, it is probably the strangest character, because it is the only one which comes with a build-in, screened, long-range interaction. This makes it particularly fun to think about their cosmological imprints on e.g.~the CMB and about their non-trivial interactions with our particle detectors. 
 For more on this, I refer the reader to Section~\ref{ssec:pospelov} and Section~\ref{ssec:tsai}.

Milli-charged particles are strongly constrained by stellar cooling considerations, but can be a good dark matter candidate for masses $\gtrsim 10$ keV, through the freeze-in mechanism \cite{Hall:2009bx}. In this scenario the Universe starts off with a negligible dark matter abundance during radiation domination, which is then slowly populated by out-of-equilibrium interactions such $e^+e^{-}\to \chi \chi$ etc.

\subsubsection{Concluding thoughts}
At the start of this talk, we posed the question whether the all of the fabulous five are featured frequently in UV complete models which address shortcomings of the SM. This was important, because they derive \emph{all} their theory motivation from being embedded in such more complete models. By supplying an example for each of the fabulous five, I hope to have showed that answer is clearly ``yes''. A more subtle question is whether the reverse is also true: By studying the fabulous five in great detail, are we covering all possible FIPs? This is a much trickier question to answer, and I am inclined to think the answer is ``no''. A compelling counter example is that of the Twin Higgs \cite{Chacko:2005pe} and in particular its generalizations in the context of Neutral Naturalness \cite{Craig:2014aea,Craig:2015pha}. These models are themselves examples of Hidden Valleys \cite{Strassler:2006im}, which can easily exhibit a phenomenology that is much more complicated and subtle than that of the fabulous five. How to systematically probe these and other possibilities is still an open question.

\clearpage
\subsection{Feebly-interacting particles and cosmological constraints}
\label{ssec:pospelov}
{\it Author: Maxim Pospelov, <pospelov@umn.edu>} 
\subsubsection{Introduction}
\label{sssec:pospelov-intro}

Experimental developments of the last 60 years have transformed Big Bang cosmology from a hypothesis to a firmly established branch of science. Today the knowledge of the cosmological expansion history is in some instances quite detailed (Big Bang nucleosynthesis (BBN), cosmic microwave background (CMB), structure formation), allowing to measure the composition of the Universe with some accuracy. This knowledge brings about clear understanding of what is possible within the prevailing cosmological paradigm, and what is constrained, as far as the field content of the Standard Model (SM) and its extensions are concerned. The sensitivity of cosmological scenarios to properties of particles was realized long ago, perhaps starting from one of the earliest papers on the topic, by Alpher, Hermann and Follin \cite{Alpher:1953zz} who noted the sensitivity of the elemental yield of the BBN to the properties of neutrinos, and specifically their Dirac vs Majorana nature. 

\vskip 2mm
The discovery of dark matter and dark energy as the most important, yet not fully identifiable components in terms of their connection with the rest of the SM particles, remains to be one of the most intriguing drivers in investigating the connection between particle physics and cosmology. 
Today there are many efficient tools that one can use to constrain the properties of SM extensions using cosmology. 
It is quite clear that only certain types of new physics models can be probed. For example, new TeV-scale resonances that conserve all quantum numbers of the SM, and decay quickly to lighter SM states cannot be constrained using cosmological considerations. On the other hand, extensions of the SM that contain feebly interacting particles (FIPs) can indeed be constrained if these particles are sufficiently long-lived. It is appropriate to remark at this point that longevity often comes hand-in-hand with the smallness of couplings. 
In this talk, I will cover some aspects of the cosmological sensitivity to FIPs. 

\vskip 2mm
The most straightforward sensitivity to light new states occur in models where such states can be thermally populated and therefore contribute, at a considerable level, to Universe's energy balance, affecting the expansion rate of the Universe. As a consequence, the timing of nuclear reactions, and elemental yield of helium and deuterium can be altered. In addition, the presence of an extra component to radiation-like energy density can be severely constrained by the CMB observable $N_{eff}$ that parametrizes the neutrino contribution to the energy density of the Universe. 

\vskip 2mm
I will discuss several examples of cosmological constraints imposed on FIPs, noting that the topic is much broader than the scope of the current talk. In particular, the interesting example I will concentrate on are: 
{\em i.} Single production of extremely weakly coupled FIPs with their subsequent decay during the BBN and CMB time, {\em ii.} Pair-annihilation of fully thermalized FIPs, with the subsequent decay of left-over during the CMB, {\em iii}. Dark radiation of FIPs that never is fully thermalized with the SM, yet being able to drive photon-dark photon oscillations after the CMB epoch. 

\begin{figure*}
\centering
  \includegraphics[height=.3\textheight]{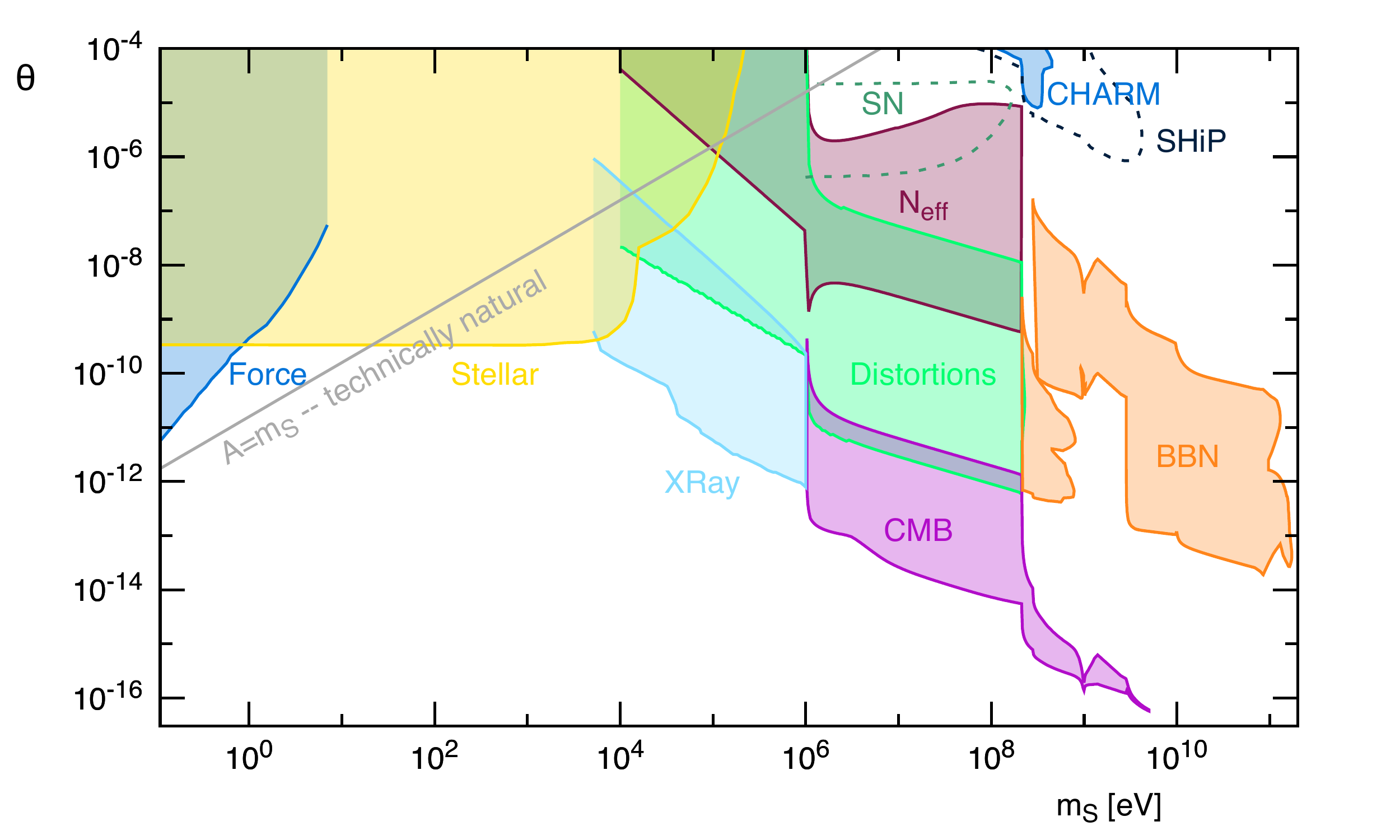}
  \caption{
  An overview of the excluded parameter space for the super-renormalizable Higgs portal scalar, including the updated constraints to the diffuse X-ray background (XRay), CMB anisotropies, spectral distortions, $N_{\rm eff}$ and BBN from Ref.~\cite{Fradette:2018hhl}.
  Constraints from new short-range forces (Force)~\cite{Kapner:2006si,Decca:2007jq,Geraci:2008hb,Sushkov:2011md} and stellar cooling (Stellar)~\cite{Hardy:2016kme} from other authors are also shown. We also display the projected SHiP sensitivity and an estimate of supernova (SN) constraints~\cite{Krnjaic:2015mbs}.
  It is easy to see that the constraints goes to very small values of $\vartheta$, and deep inside the naturalness territory.  }
\label{Fig1}
\end{figure*}

\subsubsection{Single production of FIPs with subsequent decay}

The portals to light neutral hidden sectors are an effective field theory classification of all lowest dimension operators that mediate interaction between SM and the hidden sector. Symbolically, we can write this type of contributions to the effective Lagrangian as \cite{Lanfranchi:2020crw}
\begin{equation}
\Delta {\cal L} = \sum c_O O^{\rm SM} O^{\rm hidden}.
\end{equation}
We assume that the hidden sector is not charged under the SM fields, and therefore both $O^{\rm SM} $ and $O^{\rm hidden}$ are SM singlet operators separately. Note, however, that they have to form a Lorentz singlet only in the product directly entering into $\Delta {\cal L}$, 
but separately they can transform under the Lorentz group. Given that, one can write down a set of terms of lowest dimension operators
in $\Delta {\cal L}$:
 \begin{eqnarray}
B_{\mu\nu}F'_{\mu\nu}&&~~{\rm kinetic~mixing~portal} \\  
(H^\dagger H) (A S + \lambda S^2)&&~~{\rm Higgs~portal} \\
(\bar L H)N& &~~{\rm neutrino~portal} \\
(\bar \psi_{\rm SM} \gamma_\mu \psi_{\rm SM}) A'_\mu &&~~{\rm generic~gauge~portal} \\
(\bar \psi_{\rm SM} \gamma_\mu\gamma_5 \psi_{\rm SM}) \partial_\mu a / f_a&&~~{\rm axion-like~portal}
\end{eqnarray}
In the expressions above, $B_{\mu\nu}$ is the SM hypercharge field strength, $H$ is the SM Higgs doublet, $L$ is a SM lepton doublet, and 
$\psi_{\rm SM}$ stands for a generic SM fermion. Particles from  dark sector are denoted as $A'_\mu$, $N$ and $a$. 
$F_{\mu\nu}$ stands for a usual field strength of $A_\mu$. 

\vskip 2mm
In this framework, it is easy to see that the dark sector states, $S$, $A'$, $N$, $a$ etc are not stable, on account of the existence of linear coupling to the SM. It is then clear that the small enough coupling will result in the extremely long lifetimes of these new particles, {\em and} the same coupling can be responsible for the early Universe production of these states. Thus we have, symbolically, the following situation constrained by cosmology: 
\begin{eqnarray}
{\rm Production}: ~ {\rm SM \to FIP}~{\rm at}~ T\sim T_{prod}\\
{\rm Decay}: ~ {\rm FIP \to SM}~{\rm at} ~T\sim T_{BBN}, T_{CMB}
\end{eqnarray}
The timing of the production varies depending on the type of FIPs. The production of HNL is typically peaked at $T\sim \Lambda_{QCD}$, if the mass of HNLs is in the keV range. The production of dark vectors is maximized at 
$T\sim m_{A'}$, while the production of scalar is mostly occurring during the electroweak epoch. Details of the ensuing limits can be found {\em e.g.} in these papers: \cite{Adams:1998nr,Fradette:2014sza,Vincent:2014rja,Fradette:2018hhl}. 

\vskip 2mm
The example of constraints on the parameter space of the model of a scalar particle mixed with the Higgs 
is shown in Fig. \ref{Fig1}. (Note that the mixing angle is a simple consequence of the trilinear $A H^\dagger H$ portal.) One can see extreme sensitivity to mixing angles down to values comparable to the {\em gravitational}  coupling of the electrons. (For example, a mixing angle of $10^{-17}$ times the Yukawa coupling $y_e \sim 10^{-6}$ is comparable to $m_e/M_{pl}$). It is clear that no terrestrial experiments would ever be able to probe such small couplings, underlining the importance of cosmological constraints. 

\vskip 2mm
Where does the strength of the constraints is coming from? It is because in the early Universe, despite the smallness of coupling, every particle has enough energy to produce FIPs, and the SM bath is dominated by radiation-like species. During the CMB decays, however, the expansion has diluted the radiation, making baryons and dark matter to be the dominant - but still not very numerous, $\eta_b = n_b/n_\gamma = 6\times 10^{-10}$ - species. The decay of FIPs then releases $m_{FIP}$ amount of energy, a significant fraction of which can go to ionization. Current ionization sensitivity of CMB observables allows at most $\sim 10^{-1}$ eV energy injection per baryon, while putting together the production and decay estimates 
one arrives at 
\begin{equation}
E_{per~baryon} \propto \frac{\alpha \epsilon^2 M_{pl}}{10\eta_b} \propto (\epsilon/10^{-17})^2 \times {\rm eV}
\end{equation}
for the case of vector portal FIPs. 
Much more details on this topic can be found in Ref. \cite{Fradette:2014sza}. 

\subsubsection{Higgs-mediated depletion of FIPs}
One of the more interesting in terms of the LHC application is the model of scalar FIPs with large 
quartic and small trilinear coupling:
\begin{equation}
\label{modelS}
{\cal L}_{portal} = (H^\dagger H) (A S + \lambda S^2)~{\rm at}~ (A/v) \ll \lambda.
\end{equation}
Currently, the constraints on $\lambda$ emerge due to its contribution to the Higgs boson decay via $h\to SS$. 
Today's data, however, do not allow limiting the Higgs width better than 10\% accuracy (in the assumption of the SM + portal couplings). A more hopeful approach is based on production of $S$ in the Higgs decay with subsequent sensitive detection of $S$ with a purposely-built large volume detector \cite{Chou:2016lxi}. To estimate the strength of a signal at given $\lambda$, one needs to know how large/small $A$ is allowed to be. $A$ determines the mixing angle between the Higgs and $S$, and the overall lifetime of $S$ particles. The longer $S$ lives, the weaker the signal is from its decay. 

\vskip 2mm
Cosmology comes to help, as detailed in \cite{Fradette:2017sdd}, via the {\em upper} limit on $S$ lifetime. At sizeable $\lambda$ 
there is a complete thermalization of $S$ particles in the early Universe, followed by the Higgs-mediated
$SS\to h^* \to {\rm SM}$ depletion at $T < m_S$. This depletion happens in exact analogy with WIMP annihilation. The cosmological sensitivity to the model comes from the decays of residual population of $S$ during the initial stages of BBN. These decays add charged pions (also kaons, neutrinos etc) into the primordial  fluid that affects 
neutron-proton interconversion at the critical time of $n/p$ freeze-out that determines both $^4$He and D abundances. 

\begin{figure}
\resizebox{0.48\textwidth}{!}{%
  \includegraphics{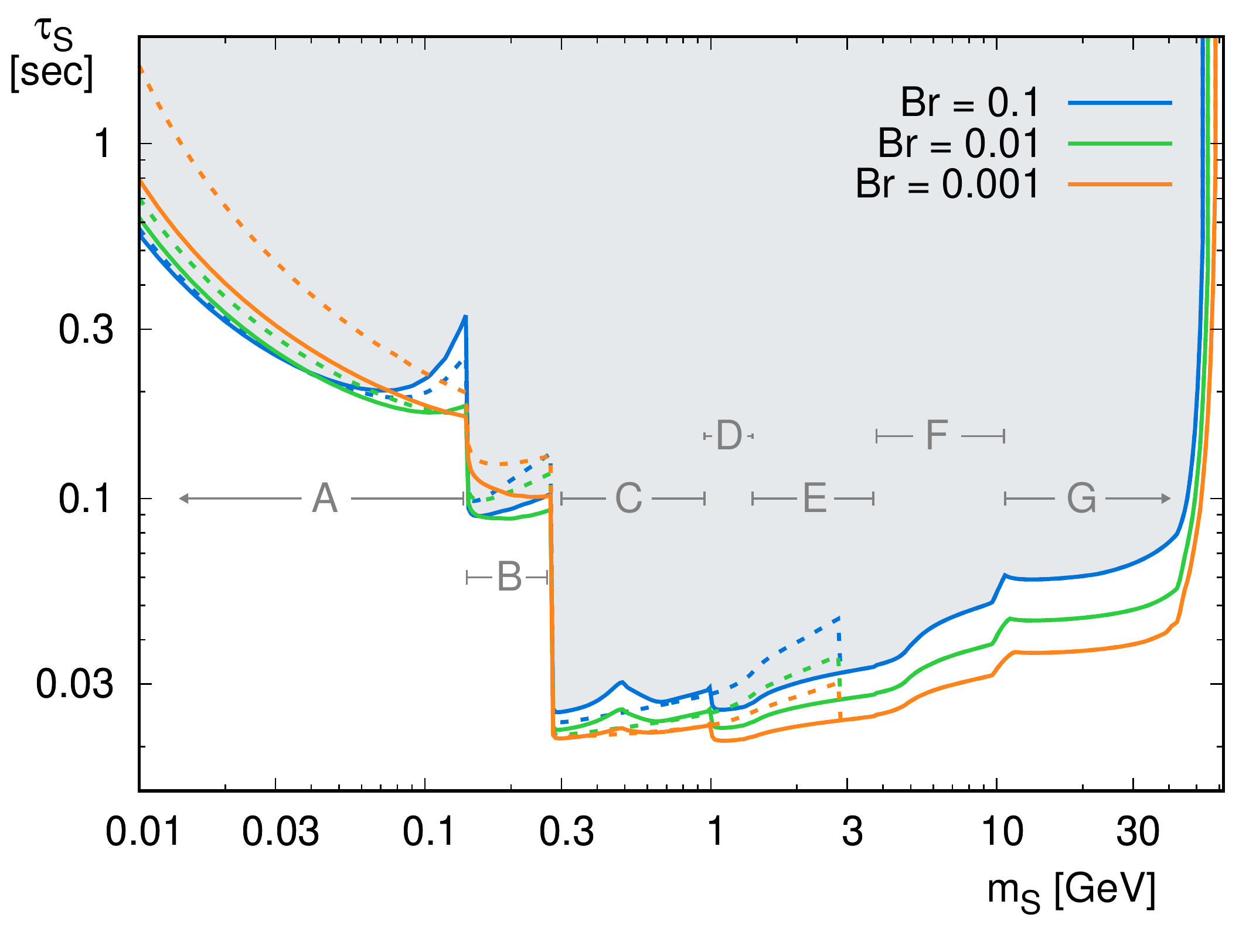}}
  \caption{Constraints on the lifetimes of the scalar $S$ with three different inputs of the $h\to 2S$ decay branching ratio.  Constraints are typically stronger than 0.1 seconds, except for lightest $m_S$.}
\label{Fig2}
\end{figure}

\vskip 2mm
The resulting constraints are summarized in Fig. \ref{Fig2}. They are expressed in terms of the physical observables, such as Higgs boson branching ratios, and the lifetime of $S$. The good news, from the point of view of building an adjacent to LHC large volume detector, is that the lifetime of the $S$ particles cannot be taken arbitrarily large, and at all points in the parameter space the lifetime has to be below 0.1 seconds. This sets an upper bound on the decay rate inside the detector with linear dimensions of $\sim 100$m, which is estimated to be $P> 10^{-6}$. 
Given copious production of the Higgs bosons at the LHC collision points, the proposal of building a large volume detector next to it would capture most of the cosmology-allowed parameter space of model (\ref{modelS}).

\subsubsection{21 cm physics and dark radiation} 
\label{sssec:pospelov-21cm}

Perhaps one of the most anticipated yet unexpected results of the last few years is the claimed detection of 
the global 21~cm absorption signal \cite{Bowman:2018yin}. While the start of the 21~cm cosmological physics has 
been widely anticipated, the results of \cite{Bowman:2018yin} taken at face value have been somewhat controversial. Indeed, the stronger than standard cosmology allows, absorption line at redshifts $O(15)$ implies serious breakdown of 
standard physics. The stronger absorption feature may imply two things in the beyond SM FIP physics: either there is a presence of some sort of non-trivial agent that cools off baryons (and consequently spin temperatures to level much below SM cosmology), or somehow the infrared content of the CMB (around 21 cm at 
corresponding redshifts) is much stronger than theory predicts. 

\vskip 2mm
The cooling off of baryons is not possible without employing an agent that is numerous, cold, and stronger coupled to baryons at later times. The only candidate that comes to mind in that respect is a millicharged FIP, that has $1/v^4$ dependence of its cross section of scattering on free charged particles \cite{Barkana:2018qrx}. Even then, the model has to be constructed 
rather carefully, as the CMB constraints limit the amount of such millicharged dark matter to be below $10^{-2}$
\cite{Kovetz:2018zan}, while its mass cannot exceed tens of MeV. Then the 21~cm preference for the coupling is at the level of $\sim 10^{-4}$. Such coupling constants can be probed indeed with the next generation of dedicated searches of millicharged FIPs (see e.g. \cite{Pospelov:2020ktu}). 

\vskip 2mm
Increasing the photon number densities at the IR tail of the CMB is another distinct possibility. It is clear that in order to achieve that the photons have to be "added" via some sort of a non-thermal process. In Ref. \cite{Pospelov:2018kdh}, the new mechanism of transferring dark radiation into soft photons was pointed out that conforms all other cosmological constraints. Imagine dark radiation in terms of the dark photons $A'$. If masses of dark photons are very small, they are not going to get thermalized by re-scattering on SM particles. Yet, the transfer of $A'$ to $A$ is possible due to a resonance, when the plasma frequency becomes equal to the dark photon mass. At that point, a very IR-enhanced population
of $A$ will develop, superimposing on the regular Planck distribution. The concrete scenario considered in \cite{Pospelov:2018kdh} was realizing the dark radiation as the product of decay of light dark matter $a$, in the following sequence:
\begin{equation}
\left. a \to A'A' \right|_{\tau_a > 10 \tau_U} ; ~ \left.A' \to A\right|_{m_{A'}=\omega_{pl}}.
\end{equation}
This process becomes of great interest for the 21 cm physics if the masses of decaying DM particles are in the range of $m_{a} \propto 10^{-4}-10^{-3}$\,eV, while the masses of dark photons are in the $m_{a} \propto 10^{-12}-10^{-8}$\,eV range. The latter insures that the resonance occur between recombination and reionization.  

\vskip 2mm
The analysis of Ref.~\cite{Pospelov:2018kdh} showed that the 21~cm physics is indeed capable of probing very soft dark radiation. (To be perfectly clear, this radiation is not detectable via measurements of $N_{eff}$). The resulting depth of the absorption can match observations (which remain highly tenuous at this point, as not confirmed by other groups yet)
if the size of the kinetic mixing is in the $\epsilon \sim 10^{-7}$ range. The particular mechanism of $A\to A'$ oscillations is difficult to probe in the lab, but perhaps not impossible in the future. 

\vskip 2mm
In conclusion, cosmology provides a unique window of FIPs. Aside of the identity of dark matter - which very well may be one of the FIP particles - it offers unique probes of unstable FIPs via modifications to the BBN and the CMB.
In this talk we have shown some application of cosmological constraints to very weakly coupled FIP states. In addition to the well understood phenomena of energy injection via decays, one could also entertain the modifications of the CMB spectrum provided by the photon-dark photon oscillations in the early Universe. With the right range of parameters such oscillations may affect late time cosmology, and even 21~cm physics - after the CMB itself the next widely anticipated chapter in the history of the Universe. 


\clearpage
\subsection{Stellar constraints on feebly-interacting particles}
\label{ssec:giannotti}
{\it Author: Maurizio Giannotti, <MGiannotti@barry.edu>} \\

Astrophysical considerations have played a quite significant role in the investigation of the physics of feebly interacting particles (FIPs)~\cite{Raffelt:1996wa},
allowing to set bounds that often exceeded the current experimental searches. 
Indeed, stars are sensitive to extremely rare processes, which are very difficult to probe in a collider. 
Very often the bounds on light, feebly interacting particles from astrophysical considerations have exceeded the bounds from direct experiments. 
Axions are a notorious example. 
Only very recently were terrestrial experiments able to test regions of the axion parameter space compatible with stellar evolution. 
The same is true for other FIPs, such as dark photons, and for non-standard neutrino properties. 

\vskip 2mm
Interestingly, having a feeble interaction may actually \textit{increase} the impact a particle has on the evolution of stars. 
Once produced, FIPs can easily escape the stellar plasma thus contributing very efficiently to the 
energy transport. 
The additional energy loss, in turn, contributes to the stellar evolution and so impacts the observed stellar populations.

\vskip 2mm
The advances in experimental astrophysics promise a significant advancement in our understanding of FIPs and 
it is likely that, if a FIP does exist and has a mass below a few keV, its impacts on stars would
be revealed before its detection in a dedicated laboratory experiment. 

\subsubsection{Bounds and hints from stellar evolution}

The idea of probing FIPs from stellar evolution is over half a century old. 
In 1963, Bernstein, Ruderman and Feinberg derived bounds on the electromagnetic properties of neutrinos from observations of the sun~\cite{PhysRev.132.1227}.
The developments of the following years were summarized in well known monographs~\cite{Raffelt:1990yz,Raffelt:1996wa} and recent years have witnessed significant advancements, as we shall discuss below.  
In Table~\ref{tab:Astro_bounds}, we present a summary of the updated bounds from stellar evolution on FIPs. 
\begin{table*}[t]
		\caption{
			Updated bounds on FIPs from stellar evolution. 
			Legend: 
			WDLF: White Dwarf Luminosity Function; 
			WDV: White Dwarf Variables; 
			RGBT: Red Giant Branch Tip;
			HB: Horizontal Branch;
			SN: Supernova; NS: Neutron Star.
			For the couplings, $ g_{ai} $ refers to the axion, $ \mu_\nu $ to the neutrino magnetic moment, and $ \varepsilon $ to the dark photon with mass $ m_V $.
			The axion bound from SN 1987A is on the effective coupling $g_{aN}=(g_{an}^2+ 0.6\, g_{ap}^2 + 0.5\, g_{an}\,g_{ap})^{1/2}$.
			In all cases, we are showing the results of the most recent analysis and, when a statistical analysis is available, we provide the $ 2\, \sigma $ bound. 
			Whenever independent studies presented different results we have shown all of them.
			This is the case of the RGBT, where there exist two independent analyses, 
			one based on 22 galactic globular clusters (GGCs) 
			and one of the NGC 4258 galaxy, 
			and of the NS, where an analysis is based on CAS A 
			and one on  HESS J1731-347. 
			We have not reported the hint from the NS in CAS A~\cite{Leinson:2014ioa} since it is in tension with the more recent bound in~\cite{Hamaguchi:2018oqw}. 
			See text for more details.}
	\begin{center}
		\begin{tabular}{ l c c}
			Star  						&  Bound  														& Reference	 		\\ \hline
			Sun \phantom{$\Big|$} 		&  $g_{a\gamma}\leq 2.7\!\times\! 10^{-10}\,{\rm GeV^{-1}}$ 	&	\cite{Vinyoles:2015aba}	\\
			\vspace{0.2cm}
			 							&  $\mu_{\nu}\leq 4\!\times\! 10^{-10}\,\mu_B$ 					&	\cite{Raffelt:1999gv}	\\
			\vspace{0.2cm}
			 							&  $ \varepsilon \cdot m_V \leq 1.8\times 10^{-12}$ eV	for $ m_V\leq 1 $ eV				&	\cite{Vinyoles:2015aba}	\\
			\vspace{0.2cm}
			 							&  $ \varepsilon \lesssim 2\times 10^{-15} $ for $ m_V\sim 300 $ eV					&	\cite{An:2014twa}	\\
			\vspace{0.2cm}
			WDLF 						&  $g_{ae}\leq 2.1 \!\times\! 10^{-13}$						 	&	\cite{Bertolami:2014wua}  	\\
			\vspace{0.2cm}
			 							&  $\mu_{\nu}\lesssim 5\!\times\! 10^{-12}\,\mu_B$				&	\cite{Bertolami:2014noa}  	\\
			\vspace{0.2cm}
			WDV (combined) 	  		 	&  $g_{ae}\leq 4.1 \!\times\! 10^{-13}$							&	 \cite{Corsico:2019nmr,Giannotti:2017hny,DiVecchia:2019ejf}	 \\
			\vspace{0.2cm}
			WDV (1351+489) 	  		 	&   $\mu_{\nu}\lesssim 7\!\times\! 10^{-12}\,\mu_B$				&	 \cite{Corsico:2014mpa}	 \\
			\vspace{0.2cm}
			RGBT	(22 GGCs)		    &  $g_{ae}\leq 1.5 \!\times\! 10^{-13}$ 				        &     \cite{Straniero:2019dtm}    \\	 
			\vspace{0.2cm}
			RGBT	(NGC 4258)          &  $g_{ae}\leq 1.6 \!\times\! 10^{-13}$ 						&	\cite{Capozzi:2020cbu}  \\
			\vspace{0.2cm}
								        &  $\mu_{\nu}\leq 1.5\!\times\! 10^{-12}\,\mu_B$ 				&	\cite{Capozzi:2020cbu}  \\
			\vspace{0.2cm}
			RGBT	(generic)           &  $\varepsilon \lesssim 10^{-15}$ for $ m_V\sim 25\, $keV		&		\cite{An:2014twa} \\
			\vspace{0.2cm}
			HB 							&  $g_{a\gamma}\leq 0.65\! \times\!10^{-10}\,{\rm GeV}^{-1}$ 	&	\cite{Ayala:2014pea,Straniero:2015nvc}\\
			\vspace{0.2cm}
										&  $\varepsilon \lesssim 10^{-15}$ for $ m_V\sim 2.6\, $keV		&		\cite{An:2014twa} \\
			\vspace{0.2cm}
			Massive Stars 				&  $g_{a\gamma}\leq 0.8\! \times\!10^{-10}\,{\rm GeV}^{-1}$ 	&	\cite{Friedland:2012hj}\\
			\vspace{0.2cm}
			 							&  $\mu_{\nu}\lesssim 5\!\times\! 10^{-11}\,\mu_B$ 	&	\cite{Heger:2008er}\\
			\vspace{0.2cm}
			SN 1987A		 		 	&  $ g_{aN}^2\lesssim 8.3 \!\times\! 10^{-19}  $ 				&	 \cite{Carenza:2019pxu}\\ 
			\vspace{0.2cm}
			NS (CAS A) 					&  $ g_{ap}^2+1.6\, g_{an}^2\lesssim 1.1\!\times\! 10^{-18} $ 	&	\cite{Hamaguchi:2018oqw} \\ 
			\vspace{0.2cm}
			NS (HESS)		 			&  $g_{an}\leq 2.8\!\times\! 10^{-10}$							&	\cite{Beznogov:2018fda} \\ 
			\hline
		\end{tabular} 

		\label{tab:Astro_bounds}
	\end{center}
\end{table*}

Furthermore, perhaps more intriguingly, a series of anomalous astrophysical observations in the last three decades have led
to speculations that new physics is at play (see,  Ref.~\cite{Giannotti:2015kwo,Giannotti:2017hny,Hoof:2018ieb,DiVecchia:2019ejf,DiLuzio:2020wdo,DiLuzio:2020jjp} for recent reviews). 
The possibility that stars were hinting at new physics was put forward in 1992, 
with the interpretation in terms of axions~\cite{Isern:1992gia} of the anomalous period change of the white dwarf variable (WDV) G117-B15A. 
Since then, more observations have been interpreted as a hint to the existence of FIPs.
These include:
a few WDV stars~\cite{Corsico:2019nmr};
red giant branch (RGB) stars~\cite{Straniero:2020iyi};\footnote{Notice, however, that Ref.~\cite{Capozzi:2020cbu} did not confirm this hint.} 
horizontal branch (HB) stars~\cite{Ayala:2014pea,Straniero:2015nvc};
the WD luminosity function (WDLF)~\cite{Bertolami:2014wua,Bertolami:2014noa,Isern:2018uce,Isern:2020non};
red clump stars~\cite{Mori:2020qqd};
helium burning intermediate mass stars~\cite{Friedland:2012hj,Carosi:2013rla}; 
and supernovae (SN) progenitors~\cite{Straniero:2019dtm} (see also discussion in Sec. 3.4.3 of Ref.~\cite{Drlica-Wagner:2019xan}).
All these hints should be taken with cautions since their significance is not very large (see Ref.~\cite{Giannotti:2015kwo}).
However, they do show a systematic problem with the present understanding of stellar cooling and 
require some further investigations. 
The axion case is especially compelling~\cite{Giannotti:2015kwo,Giannotti:2017hny} since,
contrarily to other new physics candidates, it was shown to fit particularly well all the astrophysical observations~\cite{Giannotti:2015kwo}. 

%

\vskip 2mm
Below, we review some of the most significant stellar bounds and hints on FIPs couplings to standard model fields. 

\subsubsection{Axion-photon coupling}
\label{sssec:giannotti-axion-photon}

Axions are expected to interact with photons through the Lagrangian term 
\begin{align}
	 \mathcal{L}_{a\gamma}=-\frac{1}{4}g_{a\gamma} a F\tilde{F} \,.
\end{align}
The most relevant axion production mechanism induced by this coupling is the Primakoff process, 
$ \gamma +Ze \to Ze +a$,
which consists in the conversion of an axion in the electric field of the nuclei and electrons in the stellar plasma.\footnote{It is also possible to produce 
axions from the conversion of thermal photons in the
solar magnetic field. 
In some cases, this process may even dominate over the Primakoff rate, although there are some uncertainties on the exact structure of the magnetic field 
in the solar interior. 
See Refs.~\cite{Caputo:2020quz,OHare:2020wum,Guarini:2020hps,Hoof:2021mld} for recent analyses.}
This process is efficient in the sun and is in fact expected to be one of the principal solar axion production mechanisms.
Helioseismological considerations allow to set strong bounds on the the axion-photon coupling, as originally discussed in Ref.~\cite{Schlattl:1998fz}. 
Currently, the strongest solar constraint is 
$g_{a\gamma}\leq 4.1\times 10^{-10}\,{\rm GeV^{-1}}$ at $3\,\sigma$~\cite{Vinyoles:2015aba}, 
derived from a global analysis of several solar properties. 
A somewhat weaker bound, $g_{a\gamma}\leq 7\times 10^{-10}\,{\rm GeV^{-1}}$, was inferred in Ref.~\cite{Gondolo:2008dd} from observations of neutrinos from $ ^8 $Be. 

\vskip 2mm
Hotter stars, for example helium burning low mass stars in the HB evolutionary stage, 
are significantly more efficient at producing axions through the Primakoff process, which is very sensitive to the plasma temperature. 
Effects of exotic cooling on HB stars can be properly addressed from measurements of the R-parameter,  
defined as the ratio of the number of stars in the HB and in the upper portion of the RGB: $ R=N_{\rm HB}/N_{\rm RGB} $.
From a recent analysis of 39 globular clusters in Ref.~\cite{Ayala:2014pea}, based on the sample reported in Ref.~\cite{Salaris:2004xd}, it was derived $ R=1.39 \pm 0.03 $.
In the presence of axions, this parameter is expected to be~\cite{DiLuzio:2020wdo} 
\begin{align}
	\label{eq:R-parameter}
	R=R_0(Y)-
	F_{a\gamma} \left( \frac{g_{a\gamma}}{10^{-10}{\rm GeV}^{-1}} \right) -
	F_{ae} \left( \frac{g_{ae}}{10^{-12}} \right)\,,
\end{align}
where $R_0(Y)$ is a function of the helium abundance ($Y$) in the globular clusters and the $F$ are some positive-defined functions of the axion couplings to photons and electrons.
Though it is known that both couplings can impact the R parameter~\cite{Giannotti:2015kwo}, 
surprisingly, there are no explicit stellar numerical evaluations of the R parameter which include also the axion-electron coupling. 
Instead, historically, the analysis has always been done in order to derive a bound on the axion-photon coupling under the assumption that 
$ g_{ae}=0 $~\cite{Raffelt:1985nk,Raffelt:1987yu,Ayala:2014pea,Straniero:2015nvc}.
The most recent result is $ g_{a\gamma}\leq 0.65\times 10^{-10}\,{\rm GeV^{-1}}$ at $2\,\sigma$~\cite{Ayala:2014pea,Straniero:2015nvc}.
%

A slightly weaker bound on the axion-photon coupling, $ g_{a\gamma}  \leq 0.8\times 10^{-10}\,{\rm GeV^{-1}}$ was derived from 
the observation that additional cooling during the He burning stage of intermediate mass stars could prevent the blue loop stage, whose existence is assessed by 
observation of the Cepheid stars~\cite{Friedland:2012hj,Carosi:2013rla}. 

\vskip 2mm
The bounds discussed above are derived under the assumption that axions are \textit{light}, meaning that their mass is such that their production is not hampered by the Boltzmann factor. 
In the case of HB stars, with a core mass of about 10 keV, the limits discussed above extend to 30 keV or so.  
The dependence of the bound on the axion mass was analyzed in Ref.~\cite{Cadamuro:2011fd}, and later updated in Ref.~\cite{Carenza:2020zil},
which corrected the previous result to account also for axion production through photon coalescence and for the possibility of axions decay inside the star.
The bound at a few 100 keV defines the low mass edge of the so called cosmological triangle, a triangular area in the ALP parameter space which is free from astrophysical and experimental constraints, even though in tension with considerations from standard cosmology. 

\vskip 2mm
To probe even higher masses, one has to rely on hotter stars, particularly SNe. 
Massive ALPs coupled to photons can be produced in a SN through Primakoff and through photon coalescence.
Detailed discussions can be found in~\cite{Dolan:2017osp,Carenza:2020zil,Ertas:2020xcc,Sung:2019xie,Lucente:2020whw}.
Particularly interesting, in this case, is the trapping regime, in which axions are coupled so strongly that they do not leave the star. 
Because of trapping, if the axion-photon coupling is increased above a certain value, the SN will not be cooled sufficiently and the bound would disappear. 
The exact value of the axion-photon coupling in the trapping regime is uncertain since different approaches give slightly different results 
(see discussion in Ref.~\cite{Lucente:2020whw}).
This bound defines the lower edge of the cosmological triangle.
A summary of the current bounds from stars on axions with photon-coupling is shown in Fig.~\ref{fig:large_panorama} in Section~\ref{ssec:irastorza}.

\subsubsection{Axion-electron coupling}
\label{sssec:giannotti-axion-electron}

\noindent
The axion electron coupling  is defined from the Lagrangian 
\begin{align}
	\label{eq:Lae}
	\mathcal{L}_{ae}=g_{ae}\,a\bar\psi \gamma_5 \psi\,,
\end{align}
where $ \psi $ is the electron field. 
This interaction can be best tested in high density stars, in particular in RGB and WD stars, 
where axions can be efficiently produced through the electron/ion bremsstrahlung process $e +Ze\to  e + Ze +a$.

In particular, the RGB tip (RGBT) provides the currently most efficient laboratory to test this coupling. 
The RGBT is the brightest point of the RGB, corresponding to the time when the core reaches the conditions required to ignite helium. 
Due to the high temperature dependence of the $ 3\alpha $ nuclear reaction, which produces carbon from helium, 
the exact ignition time, and thus the position of the RGBT, is very sensitive to any exotic cooling. 
The strongest bounds on $ g_{ae} $ are derived from analyses of the RGBT in several globular clusters~\cite{Straniero:2020iyi} and in the galaxy NGC 4258~\cite{Capozzi:2020cbu}.
The two analyses lead to very similar results, $ g_{ae}\lesssim 1.5\times 10^{-13} $ (see Table~\ref{tab:Astro_bounds}).

Similar bounds can also be inferred from the evolution of WDs, as shown in Table~\ref{tab:Astro_bounds}.
Notice also that the analysis of the WD luminosity function (WDLF) seems to indicate a hint to additional cooling which is, in general, compatible with the RGB bounds.
The hint is subject to several uncertainties but is confirmed in the most recent analyses (see Refs.~\cite{Isern:2018uce,Isern:2020non}).

The combination of all the hints on the axion-photon and axion-electron coupling from WDs, RGB and HB stars indicates a best fit for 
$ g_{ae} \simeq 10^{-13}$ and $ g_{a\gamma}\simeq 0.2\times 10^{-10} {\rm GeV^{-1}}$,
indicating also a preference for a finite axion electron coupling at slightly more than 2$ \sigma $.

Contrarily to the case of the axion-photon coupling, where terrestrial searches for solar axions are now competitive with~\cite{Anastassopoulos:2017ftl} 
(and may exceed~\cite{Armengaud:2019uso,Abeln:2020ywv}), 
the astrophysical bounds on the axion-electron coupling discussed above are significantly stronger than the current experimental bound,
at least in the case of solar axion searches (see, e.g., discussion in Ref.~\cite{DiLuzio:2020wdo}).
Direct experimental bounds on the axion-electron coupling from solar axion searches\footnote{for a reference, the LUX collaboration~~\cite{Akerib:2017uem} reported $g_{ae} < 3.5\times 10^{-12}$ and 
PandaX-II collaboration~\cite{Fu:2017lfc} reported $ g_{ae}<4\times 10^{-12} $.}
are, at the moment, about an order of magnitude weaker than the astrophysical bounds.
In particular, the excess found in the 2020 results from the Xenon~1T collaboration~\cite{Aprile:2020tmw}, 
which steered a lot of attention on the axion-electron coupling, cannot have a simple explanation in terms of solar axions, as originally proposed (see Ref.~\cite{DiLuzio:2020jjp} for a detailed discussion). 
The data, however, do not exclude the possibility that 
Xenon~1T observed dark matter axions with a considerably reduced coupling to electrons, compatible with astrophysical observations. 
Indeed, Ref.~\cite{Takahashi:2020bpq} proposed a possible solution in terms of dark matter axion-like particles, very weakly coupled to photons, and with a coupling to electrons in the range $ g_{ae}\sim (0.5-0.7)\times 10^{-13} $.
Remarkably, this range overlaps with the RGB best fit value in Ref.~\cite{Straniero:2020iyi}.

\subsubsection{Neutrino magnetic moment}
\label{sssec:giannotti-neutrino-magnetic-moment}

Neutrinos are a primary example of FIP and their importance for stellar evolution is well known. 
Specifically, neutrinos drive the energy loss during the RGB stage and, again, at the end of the evolution of low mass stars, during the early WD phase.
In such cases, characterized by a dense and degenerate plasma, neutrinos are mostly produced through plasmon decay $\gamma\to \nu \bar \nu $.
At higher temperature, the $e^+ + e^-\to \bar \nu \,\nu$ process becomes very important and indeed dominates after stars begin burning carbon in the core.
In all cases, a non-vanishing neutrino magnetic moment could contribute significantly to the neutrino production, since it would provide a direct coupling between neutrinos and photons. 

Early attempts to constrain the neutrino magnetic moment in stars can be traced back to the already mentioned 1963 paper by Bernstein et al.~\cite{PhysRev.132.1227}.
The analysis, based on solar observations, was revised in Ref.~\cite{Raffelt:1999gv}, which found $ \mu_\nu \leq 4\times 10^{-10} \mu_B $, where $ \mu_B $ is the Bohr magneton. 
However, considerably stronger bounds can be found from the analysis of RGB and WD stars, where neutrino processes actually dominate the stellar energy loss. 
The strongest constraint,  $\mu_{\nu}\leq 1.5\!\times\! 10^{-12}\,\mu_B$, was recently derived in~\cite{Capozzi:2020cbu} from a study of the RGBT in the galaxy NGC 4258.
Massive stars in the carbon burning and later stages provide a less stringent bound (Table~\ref{tab:Astro_bounds}) but do show some interesting implication 
of a large magnetic moment for the late evolution of massive stars and for the SN explosion~\cite{Heger:2008er}.

\subsubsection{Axion-nucleon coupling}
\label{sssec:giannotti-axion-nucleon}

The axion nucleon coupling is controlled by the Lagrangian terms 
\begin{align}
	\label{eq:LaN}
	\mathcal{L}_{aN}=g_{an}\,a\,\bar n \gamma_5 n + g_{ap}\,a\,\bar p \gamma_5 p\,,
\end{align}
where $ n,~p $ are, respectively, the neutron and proton fields. 
The most well known argument to constrain the axion interaction with protons and neutrons is the one based on the observed neutrino signal from supernova (SN) 1987A.
The signal duration depends on the efficiency of the cooling and is compatible with the assumption that SN neutrinos carry about 99\% of the  energy released in the explosion.
For a light, feebly-interacting particle, a bound can be extracted from the requirement that this observational result is not spoiled~\cite{Turner:1987by,Burrows:1988ah,Raffelt:1987yt,Raffelt:1990yz}. 

A SN, in the first few seconds after the explosion, has a core with a temperature of about 30 MeV and density of about $3\times 10^{14} \,$g\,cm$^{-3}$.
The most efficient axion production mechanisms in such environment are the nucleon-nucleon bremsstrahlung $N+N\to N+N+ a$, and the pion production process
$\pi^- + p \to a+n$, both induced by the axion coupling to nucleons. 
The first of these processes has been widely studied in the past and resulted in several strong bounds, the most recent being~\cite{Carenza:2019pxu} 
\begin{equation} 
	g_{an}^2+ 0.61\, g_{ap}^2 + 0.53\, g_{an}\,g_{ap}\lesssim 8.26 \times 10^{-19} \,.
	\label{eq:gan_gap_SN_bound}
\end{equation}
The second process, induced by pions, has been recognized for a long time as possibly competitive with the nucleon bremsstrahlung~\cite{Turner:1991ax,Raffelt:1993ix,Keil:1996ju}.
However, it received considerably less attention until, very recently, 
it was shown that the pion abundance in the early stages of a SN is much larger than 
previously expected~\cite{Fore:2019wib}.
Although a complete study of the SN evolution which includes pions self-consistently at the moment does not exist, 
Ref.~\cite{Carenza:2020cis} showed that pion induced processes may well dominate over the bremsstrahlung, and so likely reinforce the bound in Eq.~\eqref{eq:gan_gap_SN_bound}.

Other strong bounds on the axion-nucleon coupling are derived from observations of the cooling of neutron stars (NS)~\cite{Keller:2012yr,Sedrakian:2015krq,Hamaguchi:2018oqw,Beznogov:2018fda,Sedrakian:2018kdm}. 
Here, the literature presents several different analyses and results.
One of the most studied NS is the NS in CAS A.
In recent years, there has been some speculation that its 
anomalously rapid cooling 
could be a hint of axions with coupling to neutrons~\cite{Leinson:2014ioa}
$g_{an}\simeq 4\times 10^{-10}$.
However, later on the data was explained assuming a neutron triplet superfluid transition occurring  at the present time, $t\sim 320$ years, in addition to a proton superconductivity operating at $t\ll 320$ years~\cite{Hamaguchi:2018oqw}.
Under these assumptions, it was possible to fit the data well, leaving little room for additional axion cooling.
Quantitatively,
\begin{align}
	\label{eq:NS_CAS_A}
	g_{ap}^2+1.6\, g_{an}^2\leq 1.1\times 10^{-18}\,.
\end{align}
A stronger bound, though only on the axion-neutron coupling, 
\begin{align}
	\label{eq:NS_J1731}
	g_{an}\leq 2.8\times 10^{-10}\,,
\end{align}
was inferred from observations of the NS in HESS J1731-347~\cite{Beznogov:2018fda}.
However, more recently Ref.~\cite{Sedrakian:2018kdm} proposed a considerable less stringent result, 
\begin{align}
	\label{eq:NS_Sedrakian}
	g_{an}\lesssim (2.5-3.2)\times 10^{-9}\,. 
\end{align}
%

\subsubsection{Dark Photons}
\label{sssec:giannotti-dark-photon}
The relevant low energy Lagrangian describing the dark photon (DP) is (see, e.g., Ref.~\cite{Fabbrichesi:2020wbt})
\begin{align}
	\label{eq:L_DP}
	\mathcal{L} =-\frac{1}{4} F_{\mu\nu}^2+ e J^\mu A_\mu -\frac{1}{4} V_{\mu\nu}^2+\frac12 m_V^2 V_\mu V^\mu -\frac12 \varepsilon  F_{\mu\nu} V^{\mu\nu} \,,
\end{align}
where $A$ is the standard model photon field and $V$ is the DP.
The Lagrangian in Eq.~\eqref{eq:L_DP} describes a massive spin 1 field, the DP, which mixes kinematically with the standard model photon.

The kinematic mixing allows DPs to be produced in stars. 
Therefore, if they existed in certain parameter regions, they would contribute to the stellar energy loss~\cite{Redondo:2008aa,An:2013yfc,An:2014twa,An:2020bxd}.
The possibility to explain the observed cooling anomalies through DP production in stars was considered in Ref.~\cite{Giannotti:2015kwo}.
An additional DP-induced energy loss can explain well the individual observations but not easily fit all the observed anomalies.

Contrarily to the case of axions, the production of DPs depends strongly on the mass. 
In particular, it can be shown that the production of the transverse modes of the DP is resonant when the 
DP mass matches the plasma frequency of the stellar plasma (see, e.g., Ref.~\cite{Redondo:2008aa}). 
Stars in different evolutionary stages have quite different plasma frequencies. 
In the core of the sun, $ \omega_{\rm pl}\sim 300 $eV, in a RGB close to the tip $ \omega_{\rm pl}\sim 25 $keV, and in a HB star $ \omega_{\rm pl}\sim 2.6 $keV.
Thus, bounds on the DP kinetic mixing are particularly strong at considerably different masses.
The most current bounds on DP from RGB and HB stars are provided in Refs.~\cite{An:2013yfc,An:2014twa}.
In both cases, the minimal value of the kinetic mixing is about $ \varepsilon \sim 10^{-15} $, at masses equal to the plasma frequency in the core of the corresponding star. 
The limits weaken rapidly away from that mass.
The same references considered also the solar bounds, finding $ \varepsilon \lesssim 2\times 10^{-15} $ for $ m_V\sim 300 $ eV.
As with the other stars, the bound decreases rapidly away from that mass. 
At much lower masses, $ m_V\leq 1$~eV, the sun provides the strongest stellar bound $\varepsilon \cdot m_V \leq 1.8\times 10^{-12}$~eV~\cite{Vinyoles:2015aba}.

The recent excess observed in the Xenon~1T data~\cite{Aprile:2020tmw} stimulated considerably the discussion about DPs and their astrophysics.
Even in this case, the excess cannot be ascribed to solar DP, since that would require parameters in strong tension with stellar evolution considerations~\cite{An:2020bxd}.
However, the possibility that the observation was induced by dark matter DPs cannot be excluded. 
Indeed, it was shown that the DP parameters required to explain the excess overlap with those hinted by HB stars~\cite{Alonso-Alvarez:2020cdv}.

Just like for axions, strong bounds on DP can be extracted from the observed neutrino signal from SN 1987A. 
Such analysis is particularly relevant for more massive DP, $ m_V \sim 0.1 - 100\,$MeV, a range not accessible from the analysis of low mass stars. 
Currently, the strongest bound constrains the kinetic mixing parameter to $ \varepsilon \gtrsim 10^{-10}- 10^{-8} $ in the mass range $ m_V \sim 0.1 - 100\,$MeV~\cite{Chang:2016ntp} (see also Fig.~\ref{fig:DP_visible} in Section~\ref{ssec:vector-results}.)

%
%
%
%

\subsubsection{Stars as FIP factories}
Besides allowing to study FIPs through the impact they have on the stellar evolution, 
stars are also factories which may produce FIPs in large quantities. 
Such FIPs can then be detected through their interaction with the interstellar medium or because they decay in standard model particles. 
Examples are light axion-like particles converted into photons in the galactic magnetic field or heavy DP decaying into standard model photons. 
The literature about all these possibilities is too vast to be properly reviewed.
Here, we just provide a list of references to recent works.

\vskip 2mm
By far the most studied stellar FIP factory is the sun, particularly in relation to experiments aiming at a direct detection. 
Being so close to the Earth, the solar FIP flux is substantially larger than that of any other star. 
Solar axions and DPs are being studied by a variety of experiments.
Here, we mention CAST~\cite{Anastassopoulos:2017ftl}, LUX~\cite{Akerib:2017uem}, PandaX~\cite{Fu:2017lfc}, Xenon~\cite{Aprile:2014eoa,Aprile:2020tmw}, 
and the proposed IAXO~\cite{Armengaud:2019uso} and BabyIAXO~\cite{Abeln:2020ywv}.
A very recent proposal suggests that the sun may act as a \textit{basin} for axions and DPs~\cite{VanTilburg:2020jvl,Lasenby:2020goo}.
Indeed, light FIPs produced in the low energy tail of the thermal spectrum could be 
trapped in the solar gravitational field, forming a basin of extremely low energy particles, possibly detectable with next generation experiments. 

\vskip 2mm
Stars other than the sun have to pay a big price because their distance reduces substantially the expected flux. 
However, the distance may be compensated by a higher production rate.
Several production processes are steeply dependent of the plasma temperature and the sun has a relatively cold core. 
Moreover, in certain cases the distance may be an advantage. 
The best example is the case of axions like particles converting in an external magnetic field. 
Evidently, a higher distance implies a higher conversion probability. 

\vskip 2mm
Arguably the most widely studied stellar FIP factories, besides the sun, are SNe.  
Light axions can be produced in a SN either through Primakoff process~\cite{Brockway:1996yr,Grifols:1996id,Payez:2014xsa} or, 
if they are coupled also to nucleons, through nuclear bremsstrahlung~\cite{Calore:2020tjw}. 
Either way, when produced they can leave the star and convert into photons in the galactic magnetic field. 
The non-observation of a gamma flux associated by the explosion of SN 1987A allows then to set a quite strong bound, $g_{a\gamma}\leq 5.3  \times 10^{-12} {\rm GeV^{-1}}$, 
on very light axion-like particles~\cite{Payez:2014xsa}. 
In Ref.~\cite{Meyer:2016wrm}, it was estimated that the bound might improve by over an order of magnitude by observations from Fermi LAT, in the case of a new galactic SN. 

\vskip 2mm
Refs.~\cite{Raffelt:2011ft,Calore:2020tjw} considered the diffuse axion background from the totality of the SN explosions. 
Even in this case, if axions are very light the diffuse background can be detected through conversion in the galactic magnetic field~\cite{Calore:2020tjw}.
On the other hand, the diffuse background of ALPs with masses above a few keV can be detected through their decay into photons~\cite{DeRocco:2019njg,Calore:2020tjw}.
The analogous case of DP was studied in details in Ref.~\cite{Kazanas:2014mca} and later revised and improved in Ref.~\cite{DeRocco:2019njg}.

\vskip 2mm
Conventional stars and neutron stars are also surprisingly efficient in producing observable fluxes of FIPs.
Studied examples of conventional stars include Betelgeuse~\cite{Xiao:2020pra} and Wolf-Rayet stars in superclusters~\cite{Dessert:2020lil}.
Both cases were applied to axion like particles and resulted in strong bounds on the ALP-photon coupling, comparable to the bounds from SN 1987A.
Another notable example is the observed  excess of hard X-ray 
events from a group of NS known as the Magnificent Seven (M7)~\cite{Dessert:2019dos}.
The origin of this anomaly is not  understood but it was speculated in Ref.~\cite{Buschmann:2019pfp} 
that the excess might be attributed to axion-like particles with couplings to both photons and neutrons, or even to QCD axions~\cite{Darme:2020gyx}.  


\subsubsection{Conclusions and future expectations}
Stars are efficient laboratories to test the physics of FIPs, and often complement the potential of terrestrial experiments. 
Nowadays, laboratory experiments are becoming extremely sensitive to very weak couplings and the predominance of stellar arguments witnessed in the past is diminishing. 
This is actually exciting, since finally laboratory will be able to test regions of the FIP parameter space which are not in tension with astrophysical arguments. 
Astrophysical experiments are also witnessing important developments in recent years, likely leading to significant improvements in our understanding of the physics of FIPs.


\clearpage
\subsection{Search for ultra-light scalar and pseudoscalar feebly-interacting particles: atomic physics, quantum technology}
\label{ssec:stadnik}
{\it Author: Yevgeny Stadnik, <yevgenystadnik@gmail.com>} 

\newcommand{\RN}[1]{%
  \textup{\uppercase\expandafter{\romannumeral#1}}%
}
\newcommand{\Eref}[1]{Eq.~(\ref{#1})}
\newcommand{\eref}[1]{(\ref{#1})}
\newcommand{\eps }{\varepsilon}
\newcommand{\rmd}{\mathrm{d}}

\newcommand{\appropto}{\mathrel{\vcenter{
  \offinterlineskip\halign{\hfil$##$\cr
    \propto\cr\noalign{\kern2pt}\sim\cr\noalign{\kern-2pt}}}}}
\renewcommand{\v}[1]{\boldsymbol{#1}}		

\subsubsection{Introduction}
\label{Sec:ultra-low-mass-FIPs_intro}

One of the main motivations to search for feebly-interacting bosons with very small masses ($\ll \textrm{eV}$) is the possibility that they may account for the observed dark matter content of our Universe. 
Low-mass spinless bosons may be produced non-thermally via the ``vacuum misalignment'' mechanism~\cite{Preskill:1982cy,Abbott:1982af,Dine:1982ah}
and can subsequently form a coherently oscillating classical field: 
\begin{equation}
\label{oscillating_scalar_field}
\phi (t) \approx \phi_0 \cos( m_\phi c^2 t / \hbar )  \, , 
\end{equation}
which occurs, e.g.~in the case of the harmonic potential $V(\phi) = m_\phi^2 \phi^2 / 2$ when $m_\phi \gg H$, where $m_\phi$ is the boson mass and $H$ is the Hubble parameter describing the relative rate of expansion of the Universe. 
The field in Eq.~(\ref{oscillating_scalar_field}) carries an energy density, averaged over a period of oscillation, of $\langle \rho_\phi \rangle \approx m_\phi^2 \phi_0^2 / 2$. 

\vskip 2mm
The oscillations of the field in Eq.~(\ref{oscillating_scalar_field}) are temporally coherent on sufficiently small time scales, since the feebly-interacting bosons remain non-relativistic until the present day; i.e., all of the boson energies satisfy $E_\phi \approx m_\phi c^2$. 
Nowadays, the galactic dark matter is expected to be virialised, with a root-mean-square speed of $\sim 300~\textrm{km/s}$ in our local galactic region. 
The typical spread in the boson energies is hence $\Delta E_\phi / E_\phi \sim \langle v_\phi^2 \rangle / c^2 \sim 10^{-6}$, which implies a coherence time of $\tau_\textrm{coh} \sim 2 \pi / \Delta E_\phi \sim 10^6 T_\textrm{osc}$, where $T_\textrm{osc} = 2 \pi / m_\phi$ is the period of oscillation. 
In other words, the oscillations of the bosonic field are practically monochromatic, with an associated quality factor of $Q \sim 10^6$. 
On time scales exceeding the coherence time, the amplitude of oscillation $\phi_0$ fluctuates in a stochastic manner. 

\vskip 2mm
The classical nature of the field in Eq.~(\ref{oscillating_scalar_field}) follows from the fitting of $\gg 1$ bosons into the reduced de Broglie volume, $n_\phi [\lambda_{\textrm{dB},\phi}/(2\pi)]^3 \gg 1$, which for the local galactic dark-matter energy density of $\rho_\textrm{DM,local} \approx 0.4~\textrm{GeV/cm}^3$~\cite{Zyla:2020zbs}
is satisfied for $m_\phi \lesssim 1~\textrm{eV}$. 
If very-low-mass bosons are to account for the observed dark matter, then their mass cannot be arbitrarily light, since such bosonic fields would suppress the formation of structures on length scales below the ground-state de Broglie wavelength of the bosons~\cite{Khlopov:1985jw,Hu:2000ke}, which becomes astronomically large for sufficiently low-mass bosons. 
The analysis of structures in Lyman-$\alpha$ forest data provides a lower mass bound of $m_\phi \gtrsim 10^{-21}~\textrm{eV}$~\cite{Irsic:2017yje,Nori:2018pka}, 
which is comparable to bounds from other astrophysical observations~\cite{Marsh:2018zyw,Schutz:2020jox}.

\vskip 2mm
The interesting range of dark-matter particle masses $10^{-21}~\textrm{eV} \lesssim m_\phi \lesssim 1~\textrm{eV}$ corresponds to oscillation frequencies in the range $10^{-7}~\textrm{Hz} \lesssim f \lesssim 10^{14}~\textrm{Hz}$. 
From an experimental point of view, this range of frequencies is convenient and physically accessible, with the lower end corresponding to periods of oscillation of the order of a month and the upper end corresponding to the infra-red region of the electromagnetic spectrum. 
Searching for possible particle-like signatures of very-low-mass bosonic dark matter is practically impossible, since the kinetic energies of non-relativistic very-low-mass particles are extremely small. 
Instead, one may take advantage of possible wave-like signatures due to the large number density of bosons. 

\vskip 2mm
Searching for wavelike signatures of the oscillating classical field in Eq.~(\ref{oscillating_scalar_field}) via its gravitational effects, e.g.~using pulsar timing methods, is limited to the lowest possible dark-matter particle masses~\cite{Khmelnitsky:2013lxt,Porayko:2018sfa}.
It is possible, however, to probe much larger ranges of dark-matter particle masses if the boson field additionally interacts non-gravitationally with the standard-model sector. 
For spinless bosons, the non-gravitational interactions are distinguished according to the underlying parity:~scalar-type interactions are nominally associated with an even-parity spinless field, while pseudoscalar-type interactions are associated with an odd-parity spinless field. 
Scalar-type interactions have previously been considered in the context of dark-energy-type models (see, e.g.,~\cite{Bekenstein:1982vfc,Peccei:1987vfc}). 
The axion, perhaps the most widely studied type of feebly-interacting pseudoscalar boson, may solve the strong CP problem of particle physics (see, e.g.,~\cite{Kim:2008hd} for a review). 

\subsubsection{Scalar portal:~Apparent variations of the fundamental ``constants''}
\label{Sec:ultra-low-mass-FIPs_scalars}

Consider the following linear-in-$\phi$ scalar-type interactions: 
\begin{equation}
\label{linear_scalar_couplings}
\mathcal{L} = \frac{\phi}{\Lambda_\gamma} \frac{F_{\mu \nu} F^{\mu \nu}}{4} - \frac{\phi}{\Lambda_f} m_f \bar{f} f  \, , 
\end{equation}
where the first term represents the interaction of the scalar field with the electromagnetic field tensor $F$, and the second term represents the interactions of the scalar field with the standard-model fermion fields $f$. 
Here $m_f$ is the ``standard'' mass of the fermion, $\bar{f} = f^\dagger \gamma^0$ is the Dirac adjoint, and the parameters $\Lambda_{\gamma,f}$ denote the effective new-physics energy scales of the underlying model. 
For the oscillating field in Eq.~(\ref{oscillating_scalar_field}), the interactions in Eq.~(\ref{linear_scalar_couplings}) result in the following apparent oscillations of the electromagnetic fine-structure constant $\alpha$ and the fermion masses~\cite{Stadnik:2014tta}: 
\begin{equation}
\label{linear_VFCs}
\frac{\delta \alpha}{\alpha} \approx \frac{\phi_0 \cos (m_\phi t)}{\Lambda_\gamma} \, , \, \frac{\delta m_f}{m_f} \approx \frac{\phi_0 \cos (m_\phi t)}{\Lambda_f}  \, . 
\end{equation}

\begin{figure*}[t!]
\centering
\includegraphics[width=12cm]{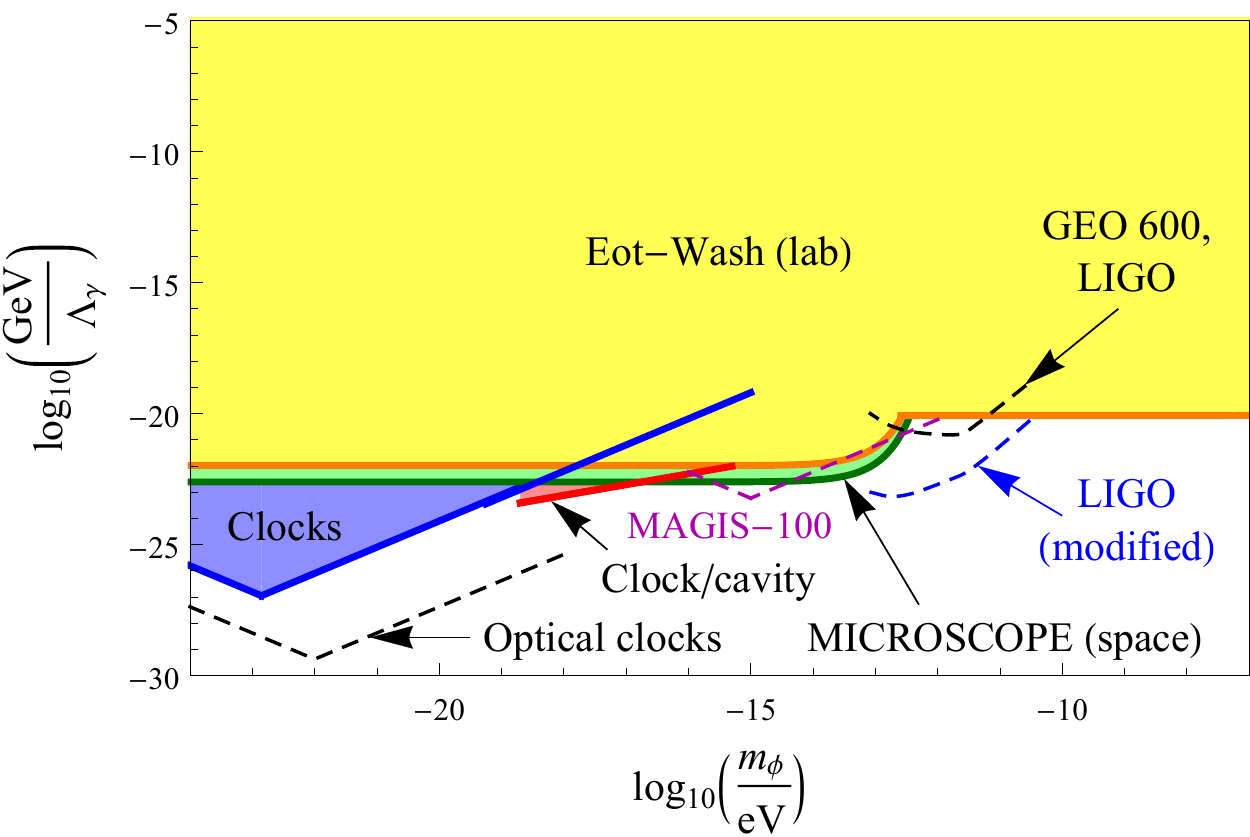}
\hspace{5mm}
\vspace{3mm}
\includegraphics[width=12cm]{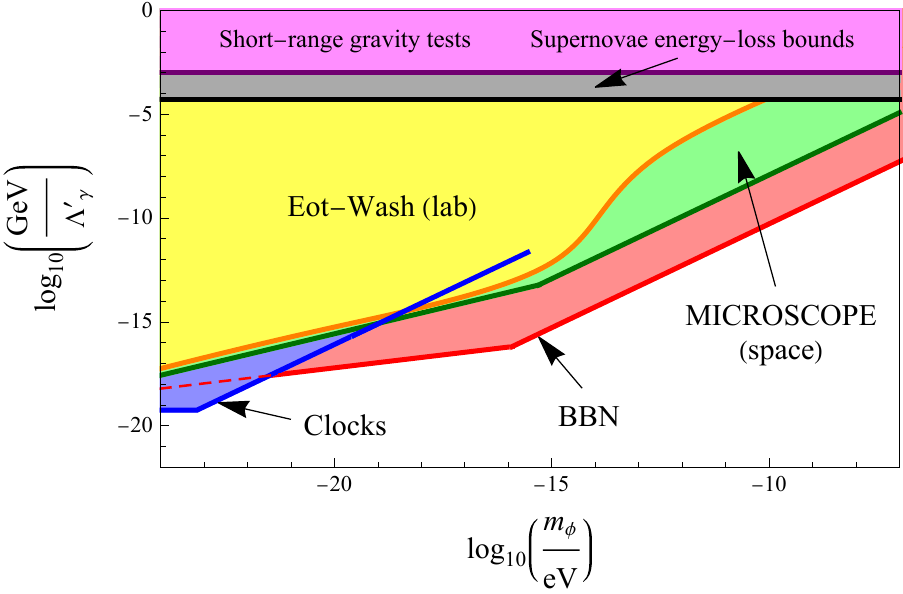}
\caption{ \normalsize 
Parameter spaces for an oscillating scalar dark-matter field interacting with the electromagnetic field via a linear-in-$\phi$ coupling (top subfigure) and via a quadratic-in-$\phi$ coupling (bottom subfigure). 
The coloured regions denote regions of parameter space ruled out by previous searches, while the dashed lines denote the projected sensitivities for existing and future experiments. 
The bounds on the $\phi^2$ coupling assume the sign of the first term in Eq.~(\ref{quadratic_scalar_couplings}), for which the scalar dark-matter field is screened near Earth's surface. 
}
\label{Fig:Scalar_limits}
\end{figure*}

Oscillations of the fundamental constants would induce oscillations of atomic transition frequencies~\cite{Stadnik:2014tta,Arvanitaki:2014faa} and of lengths of solids \cite{Stadnik:2014tta,Stadnik:2015xbn}, which can be sought with a variety of precision techniques, including (in order of increasing dark-matter particle masses):~atomic clock spectroscopy~\cite{Stadnik:2014tta,Arvanitaki:2014faa},
cavity comparisons~\cite{Stadnik:2014tta,Stadnik:2015xbn}, atom interferometry~\cite{Arvanitaki:2016fyj} and optical interferometry~\cite{Grote:2019uvn}. 
Bounds on linear-in-$\phi$ interactions from atomic clock spectroscopy experiments~\cite{VanTilburg:2015oza,Hees:2016gop} and clock-cavity comparison measurements~\cite{Kennedy:2020bac} are shown in Fig.~\ref{Fig:Scalar_limits}. 
Also shown (as dashed lines) are the projected sensitivities for various existing and future experiments. 
In this case, searches for oscillating fundamental constants give more stringent bounds over a range of dark-matter particle masses compared with traditional torsion-pendulum experiments~\cite{Wagner:2012ui,Touboul:2017grn}, which search for static equivalence-principle-violating forces mediated by virtual bosons~\cite{Hees:2018fpg}. 

\vskip 2mm
Besides the linear-in-$\phi$ interactions in Eq.~(\ref{linear_scalar_couplings}), quadratic-in-$\phi$ interactions are also possible, e.g.~if the underlying model admits a $Z_2$ symmetry under the $\phi \to -\phi$ transformation. 
In this case, Eqs.~(\ref{linear_scalar_couplings}) and (\ref{linear_VFCs}) are modified according to: 
\begin{equation}
\label{quadratic_scalar_couplings}
\mathcal{L} = \frac{\phi^2}{(\Lambda'_\gamma)^2} \frac{F_{\mu \nu} F^{\mu \nu}}{4} - \frac{\phi^2}{(\Lambda'_f)^2} m_f \bar{f} f  \, , 
\end{equation}
\begin{equation}
\label{quadratic_VFCs}
\frac{\delta \alpha}{\alpha} \approx \frac{\phi_0^2 \cos^2 (m_\phi t)}{(\Lambda'_\gamma)^2} \, , \, \frac{\delta m_f}{m_f} \approx \frac{\phi_0^2 \cos^2 (m_\phi t)}{(\Lambda'_f)^2}  \, . 
\end{equation}
In the case of $\phi^2$ interactions, there are not only apparent oscillations of the fundamental constants, but also slow ``drifts'' in the fundamental constants that are correlated with changes in the dark-matter density over time~\cite{Stadnik:2014tta,Stadnik:2015kia}. 
Additionally, in models of scalar dark matter with $\phi^2$ interactions, the dark-matter field can be strongly (anti)screened near the surface of massive bodies such as Earth, leading to equivalence-principle-violating forces directed along the resulting dark-matter density gradients~\cite{Hees:2018fpg}.
This (anti)screening phenomenon is caused by the $\phi^2$ interactions acting as a potential term in the classical equation of motion for the scalar field, in contrast to the linear-in-$\phi$ interactions, which instead provide a classical source term in the scalar-field equation of motion and give rise to Yukawa-type equivalence-principle-violating forces mediated by virtual bosons. 
For the signs appearing in Eq.~(\ref{quadratic_scalar_couplings}), the scalar dark-matter field is screened near Earth's surface. 

Bounds on quadratic-in-$\phi$ interactions from atomic clock spectroscopy~\cite{Stadnik:2015kia,Stadnik:2016zkf}, big bang nucleosynthesis~\cite{Stadnik:2015kia}, and torsion-pendula measurements~\cite{Hees:2018fpg} are shown in Fig.~\ref{Fig:Scalar_limits}. 
In this case, searches for dark-matter-induced variations of the fundamental constants give more stringent bounds over a broad range of dark-matter particle masses compared with short-range tests of gravity and supernova energy-loss bounds~\cite{Olive:2007aj}.
We remark that the quadratic-in-$\phi$ interaction parameters may be below the Planck mass-energy scale, in contrast to the linear-in-$\phi$ interaction parameters which are already constrained to be above the Planck scale for a very-low-mass scalar boson.

\subsubsection{Pseudoscalar portal:~Time-varying spin-dependent effects}
\label{Sec:ultra-low-mass-FIPs_pseudoscalars}

Traditional detection methods for pseudoscalars (axions and axion-like particles) have generally focused on their possible electromagnetic coupling of the form $aFF$ (see Section~\ref{sec:pseudoscalar} for a full discussion about models and detection techniques for axions and axion-like particles).
Let us instead consider the following non-electromagnetic pseudoscalar-type interactions: 
\begin{equation}
\label{pseudoscalar_couplings}
\mathcal{L} = \frac{g^2}{32 \pi^2} \frac{C_G}{f_a} a G\tilde{G} - \frac{C_f}{2 f_a} \partial_\mu a \, \bar{f} \gamma^\mu \gamma^5 f  \, , 
\end{equation}
where the first term represents the interaction of the pseudoscalar field with the gluonic field tensor $G$ and its dual $\tilde{G}$, and the second term represents the derivative interactions of the pseudoscalar field with the standard-model fermion fields $f$. 
Here $g^2 / (4 \pi)$ is the strong coupling constant, $\bar{f} = f^\dagger \gamma^0$ is the Dirac adjoint, $f_a$ is the axion decay constant (with dimensions of energy), and $C_{G,f}$ are model-dependent dimensionless parameters. 

\vskip 2mm
For the oscillating field in Eq.~(\ref{oscillating_scalar_field}), the gluon coupling in Eq.~(\ref{pseudoscalar_couplings}) induces oscillating electric dipole moments of nucleons~\cite{Graham:2011qk},
as well as atoms and molecules~\cite{Stadnik:2013raa,Flambaum:2019ejc}. 
In the laboratory frame of reference, the orbital motion of the Solar System through the practically-at-rest galactic dark-matter halo results in the oscillating dark-matter field acquiring a motional gradient term according to: 
\begin{equation}
\label{oscillating_pseudoscalar_field_motional_gradient}
a ( t, \boldsymbol{x} ) \approx a_0 \cos ( m_a t - \boldsymbol{p}_a \cdot \boldsymbol{x} )  \, , 
\end{equation}
where $\boldsymbol{p}_a$ is the linear momentum of the dark-matter ``wind'' as seen in the laboratory reference frame. 
In the non-relativistic limit, the derivative interaction of the field in Eq.~(\ref{oscillating_pseudoscalar_field_motional_gradient}) with a fermion field according to Eq.~(\ref{pseudoscalar_couplings}) gives rise to the following time-dependent Hamiltonian: 
\begin{equation}
\label{axion-wind_effect}
H(t) \approx \frac{C_f a_0}{2 f_a} \sin (m_a t) \, \boldsymbol{\sigma}_f \cdot \boldsymbol{p}_a  \, , 
\end{equation}
which can be interpreted as the precession of fermion spins about a time-varying pseudo-magnetic field\footnote{V.~V.~Flambaum, in \textit{Proceedings of the 9th Patras Workshop on Axions, WIMPs and WISPs, Mainz, Germany, 2013}, {http://axion-wimp2013.desy.de/e201031/index\_eng.html}. See also \cite{Stadnik:2013raa}.}. 

\vskip 2mm
These time-varying spin-dependent effects can be sought with co-magnetometry~\cite{Stadnik:2013raa,Stadnik:2017mid}
and $g$-factor measurements~\cite{Smorra:2019qfx}.
By analogy with Sec.~\ref{Sec:ultra-low-mass-FIPs_scalars}, these measurements can be interpreted in terms of apparent temporal variations of physical ``constants'', including electric dipole moments, magnetic dipole moments and $g$-factors of particles. 
In the case of the electron coupling, the ``axion wind'' spin-precession effect in Eq.~(\ref{axion-wind_effect}) can also be sought using spin-polarised torsion pendula~\cite{Stadnik:2013raa,Stadnik:2017mid}.
These types of searches can be extended to higher dark-matter particle masses by using resonant narrow-band techniques, including nuclear magnetic resonance (NMR) for the gluon and nucleon couplings~\cite{Budker:2013hfa}, and electron spin resonance (ESR) for the electron coupling~\cite{Krauss:1985ph,Raffelt:1985pvi,Barbieri:1985cp}.

\begin{figure*}[t!]
\centering
\includegraphics[width=7.5cm]{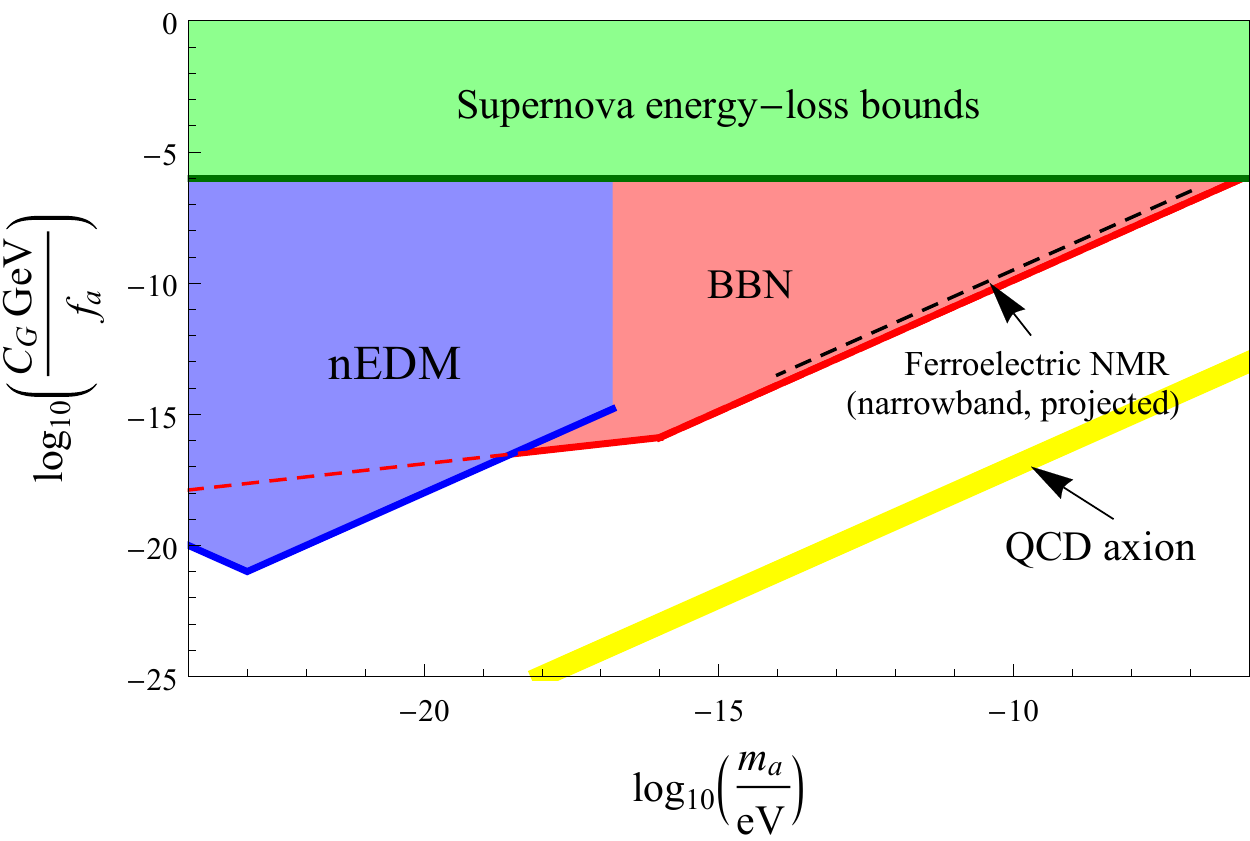}
\includegraphics[width=7.5cm]{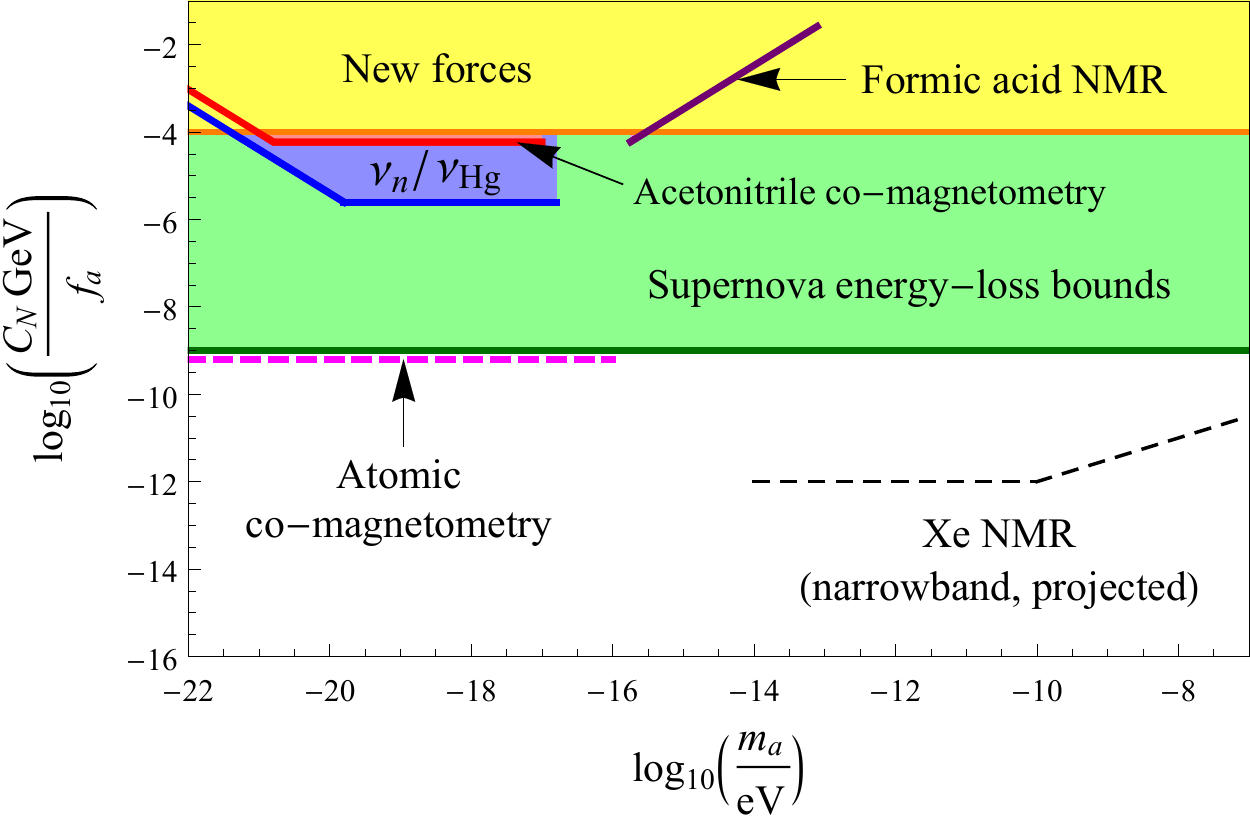}
\includegraphics[width=7.0cm]{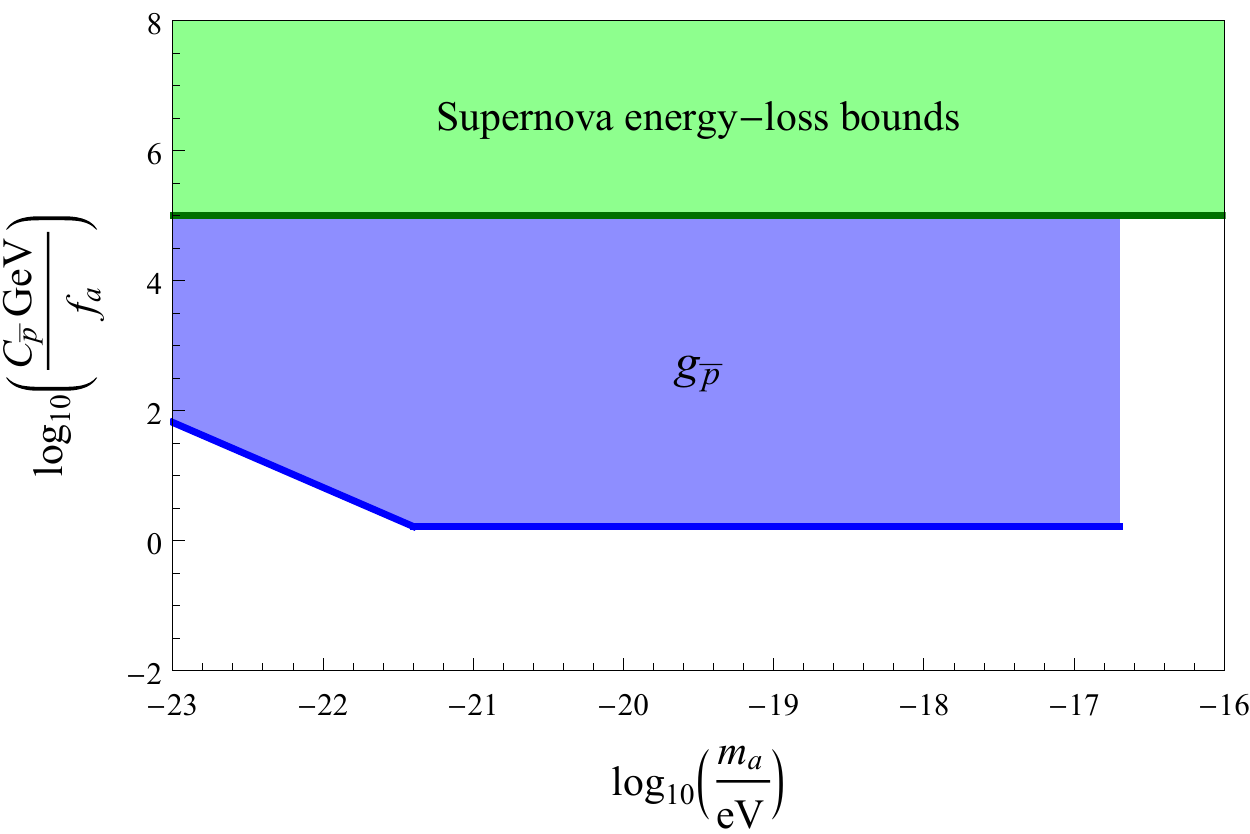}
\hspace{5mm}
\includegraphics[width=7.0cm]{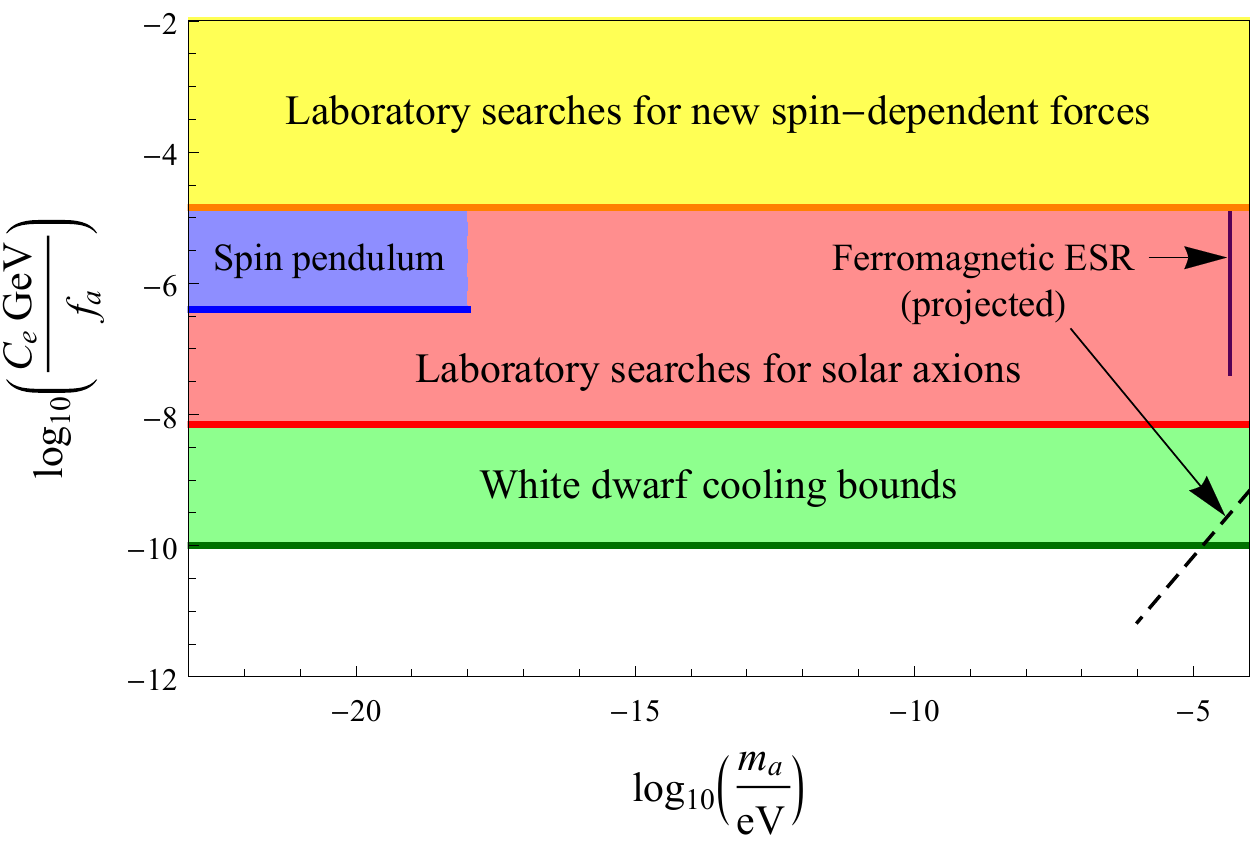}
\caption{ \normalsize 
Parameter spaces for an oscillating pseudoscalar (axion-like) dark-matter field interacting with the gluons (top-left subfigure), the nucleons (top-right subfigure), the antiproton (bottom-left subfigure) and the electron (bottom-right subfigure). The coloured regions denote regions of parameter space ruled out by previous searches, while the dashed lines denote the projected sensitivities for existing and future experiments. 
}
\label{Fig:Pseudoscalar_limits}
\end{figure*}

Bounds on the various pseudoscalar interactions are shown in Fig.~\ref{Fig:Pseudoscalar_limits}, including bounds from neutron EDM experiments and neutron-mercury co-magnetometry~\cite{Abel:2017rtm},
acetonitrile co-magnetometry~\cite{Wu:2019exd},
formic acid NMR~\cite{Garcon:2019inh},
anti-proton $g$-factor measurements~\cite{Smorra:2019qfx},
spin-polarised torsion-pendulum measurements~\cite{Terrano:2019clh},
and ferromagnetic ESR~\cite{Alesini:2020tjk,Flower:2018qgb,Crescini:2020cvl}.
In some cases, searches for dark-matter-induced time-varying spin-dependent effects already give more stringent bounds than astrophysical or different types of laboratory signatures over a range of dark-matter particle masses. 
The pseudoscalar interaction parameters may generally be below the Planck scale (compare with Sec.~\ref{Sec:ultra-low-mass-FIPs_scalars}).

\subsubsection{Concluding Remarks}
\label{Sec:ultra-low-mass-FIPs_Conclusion}

Atomic-physics and quantum-technology experiments provide a sensitive and promising avenue to search for ultra-light feebly-interacting particles. 
Rapid recent advances in these technologies have already paved the way for some of the most stringent bounds at present, courtesy of these ``unconventional'' approaches. 
Further technological advances are expected to extend the reach of these methods into new and uncharted territory.

\clearpage
\subsection{Search for FIPs with experiments at extracted beams}
\label{ssec:echenard}
{\it Author: Bertrand Echenard, <echenard@caltech.edu>} \\ 

Accelerator based searches are playing an essential role in probing light FIPs. In particular, beam dump and fixed target provide unique sensitivity to (very) small coupling at and below the QCD scale, ideally complementing the reach of collider searches. These experiments are also very versatile, as a wide variety of beams and techniques can be used to probes various couplings and final states. For example, protons produce many different final states, while muon beams open a unique window on the second generation of leptons. Finally, many dark sector searches can be performed at existing facilities with minor modifications, leveraging current infrastructures and previous investments. 

\subsubsection{Experimental techniques}
Before discussing the experimental situation, we will first review the different FIP search techniques, which can be broadly classified into the following four categories:

\begin{itemize}
\item{\it visible and semi-visible decays:} FIP decaying to visible final states can be fully reconstructed in the detector. Short lived FIPs are typically identified as narrow resonances over large irreducible backgrounds in detectors placed next to production targets, while long-lived FIPs are reconstructed in detectors placed farther away from electron or proton beam dumps. The constraints derived for long-lived particles usually exhibit a "triangular" shape characteristic of the beam intensity and the decay length between the dump and the detector. This technique requires large intensities, as FIP production is at least suppressed by the square of the coupling constant, and excellent detector performances to extract the elusive signals. Semi-visible final states can also arise from more complicated decay chains, multiple lifetimes and mass scales. Examples include inelastic DM, strongly interacting massive particle, dark sector showers,...

\item{\it re-scattering off detector material:} light DM produced by beams dumped in a stopping target can travel across shielding and scatter of electrons or protons in a detector far downstream. This technique requires a very large electron/proton flux to compensate the small production and scattering probabilities, overall suppressed by the fourth power of the coupling between the SM and dark sector, but it has the advantage of being sensitive to the dark sector coupling constant. The neutrino background, usually a limiting factor, can be reduced by using timing information and a thicker target to defocus the secondary neutrino beam. A good understanding of electromagnetic or hadronic shower development in material is also needed to correctly predict the DM flux.

\item{\it missing mass:} this technique is used to infer the mass of an invisible particle in reactions where the four-momenta of all remaining particles are well measured. Examples include $e^+e^- \rightarrow \gamma A'$ in $e^+e^-$ collider of positron beam dump where the photon is detected, and meson decays such as $K \rightarrow \pi X$ and $\pi^0 \rightarrow \gamma A'$. The FIP is typically reconstructed as a narrow resonance over a large background in the missing mass distribution. This technique is limited by backgrounds in which additional particles escape undetected.

\item{\it missing momentum / energy:} long-lived or invisibly decaying FIPs are emitted in fixed target reactions (e.g. $eZ \rightarrow eZ A'$), and identified through the missing momentum/energy carried away by the escaping particle(s). The missing momentum approach requires the ability to measure individual incoming beam particles to determine the momentum change. By contrast, the initial particle is entirely absorbed inside an active calorimeter for the missing energy case. For electron beams, the missing energy technique has a larger signal yield per electron on target (EoT), but neutrino induced backgrounds become challenging beyond a certain level. The missing momentum method remains essentially background free up to a larger luminosity thanks to electron/photon identification, and the transverse momentum spectrum contains information about the mass of the underlying FIPs. Both methods need excellent detector performance and hermiticity to ensure than no additional particle was produced. They typically offers a better signal yield than re-scattering experiments for a similar luminosity, as the FIP does not need to be detected and the sensitivity scales as the SM-mediator coupling squared.
\end{itemize}

\subsubsection{Experiments at extracted beams}
Equally diverse is the landscape of experimental efforts probing FIPs at extracted beams, including searches at electron, positrons and proton beams, as well as meson factories and neutrino experiments. A large fraction of these experiments can be performed at existing facilities with minor modifications, and a few dedicated efforts are already operating or under construction. A brief overview of these projects is discussed in the following (collider experiments are reviewed in another contribution to these proceedings). Further information can be found, for example, in References~\cite{Beacham:2019nyx,Lanfranchi:2020crw}.\\

\noindent
{\bf Experiments at electron beams}
\begin{itemize}
\item APEX (JLab)~\cite{Abrahamyan:2011gv}\\
Resonance search for visible $A' \rightarrow e^+ e^-$ decays using a high-resolution spectrometer. Status: completed.
 
\item BDX (JLab)~\cite{Battaglieri:2014qoa}\\
Search for DM scattering with homogeneous CsI calorimeter and active veto at 11 GeV beam. Sensitivity limited by the irreducible neutrino background. Status: active, BDX-Mini prototype collected $2\times10^{21}$ EOT.

\item DARKLIGHT (JLab)~\cite{Balewski:2014pxa}\\
Originally proposed to search for $e^+ \rightarrow e^+ A'(\rightarrow e^+ e^-)$ with full final state reconstruction. Very challenging concept, re-purposed as a dual spectrometer operating at the CEBAF injector. Status: under review. 

\item DARKMESA (MESA)~\cite{Christmann:2020qav}\\
Search for DM scattering with $\rm PbF_2$ and Pb-glass scintillator at 155 MeV electron beam. Very low background since the beam energy is below the $e^+\rightarrow e^+\pi^+$ production threshold. Status: under construction. 

\item HPS (JLab)~\cite{Celentano:2014wya}\\ 
Low-mass, high-rate silicon vertex tracker to search for prompt and visible dark photon decays and strongly interacting massive particles (SIMP). Status: active. 

\item LDMX (SLAC)~\cite{Akesson:2018vlm}\\
Missing momentum (and energy) experiment proposal to collect up to $10^{16}$ EOT at 4-8 GeV at the LESA beamline with high-granularity calorimeter. Status: proposal. 

\item MAGIX (MESA)~\cite{Doria:2018sfx}\\
Dual arm spectrometer with windowless gas stream target. Search for visible dark photon decays and three-body missing-mass (with recoil nucleus reconstruction). Status: under construction 

\item NA64 (CERN)~\cite{Gninenko:2019qiv}\\
Missing energy experiment at the 100 GeV secondary electron beam line at SPS with ECAL as active target to produce dark photons. Status: active. 
\end{itemize}

\noindent
{\bf Experiments at positron beams}

\begin{itemize}
\item MMAPS (CORNELL)~\cite{cornell}\\
Search for dark photon with missing mass technique at 5.5 GeV positron beam with the goal of collecting $10^{17}-10^{18}$ EOT/year. Status: proposal.

\item PADME (LNF)~\cite{Raggi:2015gza}\\
Search for dark photon with missing mass technique at 550 MeV positron beam from LNF linac. Status: active.

\item VEPP3 (BIN)~\cite{Wojtsekhowski:2012zq}\\
Search for dark photon with missing mass technique at 500 MeV positron beam with the goal of collecting $10^{15}-10^{16}$ EOT/year. Status: proposal.
\end{itemize}

\noindent
{\bf Experiments at proton beams}
\begin{itemize}
\item DarkQuest (FNAL)~\cite{Berlin:2018pwi}\\
Fixed target experiment at the 120 GeV proton beam with a short decay volume. Sensitive to many dark sector signatures (dark photon, dark scalar, inelastic DM, SIMP,...). Status: proposed upgrade of SeaQuest experiment (2023+). 

\item SHIP (CERN)~\cite{Anelli:2015pba}\\
Proton beam dump experiment to search for dark sector signature and study neutrino physics. Two detectors for complementary search strategies: scattering (LDM and neutrino) and visible decays (dark sector particles). Status: proposal.
\end{itemize}

\noindent
{\bf Neutrino experiments}
\begin{itemize}
\item COHERENT (ORNL)~\cite{Akimov:2015nza}\\
Series of detectors $\sim$20m from target at 1 GeV proton beam. First observation of Coherent Elastic Neutrino Nucleus Scattering (CEvNS). Sensitivity study to sub-GeV DM for planned 750 kg liquid argon scintillation detector. Status: active.

\item Dune (FNAL)~\cite{Berryman:2019dme}\\
Multi-purpose neutrino experiment with sensitivity to several FIPs, including DM, dark sector particles and HNLs. Status: under construction.

\item LSND (LANL)~\cite{deNiverville:2011it}\\
Reinterpret electron-neutrino elastic scattering measurement as constraints on light dark matter (re-scattering process) with 170 ton mineral oil detector. Status: completed.

\item MicroBooNE (FNAL)~\cite{Antonello:2015lea}\\
Multi-purpose neutrino experiment at BNB beamline. Search for long-lived dark scalar produced at rest in the beam dump. Status: active.

\item MiniBooNE-DM (FNAL)~\cite{AguilarArevalo:2008qa}\\
Search for DM with dedicated beam dump run with 800 ton mineral oil detector. Status: completed.

\item T2K/ND280 (J-PARC)~\cite{Abe:2019whr}\\
Multi-purpose neutrino experiment with dataset corresponding to a few $10^{21}$ POT. Search for HNLs in kaon decays with near detector, setting limits on mixing elements down to $\sim 10^{-9}$. Status: active.
\end{itemize}

\vspace{0.5cm}
\noindent
{\bf Experiments at meson factories}
\begin{itemize}
\item KOTO (J-PARC)~\cite{Ahn:2018mvc}\\
Rare neutral kaon decay experiment with sensitivity to scalar and vector portals, axions and HNLs. Status: active.

\item NA62 (CERN)~\cite{NA62:2017rwk}\\
Rare charged kaon decay experiment with sensitivity to scalar and vector portals, axions and HNLs. Possibility to run in beam dump mode to increase FIP production rate. Status: active.

\item REDTOP (CERN)~\cite{Gatto:2019dhj}\\
Intense $\eta / \eta'$ factory with sensitivity to several FIPs, including dark photon, dark scalar, baryonic dark force and axions. Status: proposal.
\end{itemize}

\vspace{2cm}
\noindent
{\bf Experiments at muon beams}
\begin{itemize}
\item Mu3e (PSI)~\cite{Arndt:2020obb}\\
Experiment searching for charged lepton flavor violating $\mu \rightarrow eee$ decays with sensitivity to visible dark photon decays.

\item M3 (FNAL)~\cite{Kahn:2018cqs}\\
Experimental concept similar to LDMX with muon beams to probe couplings to second generation and light thermal DM. Status: proposal.

\item NA64-mu (CERN)~\cite{Gninenko:2019qiv}\\
Experimental concept similar to NA64 with muon beams to probe couplings to second generation and light thermal DM. Status: proposal.
\end{itemize}

The current status for the existing bounds and future projections for different portals are shown in the plots contained in Sections~\ref{ssec:vector-results},~\ref{ssec:axion-results},~\ref{ssec:scalar-results} and ~\ref{sssec:hnl-results}. A significant fraction of the parameter space has already been explored, and future experiments will improve the sensitivity by orders of magnitude. In particular, most of the light thermal freeze-out DM below a few hundreds MeV will be explored, as well as important portions of the (extra heavy) QCD axion parameter space. Searches for HNLs will be conducted by several experiments, probing a currently largely unexplored region of masses and couplings.

\vskip 2mm
In summary, a vibrant program of fixed target and beam dump experiments is crucial to comprehensively explore feebly interacting particles.  A wide variety of beam and techniques can be used to probe many couplings and final states, offering unique sensitivity to a large fraction of the FIP parameter space. Most interestingly, many of the current and future proposals can exploit existing infrastructure and facilities, or would only require small modifications. This synergistic approach has played a central role in maximizing the discovery potential of dark sector searches so far, and will continue to be instrumental in defining an exciting future program.

\clearpage
\subsection{Search for FIPs with collider-based experiments}
\label{ssec:russell}
{\it Author: Heather Russell, <heather.russell@cern.ch>}  
\subsubsection{Overview}

Both proton-proton and electron-positron collisions provide the opportunity to discover feebly-interacting particles (FIPs). 
In such collisions, FIPs can be produced either directly in the collision or through the decay of a known particle.  General-purpose detectors, such as ATLAS and CMS, can search for many different flavours of FIPs in many different ways. However, FIPs are generally light particles with very low production cross-sections. Standard model (SM) backgrounds are overwhelming to many proposed models, though this can be counteracted by looking in rare decays, where SM backgrounds are significantly smaller. Additionally, SM backgrounds can be dramatically reduced by searching for FIPs with signatures that are uncharacteristic of the SM, such as displaced vertices from long-lived, neutral particles (LLPs).

\vskip 2mm
While the breadth of searches for FIPs at the LHC general-purpose detectors has dramatically increased over the past few years, the detectors were not built with FIPs in mind. Thus, such searches remain limited in reach by the detectors themselves, in terms of both the FIP mass and lifetime. Consequently, a variety of new experiments have been proposed to supplement the existing LHC experiments. Some proposals focus on improving the reach for LLPs, while others focus on (meta-)stable milli-charged particles. These new experiments are in varying states of approval and construction, with the earliest aiming to start their physics program during Run 3 of the LHC in 2022. 

\vskip 2mm
Experiments at $e^+e^-$ colliders provide complementary sensitivity to LHC experiments. Searches are performed both in a continuum with the $e^+e^- \rightarrow \gamma X$ process and through searches for rare decays in quarkonia samples. 

\subsubsection{Collider experiments}

\noindent
\textbf{General-purpose detectors at the LHC}

\begin{itemize}
\item ATLAS~\cite{Aad:2008zzm} and CMS~\cite{Chatrchyan:2008aa} (CERN) 

In general, ATLAS and CMS have similar sensitivity to a wide range of proposed new particles. However, in the corners of phase-space that some searches focus on, differences start to appear. For example, due to the larger diameter of ATLAS, there is a larger fiducial volume that LLPs can decay in and thus ATLAS and CMS are sensitive to complementary LLP lifetime ranges~\cite{Sirunyan:2020cao,Aad:2019xav}. 

One of the main impediments to searching for low-mass FIPs are the overwhelming SM backgrounds. The high LHC collision rate means that hardware and software triggers must rapidly select interesting events to record, with limitations of 100 kHz for hardware trigger acceptance and ~ 1 kHz for full-events accepted by the software trigger. The latter limitation comes largely from offline processing and storage requirements, and thus can be bypassed by reducing the event size. This technique is used by CMS to search for inclusive production of a dimuon resonance in the range 11.5--45~GeV, where traditional searches would have limited sensitivity~\cite{Sirunyan:2019wqq}.

\vskip 2mm
Many ATLAS and CMS searches focus on FIPs produced in SM Higgs boson decays, \emph{i.e.} through a Higgs portal. FIPs are searched for both in leptonic and hadronic decay channels, and with $H \rightarrow XX$ and $H \rightarrow ZX$ decays~\cite{Aaboud:2018fvk,Sirunyan:2020eum,Aad:2019tua,Aaboud:2018jbr,Aaboud:2018arf,Aaboud:2019opc,Sirunyan:2020zow,Sirunyan:2019xst,Sirunyan:2018mot,Sirunyan:2018mgs,Aad:2020rtv,Aaboud:2019opc,Aad:2020hzm,Aaboud:2018aqj,Sirunyan:2020cao,Aad:2019xav}. For the case where $X$ is a dark photon, $Z_d$, the latter decay is a more generic process to search for. This is because $H \rightarrow Z Z_d$ can be achieved by simply adding a dark gauge symmetry U(1)$_d$ to the SM, while for $H\rightarrow Z_d Z_d$ there must be additional mixing between a dark sector and the Higgs boson. Many of these searches are performed for both prompt and long-lived $X$.

\vskip 2mm
A result from ATLAS~\cite{Aaboud:2018fvk} showed that the fiducial acceptance for promptly-decaying pseudoscalar particles ($H\rightarrow aa$) events was significantly different than for promptly-decaying vector particles ($H\rightarrow Z_d Z_d$). There are no public searches addressing differences in fiducial acceptance between scalar, pseudoscalar, and vector particles that are long-lived.

\vskip 2mm
Higgs boson decays to FIPs can also be constrained indirectly through precise studies of SM decays or direct measurements of invisible Higgs boson decays~\cite{ATLAS-CONF-2020-008,Sirunyan:2018owy}. The strongest limits come from a combined measurement of all Higgs boson couplings~\cite{ATLAS-CONF-2020-027,Sirunyan:2018koj}. Long-lived particles are not usually reconstructed by standard reconstruction algorithms and will appear as invisible decays in many cases, so these searches also set constraints on longer-lifetime LLPs.

Both ATLAS and CMS also have searches for heavy neutral leptons (HNLs), with similar sensitivity for prompt HNLs~\cite{Aad:2019kiz,Sirunyan:2018mtv}. ATLAS has also addressed the possibility for the HNL to be long-lived, increasing the sensitivity for low-mass, small-coupling HNLs~\cite{Aad:2019kiz}.

\vskip 2mm
In addition to proton-proton collisions, the LHC can also act as a photon collider. During both proton and lead ion collisions, particles from both beams can emit photons and those photons can inelastically collide. In these collisions the beam particles remain intact, and thus there is very low detector activity. ATLAS and CMS have published searches for axions in $\gamma\gamma\rightarrow\gamma\gamma$ events, providing the strongest constraints in the axion mass region $5 < m_{a} < 100$~GeV~\cite{Aad:2020cje,Sirunyan:2018fhl}.

\item LHCb (CERN)~\cite{Alves:2008zz}

The LHCb detector provides coverage in the forward angular region $10 < \theta < 300$~mrad, and is designed to detect both prompt and displaced light resonances. LHCb searches for promptly-produced FIPs and FIPs produced in secondary decays.

A recent search for prompt and displaced dimuon resonances was designed to be as generic as possible~\cite{Aaij:2020ikh}. Although the search is less sensitive than dedicated dark photon searches~\cite{Aaij:2019bvg}, the simplicity of the analysis allows for easy reinterpretation and broad applicability to other models.

LHCb has also performed searches for new scalar particles produced in $B$-decays ($B^0\rightarrow K^{*0}\chi(\mu^+\mu^-)$,\,$B^+\rightarrow K^{+}\chi(\mu^+\mu^-)$)~\cite{Aaij:2015tna,Aaij:2016qsm}. The scalar particles are required to decay to muons, and the backgrounds are significantly constrained by requiring the secondary vertex to be consistent with the $B$-mass. These searches are sensitive to scalar masses as low as $2 m_{\mu}$.

One difficulty with many searches for new resonances is the possibility that they are hiding under known resonances. Often, for simplicity, these regions will be excluded from searches. LHCb has performed a dedicated search for a dimuon resonance $X$ in the $\Upsilon$ mass region $5.5 < m_X < 15$~GeV, though the peak regions are still excluded from the search~\cite{Aaij:2018xpt}. For a detailed discussion about LHCb results see Section~\ref{sec:non-minimal}.
\end{itemize}.

\noindent \textbf{Experiments dedicated to FIPs}

All of the collider-based experiments dedicated to discovering FIPs that are currently-proposed or ongoing are based at the LHC, operating parasitically at one of the four main collision points. 

\begin{itemize}
\item AL3X (CERN)~\cite{Gligorov:2018vkc}\\
AL3X is a newly-proposed detector that would be located in the ALICE/L3 cavern during LHC Run 5 and beyond. It would comprise a large tracking chamber located within the L3 solenoid, designed to detect LLP decays. The ALICE time projection chamber (TPC) detector could be repurposed as the AL3X tracking chamber, though a larger, dedicated tracker filling the entire solenoid would have optimal sensitivity.

Status: Proposed for LHC Run 5 and beyond if ALICE does not continue their program beyond Run 4.

\item ANUBIS (CERN)~\cite{Bauer:2019vqk}\\
ANUBIS is a proposed LLP detector that would be located in the 18~m wide, 56~m high access shaft (PX14) located directly above the ATLAS detector. Although the access shaft remains closed during LHC operations, ANUBIS would need to be removed and stored on the surface during LHC shutdowns. The detector would be made of four tracking stations, each with two triplet layers. Because ANUBIS would be located directly above ATLAS, ATLAS could be used both to trigger ANUBIS and to provide an active veto of SM activity. 

Status: Proposed, demonstrator planned for LHC Run 3.

\item CODEX-b (CERN)~\cite{Aielli:2019ivi}\\
The CODEX-b detector would be located fully underground in the DELPHI/UXA cavern, 25~m away from the LHCb interaction point IP8. The UXA cavern is separated from IP8 by a 3~m wide concrete wall, which would provide passive shielding. CODEX-b targets low-mass LLPs that are produced transversely and decay to charged particles, though the addition of calorimetry to detect photon decays is being considered. The detector would be 10~m$^3$, with six tracking layers on each wall to reconstruct the LLP decay vertex, with five tracking triplet layers evenly spaced throughout the volume of the detector. 

Status: Proposed, CODEX-$\beta$ demonstrator planned for LHC Run 3. Full detector planned for HL-LHC.

\item FASER (CERN)~\cite{Ariga:2018pin}\\
FASER is very forward LLP detector that will be located 480~m from the ATLAS interaction point, IP1, in an unused service tunnel (TI12). FASER has a 1.5~m decay volume aligned directly with the beam axis, and targets light LLPs decaying to charged particles or photons. 

Status: Under construction, full detector planned for LHC Run 3.

\item MATHUSLA (CERN)~\cite{Alpigiani:2020iam}\\
MATHUSLA is a proposed LLP detector that would be located on the surface, 60~m above and 70~m horizontally displaced from the CMS interaction point, IP5. Like CODEX-b and ANUBIS, it targets transversely-produced LLPs that decay to charged particles. CMS could be used as an active veto.

Status: Proposed, demonstrator planned for LHC Run 3. Full detector planned for HL-LHC.

\item milliQan (CERN)~\cite{Ball:2016zrp}\\
MilliQan is a detector designed to search for stable, milli-charged particles, positioned in PX56, a small drainage gallery above CMS/IP5. The relatively small 1~m$\times$1~m$\times$3~m detector would be comprised of three scintillating layers. During 2018, a demonstrator already showed competitive sensitivity to existing experiments~\cite{Ball:2020dnx}.

Status: Demonstrator collected 35~fb$^{-1}$ of data during 2018. Full detector planned for LHC Run 3.

\item MoEDAL-MAPP (CERN)~\cite{Pinfold:2019zwp}\\
MoEDAL is a detector located at IP8 designed to detect magnetic monopoles or other highly-ionizing particles. Two extensions to MoEDAL have been proposed: MAPP-mQP, to search for milli-charged particles, and MAPP-LLP, to search for neutral LLPs.

Status: Successful mQP prototype tested during LHC Run 2. Full detector planned for LHC Run 3.

\end{itemize}

\noindent \textbf{Experiments at $e^+e^-$ colliders}
\begin{itemize}
\item Belle II (KEK)~\cite{Abe:2010gxa}\\
Belle II has just started its physics programme, and has already released the results of two searches for FIPs. The exact momenta of incoming particles is known in $e^+e^-$ collisions, which allows for searches for processes not possible at the LHC. A search for a new vector boson $Z'$ decaying invisibly provides sensitivity to $Z'$ masses below $2m_{\mu}$, a region often excluded due to searches focusing on $Z'\rightarrow \mu\mu$ decays~\cite{Adachi:2019otg}. Belle II has also performed a fully data-driven search for a narrow resonance consistent with an axion in the resolved, di-photon final state, $e^+e^- \rightarrow \gamma a(\gamma \gamma)$~\cite{BelleII:2020fag}.

Status: Active.

\item BES III (IHEP)~\cite{Ablikim:2009aa}\\
The BESIII detector at BEPCII operates with variable beam energies in the $\tau$-charm region. BESIII has collected a massive sample of quarkonia, and has performed multiple direct searches for FIPs and searches for rare decays that could be enhanced by FIPs~\cite{Ablikim:2020qtn,Ablikim:2017aab,Ablikim:2018bhf,BESIII:2020soy}, including $\psi(3686)\rightarrow\pi^+\pi^-J/\psi (\gamma X)$, where $X$ decays invisibly. 

Status: Active.

\item KLOE-2 (LNF)~\cite{AmelinoCamelia:2010me}\\
The KLOE-2 experiment at DA$\phi$NE collected $e^+e^-$ collision data at the $\phi(1020)$ mass. They have performed many searches for FIPs, including searches for a new boson $U$ both rare $\phi$ decays, such as $\phi \rightarrow \eta U$, and continuum resonance searches in $e^+e^- \rightarrow U \gamma$ events, with $U$ decaying to muons or pions~\cite{Babusci:2012cr,Anastasi:2018azp}.

Status: Data-taking completed in 2018, analysis ongoing.

\end{itemize}

\subsubsection{Summary}

There is a large variety of searches for both prompt and long-lived FIPs with the current available set of detectors. Despite the general purpose detectors not being designed with long-lived particles in mind, many searches are successfully probing a wide range of FIP phase-space. However, there are limitations to this sensitivity, and a range of new experiments have been proposed, each of which would dramatically increase sensitivity to FIPs. Additionally, the milliQan and MoEDAL MAPP-mQP detectors have been proposed to increase sensitivity to milli-charged particles. Experiments at $e^+e^-$ colliders provide a complementary sensitivity to searches for FIPs at the LHC, especially in the very low-mass regime.

\clearpage

\section{Feebly-interacting vector particles as mediators for sub-GeV DM with thermal origin}
\label{sec:vector}

\subsection{Theory overview of DM models in the (1~MeV--10~GeV) mass range}
\label{ssec:berlin}
{\it Author: Asher Joseph Berlin, <ajb643@nyu.edu>} 
\newcommand{\x}{\chi}
\newcommand{\Ap}{A^\prime}
\newcommand{\mAp}{m_{A^\prime}}
\newcommand{\epsi}{\epsilon}
\newcommand{\p}{\prime}
\newcommand{\mdm}{m_{_\text{DM}}}
\newcommand{\rhodm}{\rho_{_\text{DM}}}
\newcommand{\eV}{\rm eV}
\newcommand{\MeV}{\rm MeV}
\newcommand{\GeV}{\rm GeV}
\newcommand{\TeV}{\rm TeV}

\subsubsection{Dark Matter Theory Overview}
\label{sssec:berlin-DMintro}

The theory space of DM is vast. Various attempts to provide some organizational guidance often categorize the landscape of theories according to: (1) particle content, (2) phenomenological signatures, or (3) DM's cosmological origin.  Although each has its merits, the last stands out as particularly compelling. This is because the most detailed understanding of DM arises from cosmological observations (its precise abundance is inferred from the acoustic peaks in the CMB angular anisotropies~\cite{Aghanim:2018eyx}). In organizing the theory space around DM's cosmological origin, we note that the era in which the SM bath was in thermal equilibrium is well understood in many respects. It is therefore natural to frame our questions around whether or not DM was also part of such a thermal bath. 

\begin{figure*}\centering
\includegraphics[width=1\textwidth]{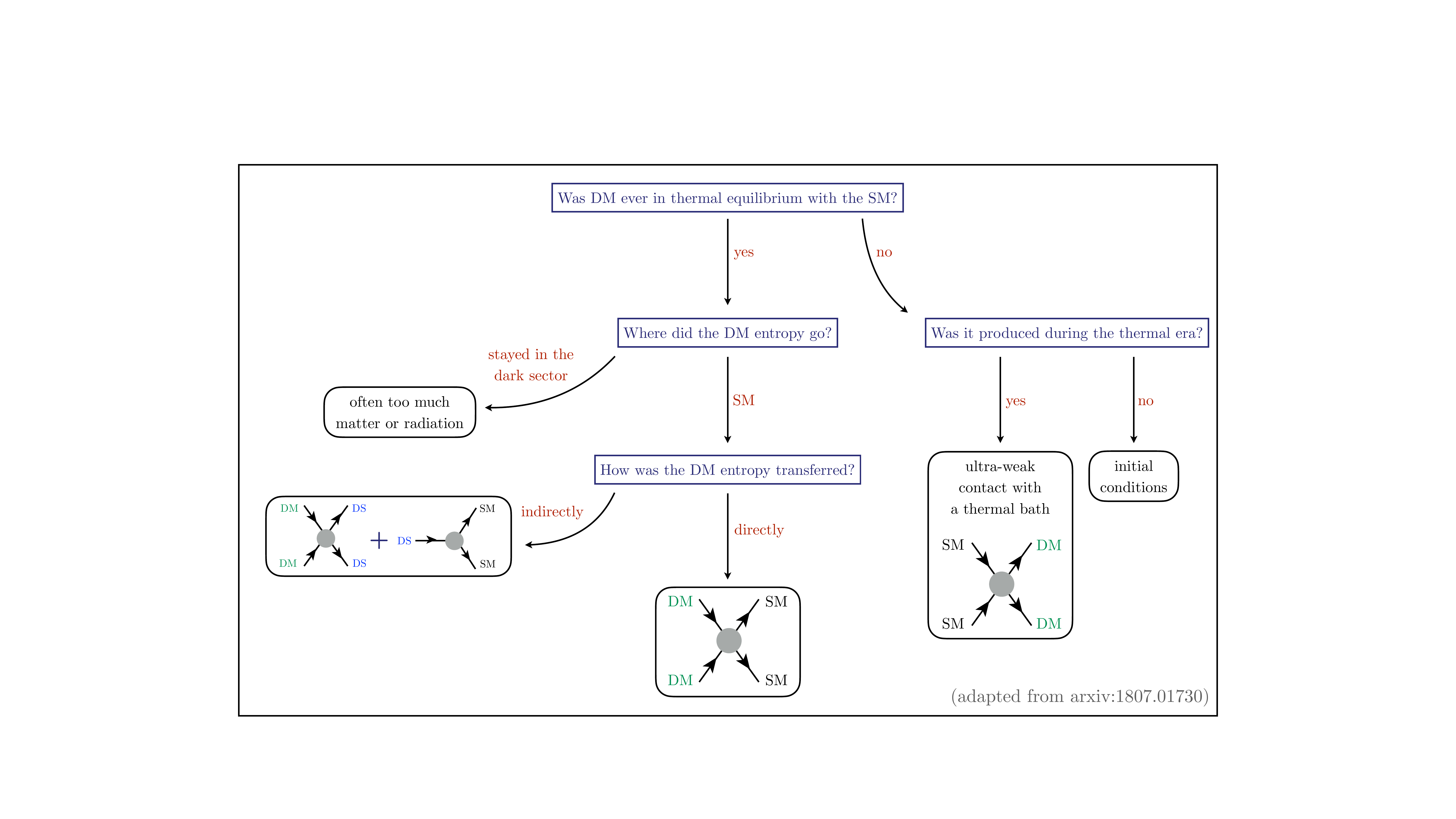}
\caption{Organizational flowchart of the DM theory space, adapted from Ref.~\cite{Berlin:2018bsc}. {\em DS} is an abbreviation for a dark sector particle. \label{fig:DMflowchart}}
\end{figure*}

\vskip 2mm
An organizational flowchart, using cosmologically-based questions as a guide, is shown in Fig.~\ref{fig:DMflowchart}. Stepping through from the top of the flowchart, if DM was in thermal equilibrium with the SM bath at early times, thermodynamics dictates that it possessed a large amount of thermal entropy. DM could not have decoupled from the SM with the entirety of this entropy, i.e., it is not a ``hot relic"~\cite{Kolb:1990vq}. As a result, this entropy was either transferred to other hidden sector degrees of freedom or to the SM bath. Instead, if DM was never in thermal contact with the SM at early times, its density may have been set by initial conditions or ultra-weak contact with thermalized particles. These possibilities encompass a vast array of cosmologies. Of the various branches shown in Fig.~\ref{fig:DMflowchart}, in this section we will focus on cosmologies in which DM directly transferred its thermal entropy to the SM, since these are especially economical and predictive. 

\subsubsection{Cosmology}
\label{sssec:berlin-DMcosmo}

There is strong motivation to consider such models of ``thermal relics." This class of cosmologies is minimal in the sense that the evolution of the DM abundance has reduced sensitivity to initial cosmological conditions and it is directly related to dynamics that are closely tied to the physics of the known universe, such as BBN. In the minimal case that thermal freeze-out occurs through DM annihilations directly into SM species, matching to the observed energy density requires an annihilation cross section of $\sigma v (T \sim \mdm) \sim 1/ (10 \ \TeV)^2$ at temperatures comparable to the DM mass $\mdm$, where $v$ is the relative DM velocity. As a result, thermal relics that only couple to the SM through the electroweak force must be heavier than the GeV-scale in order to not overclose the universe, as required by perturbativity~\cite{Lee:1977ua}. 

\vskip 2mm
The possible mass range of thermal relic DM is much wider than that typically considered for conventional WIMPs. In fact, thermal DM that is neutral under the SM gauge groups is generically viable down to MeV-scale masses, provided that its interactions with the SM are mediated by a new force-carrier parametrically lighter than the electroweak scale. This lower mass boundary is motivated by the successful predictions of BBN and observations of the CMB~\cite{Boehm:2012gr,Nollett:2013pwa,Nollett:2014lwa,Cyburt:2015mya}, although natural exceptions exist if the DM predominantly couples to neutrinos~\cite{Berlin:2017ftj,Berlin:2018ztp,Berlin:2019pbq}. In this sense, the cosmology of sub-GeV thermal relics directly motivates the consideration of new light mediators. 

\vskip 2mm
The paradigm of thermal freeze-out does much more than simply motivate the presence of light mediators, it explains why we haven't yet detected such new forces. For instance, if the DM-mediator couplings are $\order{1}$, then thermal freeze-out motivates SM-mediator couplings of size $\sim 10^{-5} \times (m_\text{force} / \mdm)^2 \, (\mdm / 100 \ \MeV)$, where $m_\text{force}$ is the mediator mass. The existence of new light forces that couple to the SM at the $10^{-5}$ level has not yet been definitely tested, but the discovery prospects in upcoming terrestrial experiments are promising in many scenarios. 

\vskip 2mm
Annihilations of thermal DM are typically no longer able to significantly modify its cosmological abundance at temperatures much below its mass. Regardless, annihilations often do proceed at much later epochs, such as near recombination (when the temperature of the universe was $T \sim {\rm eV}$). For annihilations into electromagnetically-coupled particles, measurements of the CMB constrain the quantity $\sigma v (T \sim {\rm eV}) / \mdm$, such that
\begin{equation}
\label{eq:DMCMB}
\sigma v (T \sim eV) \lesssim 10^{-3} \times \sigma v (T \sim \mdm) \, (\mdm / 100 \ {\rm MeV})
~, 
\end{equation}
where $\sigma v (T \sim \mdm) \sim 1 / (10 \ {\rm TeV})^2$ is fixed by thermal freeze-out. In other words, for small masses the annihilation rate to visible final states must be considerably suppressed compared to that near freeze-out. 

Since Eq.~(\ref{eq:DMCMB}) is a statement about the same process but at parametrically different momentum transfers, an investigation of how easily this bound is satisfied requires specifying concrete models and interaction structures. As a case example, let's consider scenarios in which DM annihilates directly to the SM through the exchange of a spin-0 or spin-1 mediator. We can then categorize such toy-models by additionally specifying the DM spin (0 or 1/2), leading to four distinct interaction structures within this model subspace. As it turns out, only one of these four possibilities (scalar DM coupled to a scalar mediator) predicts an annihilation rate that is unsuppressed at late times~\cite{Kumar:2013iva}. The other interaction structures naturally evade the bound of Eq.~(\ref{eq:DMCMB}) since they predict rates that are parametrically smaller near recombination due to the small late-time DM velocity ($v \ll 1$), a suppression that is alleviated near freeze-out when $v \sim \order{1}$. 

\subsubsection{A Concrete Model}
\label{sssec:berlin-DMmodels}

We now turn to discussing a concrete example of such models. It is important to stress from the beginning that although the model presented below is \emph{simple} in its particle content and interaction structure, this does not imply that it is a \emph{simplified} version of a necessarily more involved framework. As we will discuss, there exist both simple and theoretically consistent models of thermal DM, thus giving a complete answer to the question regarding its cosmological origin.  

\vskip 2mm
The most minimal and viable way that a new light force can couple to the SM is through mixing with the photon. This can be realized if a massive dark sector gauge boson $\Ap$ kinetically mixes with electromagnetism~\cite{Holdom:1985ag},
\begin{equation}
\label{eq:kinmix}
\mathcal{L} \supset \frac{\epsi}{2} \, F^\p_{\mu \nu} F^{\mu \nu} + \frac{1}{2} \, \mAp^2 \, \Ap_\mu A^{\p \, \mu} + e \, A_\mu \, \mathcal{J}_\text{em}^\mu + e_D \, \Ap_\mu \, \mathcal{J}^{\p \, \mu}
~.
\end{equation}
In Eq.~(\ref{eq:kinmix}), $\epsi$ is the kinetic mixing parameter, $\Ap_\mu$ is the dark photon field associated with a broken $U(1)_D$ symmetry, $A_\mu$ is the SM photon, $F^\p_{\mu \nu}$ and $F_{\mu \nu}$ are the dark photon and electromagnetic field-strengths, $\mathcal{J}^{\p \, \mu}$  and $\mathcal{J}_\text{em}^\mu$ are the dark and electromagnetic currents, $e_D$ and $e$ are the dark and electromagnetic gauge couplings, and $\mAp$ is the dark photon mass. Through the mixing term, SM particles of electromagnetic charge $e q$ acquire a millicharge under $U(1)_D$ given by $\epsi \times e q$. The natural size of $\epsi$ is $10^{-3}$ ($10^{-6}$) if it is generated radiatively at one (two) loop(s) by any particles charged under both electromagnetism and $U(1)_D$. Importantly, such small mixing parameters are consistent with small dark sector mass scales in theoretical constructions where electroweak symmetry breaking triggers $U(1)_D$ breaking at the scale $\sim \epsi \times \TeV \sim {\rm MeV} - {\rm GeV}$~\cite{Cheung:2009qd}.

\vskip 2mm
If the constituent of DM is a particle $\x$ that is directly charged under $U(1)_D$ and $\mAp \gtrsim m_\x$, then annihilations into pairs of dark photons is kinematically suppressed, and instead thermal freeze-out proceeds through direct annihilations into electromagnetically-charged SM states. For $\mAp \gtrsim \text{few} \times m_\x$, the annihilation rate near freeze-out is proportional to $\sigma v \propto \alpha_\text{em} \, y  / m_\x^2$ where
\begin{equation}
\label{eq:ydef}
y \equiv \epsi^2 \, \alpha_D \, (m_\x / \mAp)^4
~,
\end{equation}
and $\alpha_D \equiv e_D^2 / 4 \pi$~\cite{Izaguirre:2015yja}. In this case, the abundance of $\x$ is consistent with the observed DM energy density provided that
\begin{equation}
y \sim 10^{-10} \times \left( \frac{m_\x}{100 \ \MeV} \right)^2
~.
\end{equation}
Although this cosmologically-motivated range for $y$ is independent of $\mAp / m_\x$ in the limit that $\mAp / m_\x \gtrsim \text{few}$, DM observables at low-energy accelerators scale with $m_\x$, $\mAp$, and $\alpha_D$ differently, depending on the relative hierarchy between the collisional center of mass energy and $\mAp$~\cite{Berlin:2020uwy}. As a result, when presenting existing or projected sensitivities to these models in the $y-m_\x$ plane, $\alpha_D$ and $\mAp / m_\x$ are often fixed to the representative choices $\alpha_D = 0.5$ ($\alpha_D = 0.1$ is also commonly used) and $\mAp / m_\x = 3$. The motivation for this particular choice of parameters arises from the fact that large $\alpha_D$ weakens existing constraints on the cosmologically-motivated regions of parameter space, while $\mAp \gg m_\x$ maintains the predictive structure of Eq.~(\ref{eq:ydef}). For investigations that explore the consequences of relaxing these standard assumptions, see the discussion in, e.g. Refs.~\cite{Berlin:2018bsc,Berlin:2020uwy}.

\vskip 2mm
The sections below are dedicated to various experimental signatures of these models, with a particular emphasis on the ability of upcoming experiments to test the regions of parameter space that are relevant for thermal DM. Similar to the organizational structure of the theory landscape in Fig.~\ref{fig:DMflowchart}, it is also useful to provide a simple categorization of experimental signals, using DM's cosmological origin as a guide. As discussed above, cosmologies in which DM is a thermal relic that freezes-out through direct annihilations to the SM are especially interesting, since they are predictive in regards to the DM-SM interaction strength. However, such statements can only be made about the nature of DM-SM interactions at momentum-transfers comparable to the DM mass, $q \sim \mdm$, since this scale is parametrically tied to the temperature at freeze-out. To meaningfully compare different DM search strategies, it is therefore useful to compare the momentum scales probed by experiments to those that are  relevant in the early universe. 

\vskip 2mm
Accelerator searches for thermal DM are less sensitive to model variations (provided the model parameters remain consistent with standard thermal freeze-out). This is because accelerator production of DM is a probe of the interaction structure at momentum transfers $q \gtrsim \mdm$. As a result, such signatures are robust tests within a regime of momentum transfer that is directly dictated by the thermal cosmology. It is useful to contrast this to the situation for direct detection experiments. In this case, DM scattering is dictated by the small characteristic velocity in the galactic halo ($v \sim 10^{-3}$),  such that $q \ll \mdm$. Such experiments therefore probe the same couplings relevant for thermal freeze-out but in a parametrically different regime of momentum transfer. In this case, it is not surprising to note that slight variations to the interaction structure, although of little consequence to freeze-out, can lead to parametric suppressions (or enhancements) at low momentum transfer. This closely parallels our previous discussion regarding the suppression in the annihilation rate near the time of recombination.

\clearpage
\subsection{Status and prospects of DM direct detection experiments in the (1~MeV--10~GeV) mass range}
\label{ssec:cebrian}
{\it Author: Susana Cebrian, <scebrian@unizar.es}>  

\subsubsection{Introduction}
Different experimental approaches are followed for the  direct detection of the dark matter (DM) particles which can be pervading the galactic halo \cite{Schumann:2019eaa}. 

The direct detection has been typically focused on the elastic scattering off target nuclei of Weakly Interacting Massive Particles (WIMPs), either spin independent (SI) or dependent (SD); it is challenging as it is a rare signal, concentrated at very low energies with a continuum energy spectrum entangled with background. Therefore, ultra low background conditions and very low energy threshold are a must and the identification of distinctive signatures, like the annual modulation of the interaction rates and the directionality of the signal, would be extremely helpful to assign a DM origin to a possible observation. To probe specifically DM candidates with a mass below the GeV scale, lighter targets are required (to keep the kinematic matching between nuclei and DM particles), lower thresholds are necessary (to detect even smaller signals) and different search channels are being considered. Light WIMPs cannot transfer sufficient momentum to generate detectable nuclear recoils; therefore,  absorption or scattering off by nuclei (NR) or electrons (ER) are also being investigated, which requires the development of new technologies \cite{Battaglieri:2017aum}. The proposed Migdal effect \cite{Ibe:2017yqa}, where the WIMP-nucleus interaction can lead to excitation or ionization of the recoiling atoms, provides an additional signal which indeed is larger for low mass DM. DM searches in the low mass range will be discussed here (Sec.~\ref{seclowdm}) together with those of distinctive signatures (Sec.~\ref{secsign}).

\begin{figure*}[ht]
\centering
\includegraphics[width=0.9\linewidth]{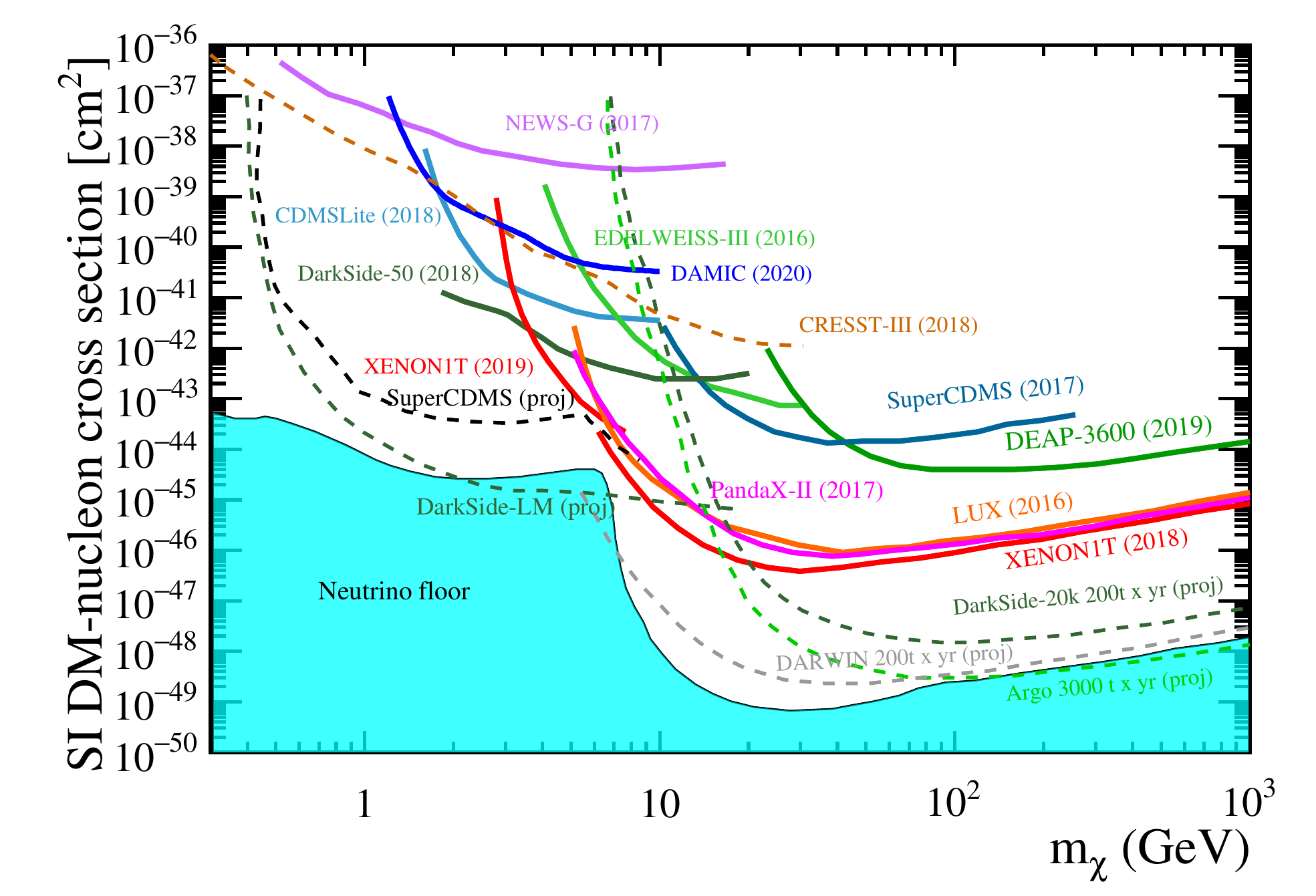}
\caption{ \normalsize Current bounds and future projections for DM direct detection  via SI elastic scattering off target nuclei.}
\label{fig:DM_SI}
\end{figure*}

\vskip 2mm
Figure~\ref{fig:DM_SI} show the current bounds and future projections for DM direct detection  via SI elastic scattering off target nuclei in the mass range (0.3-1000)~GeV.

\subsubsection{Search for low mass DM: Experimental techniques}
\label{seclowdm}
Different detection techniques are being used in the direct detection of DM; the basics of each technique and the status and latest results of main experiments implementing them will be summarized  here.

\vskip 2mm
\noindent
{\bf Bolometers}
In solid-state cryogenic detectors, phonons are measured by the tiny temperature increase induced, which requires operation at temperatures of tens of mK at most; a simultaneous measurement of ionization or scintillation allows discrimination of ER and NR. Very low mass crystals have reached low thresholds even below 100~eV$_{nr}$, and then, this type of detectors are producing leading results in GeV and sub-GeV regions.

\vskip 2mm
{EDELWEISS-III} operated at the Laboratoire Souterrain de Modane (LSM) in France 24 Ge bolometers (870 g each), presenting very good results at 5-30 GeV and limits also on Axion-Like Particles (ALPs) \cite{Hehn:2016nll}. EDELWEISS is using now much smaller Ge bolometers (33 g), having achieved a 60~eV threshold, exploring DM masses down to 45~MeV (considering the Migdal effect) and electron interactions and dark photons \cite{Armengaud:2019kfj,Arnaud:2020svb}.

\vskip 2mm
SuperCDMS operated Ge and Si bolometers of hundreds of grams at the Soudan Underground Laboratory in US. A 70~eV threshold was achieved exploiting the Neganov-Trofimov-Luke (NTL) effect at high bias voltage (HV) to convert charge into heat; results down to 1.5~GeV were presented from different analyses \cite{Agnese:2017njq,Agnese:2018gze}. Very small Si detectors (0.93 and 10.6~g) have been recently operated on surface considering nucleon and electron scattering and dark photons \cite{Amaral:2020ryn,Alkhatib:2020slm}. The operation of SuperCDMS at SNOLAB in Canada with different types of Ge and Si detectors (up to $\sim$30~kg) is expected to start in mid-2021.

\vskip 2mm
CRESST-III is operating at Laboratori Nazionali del Gran Sasso (LNGS) in Italy 10 CaWO$_4$ scintillating bolometers (24~g each), after using larger crystals too in CRESST-II \cite{Angloher:2015ewa}. A 30~eV threshold has been achieved, producing the best limits for WIMP-nucleus SI interaction down to 160~MeV \cite{Abdelhameed:2019hmk} and also for SD scattering (neutron-only case) \cite{cresstsd2019}. The use of up to 100 crystals is foreseen, with the goal of lowering the threshold to 10~eV.

\vskip 2mm
\noindent
{\bf Liquid Xe and Ar detectors}
Noble gases like Xe or Ar scintillate and can be ionized. In liquid state, they can form massive DM targets, which have produced leading results in the high mass range from a few GeV to TeV. A dual-phase detector is a TPC which can measure both the primary scintillation (S1) and the secondary scintillation from drifted electrons (S2); the ratio of these signals allows to distinguish ER and NR, with an energy threshold at the level of 1~keV$_{ee}$\footnote{Electron-equivalent energy.}. 

\vskip 2mm
Three Xe-based experiments have released over the last years the strongest constraints on WIMP-nucleus interaction above 6 GeV: XENON1T at LNGS \cite{Aprile:2018dbl}, LUX at the Sanford Underground Research Facility (SURF) in US \cite{Akerib:2016vxi} and PANDAX-II at Jinping Underground Laboratory in China \cite{Cui:2017nnn,pandaxiifull}. Considering only the S2 signal, above 0.4~keV$_{ee}$, XENON1T has also derived results for electron scattering \cite{Aprile:2019xxb}. Searches for light DM, considering the Migdal effect and Bremsstrahlung looking for ER, have been performed by XENON1T \cite{Aprile:2019jmx} and LUX \cite{Akerib:2018hck}. Many different DM models have been considered in the analyses (inelastic scattering, effective interactions, bosonic Super-WIMPs, light mediators, dark photons $\dots$); axions have been proposed as an explanation of the excess of ER observed by XENON1T \cite{Aprile:2020tmw}. Extensions of these experiments with several tonnes of active mass are being built or commissioned for operation over the next five years: XENONnT at LNGS, LUX-ZEPLIN (LZ) at SURF and PANDAX-4T at Jinping. In a longer term, the DARWIN observatory is expected to start at the end of the decade using 40~tonnes.

\vskip 2mm
In Ar-based detectors, the different scintillation pulse shape of ER and NR provides a very efficient discrimination method. DEAP-3600 is a single-phase detector operated at SNOLAB that has shown an outstanding background rejection factor \cite{Ajaj:2019jk}. DarkSide-50, at LNGS, has used a dual-phase detector filled with underground Ar having a reduced (by a factor $\sim$1400) $^{39}$Ar content \cite{Agnes:2018ep}; a low mass DM search, detecting S2 only with a 100~eV$_{ee}$ threshold, produced leading sensitivity at 1.8–3.5~GeV \cite{Agnes:2018fg} and results on electron scattering \cite{Agnes:2018ft}. The Global Argon DM Collaboration (GADMC) has been formed to work on the procurement of underground Ar from the Urania and Aria facilities in Colorado (US) and Sardinia (Itay), respectively, and to develop SiPMs. The DarkSide-20k project at LNGS will be the first step, with a smaller version using a 1~tonne target optimized for low mass DM searches (DarkSide-LowMass). For the end of the decade, the ARGO detector, with 300~tonnes at SNOLAB, is foreseen.

\vskip 2mm
\noindent
{\bf Ionisation detectors} 
Purely ionization detectors, based on semiconductors or gas chambers, are producing also very interesting results in the search for low mass DM.

\vskip 2mm
Point-Contact Ge detectors can reach sub-keV thresholds thanks to a very small capacitance in combination with a large target mass, as shown by CoGENT \cite{Aalseth:2012if}. Detectors with a mass of $\sim$1~kg each and a threshold of 160~eV$_{ee}$ are being operated at Jinping; CDEX-1, with one detector, set constraints on WIMP-nucleon SI and SD couplings and including the Migdal effect \cite{Liu:2019kzq}. CDEX-10, using a 10~kg detector array immersed in liquid N$_2$ has already presented results \cite{Jiang:2018pic} and larger set-ups are in preparation.

\vskip 2mm
Silicon charge-coupled devices (CCDs) offer 3D position reconstruction and effective particle identification for  background rejection. DAMIC, operating 7 CCDs (6 g each) at SNOLAB with a threshold 50~eV$_{ee}$ has presented results on electron scattering and hidden photon DM \cite{Aguilar-Arevalo:2016zop,PhysRevLett.123.181802} and on nucleon scattering \cite{damic3}. The commissioning of DAMIC-M, at LSM, is foreseen for 2023 using 50 larger CCDs with Skipper readout, where the multiple measurement of the pixel charge allows to reduce noise and achieve single electron counting with high resolution. This technology is already being used in the SENSEI experiment; small prototypes (0.0947 and 2 g) have been operated at Fermilab, setting leading constraints on electron scattering and hidden-sector candidates \cite{sensei,sensei2}. SENSEI plans to install a 100-g detector at SNOLAB. 

\vskip 2mm
A Spherical Proportional Counter is able to achieve very low threshold thanks to a very low capacitance. The SEDINE detector filled with Ne-CH$_4$ reached a 50~eV$_{ee}$ threshold at LSM \cite{newsg1}. NEWS-G is installing a larger sphere at SNOLAB, after taking commissioning data with CH$_4$ at LSM. Other gas detectors are being considered, like a high pressure TPC equipped with Micromegas readouts in the TREX-DM experiment being commissioned at LSC \cite{trexdmbkg}.

\vskip 2mm
\noindent
{\bf Bubble chambers}
Bubble chamber use target liquids in metastable superheated state; sufficiently dense energy depositions start the formation of bubbles, read by cameras. Although no direct measurement of the recoil energy is obtained, they are almost immune to ER background sources and, as most of the considered targets contain $^{19}$F, they offer the highest sensitivity to SD proton couplings. PICO has operated a series of bubble chambers at SNOLAB. PICO-60, with a 2.45~keV$_{nr}$ threshold, released the best SD limits \cite{Amole:2019fdf}, PICO-40L is already taking commissioning data while preparation of PICO-500 is ongoing. 

\vskip 2mm
In summary, for SI WIMP-nucleon cross section, above 10~GeV the best limits come presently from XENON1T but at lower mass results from different technologies, including liquid Xe and Ar detectors, bolometers and ionization detectors, must be considered. The huge projects like DARWIN and ARGO are expected to reach the neutrino floor. For SD interactions, also different experiments using several targets are giving the best limits at different mass ranges when considering both proton and neutron-only interactions. Concerning WIMP-electron scattering, SENSEI and XENON have presented the lowest limits for the interaction cross section below and above $\sim$10~MeV, respectively.

\subsubsection{Searches for distinctive signatures}
\label{secsign}

\vskip 2mm
\noindent
{\bf Annual modulation} 
The movement of the Earth around the Sun makes the relative velocity between DM particles and the  detector change in time, producing an annual modulation in the expected DM interaction rate with well-defined features \cite{freese1988}. The DAMA/LIBRA experiment, using 250~kg of NaI(Tl) scintillators at LNGS, has observed a modulation with all the proper features at 12.9$\sigma$ C.L. over 20~years and model-dependent corollary analyses have been carried out \cite{bernabeippnp2020}. No annual modulation signal has been found with different targets like Xe  \cite{Aprile:2017yea,Akerib:2018zoq,Kobayashi:2018jky} or Ge \cite{cdexmod} and the exclusion results obtained are in tension when interpreting the DAMA/LIBRA anual modulation as DM even assuming different halo or interaction models. Therefore, several projects are attempting a model-independent proof or disproof with the same NaI(Tl) target. COSINE-100 and ANAIS-112 are both taking data with $\sim$100~kg of NaI(Tl) in the Yangyang Underground Laboratory in South Korea and at LSC, respectively; COSINE has presented a limit for SI cross sections \cite{cosinenature} and the first annual modulation analysis is compatible with both null hypothesis and the DAMA/LIBRA signal \cite{cosinemod}. From the analysis of 3~y data of ANAIS, the deduced modulation amplitude is incompatible with DAMA/LIBRA results at 2.6 (3.3)~$\sigma$ for the 2-6 (1-6)~keV$_{ee}$ region \cite{anaismod,anais3y}. COSINUS is developing at LNGS NaI scintillating bolometers with capability to discriminate ER and NR \cite{cosinus} and SABRE is working on more ultrapure crystals to operate twin detectors in northern and southern hemispheres at LNGS and in Australia \cite{sabre}.

\vskip 2mm
\noindent
{\bf Directionality}
The average direction of the DM particles moving through the solar system comes from the constellation of Cygnus, as the Sun is moving around the Galactic center. A measurement of the direction of NR tracks could be used to distinguish a DM signal from background events (expected to be uniformly distributed) \cite{Spergel:1987kx,Mayet:2016zxu,ohare2021}. Track  reconstruction is challenging as they are very short ($\sim$1~mm in gas, $\sim$0.1~$\mu$m in solids) for keV scale NR. The identification of at least a head-tail asymmetry would be useful. Low pressure gas targets in TPCs with different amplification devices and readouts (Multi-wire proportional chambers (MWPC), Micro pattern gaseous detectors (MPGDs), optical readouts) \cite{battat} together with nuclear emulsions are being considered as directional DM detectors. Gas detectors  are mostly based on mixtures with $^{19}$F. DRIFT used MWPC at the Boulby Underground Laboratory in UK \cite{battat3}; MIMAC is using Micromegas at LSM \cite{tao}; NEWAGE develops a micro-pixel chamber ($\mu$-PIC) at Kamioka in Japan \cite{yakabe}; and DMTPC is working on optical readouts at the WIPP facility in US \cite{deaconu}. These experiments have measured directional NR, confirmed the head-tail effect and set limits for SD WIMP-proton interaction. A proto-collaboration, named CYGNUS \cite{cygnus}, has been formed to develop a multi-site, multi-target observatory of DM with directionality; first prototypes are already in operation. Regarding nuclear emulsions, the NEWS-dm project~\cite{newsdm} is working on a 10~g prototype at LNGS after showing excellent spatial resolution.

\subsubsection{Summary and Outlook}

Focused on the detection of smaller and smaller energy depositions and on the background suppression, advanced detectors based on recently developed technologies have been proposed or are at R\&D phase using, for instance: scintillating bubble chambers \cite{scintbubcham}; superfluid helium \cite{PhysRevD.100.092007}; supercooled water \cite{snowmass}; diamond detectors \cite{PhysRevD.99.123005}; or paleo-detectors \cite{PhysRevD.99.043014}. Specific ideas are being considered also to obtain the recoil direction, based on columnar recombination, anisotropic scintillators, crystal defect spectroscopy \cite{PhysRevD.96.035009} or planar targets like graphene \cite{graphene}.

Presently, the best limits from the direct detection of low mass DM, on cross sections for SD and SI interactions with nuclei and for electron scattering, come from several experiments based on very different detection techniques: cryogenic detectors, liquid noble detectors operated for charge collection only, purely ionization detectors and bubble chambers. The use of light targets, achieving extremely low energy thresholds and/or searching for different interaction channels are the strategies followed and new detection  technologies are in development. The proposed Migdal effect helps to enhance sensitivity particularly for low mass DM. Concerning the identification of distinctive signatures of DM, important results (even if still with low significance) from NaI(Tl) experiments have been presented to solve the long standing conundrum of the DAMA/LIBRA annual modulation result; work for the construction of a directional DM detector is underway, considering low-pressure TPCs and nuclear emulsions.


\clearpage

\subsection{Searches for light DM and vector mediators at FASER, CODEX-b, MATHUSLA, SHiP}
\label{ssec:salfeld}
{\it Author: Jakob Salfeld-Nebgen, <jakob.salfeld@cern.ch>} 

\subsubsection{Introduction and Model}
\label{sssec:salfeld-intro}
As a minimal extension of the Standard Model of particle physics (SM) the vector portal is very well motivated and resembles one of the four gauge and Lorentz invariant portals into the Dark Sector~\cite{Holdom:1985ag,Fayet:1980rr,Alexander:2016aln,Beacham:2019nyx}. Assuming an additional broken $U_{X}(1)$ gauge symmetry one can in general extent the SM Lagrangian with additional operators
\begin{equation}
  \mathcal{L}=\mathcal{L}_{\mathrm{SM}}+\mathcal{L}_{\mathrm{DS}}+\frac{1}{2}m_X^2 X^{\mu}X_{\mu}-g_X j^{X}_{\mu}X^{\mu}-\frac{\epsilon}{2 \mathrm{cos \theta_W}}B_{\mu\nu}X^{\mu\nu}.
\end{equation}
The kinetic mixing term of the $U(1)_X$ gauge field-strength tensor $X^{\mu\nu}$ and the SM hypercharge field-strength tensor $B_{\mu\nu}$ is parametrised by the kinetic mixing parameter $\epsilon$ and describes the coupling of SM particles with hypercharge to the new vector mediator. In its minimal version, the new model has only two parameters $(m_X,\epsilon)$ and is often referred to as Dark Photon model and the decay width (and lifetime) is proportional to $\epsilon^2$.
Possibly flavour dependent couplings of the new vector mediator $g_X$ can arise in case one of the global SM symmetries, such as $U(1)_{L_i-L_j}$ or $U(1)_{B-L}$, as studied in e.g.~\cite{Bauer:2018onh}, are gauged.

An independent Dark Sector Lagrangian $\mathcal{L}_\mathrm{DS}$ describes the dynamics in the Dark Sector with possible coupling of the new vector boson $X$ to the Dark Sector particle content. In its minimal form one can assume a dark fermion $\chi$  with dark vector mediator coupling $a_D$ and mass $m_\chi$:
\begin{equation}
\label{eq:dm}
\mathcal{L}_{DS} \supset \bar{\chi}(i\slashed{D}-m_\chi )\chi,
\end{equation}
with covariant derivative $D_\mu=\partial_\mu - i \alpha_D X_\mu$. Another less minimal example with phenomenological impact at colliders is the inelastic dark matter model as discussed in~\cite{Berlin:2018jbm,TuckerSmith:2001hy,Izaguirre:2015zva} with operator
\begin{equation}
\label{eq:idm}
\mathcal{L}_{DS} \supset i \alpha_D X_{\mu}\bar{\chi}_1 \gamma^{\mu}\chi_2,
\end{equation}
where $\chi_i$ denote the two almost mass degenerate states of Dark Matter particles.

\vskip 2mm
Finally, a mass term with vector mediator mass parameter $m_X$ can be generated for example via a spontaneous symmetry breaking mechanism of the $U(1)_X$ gauge symmetry. In case of the latter, a dark Higgs boson $S$ can give rise to additional production mechanisms for a dark vector mediator at high-energy colliders via its mixing $\kappa$ with the SM Higgs field, as discussed in e.g.~\cite{Curtin:2014cca}. The relevant operators in the Lagrangian are then:
\begin{equation}
\label{eq:hig}
\mathcal{L}\supset |D_\mu S|^2 - \kappa |S|^2|H|^2,
\end{equation}
and dark photons from decays of dark Higgs bosons can still be observed even for very small kinetic mixing parameter for suitably high $\kappa$.
%

\subsubsection{The MATHUSLA, CODEX-b, FASER, and SHiP experiments} 
The MATHUSLA~\cite{Chou:2016lxi,Lubatti:2019vkf,Alidra:2020thg}, CODEX-b~\cite{Aielli:2019ivi}, FASER~\cite{Feng:2017uoz,Ariga:2018pin,Ariga:2018uku} and SHiP~\cite{Anelli:2015pba} experiments are new detectors proposed to exploit the full physics potential of the LHC and CERN North area accelerator complex. The search for feebly interacting particles motivates dedicated detectors significantly displaced from the interaction points and highly shielded from irreducible radiation and background processes. The detectors vary in their location and pseudo-rapidity coverage which specifies the detector designs (for details see the please respective citations). Table~\ref{tab:comp} compares some of the characteristics of the four experiments. While the SHiP experiment is a fixed-target experiment using the SPS proton beam of 400 GeV (0.028 TeV centre-of-mass energy), the MATHUSLA, CODEX-b and FASER experiments make use of particle production in $\sqrt{s}=14$ TeV $pp$ collisions at the IP5, IP8 and IP1, respectively
(see also Section~\ref{ssec:russell}).

\begin{table*}[h]
\begin{center}
\caption{Comparison of some of the characteristics of the four experiments. The pseudorapidity range for SHiP is omitted because it is a fixed target experiment and the decay volume is roughly computed based on the information given in~\cite{Chou:2016lxi,Aielli:2019ivi,Feng:2017uoz,Anelli:2015pba}. The distance and location of the detectors defines their acceptance to different vector mediator production modes.}
\label{tab:comp}
\begin{tabular}{p{5cm}cccc}
\hline\hline
\textbf{Experiment}      & \textbf{$\sqrt{s}$} & \textbf{$\eta$ - range} & IP-distance & decay volume\\
\hline
FASER   		 & 14 TeV   			& $>$9 					&		480 m	& 0.06 m$^3$\\
CODEX-b      & 14 TeV             & 0.13 - 0.54  	&		25 m	& 1k m$^3$\\
MATHUSLA    & 14 TeV            & 0.9 - 1.5 			&		$\approx$ 150 m	& 800k m$^3$\\
SHiP   			 & 0.028 TeV        & -- 					&		 70 m	& 10k m$^3$\\
\hline\hline
\end{tabular}
\end{center}
\end{table*}

\vskip 2mm
MATHUSLA and CODEX-b, being located at more central pseudorapidities ($|\eta| < 1.5$), are of significant size to achieve good acceptance. CODEX-b is located closer to the IP than MATHUSLA and achieves good sensitivity for particle decays with significant lifetime for a $\mathcal{O}(100)$ times smaller detector size. FASER and SHiP are both primarily targeting low mass vector mediators. While SHiP is expected to have excellent sensitivity to a full suite of long-lived particles, good sensitivity can already be achieved with 10$^5$ times smaller detector such as FASER. 

\vskip 2mm
The FASER detector is located about 480 m away from IP1 and centered around the tangential extrapolation of the beam-axis at IP1. As such it can benefit from the highly collimated beam of mesons produced along the beam-axis and the decay volume can be small, 1.5 m in length and $0.2\times 0.2$ m$^2$ in transverse dimensions. The detector features 3 tracking stations interleaved with 3 magnets of 0.55 T field-strength to bend the charged particle tracks, and followed by a calorimeter for proper energy measurements. As of now all detector components are assembled, fully commissioned and final installation has started with full completion expected before LHC Run-3.

\subsubsection{Vector Mediator Production Modes} 
For the framework mentioned in Sec.~\ref{sec:intro}, there are multiple relevant production modes at the LHC providing sensitivity with the proposed detectors. In its minimal version with parameter space $(m_X,\epsilon)$ and $m_X<m_\eta ,m_{\pi^0}$, the vector mediator is produced via virtual photon conversion from meson decays, $\pi^0\rightarrow \gamma+X$ or $\eta\rightarrow \gamma+X$. Mesons with momenta above 1 TeV are copiously produced at high pseudorapidities at the LHC, motivating the location of detectors in the far forward region of the LHC interaction points. In this case, vector mediators with kinetic mixing parameter in the range $10^{-6}<\epsilon<10^{-4}$ can travel $\mathcal{O}(100)$m before decaying into SM particles, see e.g.~\cite{Ariga:2018uku}.

\vskip 2mm
For masses $m_\eta ,m_{\pi^0} \lesssim m_X$, vector mediators are predominantly produced from Bremsstrahlung via virtual photon exchange, $pp\rightarrow pp+X$, or hard scattering process, via e.g. quark anti-quark annihilation $q\bar{q}\rightarrow X$ or t-channel quark exchange $qg\rightarrow q+X$. Proton parton distribution functions for small $x$ and $Q^2$ exhibit significant theoretical uncertainties and are experimentally insufficiently measured, hence predictions for hard scattering production rates for large pseudorapidities at the LHC are difficult. For fixed-target experiments such as SHiP, however, vector mediator production via parton interactions can be well predicted~\cite{CERN-SHiP-NOTE-2016-004}.

\vskip 2mm
For $|\eta|\lesssim 5$, $m_X \approx \mathcal{O}(1)$ GeV and HL--LHC integrated luminosities of $\mathcal{O}(1)$ ab$^{-1}$ the expected number of vector mediators produced at the LHC via hard scattering is too low to observe a potential decay significantly displaced ($\mathcal{O}(10)$ $\mu$m) from the $pp$ interaction point.

However, in case the vector mediator mass is generated via Higgs mechanism the production through the scalar portal opens new discovery possibilities for sufficiently large mixing parameter $\kappa$ (c.f. Eq~\ref{eq:hig}). The ATLAS and CMS collaborations are expected to measure the Higgs total decay width only with up to $\mathcal{O}(1)$\% precision at the HL-LHC~\cite{Cepeda:2019klc}, and hence significant branching ratios into beyond the standard model particles remain non-excluded.  In particular, detectors positioned centrally ($|\eta| \lesssim 2$) and distant with respect to the interaction point can observe enough events to discover vector mediators produced via decays of dark Higgs bosons~\cite{Chou:2016lxi,Curtin:2014cca,Aielli:2019ivi}. The inelastic dark matter model can result in additional sensitivity to new vector mediators from $\chi_2$ decays ($\chi_2 \rightarrow \chi_1 X$, c.f. Eq.~\ref{eq:idm}) which are in turn produced in $pp$ collisions at the interaction point for example via dark photon decay~\cite{Berlin:2018jbm}.

\vskip 2mm
Secondary production modes of vector mediators at the LHC can increase the sensitivity in certain non-minimal extensions of the vector mediator model and are discussed in e.g.~\cite{Jodlowski:2019ycu}. In case of a \textit{secluded} dark Higgs model with $m_S < m_X$ and $\kappa<10^{-6}$ dark Higgs boson scatterings off nuclei via t-channel vector mediator exchange can result in additional vector mediator production. Similarly, in case of the inelastic dark matter model up-scatterings of $\chi_1$ to $\chi_2$ via vector mediator interactions with nuclei can result in additional vector mediator production from subsequent decays $\chi_2 \rightarrow \chi_1 X$. Finally, Dark bremsstrahlung from Dark Matter particles traversing dense material provide an additional source for vector mediator production.

\subsubsection{Sensitivity within a minimal vector model} 
All four detectors exploit visible decays of the vector mediator and are hence sensitive to vector mediator masses above 1 MeV (twice the electron mass). Only FASER and SHiP are expected to be sensitive to vector mediator production in the minimal Dark Photon model. For only $\mathcal{O}(10)\mu$m displacement and dark photon masses at about 5 GeV, kinetic mixing parameter values $\epsilon<10^{-5}$ would be needed, resulting in too low production rates via hard scattering for centrally located detectors. 

\begin{figure}
 \begin{center}
 \vspace{-10pt}
 \resizebox{0.45\textwidth}{!}{%
    \includegraphics[width=0.5\textwidth]{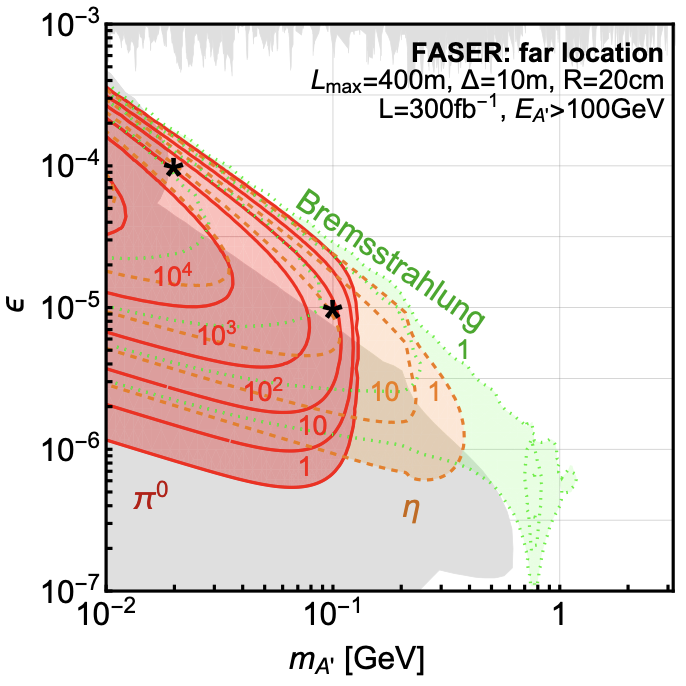}}
  \caption{
  Sensitivity for a FASER-like detector subdivided into production modes via $\pi^0 /\eta$-mesons decays and Bremsstrahlung~\cite{Feng:2017uoz}. The grey area indicates the parameter space already excluded by previous searches.}
\label{fig:dpsensitivity}
 \end{center}
\end{figure}

\vskip 2mm
With copious production of $\pi^0 /\eta$-mesons in direction of the SHiP and FASER decay volumes, these detectors will be able to probe so far non-excluded parameter space. The sensitivity is shown in Fig.~\ref{fig:DP_visible} assuming 150 fb$^{-1}$ of $pp$ collisions for the FASER and $10^{20}$ protons on target for the SHiP experiments. The expected sensitivity and yield for the FASER experiment subdivided into Dark Photon production mode is shown in Fig.~\ref{fig:dpsensitivity}. The leading contribution in the production cross section times branching ratio at $\epsilon$ above 10$^{-4}$ is roughly $\sigma A \propto e^{-m^2\epsilon^2}$, resulting in exponential increase for decreasing kinetic mixing parameter due to the acceptance of the FASER experiment~\cite{Feng:2017uoz}. The production cross section times branching ratio then decreases according to $\sigma A \propto m^2\epsilon^4$ for $\epsilon$ below roughly 10$^{-4}$, which leads to the distinct shape in parameter space coverage. 

\subsubsection{Sensitivity within non-minimal vector models}
\begin{figure*}
 \begin{center}
 \vspace{-10pt}
    \raisebox{20pt}{\includegraphics[width=0.49\textwidth]{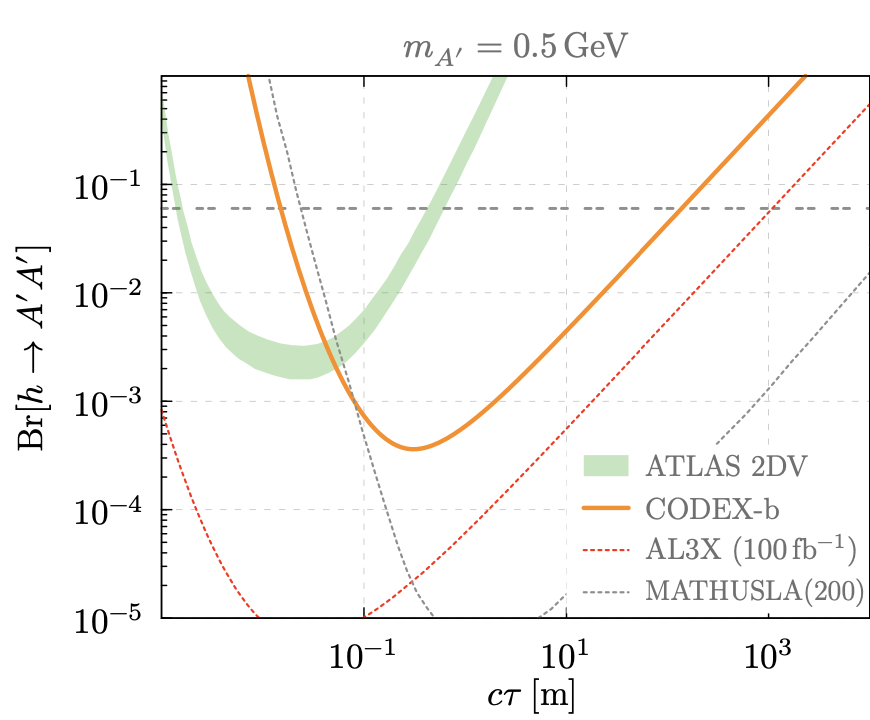}}
    \includegraphics[width=0.5\textwidth]{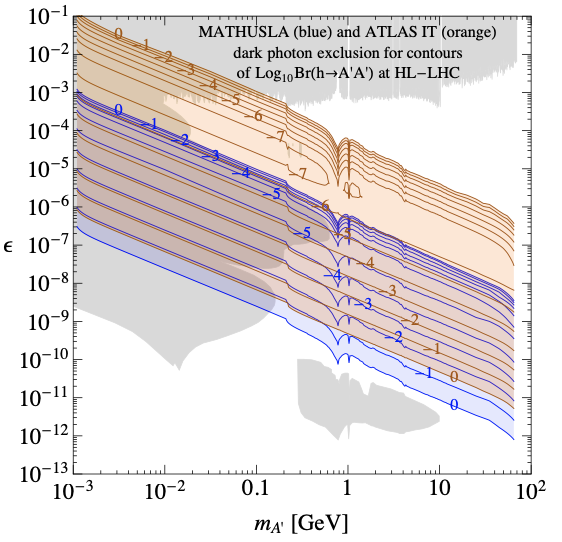}
  \caption{Left: Sensitivity for CODEX-b and MATHUSLA for vector mediators from $\mathrm{BR}(H\rightarrow XX)$ decays as a function of decay length ~\cite{Aielli:2019ivi}. Also shown are the sensitivities of the AL3X and ATLAS experiments. Right: 
  Sensitivity of the MATHUSLA experiment for vector mediators from $\mathrm{BR}(H\rightarrow XX)$ in the $(m_X,\epsilon)$ parameter plane. The contours and numbers indicate the sensitivities for different exponents for $\mathrm{BR}(H\rightarrow XX)$. The grey area indicates the parameter space already excluded by previous searches.~\cite{Curtin:2018mvb}. Also shown is the expected sensitivity for searches by the ATLAS Collaboration.
  }
   \label{fig:dhiggs}
 \end{center}
\end{figure*}

The sensitivity of the proposed MATHUSLA and CODEX-b experiments to vector mediators from decays of dark Higgs bosons is shown in Fig.~\ref{fig:dhiggs}. Also shown are possible sensitivities of searches performed with the ATLAS detector. Detectors such as FASER and SHiP have no sensitivity to this production mode due to the low Higgs boson production rate at high pseudorapidities and low center of mass energy, respectively. Due to the distant placement and highly shielded detector environments the MATHUSLA and CODEX-b experiments can extent the sensitivity reach up to vector mediator decay lengths of $\mathcal{O}(1)$km and $\mathcal{O}(10)$m at $\mathrm{BR}(H\rightarrow XX)=0.06$, respectively, where the decay of $X$ to SM particles is set to be 100\%. Likewise, on the right in Fig.~\ref{fig:dhiggs} the lifetime is converted to values of kinetic mixing parameter $\epsilon$ for different Higgs branching fractions in case of the MATHUSLA experiment. In this model, MATHUSLA would be able to probe kinetic mixing parameter values as low as about $\epsilon=10^{-10}$ at $m_X=10$ GeV and $\mathrm{BR}(H\rightarrow XX)=0.10$. Different contours and the numbers indicate the coverage for different exponents for $\mathrm{BR}(H\rightarrow XX)$, relating the parameter space to different scenarios of measurements of the Higgs total decay width.

\begin{figure*}
 \begin{center}
 \vspace{-20pt}
    \includegraphics[width=0.48\textwidth, height=5.5cm]{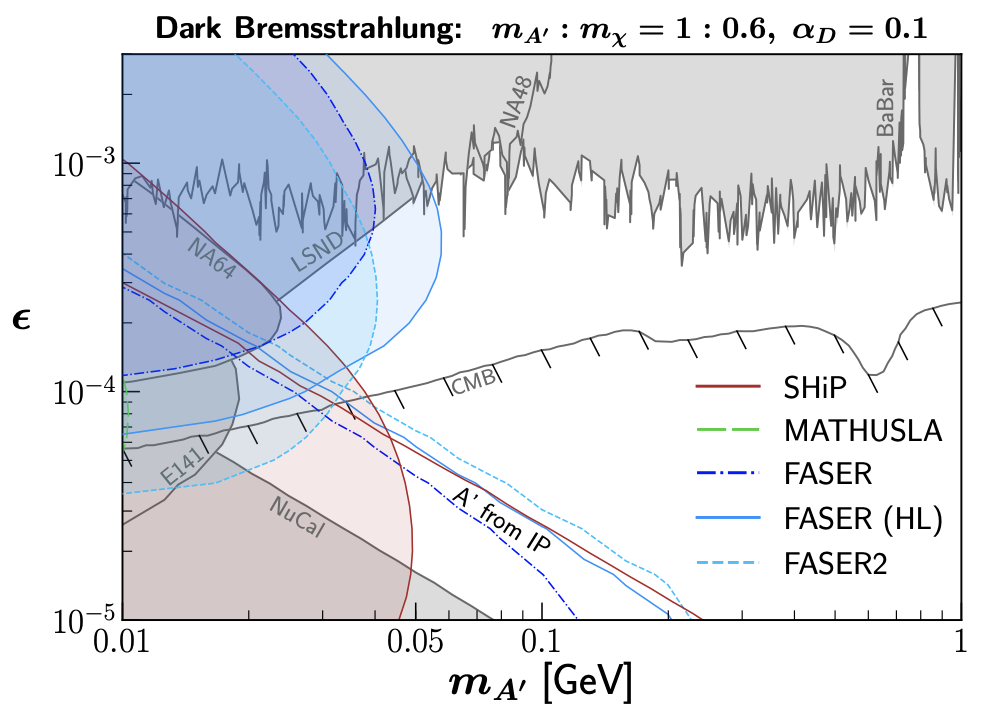}
    \includegraphics[width=0.48\textwidth,height=5.5cm]{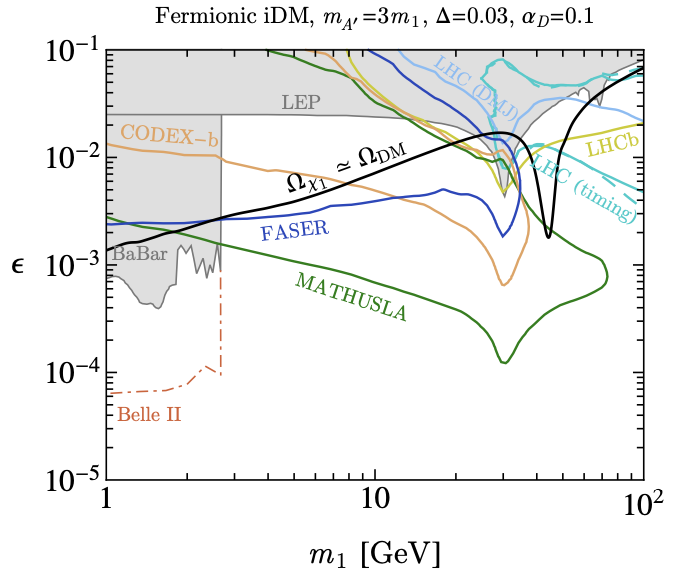}
  \caption{Left: Sensitivity regions in the $(m_X,\epsilon)$-plane of FASER, MATHUSLA and SHiP (filled areas with different colours) for vector-mediator detection via Dark Bremsstrahlung production mode. Also shown is the constraint from Cosmic Microwave Background measurements~\cite{Jodlowski:2019ycu}. Right: Sensitivity lines in particular for CODEX-b, FASER and MATHUSLA in the $(m_{\chi_1},\epsilon)$ plane for vector mediators from $\chi_2\rightarrow \chi_1 X$ decays in the inelastic dark matter model. Also shown is the line indicating the correct relic abundance for the parameters assumed~\cite{Berlin:2018jbm}.}
   \label{fig:nonmin}
 \end{center}
\end{figure*}

\vskip 2mm
In principle, all of the four experiments have additional sensitivity in case further assumptions about the Dark Sector are made, as can be seen in Fig.~\ref{fig:nonmin}. A light dark matter particle ($m_\chi=1$ GeV) produced at the IP can radiate off the vector mediator away from the IP according to the dark coupling constant $\alpha_D$. This way, in particular with FASER and SHiP, larger kinetic mixing parameters can be probed for small $m_X$ when compared to the production mode via meson decays dominant in the minimal model. In this case, FASER can probe yet non-excluded phase space for $m_X<0.05$ GeV and $10^{-4}<\epsilon<10^{-3}$. In the inelastic dark matter model in particular large vector mediator masses can be probed also with a forward detector such as FASER. Sensitivity lines are shown in Fig~\ref{fig:nonmin} right, with $\frac{m_X}{m_{\chi_1}}=3$, $\alpha_D=0.1$ and $m_{\chi_2}-m_{\chi_1}=0.03$. Assuming the main production mode $X\rightarrow \chi_1\chi_2$ the FASER detector reaches good sensitivity in the $(m_X,\epsilon)$ plane for large kinematic mixing parameter, due the acceptance favouring $\chi_2$ production with large boost. MATHUSLA and CODEX-b have qualitatively similar phase space coverage. However, MATHUSLA has better reach due to its increased acceptance. For these centrally located detectors, the on average smaller transverse boost leads to sensitivity to mostly smaller kinetic mixing parameter, or likewise, small mass splitting $\Delta$. 

\clearpage
\subsection{Searches for light DM and vector mediators at NA62, NA64, MESA, PADME}
\label{ssec:gninenko}
{\it Author: Sergei Gninenko, <sergei.gninenko@cern.ch>} 
\newcommand{\ainv}{\ensuremath{A'}\to\mathrm{invisible}}
\newcommand{\aee}{\ensuremath{A'}\to\mathrm{e}^+\mathrm{e}^-}

\subsubsection{Introduction}
\label{sec:intro}
Despite the intensive experimental searches  dark matter (DM) still is a great
puzzle. The difficulty so far is that DM can be probed only through its gravitational interaction with visible matter (VM).
An  exciting possibility is that in addition  to gravity, a new interaction  between DM and VM  through a vector portal might exists.
The new force could be transmitted  by a new gauge  boson,  $A'$ , called dark photon. 
The $A'$  couples to  the standard model (SM)  via kinetic mixing  with the SM photon,  described by the  term  $\frac{\epsilon}{2}F'_{\mu\nu}F^{\mu\nu}$ and  parameterized by the mixing strength  $\epsilon$.  The $A'$ could have a mass, e.g.  $m_{A'}\lesssim 1$ GeV,  associated with the  spontaneously broken $U_D(1)$ gauge group, and a coupling  $e_D$  of the $U(1)_D$ interactions to the Dark scalars or fermions.
   The mixing term   results in the interaction $\mathcal{L}_{int}= \epsilon e A'_{\mu} J^\mu_{em}$ of $A'$s with the electromagnetic current $J^\mu_{em}$ with a mixing strength $\epsilon e$,
where $e$ is the electromagnetic coupling and $\epsilon \ll 1$. 

\vskip 2mm
\par Since there are no firm predictions for  the $A'$, its experimental searches  have been performed   over a wide range of $A'$ masses and decay modes.
If the $A'$ is the lightest state in the dark sector,  then it would decay mainly visibly, i.e., typically to SM leptons 
or hadrons which can be used to detect it. 
 Such dark photons in the  mass region below a few $\rm GeV$ has been mainly  searched for in  beam dump, fixed target, collider and rare meson decay  experiments, which already put stringent  limits  on  the  mixing strength $\epsilon^2 \lesssim 10^{-7}$. However, in the presence of light dark states $\chi$,  in particular, DM with the masses $m_\chi<m_{A'}/2$, the $A'$  would predominantly  decay invisibly into those particles, i.e. $\Gamma (A'\rightarrow \bar{\chi}\chi)/\Gamma_{tot} \simeq 1$,    provided that  coupling $g_D >\epsilon e$.  Various dark sector models motivate  sub-GeV scalar and Majorana or pseudo-Dirac fermion DM coupled to dark photons 
\cite{Battaglieri:2017aum,Beacham:2019nyx,Berlin:2018bsc}. 

\vskip 2mm
To interpret the observed  abundance of thermal relic density,  the requirement of the  thermal freeze-out of DM annihilation into visible matter through $\gamma-A'$ kinetic mixing allows one to derive a relation among the parameters  
\begin{equation}
\alpha_D \simeq 0.02 f \Bigl( \frac{10^{-3}}{\epsilon}\Bigr)^2\Bigl(\frac{m_{A'}}{100~ MeV}\Bigr)^4
\Bigl( \frac{10~MeV}{m_\chi}\Bigr)^2
\label{alphad}
\end{equation}
where $\alpha_D = e_D^2/4\pi$, $f \lesssim 10$ for a 
scalar \cite{deNiverville:2011it}, and $f\lesssim 1$ for a  fermion 
\cite{Izaguirre:2014bca}. This 
 prediction combined with the fact  that  the intrinsic scale of the dark sector could be  smaller than, or comparable to, that of the visible sector, provide  an important target for the ($\epsilon, ~m_{A'}$) parameter space that can be probed at energies currently attainable at the current CERN SPS and other facilities. 
This has motivated a worldwide experimental and theoretical effort  towards  dark forces and other portals between the visible and dark sectors, see Refs. \cite{Essig:2013lka,Alexander:2016aln,Battaglieri:2017aum,Zyla:2020zbs} for a review.

\vskip 2mm
  If such  $A'$ exists, many crucial questions about its mass scale, coupling constants, decay modes, etc. arise. One possible way to answer these questions, is  to search  for the invisible $A'$ decays in high-intensity accelerator experiments. The $A'$s  could be 
produced in a beam dump  and generate a flux of DM particles  through their decays,  which can be detected through the  scattering off electrons in a detector target.  In this case, the signal event rate in the detector  scales as $\epsilon^2 y \propto \epsilon^4 \alpha_D$, where  the parameter $y$ is  defined as 
$y = \epsilon^2 \alpha_D \Bigl(\frac{m_\chi}{m_{A'}}\Bigr)^4$. If the $A'$ is the lightest state in the dark sector,  then it would decay mainly visibly  to SM leptons or hadrons. 

\vskip 2mm
Another approach, proposed in Refs.~\cite{Gninenko:2013rka,Andreas:2013lya}, 
 is based on the detection of the large missing energy, carried away by the energetic $A'$
 produced in the interactions of high-energy electrons in the active beam dump target. The advantage of this technique is that the sensitivity is proportional to the mixing strength squared, $\epsilon^2$,  associated with the $A'$ production  in the primary reaction and its subsequent prompt invisible decay, while in the former case it is proportional to $\epsilon^4\alpha_D$, with $\epsilon^2$ associated with  the $A'$ production in the beam dump and  $\epsilon^2 \alpha_D$ coming from the $\chi$ particle interactions in the detector.  The accessibility of $\epsilon, \alpha_D, y$
  parameter space that could explain the observed abundance of the thermal relic DM density at accelerator experiments,  has motivated a worldwide effort towards  dark forces and other portals between the visible and dark sectors; see Refs. \cite{Essig:2013lka,Alexander:2016aln,Battaglieri:2017aum,Beacham:2019nyx,Berlin:2018bsc,Lanfranchi:2020crw} for a review. Below  we report new results and prospects on the search for  the $A'$ and light DM in the fixed-target  experiment NA64, NA62, MESA and PADME.

\subsubsection{NA64}
NA64 is  specially designed as a hermetic general purpose detector  to search for dark sector physics in missing energy events from  high-energy electron and muon scattering off nuclei in an active dump.  
The main aim of the NA64 experiment (North Area experiment 64) at the CERN SPS  is the detection of Light Dark Matter (LDM) motivated by freeze-out  DM models. 
The method of the search can be illustrated by considering, as an example,  the search for the vector mediator, e.g. the dark photon $A'$,  of Dark Matter ($\chi$)  production in invisible decay mode, $ A' \to \chi \overline{\chi}$ \cite{Gninenko:2013rka,Andreas:2013lya}. If the $A'$ exists it could be produced via the kinetic mixing with bremsstrahlung photons  in the reaction  of high-energy electrons absorbed in  an active beam dump (target)  and exhibited itself as an event with a large missing energy due to its prompt invisible decay into DM particles  in  a hermetic detector \cite{Gninenko:2016kpg,Gninenko:2017yus}.

\vskip 2mm
NA64 exploits the high luminosity achievable at the extracted 100~GeV $e^-$ beam serving the experiment 
to look with high sensitivity to manifestations of the $A'$. The long period (4.8 s long beam spills) over which the beam particles arrive  provides a relatively clean experimental environment with acceptable rate of events overlapping in time (pile-up). This allows the clean detection of events with large missing energy on an event-by-event basis. For the search of the $\ainv$ decay mode, the NA64 experiment looks for events where a 100 GeV electron with intensity up to $\sim 10^7~e^-$ per SPS spill  deposits less than about 50 GeV in a hermetic electromagnetic (em) calorimeter served as an active dump. The missing energy would be carried away by the $A'$, and the occurrence of $\ainv$ decays  would appear as an excess of events with  isolated single em showers  accompanied by a large missing energy above those expected from  standard $e^-$ interactions.  
NA64 aims  for a single event sensitivity of better than $10^{-12}$ per incoming electron (EOT). Such sensitivity requires an extremely precise understanding of the incoming beam, where contamination by other particles, such as pions, kaons, muons, and antiprotons, must be controlled at the $10^{-6}$ level, whereas the typical hadron contamination present in the beam can be at the $\simeq 1\%$ level. This is achieved by analyzing the synchrotron radiation emitted by the electrons in the beam, combined with precise tracking.

\vskip 2mm
With small modifications to the setup, the NA64 experiment will also search for the so-called visible decay mode $A' \to e^+ e^-$ . As a first step, NA64 started a search for a recently reported signal 
at 17~MeV mass in a $^{8}$Be anomaly (see discussion below). For a second phase, searches with other beam particles, such as pions or muons, are considered.
The experimental signature of events from the $A' \to e^+ e^- $ decays is clean and they can be selected with small background due to the excellent capability of NA64 for the  precise identification and measurements of the 
initial electron state.

\vskip 2mm
\par
\begin{figure*}[tbh!]
\centering
\includegraphics[width=0.48\textwidth]{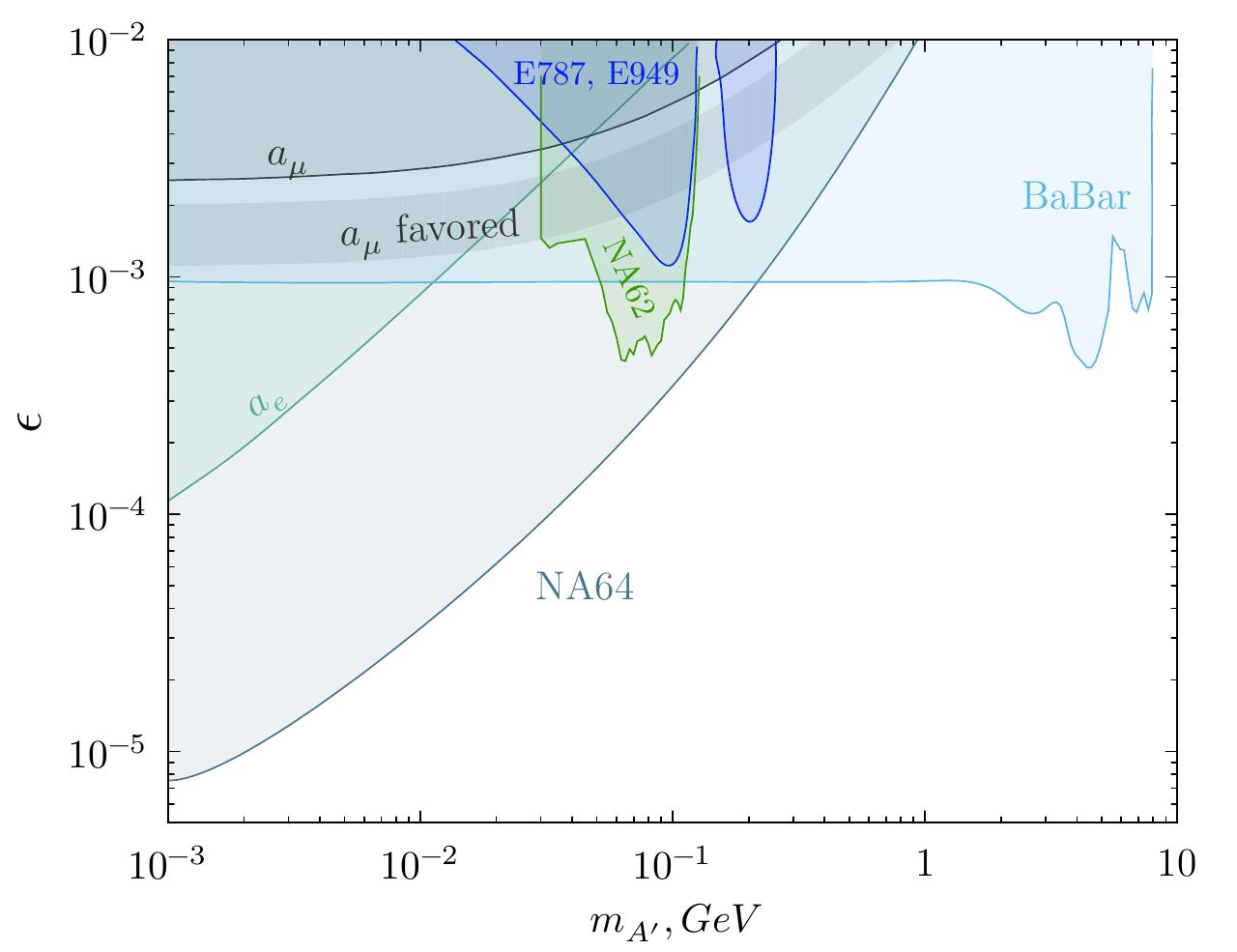}
\includegraphics[width=0.48\textwidth]{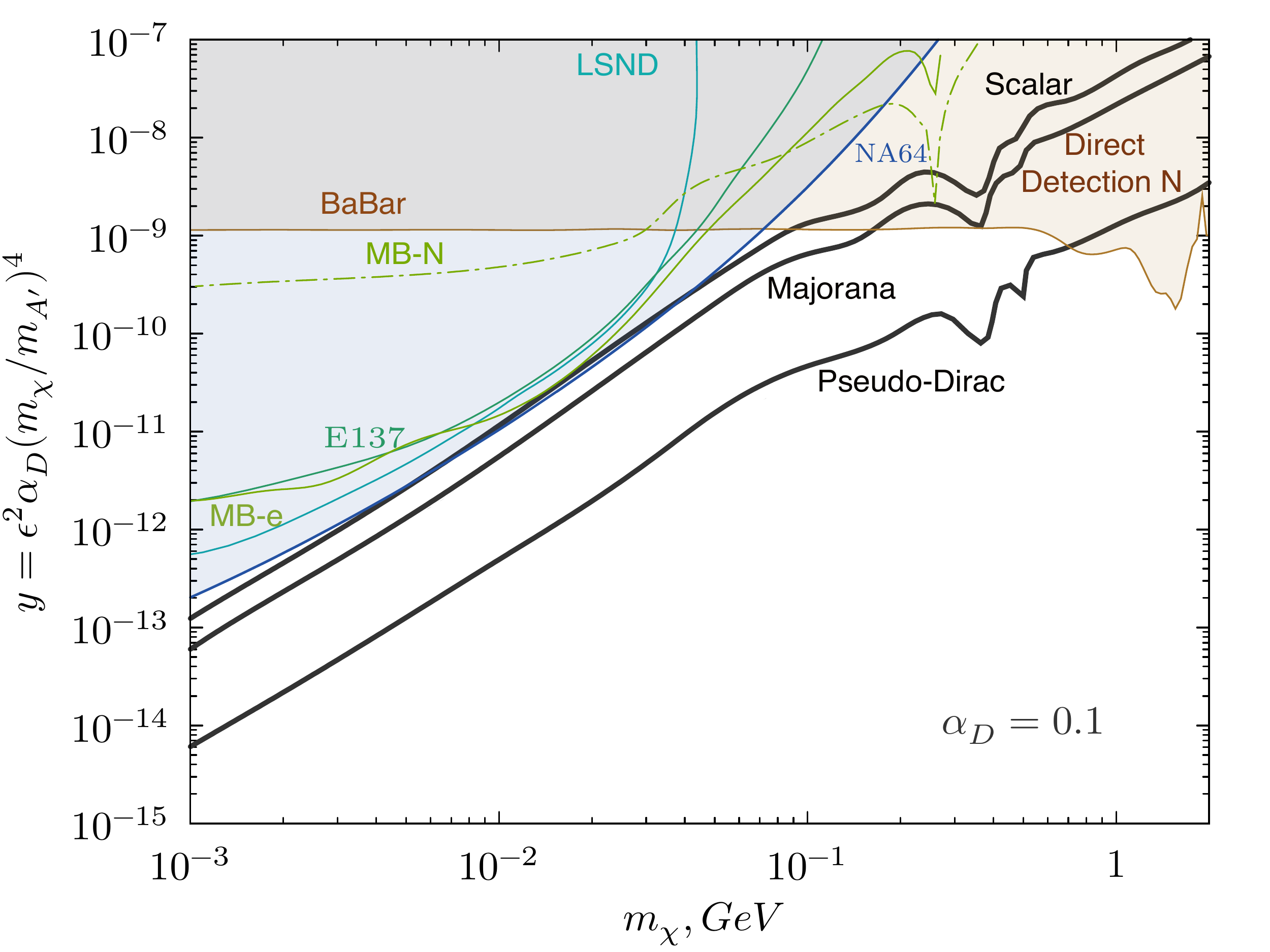}
\includegraphics[height=0.38\textwidth] {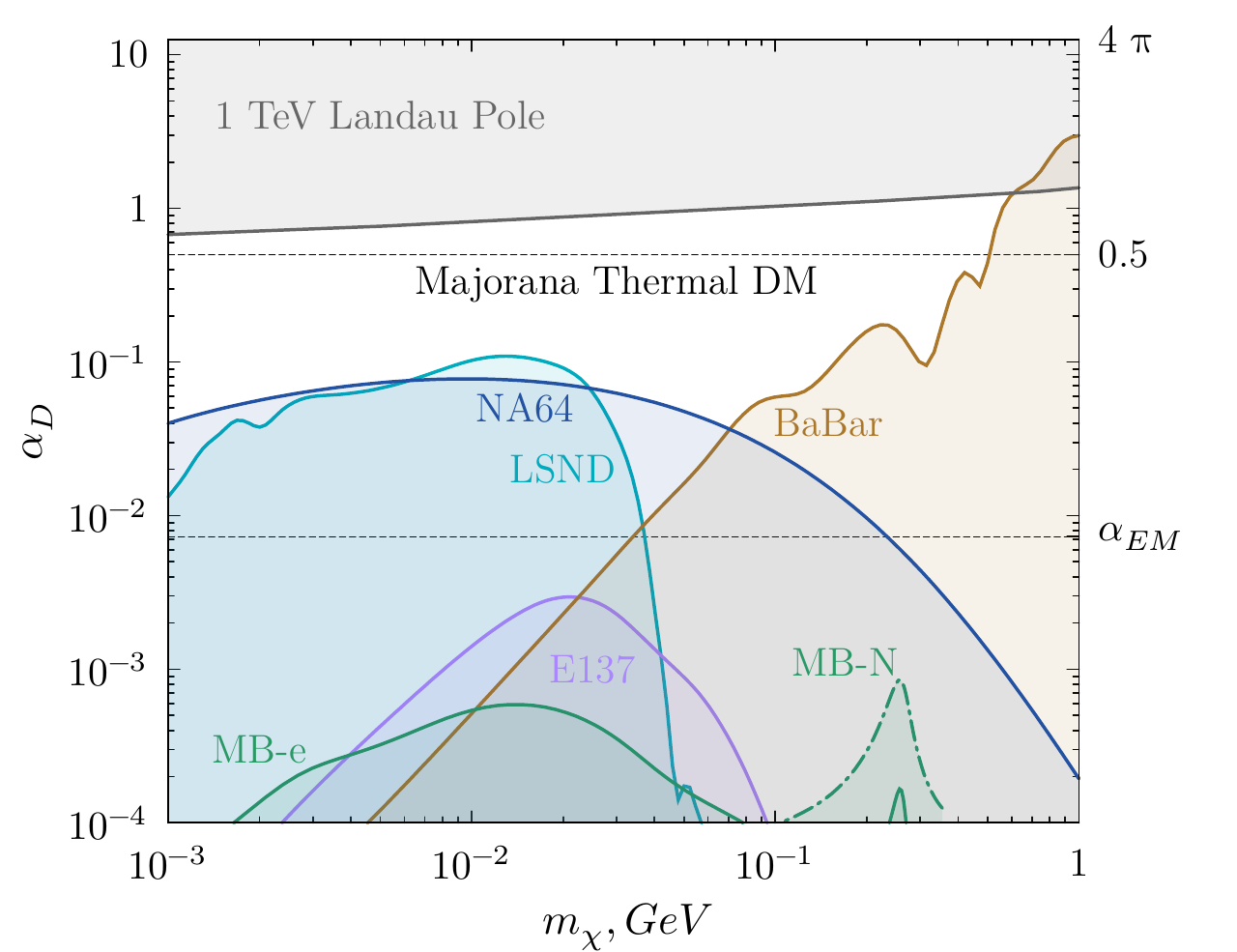}
\includegraphics[width=0.48\textwidth]{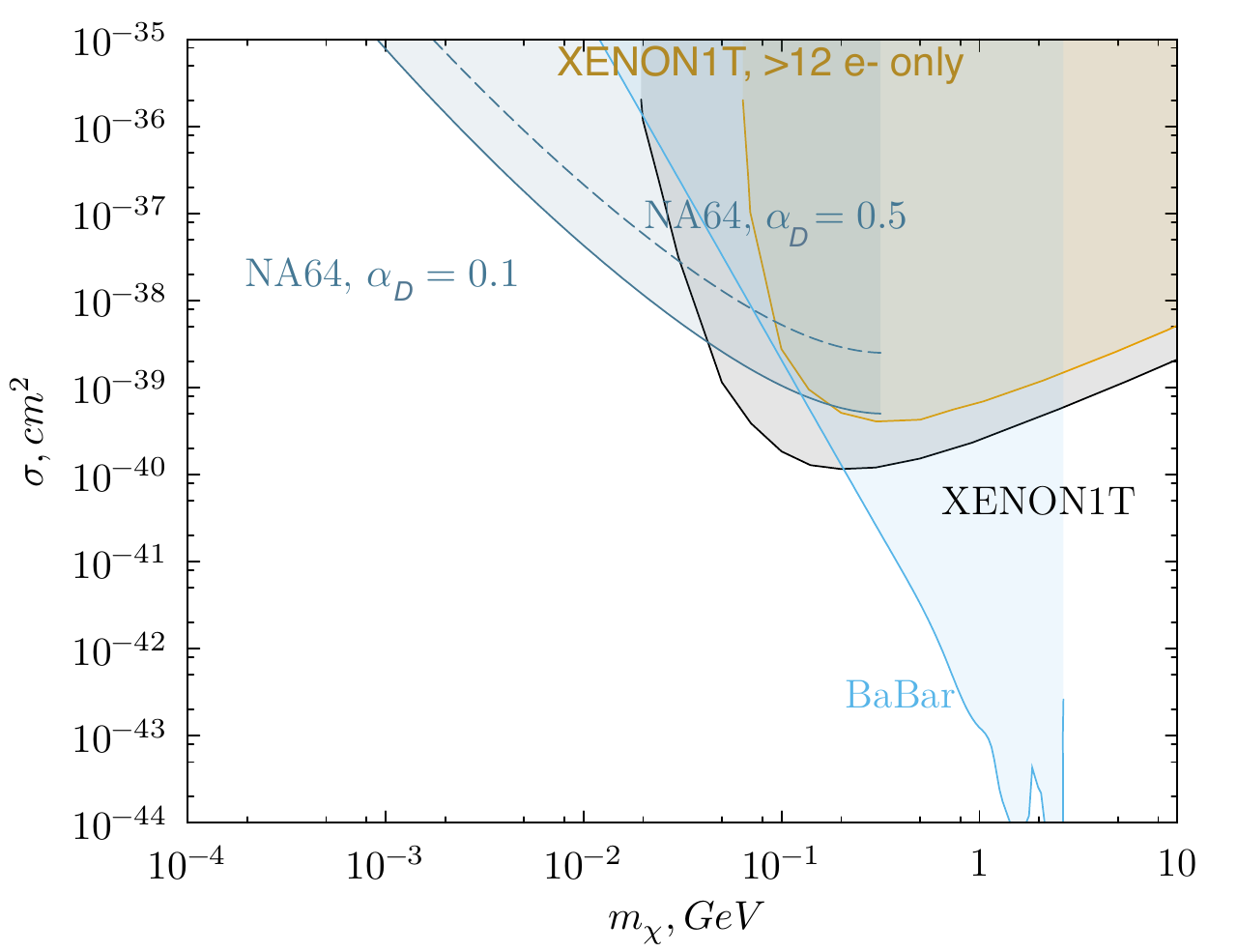}
\caption{The {\it top left panel} shows the NA64 90\% C.L. exclusion regions  in the ($m_{A'}, \epsilon$) 
plane.   Constraints from the   E787 and  E949  \cite{Davoudiasl:2014kua,Essig:2013vha}, $BABAR$ \cite{Lees:2017lec} and recent NA62 \cite{CortinaGil:2019nuo} experiments,  as well as the muon  $\alpha_\mu$ favored area are also shown. The {\it top right panel} presents limits  in the (y;$m_{\chi}$) plane obtained for $\alpha_D=0.1$ and $m_{A'} = 3 m_\chi$. The  favored parameters to account for the observed relic DM density for the  scalar, pseudo-Dirac  and Majorana  type of light DM are shown as the lowest solid line; see, e.g. \cite{Berlin:2018bsc}.
The {\it bottom left panel} shows  an example of the NA64 constraints in the ($\alpha_D$;$m_{\chi}$) plane  on Majorana DM.  The limits are shown in  
 in comparison with bounds obtained in Refs.\cite{Alexander:2016aln,Battaglieri:2017aum,Izaguirre:2014bca}  from  the results of the $BABAR$ \cite{Lees:2017lec}, 
LSND~\cite{deNiverville:2011it,Batell:2009di},   E137 \cite{Batell:2014mga}, and MiniBooNE \cite{Aguilar-Arevalo:2017mqx} experiments. 
 The {\it bottom right panel} shows the comparison of 90 \% C.L. upper limits on LDM-electron scattering cross-sections calculated in Ref. \cite{Gninenko:2020hbd} by using NA64  \cite{NA64:2019imj} and 
BaBar  \cite{Lees:2017lec} constraints on kinetic-mixing with results from LDM direct searches by XENON1T \cite{Aprile:2019xxb}. The  blue curves are calculated for $\alpha_D=0.1$, while the dashed blue 
for $\alpha_D=0.5$. The yellow curve shows  XENON1T limits obtained without considering signals with $<$ 12 produced electrons \cite{Aprile:2019xxb}.}
\label{fig:excl}
\end{figure*}

  In the 2016-18 runs, the NA64 experiment  has accumulated about $\sim 3\times10^{11}$ EOT  and successfully performed  sensitive  searches  for LDM 
production mediated though the vector portal and other rare processes  in  missing energy events. 
The combined 90\% C.L. exclusion limits on the mixing  $\epsilon$  as a function of the $A'$ mass,  calculated by taken into account the expected backgrounds and estimated systematic errors, can be seen  in Fig.~\ref{fig:excl} (top left panel).
The limits are obtained with a full NA64 detector simulation \cite{Gninenko:2016kpg} and by using the $A'$ production cross section obtained with the exact tree-level calculations \cite{Gninenko:2017yus}.  
The derived bounds are currently  the best for the mass range $0.001\lesssim m_{A'} \lesssim 0.2 $ GeV  \cite{Zyla:2020zbs}.

\vskip 1mm
\par Using  constraints on the cross section of  the DM annihilation freeze-out and obtained limits on mixing 
strength, one  can derive constraints on the LDM models, which are shown  in the ($y$;$m_{\chi}$) and ($\alpha_D$;$m_{\chi}$)  planes in Fig.~\ref{fig:excl}, top right and bottom left panels, respectively,  for masses  $m_{\chi} \lesssim 1$~GeV. 
  On the ($y$;$m_{\chi}$) plot one can also see   the favored $y$ parameter curves  for scalar,  pseudo-Dirac (with a small splitting)  and Majorana  scenario  of  LDM  obtained by  taking into account  the observed relic DM density \cite{Berlin:2018bsc}. The limits on the variable $y$  are calculated 
   under the convention $\alpha_D= 0.1$ and  $m_{A'}=3 m_{\chi}$ \cite{Battaglieri:2017aum,Beacham:2019nyx} and shown also for comparison with bounds  from other experiments. 
    One can see that using the NA64 approach allows us to obtain  more stringent bounds on $\epsilon,~y, ~ \alpha_D$ 
   for the mass range $m_{\chi} \lesssim 0.1$ GeV than the limits obtained from the results of  classical beam dump experiments, thus,  demonstrating its power for the dark matter search. 
 Further improving of the sensitivity  is  expected after  the   NA64 detector upgrade.

\begin{figure}
\resizebox{0.45\textwidth}{!}{
    \includegraphics[width=1.0\textwidth]{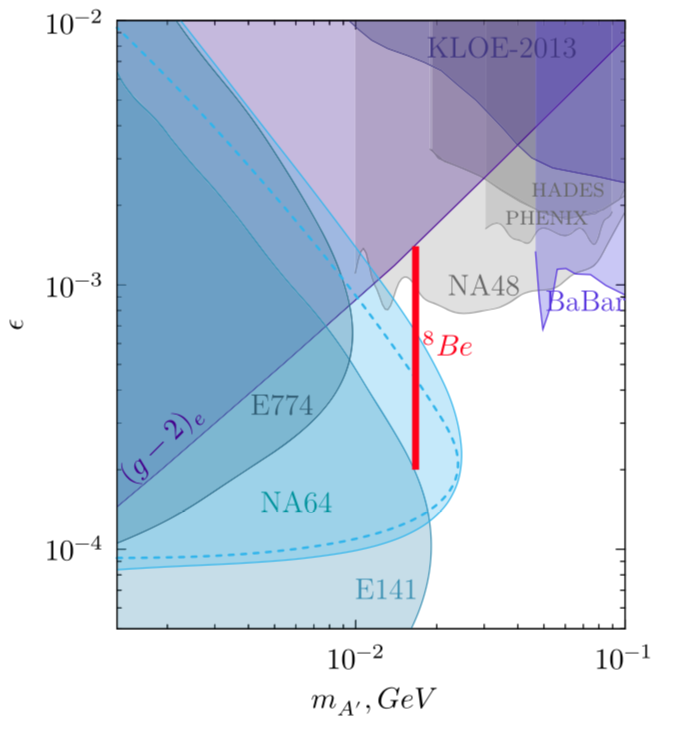}}
\caption{The  90\% C.L.\ exclusion areas  in the ($m_{X}; \epsilon$) plane from 
the NA64 experiment (shaded blue area) using 2017 data only (dashed line) and 2017-2018 data combined. For the mass of 16.7 MeV, the
$X-e^-$ coupling region excluded by NA64  is  $1.2\times 10^{-4}< \epsilon_e < 6.8~\times~10^{-4}$.
The NA48 limits only apply to dark photons, but not the X boson because differently from the dark photon it has nonuniversal couplings to u, d quarks 
 allowing to explain the anomaly \cite{Feng:2016jff}.
 The full allowed range of $\epsilon_e$ for the X17 boson, $2\times 10^{-4}< \epsilon_e < 1.3~\times~10^{-3}$  is shown as a vertical red bar. The constraints on the mixing strength from the experiments E774 and E141\cite{Andreas:2012mt,Bross:1989mp,Riordan:1987aw}, BaBar \cite{Lees:2014xha}, KLOE \cite{Anastasi:2016ktq}, HADES \cite{Agakishiev:2013fwl}, PHENIX \cite{Adare:2014mgk}, NA48 \cite{Batley:2015lha} and bounds from the electron anomalous magnetic moment
(g-2)$_2$  see e.g. \cite{Krasnikov:2019dgh}, are also shown.  }\label{fig:x17}
\end{figure}

\vskip 2mm
\par Recently, the XENON1T collaboration has published new results on the search for direct electron LDM scattering~\cite{Aprile:2019xxb}. 
New bounds on elastic electron-LDM cross sections  were obtained for masses $m_{\chi} \geq 30~$MeV. For the 
model with dark photon mediator the use  of XENON-1T results  allows the Collaboration to  derive bound on $\epsilon^2\alpha_D$  and cross-sections of LDM scattering 
on electrons~\cite{Aprile:2019xxb}. 
Using bounds on the $\gamma -A'$ mixing strength obtained by NA64~\cite{NA64:2019imj} and BaBar~\cite{Lees:2017lec}, 
allows us also  to extract 90 \% C.L. upper limits on cross-sections of LDM-electron scattering  transmitted by the dark-photon mediator $A'$ \cite{Gninenko:2020hbd}.  
In Fig.\ref{fig:excl} (bottom right panel) the comparison of 90 \% C.L. upper limits on cross-sections of LDM-electron scattering calculated by using NA64 and BaBar bounds
and $\alpha_D = 0.1$, and  the  XENON1T bounds~\cite{Aprile:2019xxb} is shown. 
For $m_{A'} \leq 50$~MeV NA64 bounds  are stronger than those from XENON-1T. Note that 
for pseudo-Dirac fermions with not too small mass splitting $\delta = \frac{m_{\chi_2} - m_{\chi_1}}{m_{\chi_1}}$ the reaction of $\chi_2$ electro-production 
$\chi_1 ~e \rightarrow \chi_2~e$ for nonrelativistic LDM $\chi_1$  is prohibited due to energy 
conservation law, while elastic $\chi_1 ~e \rightarrow \chi_1~e$ scattering is absent at tree level. This fact 
extremely complicates the direct LDM detection for pseudo-Dirac fermions.

\vskip 2mm
The final results of the LDM analysis using the 2016-2018 dataset presented the best exploration of  the parameter space for the sub-GeV DM  as well as prospects for the searching beyond LS2. NA64 also has sensitivity to various other rare processes related to the observed experimental anomalies , as well as some hidden sector models. Below, new results of a search for a new X17 boson from the 
the $^8$Be- $^4$He (ATOMKI) anomaly are  presented. 
The ATOMKI experiment of Krasznahorkay et al. \cite{Krasznahorkay:2015iga} has reported the observation of a 6.8 $\sigma$ excess of events
in the invariant mass distributions of $e^+ e^-$ pairs produced in the nuclear transitions of excited  $^8Be^*$ to its ground state via
internal pair creation. This observation has been recently confirmed by the same group from the measurements with $^4$He nuclei \cite{Krasznahorkay:2019lyl}. 
The anomaly can be interpreted as the emission of a new $X17$ boson with a mass of $\sim$17 MeV followed by its $e^+ e^-$ decay.
The X17 could be a scalar, pseudo-scalar, vector,  or an axial-vector particle. Assuming that the $X$ is a vector with non-universal coupling to quarks, coupling to electrons is in the range $2\times 10^{-4} \lesssim \epsilon_e \lesssim 1.4\times 10^{-3}$ and the lifetime 
$10^{-14}\lesssim \tau_X \lesssim 10^{-12}$~s \cite{Feng:2016jff}. 
 The NA64 90\% C.L. exclusion limits on the mixing  $\epsilon$ as a function of the $A'$ mass obtained from the 2017-2018 search for the $A'(X17)\to ee$ decays 
  by using the $A'(X17)$ production  and subsequent decay in the reaction $e^- + Z \to e^- + Z + A'(X17)   ;~ A'(X17)\to e^+ e^-$  are shown in Fig.~\ref{fig:x17} together 
with the current constraints from other experiments. The NA64 results exclude the X17 as an explanation for the ATOMKI anomaly
for the $X17-e^-$ vector coupling  $ \epsilon_e  \lesssim 6.8 \times 10^{-4}$ and mass value of 16.7 MeV \cite{Banerjee:2019hmi}, leaving the still unexplored  region
$6.8~\times  10^{-4} \lesssim  \epsilon_e  \lesssim 1.4 \times 10^{-3}$ as  quite an exciting prospect for future searches \cite{Depero:2020zfy}. 

New preliminary limits on the existence of a new light $Z'$
able to explain the XENON1T and KOTO anomalies and new results on searching for ALP and scalar particles \cite{Banerjee:2020fue} were also presented at the workshop but not reported in these proceedings.

\vskip 2mm
\par Although NA64 has already produced significant results on FIPs physics, it has just begun to exploit physics potential of the proposed experimental techniques. The experiment is expected to continue in 2021 aiming to accumulate a few $10^{12}$ EOT  before the CERN Long Shutdown Period 3 (LS3)  and to improve the sensitivity reaches. 

\subsubsection{NA62}
  The primary goal of the NA62 experiment at the CERN SPS is to measure the "golden" rare decay modes is $K^+ \to \pi^+ \nu \overline{\nu} $, which has a SM branching ratio of 
$(8.4 \pm1.0)\times  10^{-11}$,  with  a 10\% accuracy, based on 100 events with very small backgrounds. 
The NA62 detector took data  in 2016-2018 runs resulting in observation of a few signal events. 
After the LS2 the detector must be able to collect many kaon decays and thus have very high rate capability as well as excellent timing resolution to disentangle overlapping events that are very close in time. 
Backgrounds from other decays are rejected by a combination of severe kinematic cuts, combined with particle identification of the kaons in the beam (about 6\% of the beam) and of the outgoing pion.
 A very hermetic and efficient veto system 
 completes the setup.
 
 \vskip 2mm
\par The high rate capability and excellent resolution and particle identification allow, as a by-product, the measurement of or search for many other rare decays, including the
dark photons, axion-like particles, and heavy neutral lepton (HNL) decays into SM lepton or hadron pairs. 
The NA62 limits on the mixing $\epsilon$ derived from a search for $\ainv$ decays by using the decay chain $K^+ \to \pi^+ + \pi^0; \pi^0 \to g + A'; A' \to {\rm invisible}$  
as a source of $A'$s,  are shown in  top left panel  of Fig. \ref{fig:excl}. The NA64 constraints currently surpass the NA62  ones,  however one should take into account 
that the latter are obtained just with 1\% of full available statistics. Hence, further improvement by a factor about $\sim$10 can be expected. Another important feature making 
the further NA62 search interesting is that the experiment is also sensitive to the $A'$ predominantly coupled to quarks, thus making the search  complementary 
to the NA64 one.

\vskip 2mm
\par A special  search for dark sector particles, by operating the detector into beam dump mode, is also under consideration. The idea 
is to move the  Be target used for the $K^+$ production away from the beam and to allow the 400-GeV protons  impinge directly on the two 1.6 m long, beam defining, Fe/Cu collimators (about 21 nuclear interaction lengths when closed) located about 22~m downstream of the primary Be target. The long ($\sim 100$~m) decay volume  along with a good particle ID system and a tracker with excellent tracking performance make NA62 an attractive and cost-effective 
 setup to probe new physics in the beam dump mode. A high-sensitivity  search of  a wide spectrum of  feebly-interacting long-lived particles, including dark photons, scalars, ALPs, HNLs, etc. just with 
 about a few months of running, is expected.   The current sensitivity to such particles will be significantly improved.
 Projected exclusion limits for visible dark photon decays are shown in Fig.~\ref{fig:DP_visible} of Section~\ref{ssec:vector-results}.

The current status of the experiment: 
NA62 plan to resume data taking in 2021. The promising program of searching FIPs in beam dump mode, including $A'$, is expected to start before LS3.  NA62-dump plan to take 3 months run in this period.

\subsubsection{MESA}
 MESA - the Mainz Energy-Recovering Superconducting Accelerator - with its two experiments, MAGIX and DarkMESA,  is a facility dedicated to dark sector physics currently  under development in Mainz \cite{Christmann:2020qav}.

\begin{figure*}[tbh!]
\centering
\includegraphics[width=0.99\textwidth]{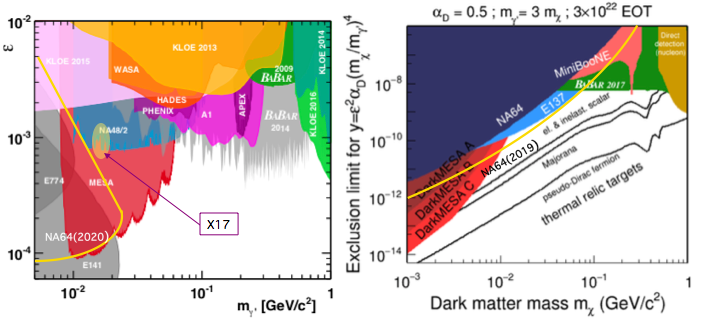}
\caption{The projected sensitivities of MAGIX (left panel) and DarkMESA (right panel)  for the searches for $A' \to e^+ e^-$ and $A' \to {\rm invisible}$ decays, respectively, in the $(\epsilon; m_{A'})$ plane (adapted from~\cite{Doria:2018sfx} and~\cite{Christmann:2020qav}).
The yellow curves corresponds to the recent NA64 results of Ref.\cite{Banerjee:2019hmi} and Ref.\cite{NA64:2019imj}, respectively.
The not yet excluded parameter space for the vector X17 boson from the ATOMKI anomaly is indicated by the arrow.}\label{fig:mesa}
\end{figure*}   

The  MAGIX (MESA Gas Internal target eXperiment)  experiment will exploit the  high-current  electron-beam on internal targets at low energies ($\lesssim$ 105 MeV, 1~mA). With its magnetic spectrometer system and a novel windowless target, MAGIX will contribute to the experimental searches for dark photons, by exploring the  reaction $eZ\to eZA'$ for the $A'$ production with the subsequent $A' \to e^+ e^-$ decay. 
\par Another  dark sector experiment DarkMESA plans to utilise the high-power beam dump facility from the P2  experiment. At MESA, 
 in the EB (Extracted Beam) operation ($\sim$150 MeV, 150 $\mu$A) the  beam electrons for the P2 experiment scatter in a 60 cm long liquid hydrogen target, losing an average energy of 17 MeV. The beam dump is 12 m behind the target and in 10,000 h of operation time approximately $3\times10^{22}$  electrons with a total charge of 5400~C will be dumped. This is 
ideally suited for  DarkMESA, which intends to use the reaction $eZ \to eZA'; ~A'\to \chi \chi$ in the dump for the production of a flux of DM $\chi$ particles, which can  be detected via its scattering off electron and protons of the downstream calorimeters of their setup. The expected sensitivity of these experiments, which are currently in preparation for running in a few years,  is shown in Fig.\ref{fig:mesa}. 

\subsubsection{PADME}
PADME is a fixed target experiment at the Beam Test Facility of Laboratori Nazionali di Frascati 
 designed to search for both, $A' \to e^+ e^-$ and $A' \to {\rm invisible}$ decay  modes of the $A'$  produced in $e^+ e^-$ annihilations by using a 550 MeV bunched positron beam \cite{Nardi:2018cxi,Gianotti:2021wiv}. It is expected to be sensitive to the mixing   $\epsilon \gtrsim 10^{-4}$ for the $A'$ masses $m_{A'} \lesssim 24$ MeV  after about one year of running.
PADME plan to search  for the $A'$  by using the missing mass technique in events accompanied with only one photon in the final state. This technique has the advantage to be not controversial, if a signal bump above a continuous background of standard electromagnetic processes is observed.
An excellent missing mass resolution is obtained with a BGO calorimeter, which is further improved by using a narrow positron beam and an active target to determine the beam position bunch-by-bunch.

\begin{figure}
  \begin{minipage}[c]{0.5\textwidth}
    \includegraphics[width=.9\textwidth]{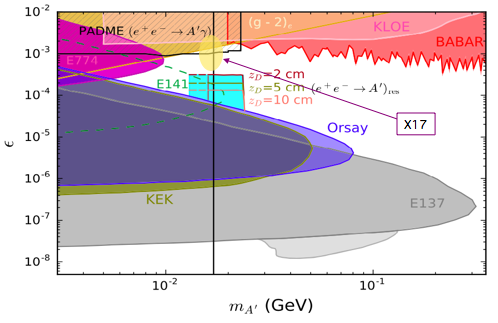}
  \end{minipage}\hfill
  \begin{minipage}[c]{0.5\textwidth}
\caption{The projected sensitivities of PADME for the searches for $A' \to e^+ e^-$  decays, respectively (adapted from~\cite{Nardi:2018cxi}).
The yellow spot indicates the not yet excluded parameter space for the vector X17 boson from the ATOMKI anomaly.}\label{fig:padme}
\end{minipage}
\end{figure}

Another aspect of PADME is the use of a bunched positron beam to increase the number of positron on target (POT) and improve the sensitivity. This choice leads to an increase of events occurring in the same positron bunch which must be disentangled. The strategy to cope with pile-up events foresees to use fast detectors, to veto on charged particles and small angle photons, to digitize all detector waveforms, by a powerful data acquisition, and to develop robust multi-hit reconstruction algorithms.

The current status of the experiment: PADME is the first experiment to search for a dark photon with the missing mass technique and a pulsed positron beam.
 In  the first data taking period from September 2018 to February 2019 about  $7.5\times 10^{12}$ POT were collected with stable positrons beams during the experiment commissioning runs. The next steps are to finalize the detectors absolute calibration, to measure bremsstrahlung and annihilation cross-sections with the collected data, minimize the beam background, and, finally, to collect up to about  $10^{13}$ POT in the physics run in 2020.
 
 \vskip 2mm
\par In conclusion the NA62, NA64, MESA and PADME experiments are complementary to each other providing a remarkable sensitivity to the sub-GeV light dark matter  parameter space, including areas difficult to access by others. They are also complementary to direct searches for DM, and others at LHC and in astrophysics, can efficiently probe large part of the remaining parameter space of sub-GeV LDM model and variety of other models and experimental anomalies.

\clearpage
\subsection{Accelerator probes of milli-charged particles}
{\it Author: Yu-Dai Tsai, <yt444@cornell.edu>}
\label{ssec:tsai}

\subsubsection{Introduction}
The study of the millicharged particle (MCP) is linked to several fundamental mysteries in particle physics.
To begin with, it is connected to the test of the empirical electric charge quantization \cite{Dirac:1931kp} and the related theories \cite{Pati:1973uk,Georgi:1974my,Shiu:2013wxa}. It is also considered as a low-energy consequence of well-motivated dark-sector models \cite{Holdom:1985ag}.
MCP is proposed as a potential dark matter candidate \cite{Brahm:1989jh,Feng:2009mn,Cline:2012is}, and has recently been considered as a solution to the anomaly of 21 cm absorption spectrum reported by the EDGES collaboration \cite{Bowman:2018yin,Barkana:2018lgd,Berlin:2018sjs,Slatyer:2018aqg,Liu:2019knx}.

\vskip 2mm
We consider MCP, labelled $\chi$, with electric charge $Q_\chi$ and define $\epsilon \equiv Q_\chi/e$. This can arise if $\chi$ directly has a small charge under standard model $U(1)$ hypercharge, or if $\chi$ is coupled to a massless kinetic mixing dark photon \cite{Holdom:1985ag}.

\vskip 2mm
MCPs are heavily explored in terrestrial experiments, and their signatures as dark matter is also studied in astrophysical/cosmological observations \cite{Dobroliubov:1989mr,Prinz:1998ua,Davidson:2000hf,Golowich1987,Babu:1993yh,Gninenko:2006fi,Agnese:2014vxh,Haas:2014dda,Ball:2016zrp,Alvis:2018yte,Magill:2018tbb,Kelly:2018brz}. Our focus here is to briefly describe and classify the accelerator-based probes. 
The MCPs are usually produced when the beam collides with another beam or impacts a target.
They can be produced either directly, or through secondary mesons decay. The experimental signature can be roughly classified as tracking ($dE/dx$ signature), hard scattering (to detect the electron recoil), or missing momentum/energy.
The electron-scattering signatures have been one of the main focus to study MCPs. When studying such signatures, since there is a $1/E$ enhancement in the scattering cross-section ($E$ here is the electron-recoil energy), experiments with sensitivity to low-energy recoil or scintillation are often preferred as MCP probes \cite{Magill:2018tbb}.

\subsubsection{Existing bounds and future projections}
In the following paragraphs, we roughly classify the accelerator probes of MCPs.

\vskip 2mm
\noindent
{\bf Colliders - } Searches for MCPs at the Large Hadron Collider (LHC) and the Tevatron have delivered the strongest constraints in the mass region above $100$ MeV. These consist of bounds from trident process searches, the invisible width of the $Z$ boson as well as direct searches for particles with fractional charges at LEP \cite{Davidson:2000hf} and low ionizing particles at CMS \cite{CMS:2012xi,Jaeckel:2012yz}, with focus on $\pm 2e/3$ and $\pm e/3$ . In addition, new sensitivity is achieved recently by milliQan (a prototype scintillator-based detector) for masses larger than a few hundred MeV \cite{Ball:2020dnx}. The proposed electron collider such as Beijing Electron-Positron Collider \cite{Liu:2018jdi} could also improve the sensitivity to MCPs.

\vskip 1mm
\noindent
{\bf Proton fixed-target and neutrino experiments - }
In the fixed-target neutrino experiments category, the Liquid Scintillator Neutrino Detector (LSND) and MiniBooNE experiments are found to provide new constraints in certain MCP mass windows, $5-35$ MeV and $100-180$ MeV respectively \cite{Magill:2018jla}. Using existing data from Liquid Argon detectors in neutrino beams at Fermilab, ArgoNeuT has also constrained new regions of the MCP parameter space \cite{Acciarri:2019jly}. 
Sensitivity projections for MCPs over a range of masses $5$ MeV to $5$ GeV has been analyzed recently \cite{Magill:2018jla}, considering the proposed upcoming neutrino experiments and proton fixed-targets such as MicroBooNE, the Short-Baseline Neutrino Program (SBND) and the Deep Underground
Neutrino Experiment (DUNE) at Fermilab, as well as the Search for Hidden Particles (SHiP) at CERN. 

\vskip 2mm
\noindent
{\bf Lepton fixed-target experiments - } In the low-mass region, the most sensitive constraints on MCPs were placed by electron fixed-target experiments, \textit{e.g.} SLAC mQ experiment \cite{Prinz:1998ua} with the leading sensitivity for $m_{\rm MCP}<100$ MeV. Despite the mass reach limit due to the beam energy, further sensitivity enhancement to MCP coupling can be reached by future electron beam dump facilities using missing energy and momentum techniques, \textit{e.g.} LDMX \cite{Berlin:2018bsc} and NA64$\mu$ \cite{Gninenko:2018ter} with $10^{16}$ electron-on-target and $5\times 10^{13}$ muon-on-target, respectively.

\begin{figure}[t]
\centering
\includegraphics[width=\linewidth]{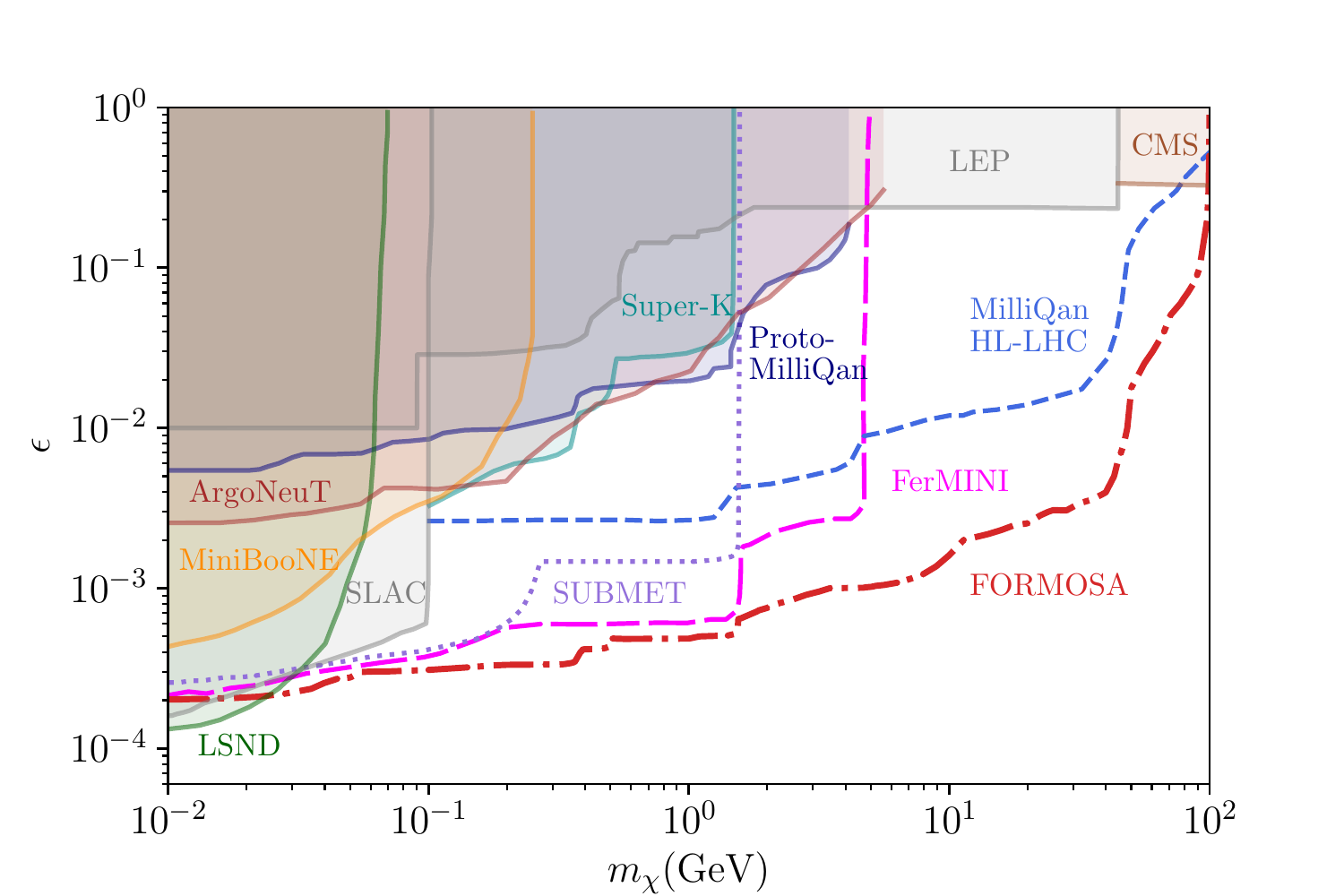}
\caption{Expected $95\%$ confidence level sensitivity to millicharged particles shown in the plane of the fractional charge versus MCP mass. Exclusions from direct searches including SLAC \cite{Prinz:1998ua}, Colliders \cite{Davidson:2000hf,Haas:2014dda}, CMS \cite{CMS:2012xi,Jaeckel:2012yz}, MiniBooNE and LSND \cite{Magill:2018tbb}, ArgoNeuT at Fermilab \cite{Acciarri:2019jly}, recent search by milliQan at LHC \cite{Ball:2020dnx}, the diffuse supernova neutrino background search in Super-K \cite{Plestid:2020kdm}, as well as the projections for milliQan HL-LHC \cite{Haas:2014dda} (dashed blue), FerMINI \cite{Kelly:2018brz} at DUNE (dashed pink), SUBMET \cite{Choi:2020mbk} (dashed purple), and FORMOSA \cite{Foroughi-Abari:2020qar} (dot-dashed red) are shown for comparison (see the text for details).}
\label{fig:exsisting_constraints}
\end{figure}

\vskip 2mm
\noindent
{\bf milliQan/FerMINI: dedicated detectors - }
Dedicated MCP detectors were proposed at the LHC, proton fixed-target, and neutrino experiments, e.g. milliQan \cite{Ball:2016zrp} and FerMINI \cite{Kelly:2018brz}.
The detectors consist of layers of scintillator arrays, where MCPs traversing the scintillators would produce a few photo-electrons in each layer. The idea is to use the multiple-coincidence scintillation as an experimental signature within a short time window. 
The milliQan detector is located in the transverse region with respect to the LHC beam-line. The milliQan demonstrator, benefiting from the high beam energy of the LHC, could deliver sensitivity for the high mass region. Recently, new sensitivity has been achieved by milliQan demonstrator for masses larger than a few hundred MeV \cite{Ball:2020dnx}.

\vskip 1mm
The FerMINI detector consists of a
milliQan-type detector and is located downstream of the proton target of a neutrino experiment. Impressive sensitivity to $\epsilon$ below $10^{-3}$ and up to about $m_{\chi} \sim 5$ GeV could be provided by FerMINI proposal due to the higher flux of MCP produced in the high-luminosity proton beam on targets. SUBMET is also proposed at the J-PARC proton fixed-target facilities having a similar sensitivity as FerMINI for masses below 1 GeV \cite{Choi:2020mbk}. As another example of dedicated detectors, MAPP is a new MoEDAL sub-detector in LHC’s RUN-3, which is being designed to search for fractionally charged particles \cite{Staelens:2019fan}.
Recently, a new  experimental setup, FORMOSA  \cite{Foroughi-Abari:2020qar}, has been proposed to place a milliQan-like detector at the LHC forward region to study MCP .

\begin{figure}[t]
\centering
\includegraphics[width=\linewidth]{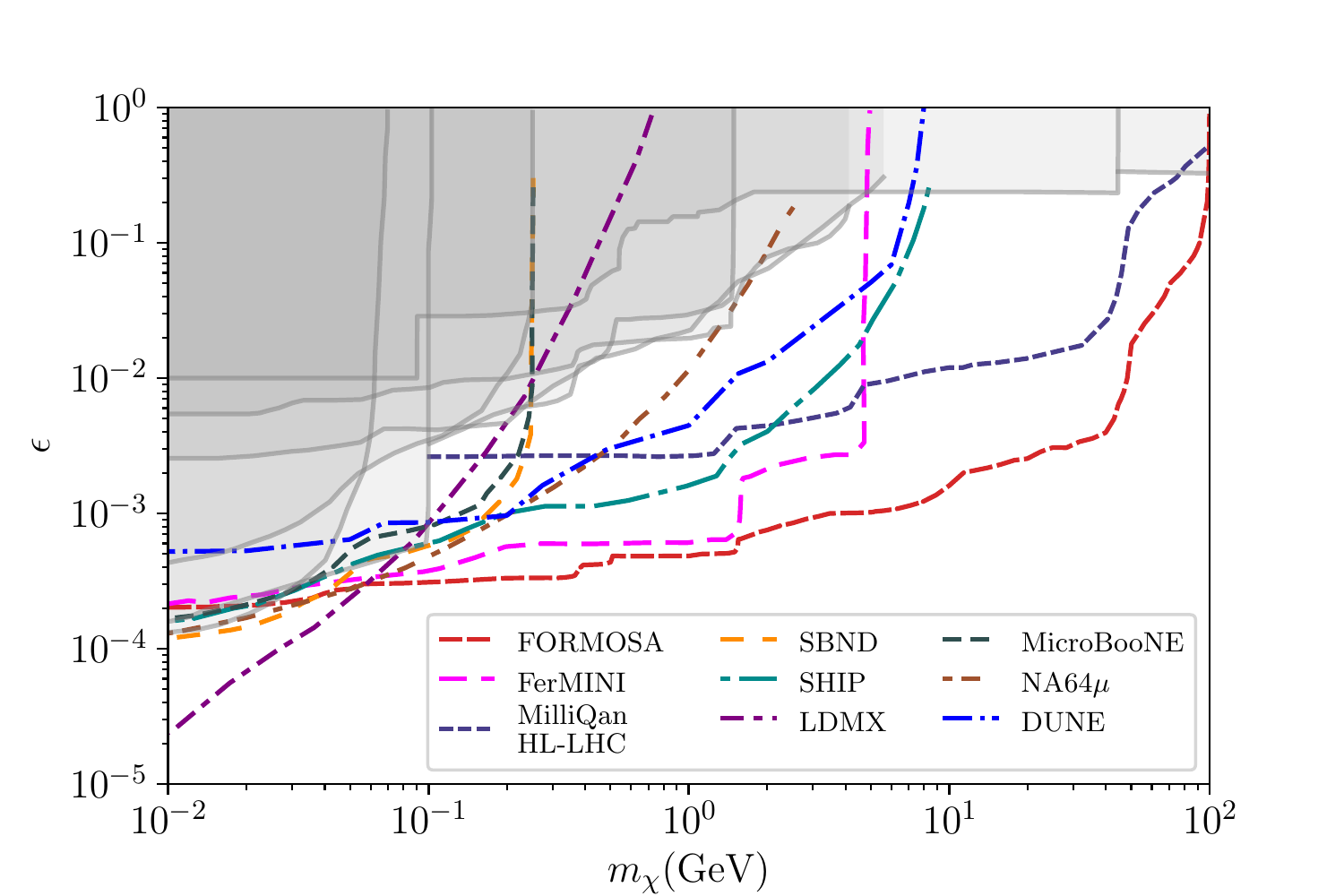}
\caption{Shaded areas (in gray) are the existing bounds excluded by different source mentioned in the text. The projected sensitivities including FerMINI \cite{Kelly:2018brz} at DUNE, FORMOSA \cite{Foroughi-Abari:2020qar}, milliQan HL-LHC \cite{Haas:2014dda}, NA$64\mu$ \cite{Gninenko:2018ter} with $5\times 10^{13}$ muon-on-target at CERN and LDMX \cite{Berlin:2018bsc} with $10^{16}$ electron-on-target are shown by dashed curves for comparison. The reach of neutrino experiments such as $\mu$BooNE, SBND, SHIP are taken from \cite{Magill:2018jla}, and DUNE is taken from \cite{Harnik:2019zee}.}
 \label{fig:future_projections}
\end{figure}
 
 \vskip 2mm 
 \noindent
{\bf Cosmic-ray accelerator - }Another interesting and important probe of the millicharged particles is through the production of MCPs from cosmic ray hitting the atmosphere. Using large underground neutrino detectors such as Super-K, a recent study has set new limits on MCPs for the mass range 0.1 GeV to 1.5 GeV \cite{Plestid:2020kdm}.

\vskip 2mm
In summary, the existing $95 \%$ CL sensitivity limits to millicharged particles are shown in Fig. \ref{fig:exsisting_constraints}. The future sensitivity reaches of different accelerator-based experiments are shown in Fig. \ref{fig:future_projections}.

\clearpage
\subsection{Results and remarks for light DM and vector mediator in minimal models}
\label{ssec:vector-results}

\subsubsection{Results}
\label{sssec:vector-results}
\begin{figure*}[ht!]
\centering
\includegraphics[width=0.8\linewidth]{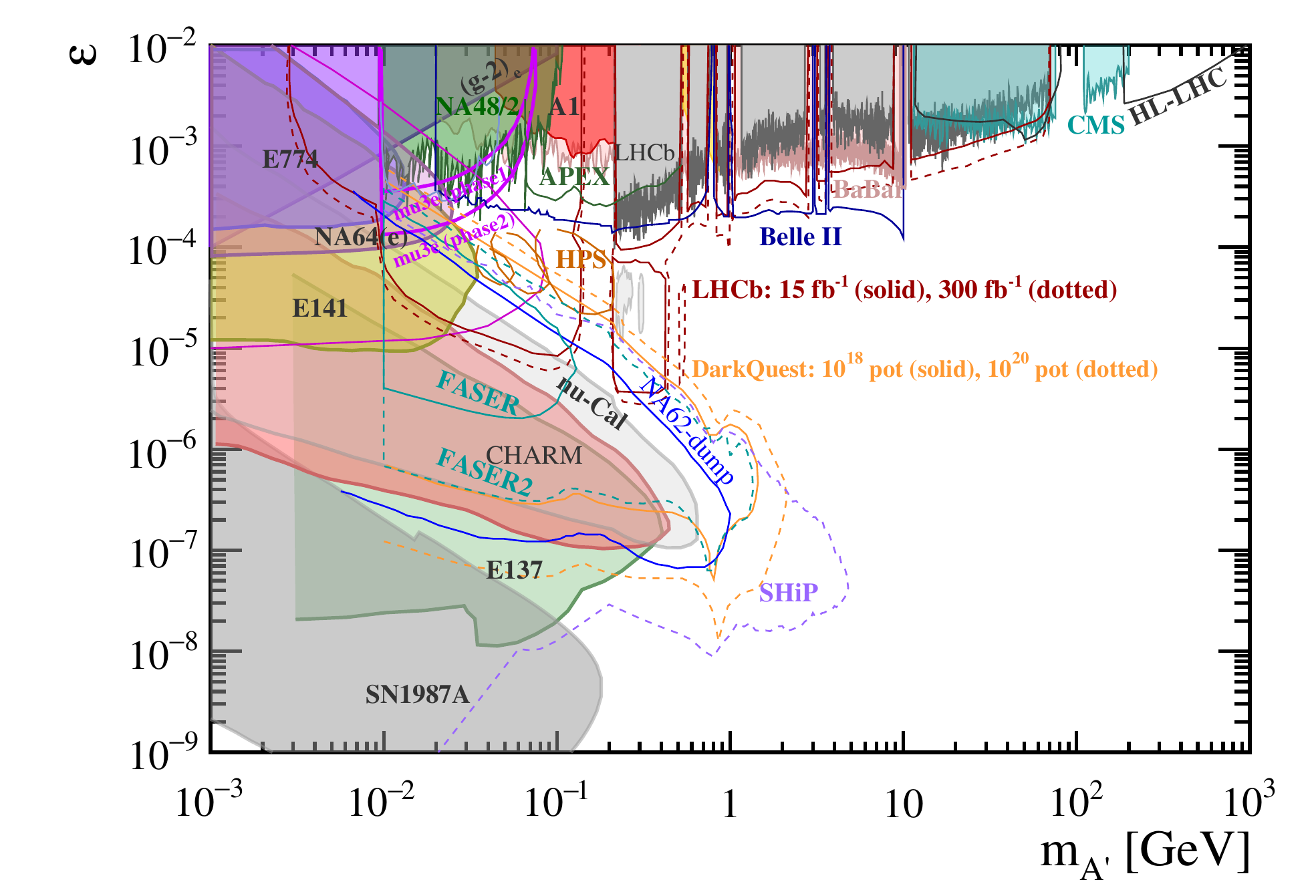}
\caption{ {\bf Dark photon into visible final states:} $\varepsilon$ versus $m_{A'}$. Filled areas are existing limits from searches at experiments at collider/fixed target (A1~\cite{Merkel:2014avp}, LHCb~\cite{Aaij:2019bvg},CMS~\cite{CMS:2019kiy},BaBar~\cite{Lees:2014xha}, KLOE~\cite{Archilli:2011zc,Babusci:2012cr,Babusci:2014sta,Anastasi:2016ktq}, and NA48/2~\cite{Batley:2015lha}) and old  beam dump:  E774~\cite{Bross:1989mp}, E141~\cite{Riordan:1987aw}, E137~\cite{Bjorken:1988as,Batell:2014mga,Marsicano:2018krp}), $\nu$-Cal~\cite{Blumlein:2011mv,Blumlein:2013cua}, CHARM (from~\cite{Gninenko:2012eq}), and BEBC (from~\cite{Marocco:2020dqu}).Bounds from supernovae~\cite{Chang:2016ntp} and $(g-2)_e$~\cite{Pospelov:2008zw} are also included. Coloured curves are projections for existing and proposed experiments: Belle-II~\cite{Kou:2018nap}; LHCb upgrade~\cite{Ilten:2016tkc,Ilten:2015hya}; NA62 in dump mode~\cite{NA62:dump} and NA64(e)$^{++}$~\cite{Gninenko:2013rka,Andreas:2013lya}; FASER and FASER2~\cite{Ariga:2018uku}; seaQUEST~\cite{Berlin:2018pwi}; HPS~\cite{Adrian:2018scb};  Dark MESA~\cite{Doria:2019sux}, Mu3e~\cite{Echenard:2014lma}, and  HL-LHC~\cite{Curtin:2014cca}. Figure revised from Ref.~\cite{Lanfranchi:2020crw}.
}
\label{fig:DP_visible}
\end{figure*}

\begin{figure*}[ht!]
\centering
\includegraphics[width =0.8\linewidth]{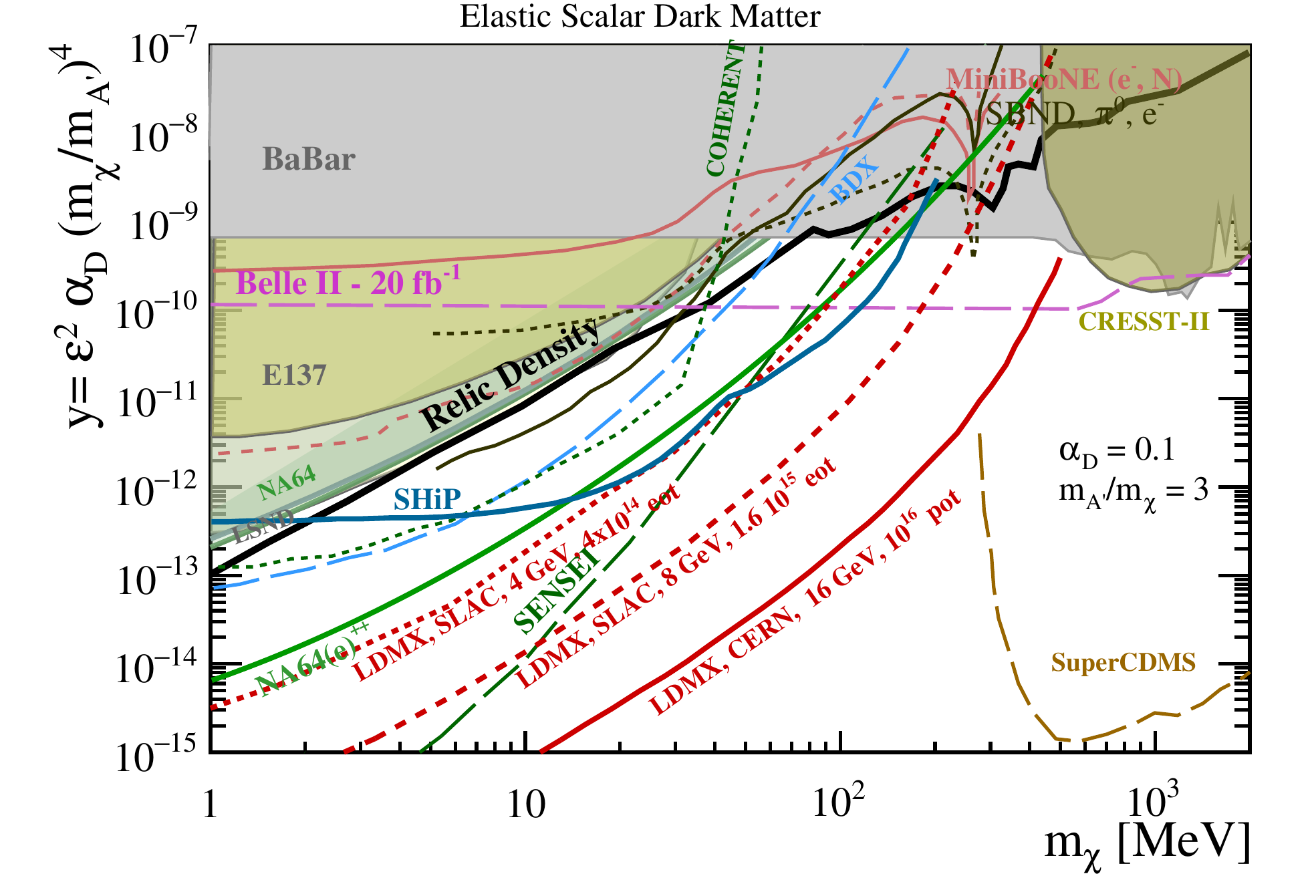}
\includegraphics[width = 0.8\linewidth] {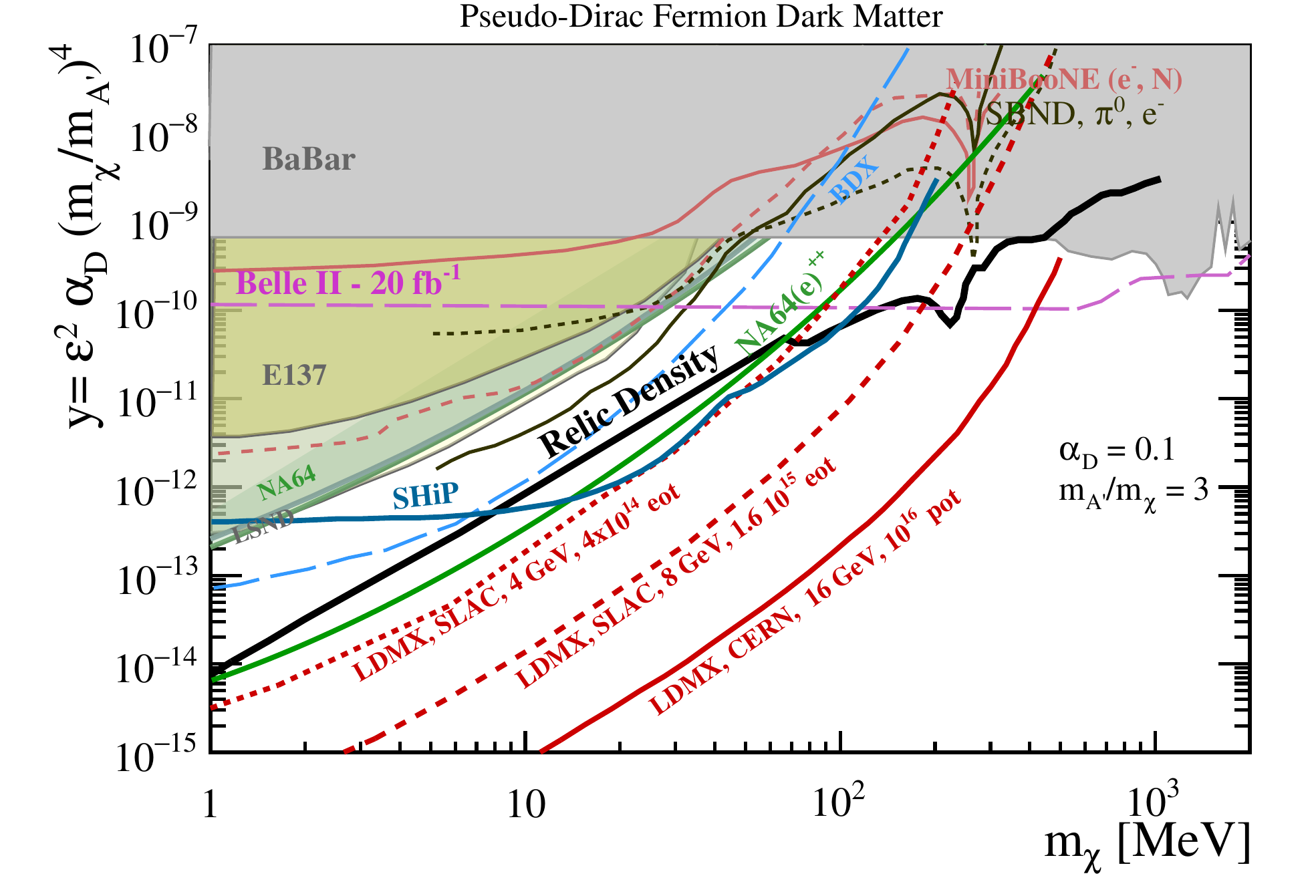}
\caption{\small 
  Existing limits ( filled areas) and future sensitivities of existing or proposed experiments (coloured curves) to light dark matter production through a dark photon in the plane defined by the yield variable $y$ as a function of DM mass $m_{\chi}$ for a specific choice of $\alpha_D = 0.1$ and $m_{A'}/m_{\chi} = 3$. The DM candidate is assumed to be an elastic scalar (top) or a pseudo-Dirac fermion (bottom).  Limits shown as filled areas are: BaBar~\cite{Lees:2017lec}; NA64$_e$~\cite{NA64:2019imj}; reinterpretation of the data from E137~\cite{Batell:2014mga} and LSND~\cite{deNiverville:2011it};  result from MiniBooNE~\cite{Aguilar-Arevalo:2018wea}; inferred sensitivity from the direct detection DM experiment CRESST-II~\cite{Angloher:2015ewa}. The projected sensitivities, shown as solid or dashed lines, come from: SHiP~\cite{Anelli:2015pba}, BDX~\cite{Battaglieri:2016ggd}, SBND~\cite{Antonello:2015lea}, COHERENT~\cite{deNiverville:2015mwa}, LDMX@SLAC~\cite{Akesson:2018vlm,Raubenheimer:2018mwt}, LDMX@CERN~\cite{LDMX:CERN}, Belle-II~\cite{Kou:2018nap}; SENSEI~\cite{Battaglieri:2017aum} and SuperCDMS~\cite{Agnese:2016cpb}. Figure revised from~\cite{Lanfranchi:2020crw}.}
\label{fig:DP_y_scalar}
\end{figure*}

\begin{figure*}[ht!]
\centering
  \includegraphics[width=0.8\linewidth]{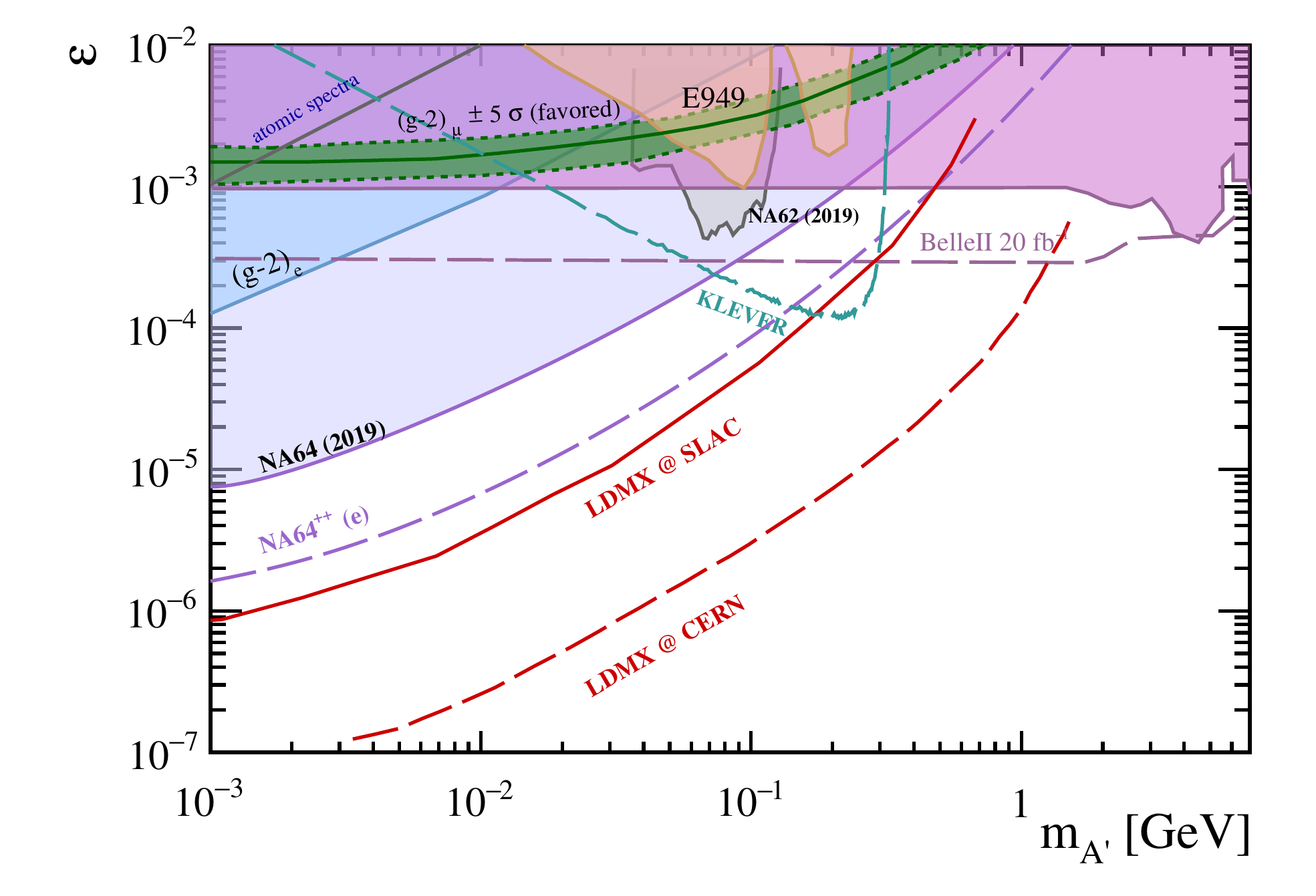}
\caption{\small 
  \label{fig:massive31}
  Existing limits  and future sensitivities for  a massive dark photon going to invisible final states ($\alpha_{\dd}>> \alpha \varepsilon^2 $). Existing limits from Kaon decay experiments (E787~\cite{Adler:2001xv},  E949~\cite{Artamonov:2009sz}, NA62~\cite{CortinaGil:2019nuo}), BaBar~\cite{Lees:2017lec}, and NA64(e)~\cite{Gninenko:2019qiv}. The constraints from $(g-2)_{\mu}$~\cite{Bennett:2006fi} and $(g-2)_e$ are also shown. Future sensitivities for NA64(e)$^{++}$~\cite{NA64:eplus}, Belle II~\cite{Kou:2018nap}, KLEVER~\cite{Ambrosino:2019qvz}, PADME~\cite{Raggi:2015gza}, LDMX@SLAC~\cite{Akesson:2018vlm,LDMX:CERN}, and LDMX@CERN~\cite{Akesson:2018vlm,Raubenheimer:2018mwt}. The sensitivity curves for LDMX@SLAC and LDMX@CERN assume $10^{14}$~electrons-on-target and $E_{\rm beam} = 4$~GeV and $10^{16}$~electrons-on-target and $E_{\rm beam} = 16$~GeV, respectively.}
\end{figure*}

The current status and future projections for searches for dark photons in the minimal model are shown in 
Figures~\ref{fig:DP_visible},~\ref{fig:DP_y_scalar} and ~\ref{fig:massive31}.
These three plots correspond to the established PBC benchmarks $BC1$ and $BC2$, as reported in~\cite{Beacham:2019nyx}
and summarised here below for convenience.

\begin{itemize}

\item {\em BC1, Minimal dark photon model}: in this case the SM is augmented by a single new state $A'$. DM is assumed to be either 
heavy or contained in a different sector. In that case, once produced, the dark photon decays back to the SM states. The parameter 
space of this model is then $\{ m_{A'}, \epsilon \}$.

\item{\em BC2, Light dark matter coupled to dark photon}: this is the model where minimally coupled viable WIMP dark matter 
  model can be constructed.
  The preferred values of dark coupling $\alpha_{D} = g_D^2/(4\pi)$ are such that the decay of $A'$ occurs predominantly 
  into $\chi\chi^*$ states. These states can further rescatter on electrons and nuclei due to $\epsilon$-proportional
  interaction between  SM and DS states mediated by the mixed $AA'$ propagator.
  
  The parameter space for this model is  $\{ m_{A'}, \epsilon, m_\chi, \alpha_D \}$ with further model-dependence associated
  with properties of $\chi$ (boson or fermion). The suggested choices for the PBC evaluation are
  1. $\epsilon$ vs $m_{A'}$ with $\alpha_D \gg \epsilon^2 \alpha$ and $2m_\chi <m_{A'}$, 2. $y$ vs. $m_\chi$ plot
  where the {\it yield} variable $y$,  $y = \alpha_D \epsilon^2 (m_\chi/m_{A'})^4$, is argued
  to contain a combination of parameters relevant for the freeze-out and 
  DM-SM particles scattering cross section. One possible choice is $\alpha_D = 0.1$ and $m_{A'}/m_\chi = 3$. 

\end{itemize}

\clearpage
\subsubsection{Remarks about comparison between accelerator-based and direct detection dark matter searches in the sub-GeV range}
\label{sssec:vector-recommendations}

This Section aims at clarifying a few aspects in the formalism of light DM searches in the vector portal framework that are important to properly compare results between accelerator-based and direct detection DM experiments.

\noindent 
What we know:
\begin{enumerate}
    \item {\bf Lee-Weinberg bound:} \\
Thermal WIMPs interacting solely through the electroweak force must be heavier than 1-2 GeV~\cite{Lee:1977ua}. Hence:  sub-GeV thermal DM requires new forces with light mediators~\cite{Boehm:2003hm}. \\
     
     \item {\bf CMB bound:} \\
     If DM annihilates during CMB era, strong constraints exist on the energy injected in the photon plasma during recombination. 
     The CMB bounds are based on visible energy injection at T$\sim$~eV, which reionizes the newly recombined hydrogen and thereby modifies the ionized fraction of the early universe. Hence: for s-wave annihilating DM, CMB bounds rule out $m_{\rm DM} < 10$~GeV. \\
     
     \item {\bf Viable options: }\\
     Given bounds from 1. and 2. viable options for thermal DM in the MeV-GeV range are: 
     \begin{itemize} 
     \item[i)] DM annihilates in $p-$wave ($\sigma v$ is $v^2$ suppressed, hence smaller at low temperature): assuming a vector mediator, DM can be a scalar particle (Figure~\ref{fig:DP_y_scalar}, left).
   \item[ii)] Presence of a mechanism that cuts off late time annihilation, as e.g.\ mass splitting in the $\chi - \overline{\chi}$ system ($\chi$ is the DM): DM can be a pseudo-Dirac fermion (Figure~\ref{fig:DP_y_scalar}, right). Similar considerations apply to DM as Majorana particle.
   \end{itemize}
  These two cases are discussed in the following.
  \end{enumerate}
   
   \begin{enumerate}
   \item {\bf Vector mediator and DM as elastic scalar particle:}
   
   \begin{itemize}
   \item {\it How to compare accelerator-based results with direct detection DM cross sections. }\\
  Accelerator-based results are usually shown in terms of $y$ variable versus $m_{\chi}$ (being $\chi$ the DM candidate). This is convenient because the results can be  compared against the non-relativistic cross section for DM annihilation to a pair of light ($m_{\ell} << m_{\chi}$) SM leptons:

\begin{eqnarray*}
    \sigma v (\chi \chi^* \to A'^{*} \to \ell^+ \ell^-) & \simeq & {8 \pi \over 3} {\alpha_{em} \varepsilon^2 \alpha_D m_{\chi}^2 v^2 \over (4 m^2_{\chi} - m^2_{A'})^2 }  \nonumber \\
    & \simeq & { 8 \pi \alpha_{em} v^2 \over 3 } {y \over m^2_{\chi}}
\end{eqnarray*}

where:
\begin{equation} 
y = \varepsilon^2 \alpha_D  {m^4_{\chi} \over m^4_{A'}}, 
\label{eq:y}
\end{equation}

and $v \sim 10^{-3}$ is the relative DM velocity. We assume that $m_{A'} >> m_{\chi}$. 

Accelerator-based results usually show results in the plane $y$ versus $m_{\chi}$. In doing that they assume $\alpha_D = 0.1$  (also $\alpha_D = 0.5$ is commonly used) and $m_{A'}/m_{\chi} = 3$. This is a conservative choice of the parameters. The scaling of the bounds as a function of $\alpha_D$ and $m_{A'}/m_{\chi}$ are studied in Ref.~\cite{Berlin:2020uwy}.

In this model, the (non-relativistic) electron scattering cross-section measured by the DM direct detection experiments is:
\begin{eqnarray}
   \sigma_{\chi e}  & = & \frac{16\pi \alpha \alpha_D\epsilon^2 \mu_{e\chi}^2}{m_{A'}^4} \simeq \frac{16\pi \alpha m_e^2}{m_\chi^4} \times y \nonumber \\
    & \simeq &  3.7\times 10^{-27}\,{\rm cm}^2 \cdot (10\,{\rm MeV}/m_\chi)^4\cdot y,
\label{eq:dm_elastic_scattering}
\end{eqnarray}
where $\mu_{e\chi}$ is reduced mass of the $\{e,\chi\}$ system and a similar formula applies for the scattering on nuclei. Here $y$ is the same yield parameter that largely controls the abundance (see Eq.~\ref{eq:y}). Consequently, low threshold direct detection experiments probing DM-nucleus scattering can rule out significant fractions of parameter space for this model ({\em e.g.} $m_\chi > 400$\,MeV is disfavoured by CRESST, see Figure~\ref{fig:DP_y_scalar} (top), while future planned DM-electron experiments have yet to achieve the levels of sensitivity relevant for sub-GeV freeze-out DM.

{\it Equation~\ref{eq:dm_elastic_scattering}
can be used to compare accelerator-based  with DM direct detection results within this model}.\\

\item
{\it How to scale direct detection DM cross sections above the thermal relic target}\\
Above the "thermal relic target line", if there are no other ingredients in the theory, DM is under-abundant. Below this line it is over-abundant and the model is excluded unless some additional ingredients are added to the model. 

Let's address the re-scaling when the experimental results are above the thermal relic target line.
In that case we need to account for the fact that experimental collaborations quote their limits saturating the observed DM abundance. They typically have:
\begin{equation}
    \sigma_0 (m_{\chi}) \cdot y < \sigma_{\rm limit} (m_{\chi}) \;\;\; {\rm at} \;\;\; \Omega_{\chi} = \Omega_{\rm DM, obs}
\end{equation}

where $\sigma_{\rm limit}$ is the limit of their sensitivity in terms of cross-section.
To re-scale, we need to take into account that:
\begin{equation}
    {\Omega_{\chi} \over \Omega_{\rm DM, obs}} = {y_0 \over y}
\end{equation}
where $\{ m_{\chi}, y\}$ is a point in the parameter space above the thermal relic target and 
$\{ m_{\chi}, y_0\}$ is the point for the same mass value on the thermal relic target. Then the true bound re-scaled from direct detection experiments in our parameter space should be given by:
\begin{equation}
    { y_0 \over y} \cdot \sigma_0(m_{\chi}) \cdot y = \sigma_0(m_{\chi}) \cdot y_0 < \sigma_{\rm limit} (m_{\chi}) \;\;\; (y_0 < y). 
\end{equation}

{\it Notice that the result is independent of $y$}.
That means the following: {\it if the direct detection experimental limit does not reach the thermal relic target then any value of $y$ is allowed}. This is because the larger the $y$, the less abundance we have, and this compensates the growth of the cross section. Therefore the actual signal in DD that is given by abundance $\times$ cross section stays constant.

{\it We advocate to show DM direct detection bounds in the $y$ versus mass plane only if they stay below or at the thermal relic curve (as done in Figure~\ref{fig:DP_y_scalar}, top)}.

\end{itemize}
\item {\bf Vector mediator and DM as a pseudo-Dirac particle}\\
It can happen that the $U(1)_{A'}$ breaking mass term $\Delta m^2\chi^2$ generates a mass splitting $\Delta m$ between the real components of $\chi = 2^{-1/2}(\chi_1+i\chi_2)$ that is easily larger than the kinetic energy of DM today ($E_{\rm kin} \sim \frac12 m_\chi c^2 (v_{\rm SM}/c)^2 \sim 10^{-5}{\rm eV}\times m_\chi/20\,{\rm MeV}$). In this case, $s$-wave elastic direct detection scattering is completely quenched, while having a negligible effect on the primordial abundance. Similar arguments apply to fermionic DM as well.

{\it In this case accelerator-based experiments would see a signal while DD would see none. This has been shown in Figure~\ref{fig:DP_y_scalar} (bottom)}.
   
\end{enumerate}

\noindent{\bf Final considerations}\\
It is important to stress the underlying physical complementarity between direct detection and accelerator probes, especially when comparisons are made. Ultimately, direct detection probes DM deep in the non-relativistic limit, whereas accelerator experiments probe DM interactions in the relativistic limit, and this has profound consequences on the phenomenology. 

For example, different choices of DM spin can lead to direct detection rates suppressed by multiple powers of DM halo velocity $v\sim 10^{-3}$ among different models, whereas accelerator rates are only mildly affected by changing the spin because production occurs near $v\sim 1$. Additionally, direct detection rates are extremely sensitive to even small perturbations of the DM mass terms (as shown for the Pseudo-Dirac case).

The examples above show that the direct detection program alone cannot be viewed as a {\it universal substitute} to the accelerator-derived probes of the vector portal DM models. Likewise, accelerator probes alone cannot verify that the particles produced are indeed cosmologically long-lived. Both types of experiments are crucially important. 

\clearpage

\section{Feebly-interacting pseudoscalar particles: Axions and ALPs}
\label{sec:pseudoscalar}

\subsection{Phenomenology of axions/ALPs as DM candidates or DM mediators}
\label{ssec:ringwald}
{\it Author: Andreas Ringwald, <andreas.ringwald@desy.de>}  \\ 

Axions, or more generally ALPs, $a$, are pseudoscalar pseudo-Nambu-Goldstone bosons arising from approximate Abelian global symmetries beyond the SM which are broken spontaneously at a scale $f_a$ much greater than the electroweak scale $v$.  At the latter, 
their most general interactions with the SM fermions $\psi_F$ and the SM 
$SU(3)_c$, $SU(2)_L$ and $U(1)_Y$ gauge field strengths $G_{\mu\nu}^a$, $W_{\mu\nu}^A$ and $B_{\mu\nu}$, respectively, and their 
duals (denoted by a tilde) are summarised by the following low-energy effective Lagrangian 
\cite{Georgi:1986df,Bauer:2020jbp}:
\begin{equation}
\begin{aligned}
   {\cal L}_{\rm eff}
   &= \frac12 \left( \partial_\mu a\right)\!\left( \partial^\mu a\right)  - \frac{m_{a,0}^2}{2}\,a^2
    +  \frac{\partial^\mu a}{f_a}\,
\sum_F \,\bar\psi_F \gamma_\mu C_F \psi_F
\\
   &\quad\mbox{}  -C_{aGG}\,\frac{\alpha_s}{8\pi}\,\frac{a}{f_a}\,G_{\mu\nu}^a\,\tilde G^{\mu\nu,a} 
    -C_{aWW}\,\frac{\alpha_2}{8\pi}\,\frac{a}{f_a}\,W_{\mu\nu}^A\,\tilde W^{\mu\nu,A} \\
    &\quad\mbox{} -C_{aBB}\,\frac{\alpha_1}{8\pi}\,\frac{a}{f_a}\,B_{\mu\nu}\,\tilde B^{\mu\nu}  .
\end{aligned}
\label{Leff_a}
\end{equation}
Here, $\alpha_s=g_s^2/(4\pi)$, $\alpha_2=g^2/(4\pi)$ and $\alpha_1=g^{\prime\,2}/(4\pi)$ are the respective SM gauge coupling parameters, and 
$F$ denotes the left-handed fermion multiplets in the SM. $C_F$ is a  
Hermitian matrix in generation space with dimensionless entries which, together with the dimensionless coefficients $C_{aGG}$, $C_{aWW}$ and
$C_{aBB}$, depend on the specific UV completion featuring the Abelian global symmetry. 
Since all the interactions with the SM in Eq.~\eqref{Leff_a} are inversely proportional to $f_a$, 
axions and ALPs with $f_a\gg v$ are indeed feebly-interacting pseudoscalar particles.  

The low-energy effective 
Lagrangian \eqref{Leff_a} realises the 
approximate Abelian global symmetry non-linearly through an approximate symmetry under constant shifts, 
$a\to a+\kappa f_a$, broken only by the bare mass term $\propto m_{a,0}^2 a^2$, which parametrises the effect of a possible explicit breaking of the global symmetry, and by the coupling to the topological charge density, $\frac{\alpha_s}{8\pi}\,G_{\mu\nu}^a\,\tilde G^{\mu\nu,a}$, of the $SU(3)_c$ gauge fields.\footnote{For the couplings of $a$ to the $SU(2)_L$  
the additional term arising from a constant shift $a\to a+\kappa f_a$ can be removed by field redefinitions~\cite{Perez:2014fja}. For $U(1)_Y$ it is unobservable in the SM.} 

The axion~\cite{Weinberg:1977ma,Wilczek:1977pj}, arising from the Peccei-Quinn (PQ) solution of the strong CP problem~\cite{Peccei:1977hh}, is defined via $C_{aGG}\neq 0$ and $m_{a,0}=0$.
In fact, in this case, one can eliminate the CP-violating QCD $\bar\theta$-term, $\bar\theta\, \frac{\alpha_s}{8\pi}\,G_{\mu\nu}^a\,\tilde G^{\mu\nu,a}$,  
by the shift $a \to a + \bar\theta f_a/C_{aGG}$. Non-perturbative topological fluctuations of the gluon fields in QCD induce a
potential for the shifted field whose minimum is at zero field value, thereby ensuring the CP conservation of strong interactions~\cite{Vafa:1983tf}.\footnote{In order for the axion vacuum expectation value to be relaxed to zero, the PQ global symmetry has to be preserved from explicit breaking to a great degree of accuracy, such that $m_{a,0}\approx 0$ to a precision compatible with the 
upper bound on $\bar\theta$ arising from the non-observation of the neutron electric dipole moment.}
The second derivative of the dynamically induced potential gives the axion mass. It is inversely proportional to the scale $f_a$ and proportional to 
the square root of the topological susceptibility $\chi$ of QCD, which can be determined either using 
chiral effective field theory (for recent evaluations, see Ref.~\cite{diCortona:2015ldu,Gorghetto:2018ocs,Lu:2020rhp}) or lattice 
QCD~\cite{Borsanyi:2016ksw}, resulting in the prediction\footnote{In the literature about axions, it is customary to absorb the coefficient $C_{aGG}$ in the decay constant, $f_A=f_a/C_{aGG}$.} 

\begin{equation}
\label{canonical_axion_mass}
m_a = C_{aGG} \frac{\sqrt{\chi}}{f_a} \simeq 5.7\,  \left(\frac{10^{9}\,{\rm GeV}}{f_a/C_{aGG} } \right) {\rm meV}\,.
\end{equation}
Correspondingly, all the axion couplings to the SM grow proportional to the axion mass, while their magnitude depend on the 
specific model-dependent coefficients of the relevant operators in the effective Lagrangian. This gives rise to the so called axion bands  
in plots of coupling constant versus mass, see for example Figs.~\ref{fig:gagg}, \ref{fig:large_panorama}, and \ref{fig:helioscopes}. Those 
display the photon coupling, 
\begin{equation}
\label{axion:eq:eq4}
  {\cal L}_{\rm eff}
 \supset -\frac{{g_{a\gamma}}}{4}\,a\,F_{\mu\nu} \tilde F^{\mu\nu}
 =g_{a\gamma}\,a\,{\bf E}\cdot{\bf B}\,,
\end{equation}
which, at energy scales below the confinement scale of QCD and for $m_a\ll m_\pi$,\footnote{At larger values of $m_a$, additional contributions to the photon coupling 
have to be taken into account, see 
Refs.~\cite{Bauer:2017ris,Bauer:2020jbp}.}, is predicted in terms of the coefficients in the effective Lagrangian \eqref{Leff_a} as \cite{diCortona:2015ldu,Lu:2020rhp},
\begin{equation}\label{axion:eq:eq5}
g_{a\gamma}
=\frac{\alpha}{2\pi f_a}\,
  \left[ C_{aWW}+C_{aBB} -1.92(4) C_{aGG}\right]\,,
\end{equation}
for particular UV completions of the axion model, notably for the rather minimalist KSVZ~\cite{Kim:1979if,Shifman:1979if} and  
DFSZ~\cite{Zhitnitsky:1980tq,Dine:1981rt} models and variants. 

Recently, there have been proposals to enlarge the axion bands by considering less minimal UV completions which 
result in larger values of the coefficients of specific axion-SM interactions, e.g.  of the one for two-photon interactions, 
$C_{a\gamma\gamma}=C_{aWW}+C_{aBB}$. Such proposals are discussed in Sec.~\ref{ssec:agrawal}. 
They allow the axion to populate new regions of parameter space 
by moving vertically the axion band in the mass versus coupling plane (for examples, see Fig.~\ref{fig:gagg}). 
The results are then “channel specific”: different couplings are modified differently.

The axion bands can be alternatively changed enlarging the confining sector beyond QCD. New contributions of 
topologically non-trivial gauge field fluctuations give then additional contributions to the right-hand side of 
Eq.~\eqref{canonical_axion_mass}. 
The practical consequence is a universal modification of the parameter space of all axion couplings at a given mass, 
at variance with the vertical displacement scenarios discussed above.
Examples of horizontal enlargement of the parameter space towards the right of the
canonical axion band are heavy axion models that solve the strong CP problem
at low scales (e.g. $f_a\sim {\rm TeV}$)
~\cite{rubakov:1997vp,Berezhiani:2000gh,Gianfagna:2004je,Hsu:2004mf,Hook:2014cda,Fukuda:2015ana,Chiang:2016eav,Dimopoulos:2016lvn,Gherghetta:2016fhp,Kobakhidze:2016rwh,Agrawal:2017ksf,Agrawal:2017evu,Gaillard:2018xgk,Buen-Abad:2019uoc,Hook:2019qoh,Csaki:2019vte,Gherghetta:2020ofz}. 
More recently, there has been also a proposal producing instead a left horizontal shift, that is 
an axion solving the strong CP problem with $m_a\,f_a\ll C_{aGG} \sqrt{\chi}$~\cite{Hook:2018jle}. It is based on a mirror 
extension of the SM:
$N$ mirror and degenerate worlds are assumed to coexist, linked by an axion field which implements non-linearly a 
$Z_N$ symmetry. 
All the confining sectors contribute now to the right-hand side of Eq.~\eqref{canonical_axion_mass}, conspiring by symmetry to suppress the axion mass exponentially by a factor~\cite{DiLuzio:2021pxd}
\begin{align}
\label{eq:ratio-holo-mass} 
\bigg(\frac{m_a f_a}{C_{aGG} \sqrt{\chi}}\bigg)^2 \simeq \frac{1}{\sqrt{\pi}}\sqrt{1-z^2} (1+z)\, N^{3/2} z^{N-1}\,,
\hspace{1ex} {\rm for}\ N\geq 3\,,
\end{align}
without spoiling the solution to the strong CP problem. Here $z \equiv m_u / m_d \approx 0.48$, such that the above ratio goes 
as $\sim 2^{-N}$. 
The resulting universal enhancement of all axion interactions relative to those of the canonical QCD axion has a strong impact on the prospects of experiments searching for sub-eV mass ALPs, see Sec.~\ref{ssec:irastorza}.  A pseudoscalar observed by
the light-shining-through-walls experiment ALPS II or by the helioscope (Baby)IAXO can indeed solve the strong CP problem 
and explain the hints of excessive stellar cooling (cf. Sec.~\ref{ssec:giannotti}) in one stroke~\cite{DiLuzio:2021pxd}.

Sub-eV mass axions or ALPs with large decay constants are excellent dark matter (DM) candidates. They are produced non-thermally in the 
early universe via the misalignment mechanism \cite{Preskill:1982cy,Abbott:1982af,Dine:1982ah} and, in case of post-inflationary breaking of the Abelian global symmetry, also by the decay of topological defects such as cosmic strings and domain walls. Further mechanisms of their production such as kinetic misalignment~\cite{Co:2019jts} and trapped misalignment~\cite{DiLuzio:2021gos} have also been explored recently.  The relevant parameters 
for the relic density are the mass $m_a$ and the decay constant $f_a$ and for pre-inflationary symmetry breaking also the 
initial value of the ALP field in units of $f_a$, $-\pi\leq \theta_i\leq \pi$, 
attained in the causally connected region which evolved into today's observable universe.
For example, 
for an ALP with temperature independent mass $m_a$, the fractional energy density of ALP from the misalignment mechanism is
predicted to go like \cite{Arias:2012az}
\begin{equation}
\Omega_a h^2 
 \approx
 0.12\,
\left(\frac{m_a}{{\rm 10\, \mu eV}}\right)^{1/2}
\left(\frac{f_a}{4.7\times 10^{12}\,{\rm GeV}}\right)^{2}\,
\theta_i^2\,,
\end{equation}
while, for the canonical axion, which obtains its mass from the temperature dependent topological susceptibility~\cite{Borsanyi:2016ksw},
\begin{equation}
\label{axion:eq:eq16}
\begin{aligned}
\Omega_a h^2 
 \approx
 0.12\,\left(\frac{f_a}{9\times 10^{11}~{\rm GeV}}\right)^{1.165}\,\theta_i^2
\approx 0.12\,\left(\frac{6~\mu{\rm eV}}{m_A}\right)^{1.165}\,\theta_i^2\,.
\end{aligned}
\end{equation}
Generically, a wide range in $f_a$ and $m_a$ can be consistent with axions or ALPs being the main component 
of dark matter. 
Heavy axions or ALPs with too large two-photon coupling, however, cannot be the dark matter, since their lifetime,  
\begin{equation}
\label{axion:eq:eq6}
\tau_{a\to\gamma\gamma}
 =\frac{64\,\pi}{{g_{a\gamma}^2m_a^3}}
=1.3\times 10^{16}\,{\rm s} 
\left(  \frac{10^{-10}\,{\rm GeV}^{-1}}{g_{a\gamma}}\right)^2
\left(  \frac{{\rm keV}}{m_a}\right)^3\,,
\end{equation}
tends to be smaller than the age of the universe.

\clearpage
\subsection{Axions beyond the axion-band}
\label{ssec:agrawal}
{\it Author: Prateek Agrawal, <prateek.agrawal@physics.ox.ac.uk>} 
\subsubsection{The QCD axion}
\label{sssec:agrawal-qcd-axion}

The QCD axion is a very well motivated candidate for new physics. It elegantly explains why the strong-CP angle is small, and at the same time is an excellent dark matter candidate. Thus, it solves two puzzles in the standard model in an economical framework. 
Additionally, the QCD axion model is highly predictive, and there is a narrow range of expected parameters.

\vskip 2mm
The solution of the strong CP problem relies on the fact that the axion receives almost all of its potential from non-perturbative QCD effects. In the simplest models, this fixes the axion mass in terms of its decay constant.
The QCD axion mass is generated by non-perturbative QCD dynamics, and is determined by its coupling to gluons,
\begin{align}
  \mathcal{L}
  &=
  \frac{\alpha_s}{8\pi} \frac{a}{f_a} 
  G^a_{\mu \nu} \widetilde{G}^{a\mu\nu}
  \\
  \Rightarrow
  m_a^2 &\simeq \frac{f_\pi^2 m_\pi^2}{f_a^2}
  \,.
\label{eq:QCDcouplingdef}
\end{align}

The DM abundance of the QCD axion is also quite predictive. It depends on a few discrete choices -- here we will assume that the PQ breaking happens before inflation. In that case, it depends on the initial misalignment angle of the axion at the end of inflation,
\begin{align} 
  \Omega_{a} h^2 
  \sim
  0.1
    \left(
    \frac{f_a} {10^{12}~{\rm GeV}} 
    \right)^{7/6} \theta_i^2 
  \label{eq:axion-abundance}
  \,.
\end{align} 
(In the region $f_a > 10^{15}$~GeV, the expression is modified slightly.)
Thus, the QCD axion has an extremely interesting and relatively narrow target space. We have assumed that
the PQ symmetry is broken before inflation, but there is
an analogous target space for QCD axions where PQ symmetry is broken after inflation.

\vskip 2mm
While it is fortunate the axion provides us with a narrow
target, it is worth exploring the boundaries of this statement -- how difficult is it for extended models to modify some of these predictions? We will list below a few possibilities that qualitatively change the prediction for axion phenomenology.

\subsubsection{Dark matter at higher $f_a$}
\label{sssec:agrawal-dm}
\begin{figure}[htbp]
  \centering
  \includegraphics[width=0.47\textwidth]{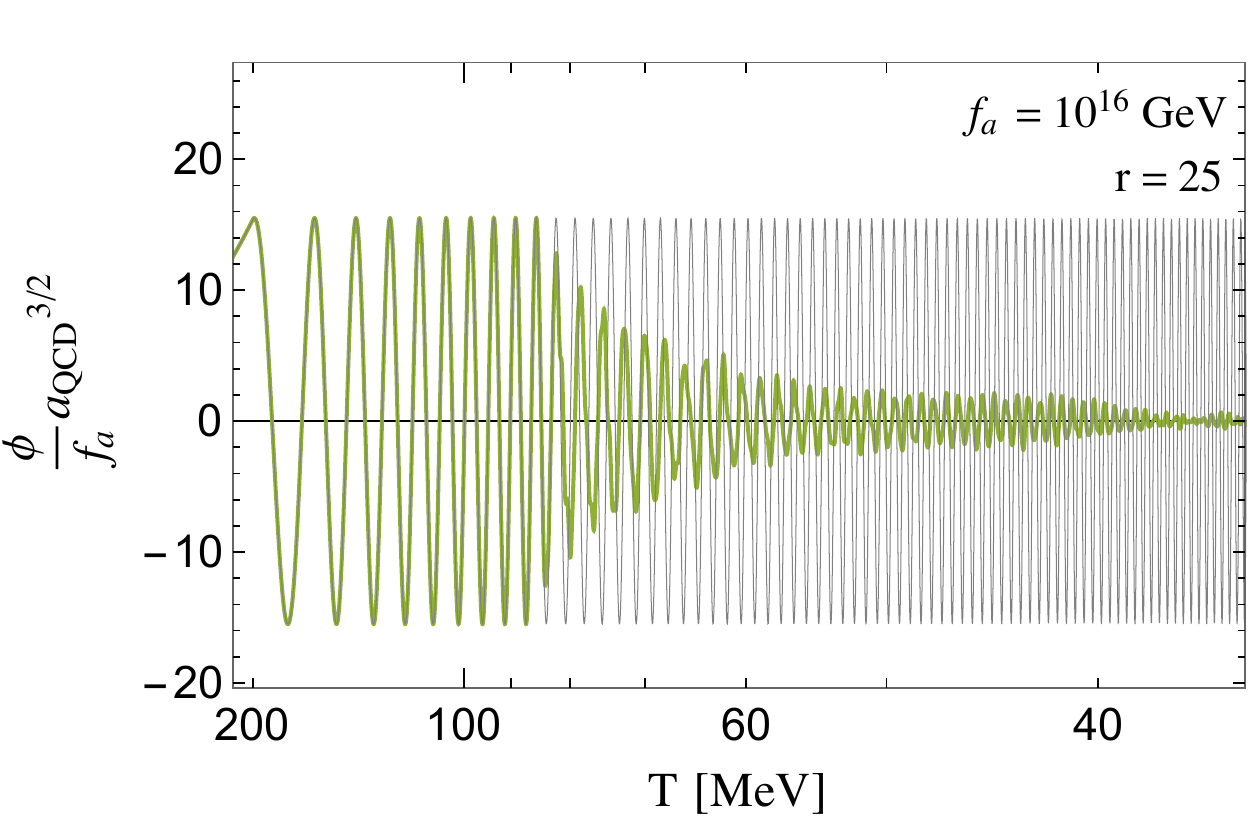}
  \qquad
  \includegraphics[width=0.45\textwidth]{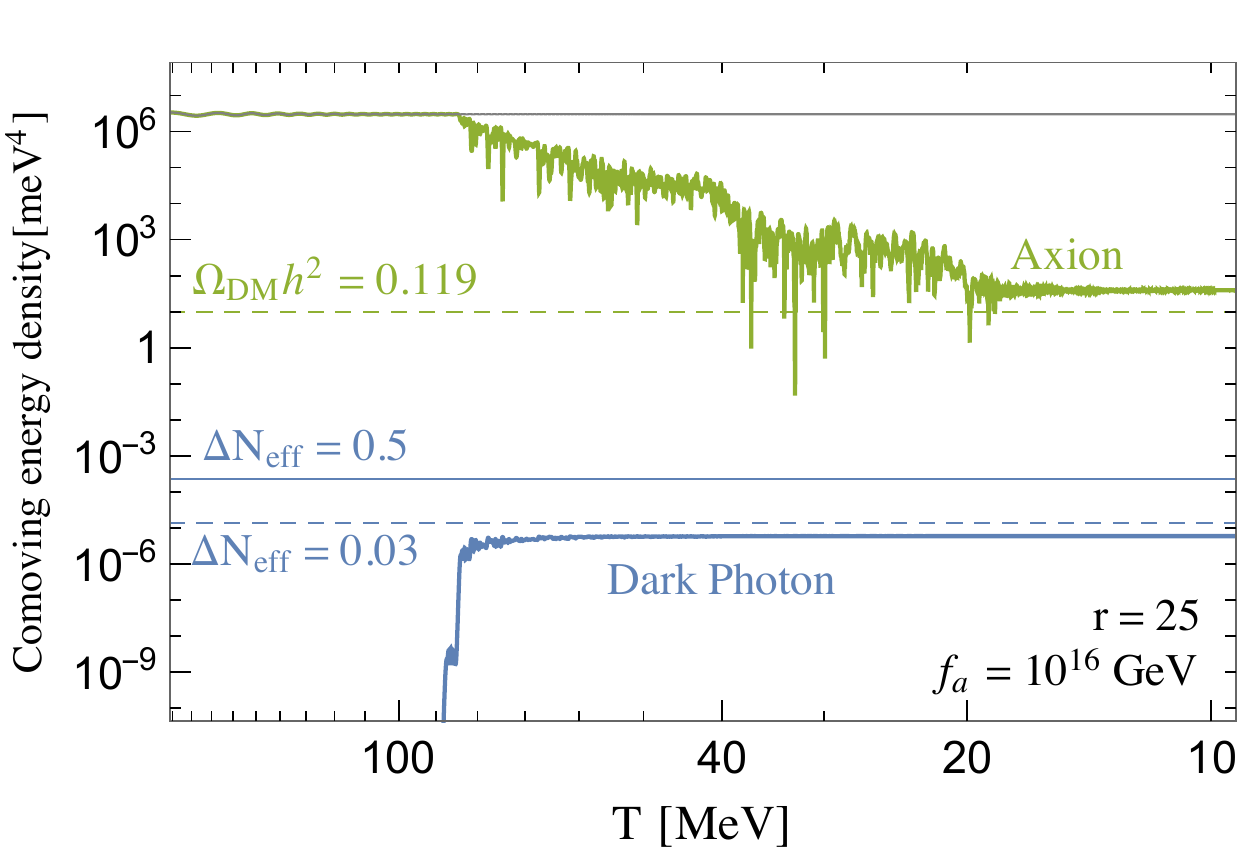}
  \caption{Cosmological evolution of the axion field (denoted $\phi$
    here) and the rescaled comoving energy density of the axion and the
    dark photon field, which are defined to scale with the scale factor as
    $\rho a^3$ and $\rho a^4$ respectively. Backscattering effects have been neglected for this plot, see~\cite{Agrawal:2017eqm,Kitajima:2017peg} for more details. 
  }
  \label{fig:tachyon}
\end{figure}
In many theoretical models the decay constant of the axion is closer
to the GUT scale, $f_a \sim 10^{16}$ GeV. In this case, generically there would
be too much dark matter, unless the initial misalignment angle is tuned. Additionally, there are experimental techniques which are relevant to find axions in the mass range dictated by large $f_a$.
Thus, it is important to understand any mechanisms that lead to a viable cosmology at large $f_a$. 

We will focus on a particular mechanism of particle production. The
axion is assumed to couple to a dark photon, 
\begin{align}
  \mathcal{L}
  &=
  \frac{\alpha_s}{8\pi f_a} \phi \, G^{a,\mu\nu} \widetilde{G}^a_{\mu\nu}
  +
  \frac{\alpha_d r }{8\pi f_a} \phi  \,F_D^{\mu\nu} (\widetilde{F}_D)_{\mu\nu}
  \label{eq:axionefflagrangian}
  \,,
\end{align}
with a coupling that is $r \sim
\mathcal{O}(10)$
times larger than its coupling to the standard model particles. 
This coupling is responsible for depleting the axion abundance.

As the axion starts oscillating, the dark photon develops an instability in the
oscillating axion background, leading to copious production of dark photons in a short duration of time.
Treating the homogeneous axion field as a classical field, the equation of motion for the gauge field is,
\begin{align}
  A_\pm'' + \left(k^2 \pm \frac{k \phi'}{f_d}\right) A_\pm &=0\,,
\end{align}
where $A_\pm$ are the helicities of the gauge field. 
If the effective frequency $
  \omega^2(k, \phi') 
  = 
  k^2 \pm \frac{k \phi'}{f_d}
$ of a given mode becomes
negative, the corresponding mode becomes tachyonic which can lead to
an exponential growth of the corresponding ``mode function''.

The exponential growth extracts energy from the axion field
effectively, and depletes the axion abundance substantially. This can
make higher-$f_a$ axion models viable.

In figure~\ref{fig:tachyon} we show a benchmark parameter point where
a dramatic reduction of the axion abundance can be achieved. This
calculation ignores the effect of backscattering of the dark photons
to produce high-$k$ axions; this effect was incorporated in a lattice
study in~\cite{Kitajima:2017peg}.

\subsubsection{Enhanced coupling to photons}
\label{sssec:agrawal-photon-coupling}
The mass and the decay constant of the axion set much of its phenomenology. An important coupling of the axion is to the photon,
for experiment. This coupling is given as
\begin{align}
  \mathcal{L}_{a\gamma\gamma}
  &=
  \frac{e^2}{32 \pi^2 f_a} 
  \left(\frac{E}{N}-1.92\right)
  a F_{\mu\nu} \widetilde{F}^{\mu\nu}
  \equiv
  \frac14 g_{a\gamma\gamma}
  \frac{a}{f_a} F_{\mu\nu} \widetilde{F}^{\mu\nu}
  \label{eq:gagg}
  \,,
\end{align}
where $E,N$ are anomaly integers and the correction arises from
axion-meson mixing. In the simplest models, $E/N \sim 1$.

The clockwork mechanism or the Kim-Nilles-Peloso mechanism 
provides an interesting way to enhance the
coupling to photons. 
The KNP alignment mechanism has been proposed and studied extensively
in the natural inflation context~\cite{Kim:2004rp,Choi:2014rja,Tye:2014tja,Kappl:2014lra,Ben-Dayan:2014zsa,Bai:2014coa,delaFuente:2014aca}. 
In the KNP scenario, we have
two axion fields $a(x)$ and $b(x)$, one of which couples to photons
and the other to the QCD gluons. The basic
mechanism could be described by the Lagrangian
\begin{align}
\mathcal{L} 
= 
\frac{a}{f_a}
\frac{\alpha}{8\pi}
F \widetilde{F}
+
\frac{a}{f_a}
\frac{\alpha_s}{8\pi}
G^a \widetilde{G}^a
+
\Lambda^4 \cos\left(\frac{a}{f_a} + \frac{Q b}{f_b} \right)\,.
\end{align}
The light eigenstate is
\begin{align}
  \phi
  &=
  \frac{-Q f_a a + f_b b}{\sqrt{f_b^2 + Q^2 f_a^2}}\,,
\end{align}
and couples to photons with an enhanced coupling,
\begin{align}
  \mathcal{L}
  &=
  \frac{Q\phi}{f_{\rm eff}}
  \frac{\alpha}{8\pi}
  F\widetilde{F}
  +
  \frac{\phi}{f_{\rm eff}}
  \frac{\alpha_s}{8\pi}
  G^a\widetilde{G}^a\,,
\end{align}
where $f_{\rm eff} = \sqrt{Q^2 f_a^2 + f_b^2}$.

The clockwork mechanism operates on a series of aligned potentials
with $n$ axions~\cite{Choi:2015fiu,Kaplan:2015fuy,Farina:2016tgd},
\begin{align}
  \mathcal{L}
  &=
  \frac{a_1}{f}\frac{\alpha}{8\pi}
  F\widetilde{F}
  +\Lambda^4\left(\frac{a_1}{f} + \frac{Q a_2}{f} \right) 
  +\ldots \\ \nonumber
  & +\Lambda^4\left(\frac{a_{n-1}}{f} + \frac{Q a_n}{f} \right)
  +\frac{a_n}{f}\frac{\alpha_s}{8\pi} G^a\widetilde{G}^a\,.
\end{align}
The light eigenstate in this case is,
\begin{align}
  \phi
  \approx
  a_1 + \frac{a_2}{Q}+\ldots+\frac{a_n}{Q^{n-1}}\,.
\end{align}
This leads to an exponential enhancement of the coupling of the light
eigenstate to the photon,
\begin{align}
  \mathcal{L}
  &=
  \frac{Q^{n-1} \phi}{f_{\rm eff}}
  \frac{\alpha}{8\pi} F \widetilde{F}
  +
  \frac{\phi}{f_{\rm eff}}
  \frac{\alpha_s}{8\pi} G^a \widetilde{G}^a\,.
\end{align}
Thus, these extended models motivate experiments looking for the QCD
axion far from the standard QCD band.

\begin{figure*}[t]
  \centering
  \includegraphics[width=0.65\textwidth]{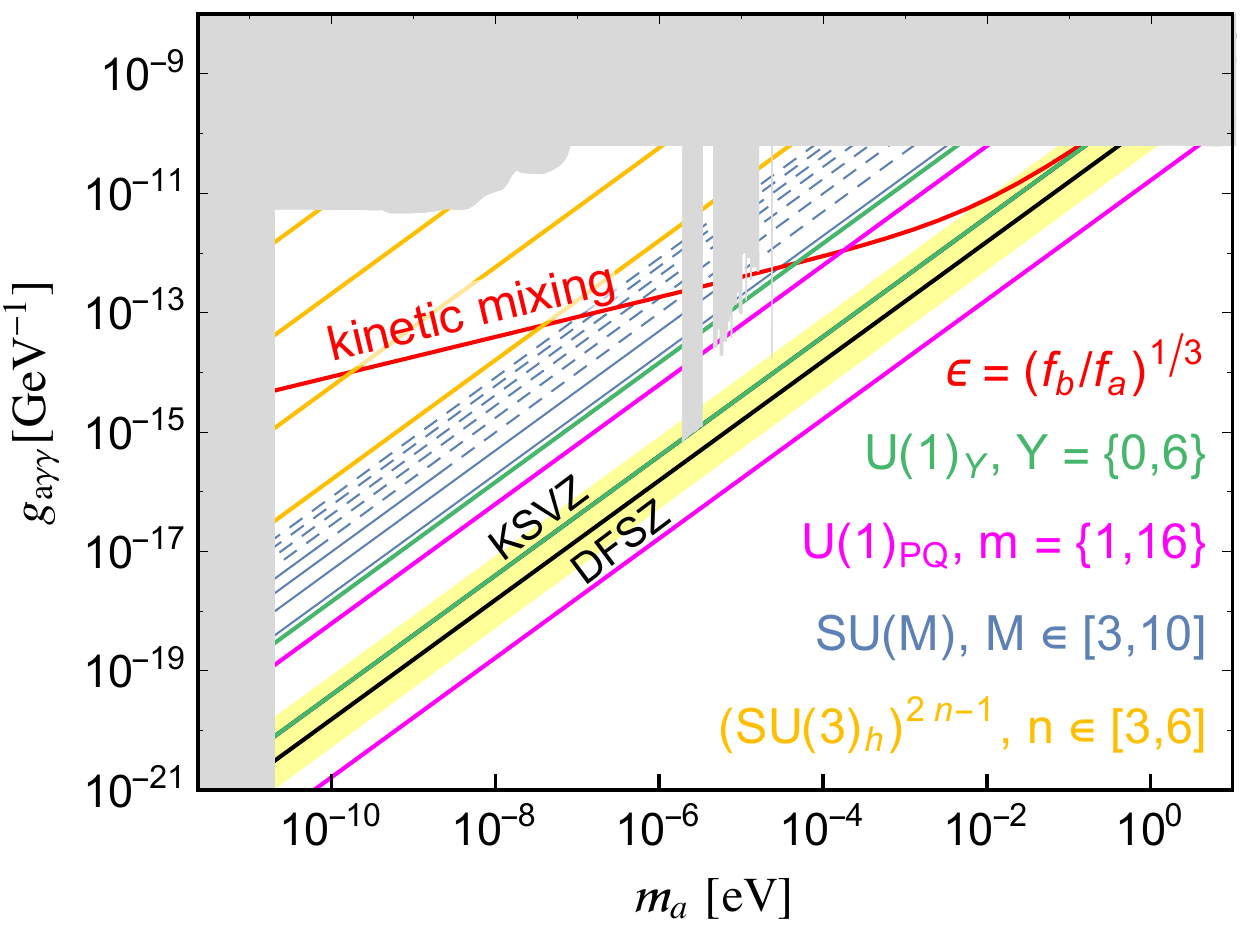}
  \caption{Enhancement of the photon axion coupling in a number of
  extended QCD axion models. Details of the models can be found
in~\cite{Agrawal:2017cmd}. The clockwork mechanism (yellow solid
lines) provides a dramatic
enhancement, making the QCD axion viable in the entire parameter space.}
  \label{fig:gagg}
\end{figure*}

\vskip 2mm
There are a variety of other mechanisms that have been used to enhance the photon-axion coupling.
In figure~\ref{fig:gagg}, we show the enhancement of the axion-photon coupling relative to the standard QCD axion bands in various cases (details can be found in~\cite{Agrawal:2017cmd}).
We have focused on the clockwork mechanism in this note since that is the mechanism that provides a model basis for the entire
$m_a$--$g_{a\gamma\gamma}$ plane.

\subsubsection{Heavier QCD axions}
\label{sssec:agrawal-heavy-axions}
In addition to changing the couplings and the decay constants of the axion, it is also possible to change the relationship between the mass and the decay constant of the QCD axion.
This occurs for an extension of the QCD axion framework where instead of a single axion relaxing the vacuum to $\theta = 0$, two or more axions naturally cooperate to solve the strong CP problem. Each of these axions can have a mass much larger than the standard QCD axion, motivating searches for axions in the $(m_a, f_a)$ plane outside of the
QCD axion window.  

\vskip 2mm
This multi-axion solution
to the strong CP problem arises in models that extend the low-energy $SU(3)_c$ gauge group to be the diagonal subgroup of a parent
$SU(3) \times SU(3) \times \ldots$ product gauge group, which is broken down to $SU(3)_c$ at some high scale $M$. 
As an example, consider a simple extension of the SM where the QCD gauge group emerges from Higgsing a product group 
\begin{align}
SU(3)_{1}\times SU(3)_{2} \to SU(3)_c 
\,. 
\end{align}
at a scale $M \gg$ TeV.
The SM quarks are charged only under the $SU(3)_1$ gauge factor, and there are no
fermions charged under $SU(3)_2$. 
We take the theory to have two spontaneously broken anomalous $U(1)_{PQ}$ symmetries at scales $f_i > M$, giving an axion
$a_{1,2}$ in each $SU(3)$ sector:
\begin{align}
  \mathcal{L} 
  &= 
  -\frac14 (G_1)^a_{\mu\nu}\ (G_1)^{a,\mu\nu}
  +\frac{g^2_{s1}}{32\pi^2}
  \left( \frac{a_1}{f_1} - \theta_1 \right)
  (\widetilde{G}_1)^a_{\mu\nu}\ (G_1)^{a,\mu\nu}
  \nonumber\\
  &-\frac14 (G_2)^a_{\mu\nu}\ (G_2)^{a,\mu\nu}
  +\frac{g^2_{s2}}{32\pi^2}
  \left( \frac{a_2}{f_2} - \theta_2 \right)
  (\widetilde{G}_2)^a_{\mu\nu}\ (G_2)^{a,\mu\nu}\,.
\end{align}
The gauge couplings $g_{s{1,2}}$ and theta angles $\theta_{1,2}$ are independent parameters not related to each other by any symmetries.
The presence of two independent axion degrees of freedom will allow both physical $\theta$-angles to be dynamically removed. 

The theory will match to the SM at a scale $M$ where the $SU(3)_1 \times SU(3)_2$ gauge group is Higgsed to a diagonal $SU(3)_{c}$. 
The effective Lagrangian generated by
short-distance non-perturbative effects in the individual $SU(3)_1$ and $SU(3)_2$
factors is~\cite{tHooft:1976snw,Callan:1977gz,Andrei:1978xg}
\begin{align}
  \mathcal{L}_a
  &=
  \Lambda_1^4 \cos \left(\frac{a_1}{f_1}  - \bar\theta_1\right) +
  \Lambda_2^4 \cos \left(\frac{a_2}{f_2} - \bar\theta_2\right) +
  \nonumber \\
  & \frac{g_s^2}{32\pi^2}
  \left(\left( \frac{a_1}{f_1} - \bar{\theta}_1\right) 
  +\left(\frac{a_2}{f_2} - \bar{\theta}_2\right) \right) G\tilde{G}\,.
  \label{eq:LUV}
\end{align}

The new short-distance non-perturbative contributions to the axion potentials are exactly aligned to remove the effective $\theta$-angle by a generalization of the Vafa-Witten argument \cite{Vafa:1984xg}. 
We can estimate how the mass of the two axions is affected by these UV non-perturbative contributions to the potential.

\vskip 2mm
For brevity, we provide rough estimates for the mass scales. In a particular model, precise calculation for the modifications to the axion mass can be calculated in e.g.~the constrained instanton formalism (see \cite{Fuentes-Martin:2019bue,Csaki:2019vte}).
In the $SU(3)_2$ factor, where no colored fermions are present, the axion potential is suppressed only by the non-perturbative instanton action. The scale can be estimated as~\cite{tHooft:1976snw,Callan:1977gz,Andrei:1978xg}
\begin{align}
  \Lambda_2^4 
  &\simeq
  D[\alpha_{s_2}(M)] M^4 \,,
\end{align}
where the dimensionless instanton density depends non-perturbatively on the running gauge coupling as
\begin{align}
  D[\alpha] 
  &= 0.1
  \left(\frac{2\pi}{\alpha}\right)^6
  e^{-\frac{2\pi}{\alpha}}\,,
\end{align}
with $\alpha$ denoting the running coupling evaluated at the scale $M$. We have
made the approximation that the contribution is dominated by the instantons at the scale $M$. In the $SU(3)_1$ factor, there is a further suppression due to the Yukawa couplings and Higgs loops, which can be estimated as~\cite{Flynn:1987rs,Choi:1998ep}
\begin{align}
  \Lambda_1^4 
  \approx 
  K D[\alpha_{s_1}(M)] M^4\,,
  \label{eq:Lambda1}
\end{align}
where $K$ is a chiral suppression factor capturing the breaking of the $U(1)^6$ axial symmetry of the individual quarks by the Yukawa couplings,
\begin{align}
  K 
  &= 
  \left(\frac{y_u}{4\pi}\right)
  \left(\frac{y_d}{4\pi}\right)
  \left(\frac{y_c}{4\pi}\right)
  \left(\frac{y_s}{4\pi}\right)
  \left(\frac{y_t}{4\pi}\right)
  \left(\frac{y_b}{4\pi}\right) 
  \approx 10^{-23}\,.
\end{align}
Of course, $\alpha_{s_1}$ and $\alpha_{s_2}$ are related by the matching condition to the QCD coupling at the scale $M$,
\begin{align}
  \frac{1}{\alpha_{s_1}}
  +
  \frac{1}{\alpha_{s_2}}
  =
  \frac{1}{\alpha_{s}}
  \,.
  \label{eq:GaugeMatching}
\end{align}

The sensitivity to the large scale $M$ can overcome the non-perturbative and
chiral suppression factors. The relation  equation~\eqref{eq:GaugeMatching}
between $\alpha_{s_1}$ and $\alpha_{s_2}$  implies that
$\Lambda_1$ grows as $\Lambda_2$ shrinks. 
The most interesting regime is when $\Lambda_1, \Lambda_2  \gg
\Lambda_0$ -- in this case there will be no light state resembling the standard QCD axion. This regime generally corresponds to $\alpha_{s_1} > \alpha_{s_2}$ to compensate for the extra chiral suppression factor in $\Lambda_1$.  For gauge groups with additional $SU(3)$ factors, the gauge couplings in each gauge factor is larger, and the masses of each of the axions can easily be parametrically larger than the QCD axion.


\clearpage
\subsection{Searches for axions/ALPs in the sub-eV range: Experimental review}
\label{ssec:irastorza}
{\it Author: Igor Irastorza, <Igor.Irastorza@cern.ch>} 
\subsubsection{Introduction}
\newcommand{\gagamma}{g_{a\gamma}}

Axion-like particles (ALPs) appear in many extensions of the Standard Model (SM), typically those with the spontaneous breaking of one or more global symmetries at high energies. ALP models are invoked in attempts to solve shortcomings of the SM, but also of cosmological or astrophysical unexplained observations. Most relevantly, ALPs are ideal dark matter candidates. In addition, and not exhaustively, ALPs have been invoked to solve issues as diverse as the hierarchy problem in the SM, the baryon asymmetry of the Universe, inflation, dark energy, dark radiation, or to explain the anomalous cooling observed in several types of star (see recent reviews \cite{Irastorza:2018dyq,DiLuzio:2020wdo} for references and also section~\ref{ssec:giannotti} for a brief discussion). The QCD axion is the prototype particle of this category, proposed long ago to solve the strong-CP problem of the SM. Still the most compelling solution to this problem, it  remains maybe the strongest theoretical motivation for the pseudoscalar portal. 

\vskip 2mm
Typical axion models are constrained to very small masses below $\sim$1~eV. Because of that, signatures of these particles are not expected at accelerators, and novel specific detection techniques are needed. The particular combination of know-hows needed for these experiments, some of them not present in typical HEP groups (and including, among others, high-field magnets, superconducting magnets, RF techniques, X-ray optics  \& astronomy, low background detection, low radioactivity techniques, quantum sensors, atomic physics, etc...), and their effective interplay with axion particle physicists is an important challenge in itself. We will focus here on the detection efforts of these low-energy axions\footnote{In this section, the term axion will be used to refer to ALPs too.}. 

\subsubsection{Detecting low energy axions}
\vskip 2mm
Most (but not all) of the axion detection strategies rely on the axion-to-photon coupling $\gagamma$. This is due to the fact that this coupling is generically present in most axion models, as well as that coherence effects with the electromagnetic field are easy to exploit to increase experimental sensitivity. The different experimental approaches can be categorized in three groups, depending on the source of axions invoked:

\begin{itemize}
    \item Experiments looking for axion or axion-induced effect purely in the laboratory.
\item Haloscopes looking for axions as local dark matter constituents.
\item Helioscopes searching for axions emitted by the sun.

\end{itemize}

\vskip 2mm
Purely laboratory-based experiments constitute the most robust search strategy, as they do not rely on astrophysical or cosmological assumptions. However, their sensitivity is hindered by the low probability of photon-axion-photon conversion in the lab. Haloscopes and helioscopes take advantage from the enormous flux of axions expected from extraterrestrial sources. Because of this, they are the only techniques having reached sensitivity down to QCD axion couplings. Haloscopes rely on the assumption that 100\% of the dark matter is in the form of axions, and in the case of a subdominant axion component their sensitivity should be rescaled accordingly. Helioscopes rely on the Sun emitting axions, but in its most conservative channel (Primakoff conversion of solar plasma photons into axions) this is a relatively robust prediction of most models, relying only on the presence of the $\gagamma$ coupling. 

\vskip 2mm
In the following section, we briefly review the status of these three experimental ``frontiers''. Figure~\ref{fig:large_panorama} shows the overall panorama of experimental and observational bounds on the $\gagamma$-$m_a$ plane. For a description of the latter, we refer to~\cite{Irastorza:2018dyq}.

\subsubsection{Laboratory experiments}

\vskip 2mm
\noindent
{\bf LSW} 

The most well-known laboratory technique to search for axions is photon regeneration in magnetic fields, colloquially known as \textit{light-shining-through-walls} (LSW). A powerful source of photons (e.g.\ a laser) is used to create axions in a magnetic field. Those axions are then reconverted into photons after an optical barrier. 

\vskip 2mm
A number of LSW experiments has been carried out in the past \cite{Redondo:2010dp},  
all of them producing limits to $\gagamma$ in the ballpark of 10$^{-6}$--10$^{-7}$~GeV$^{-1}$. 
Currently two active collaborations are working on LSW experiments and have produced the most competitive bounds below 10$^{-7}$~GeV$^{-1}$: The ALPS~\cite{Ehret:2010mh} experiment at DESY and the OSQAR~\cite{Ballou:2015cka} experiment at CERN, both making use of powerful accelerator dipole magnets, from HERA and LHC accelerators respectively. 
ALPS features power build-up in the production region, while OSQAR has slightly higher magnet and laser parameters.

\begin{figure*}[t]
\centering
\includegraphics[width=0.7\linewidth]{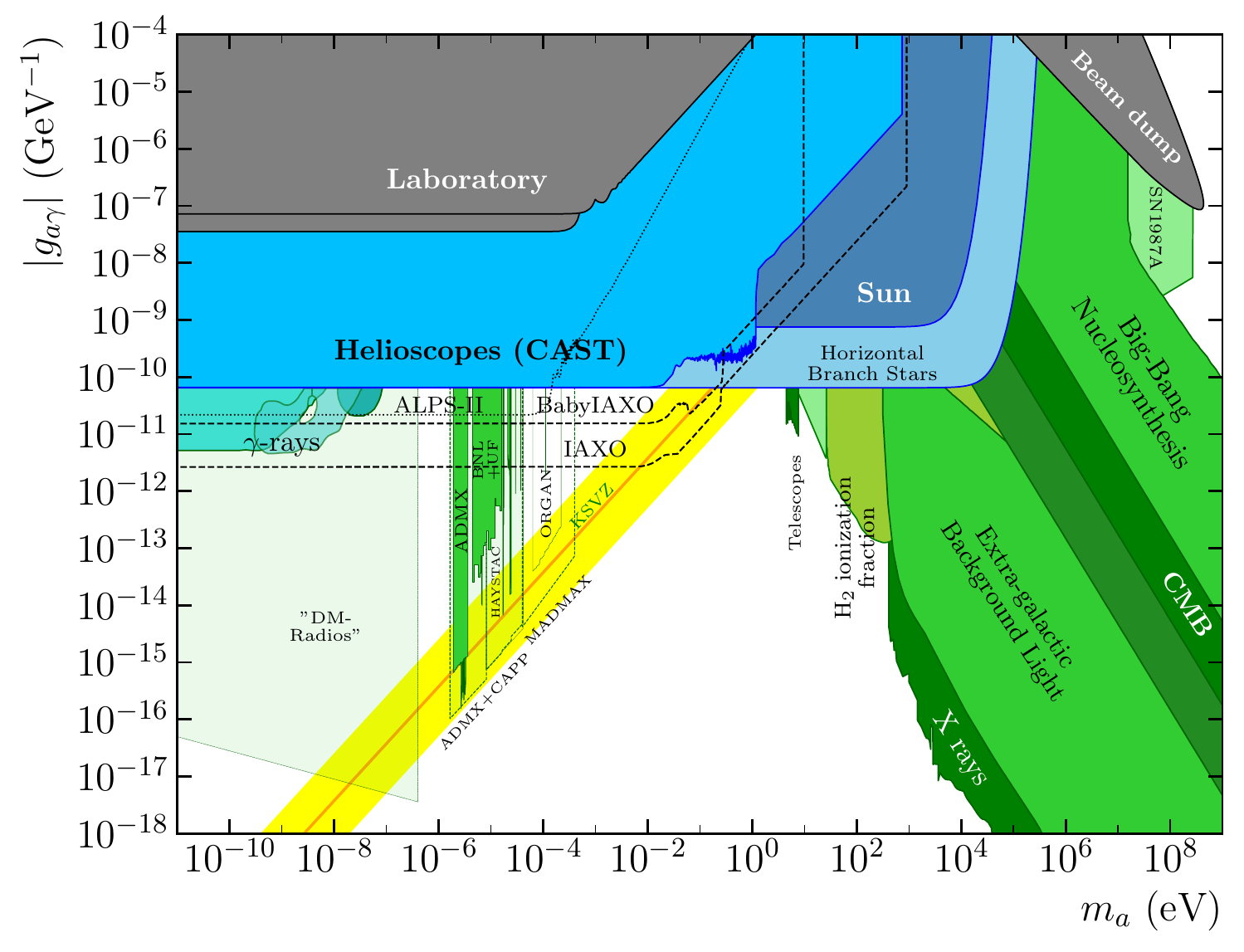}
\caption{Overall panorama of current bounds (solid area) and future prospects (semi-transparent areas or dashed lines) in the $\gagamma$-$m_a$ plane. See \cite{Irastorza:2018dyq} for details on the different lines}.
\label{fig:large_panorama}
\end{figure*}

These results can be improved by implementing resonant regeneration schemes. If adequately matched Fabry-Perot resonators are used in both the generation and conversion parts, the conversion probability can be boosted several orders of magnitude. However, it poses challenging requirements on the optical system.  The ALPS-II experiment~\cite{Bahre:2013ywa}, currently finishing construction at DESY, will be the first laser LSW using resonant regeneration in a string of 2$\times$10 HERA magnets (i.e.\ a length of 2$\times$100 m) for the production and the conversion regions. The expected sensitivity of ALPS-II goes down to $\gagamma < 2\times 10^{-11}$~GeV$^{-1}$ for low $m_a$, and will be the first laboratory experiment to surpass current astrophysical and helioscope bounds on $\gagamma$ at low $m_a$, partially testing ALP models hinted by the excessive transparency of the Universe to ultrahigh-energy photons (see Fig.~\ref{fig:large_panorama}).

\vskip 2mm
A more ambitious extrapolation of this experimental technique is conceivable, for example, as a byproduct of a possible future production of a large number of dipoles like the one needed for the Future Circular Collider (FCC). This is the idea behind JURA, a long-term possibility discussed in the Physics Beyond Colliders study group~\cite{PBC}. JURA contemplates a magnetic length of almost 1 km, and would suppose a further step in sensitivity of more than one order of magnitude in $\gagamma$ with respect to ALPS-II. 

\vskip 2mm
LSW experiments with photons at frequencies other than optical have also been performed. The most relevant result comes from the CROWS experiment at CERN~\cite{Betz:2013dza}, a LSW experiment using microwaves~\cite{Hoogeveen:1992nq}. Despite the small scale of the experiment, its sensitivity approached that of ALPS or OSQAR, thanks to the resonant regeneration, more easily implemented in microwave cavities. A large-scale microwave LSW experiment has been discussed in the literature~\cite{Capparelli:2015mxa}. LSW experiments have also been performed with intense X-ray beams available at synchrotron radiation sources \cite{Battesti:2010dm,Inada:2013tx}. However, due to the relative low photon number available and the difficulty in implementing high power build-ups at those energies,  X-ray LSW experiments do not reach the sensitivity of optical or microwave LSW.

\vskip 2mm
\noindent
{\bf Polarization experiments}

Laser beams traversing magnetic fields offer another opportunity to search for axions. The photon-axion oscillation in the presence of the external $B$-field has the effect of depleting the polarization component of the laser that is parallel to the $B$-field (dichroism), as well as phase-delaying it (birefringence), while leaving the perpendicular component untouched. The standard Euler-Heisenberg effect in QED (also dubbed \textit{vacuum magnetic birefringence}) would be a (still unobserved) background to these searches. The most important experimental bound from this technique comes from the PVLAS experiment in Ferrara~\cite{DellaValle:2015xxa}, reaching a sensitivity only a factor of $\sim$8 away from the QED effect~\cite{DellaValleTrento}. The BMV collaboration in Toulouse~\cite{Hartman:2017nez}, as well as OSQAR at CERN have reported plans to search for the QED vacuum birefringence. Recently, efforts towards an enhanced experiment of this type, dubbed VMB@CERN, are being discussed in the context of the Physics Beyond Colliders initiative at CERN. 

\vskip 2mm
\noindent
{\bf Fifth force experiments}

Although very different from the above examples, experiments looking for new macroscopic forces (e.g.\ torsion balance experiments, among many others) could in principle be sensitive to axion effects in a purely laboratory setup. Axion-induced forces via e.g.\ a combination of axion-fermion couplings, could compete with gravity at $\sim 1/ m_a$ scales. However, the interpretation of current bounds in terms of limits to axion couplings are typically not competitive with astrophysical bounds or electric dipole moment (EDM)  limits on CP-violating terms (see \cite{Irastorza:2018dyq} for a recent discussion). The recently proposed ARIADNE experiment intends to measure the axion field sourced by a macroscopic body using nuclear magnetic resonance (NMR) techniques~\cite{Arvanitaki:2014dfa} instead of measuring the force exerted to another body. Very relevantly, ARIADNE could be sensitive to CP-violating couplings well below current EDM limits, in the approximate mass range 0.01 to 1~meV. Therefore, it could be sensitive to QCD axion models with particular assumptions; most importantly, they should include beyond-SM physics leading to a CP-violating term much larger that the expected SM contribution. Because of this assumption, ARIADNE would not allow for a firm model-independent exclusion of the axion in this
mass interval.

\subsubsection{DM axion experiments}
If our galactic dark matter halo is totally made of axions, the number density of these particles around us would be huge. In addition, these dark matter axions would be non-relativistic particles (with a velocity roughly equal to the virial velocity inside our galaxy). Therefore this population behaves like a classical field that is
coherent over a length approximately equal to their de~Broglie wavelength, which for typical axion
masses is at the meter or even larger scales (see  Section~\ref{ssec:stadnik} for an in depth discussion). This 
allows to employ coherent techniques for their detection, further enhancing sensitivities. In the following a brief review of the current efforts is presented (see~\cite{Irastorza:2018dyq} for a more detailed review).

\vskip 2mm
\noindent
{\bf Classical haloscopes}

\begin{figure*}[t]
\centering
\includegraphics[scale=0.7]{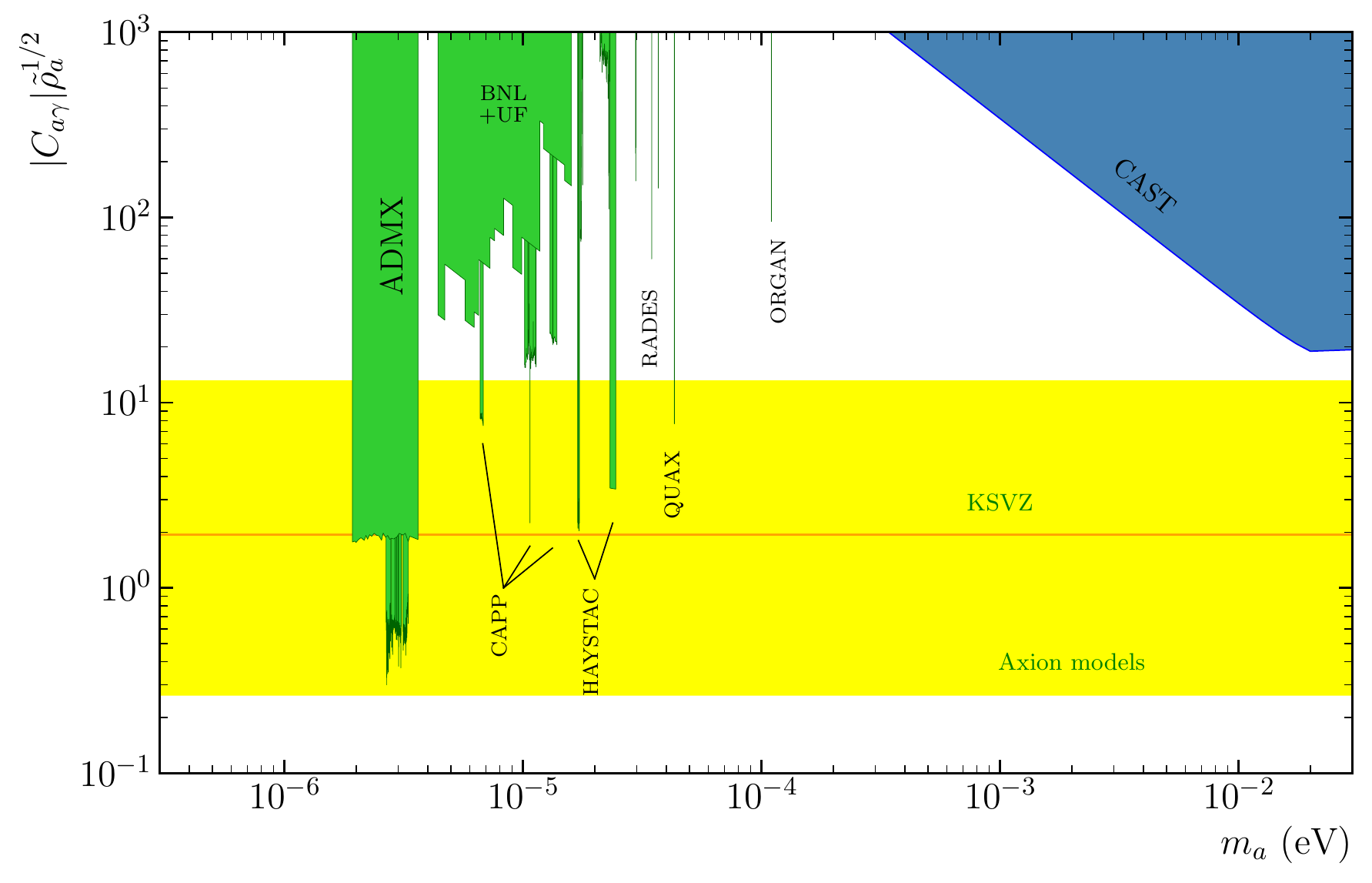}
\caption{Zoom-in of the region of parameters where most axion dark matter experiments are active (in green). The $y$-axis shows the adimensional coupling $C_{a\gamma} \propto \gagamma / m_a$ (scaled with the local DM density $\tilde{\rho}_a$, to stress that these experiments produce bounds that are dependent on the assumed fraction of DM in the form of axions). Thus, the yellow region, where the conventional QCD axion models are, appears now as a horizontal band, but is the same yellow band shown in the other plots of this section.}
\label{fig:haloscopes}
\end{figure*}

The conventional axion haloscope technique~\cite{Sikivie:1983ip} involves a high quality factor $Q$ microwave cavity inside a magnet. Since dark matter axions are non-relativistic their energy
is essentially equal to their mass and they  convert into almost monochromatic photons, with a very small spread in frequency $\sim 10^{-6}$. 
If the axion energy matches the cavity frequency, the conversion is enhanced by a factor $Q$. This enhancement allows to reach sensitivity down to very low coupling values, but only for a very thin interval of masses at the resonant frequency. Given that the axion mass in unknown, this technique requires the implementation of a scanning procedure (usually involving very precise mechanical modification of the cavity geometry). Covering a wide mass range poses a experimental challenge.

\vskip 2mm
Efforts in this direction have been led by the ADMX collaboration, which has pioneered many
relevant technologies (high-Q cavities inside magnetic fields, RF detection close to the quantum limit, and others). The main ADMX setup includes a 60~cm diameter, 1~m long cavity inside a solenoidal $\sim$8~T magnet. The cavity can be tuned in frequency by the precise movement of some dielectric rods. With this setup, ADMX has achieved sensitivity to axion models in the $\mu$eV range~\cite{Asztalos:2009yp}, and
is currently taking data to extend this initial result at lower temperatures (and thus lower noise). The latest results  released~\cite{Du:2018uak,Braine:2019fqb} include sensitivity down to pessimistically coupled axions in the 2.66--3.31~$\mu$eV range (see figure~\ref{fig:haloscopes}). Preliminary results exteding this range up to 4.2~$\mu$eV, with sensitivity at roughly half the DFSZ coupling, have been released this very year~\cite{Patras2021}.

\vskip 2mm
In recent years, a number of new experimental efforts are appearing, some of them implementing variations of the haloscope concept, or altogether novel detection concepts, making this subfield one of the most rapidly changing in the axion experimental landscape. Figure~\ref{fig:haloscopes} shows the current situation, with a number of new players accompanying ADMX in the quest to cover different axion mass ranges.
Applying the haloscope technique to frequencies considerably higher or lower than the one ADMX is targeting is challenging for different reasons. Lower frequencies imply proportionally larger cavity volumes and thus bigger, more expensive, magnets, but otherwise they are technically feasible. The use of large existing (or future) magnets has been proposed in this regard, e.g.\ the KLASH~\cite{Alesini:2019nzq} proposal at LNF,  ACTION~\cite{Choi:2017hjy} in Korea, or a possible haloscope setup in the the future (Baby)IAXO helioscopes~\cite{redondo_patras_2014}). 

\vskip 2mm
Higher frequencies imply lower volumes and correspondingly lower signals and sensitivity. This could be in part compensated by enhancing other experimental parameters (more intense magnetic fields, higher quality factors, noise reduction at detection, etc.). The HAYSTAC experiment~\cite{Kenany:2016tta} at Yale, born in part out of developments initiated inside the ADMX collaboration~\cite{Shokair:2014rna} has implemented a scaled-down ADMX-like setup, and has  been the first experiment proving sensitivity to QCD axion models in the decade in $m_a$ above ADMX, in particular in the mass range 23.15--24.0~$\mu$eV~\cite{Brubaker:2016ktl}. More recently, HAYSTAC has reported the first axion DM search with squeezed photon states~\cite{Backes:2020ajv}, which effectively allows to push the noise below the quantum limit~\cite{Malnou:2018dxn}, reaching sensitivity almost to the KSVZ coupling in the 16.96--17.28~$\mu$eV range. HAYSTAC is also pioneering analysis methodology~\cite{Brubaker:2017rna,Palken:2020wgs} in this type of searches. The QUAX-$a\gamma$ experiment in Frascati has also recently reported~\cite{Alesini:2020vny} a first result with a cavity resonating at an axion mass of 43~$\mu$eV, cooled down to 200 mK and read out with a Josephson parametric amplifier whose noise fluctuations are at the SQL, making this the axion haloscope having reached sensitivity to the QCD axion at the highest mass point.

\vskip 2mm
Another similar program is CULTASK, the flagship project of the recently created Center for Axion and Precision Physics (CAPP) in Korea. CAPP also hosts several other projects and R\&D lines, with the general long-term goal of exploring DM axions in the mass range 4--40~$\mu$eV. The first result from this line is the CAPP-PACE pilot experiment~\cite{Lee:2020ygx}, that has recently produced an exclusion in the 10.1--11.37~$\mu$eV range. A more recent result, from a larger setup CAPP-8TB~\cite{Lee:2020cfj} has produced another excluded region in the 6.62--6.82~$\mu$eV mass range with sensitivity down to the upper part of the QCD band. 
Even higher frequencies are targeted by the ORGAN~\cite{McAllister:2017lkb} program, recently started in the University of Western Australia. A first pathfinder run has already taken place~\cite{McAllister:2017lkb}, at a fixed frequency of 26.531 GHz, corresponding to  $m_a=110~\mu$eV. Several groups explore the possibility to increase of the cavity $Q$ by coating the inside of the cavity using a superconducting layer. In particular, this strategy has been implemented in the QUAX-$a\gamma$ setup, with which the QUAX collaboration has proven an improvement of a factor of 4 with respect a copper cavity, and has performed a single-mass axion search at about $\sim$37~$\mu$eV~\cite{Alesini:2019ajt}. Another strategy to reach higher frequencies is to select a higher order mode of the cavity than the one that couples with the axion field, albeit with a lower geometric factor. This has been done by the ADMX ``Sidecar'' setup, a testbed experiment living inside of and operating in tandem with the main ADMX experiment~\cite{Boutan:2018uoc}. 

\vskip 2mm
Higher frequencies eventually require to increase the instrumented volume, either by combining many similar phase-matched cavities, or by implementing more complex extended resonant structures that effectively decouple the detection volume $V$ from the resonant frequency. 
The former has already been done long ago for four cavities within the ADMX R\&D~\cite{Kinion:2001fp}, but going to a much larger number of cavities has been considered not feasible in practice. More recently, the CAST-CAPP project~\cite{Desch:2221945} is operating several long-aspect-ratio rectangular (i.e.\ waveguide-like) cavities inserted in CAST dipole magnet at CERN. The option of subdividing the resonant cavity is investigated by the RADES project~\cite{Melcon:2018dba}, also implemented in the CAST magnet. RADES is exploring the use of arrays of many small rectangular cavities connected by irises, carefully designed to maximally couple to the axion field for a given resonant mode. Data with a 5-subcavity prototype have already been obtained~\cite{Melcon:2020xvj}, and a first result at an axion mass of 34.67~$\mu$eV has been obtained~\cite{CAST:2021add}. A larger, 30-subcavity model is under operation. A similar concept, better adapted to a solenoidal magnet, is being followed at CAPP, with the concept of a sliced-as-a-pizza cavity~\cite{Jeong:2017hqs}, which consists on dividing the cylindrical cavity in sections connected by a longitudinal iris along the cylinder's axis of symmetry. A first version with two subcavities has been recently used~\cite{Jeong:2020cwz} to perform a search in the 13.0--13.9~$\mu$eV mass range. Finally, resonance to higher frequencies with a relatively large resonator can also be achieved by filling it with individually adjustable current carrying wire planes. R\&D is ongoing in this direction by the ORPHEUS experiment \cite{Rybka:2014cya}.

\vskip 2mm
\noindent
{\bf Dish antennas and dielectric haloscopes}\\

Going to even higher frequencies requires altogether different detection concepts. Most relevant is the concept of the  \textit{magnetized dish antenna} and its evolution, the \textit{dielectric haloscope}. A dielectric interface (e.g.\ a mirror, or the surface of a dielectric slab) immersed in a magnetic field parallel to the surface should emit electromagnetic radiation perpendicular to its surface, due to the presence of the dark matter axion field~\cite{Horns:2012jf}. This tiny signal can be made detectable if the emission of a large surface is made to concentrate in a small point, like e.g.\ in the case of the surface having a spherical shape. This technique has the advantage of being broad-band, with sensitivity to all axions masses at once\footnote{In practice this is limited by the bandwidth of the photon sensor being used.}. This technique is being followed by the BRASS~\cite{BRASSweb} collaboration at U. of Hamburg, as well as G-LEAD at CEA/Saclay~\cite{PC_brun}. 

\vskip 2mm
Given that no resonance is involved in this scheme, very large areas are needed to obtain competitive sensitivities. Dielectric haloscopes are an evolution of this concept, in which several dielectric slabs are stacked together inside a magnetic field and placed in front of a metallic mirror. This increases the number of emitting surfaces and, in addition, constructive interference among the different emitted (and reflected) waves can be achieved for a frequency band if the disks are adjusted at precise positions. This effectively amplifies the resulting signal. The MADMAX collaboration~\cite{TheMADMAXWorkingGroup:2016hpc} plans to implement such a concept, using 80 discs of LaAlO$_3$ with 1 m$^2$ area in a 10 T B-field,  leading to a boost in power emitted by the system of a $>$10.000 with respect to a single metallic mirror in a relatively broad frequency band of 50 MHz. By adjusting the spacing between the discs the frequency range in which the boost occurs can be adjusted, with the goal of scanning an axion mass range between 40 and 400~$\mu$eV (Fig.~\ref{fig:large_panorama}). The experiment is expected to be sited at DESY. A first smaller-scale demonstrating prototype will be operated in the MORPURGO magnet at CERN in the coming years, before jumping to the full size experiment. The experiment is expected to be sited at DESY. Finally, let us mention that an implementation of the dielectric haloscope concept but at even higher frequencies has been discussed in the literature, with potential sensitivity to 0.2~eV axions and above~\cite{Baryakhtar:2018doz}. 

\vskip 2mm
\noindent
{\bf DM Radios}

For much lower axion masses (well below~$\mu$eV), it may be more effective to attempt the detection of the tiny oscillating $B$-field associated with the axion dark matter field in an external constant magnetic field, by means of a carefully placed pick-up coil inside a large magnet~\cite{Sikivie:2013laa,Chaudhuri:2014dla,Kahn:2016aff}. Resonance amplification can be achieved externally by an $LC$-circuit, which makes tuning in principle easier than in conventional haloscopes. A broad-band non-resonant mode of operation is also possible ~\cite{Kahn:2016aff}.  Several teams are studying implementations of this concept~\cite{Silva-Feaver:2016qhh,Kahn:2016aff}. Two of them, the ABRACADABRA~\cite{Ouellet:2018beu} and SHAFT~\cite{Gramolin:2020ict} experiments, have recently released results with small table-top demonstrators, reaching sensitivities similar to the CAST bound for masses in the 10$^{-11}$--10$^{-8}$~eV range. Another similar implementation, that of BEAST~\cite{McAllister:2018ndu}, has obtained better sensitivities in a narrower mass range around 10$^{-11}$~eV. Similarly, the more recent result from the ADMX SLIC pilot experiment has probed a few narrow regions around $2\times10^{-7}$~eV and down to $\sim 10^{-12}$~GeV$^{-1}$~\cite{Crisosto:2019fcj}. Finally, the BASE experiment, whose main goal is the study of antimatter at CERN, has recently released a result adapting its setup to the search of axions following this concept~\cite{Devlin:2021fpq}. In general, this technique could reach sensitivity down to the QCD axion for masses $m_a \lesssim 10^{-6}$~eV, if implemented in magnet volumes of few~m$^3$ volumes and few~T fields. 

\vskip 2mm
\noindent
{\bf Other techniques}

A recent proposal to detect axion DM at even higher mass values involves the use of certain antiferromagnetic topological insulators~\cite{Marsh:2018dlj,Schutte-Engel:2021bqm}. Such materials contain axion quasiparticles (AQs), that are longitudinal antiferromagnetic spin fluctuations. These AQs have similar dynamics than the axion field, including a mass mixing with the electric field in the presence of magnetic fields. The dispersion relation and boundary conditions permit resonant conversion of axion DM into THz photons in a way that is independent of the resonant frequency. An advantage of this method is the tunability of the resonance with applied magnetic field. The technique could be competitive in the search for DM axions of masses in the 1 to 10 meV range.

At the very low masses, DM axions can produce an oscillation of the optical linear polarization of a laser beam in a bow-tie cavity. The DANCE experiment has already provided proof-of-concept results~\cite{Oshima:2021irp} with a table-top setup, while large potential for improvement exist in scale-up projections.

The techniques presented above are all based on the axion-photon coupling. If the axion has relevant fermionic couplings, the axion DM field would couple with nuclear spins like a fictitious magnetic field and produce the precession of nuclear spins. Moreover, by virtue of the same Peccei-Quinn term that solves the strong CP problem, the DM axion field should induce oscillating electric-dipole-moments (EDM) in the nuclei. Both effects can be searched for by nuclear magnetic resonance (NMR) methods. The CASPEr project ~\cite{Budker:2013hfa,JacksonKimball:2017elr} is exploring several NMR-based implementations to search for axion DM along this directions. The prospects of the technique may reach relevant QCD models for very low axion masses ($\lesssim 10^{-8}$~eV). A conceptually similar is done by the QUAX experiment, but invoking the electron coupling using magnetic materials~\cite{Barbieri:2016vwg}. In this case, the sample is inserted in a resonant cavity and the spin-precession resonance hybridises with the electromagnetic mode of the cavity. The experiment focuses in a particular axion mass $m_a\sim 200~\mu$eV, but sensitivity to QCD models will require lowering the detection noise below the quantum limit. Another  recently proposed technique is to search for the axion/ALP induced EDM in the future proton storage ring developed to measure the static proton EDM~\cite{Chang:2017ruk}. 

\vskip 2mm
DM axions can produce atomic excitations in a target material to levels with an energy difference equal to the axion mass. This can again happen via the axion interactions to the nuclei or electron spins. The use of Zeeman effect has been proposed~\cite{Sikivie:2014lha} to split he ground state of atoms to effectively create atomic transition of energy levels that are tunable to the axion mass, by changing the external magnetic field. 
The AXIOMA~\cite{1367-2630-17-11-113025,Braggio:2017oyt} project has started feasibility studies to experimentally implement this detection concept. Sensitivity to axion models (with fermion couplings) in the in the ballpark of $10^{-4}$--10$^{-3}$~eV could eventually be achieved if target material of $\sim$kg mass are instrumented and cooled down to mK temperatures

For a more thorough review of the possibilities that atomic physics offer to axion physics we refer to Section~\ref{ssec:stadnik}. 

\vskip 2mm
Before concluding this section, let us mention that a DM axion with keV mass (or higher) and with sufficiently strong coupling to electrons would show up in low background massive detectors developed for WIMP searches~\cite{Ahmed:2009ht,Aprile:2014eoa,Armengaud:2013rta}, as a non-identified peak at an energy equal to the mass, by virtue of the axioelectric effect. The recent XENON1T low-energy electronic recoil event excess~\cite{Aprile:2020tmw} could be interpreted as such a signal, with a favoured value for $m_A \sim 2.3$ keV.

\subsubsection{Solar axion experiments}

\begin{figure*}[t]
\centering
\includegraphics[scale=0.8]{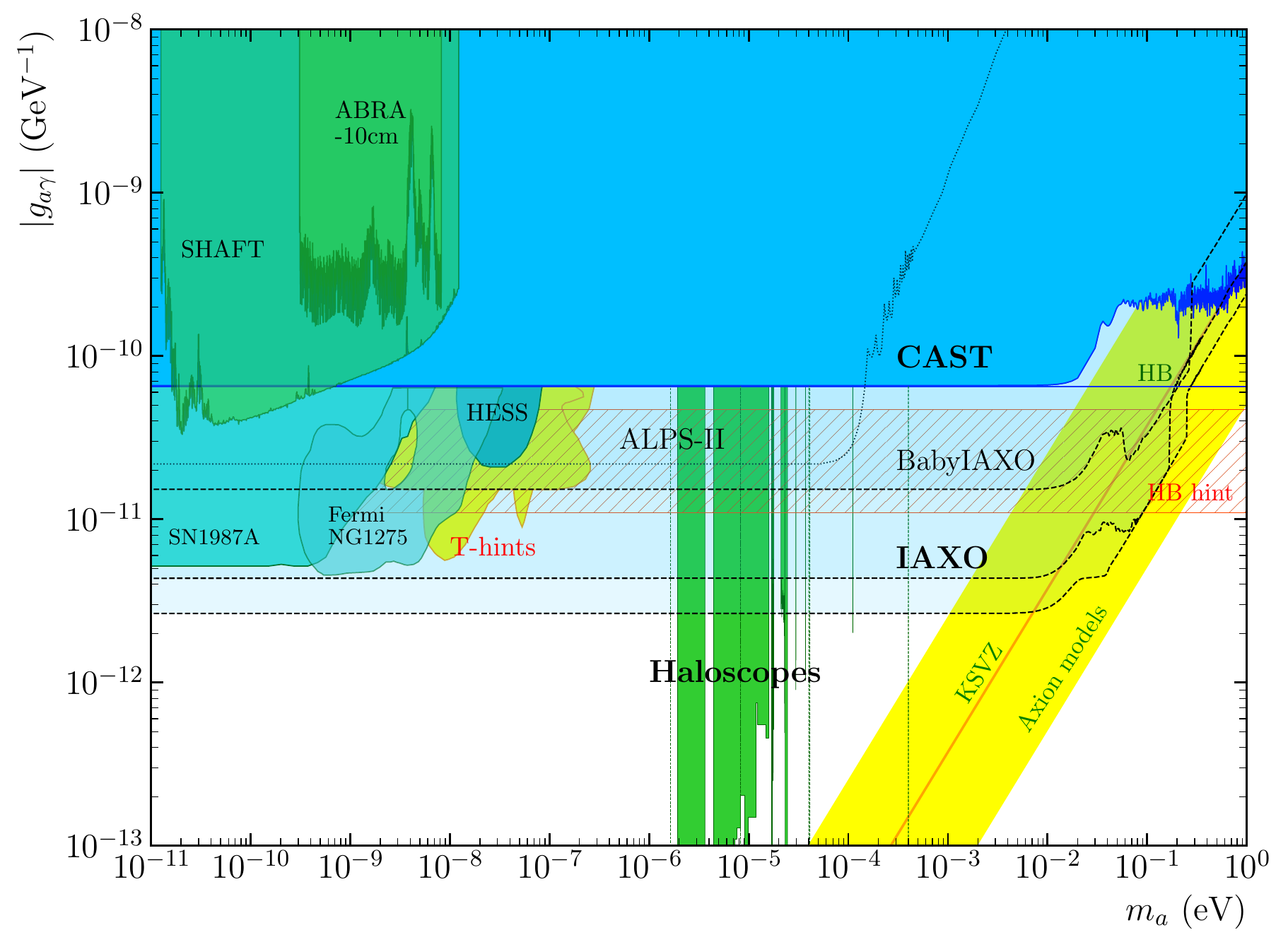}
\caption{Panorama of searches in the $\gagamma$-$m_a$ plane.}
\label{fig:helioscopes}
\end{figure*}

If axions exist, they would be produced in large quantities in the solar interior. Photons from the solar plasma would convert into axions in the Coulomb fields of charged particles via the Primakoff axion-photon conversion. If axions couple with electrons, additional production channels are possible~\cite{Redondo:2013wwa}. Once produced, axions get out of the star unimpeded and travel to the Earth, offering a great opportunity for direct detection in terrestrial experiments. The leading technique to detect solar axions are axion helioscopes~\cite{Sikivie:1983ip}, one of the oldest concepts used to search for axions. Axion helioscopes are sensitive to a given $\gagamma$ over a very wide mass range (Fig.~\ref{fig:helioscopes}), and after several past generations of helioscopes, the experimental efforts are now directed to increase the scale and thereby push sensitivity to lower $\gagamma$ values. Contrary to the scenario described in the haloscope frontier, with a plethora of relatively small, sometimes table-top, experiments, most of the helioscope community has coalesced into a single collaboration, IAXO, to face the challenges to build a large scale next-generation helioscope. 
Indeed, the IAXO collaboration is by far the largest experimental collaboration in axion physics, currently with about 125 scientists from 25 different institutions.

\vskip 2mm
\noindent
{\bf Helioscopes} 

The axion helioscope~\cite{Sikivie:1983ip} invokes the conversion of the solar axions back to photons in a strong laboratory magnet. The resulting photons are X-rays that can be detected behind the magnet when it is pointing to the Sun. In the baseline (or phase-I) configuration, this conversion takes place in the evacuated bores of the magnet, and provides a flat sensitivity to $\gagamma$ up to a mass of about $0.01$~eV. Above this value the sensitivity drops due to lack of coherence in the axion-photon oscillation along the magnet. Phase-II operation involves the use of a pressure-varying buffer gas inside the magnet bores. This gas provides the photon with a mass and restores the coherence for a narrow window of axion masses around the photon refractive mass. This has been the strategy followed by CAST at CERN using a decommissioned LHC test magnet that provides a 9~T field inside the two 10~m long, 5~cm diameter, magnet bores. CAST has been active for more than 15 years at CERN and represents the state-of-the-art in the search for solar axions. The latest result sets an upper bound on the axion-photon coupling of $0.66 \times 10^{-10}$~GeV$^{-10}$ (Fig.~\ref{fig:helioscopes}). This value competes with the strongest bound coming from astrophysics. Advancing beyond this bound is now highly motivated, not only because it would mean to venture into regions of parameter space allowed by astrophysics, but also because a number of astrophysical observations seem to hint at precisely this range of parameters. CAST has also searched for solar axions produced via the axion-electron coupling, although the very stringent astrophysical bound on this coupling remains so far unchallenged by experiments.

\vskip 2mm
The International Axion Observatory (IAXO) is a new generation axion helioscope, aiming at the detection of solar axions with sensitivities to the axion-photon coupling $g_{a\gamma}$ down to a few 10$^{-12}$~GeV$^{-1}$, a factor of 20 better than the current best limit from CAST (a factor of more than $10^4$ in signal-to-noise ratio). This leap forward in sensitivity is achieved by the realization of a large-scale magnet, as well as by extensive use of X-ray focusing optics and low background detectors. 

The main element of IAXO is thus a new dedicated large superconducting magnet, designed to maximize the helioscope figure of merit. The IAXO magnet will be a superconducting magnet following a large multi-bore toroidal configuration, to efficiently produce an intense magnetic field over a large volume. The design is inspired by the ATLAS barrel and end-cap toroids, the largest superconducting toroidal magnets ever built and presently in operation at CERN. Indeed the experience of CERN in the design, construction and operation of large superconducting magnets is a key aspect of the project. 

X-ray focusing relies on the fact that, at grazing incident angles, it is possible to realize X-ray mirrors with high reflectivity. IAXO envisions newly-built optics similar to those used onboard NASA's NuSTAR satellite mission, but optimized for the energies of the solar axion spectrum. Each of the eight $\sim$60~cm diameter magnet bores will be equipped with such optics. At the focal plane of each of the optics, IAXO will have low-background X-ray detectors. Several detection technologies are under consideration, but the most developed one are small gaseous chambers read by pixelized microbulk Micromegas planes. They involve low-background techniques typically developed in underground laboratories, like the use of radiopure detector components, appropriate shielding, and the use of offline discrimination algorithms. Alternative or additional X-ray detection technologies are also considered, like GridPix detectors, Magnetic Metallic Calorimeters, Transition Edge Sensors, or Silicon Drift Detectors. All of them show promising prospects to outperform the baseline Micromegas detectors in aspects like energy threshold or resolution, which are of interest, for example, to search for solar axions via the axion-electron coupling, a process featuring both lower energies than the standard Primakoff one, and monochromatic peaks in the spectrum.

\vskip 2mm
An intermediate experimental stage called BabyIAXO~\cite{Abeln:2020ywv} is the near term goal of the collaboration. BabyIAXO will test magnet, optics, and detectors at a technically representative scale for the full IAXO and, at the same time, it will be operated and will take data as a fully-fledged helioscope experiment, with sensitivity beyond CAST (see Fig.~\ref{fig:helioscopes}). It will be located at DESY, and it is expected to be built in 2--3 years.

The physics reach of IAXO is highly complementary to all other initiatives in the field, with sensitivity to motivated parts of the axion parameter space that no other experimental technique can probe. Most relevantly, IAXO will probe a large fraction of QCD axion models in the meV to eV mass band. This region is the only one in which astrophysical, cosmological (DM) and theoretical (strong CP problem) motivations overlap. As solar axion emission is a generic prediction of most axion models, solar axion searches represent the only approach that combines relative immunity to model assumptions plus a competitive sensitivity to parameters largely complementary to those accessible with other detection techniques. Finally, IAXO will also constitute a generic infrastructure for axion/ALP physics with potential for additional  search strategies (e.g.\ the option of implementing RF cavities to search for DM axions).

\vskip 2mm
\noindent
{\bf Other techniques} 

A variant of the helioscope technique, dubbed AMELIE~\cite{Galan:2015msa}, can be realized in a magnetized large gaseous detector (e.g.\ a time projection chamber). In this configuration, the detector gaseous volume plays both the roles of buffer gas where the Primakoff conversion of solar axions takes place, and X-ray detection medium. Contrary to standard helioscopes, in which the resulting X-rays need to cross the buffer gas to reach the detectors, here high photoabsorption in the gas is pursued. Therefore, high pressures or high-$Z$ gases are preferred. Due to the short range of the X-rays in the gas, the coherence of the conversion is lost, there is no privileged direction, and moving the magnet to track the Sun is no longer necessary. Still the signal depends on the $B$ field component perpendicular to the axion incident direction, and therefore even in a stationary magnet a daily modulation of the signal is expected, which give a useful signal signature. 
The technique could have some window of opportunity at higher masses $\gtrsim 0.1 $~eV where buffer gas scanning in helioscopes is increasingly difficult. 

\vskip 2mm
Axion-photon conversion (and vice versa) can also happen in the atomic electromagnetic field inside materials. In the case of crystalline media, the periodic structure of the field imposes a Bragg condition, i.e., the conversion is coherently enhanced if the momentum of the incoming particle matches one of the  Bragg angles~\cite{Buchmuller:1989rb}. This concept has been applied to the search for solar axions with crystalline detectors~\cite{Paschos:1993yf,Creswick:1997pg}. The continuous variation of the relative incoming direction of the axions with respect to the crystal planes, due to the Earth rotation, produces very characteristic sharp energy- and time-dependent patterns in the expected signal in the detector, which can be used to effectively identify a putative signal over the detector background. This technique has been used as a byproduct of low-background underground detectors developed for WIMP searches~\cite{Avignone:1997th,Morales:2001we,Bernabei:2001ny,Ahmed:2009ht,Armengaud:2013rta,Li:2015tsa,Xu:2016tap}. However, in the mass range where helioscopes enjoy full coherent conversion of axions, the prospects of this technique are not competitive~\cite{Cebrian:1998mu,Avignone:2010zn}. 

\vskip 2mm
Finally, solar axions could also produce visible signals in ionization detectors by virtue of the axioelectric effect~\cite{Ljubicic:2004gt,Derbin:2011gg,Derbin:2011zz,Derbin:2012yk,Bellini:2012kz}, most relevantly, in  large liquid Xe detectors~\cite{Abe:2012ut,Aprile:2014eoa,Fu:2017lfc,Akerib:2017uem}. However, the sensitivity to $g_{ae}$ is still far from the astrophysical bound. Interactions via nucleon coupling can also be used. For monochromatic solar axions emitted in M1 nuclear transitions, a reverse absorption can be invoked at the detector, provided the detector itself (or a component very close to it) contains the same nuclide, as e.g.\ in Fe$^{57}$~\cite{Moriyama:1995bz,Krcmar:1998xn}, Li$^7$~\cite{Krcmar:2001si} or Tm$^{169}$~\cite{Derbin:2009jw}. The upper limits to the nucleon couplings obtained by this method are however larger than the bounds set by astrophysics.
As a final comment, a combination of different couplings at emission and detection can also be invoked. The recent XENON1T excess~\cite{Aprile:2020tmw}, mentioned in a previous section, has also been interpreted as a signal of solar axions via a combination of couplings at emission and detection, including axion-photon conversion in the atomic field of the Xe atoms~\cite{Gao:2020wer} (this time with no Bragg-like effect). In all cases, the values of the couplings are already excluded by CAST or by astrophysical bounds~\cite{DiLuzio:2020jjp}.

\subsubsection{Conclusions and prospects}

Axions and axion-like particles at the low mass frontier appear in very motivated extensions of the SM. For long considered ``invisible'', very light axions are within reach of current and near-future technologies in different parts of the viable parameter space. The field is now undergoing a blooming phase. As shown in this review, the experimental efforts to search for axions are rapidly growing in intensity and diversity. Novel detection concepts and developments are recently appearing and are being tested in relatively small setups, yielding a plethora of new experimental initiative. In addition to this, consolidated detection techniques are now facing next-generation experiments with ambitious sensitivity goals and  challenges related to large-scale experiments and collaborations. As an example of the importance that this subfield is getting, let us mention that axion searches are explicitly mentioned in the last Update of the European Strategy for Particle Physics~\cite{Strategy:2019vxc}. The near and mid-term sensitivity prospects show promise to probe a large fraction of the axion parameter space, and a discovery in the coming years is not excluded. Such a result would  be a breakthrough discovery that could reshape the subsequent evolution of Particle Physics, Cosmology, and Astrophysics.

\clearpage
\subsection{Axions/ALPs phenomenology at accelerator-based experiments}
\label{ssec:kahloefer}
{\it Author: Felix Kahlhoefer, <kahlhoefer@physik.rwth-aachen.de>} \\ 

Pseudo-Goldstone bosons arise naturally in many extensions of the Standard Model with spontaneously broken approximate global symmetries. The most well-known example for such a particle is the QCD axion, which arises from the breaking of Peccei-Quinn symmetry. QCD axion models typically predict specific coupling patterns and relations between the coupling strength and the axion mass. In a model-independent approach one instead considers particles with the same types of effective interactions, but without imposing specific relations between the various parameters. These so-called axion-like particles (ALPs) can in general couple to any combination of Standard Model gauge bosons, fermions and Higgs bosons.

\vskip 2mm
The most frequently studied case is the one where the ALPs couple to hypercharge gauge bosons
\begin{equation}
\mathcal{L} \supset g'^2 C_{BB} \frac{a}{\Lambda} B_{\mu\nu} \tilde{B}^{\mu\nu}  \; ,
\end{equation}
so that their low-energy phenomenology is dominated by the effective ALP-photon interaction
\begin{equation}
 \mathcal{L} \supset -\frac{g_{a\gamma\gamma}}{4} a F_{\mu\nu} \tilde{F}^{\mu\nu} \,,
\end{equation}
with $g_{a\gamma\gamma} = 4 e^2 C_{BB} / \Lambda$~\cite{Bauer:2017ris}.
For ALPs below the MeV-scale, astrophysical constraints (in particular from horizontal branch stars) require the coupling $g_{a\gamma\gamma}$ to be tiny, such that ALPs are nearly stable and may constitute some or all of dark matter. For heavier ALPs, on the other hand, these constraints are absent and ALPs can have much larger couplings and shorter lifetimes, leading to interesting implications for particle physics and cosmology (see e.g.~\cite{Depta:2020wmr} for a recent discussion of constraints on ALPs from Big Bang Nucleosynthesis).

\vskip 2mm
Indeed, for an ALP mass $m_a \gtrsim 100~\mathrm{MeV}$ and effective couplings $g_{a\gamma\gamma} \gtrsim 10^{-4}~\mathrm{GeV^{-1}}$ ALPs decay promptly on typical detector scales and can be searched for at collider experiments, for example using the process $e^+ e^-\to \gamma a (\to \gamma \gamma)$, where the two photons from the ALP decay can be separately reconstructed, allowing to suppress backgrounds by searching for a peak in the invariant mass distribution~\cite{Dolan:2017osp}. A recent search for this signature at Belle II places world-leading limits on GeV-scale ALPs~\cite{BelleII:2020fag}.

For smaller masses and smaller couplings the ALP decay length can be macroscopic and one of the most promising strategies is to look for ALPs in beam dump experiments. Conventionally, the ALP yield in these experiments is predicted using the equivalent photon approximation~\cite{Dobrich:2015jyk}, but it was recently shown that for proton beams the dominant production mechanism is Primakoff production from photons produced in meson decays in the hadronic shower~\cite{Dobrich:2019dxc}.

\vskip 2mm
Analogous to the coupling to hypercharge gauge bosons, ALPs can also couple to $SU(2)_L$ gauge bosons:
\begin{equation}
\mathcal{L} \supset g^2 C_{WW} \frac{a}{\Lambda} W^a_{\mu\nu} \tilde{W}^{\mu\nu,a}  \; .
\label{eq:CWW_definition}
\end{equation}
The resulting phenomenology is however very distinct, because ALPs can now also be produced in rare meson decays, such as $K^+ \to \pi^+ a$~\cite{Izaguirre:2016dfi}. This specific process is of great interest experimentally, as it can be searched for using experiments like NA62 that aim to measure the ultra-rare Standard Model process $K^+ \to \pi^+ \nu \bar{\nu}$. Indeed, NA62 has very recently published first results from a search for a peak in the missing mass distribution, achieving sensitivity to branching ratios at the level of $10^{-10}$~\cite{CortinaGil:2020fcx}.

\vskip 2mm
We note that an additional contribution to rare decays involving ALPs arise from ALP couplings to heavy quarks and from ALP-meson mixing.  Unfortunately, these contributions suffer from substantial theory uncertainties as well as an unavoidable sensitivity of the predictions to the specific UV completion of the effective interactions~\cite{Batell:2009jf}. The contribution arising from $SU(2)_L$ gauge bosons, on the other hand, is theoretically clean and exhibits no UV dependence. Nevertheless, the case of ALPs coupled to $SU(2)_L$ gauge bosons has up to now received comparably little attention in the literature. 

\vskip 2mm
Another interesting albeit less studied case is that ALPs couple to gluons in a similar way as the QCD axion~\cite{Aloni:2018vki}:
\begin{equation}
 \mathcal{L} \supset g_s^2 C_{GG} G_{\mu\nu}^a \tilde{G}^{\mu\nu,a} \; .
\end{equation}
At low energies these interactions induce effective couplings to photons as well as flavour changing processes resulting from ALP-meson mixing (in particular ALP-$\eta^{(\prime)}$ mixing). Moreover, effective ALP-nucleon interactions lead to a significant suppression of the otherwise very strong constraints from SN1987A, because ALPs are trapped in the supernova core and cannot contribute to its cooling~\cite{Chang:2018rso}. As a result there are large swathes of allowed parameter space around $C_{GG} / \Lambda \sim 10^{-4} \text{--} 10^{-3}~\mathrm{GeV^{-1}}$, which may be explored with laboratory experiments.

A particularly promising strategy to probe these parameter regions are proton beam dump experiments. A recent study has for example pointed out the great potential of searching for ALPs coupled to gluons at the DUNE near detector~\cite{Kelly:2020dda}. However, at the moment there is no consensus on how to calculate the production of ALPs coupled to gluons in proton beam dump experiments and different groups include different effects in their calculation. At the same time, there are sizeable uncertainties in the ALP decay width, which enters exponentially in the predicted number of events. There is hence great need for collaboration between theorists and experimentalists to obtain precise and comprehensive predictions for ALPs with gluon couplings.

Another effective interaction, which is often neglected, is the one between ALPs and Higgs bosons, which arises at dimension 6 and 7 in the ALP effective theory~\cite{Bauer:2017ris}. These interactions can give rise to striking processes, such as the decay $h \to a a$ or the decay $h \to Z a$, which can be searched for with great precision at the LHC. A detailed discussion of the resulting constraints is however beyond the scope of this contribution.

So far, we have adopted the common strategy to consider one type of ALP interactions at a time (referred to in Ref.~\cite{Beacham:2019nyx} as \emph{photon dominance}, \emph{fermion dominance} and \emph{gluon dominance}). This approach is based on the naive intuition that ALPs with several interactions will be more tightly constrained than those with a single type of interactions. Hence, setting all but one of the couplings to zero would yield conservative predictions. However, it turns out that there are several constraints that can be suppressed if several couplings are present at the same time~\cite{Ertas:2020xcc}:
\begin{itemize}
 \item Constraints from SN1987A are strongly suppressed as soon as effective ALP-nucleon couplings are included.
 \item Constraints from beam-dump experiments are suppressed if additional decay modes reduce the ALP decay length.
 \item Accidental cancellations in the effective ALP-photon coupling can suppress all constraints that rely on this interaction.
\end{itemize}
Hence, it is important to consider the case where several couplings have a comparable magnitude. This approach has been coined \emph{co-dominance} in Ref.~\cite{Kelly:2020dda}, and it may reveal a number of interesting opportunities that would otherwise be missed.

\begin{figure*}[!t]
\centering
\includegraphics[width=0.45\textwidth]{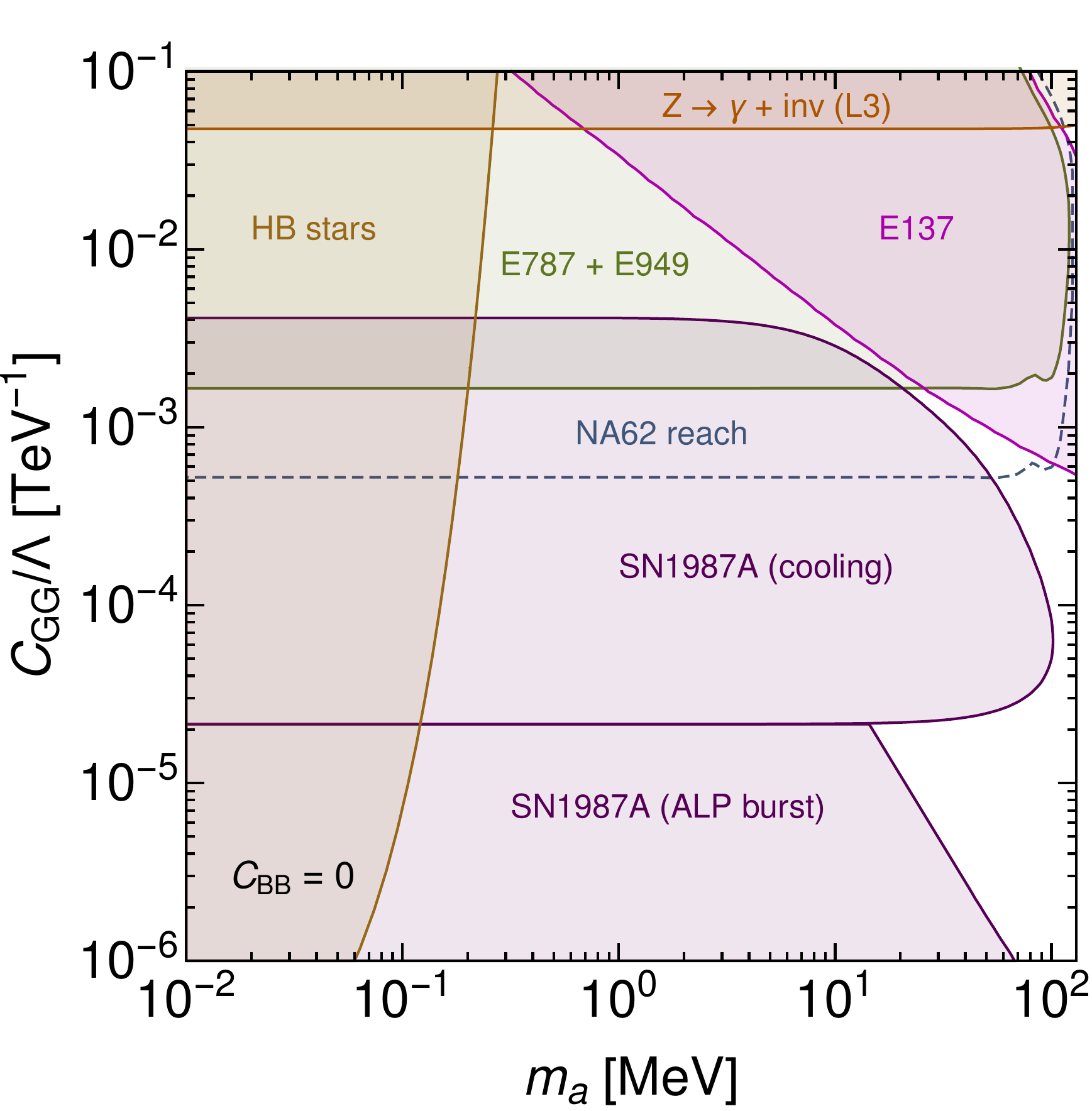}\hspace{0.2cm}
\includegraphics[width=0.45\textwidth]{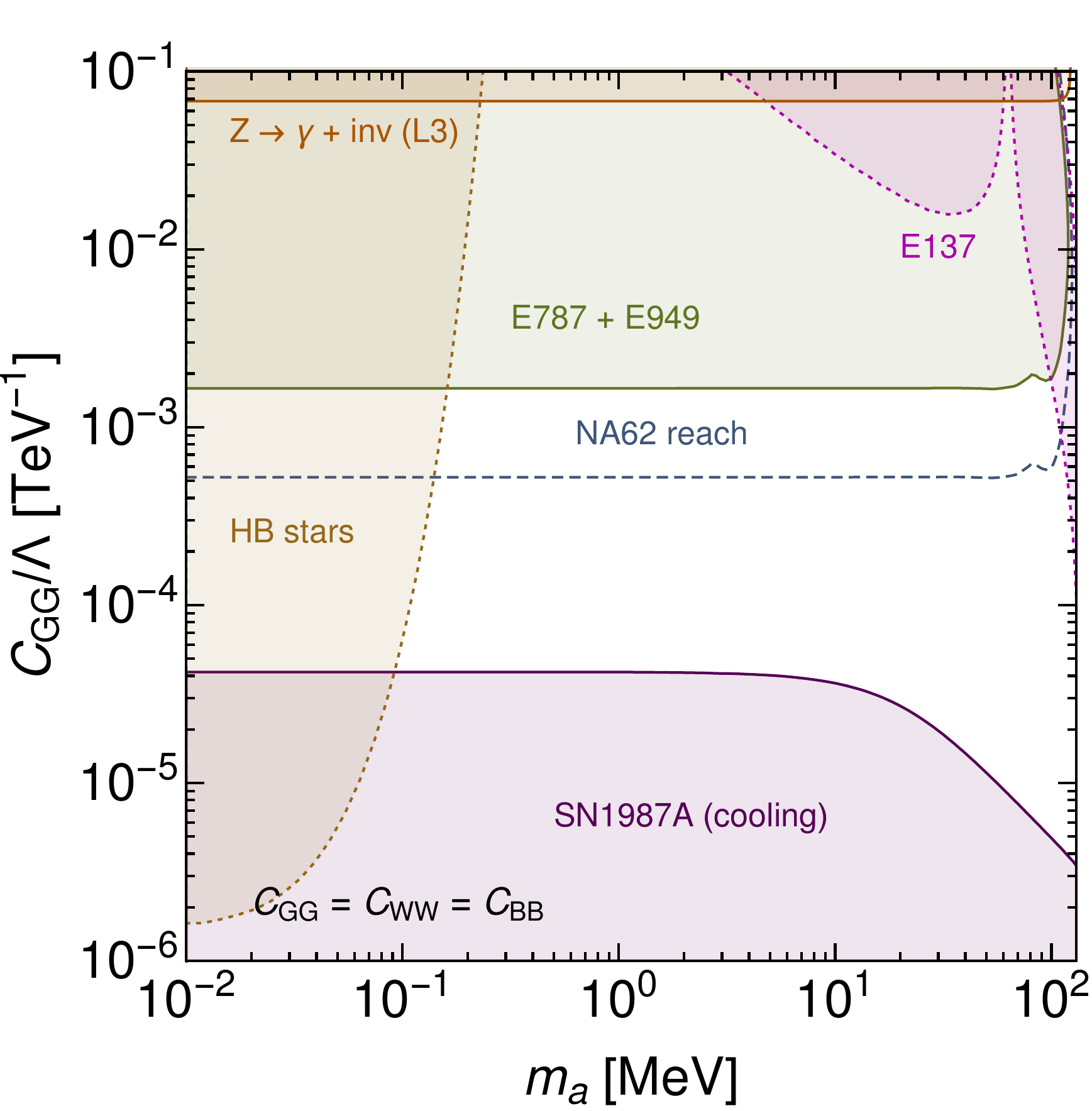}
\hspace*{1cm}
\caption{Experimental constraints and prospects on ALPs interacting dominantly with Standard Model gauge bosons. The left panel considers interactions only with $SU(2)_L$ gauge bosons, while in the right panel all gauge bosons are included, setting $C_{WW} = C_{BB} = C_{GG}$. This \emph{co-dominance} scenario reveals large regions of allowed parameter space. Figure taken from Ref.~\cite{Ertas:2020xcc}.}
\label{fig:1}	
\end{figure*}

This is illustrated in Fig.~\ref{fig:1}, which shows the case that ALPs couple only to $SU(2)_L$ gauge bosons (left panel) and the case that ALPs couple with the same coefficient to all SM gauge bosons ($C_{WW} = C_{BB} = C_{GG}$, right panel). In this case, the supernova trapping regime is substantially extended and hence the bound from SN1987A is shifted to smaller couplings. Moreover, since $g_{a\gamma\gamma} \approx 4 e^2 (C_{WW} + C_{BB} - 1.92 C_{GG}) / \Lambda$, constraints that rely on the ALP-photon coupling are weakened. Whereas the left panel would lead one to conclude that no interesting findings can be obtained from NA62, the right panel reveals that the sensitivity of NA62 is not nearly sufficient to probe all of the interesting parameter space for these models. 

\vskip 2mm
The study of ALPs reveal many exciting opportunities for future discoveries that do not require ambitious high-energy colliders. Indeed, many searches for ALPs can be performed using experiments already at our disposal, or future facilities that are being constructed for very different reasons. At the same time, ALPs may be connected to various open problems of particle physics. For example, they can act as the mediator responsible for communicating the interactions of dark matter with the Standard Model~\cite{Dolan:2014ska,Dolan:2017osp}.
The intriguing interplay between constraints derived from fixed-target experiments, rare decays,
and $e^+ e^-$ colliders as well as astrophysical bounds will enable us to map out these possibilities in more detail and explore one of the most well-motivated candidates for new physics below the GeV scale.


\clearpage
\subsection{Search for axions/ALPs at kaon factories}
\label{ssec:tobioka}
{\it Author: Kohsaku Tobioka, <ktobioka@fsu.edu>}
\subsubsection{Introduction}

The current kaon experiments have the unique potential to probe new physics beyond the standard model (BSM). A high-intensity beam of charged kaons $K^+$ is produced at the NA62 experiment at CERN, and the neutral counterpart for $K_L $ exists at the KOTO experiment at J-PARC. The main channels at NA62 and KOTO are rare kaon decay, $K^+\to \pi^+ \nu\bar\nu$ and $K_L\to \pi^0 \nu\bar\nu$, respectively, and both experiments aim for the extremely high precision measurements, at the level of $10^{-11}$ in their branching ratios. 

The isospin relation with $\Delta I=1/2$ dominance leads to so-called Grossman-Nir bound~\cite{Grossman:1997sk}, 
\begin{align}
{\cal B}(K_L \to \pi^0 \nu\bar\nu)\leq 4.3 {\cal B}(K^+\to \pi^+ \nu\bar\nu) . 
\label{eq:GN1}
\end{align}
It is well-known that this relation holds even if there are significant BSM contributions. Furthermore, one can replace the neutrinos with a light BSM particle(s) $X$ and the generalized GN generically holds~\cite{Kitahara:2019lws}, 
\begin{align}
{\cal B}(K_L \to \pi^0 X)\leq 4.3 {\cal B}(K^+\to \pi^+ X) \,,
\label{eq:GN2}
\end{align}
where $X$ does not carry any SM charge, and it is thus experimentally invisible.

In this section, we discuss how the generalized Grossman-Nir bound is violated with BSM physics. This theoretical investigation is partially motivated by four events\footnote{KOTO collaboration later confirmed that one of the four events is mistakenly kept in the signal region, and also the KOTO found several new sources of background that make this excess statistically insignificant~\cite{Ahn:2020opg}.} that the KOTO experiment reported in 2019~\cite{KOTOslides}. However, the idea remains useful for any future kaon data. 
Also, we briefly summarize BSM scenarios for which NA62 and KOTO have unique sensitivities thanks to their enormous number of kaons. We emphasize axion-like particles in this context. 

\subsubsection{BSM scenarios violating Grossman-Nir bound}
In the SM and most of BSM scenarios, we always expect an excess of $K^+\to \pi^+ \nu\bar\nu$ if an excess of $K_L \to \pi^0 \nu\bar\nu$ is observed. Otherwise, the GN bound is violated, and this situation was realized in 2019 when NA62 reported the result consistent with the SM expectation while KOTO reported anomalous events (which later became compatible with the SM after examining the unblinded data). 
It is interesting to investigate scenarios violating the GN bound because we can nail down particular BSM scenarios if excess is seen only in $K_L$ decay. 

\vskip 2mm
Refs.~\cite{Kitahara:2019lws,Fuyuto:2014cya} pointed out possibilities to violate the GN bound by the experimental setup difference. In particular, there is a gap around $m_X\simeq m_{\pi^0}$ in the search of $K^+\to \pi^+ X$ ($X$ is invisible) due to the large background of $K^+\to \pi^+ \pi^0$~\cite{Fuyuto:2014cya}. It is easily realized in the minimal Higgs portal model~\cite{Egana-Ugrinovic:2019wzj}.  
The GN bound is violated in another way~\cite{Kitahara:2019lws}: if $X$ has a finite lifetime, $\tau_X\sim 10^{-10}$~s, the produced $X$ decays inside the NA62 setup, while some of $X$ escape the KOTO detector mimicking an invisible particle. This possibility exploits the effective detector size, $L_{\rm NA62}/p_{\rm NA62}>L_{\rm KOTO}/p_{\rm KOTO}$ where $L$ is the detector length and $p$ is momentum of $X$ in the lab frame. 

\vskip 2mm
While the GN bound is not violated at the fundamental level in the above scenarios, Refs.~\cite{Ziegler:2020ize,Gori:2020xvq,Hostert:2020gou} pointed out the fundamental violation is realized if $X$ includes multiple particles. The point is that the new particles in the dark sector are generically neutral under the SM gauge groups, so neutral particles such as neutral kaons could decay to the dark sector particles directly, but the similar decay from charged particles would be suppressed even forbidden. For example, if there is a dark sector coupling $\bar{s}d X_1 X_2$ where $X_2$ further decays $\gamma\gamma$ or $X_1\pi^0$,  neutral kaons decay to the dark sector particles, $K_L\to X_1 X_2\to X_1 \gamma\gamma, \ X_1 X_1 \pi^0(\to\gamma\gamma)$ mimicking the SM rare decay $K_L \to \pi^0 \nu\bar\nu$. On the other hand, decays of charged kaon by this operator are suppressed because the charge conservation requires an extra charged particle in the final state as $K^+ \to X_1 X_2 \pi^+$. Or even this channel is kinematically forbidden depending on the $X_{1,2}$ mass. This type of scenario is realized if the dark sector particle carries {\it strangeness}~\cite{Gori:2020xvq}, and the operators are  
\begin{align}
 H \bar{Q}_1 s \phi^2 /\Lambda^2 \ , \quad 
 H \bar{Q}_2 d \phi^2 /\Lambda^2 \,,
\end{align}
where $\phi$ is a complex scalar with $1/2$ strangeness and $\Lambda \ge 10^7$~GeV. We need a small breaking of strangeness, $\chi F\tilde{F}/\Lambda_\chi$, that induces a decay $\chi\to\gamma\gamma$ where $\chi$ is the imaginary component of $\phi$. In Ref.~\cite{Hostert:2020gou}, other realizations in UV completions with $Z'$ or Higgs portals are discussed. 

\vskip 2mm
There is another way to violate the GN bound using mixing among charge-neutral particles~\cite{Ziegler:2020ize}.
Suppose dark sector particles, $X_1$ and $X_2$, have couplings, 
\begin{align}
g_{dd} X_2 \bar{d}d+g_{sd} X_2 \bar{s}d +\lambda m_X X_2^2 X_1 +h.c. \ ,
\end{align}
where $X_2$ is heavier than $m_K$ while $X_1$ is lighter such that $K\to \pi X_1$ is kinematically allowed. 
In this case, the neutral kaons have a mixing contribution, $K_{L}\to X_2^*\to \pi^0 X_1$, but the charged kaon does not have the corresponding effect due to the charge conservation. Therefore, this scenario directly violates the generalized GN bound (Eq.~\eqref{eq:GN2}). 

\begin{figure*}[t!]
\begin{center}
\includegraphics[width=0.43\textwidth]{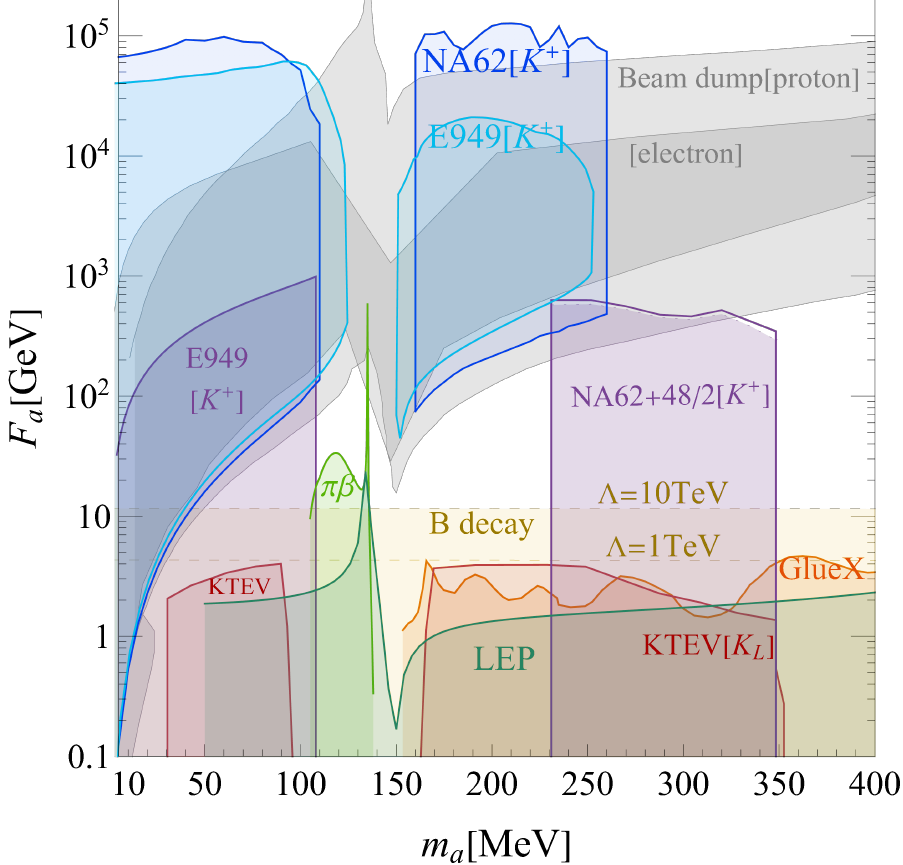}~~~
\includegraphics[width=0.43\textwidth]{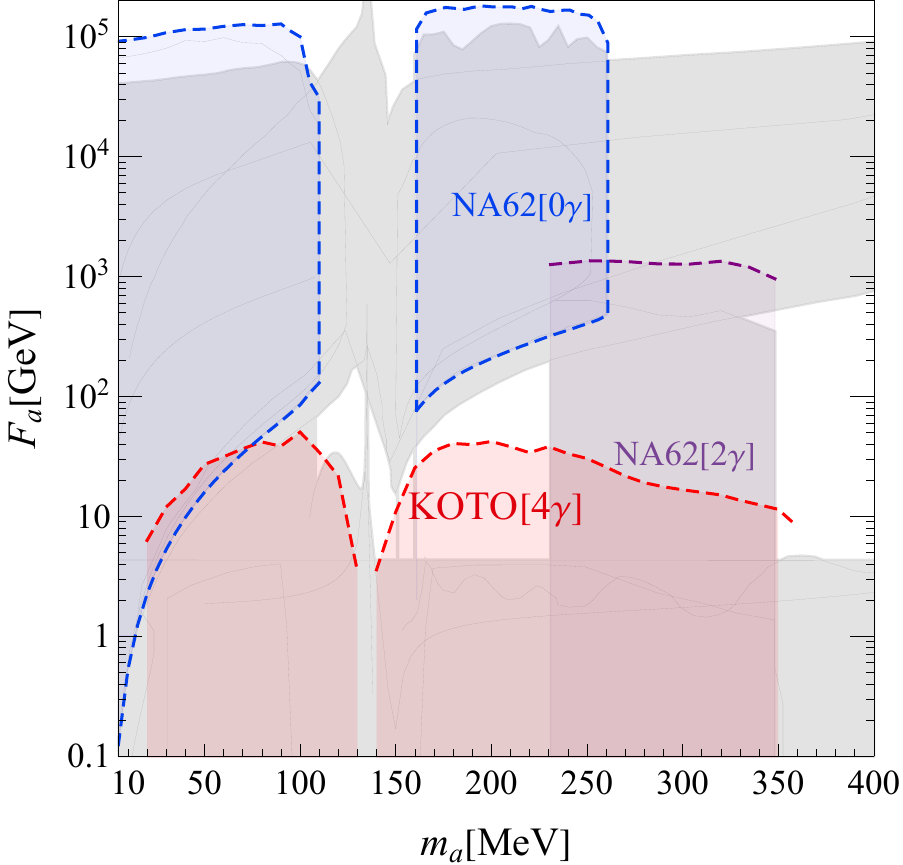}
\end{center}
\caption{Left panel: Present bounds on the parameter space of the $G\tilde G$-coupled ALP benchmark, as a function of the ALP mass, $m_a$, and of its decay constant, $F_a$. Right panel: Present and future bounds on the parameter space. In gray, we present the present bound (as shown in the left panel). The bound of $B$ decay depends on the EFT cutoff $\Lambda$, and we include the bound with $\Lambda=$1~TeV. In red, purple, and blue we present the future bounds at KOTO ($4\gamma$ proposed search),  at NA62 ($\pi^++2\gamma$ signature), and at NA62 ($\pi^++missing$ signature), respectively. 
}
\label{fig:aGG}
\end{figure*}

\subsubsection{Other BSM scenarios for future data of NA62 and KOTO}
Since NA62 and KOTO will take data with unprecedented kaon decays, we can explore many exciting BSM scenarios. One of the leading examples is a heavy axion or ALP in the MeV mass range. A reasonable parameter space with a low decay constant is unexplored in the mass range from MeV to 10~GeV, and the kaon experiments have unique sensitivities for $m_a<350$~MeV.  
In particular, we investigate two effective field theories (EFTs) in~\cite{Gori:2020xvq}. One is a heavy axion which dominantly couples to gluons, 
\begin{align}
\frac{1}{2}m_a^2 a^2 + \frac{\alpha_s}{8\pi F_a} G_{\mu\nu}^a \tilde{G}_{\mu\nu}^a \ ,
\end{align}
where the standard relation $m_a\simeq m_\pi f_\pi/F_a$ is not imposed. 
The other EFT is an ALP dominantly coupling to SU(2) gauge bosons~\cite{Izaguirre:2016dfi}, 
\begin{align}
\frac{1}{2}m_a^2 a^2 + \frac{g_{aW}}{4} W_{\mu\nu}^a \tilde{W}_{\mu\nu}^a \,. 
\end{align}

We consider $K\to \pi +missing$ from E949, NA62, and KOTO for bounds~\cite{Artamonov:2009sz,CortinaGil:2018fkc,talkNA62pheno,Ahn:2018mvc}, and we study the projected sensitivities assuming $2.5 \times10^{11}~K^+$ decays at NA62 and $2.1 \times 10^{14}~K_L$ at the beam entrance of KOTO. In addition, we point out that $K\to \pi +\gamma\gamma$~\cite{Artamonov:2005ru,Ceccucci:2014oza,Abouzaid:2008xm} takes an essential role in the search for ALPs since the diphoton decays are  expected in the relevant parameter space. Other colored regions comprise bounds from \cite{Bjorken:1988as,Aloni:2018vki,Gori:2020xvq,Aloni:2019ruo,Bergsma:1985qz,Abbiendi:2002je,Knapen:2016moh,Chakraborty:2021wda} (see also Sec.~\ref{ssec:soreq}).  
 The left panel of Figs.~\ref{fig:aGG} shows the updated bounds for the scenario of gluon coupling dominance, and the right one shows the projections at KOTO and NA62. 
 The projections shown in the right panel of Fig.~\ref{fig:aGG} show that $K\to \pi +\gamma\gamma$ channels will cover a large parameter space.  
For the scenario of $W$-coupling dominance, see Fig.~9 of \cite{Gori:2020xvq}. 



\clearpage
\subsection{Search for axions/ALPs with photon beams}
\label{ssec:soreq}
{\it Author: Yotam Soreq, <soreqy@physics.technion.ac.il>}  \\

Axion-like-particles~(ALPs), denoted as $a$, appear in many extensions of the standard model~(SM).  
Originally, the axion is predicted as part of the solution for the strong-CP problem~\cite{Peccei:1977hh,Peccei:1977ur,Weinberg:1977ma,Wilczek:1977pj} and, more recently, SM hierarchy problems~\cite{Graham:2015cka}.
ALPs are valid dark matter candidates~(DM)~\cite{Abbott:1982af,Preskill:1982cy,Dine:1982ah} or can be a portal to DM and/or dark sector~\cite{Nomura:2008ru,Freytsis:2010ne,Dolan:2014ska,Hochberg:2018rjs}. 
The ALPs are pseudoscalars and pseudo-Goldstone modes, thus, their mass can be much smaller than the scale that suppresses their interaction with SM particles, i.e.\ $m_a \ll \Lambda$.

Here, we mainly focus on ALPs with mass at the MeV-to-GeV scale in photon beam experiments, such as as {\sc PrimEx} and {\sc GlueX}, and follow Refs.~\cite{Aloni:2018vki,Aloni:2019ruo}.
The effective interaction of ALP to photons and gluons is given by
\begin{align}
	\mathcal{L}_{\rm eff}
=	-\frac{4\pi \alpha_s c_g}{\Lambda} a G^{\mu\nu}\tilde{G}_{\mu\nu}
	+\frac{c_\gamma}{4\Lambda}a F^{\mu\nu}\tilde{F}_{\mu\nu} \, ,
\end{align}
where $F_{\mu\nu}\,(G_{\mu\nu})$ is the photon\,(gluon) field strength and the their duals are denoted with tilde. 
Note that $\frac{c_g}{\Lambda}=\frac{C_{aGG}}{32\pi^2  f_a}$ and $\frac{c_\gamma}{\Lambda}=-g_{a\gamma}$ in the notation of Eqs.~\eqref{Leff_a} and \eqref{axion:eq:eq4}, respectively.

\vskip 2mm
We consider a photon beam with energy of $k_\gamma\approx 5$--10~GeV colliding with a fixed thin target, $N$, made of protons, C, Si, or Pb. 
In case of $c_\gamma\ne0$ and $c_g=0$, sub-GeV ALP are produced via the coherent Primakoff production, $\gamma \, N \to a \, N$.
In the opposite case, where $c_g\ne0$ and $c_\gamma=0$, the ALP is produced via photoproduction of photon-vector meson mixing and $t$-channel vector meson exchange, where for simplicity we consider only proton target.
In both cases, the ALP promptly decays into SM final states, such as photons or hadrons (if kinematically available). 

\begin{figure*}[!t]
\begin{center}
\raisebox{20pt}{\includegraphics[width=0.52\textwidth]{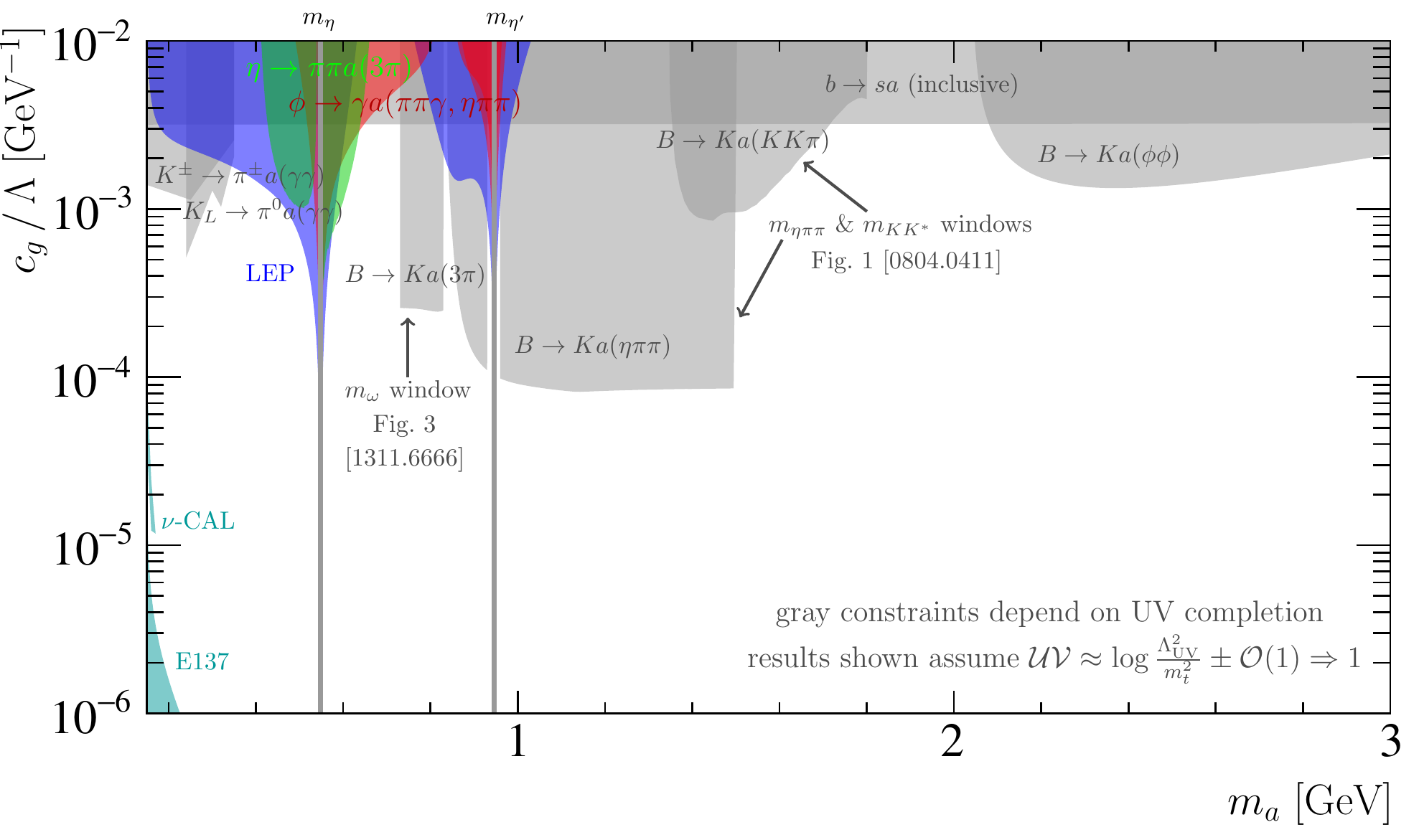}}
\includegraphics[width=0.46\textwidth]{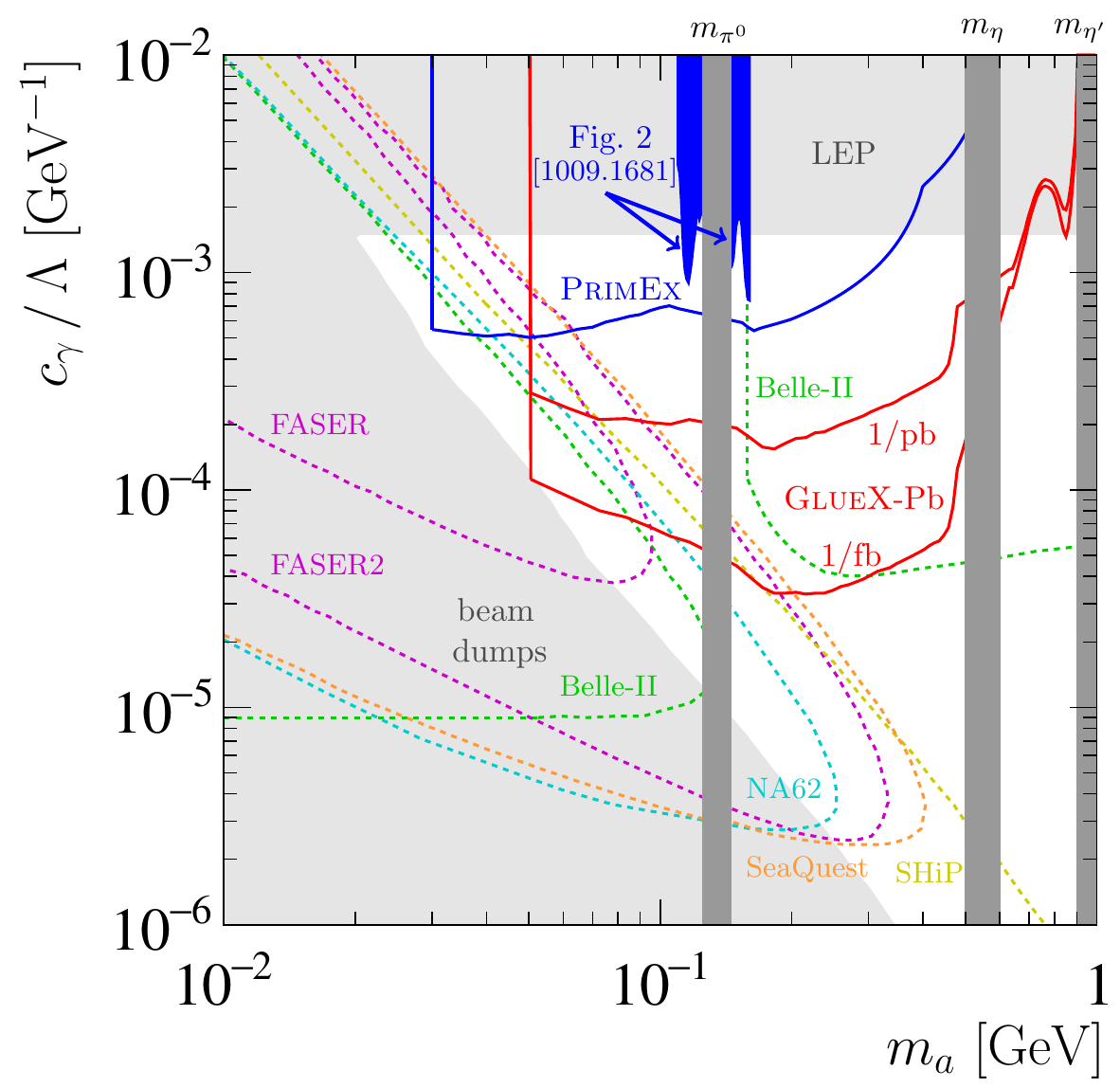}
\end{center}
\caption{
Left: New limit set (dark blue shaded regions) using the published $m_{\gamma\gamma}$ spectrum from one angular bin of carbon-target {\sc PrimEx} data from Fig.~2 of Ref.~\cite{Larin:2010kq}.
Projection for {\sc PrimEx}\,(blue) and {\sc GlueX}\,(red) for the $a$-$\gamma$ coupling compared to the current bounds~\cite{Bjorken:1988as,Blumlein:1990ay,Abbiendi:2002je,Knapen:2016moh} and projections of NA62, SeaQuest, Belle-II, SHiP, and FASER~\cite{Feng:2018noy,Berlin:2018pwi,Dobrich:2015jyk,Dolan:2017osp}.
The plot is taken    from Ref.~\cite{Aloni:2019ruo}.
Right: Constraints on the ALP-gluon coupling from Ref.~\cite{Aloni:2018vki}.
 }
\label{fig:prim_lims}
\end{figure*}

\vskip 2mm
The ALP Primakoff production is similar to that of $\pi^0$ and $\eta$ mesons, up to known kinematic corrections.
Thus, the expected ALP yield can be normalized to the number of observed $\pi^0$'s and $\eta$'s in the data~\cite{Aloni:2019ruo}. 
Therefore, many uncertainties such as the photon flux and the nuclear form factors are cancelled in the ratio. 
By recasting the {\sc PrimEx} data from~\cite{Larin:2010kq}, we derive new bounds on the $a$-$\gamma$ interaction, $c_\gamma$, for $m_a$ around the pion mass.  
In addition, we project the sensitivity of the full {\sc PrimEx} data and future {\sc GlueX} runs to probe this coupling.
Our results are summarized in the left panel of Fig.~\ref{fig:prim_lims} can compared to current bounds from LEP and beam dumps~\cite{Bjorken:1988as,Blumlein:1990ay,Abbiendi:2002je,Knapen:2016moh}, as well as future projections of Belle-II, FASER, NA62, SeaQuest, and SHiP~\cite{Feng:2018noy,Berlin:2018pwi,Dobrich:2015jyk,Dolan:2017osp}. 

In case of ALP gluon coupling, $c_g\ne0$ and $c_\gamma=0$, we consider the photon beam on proton target and extract the bound from the public {\sc GlueX} data~\cite{AlGhoul:2017nbp}, as well as give projection for future runs (see Fig.~3 of~\cite{Aloni:2019ruo}). 
This scenario can be also probed by the KOTO experiment via $K_L\to\pi^0 a\to 4\gamma$ process or in NA62~\cite{Gori:2020xvq}. 

One challenge is to estimate the ALP hadronic decay rate for $1\,{\rm GeV} \lesssim m_a \lesssim 3\,{\rm GeV}$, where both chiral perturbation theory and perturbative QCD fail. 
These rates can be estimated by using data-driven method~\cite{Aloni:2018vki}, which is based on $e^+e^-$ and $U(3)$ flavor symmetry.
For instance, the decay amplitude of ALP into two vector mesons, $V_{1,2}$, $a\to V_1V_2$, can be extracted from $e^+e^-\to V_1^*\to V_2 P$ cross section ($P$ is a meson) by using $a$-$P$ mixing and exchanging initial and final states.
This method can be applied to different hadronic final states. 
As a cross check, the $\eta_c$ meson, with $U(3)$ flavor quantum numbers of an ALP, has branching ratios into $V_1V_2$ that can be predicted with a good success~\cite{Aloni:2018vki}. 
Finally, we use different $B\to K$ and $K\to\pi$ decays to constrain the ALP-gluon coupling, $c_g/\Lambda$ (Fig.~\ref{fig:prim_lims}, right).

Rare kaon decays, $K\to \pi a$, can be also used to place strong bounds on the $aWW$ coupling, $g_{aW}$, via the one-loop FCNC diagram, similar to the SM one~\cite{Izaguirre:2016dfi,Gori:2020xvq}.
For $m_a<m_K$, one can probe $g_{aW}$ smaller than $10^{-4}\,{\rm GeV}^{-1}$ with the current data.
Similarly, $B\to K a$ can probe $g_{aW}$ at the level of $10^{-5}\,{\rm GeV}^{-1}$~\cite{Izaguirre:2016dfi}.  
We note also that higher ALP masses, $m_a>5\,{\rm GeV}$, can be probed in $\gamma\gamma$ resonance searches in Pb-Pb~\cite{Knapen:2016moh,Sirunyan:2018fhl,Aad:2020cje} and p-p~\cite{Mariotti:2017vtv,CidVidal:2018blh} collisions (see Section~\ref{ssec:denterria}).
For details on ALP searches at future colliders, see~\cite{Bauer:2017ris,Bauer:2018uxu}.

\clearpage
\subsection{Search for axions/ALPs at the LHC}
\label{ssec:denterria}
{\it Author: David d'Enterria \\ (for the ATLAS and CMS Collaborations),\\
<dde@cern.ch>}  
\newcommand{\mrm}[1]{\mathrm{#1}}

\def\GeV{{\text{ }\mathrm{GeV}}}
\def\MeV{{\text{ }\mathrm{MeV}}}
\def\keV{{\text{ }\mathrm{keV}}}
\def\TeV{{\text{ }\mathrm{TeV}}}
\newcommand{\Lumi}{\mathcal L}
\newcommand{\pp}{\mathrm{pp}}
\newcommand{\ppbar}{\mathrm{p\bar{p}}}
\newcommand{\PbPb}{\mathrm{PbPb}}

\newcommand{\epem}{e^+e^-}
\newcommand{\gaga}{\gamma\gamma}
\newcommand{\alphas}{\alpha_{S}}
\newcommand{\pt}{p_{_\perp}}
\newcommand{\pT}{p_{\rm T}}
\newcommand{\ET}{E_{\rm T}}
\newcommand{\alphasmZ}{\alphas(m_\mathrm{Z})}
\newcommand{\lqcd}{\Lambda_{_\mathrm{QCD}}}
\newcommand{\MSbar}{\overline\mathrm{MS}}
\newcommand{\sqrts}{\sqrt\mathrm{s}}
\newcommand{\sqrtsnn}{\ensuremath{\sqrt{s_{_{NN}}}}}

\newcommand{\Mt}{m_\mathrm{t}}
\newcommand{\MW}{m_\mathrm{W}}
\newcommand{\MZ}{m_\mathrm{Z}}

\newcommand{\starlight}{\textsc{Starlight}}
\newcommand{\superchic}{\textsc{SuperChic}}

\providecommand{\qqbar}{\ensuremath{q\overline{q}}}
\providecommand{\ccbar}{c\overline{c}}
\providecommand{\bbbar}{b\overline{b}}

\newcommand*{\cm}{c.m.\@\xspace}
\newcommand*{\eg}{e.g.\@\xspace}
\newcommand*{\ie}{i.e.\@\xspace}

\def\cO#1{{{\cal{O}}}\left(#1\right)}

\subsubsection{Introduction}
The existence of fundamental scalar particles in nature has received very strong evidence with the discovery of the Higgs boson. Searches for additional (pseudo)scalar particles have thereby been attracting an increasing interest in the last years in collider studies of physics beyond the Standard Model (BSM)~\cite{Jaeckel:2012yz,Mimasu:2014nea,Jaeckel:2015jla,Bauer:2017ris}, and have become key ingredients of the ``intensity frontier'' physics program aiming at discovering new feebly interacting particles~\cite{Essig:2013lka,Lanfranchi:2020crw}. Axion-like particles (ALPs) are hypothetical pseudoscalar bosons that appear naturally in many extensions of the SM, often as pseudo Nambu--Goldstone bosons (pNGBs) arising in the spontaneous breaking of a new global approximate symmetry~\cite{Ringwald:2014vqa}
(similarly as the neutral pion does for the chiral symmetry of quantum chromodynamics, QCD).
ALPs are ubiquitous in a variety of particle physics and cosmology BSM scenarios such as
\begin{description}
\item (i) the original axion, a pNGB of the Peccei--Quinn symmetry broken by the axial anomaly of QCD, introduced to solve the absence of observed charge-parity (CP) violation in the strong interaction (``strong CP problem'')~\cite{Peccei:1977hh,Weinberg:1977ma,Wilczek:1977pj};
\item (ii) light pseudoscalars proposed as cold dark matter (DM) candidates particles~\cite{Duffy:2009ig,Marsh:2015xka}, or dark-sector mediators~\cite{Nomura:2008ru,Dolan:2014ska,Kozaczuk:2015bea}; 
\item (iii) generic pNGBs arising from the spontaneous breaking of a new U(1) global symmetry at some large energy scale $\Lambda$, like \eg\ the R-axion (pNGB of the broken R-symmetry in SUSY)~\cite{Bellazzini:2017neg}, pseudoscalar states in Higgs compositeness models (from the broken U(1) symmetry acting on all underlying fermions of the theory)~\cite{Cacciapaglia:2019bqz}, or in cosmic inflation~\cite{Freese:1990rb}; 
\item (iv) generic pseudoscalar bosons that appear \eg\ in extended Higgs sectors~\cite{Branco:2011iw}, 
in phenomenological realizations of string theory~\cite{Ringwald:2012cu}, in models where they contribute to lepton dipole moments (thereby solving the $(g-2)_\mu$ problem)~\cite{Chang:2000ii,Marciano:2016yhf} or to address the electroweak hierarchy problem~\cite{Graham:2015cka}.
\end{description}

\vskip 2mm
Since light pseudo-scalars naturally couple to photons (due to spin selection rules and mass\footnote{For a very light ALP with $m_a < 2m_e$, the diphoton is the only SM decay mode allowed.} constraints), ALP searches from cosmology, astrophysical, and low-energy accelerator studies have mostly exploited the (inverse) Primakoff effect~\cite{Halprin:1966zz}, $\gaga^{(*)}\to a \to \gaga^{(*)}$, that converts photons into axions (and vice versa) in collisions with other photons or, equivalently, electromagnetic fields ($\gamma^*$). Collider searches have thereby predominantly focused on initial and/or final states with photons that, first, allow a direct connection with the vast range of all other existing searches and, secondly, are easier to identify over intrinsically large hadronic backgrounds. This type of searches thereby correspond to the benchmark point referred to as \emph{photon dominance} (BC9) in~\cite{Beacham:2019nyx} (see the discussion in Sect.~\ref{sssec:axion-results}).

\begin{figure}
\centering
\includegraphics[width=0.9\columnwidth]{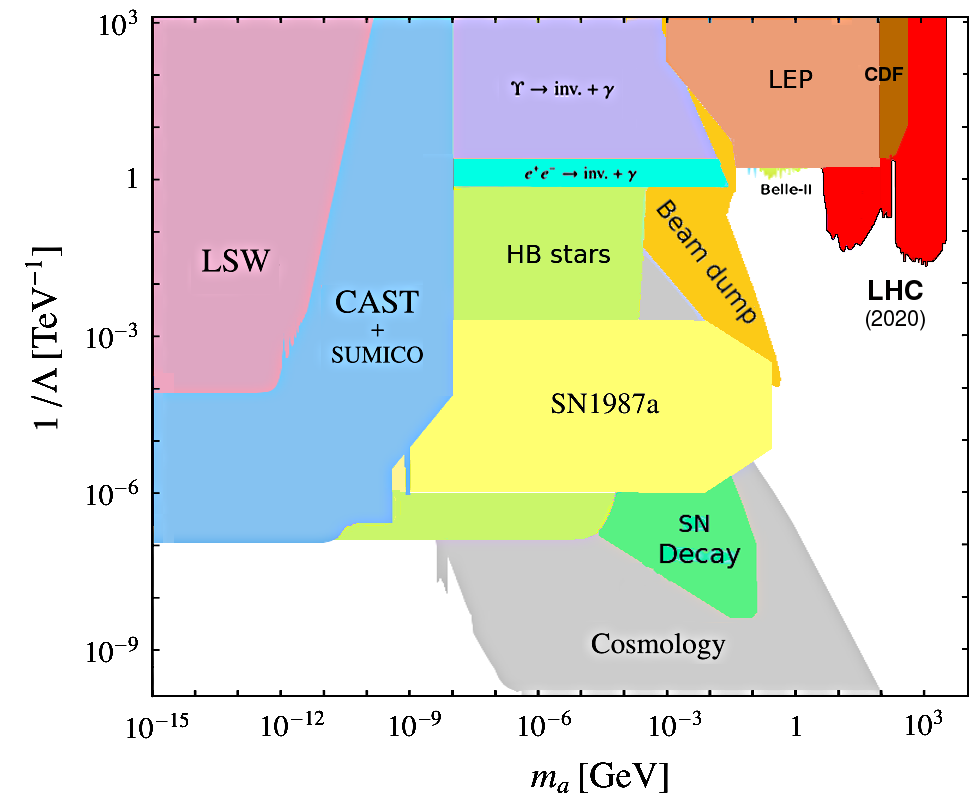}
\caption{Limits on the ALP-photon coupling vs.\ ALP mass from cosmological, astrophysical, and accelerator searches over the mass range $m_a = 1~\mu$eV--10~TeV. The top-right red contour depicts the approximate area currently covered by the latest LHC studies summarized here. Figure adapted from~\cite{Jaeckel:2015jla,Bauer:2017ris}.}
\label{fig:limits_preLHC}
\end{figure}

\vskip 2mm
Focusing on the ALP-$\gamma$ coupling, the effective Lagrangian reads (see Eq.~(\ref{axion:eq:eq4})):
\begin{equation}
\mathcal{L} \supset -\frac{1}{4}g_{a\gamma}\,a\,F^{\mu\nu}\Tilde{F}_{\mu\nu},\, \mbox{ with } g_{a\gamma} = C_{\gaga}/\Lambda,
\label{eq:L_agamma}
\end{equation}
where $a$ is the ALP field, $F^{\mu\nu}$ ($\Tilde{F}_{\mu\nu}$) is the photon field strength (dual) tensor, and the dimensionful ALP-photon coupling strength $g_{a\gamma}$ is related\footnote{For the QCD axion, decay constant and mass are directly related and $g_{a\gamma}$ is proportional to $m_a$ with an essentially known fixed constant
(thereby populating a linear band in Fig.~\ref{fig:limits_preLHC}), but such a relationship is relaxed for generic ALPs, which do not necessarily couple to gluons, and for which $m_a$ and $g_{a\gamma}$ are essentially free parameters.} 
to the high-energy scale $\Lambda$ associated with the broken symmetry in the ultraviolet (the effective dimensionless coefficient $C_{\gaga}$ scales appropriately the $a$-$\gamma$ coupling whenever the ALP couples to other SM particles, most often $C_{\gaga}=1$ is considered hereafter).
The whole ALP phenomenology (production, decay, interaction with magnetic fields) is thereby fully defined in the $(m_a,\,g_{a\gamma})$ parameter space.

\vskip 2mm
Figure~\ref{fig:limits_preLHC} shows the current limits over about 19 orders of magnitude in ALP mass (up to the largest masses explored so far) and 13 orders in ALP-$\gamma$ coupling. In the very light mass range ($m_a\lesssim 10$~eV), the most stringent limits are provided by light-shining-through-a-wall (LSW) experiments, as well as solar photon instruments like the Tokyo Axion Helioscope (SUMICO) and the CERN Axion Solar Telescope (CAST)~\cite{Graham:2015ouw}  (see Section~\ref{ssec:irastorza}). 
%
%
A combination of cosmological constraints (number of effective degrees of freedom, primordial big-bang nucleosynthesis, distortions of the cosmic microwave-background spectrum, diffuse extragalactic photon measurements) leads to the grey region at the lowest $g_{a\gamma}$ couplings, over $m_a \approx 1$~keV--10~GeV~\cite{Cadamuro:2011fd,Millea:2015qra}. For ``intermediate'' masses and couplings, astrophysical bounds have been derived from energy loss of stars through ALP radiation, which are constrained by the ratio of red giants to younger stars in the so-called horizontal-branch (HB, light green), from the measurement of the length of the neutrino burst from Supernova SN1987A (yellow)~\cite{Chang:2018rso}, and from the absence of SN photon bursts from decays of produced ALPs (green)  (see Section~\ref{ssec:giannotti}).

\begin{figure}[htbp!]
\centering
\includegraphics[width=0.8\columnwidth]{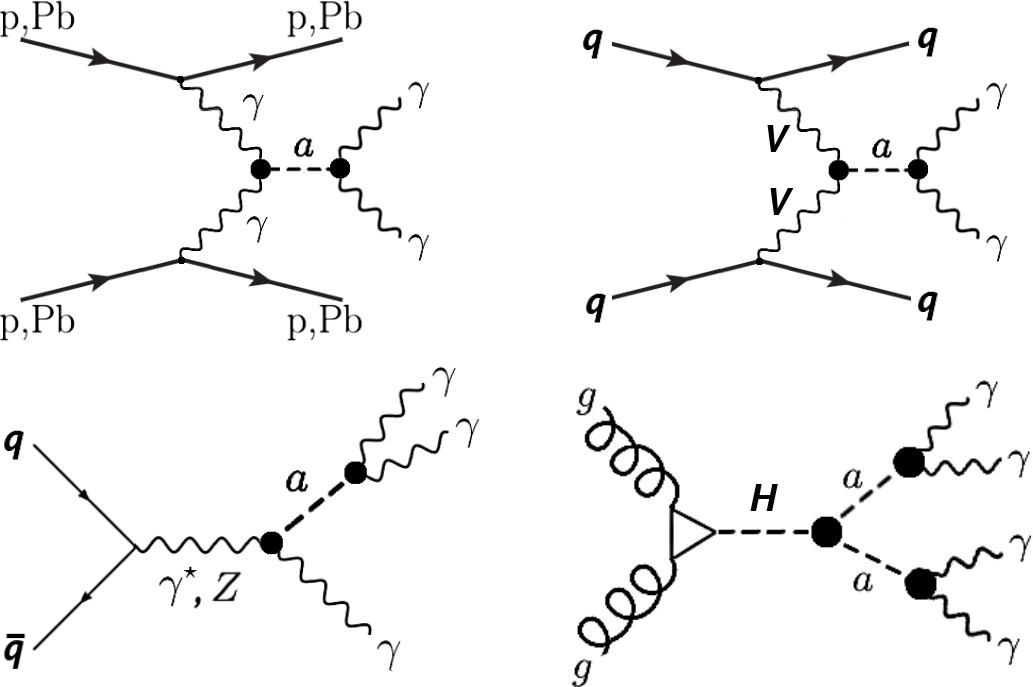}
\caption{Representative diagrams of ALPs searches at the LHC via exclusive diphotons (top left), diphotons from vector-boson-fusion (with $V = \gamma, \mathrm{Z}$, top right), exotic triphoton Z-boson decays (bottom left), and exotic 4-photon Higgs-boson decays (bottom right).}
\label{fig:diags}
\end{figure}

\vskip 2mm
Experimental attention to ALPs above the MeV scale at particle accelerators is relatively recent. Since the $a \to \gaga$ decay rate scales with the cube of the ALP mass, 
\begin{equation}
 \Gamma_{a\to\gamma\gamma} \propto \frac{g_{a\gamma}^2 m_a^3}{64\pi} \; ,
\end{equation}
for decreasing values of $m_a$ and/or $g_{a\gamma}$ the decay rate becomes so small that the ALP leaves the detector and appears as an invisible particle. In fixed-target proton and electron experiments (orange ``beam dump'' band), photons copiously produced via meson decays, or via bremsstrahlung off electron beams, can convert into ALPs via the Primakoff process off a nuclear target, and ALPs are searched for in final states with single photons (inverse Primakoff scattering process) or diphotons (from $a\to\gaga$ decays) emitted in the forward direction~\cite{Dobrich:2015jyk,Banerjee:2020fue}. The top area of Fig.~\ref{fig:limits_preLHC} is dominated by searches at $\epem$ colliders via mono-photon final states with missing energy from long-lived ``invisible''\footnote{Such final states include also the case where the ALP is a short-lived dark-matter mediator that decays into invisible DM particles.} ALPs via radiative $\Upsilon\to\gamma a$ decays (violet area) and $\epem\to \gamma a$ (light blue area) at CLEO and BaBar~\cite{Balest:1994ch,delAmoSanchez:2010ac}, and diphoton and triphoton final states ($\epem\to 2\gamma, 3\gamma$), for $m_a=50$~MeV--8~GeV and $m_a=20$--100~GeV respectively, at LEP-I and II~\cite{Acciarri:1994gb,Abbiendi:2002je} (light brown) derived in~\cite{Jaeckel:2015jla,Knapen:2016moh}, as well as similar searches at Belle-II~\cite{BelleII:2020fag}, PrimEx~\cite{Aloni:2019ruo}, and CDF~\cite{Aaltonen:2013mfa}. As one can see from Fig.~\ref{fig:limits_preLHC}, ALP limits for masses above $m_a \approx 1$~GeV were very scarce (and nonexistent above $m_a\approx 300$~GeV) before the LHC, and this is the region where the energy-frontier collider plays a unique role. Searches for axion-like particles at the LHC share the following properties:
\begin{description}
\item (1) Final states with photons plus, in some cases, Z and H bosons, are considered;
\item (2) if ALPs are long-lived (usually for $m_a \lesssim 1$~GeV), bounds can be derived from mono-photon final-states with missing transverse energy (from the escaping $a$) in the LHC detectors;
\item (3) if ALPs are short-lived (usually for $m_a \gtrsim 1$~GeV), bounds are set from their decays into photons inside the LHC detectors volume.
\end{description}
For case (2) above, LHC limits on ALPs with masses below 1~GeV in mono-photon topologies have been discussed in detail in~\cite{Mimasu:2014nea} and, since they basically overlap with the beam-dump bounds shown in Fig.~\ref{fig:limits_preLHC}, will not be discussed further. Instead, for case (3) above, competitive searches have been set for masses $m_a > 5$~GeV in 2-, 3-, 4-photon final states via searches for exclusive and inclusive $\gaga$ resonances, and exotic Z or Higgs boson decays, with the corresponding diagrams shown in Fig.~\ref{fig:diags}. 
ALP bounds have been set in some cases directly by the LHC experiments themselves, but mostly by phenomenological reinterpretations of experimental data from ATLAS~\cite{Aad:2008zzm} and CMS~\cite{Chatrchyan:2008aa} based on generic spin-0 $\gaga$ resonance searches, as discussed case-by-case below.

\subsubsection{ALP searches in exclusive $\gaga$ final states}

Proton (p) and lead ions (Pb) accelerated at LHC energies generate strong electromagnetic (e.m.) fields that, in the equivalent photon approximation~\cite{Budnev:1974de}, can be identified as quasi-real photon fluxes with very low virtualities $Q^{2} < 1/R_A^{2}\approx 0.08,\,8\cdot10^{-4}$~GeV$^2$ and longitudinal energies as large as $\omega_{\rm max} = \gamma_L/R_A \approx 2.5$~TeV, 80~GeV, respectively\footnote{Taking $R_A\approx 0.7,7$~fm as p, Pb charge radii, and $\gamma_L=E_{\rm beam}/m_{p,N}$ for their beam Lorentz factors.}~\cite{Baltz:2007kq,dEnterria:2013zqi}. On the one hand, in the Pb-Pb case, since the photon flux scales as the square of the charge of each colliding particle, $\gaga$ cross sections are enhanced by up to a factor of $Z^4 \approx 5\cdot 10^{7}$ compared to p-p (or $\epem$) collisions, although they reach relatively moderate maximum centre-of-mass (\cm) energies of the order of $\sqrt{s^{\rm max}_{\gaga}}\approx 200$~GeV. On the other, photon-photon processes in p-p collisions, although featuring much lower $\gamma$ fluxes (compensated, in part, by the much larger beam luminosities compared to heavy ions), can reach much higher $\sqrt{s^{\rm max}_{\gaga}}\approx 4.5$~TeV values, thanks to the harder proton $\gamma$ fluxes. Exploiting such interesting photon-photon possibilities in these so-called ``ultraperipheral'' collisions to search for axions at the LHC via the diagram shown in Fig.~\ref{fig:diags} (top left) was suggested in~\cite{Knapen:2016moh,Knapen:2017ebd}. The final state of interest is that of an exclusive diphoton event, \ie\ without any other particle produced, with the two photons back-to-back (the ALP is produced basically at rest due to the low virtuality of the colliding photons) and featuring a peak on top of the invariant mass ($m_{\gaga}$) distribution of the light-by-light (LbL) scattering $\gaga\to\gaga$ continuum~\cite{dEnterria:2013zqi}. In p-p collisions, the presence of very large pileup events hinders the observation of this process unless one can tag one or both protons in very forward spectrometers, such as the CMS-TOTEM PPS and ATLAS AFP ones~\cite{Royon:2015tfa}, to remove overwhelming hadronic backgrounds. The absence of pileup in Pb-Pb collisions makes this colliding mode the most competitive one for exclusive ALP searches up to $\sqrt{s^{\rm max}_{\gaga}}\approx 100\,(300)$~GeV compared to p-p running with proton taggers located at 420 (220)~m from the interaction point (IP)~\cite{Bruce:2018yzs}.

\vskip 2mm
The first exclusive diphoton search at the LHC was carried out by CMS in p-p collisions at 7~TeV with the first 36-pb$^{-1}$ of integrated luminosity~\cite{Chatrchyan:2012tv}. The event selection consisted of two photons with transverse energy $\ET>2$~GeV and pseudorapidity $|\eta|<2.5$ with no other hadronic activity over $|\eta|<5.2$. The lack of observed events imposes an upper limit cross section of $\sigma(\pp\to\rm{p}\gaga\rm{p})>1.18$~pb at 95\% confidence level (CL). Such a result was subsequently recast into the $(m_a,g_{a\gamma})$ plane (red area in Fig.~\ref{fig:limits_VBF}, right)~\cite{Knapen:2016moh,Knapen:2017ebd}, overlapping with the LEP-II limits over $m_a \approx 5$--90~GeV. A recent search of the same exclusive final state has been carried out by CMS\,+\,TOTEM in p-p collisions at 13~TeV with forward proton tagging, requiring two photons with $\ET>75$~GeV over $|\eta|<2.5$, with $m_{\gaga}> 350$~GeV and low acoplanarity~\cite{CMS:2020rzi}. Matching the diphoton mass and pseudorapidity measured in the central detectors with those of the CT-PPS proton spectrometers allows the removal of p-p pileup events. No exclusive $\gaga$ event is found above expected backgrounds, leading to an upper limit cross section of $\sigma(\pp\to\rm{p}\gaga\rm{p})>3.0$~fb at 95\% CL, that has been recast into limits on anomalous quartic photon couplings over $m_{\gaga}\approx 0.4$--2~TeV~\cite{CMS:2020rzi}. However, in terms of ALP bounds in this mass range, such a measurement is not yet competitive compared to other $\gaga$ final states discussed below.\\

\vskip 1mm
The most stringent limits on ALPs over $m_a\approx 5$--100~GeV have been set by searches in ultraperipheral Pb-Pb collisions by CMS and ATLAS~\cite{Sirunyan:2018fhl,Aad:2020cje}. The final state of interest corresponds to two photons with $\ET>2,3$~GeV, $|\eta|<2.5$, $m_{\gaga}>5$~GeV with no other hadronic activity over $|\eta|<5$. Extra photon-pair kinematic criteria are applied to remove any remaining background (mostly from misidentified $\gaga\to\epem(\gamma)$ events): $\pT^{\gaga}<1$~GeV, and acoplanarity $\Delta\phi_{\gaga}<0.01,0.03$, which enhance the photon-fusion production characterized by two back-to-back photons. Both ATLAS and CMS observe a number of exclusive diphoton counts consistent with the expectations of the very rare LbL scattering process~\cite{dEnterria:2013zqi}. For ALP searches, both experiments have subsequently injected into their measured $m_{\gaga}$ distribution, simulated $\gaga\to a\to\gaga$ peaks generated with \starlight~\cite{Klein:2016yzr} with the ALP-$\gamma$ coupling implemented as described in~\cite{Knapen:2016moh}.

\begin{figure*}[htbp!]
\centering
\includegraphics[width=0.45\linewidth]{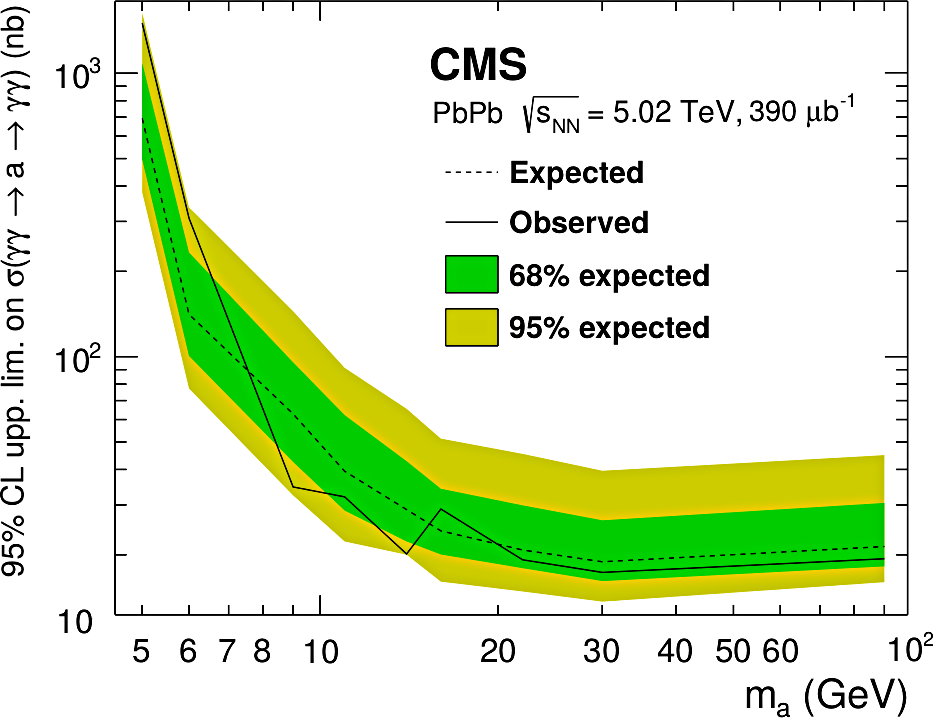}\hspace{0.5cm}
\includegraphics[width=0.45\linewidth]{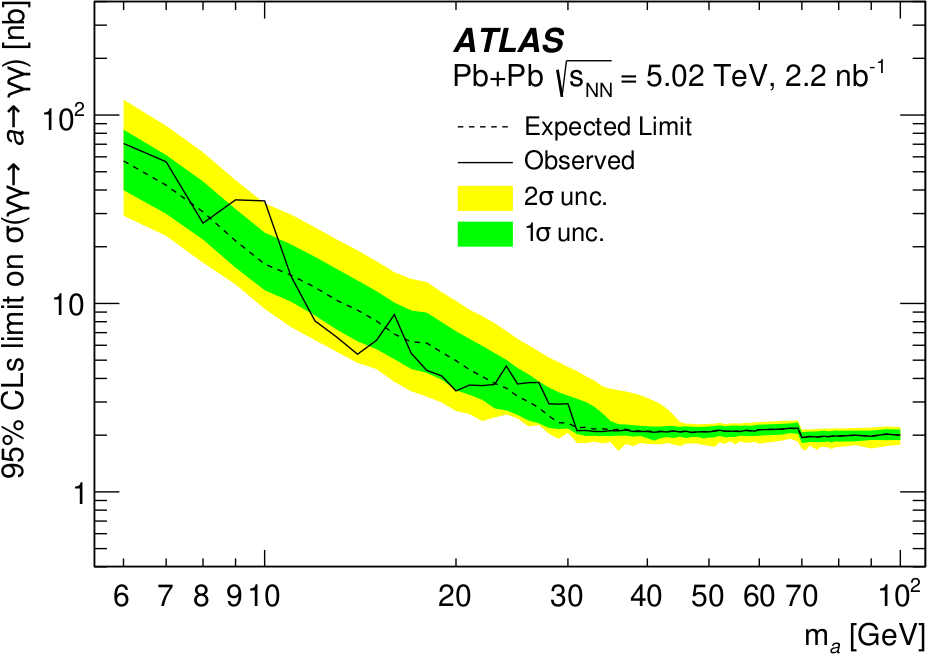}
\caption{95\% CL upper limits on the ALP cross section $\sigma(\gaga\to a\to\gaga)$ as a function of ALP mass $m_a$ measured in ultraperipheral Pb-Pb collisions at $\sqrtsnn = 5.02$~TeV in CMS (left)~\cite{Sirunyan:2018fhl} and ATLAS (right)~\cite{Aad:2020cje}. The observed (expected) upper limits are shown as solid (dashed) black curves (with $\pm1$ and $\pm2$ standard deviation bands).}
\label{fig:sigma_UPC_ALPs}
\end{figure*}

\vskip 2mm
The diphoton trigger and reconstruction efficiencies are found to be $\sim$20\% (50\%) at $m_a=5$ (50)~GeV, and the assumed ALP width is dominated by the experimental $\gaga$ mass resolution. 
It is worth to note that such experimental effects are not always taken appropriately into account by phenomenological works, which often use very simplified detector inefficiencies when reinterpreting existing data to extract new ALP limits.\\
The absence of any visible peak above the LbL continuum is used to set the expected and observed limits on the ALPs production cross section as a function of mass shown in Fig.~\ref{fig:sigma_UPC_ALPs}. The ATLAS results (right plot) exploit the largest integrated luminosity of the 2018 Pb-Pb run, and allow one to exclude ALP integrated cross sections above 2 to 70~nb at the 95\% CL over the $m_a = 6$--100 GeV range. The cross section limit from CMS (left plot) in the bin $m_a = 5$--6~GeV provides the lowest ALP mass probed so far at the LHC. It is unlikely that ATLAS or CMS will be able to go below this mass value due to the difficulties in triggering and reconstructing very low $\ET\lesssim 2$~GeV photons. However, LHCb and ALICE have good photon reconstruction capabilities down to $\sim$1~GeV (or lower, exploiting $\gamma\to\epem$ conversions) and could improve those limits in a mass region $m_a \approx 0.5$--5~GeV of otherwise difficult experimental access, thereby bridging with the beam-dump and Belle-II constraints in the ``wedge'' area shown in Fig.~\ref{fig:limits_preLHC}. Experimental challenges in this mass region are the $m_{\gaga}$ resolution, and the presence of multiple spin-0 and 2 resonances with visible diphoton decays~\cite{Klusek-Gawenda:2019ijn}.

\vskip 2mm
The limits on the ALP cross section (with signal strength $\sqrt{\mu_{\rm CLs}}$) are transformed into the corresponding limits for the ALP-$\gamma$ coupling $g_{a\gamma}\equiv 1/\Lambda$, assuming a BR$(a\to\gaga)=100\%$ branching ratio, \ie\ effectively setting $C_{\gaga}=1$ in Eq.~(\ref{eq:L_agamma}), calculated from $1/\Lambda = \sqrt{\mu_{\rm CLs}}\cdot (1/\Lambda^{\rm MC})$ where the latter is the coupling used in the MC generator. The derived $1/\Lambda$ constraints range from $g_{a\gamma}\approx 1$~TeV$^{-1}$ to 0.05~TeV$^{-1}$ (violet area in Fig.~\ref{fig:limits_final}, right). It is worth to mention that the CMS Pb-Pb ALP results have been also recast into limits on the $(m_a,g_{aB})$ plane, including also the hypercharge coupling $g_{aB}$, \ie\ processes involving the Z boson discussed in Section~\ref{sec:gammaZ}. Locally, these bounds are the strongest ones over $m_a\approx 25$--60~GeV~\cite{Sirunyan:2018fhl}, above the $\pp\to 3\gamma$ ones discussed later.

\subsubsection{ALP searches in generic $\gaga$ final states}

In p-p collisions at the LHC, ALPs can also be produced in parton-parton scatterings, via gluon-fusion and vector-boson-fusion (VBF) processes, followed by their diphoton decay. Since one usually focuses on ALP couplings to electroweak bosons, and in particular to photons in order to compare them to existing cosmological and astrophysical limits, the gluon-fusion production mode will not be further considered here\footnote{Detailed discussions on exploiting the LHC data to place bounds on ALPs sensitive to both photons and gluons can be found \eg\ in~\cite{Jaeckel:2012yz,Mimasu:2014nea,Mariotti:2017vtv,CidVidal:2018blh,Gavela:2019cmq}.} and we will only discuss diphoton final states from VBF processes such as the one shown in Fig.~\ref{fig:diags} (top-right).

\begin{figure*}[htbp!]
\centering
\includegraphics[width=0.45\linewidth]{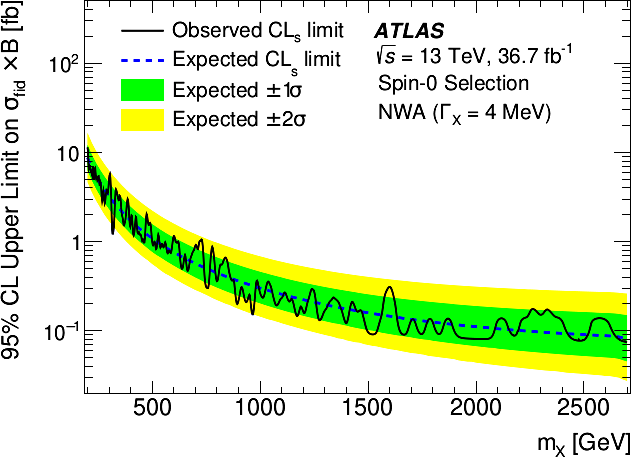}\hspace{0.1cm}
\includegraphics[width=0.45\linewidth]{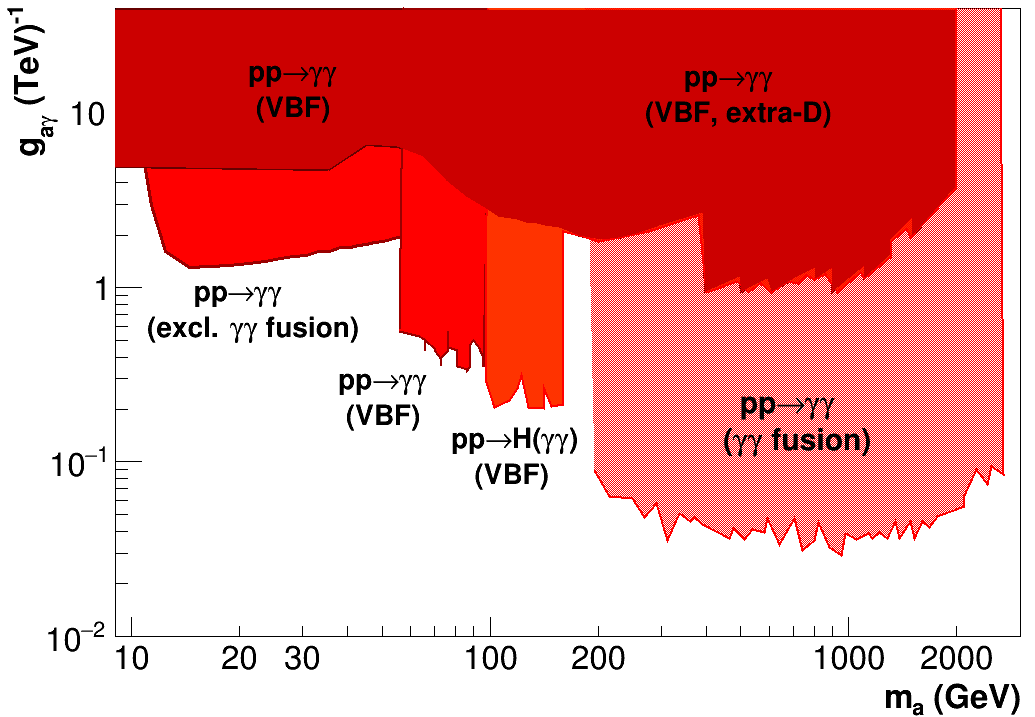}
\caption{Left: Upper limits on the fiducial cross section times diphoton branching ratio of a narrow-width ($\Gamma_X = 4$~MeV) spin-0 resonance as a function of its mass $m_X$ in p-p collisions at $\sqrts = 13$~TeV~\cite{Aaboud:2017yyg}.
Right: Derived exclusion regions on the ALP-$\gamma$ coupling vs.\ ALP mass plane from various diphoton production data sets in p-p collisions at the LHC~\cite{Jaeckel:2012yz,Jaeckel:2015jla,Knapen:2016moh,Bauer:2018uxu}.}
\label{fig:limits_VBF}
\end{figure*}

\vskip 2mm
Limits on the production of ALPs have been derived in~\cite{Jaeckel:2012yz,Jaeckel:2015jla,Knapen:2016moh,Bauer:2018uxu} based on reinterpretations of existing searches of Higgs, generic spin-0, and/or extra dimensions, as well as QCD studies, in diphoton final states measured by ATLAS and CMS over the range $m_{\gaga}\approx 10$~GeV--2.6~TeV in p-p collisions at $\sqrts = 8, 7, 13$~TeV (Fig.~\ref{fig:limits_VBF}, right). First, the ATLAS QCD diphoton data at 7~TeV~\cite{Aad:2012tba} has been used to set limits on the ALP-photon coupling of the order of $g_{a\gamma}\approx 5$~TeV$^{-1}$ over $m_a = 10$--100~GeV. The ALP diagram of Fig.~\ref{fig:diags} (top-right) is identical to Higgs boson production via VBF, accompanied by forward and backward jets from the radiating quarks, followed by its diphoton decay, with kinematic differences between scalar and pseudoscalar resonances being of the order of $\mathcal{O}$(10\%)~\cite{Jaeckel:2012yz}. Therefore, in the region $m_a = 100$--160~GeV, the VBF data from the ATLAS Higgs observation at 7 and 8~TeV~\cite{Aad:2012tfa}, establishing the maximum allowed cross sections in each mass bin, has been exploited to set the corresponding ALP bounds. For $m_a = 150$--400~GeV and $m_a = 400$--2000~GeV, the CMS~\cite{Chatrchyan:2011jx,Chatrchyan:2011fq} and ATLAS~\cite{ATLAS:2011ab} extra-dimension searches at $\sqrts = 7$~TeV respectively, have been used. 

\vskip 1mm
In addition, the work of~\cite{Knapen:2016moh} exploited the ATLAS searches for scalar diphoton resonances at $\sqrts =8$~TeV~\cite{Aad:2014ioa} to set limits over $m_{a} =65$--100~GeV (red rectangular region in Fig.~\ref{fig:limits_VBF}, right), which are basically coincident with those obtained from the OPAL $\epem\to 3\gamma$ studies at LEP~\cite{Abbiendi:2002je}. More recently, Ref.~\cite{Bauer:2018uxu} has computed the constraints on $a\to\gaga$ produced via photon-fusion in p-p collisions, employing the photon distribution functions in the proton from Ref.~\cite{Manohar:2016nzj}, to reinterpret the ATLAS upper limits on the cross sections for spin-0 resonances in 39.6 fb$^{-1}$ of p-p data collected at 13~TeV~\cite{Aaboud:2017yyg} (Fig.~\ref{fig:limits_VBF}, left). This work leads to the most competitive limits today over $m_{a} = 200$--2600~GeV (lighter red area in Fig.~\ref{fig:limits_VBF}, right). It would be appropriate, however, for LHC experiments to revisit these latter limits, as their derivation may be too optimistic in the treatment of the relevant backgrounds~\cite{Florez:2021zoo}.

\subsubsection{ALP searches in exotic Z and H boson photon decays}
\label{sec:gammaZ}

If one considers the possibility that the ALP couples not only to photons, but also to hypercharge bosons $B$, one can generalize Eq.~(\ref{eq:L_agamma}) to the following effective Lagrangian
\begin{equation}
\mathcal{L}  \supset -\frac{1}{4}g_{aB}\,a\,B^{\mu\nu}\Tilde{B}_{\mu\nu} \,.
\label{eq:L_agammaZ}
\end{equation}

\begin{figure*}[htbp!]
\centering
\includegraphics[width=0.47\linewidth]{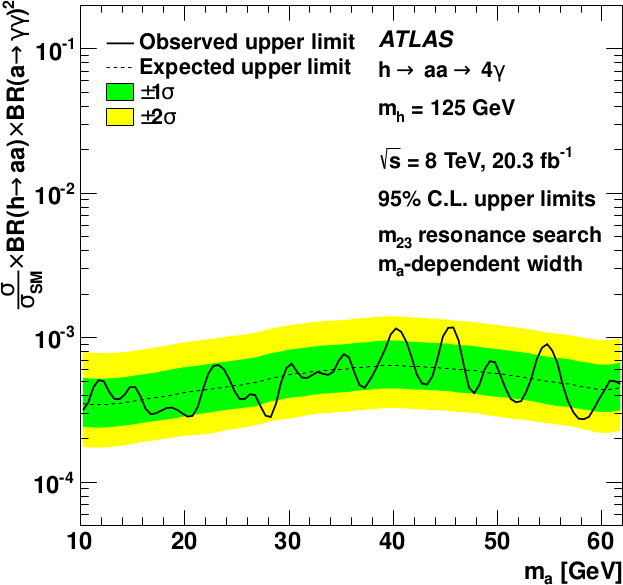}
\hspace{0.1cm}
\includegraphics[width=0.47\linewidth]{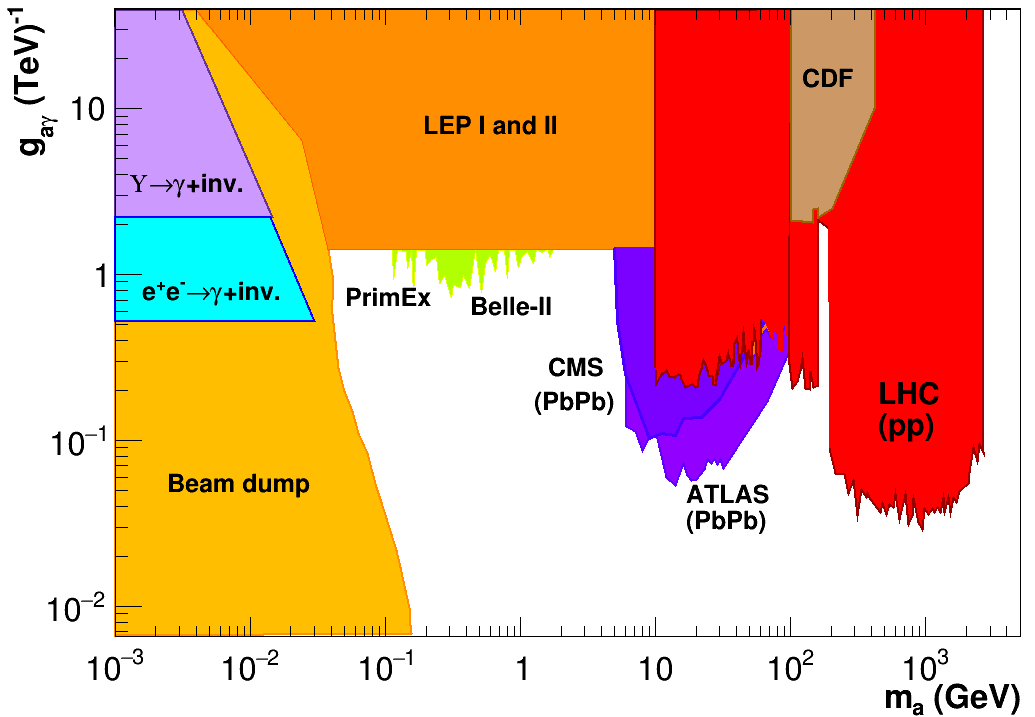}
\caption{Left: 95\% CL upper limits on $(\sigma/\sigma_{\rm SM}) \times$\,BR$(\mathrm{H}\to aa) \times$ BR$(a \to\gaga)$ with $\pm 1\sigma$ and $\pm2\sigma$ uncertainty bands from the resonance search hypothesis tests, accounting for statistical and systematic uncertainties from simulated signal samples~\cite{Aad:2015bua}.
Right: Detailed bounds in the $(m_a,g_{a\gamma})$ plane from all existing accelerator and collider ALP searches for masses $m_a\approx 1$~MeV--3~TeV. The LHC constraints summarized here are indicated by the area in orange (p-p) and violet (Pb-Pb collisions).}
\label{fig:limits_final}
\end{figure*}

The individual ALP-electroweak couplings can be related to each other through the underlying effective Wilson coefficients $C_{\gaga} = C_{WW} + C_{BB}$,  $C_{\gamma \mathrm{Z}} = c_{\theta}^2\,C_{WW} -s_{\theta}^2\,C_{BB}$, and $C_{ZZ} = c_{\theta}^4\,C_{WW}+s_{\theta}^4\,C_{BB}$, with $s_{\theta}$ ($c_{\theta}$) being the (co)sine of the weak mixing angle~\cite{Bauer:2018uxu}. The exotic decay $\mathrm{Z} \to \gamma a$ (Fig.~\ref{fig:diags}, bottom left)  is governed by the coefficient $C_{\gamma \mathrm{Z}}$. When interpreting the corresponding Z-boson experimental data in the $(m_a,g_{a\gamma})$ plane, one often assumes that the two effective coefficients $C_{\gaga}$ and $C_{\gamma \mathrm{Z}}$ are correlated, \eg\ if the ALP couples to hypercharge but not to SU(2)$_L$, then $C_{\gamma \mathrm{Z}} = -s_{\theta}^2 C_{\gaga}$ since $C_{WW} = 0$. The search for unusual triphoton decays of onshell Z bosons at colliders provides thereby sensitivity to ALPs with masses $m_a\lesssim m_Z \approx 90$~GeV~\cite{Jaeckel:2015jla,Brivio:2017ije,Bauer:2017ris}. The LEP bounds shown in Fig.~\ref{fig:limits_preLHC} include ALP searches via triphoton final states on and off the Z pole at LEP~\cite{Jaeckel:2015jla,Abbiendi:2002je} and in $\ppbar$ collisions at Tevatron (CDF)~\cite{Aaltonen:2013mfa} (brown area), assuming $C_{WW} = 0$. Given the much larger number of Z bosons measured at the LHC, significant improvements in such limits are expected from the corresponding ATLAS and CMS data sets. The most stringent constraint is set by the ATLAS measurement of BR$(\mathrm{Z} \to 3\gamma) < 2.2\cdot 10^{-6}$ at 95\% CL~\cite{Aad:2015bua}, improving by a factor of five the previous BR$<10^{-5}$ result at LEP~\cite{Acciarri:1994gb}. Different phenomenological analyses~\cite{Knapen:2016moh,Bauer:2017ris} have recast such a result into limits in the $(m_a,g_{a\gamma})$ plane, covering the region $m_a \approx 10$--70~GeV, the latter value determined from the experimental requirement of $\pT^{\gamma} > 17$~GeV (red area below the Pb-Pb bounds, in Fig.~\ref{fig:limits_final} right). It is worth noting that the phenomenological recast of the $\pp\to 3\gamma$ data is subject to intrinsic 3-photon combinatorics uncertainties, and that ideally such ALP limits extraction should be redone by the experiments themselves.

\vskip 2mm
Searches of decays of the Higgs boson to a pair of pseudoscalar bosons  (Fig.~\ref{fig:diags}, bottom-right) have been also a common BSM search channel at the LHC in the last years, motivated \eg\ by Higgs compositeness~\cite{Cacciapaglia:2019bqz} and extended Higgs sectors~\cite{Branco:2011iw} models. Whereas the Lagrangian (\ref{eq:L_agammaZ}) encoding the coupling of ALPs to electroweak gauge bosons is of dimension-5, interactions of the ALP with the Higgs boson appear only at dimension-6 and higher, with two operators mediating the decay $\mathrm{H}\to aa$ and $\mathrm{H}\to \mathrm{Z} a$ respectively~\cite{Bauer:2017ris}. Although searches for the exotic Higgs decays $\pp \to \mathrm{H} \to \mathrm{Z} a \to \mathrm{Z} \gaga$ and $\pp \to \mathrm{H} \to aa \to 4\gamma$ cannot be directly translated into constraints in the $(m_a,g_{a\gamma})$ plane, because the corresponding ALP-Higgs coefficients $C_{HZ}$ and $C_{Ha}$ are generally not related to $C_{\gaga}$, these processes probe the parameter space in which an ALP can provide the explanation of the anomalous magnetic moment of the muon~\cite{Bauer:2017ris}. The ATLAS collaboration has carried out a search of the $\mathrm{H}\to aa$ decay in p-p collisions at 8~TeV~\cite{Aad:2015bua} (Fig.~\ref{fig:limits_final}, left), which has been subsequently translated into exclusion limits on the $(m_a,g_{aH})$ plane at low ($m_a<400$~MeV) and high ($m_a=10$--62.5~GeV) ALP masses~\cite{Bauer:2017ris,Bauer:2018uxu}.

\subsubsection{Summary and outlook}

High-energy colliders provide unique possibilities to explore heavy axion-like particles (ALPs), with masses above $m_a\gtrsim 1$~GeV, that otherwise lie beyond the reach of lower energy accelerators and of existing cosmological and astrophysical constraints. ALPs with masses above $m_a\approx 1$~GeV can only be accessed via collider experiments (except for extremely weakly coupled ones, with $g_{a\gamma}<10^{-6}$~TeV$^{-1}$, excluded by cosmological observations), with the LEP and Tevatron data providing the best limits up to $m_a\approx 100,\,300$~GeV before the LHC started operation 10 years ago. Searches at the CERN LHC for ALPs decaying into two photons, $a\to\gaga$, have been summarized focusing on $a$-$\gamma$ dominant couplings. Different channels have been scrutrinized including exclusive and inclusive diphoton production in proton-proton and lead-lead collisions, $\pp,\,\PbPb \to a \to \gaga\,(+X)$, as well as exotic Z and Higgs boson decays, $\pp \to \mathrm{Z},\mathrm{H}\to a\gamma \to 3\gamma$ and $\pp \to \mathrm{H}\to aa \to 4\gamma$. The limits in the ALP-photon coupling vs.\ ALP mass plane $(m_a,g_{a\gamma})$ derived from the LHC data extend by one-order-of-magnitude the range of preceding bounds in both mass (from $m_a\approx 300$~GeV previously at Tevatron, up to 2.6~TeV now) and in $g_{a\gamma}$ for masses $m_a\approx 100$~GeV (from $g_{a\gamma} = 0.5$~TeV$^{-1}$ previously at LEP, down to 0.05~TeV$^{-1}$ now).  Exclusive diphoton searches in PbPb collisions provide currently the best ALP exclusion~limits for masses $m_a\approx 5$--100~GeV, whereas the other channels are the most competitive ones over $m_a\approx 100$~GeV--2.6~TeV. ALP searches in exotic Z boson decays have been also exploited to set bounds in the $m_a\approx 10$--70~GeV range, with specific assumptions about the relationship between the $C_{\gaga}$ and $C_{\gamma \mathrm{Z}}$ coupling operators. Searches for exotic 3- and 4-photon Higgs boson decays through intermediate ALPs provide also strong constraints on new dimension-6 and -7 operators. Although the latter cannot be recast into bounds in the standard $(m_a,g_{a\gamma})$ plane, the study of several couplings simultaneously is crucial to identify the most interesting regions in ALP parameter space.

\vskip 2mm
The LHC constraints summarized here are indicated by the orange (for p-p collisions) and violet (for Pb-Pb collisions) areas in Fig.~\ref{fig:limits_final} (right), compared to previous fixed-target and collider results. In the next 15 years, the currently uncovered (white) areas in this plot will be probed by different experiments. The ``empty'' region $m_a\approx 50$~MeV--5~GeV will be accessible to various future experiments both descending along the vertical $g_{a\gamma}$ axis alone, such as Belle-II with 50~ab$^{-1}$ (improving its current limits at $g_{a\gamma}\approx 1$~TeV$^{-1}$ by two orders of magnitude, over $m_a\approx 0.2$--8~GeV)~\cite{Dolan:2017osp} and GlueX (50 times better than the local PrimEx limit today of $g_{a\gamma}\approx 1$~TeV$^{-1}$, over $m_a\approx 50$--500~MeV)~\cite{Aloni:2019ruo} or extending the beam-dump ``wedge'' in both $g_{a\gamma}$ and $m_a$ directions at new CERN experiments searching for long-lived particles such as \eg\   SHiP~\cite{Alekhin:2015byh}, FASER~\cite{Feng:2018pew}, and MATHUSLA~\cite{Alpigiani:2018fgd}.

\vskip 2mm
The full completion of the LHC physics program will provide about 10 and 100 times more integrated luminosities in heavy-ion and p-p collisions, respectively, than exploited so far in ALP searches. The clean searches via exclusive $\gaga\to a\to\gaga$ final states will benefit, in particular, from the possibility of forward proton-tagging (as well as new ion species, including p-Pb running) that will further extend the LHC coverage in mass regions $m_a\approx 2.6$--5~TeV (and $m_a\approx 200$--400~GeV) with nonexistent (or locally weaker) bounds today~\cite{Baldenegro:2018hng,Schoeffel:2020svx}. Thus, ultimate 10--100 improvements in the ALPs bounds are expected at the LHC with ALP-photon couplings approaching $g_{a\gamma}\approx 10^{-3}$~TeV$^{-1}$ (or, ideally, ALP discovery) relatively uniformly over $m_a\approx 1$~GeV--5~TeV. Going to lower couplings for axion-like particles with masses in the multi-GeV regime, will require a next generation of $\epem$ colliders such as FCC-ee~\cite{Abada:2019zxq}, and ILC~\cite{Steinberg:2021iay}.

\clearpage
\subsection{Results and new benchmarks for axion/ALP searches}
\label{ssec:axion-results}
\subsubsection{Results}
\label{sssec:axion-results}

The following figures~\ref{fig:PS_photon},~\ref{fig:PS_gluon}, and ~\ref{fig:PS_fermion} show the current status of experimental bounds and future projections for the three benchmark models related to axion/ALP physics, namely {\it photon dominance}, {\it gluon dominance}, and {\it fermion dominance}.
These three benchmarks scenarios have been described in the Physics Beyond Colliders BSM report~\cite{Beacham:2019nyx}, and are repeated here below for the reader's convenience.

Taking a single pseudoscalar field $a$ one can write a set of its couplings 
to photons, quarks, leptons and other fields of the SM. This is shown in Eq.~(\ref{Leff_a}). In principle, the set of possible couplings is very large and only the flavour-diagonal subset is considered in this study. 

The PBC proposals have considered the following benchmark cases:

\begin{itemize}
\item  {\em BC9, photon dominance:} Assuming a single ALP state $a$, and its predominant coupling to photons, all phenomenology (production, decay, oscillation in the magnetic field) can be determined as functions on  $\{m_a, g_{a\gamma} \}$ parameter space, where the $g_{a\gamma}=f^{-1}_\gamma$ notation is used (previous sections of this report have also used $\Lambda \equiv 32 \pi^2 f$). The current status of experimental bounds and projections for future experiments are shown in Figures~\ref{fig:PS_photon} and \label{fig:PS_photon2}.

\item {\em BC10 fermion dominance:} Assuming a single ALP state $a$, and its predominant coupling to fermions, all phenomenology (production and decay) can be determined as functions of  $\{m_a, f^{-1}_l, f^{-1}_q \}$.
Furthermore, for the sake of simplicity, we take $f_q=f_l$.  The current status of experimental bounds and projections for future experiments are shown in Figure~\ref{fig:PS_fermion}.
 
 \item  {\em BC11, gluon dominance:} This case assumes an ALP coupled to gluons. The parameter space is $\{m_a, f^{-1}_G \}$. Notice that in this case, the limit of $m_a < m_{a,QCD}|_{f_a=f_G}$ is unnatural as it requires fine tuning and therefore is less motivated. The current status of experimental bounds and projections for future experiments are shown in Figure~\ref{fig:PS_gluon}.
 
\end{itemize} 

The ALP portals, $BC9-BC11$, are {\em effective} interactions, and would typically require UV completion at or below $f_i$ scales. This is fundamentally different from vector, scalar, and neutrino portals that do not require external UV completion.   Moreover, the renormalization group evolution is capable of inducing new couplings. All the sensitivity plots in this Section assume a cut-off scale of $\Lambda =1$\, TeV.


\subsubsection{Final considerations and proposal of new benchmarks} 
\label{sssec:axion-recommendations}

The existing search programme for ALPs has focused on three benchmark scenarios~\cite{Beacham:2019nyx}, namely \emph{photon dominance}, \emph{fermion dominance}, and \emph{gluon dominance}, each of which considers one specific term in the effective Lagrangian for ALPs given in Eq.~(\ref{Leff_a}).\\

{\it We recommend to keep the fermion, gluon and photon dominance scenarios as benchmarks for future experimental results}.

\vskip 2mm
However, these benchmark scenarios do not cover the case that the phenomenology of ALPs is dominated by their effective coupling to $W$ bosons. This scenario, which was first proposed in Ref.~\cite{Izaguirre:2016dfi}, differs from the three existing benchmark scenarios in that the production of ALPs is dominated by rare decays (as in the case of \emph{fermion dominance}),\footnote{We note that the theoretical predictions for rare meson decays involving ALPs with couplings to $W$ bosons are theoretically cleaner than the ones for \emph{fermion dominance} (which suffer from a sensitivity to unknown UV physics) and the ones for \emph{gluon dominance} (which are affected by large hadronic uncertainties).} but the decays are dominated by the effective ALP-photon coupling (as in the case of \emph{photon dominance}). The relevant formulas for the various rare meson decays can be found in Ref.~\cite{Izaguirre:2016dfi}, noting the convention $g_{aW} = 4 g^2 C_{WW} / \Lambda = \alpha_2 C_{aWW} / (2 \pi f_a)$, where $g$ is the $SU(2)_L$ gauge coupling and $\alpha_2 = g^2/(4\pi)$, see Eqs.~(\ref{Leff_a}) and~(\ref{eq:CWW_definition}). The ALP decay width into photons is given by
\begin{equation}
 \Gamma_{a\gamma\gamma} = \frac{4\pi\alpha^2 m_a^3}{\Lambda^2} \left|C_{WW}\right|^2 \; ,
\end{equation}
where $\alpha$ is the electromagnetic fine-structure constant.
The interplay between these effects leads to a different relation between production cross section and decay length than in any of the currently studied benchmarks. Furthermore, to constrain this scenario it will be crucial to carry out new searches, for example of $B \to K + a(\to \gamma \gamma)$.\\

{\it We therefore recommend that the case of {$W$ boson dominance} be included as a new benchmark scenario in future studies of experimental sensitivities for ALPs.}

\vskip 2mm
Including these additional interactions in the calculation of experimental predictions also paves the way for the next milestone in the study of ALP phenomenology: scenarios with more than one type of interactions (so-called \emph{codominance}). An interesting example, explored recently in Refs.~\cite{Ertas:2020xcc,Kelly:2020dda}, is the case $C_{BB} = C_{WW} = C_{GG}$ (and negligible couplings to fermions), which leads to an accidental cancellation in the effective ALP-photon coupling.  

While a detailed prescription about how to practically implement the co-dominance scenario in future experimental searches is still being worked out,
we emphasize that they naturally occur in many UV completions and that it will be essential to consider them in order to obtain a comprehensive understanding of the parameter space of ALP models (see section~\ref{ssec:kahloefer} for further details).

\vskip 2mm
Another important challenge for the near future will be to improve the calculational methods to obtain predictions for the \emph{gluon dominance} benchmark, in particular regarding the different ways in which ALPs with couplings to gluons can be produced (including production in partonic showers~\cite{Aielli:2019ivi}, production from gluon fusion~\cite{Kelly:2020dda}, and production in rare $B$ meson decays~\cite{Chakraborty:2021wda}).\\ 

{\it We recommend that further studies of these effects be carried out in order to establish a common framework for simulating ALP signals in the {gluon dominance} benchmark and to ensure that future sensitivity studies can be compared on equal footing.}

\begin{figure*}[ht!]
\centering
\includegraphics[width=0.8\linewidth]{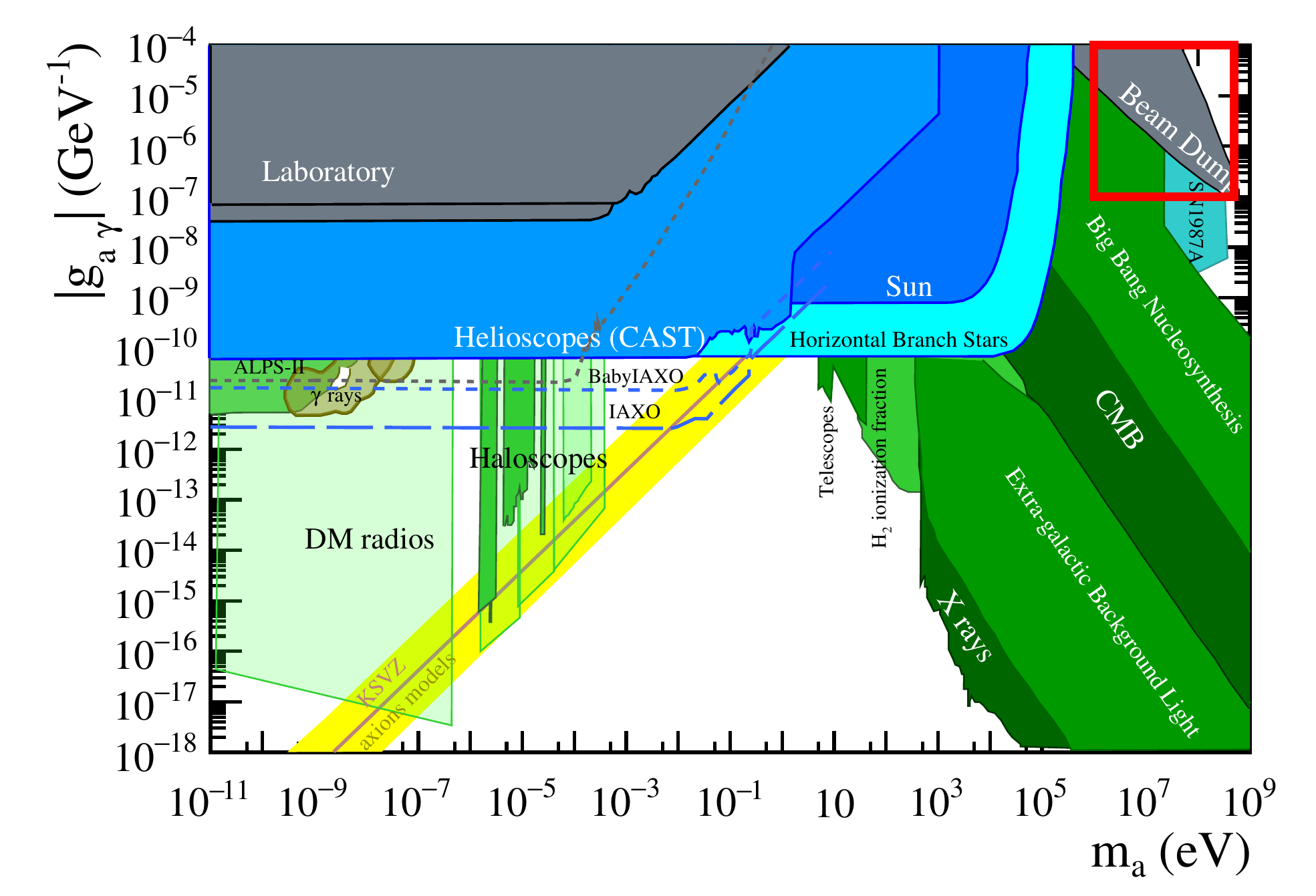}
\includegraphics[width=0.8\linewidth]{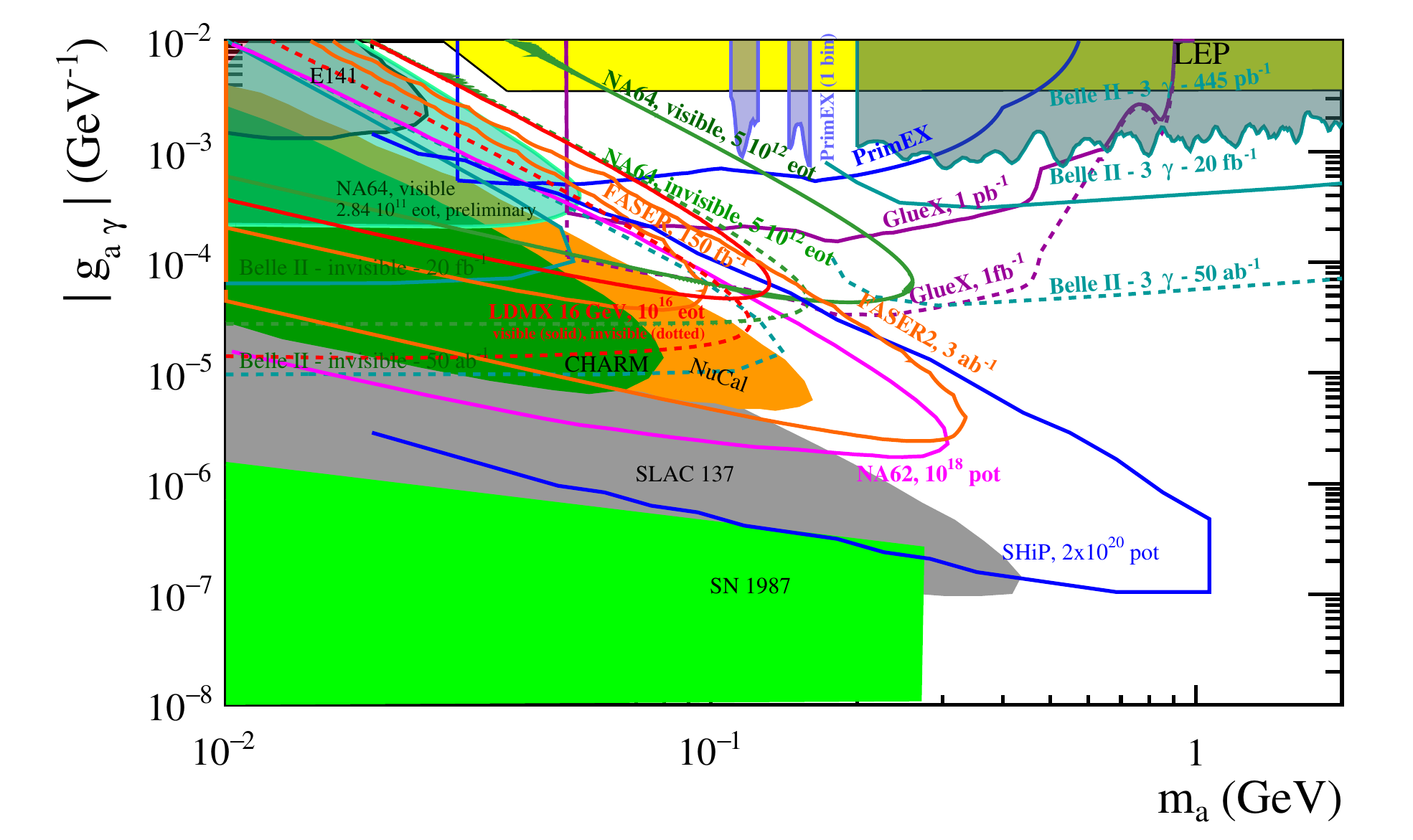}
\caption{{\bf Axions/ALPs with photon coupling.} {\it Top plot:} overall view of axion/ALP searches. Figure revised by I.~Irastorza in Ref.~\cite{Beacham:2019nyx}. {\it Bottom plot:} zoom in the region of interest for accelerator-based experiments up to a few GeV (note the different units for the mass ranges in the two plots). 
Shaded areas are excluded regions from:
LEP (data:~\cite{Acciarri:1994gb,Abreu:1991rm,Abreu:1994du,Acciarri:1995gy}; interpretation:~\cite{Jaeckel:2015jla});
Belle II~\cite{BelleII:2020fag};
E141 (data~\cite{Riordan:1987aw} and interpretation~\cite{Dobrich:2017gcm}); 
E137~\cite{Bjorken:1988as}; 
NA64~\cite{Banerjee:2020fue};
CHARM~\cite{Gninenko:2012eq}; 
NuCal~\cite{Blumlein:1990ay};
PrimEx~\cite{Aloni:2019ruo} based on~\cite{Larin:2010kq}.
Curves are projections from:
Belle II~\cite{Kou:2018nap} for 20~fb$^{-1}$ and 50~ab$^{-1}$;
SHiP~\cite{Anelli:2015pba};
FASER and FASER2~\cite{Ariga:2018uku};
NA64$_e^{++}$~\cite{NA64:eplus} in visible and invisible modes;
interpretation of the physics reach~\cite{Aloni:2019ruo}
of PrimEx~\cite{Larin:2010kq} and GlueX experiments at JLab. The Figures are revised from Ref.~\cite{Lanfranchi:2020crw}.
}
\label{fig:PS_photon}
\end{figure*}

\begin{figure*}[ht!]
\centering
\includegraphics[width=0.8\linewidth]{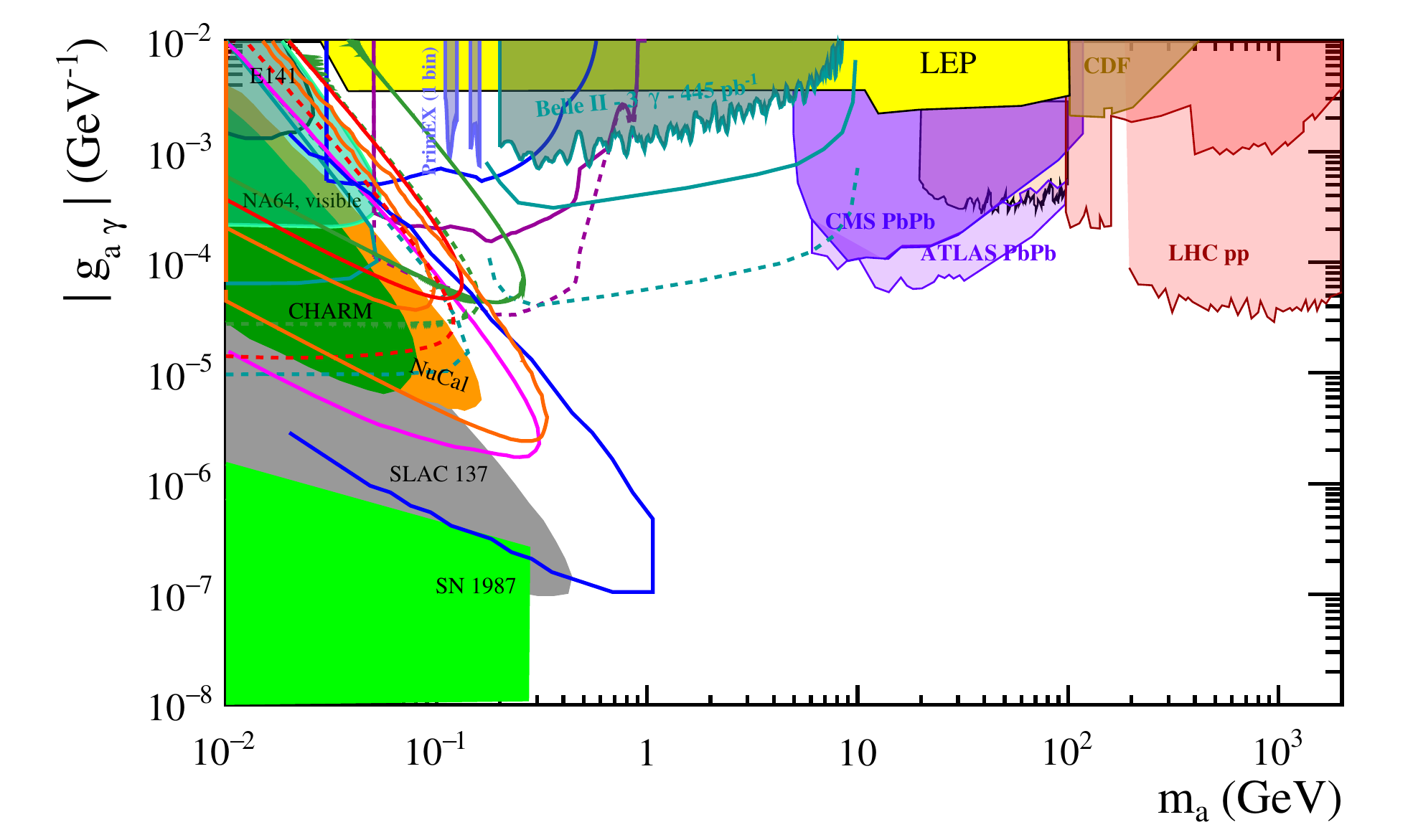}
\caption{{\bf Axions/ALPs with photon coupling.}  Region of interest for accelerator-based (including collider) experiments. The current bounds and projections below a few GeV, including LEP bounds, are explained in Fig.~\ref{fig:PS_photon}. Above a few GeV, the current limits come from analyses of data from CDF~\cite{Aaltonen:2013mfa}, and ATLAS and CMS in PbPb~\cite{Sirunyan:2018fhl,Aad:2020cje} and $pp$ collisions (details on the latter can be found in Fig.~\ref{fig:limits_VBF}, right).}
\label{fig:PS_photon2}
\end{figure*}

\begin{figure*}[ht!]
\centering
\includegraphics[width=0.8\linewidth]{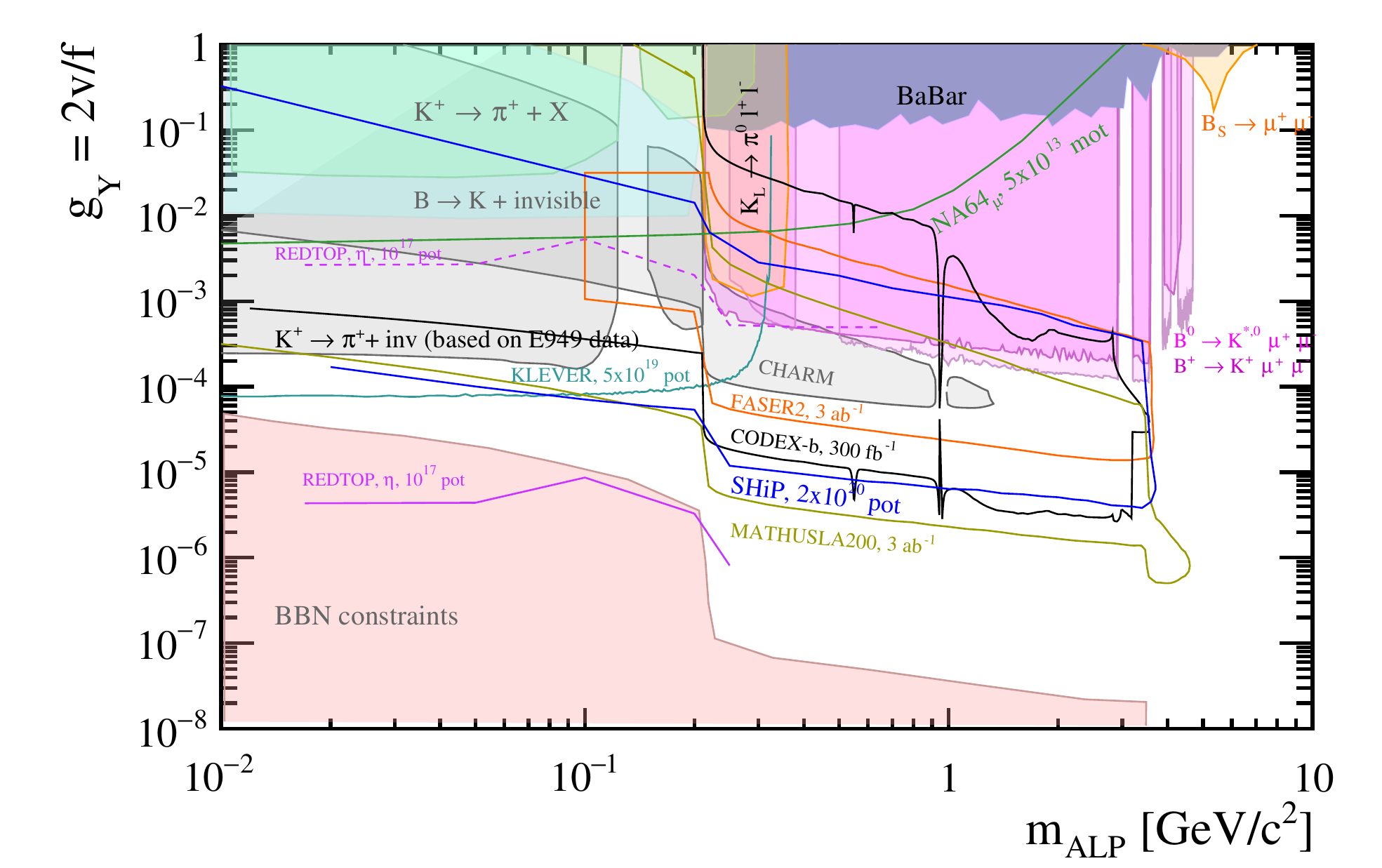}
\caption{{\bf Axions/ALPs with fermion coupling (MeV--GeV range)}: Current bounds (filled areas) and prospects (solid lines) from FASER2~\cite{Ariga:2018uku}, CODEX-b~\cite{Aielli:2019ivi}, MATHUSLA~\cite{Beacham:2019nyx}, SHiP~\cite{Beacham:2019nyx},
and REDTOP~\cite{Beacham:2019nyx}.
CHARM and LHCb filled areas have been adapted
by F.~Kahlhoefer, following Ref.~\cite{Dobrich:2018jyi}. E949 area has been computed by the KLEVER collaboration and M.~Papucci based on E949 data. All other exclusion regions have been properly recomputed by M.~Papucci, following Ref.~\cite{Dolan:2014ska}.}
\label{fig:PS_fermion}
\end{figure*}

\begin{figure*}[ht!]
\centering
\includegraphics[width=0.8\linewidth] {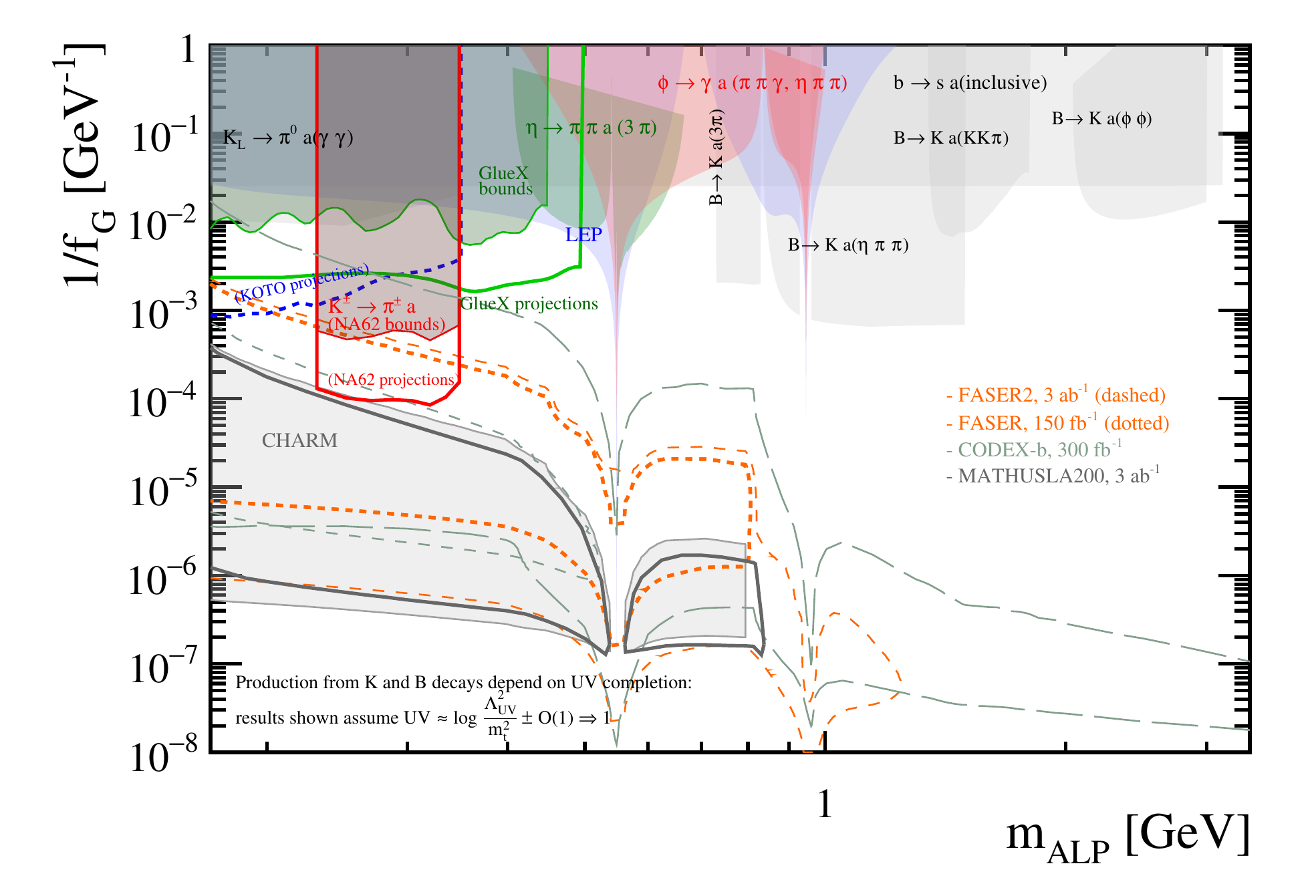}
\caption{{\bf Axions/ALPs with gluon coupling (MeV--GeV range)}. 
Current bounds (coloured filled areas) and prospects (solid and dashed lines) for FASER (FASER2)~\cite{Ariga:2018uku}, CODEX-b~\cite{Aielli:2019ivi}, and MATHUSLA200.
The CHARM grey filled area has been computed by F.~Kling, recasting the search for long-lived particles decaying to two photons performed by CHARM~\cite{Bergsma:1985qz}.
Other coloured filled areas are kindly provided by Mike Williams and revisited from Ref.~\cite{Aloni:2018vki}.
The gray areas depend on UV completion and the results shown
assume $\approx [{\rm log \Lambda^2_{UV}}/m^2_t \pm {\mathcal{O}}(1)] \Rightarrow 1$.
The NA62 bounds are based on $1.6 \cdot 10^9$ $K^+$ decays~\cite{Ceccucci:2014oza}. NA62 and KOTO projections are based on~\cite{Gori:2020xvq}.
}
\label{fig:PS_gluon}
\end{figure*}

\clearpage
\section{Feebly-interacting scalar particles}
\label{sec:scalar}
\subsection{Scalar portal and its connection to Higgs physics from a theory viewpoint}
\label{ssec:gori}
{\it Author: Stefania Gori, <sgori@ucsc.edu>} 
\newcommand{\beq}{\begin{eqnarray}}
\newcommand{\eeq}{\end{eqnarray}}
\newcommand{\bea}{\begin{eqnarray}}
\newcommand{\eea}{\end{eqnarray}}
\newcommand{\gev}{{~\rm GeV}}
\newcommand{\lsim}{\begin{array}{c}\,\sim\vspace{-26pt}\\< \end{array}}
\newcommand{\gsim}{\begin{array}{c}\sim\vspace{-26pt}\\> \end{array}}
\newcommand{\ZZ}{{\rm Z}}
\newcommand{\mZ}{m_{\rm Z}}
\newcommand{\mW}{m_{\rm W}}
\newcommand{\dPS}{{\rm dPS}}
\newcommand{\mupmum}{\mu^+\mu^-}

\newcommand{\zprime}{{\rm Z^\prime}}
\newcommand{\mzp}{m_{\zprime}}
\newcommand{\vprime}{v_\Phi}
\newcommand{\gp}{g'}

\newcommand{\CV}{C_{\rm _V}}
\newcommand{\CA}{C_{\rm _A}}
\newcommand{\thetaW}{\theta_{\rm _W} }
\newcommand{\GF}{G_{\rm F}}

\subsubsection{Introduction and Motivation}
\label{sssec:gori}
The Higgs portal interaction between the Higgs and a new scalar $\hat S$, $\frac{\epsilon}{2} \hat S^2 |H|^2$, arises in many models proposed to address some of the most important open problems in particle physics: the nature of dark matter (DM), the baryon asymmetry (for a review see e.g.~\cite{Assamagan:2016azc}), and the origin of the electroweak scale. It is one of the only three renormalizable portals that connects the Standard Model (SM) to the dark sector.

In these proceedings, we will show that the Higgs portal interaction is and will be probed by several complementary searches and measurements at accelerator experiments, both at the high-energy frontier (LHC and future colliders) and at the high-intensity frontier (beam dump experiments, flavor physics experiments, and neutrino experiments). For simplicity, we will focus on the minimal dark scalar model. We emphasize, however, that the phenomenology discussed in these proceedings is rather generic and arises in many richer dark sector models. Additional non-minimal model signatures not discussed in these proceedings include invisible and semi-visible decays of the dark scalar, as predicted by many DM models (see e.g.~\cite{Krnjaic:2015mbs,Arcadi:2021mag}), as well as decays into long-lived new physics particles, as predicted by neutral naturalness models (see e.g.~\cite{Kilic:2018sew,Alipour-fard:2018mre}). 

\subsubsection{A minimal dark scalar model}\label{sssec:gori:MinimalModel}
The most minimal model for the Higgs portal interaction contains a new dark scalar singlet under the SM gauge group, $\hat S$, interacting with the SM Higgs through the potential

\beq\label{eq:SH}
\mathcal L = \mathcal L_{kin}+ \frac{\mu_s ^ 2}{2} \hat S^2-\frac {\lambda_s} {4!} \hat S ^ 4-\frac{\epsilon}{2}\hat S ^ 2 |\hat H | ^ 2 - V(\hat H),
\eeq
where $V(\hat H)= -\mu ^ 2 |\hat H |^ 2+\lambda |\hat H | ^ 4$ is the SM Higgs potential and where we have imposed a $Z_2$ symmetry under which $\hat S\to -\hat S$ and $\hat H\to \hat H$. In the case of a broken $Z_2$ symmetry, the additional term $\hat S|\hat H | ^ 2$ can be added to the Lagrangian.
The Higgs portal interaction is the only gauge-invariant, $Z_2$-symmetric, interaction of the dark scalar with the SM fields. Extended models can contain additional interactions of the dark scalar, as e.g.\ interactions of $\hat S$ with a fermionic dark matter candidate, or with additional states of the dark sector as e.g.\ dark glue-balls in twin Higgs models.

Phenomenologically, there are two interesting regimes, depending on the sign of $K\equiv(2\lambda\mu_s^2 - \epsilon\mu^2)(2\lambda\lambda_s-3\epsilon^2)$:

\begin{enumerate}
\item If $K>0$, the dark scalar acquires a VEV, $v_s$, spontaneously breaking the $Z_2$ symmetry. This leads to mixing between $\hat S$ and the SM Higgs boson. The dark scalar mass eigenstate is $S=\hat S \cos\theta +\hat H\sin\theta$, with a mixing angle that, at the first order in $\epsilon$, is given by $\tan\theta\sim \epsilon v v_s/(m_h^2-m_s^2)$, with $m_s$ the physical mass of the dark scalar.
\item If $K\leq0$, the dark scalar does not acquire a VEV and does not mix with the SM Higgs boson.
\end{enumerate}

This minimal model can be therefore fully phenomenologically described by a small set of independent parameters:
\beq\label{eq:param}
m_s, \sin\theta,\kappa\,,
\eeq
 where 
 $\kappa$  is the $Z_2$-preserving coupling between two dark scalars and a Higgs, $\kappa S^2 h$, that can be computed in terms of the Lagrangian parameters in (\ref{eq:SH}).
 
\begin{figure*}[t]
\begin{center}
 \includegraphics[width=0.9\linewidth]{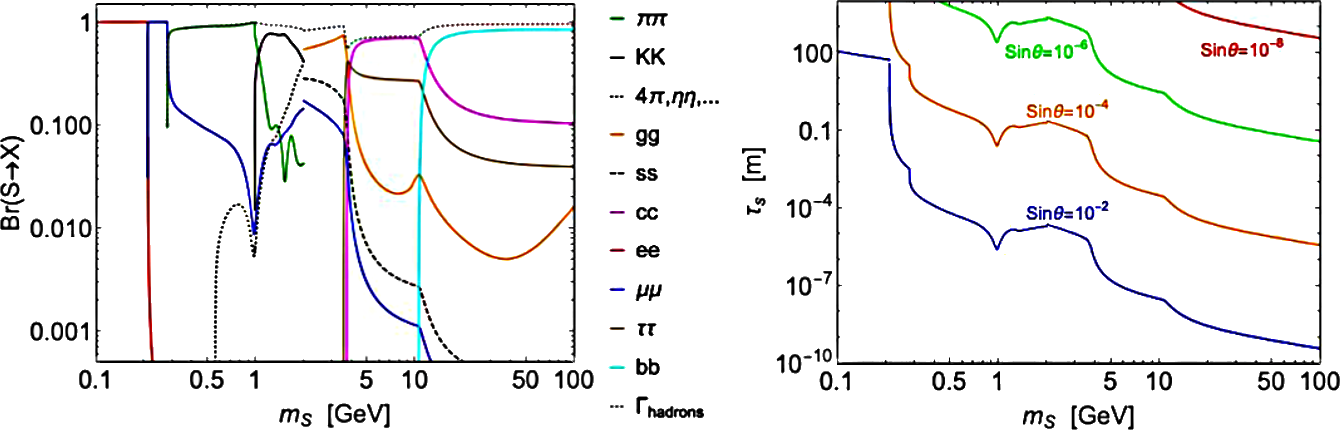}
\caption{Left: Dark scalar branching ratios as a function of the mass $m_s$. Note that the branching ratios are independent of $\sin\theta$. Right: Dark scalar lifetime in unit of meters as a function of the scalar mass, $m_s$, having fixed representative values for the mixing angle $\sin\theta$.}
\label{Fig:DecaysS}
\end{center}
\end{figure*}

In the first case, the dark scalar mass eigenstate, $S$, can be singly produced at accelerator experiments thanks to its mixing with the Higgs. It can also be pair-produced through an off-shell or on-shell Higgs boson. The dark scalar will also affect the Higgs couplings to SM particles, through its mixing with the Higgs.
The mixing term drives also the dark scalar decays. In fact, the dark scalar couples to SM fermions and gauge bosons as a SM Higgs, but with a strength reduced by $\sin\theta$. The several branching ratios of the dark scalar and its proper lifetime are presented in Fig. \ref{Fig:DecaysS}. This lifetime is computed in the minimal dark scalar model, in which $S$ can only decay back to SM particles. It is important to keep in mind that there are large theoretical uncertainties in the calculation of these decay widths for light scalars below the GeV scale (see~\cite{Winkler:2018qyg} for a review; this is the reference we use to produce Fig.~\ref{Fig:DecaysS}). In the second case, $S$ can only be pair-produced through an off-shell or on-shell Higgs boson and is stable (it can only decay to lighter dark sector particles in non-minimal models).

In the following Sections, we will discuss the complementary ways of probing the dark scalar at accelerator experiments. 

\subsubsection{Production and detection of a light scalar below the Higgs threshold}
\begin{figure*}[t!]
\begin{center}
\raisebox{11.5pt}{  \includegraphics[width=.44\linewidth]{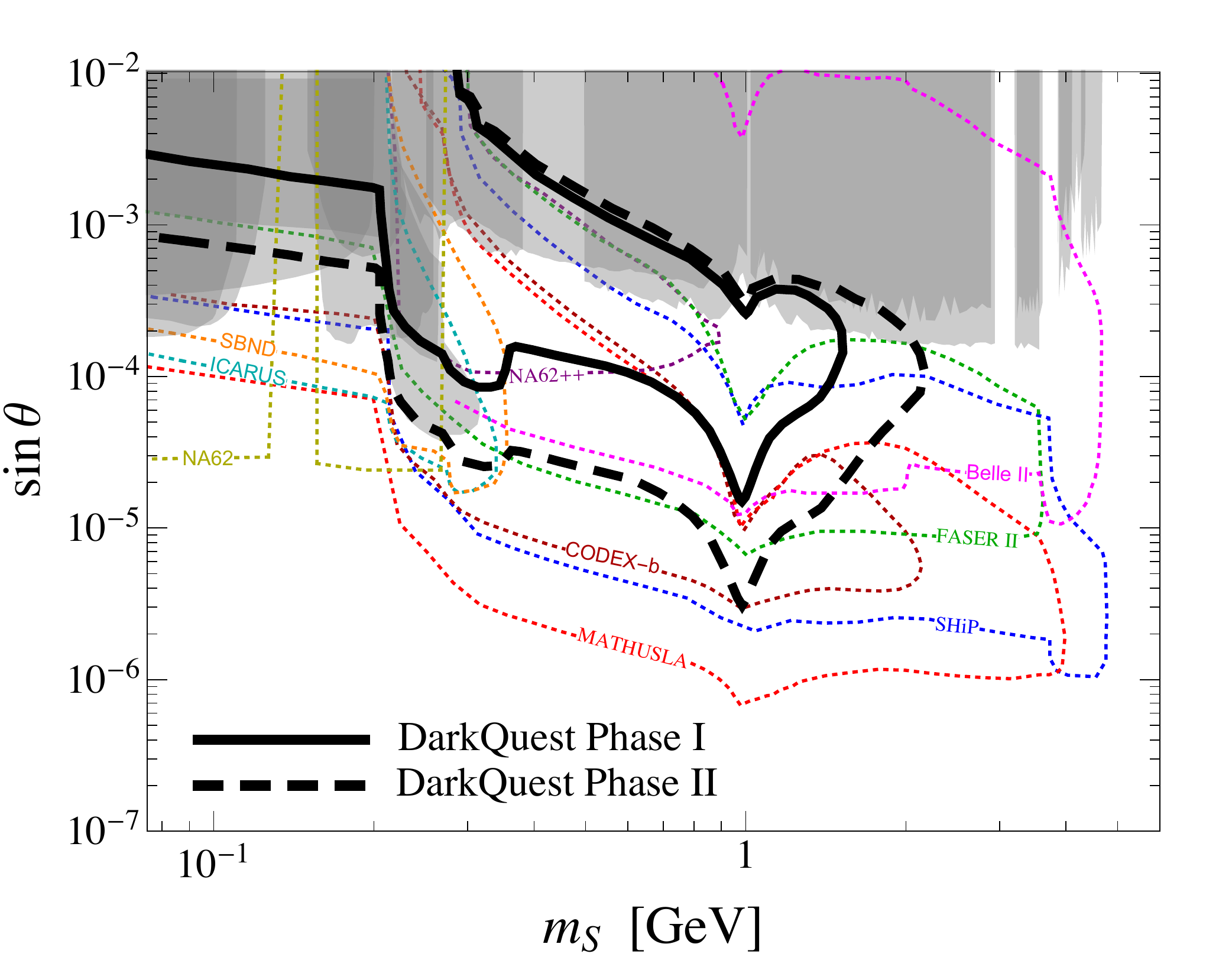}}
  \includegraphics[width=.54\linewidth]{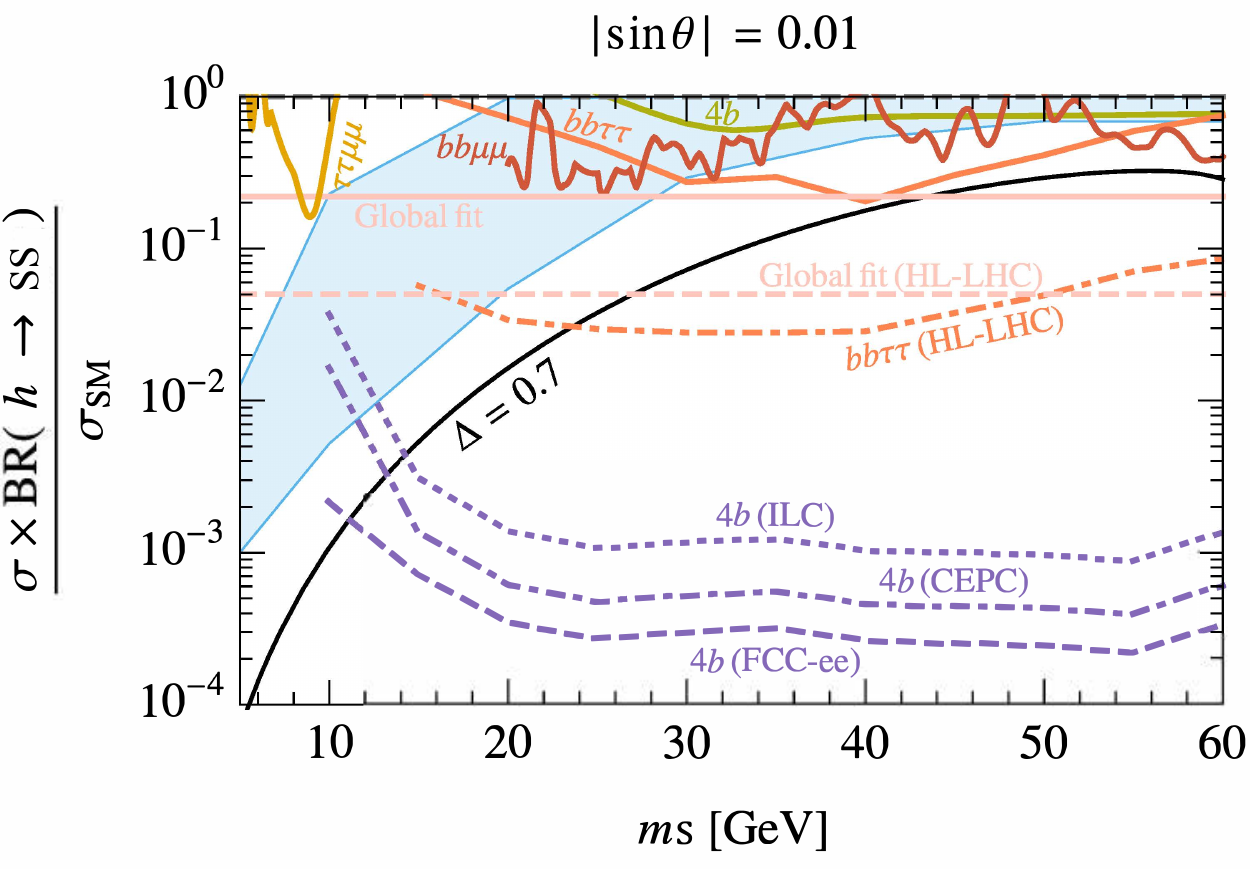}
\caption{ Left: Sensitivity of neutrino experiments (SBND and ICARUS), Kaon experiments (NA62), proposed experiments (CODEX-b, DarkQuest Phase I and Phase II, FASER, MATHUSLA, SHiP), and Belle II to the light scalar parameter space. The gray regions are regions already probed by past experiments. From~\cite{Batell:2020vqn}. Right: Bounds on the light scalar parameter space from LHC searches for Higgs exotic decays. Shown are also the future expected sensitivities on $h\to SS\to 4b$ at ILC, CEPC, and FCC-ee (purple lines).  The indirect limit on the total exotic branching fraction from global fits is indicated by the pink horizontal line, and its HL-LHC projection is shown with a pink dashed line. The blue shaded region shows points predicting a strong first-order electroweak phase transition. To obtain this region, the Higgs-scalar mixing angle is fixed to $\sin\theta=0.01$. Adapted from~\cite{Kozaczuk:2019pet}.}
\label{Fig:ProbesLightScalars}
\end{center}
\end{figure*}

Light dark scalars can be produced at high-energy and high-intensity accelerator experiments. Below the few GeV scale, dark scalars can be abundantly produced by future high-intensity accelerator experiments as CODEX-b, DarkQuest, FASER, MATHUSLA, SHiP through the bremsstrahlung production and meson decays, $K\to\pi S,~B\to KS$. The dark scalar can be detected through its decays to two charged hadrons or leptons. The corresponding reach curves are shown in the left panel of Fig. \ref{Fig:ProbesLightScalars} and are compared to the reach of past and present high-intensity experiments, CHARM, LSND, E787/E949, LHCb, and NA62 (gray region in the plot). In the figure, we also show the future reach of Belle II (through $B\to K^{(*)}S$), and of the neutrino experiments SBND and ICARUS. See~\cite{Batell:2020vqn} and references therein for the several curves. For a further discussion about the light scalar parameter space, see also the recent review article~\cite{Lanfranchi:2020crw}. It is important to keep in mind that the uncertainties in the calculation of the several scalar decay widths mentioned in the previous section particularly affect the reach of high-intensity experiments on displaced dark scalars. 

\vskip 2mm
Above the few GeV mass range, dark scalars can be exclusively produced at the LHC and future high-energy colliders. The production cross section is given by the corresponding SM Higgs cross section times $\sin^2\theta$. So far, the LHC has performed a few searches for singly produced dark scalars, decaying promptly to $\mu^+\mu^-$~\cite{Sirunyan:2019wqq} and $\tau\tau$~\cite{CMS:2019hvr} in some part of the mass range $m_s<m_h/2$ (diphoton resonance searches have been performed in the 65--110 GeV mass range~\cite{Sirunyan:2018aui}). The former signature has also been searched for by the LHCb collaboration~\cite{Aaij:2020ikh}. These searches only very weakly constrain the mixing parameter $\sin\theta$ ($\sin\theta\sim\mathcal O(1)$) in the minimal model discussed in these proceedings. It will be interesting to see the reach on singly-produced light dark scalars using future LHC(b) data.

\vskip 2mm
Higgs exotic decays can also (pair-) produce light dark scalars~\cite{Curtin:2013fra}. The corresponding cross section is proportional to $\kappa^2/m_h$ (see (\ref{eq:param})). Several searches have been done by the ATLAS and CMS collaborations for both prompt and displaced scalars (see e.g.~\cite{CMSExoDecays,ATLAS:2018wjg} for the latest summary plots for prompt decays, and~\cite{Aad:2019xav,CMS:2020idp} for displaced decays). Present searches are sensitive to exotic decay branching ratios as small as $\sim{\rm{BR}}(h\to SS)\sim \mathcal O(10^{-1})$, and put stringent bounds on the $\kappa$ parameter in (\ref{eq:param}). In the right panel of Fig.~\ref{Fig:ProbesLightScalars} we report the several bounds on the Higgs exotic branching ratios and the projections for the HL-LHC. The future bounds on $h\to SS\to 4b$ at ILC, CEPC, and FCC-ee are also shown in the figure (see purple lines). For comparison, in the figure, the authors also report the region of parameter space that leads to a strong first-order phase transition, after having set $\sin\theta=0.01$ (blue shaded region)~\cite{Kozaczuk:2019pet}. Present and future Higgs exotic decay searches will probe a large part of this parameter space.

\subsubsection{Production and detection of a heavy scalar above the Higgs threshold}
Dark scalars with a mass above the Higgs mass can also be produced at the LHC. $S$ has SM-like properties and can be searched for in diboson or difermion resonant signatures. Also, $S$ can be pair-produced through an off-shell Higgs boson exchange~\cite{Craig:2014lda}. The corresponding cross section is, however, relatively small in the interesting regions of parameters. 

\begin{figure}[h]
\begin{center}
 \includegraphics[width=.9\columnwidth]{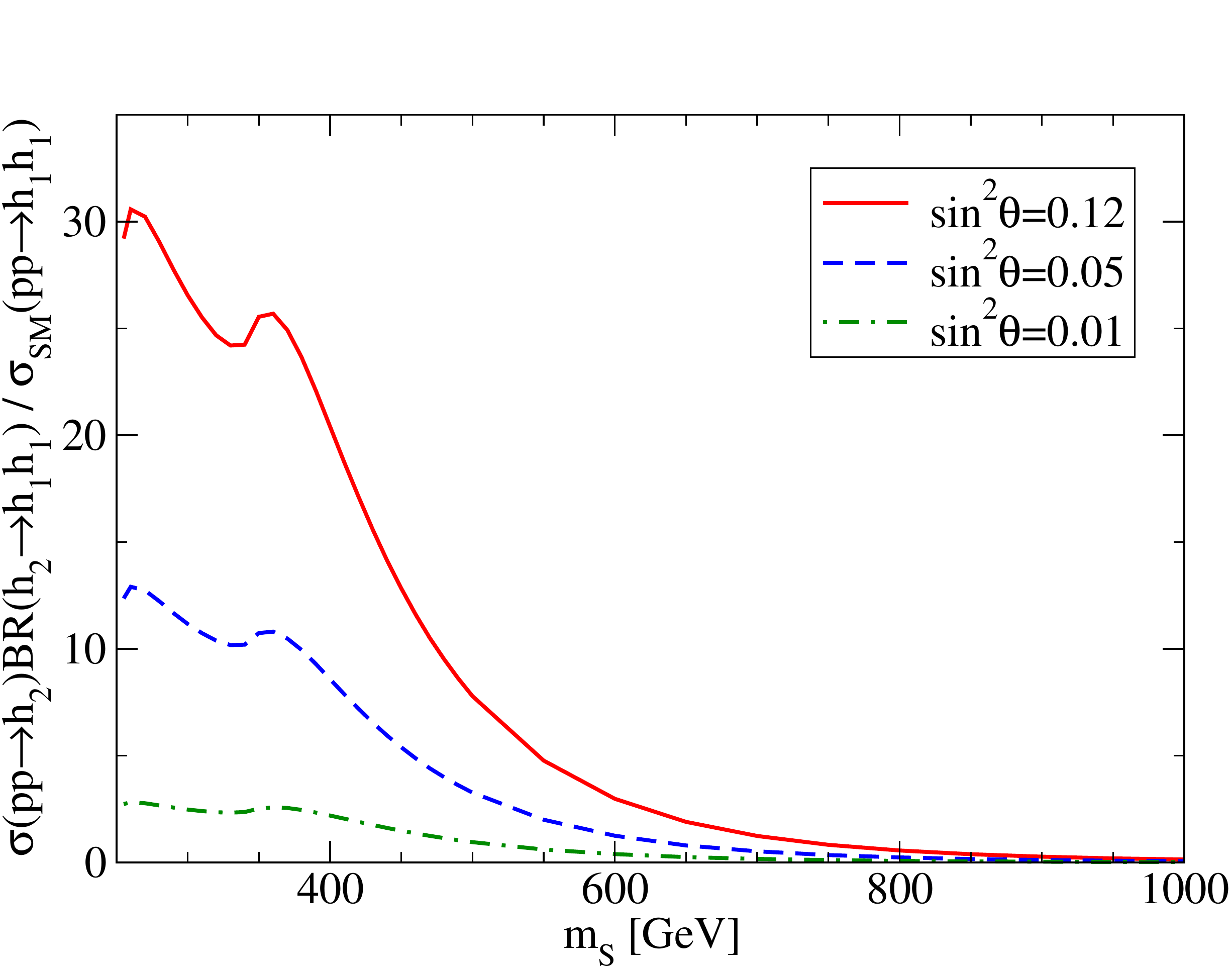}
 \caption{Maximum 13 TeV di-Higgs production cross section normalized to the SM prediction, as a function of the heavy dark scalar mass. For the plot, the quartic coupling $\lambda_s/4$ in (\ref{eq:SH}) has been fixed to $\lambda_s/4=1.05$. Adapted from~\cite{Lewis:2017dme}.
 }
\label{Fig:DiHiggs}
\end{center}
\end{figure}

Perhaps the most interesting set of signatures is the single production of a heavy dark scalar that subsequently decays into two SM Higgs bosons: $pp\to S\to hh$. This is a $Z_2$-breaking process and, depending on the value of $\sin\theta$, can enhance the LHC di-Higgs production cross section by one order of magnitude if compared to the SM prediction (see Fig. \ref{Fig:DiHiggs}). Part of the parameter space for $\sin^2\theta=\mathcal O(0.1)$ is already probed by the ATLAS and CMS searches for resonant di-Higgs production (see~\cite{Sirunyan:2018two,Aad:2019uzh} for the combination of searches using 13 TeV data).

\subsubsection{Indirect probes of a dark scalar}
The mixing between the dark scalar and the SM Higgs boson also modifies the Higgs rates at the LHC. In particular, the several SM rates will be reduced by a $\cos^2\theta$ factor. The present LHC Higgs measurements are in approximate agreement with the SM predictions (see~\cite{CMS:2020gsy,Aad:2019mbh} for the latest combinations of Higgs rate measurements). This leads to a bound at the level of $\sin\theta\lesssim\mathcal O({\rm{few}}\times 0.1)$. For some recent analysis of these constraints see e.g.~\cite{Robens:2015gla}. This bound is independent of the value of the heavy scalar mass, even if the mixing angle, $\theta$, and the scalar mass, $m_s$, are related as discussed at the beginning of Sec. \ref{sssec:gori:MinimalModel}. 

\subsubsection{Summary}
The Higgs portal is among the most motivated portals connecting the dark sector to the SM sector. In these proceedings, we have reviewed the complementarity of high-energy and high-intensity experiments to probe the minimal model for the Higgs portal. Present and future searches will be able to probe vast regions of parameter space of this minimal model.


\clearpage
\subsection{Direct searches for feebly-interacting dark scalars at the LHC central detectors}
{\it Author: Verena Ingrid Martinez Outschoorn,\\ <Verena.Martinez@cern.ch>} 
\subsubsection{Overview}

Dark scalars appear in models for naturalness, dark matter, axions, and electroweak baryogenesis, among others. Various anomalies can be explained by the presence of additional scalars, often through interactions with the Higgs boson, such as the gamma-ray excess at the center of the galaxy measured by the FermiLAT~\cite{Ackermann:2013yva} and the antiproton excess in cosmic rays measured by the AMS-02 experiment~\cite{Aguilar:2016kjl} (see for example Ref.~\cite{Hooper:2019xss}). The ATLAS, CMS and LHCb experiments at the LHC are able to probe new scalars in the mass range $\mathcal{O}(100~\mathrm{MeV}-100~\mathrm{GeV})$.

\vskip 2mm
New scalars have very rich phenomenology with many final states, mass regimes and lifetime ranges. Searches at the LHC are typically interpreted in terms of phenomenological models that guide the analysis strategies. Some of the most widely used examples include exotic Higgs boson decays~\cite{Curtin:2013fra}, where in the simplest example a new scalar is added to the SM, mixing with the Higgs boson and inheriting its Yukawa couplings. In this case, the preferred decays are to heavy particles, a recurring theme in searches. In the case of axion-like particles, exotic Higgs boson decays arise from higher dimensional operators~\cite{Bauer:2017ris}. In these models, decays to gluons and photons are favored. Note these are examples of general expectations, though detailed predictions depend on the specific model and the mass range considered, since several final states may be kinematically disfavored. 

\vskip 2mm
Searches in ATLAS, CMS and LHCb may be categorized into a few types as discussed in the following sections: inclusive searches (Section~\ref{sec:vimartin_inclusive_searches}), exotic Higgs boson decays (Section~\ref{sec:vimartin_exo_higgs_decays}) and long-lived particle searches (Section~\ref{sec:vimartin_LLP_searches}). Other related searches not discussed in this section target visible or semi-visible signals, corresponding to final states with missing transverse energy. 
This report focuses on a selection of recent results from ATLAS, CMS and LHCb. 

\subsubsection{Inclusive Searches}
\label{sec:vimartin_inclusive_searches}

Inclusive searches for a new state decaying to SM particles are prototypical analyses in high energy collider experiments and are often referred to as bump hunting. At low masses, these are challenging analyses because of trigger limitations and require dedicated strategies. 

A very powerful search that sets constraints on several new physics scenarios from LHCb targets a narrow resonance decaying to a pair of muons $X \rightarrow \mu\mu$ in the mass range $m_X \approx 1~\mathrm{GeV}-60~\mathrm{GeV}$~\cite{Aaij:2020ikh}. This search is able to achieve sensitivity at such low masses because of very soft triggers, as well as advanced techniques such as real time analysis that allows for offline-quality alignment online and the Turbo stream that selects interesting portions of events. The results probe a range of scenarios with dark scalars and different mixing angles with the Higgs boson. This search is also able to exclude a tantalizing excess from CMS in the dimuon spectrum at $28$~GeV when accompanied by one $b$-jet and one forward jet~\cite{Sirunyan:2018wim}. A cross-check by ATLAS also saw no hints of such a signal~\cite{ATLAS-CONF-2019-036}.

CMS is also able to search for similar low mass dimuon resonances $X \rightarrow \mu\mu$ in the mass range $m_X \approx 11~\mathrm{GeV}-200~\mathrm{GeV}$, excluding the $Z$ boson mass window~\cite{Sirunyan:2019wqq}. This search is also made possible by the use of innovative trigger techniques such as scouting where only a fraction of events are recorded to reduce the rates and thresholds. 

Several other searches target different final states and mass ranges. CMS has a search for diphoton resonances $X\rightarrow \gamma\gamma$ targeting the mass range $m_X \approx 70~\mathrm{GeV}-110~\mathrm{GeV}$~\cite{Sirunyan:2018aui}. The reach at low mass is primarily driven by the trigger. Using only a partial dataset, this search currently has a slight excess $m_{\gamma \gamma} \approx 95.3$~GeV with 2.8 (1.3) $\sigma$ of local (global) significance. 

\subsubsection{Exotic Higgs Boson Decays}
\label{sec:vimartin_exo_higgs_decays}

The Higgs boson is a very narrow resonance and even weakly coupled new particles may generate visible branching ratios. Since the Higgs boson is produced at the LHC, it is possible to search for exotic Higgs boson decays to new particles directly. This section focuses on prompt signatures, while the next section focuses on displaced signatures from long-lived particles. 

A major challenge for several searches for exotic Higgs boson decays are soft and overlapping decay products. This is because some of the most common signals targeted produce decay chains such as $h \rightarrow aa \rightarrow (XX)(YY)$ or $h \rightarrow Za \rightarrow (\ell\ell)(XX)$, where $a$ is a new scalar, and $(XX)$ and $(YY)$ are pairs of standard model particles from each $a$-boson decay. 

One recent result from ATLAS searches for $h\rightarrow Za \rightarrow (\ell\ell) (jj)$, where $(jj)$ is a single hadronic jet, and targets the mass range $m_a \approx 500~\mathrm{MeV}-4~\mathrm{GeV}$~\cite{Aad:2020hzm}. New dedicated multivariate techniques are used to separate $a \rightarrow (jj)$ decays from the dominant $Z+\mathrm{jets}$ background and to estimate $m_a$. The search is interpreted in terms of a decays to a pair of gluons, a signature favored in axion-like particle scenarios, setting limits on exotic Higgs boson branching ratios of around $20\%$.

Another recent result from ATLAS is the search for $h\rightarrow aa \rightarrow (bb) (bb)$, where $(bb)$ is a single jet containing the decay products of two $b$-quarks, and targets the mass range $m_a \approx 15~\mathrm{GeV}-30~\mathrm{GeV}$~\cite{Aad:2020rtv}. In this case a new dedicated multivariate algorithm is also used to separate $a \rightarrow (bb)$ decays from the dominant $Z+\mathrm{jets}$ background. The search is sensitive to scenarios with additional scalars with Yukawa couplings that decay preferentially to $b$-quarks and sets limits on exotic Higgs boson branching ratios of around $80\%$. This is the first time that such low momentum signatures are targeted with gluons or $b$-jets and there is much room for performance improvements. 

Other searches for exotic Higgs decays target final states with leptons, where signals with taus can be important below the $b$-production threshold and muons below the $\tau$-production threshold. CMS has performed a search for $h \rightarrow aa \rightarrow (\mu\tau) (\mu\tau)$ or $(\tau\tau) (\tau\tau)$ targeting the mass range $m_a \approx 4~\mathrm{GeV}-15~\mathrm{GeV}$~\cite{Sirunyan:2019gou}. In this analysis, the different decay modes cannot be distinguished because the final state is a muon with a nearby track that can be interpreted as the one-prong decay of the tau. This search sets limits on $h \rightarrow aa$ assuming a relationship between branching ratios coming from Yukawa couplings. 

Another recent search from CMS focuses on exclusive final states and searches for $h \rightarrow aa \rightarrow (\mu\mu) (\tau\tau)$ targeting the mass range $m_a \approx 3.6~\mathrm{GeV}-21~\mathrm{GeV}$~\cite{Sirunyan:2020eum}. In order to have sensitivity to the lower mass range, the analysis uses a dedicated identification algorithm to identify hadronic tau decays with a muon inside, targeting $a \rightarrow (\tau_\mu\tau_{\mathrm{had}})$. The search sets limits on exotic Higgs boson branching ratios of around $10^{-4}$. 

Finally, several searches have been performed in the four-lepton final state. The most recent result is from CMS searching for $h \rightarrow aa \rightarrow (\ell \ell) (\ell\ell)$ where $\ell$ is an electron or muon, and targeting the mass range $m_a \approx 4~\mathrm{GeV}-60~\mathrm{GeV}$~\cite{CMS:2020bni}. This final state has been interpreted in both scenarios with new light scalars with Yukawa couplings, which is particularly sensitive in the final state with muons at low mass, and in scenarios with axion-like particles, setting limits on exotic Higgs boson branching ratios of around $10^{-4}$~or~$10^{-5}$. 

\begin{figure*}
    \centering
    \begin{minipage}{0.55\textwidth}
            \centering
            \vspace{-10pt}
            \includegraphics[width=1.0\textwidth]{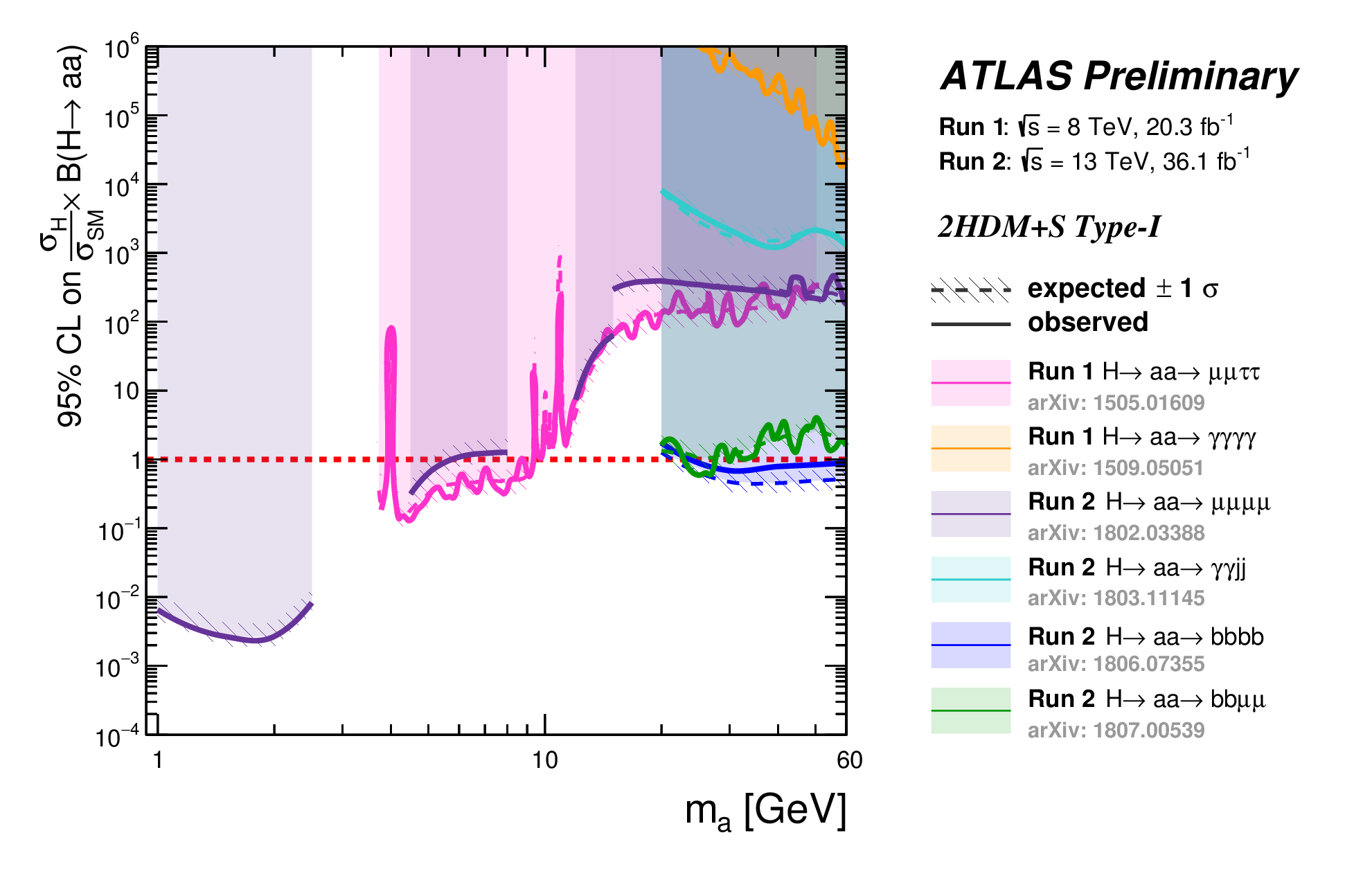}
            \label{fig:exoHdecay_ATLAS}
        \end{minipage}
        \begin{minipage}{0.05\textwidth}
        \end{minipage}
        \begin{minipage}{0.4\textwidth}
            \centering
            \vspace{-10pt}
            \includegraphics[width=1.0\textwidth]{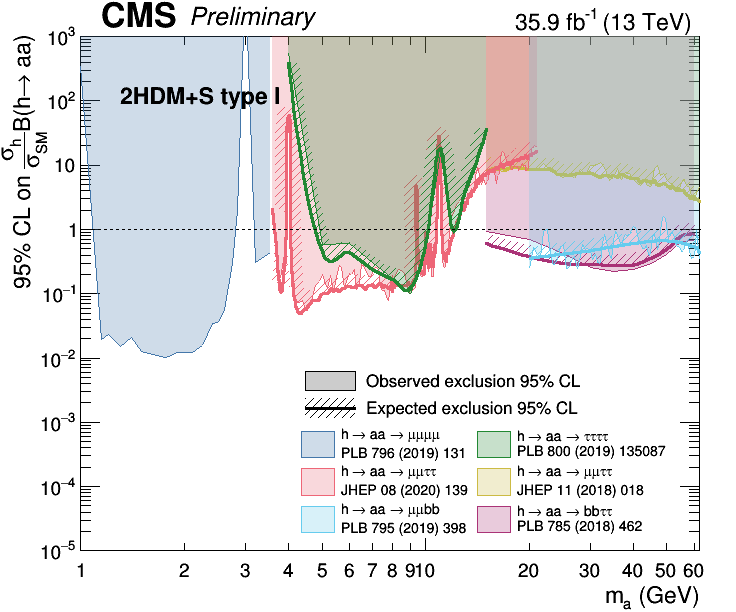}
            \label{sig:exoHdecay_CMS}
    \end{minipage}
    \vspace{-10pt}
    \caption{Observed and expected 95\% CL upper limits on $\sigma_H / \sigma_{\mathrm{SM}} \times B(h\rightarrow aa)$ in the 2HDM+S Type I model for (left) ATLAS~\cite{ATL-PHYS-PUB-2018-045} and (right) CMS. The branching fractions of the new boson to SM particles are computed following the prescriptions in Refs.~\cite{Curtin:2013fra,Haisch:2018kqx}}
    \label{fig:ExoH_summary}
\end{figure*}

\begin{figure*}
    \centering
    \includegraphics[width=1.0\textwidth]{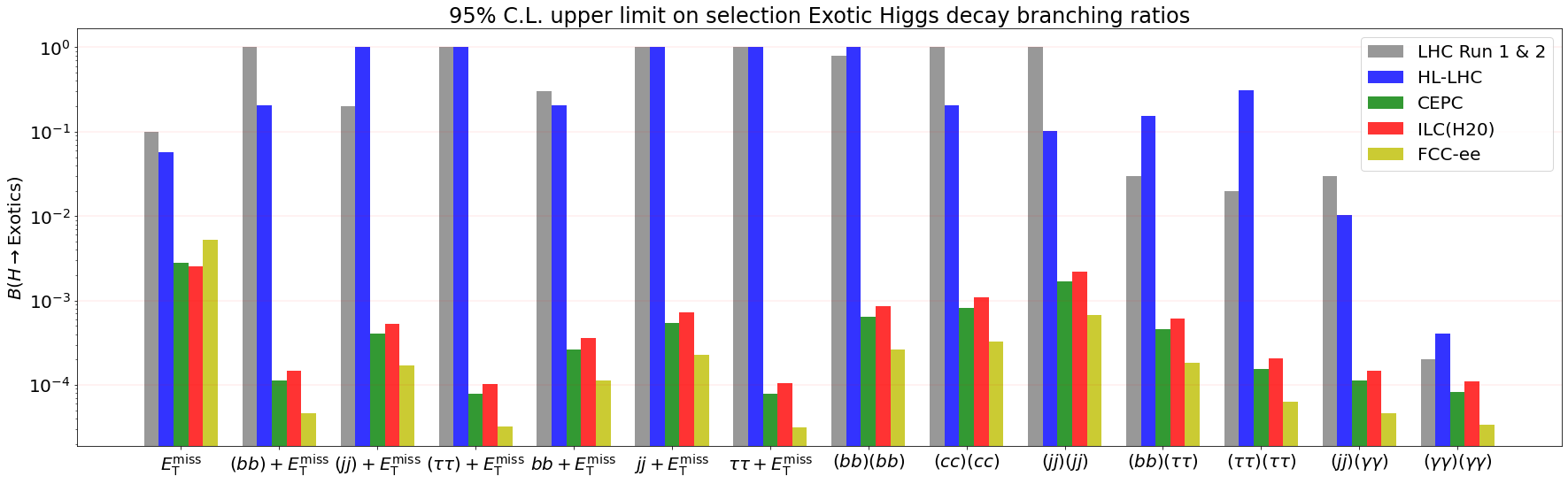}
    \caption{The 95\% C.L. upper limit on selected Higgs exotic decay branching fractions comparing results from Run 1 and 2 at the LHC to predictions for the HL-LHC, CEPC, ILC and FCC-ee for a set of benchmark parameter choices. Note in some cases current results are already more sensitive than the predictions. Future predictions are taken from~\cite{Liu:2016zki}.}
\label{fig:ExoHdecay_projection}
\end{figure*}

The results in the different final states can be compared in the context of a model to set limits on the branching ratio of the exotic Higgs boson to a new light scalar $B(h\rightarrow aa)$, as shown in Figure~\ref{fig:ExoH_summary}. Different channels offer sensitivity for different mass ranges. Projections to HL-LHC and especially to future colliders show major gains in sensitivity~\cite{Liu:2016zki}, as can be seen in Figure~\ref{fig:ExoHdecay_projection}. 

\subsubsection{Long-Lived Particle Searches}
\label{sec:vimartin_LLP_searches}

Searches for long-lived particles target new signals that travel macroscopic distances in the detector before decaying. They are very challenging analyses because they require dedicated trigger and reconstruction methods. There are many possible signatures sensitive to a broad range of lifetimes, see Ref.~\cite{Alimena:2019zri} for a comprehensive review.

The inclusive $X\rightarrow \mu\mu$ analysis from LHCb described in Section~\ref{sec:vimartin_inclusive_searches} also includes a dedicated search for long-lived decays using muons that are inconsistent with originating from any primary vertex and with a decay topology consistent with a long-lived particle~\cite{Aaij:2020ikh}. This search is particularly sensitive to the lower mass regime $m_X < 3$~GeV and low lifetimes.

CMS and ATLAS also have analyses that search for particles with very long lifetimes, comparable to the size of the detector, that may also be produced in exotic Higgs boson decays. Several searches target displaced hadronic jets, including a recent search from CMS that uses a new multivariate algorithm to identify the distinctive topology of displaced tracks and displaced vertices associated with a dijet system~\cite{Sirunyan:2020cao}. The search is sensitive to exotic Higgs boson branching ratios below 1\% for decays to long-lived scalar particles that each decay to a quark-antiquark pair for mean proper decay lengths in the range $1~\mathrm{mm}-340~\mathrm{mm}$ and masses in the range $40~\mathrm{GeV}-55~\mathrm{GeV}$.

ATLAS has also performed a search for long-lived particles decaying to displaced hadronic jets using displaced vertices in the inner detector and muon system~\cite{Aad:2019xav}. The analysis benefits from dedicated algorithms for tracking and vertexing that are able to reconstruct trajectories and decays at large radii. In combination with searches for displaced jets in the calorimeter~\cite{Aaboud:2019opc} and the muon spectrometer~\cite{Aaboud:2018aqj}, the search is sensitive to exotic Higgs boson branching ratios below $10\%$, reaching as low as $10^{-3}$, for the range of proper decay lengths $6~\mathrm{cm}-8~\mathrm{m}$ and masses in the range $8~\mathrm{GeV}-55~\mathrm{GeV}$.

\subsubsection{Summary}
\label{sec:vimartin_summary}

There is a large program of searches for dark sector scalars at the central LHC detectors. Several signatures are targeted following a range of approaches in inclusive searches, exotic Higgs boson decays and long-lived particle signatures. The signatures are motivated by a broad range of phenomenology and benchmark models are very useful to guide analyses. Searches for dark scalars are a vast unexplored territory, since in many cases signals will only be visible if new trigger and reconstruction techniques are developed. There are several uncovered channels and regions of phase space. In addition, a comparison of the sensitivity of analyses searching for prompt signals and dedicated long-lived particle searches show gaps that remain unexplored. While invisible decays are already being explored at the LHC, semi-visible decays are also largely uncovered so far. Other unexplored directions include new production channels such as the production of new scalars in association with $t\bar{t}$. There are many exciting possibilities for full Run 2 and Run 3 analyses, especially with new triggers, algorithms and analysis strategies.

\clearpage
\subsection{Search for light feebly-interacting scalar particles at extracted beams} 
\label{ssec:swallow-goudzovski}
{\it Authors: Joel Christopher Swallow, \\ <joel.christopher.swallow@cern.ch>  and  \\
Evgueni Goudzovski, <goudzovs@cern.ch>} 
\subsubsection{Introduction}
\label{sec:intro_swallow}

\noindent
Construction of generic theoretical frameworks has lead to identification four `portals' to physics beyond the Standard Model (SM), as described in section~\ref{ssec:knapen}. 
The `scalar portal' includes a new scalar $S$ which mixes with the Higgs boson (section~\ref{sec:intro_DarkHiggs}).
In a benchmark model BC4 of Ref.~\cite{Beacham:2019nyx}, the production and decay of the scalar $S$ are determined by mixing angle $\theta$. The decay width depends on the mass $m_S$ which determines the allowed decay channels, and the branching ratio to different final states varies with mass.
Moreover there is a model-dependent relationship between the mixing angle $\theta$, mass $m_{S}$, and lifetime $\tau_{S}$. 

\vskip 2mm
Searches for the dark scalar can be performed at experiments at extracted beam lines. In general, the experimental signature depends on the properties of $S$: 
\begin{enumerate}
    \item if $S$ decays to visible (SM) particles and is sufficiently short-lived to decay in the experiment's sensitive volume, a particle pair is detected with invariant mass equal to $m_{S}$;
    \item if $S$ decays to invisible particles (for example dark matter), or is long-lived, missing energy/momentum may be deduced and associated with a missing mass equal to $m_{S}$.
\end{enumerate}
The distance $S$ travels, on average, depends on: the Lorentz factor, which depends on the beam momentum; and its lifetime, $\tau_{S}$. 
Experiments measuring the momentum of beam particles may search for both signatures (due to the capability to measure missing energy/momentum/mass) while beam dump experiments can only search for visible decays.

\subsubsection{NA62}
\label{sec:NA62}
NA62 is a high-intensity $K^{+}$ decay experiment at CERN with primary objective of studying the ultra-rare $K^+\to\pi^+\nu\bar\nu$ decay~\cite{NA62:2017rwk,CortinaGil:2020vlo}. A $400\,\text{GeV}/c$ proton beam is extracted from the CERN SPS and steered onto a beryllium target to produce a 75~GeV/$c$ secondary hadron beam with a 6\% component of $K^+$, and decays of $K^{+}$ in a 75~m long fiducial volume contained in a vacuum tank (at a typical decay rate of 4~MHz) are studied.

\vskip 2mm
\noindent
\boldmath
{\bf Search for $K^{+}\to\pi^{+}S$ outside of the $\pi^0$ mass window} \\
\unboldmath
The $K^+\to\pi^+S$ and $K^+\to\pi^+\nu\bar\nu$ decays share the same signature with missing energy/momentum. Therefore the search for the $K^+\to\pi^+S$ decay, with the assumption that the $S$ is long-lived and $m_S$ is outside of a ($\pm20$ MeV/$c^{2}$) $\pi^0$ mass window, is performed as a peak search in the missing mass spectrum of the $K^+\to\pi^+\nu\bar\nu$ candidates, with the $K^+\to\pi^+\nu\bar\nu$ decay representing the main source of background~\cite{CortinaGil:2020fcx}. The squared missing mass is defined as $m_{\rm miss}^2=(P_{K^+}-P_{\pi^+})^2$, where $P_{K^+}$ and $P_{\pi^+}$ are the reconstructed 4-momenta of the beam kaon and the daughter pion. Signal regions are defined with $m_{\rm miss}^{2}$ in ranges 0--$0.01\,\text{GeV}^{2}/c^{4}$ (Region 1) and 0.026--$0.068\,\text{GeV}^{2}/c^{4}$ (Region 2).

\begin{figure*}
    \centering
    \begin{minipage}{0.45\textwidth}
            \centering
            \vspace{-10pt}
            \includegraphics[width=1.05\textwidth]{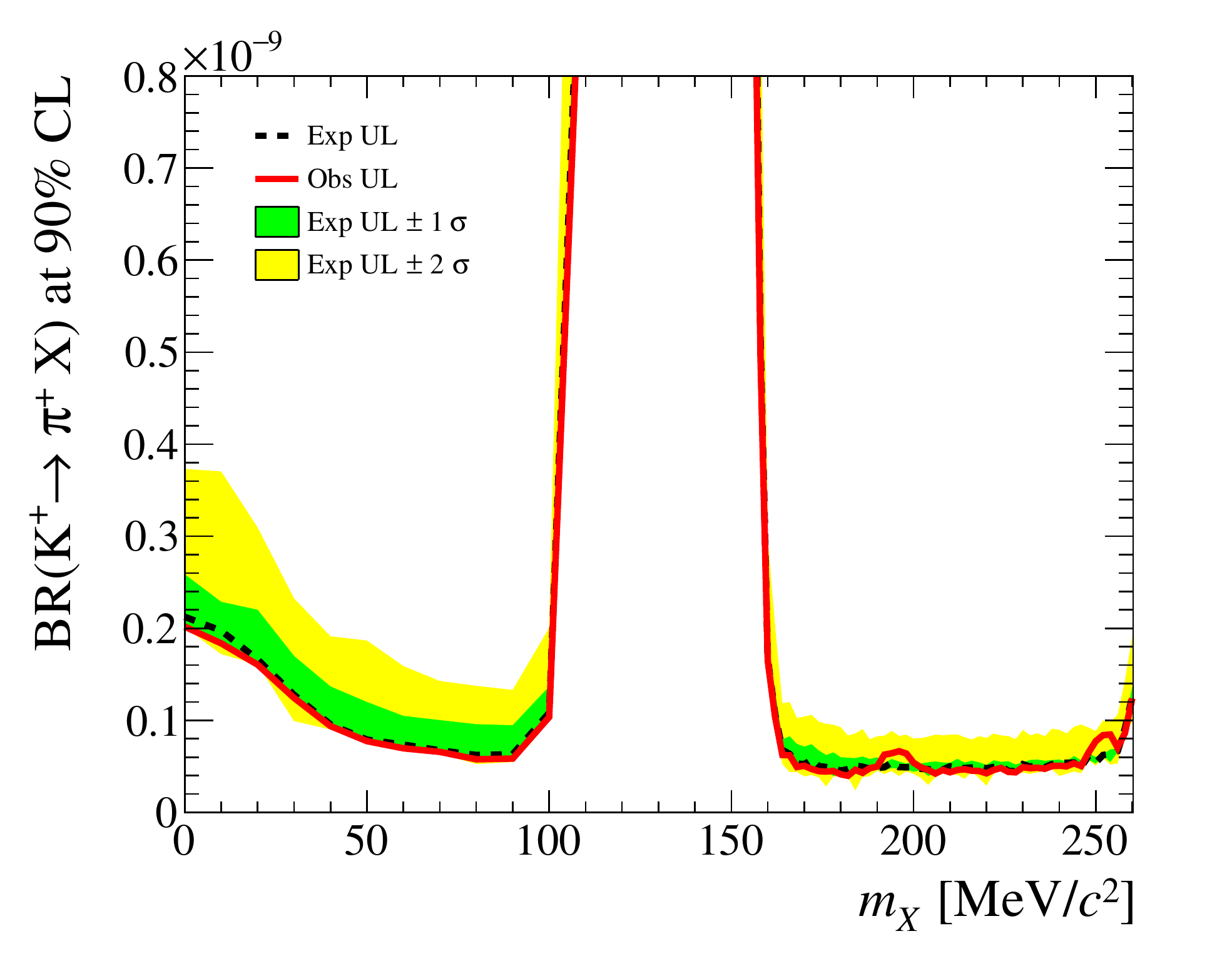}
            \label{fig:KpiX_BRUL}
        \end{minipage}
        \begin{minipage}{0.05\textwidth}
        \end{minipage}
        \begin{minipage}{0.45\textwidth}
            \centering
            \vspace{-5pt}
            \includegraphics[width=1.0\textwidth]{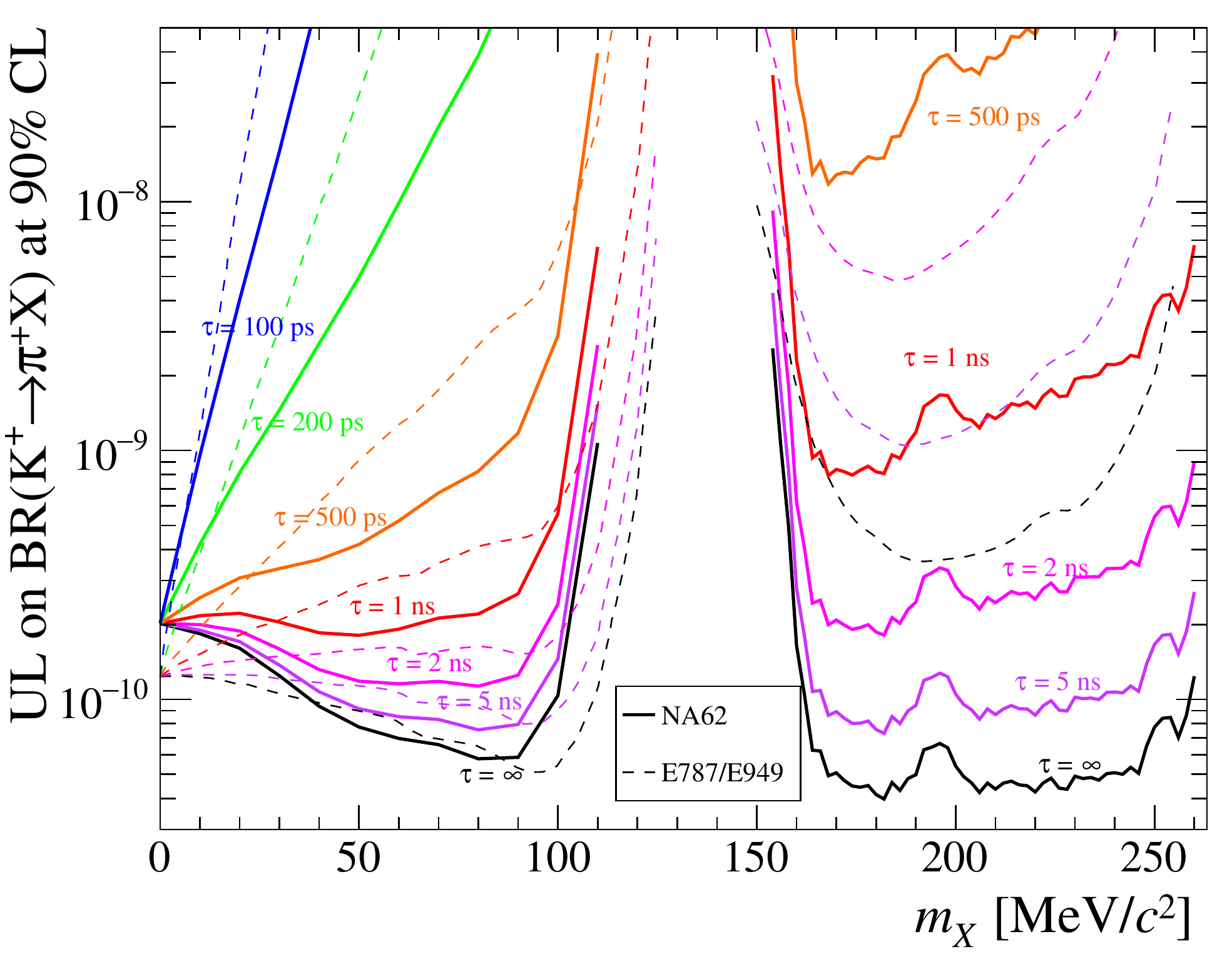}
            \label{sig:KpiX_BRUL_lifetimes}
    \end{minipage}
    \vspace{-10pt}
    \caption{Upper limits of $\mathcal{B}(K^+\to\pi^+S)$ at $90\%$ confidence level as a function of the assumed mass obtained using the NA62 2017 dataset~\cite{CortinaGil:2020fcx} under assumptions of $\tau_S>100$~ns or $S$ decaying to invisible particles (left), and $S$ decaying to visible SM particles with lifetime $\tau_{S}$ (right). The limits are compared with those from the BNL-E949 experiment~\cite{Artamonov:2009sz}.}
    \label{fig:KpiXresults}
\end{figure*}

\vskip 2mm
Two $K^+\to\pi^+\nu\bar\nu$ candidates are observed in the NA62 2017 data (both in Region 2) leading to upper limits on the branching ratio ${\cal B}(K^+\to\pi^+S)$, under the assumption of long $S$ lifetime ($\tau_{S}>100$~ns) or $S$ decaying only to invisible particles, shown in figure~\ref{fig:KpiXresults}~(left). 
Alternatively, assuming that $S$ has finite lifetime $\tau_{S}$ and decays only to visible (SM) particles, upper limits are established accounting for the probability of $S$ decaying in the sensitive volume of the experiment and of the event being rejected either by either the trigger or the $K^+\to\pi^+\nu\bar\nu$ selection. Limits on the branching ratio ${\cal B}(K^+\to\pi^+S)$ under the assumption of small lifetimes are shown in figure~\ref{fig:KpiXresults}~(right).

\vskip 2mm
Analysis of the complete NA62 dataset collected in 2016--18 has led to observation of 20 candidate $K^+\to\pi^+\nu\bar\nu$ events~\cite{pnnICHEP20}. Further large datasets for the $K^+\to\pi^+\nu\bar\nu$ measurement are to be collected by NA62 after LS2. Therefore the sensitivity of the search is expected to improve significantly in near future.

\vskip 2mm
\noindent
\boldmath
{\bf Search for $K^{+}\to\pi^{+}S$ in the vicinity of the $\pi^0$ mass}\\
\unboldmath
The search for $K^+\to\pi^+S$ decay involving a long-lived $S$ particle with a mass in the vicinity of the $\pi^0$ mass has been performed with 30\% of the NA62 2017 dataset~\cite{CortinaGil:2020zwa}. Contrary to the previous case, the dominant background to this search comes from the $K^+\to\pi^+\pi^0$, $\pi^0\to\gamma\gamma$ decay chain. Therefore measurements of photon detection inefficiency in the hermetic NA62 photon veto system are critical for the background subtraction. Such measurements have been performed in bins of photon energy using a sample of $K^+\to\pi^+\pi^0$ decays collected with the minimum-bias trigger. Signal event selection, including the range of $\pi^+$ momentum selected, is optimised to minimise the $\pi^{0}$ rejection inefficiency. The total background expected is $10^{+22}_{-8}$, and $12$ events are observed, leading to new upper bounds on the $K^{+}\to\pi^{+}S$ decay (Fig.~\ref{fig:MicroBooNEresult}). A new upper limit on the rate of the $\pi^0$ decay to the invisible final state is also obtained,
\begin{equation}
    \mathcal{B}(\pi^{0}\to{\rm invisible}) < 4.4\times10^{-9}~~~{\rm at~ 90\% ~CL},
\end{equation}
improving by a factor of 60 on the previous search~\cite{Artamonov:2005cu}.

\vskip 2mm
\noindent
{\bf Projected sensitivity in beam dump mode} \\
Collection of a dataset in beam dump mode is envisaged at NA62 shortly after LS2, with a movable collimator closed and acting as a beam dump. With 90 days of dump mode data, $10^{18}$ protons on target (POT) are expected to be collected with a minimum-bias trigger and in a low-background regime~\cite{NA62_Addendum}. The  projected sensitivity to $S$ decays with this dataset is shown in Fig.~\ref{fig:DS_1}.

\subsubsection{MicroBooNE}
\label{sec:MicroBooNE}
The MicroBooNE experiment at Fermilab is a large ($10.4\times2.5\times2.3\,\text{m}^{3}$) liquid argon time projection chamber (LArTPC) positioned downstream of extracted beams from fixed targets used for neutrino experiments~\cite{Adams:2019bzt}. The primary objectives of MicroBooNE are investigations of low energy excess events observed by the MiniBooNE experiment, low-energy neutrino cross-sections and astroparticle physics. 

\begin{figure*}
    \centering
    \begin{minipage}{0.45\textwidth}
            \centering
            \vspace{-10pt}
            \includegraphics[width=0.9\textwidth]{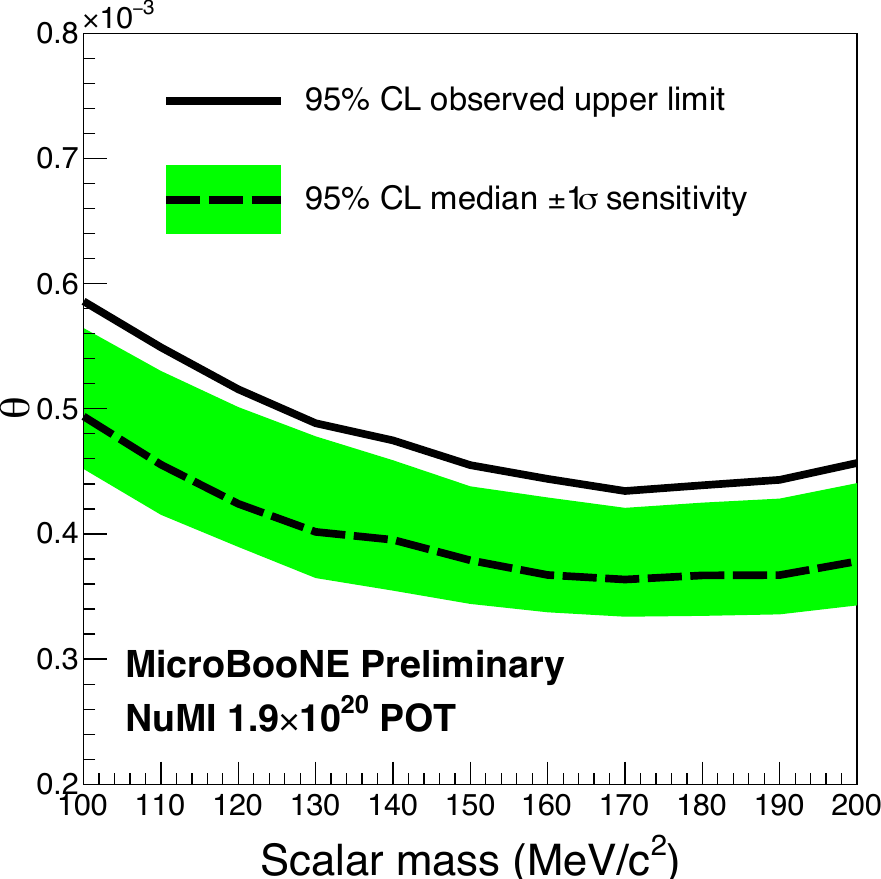}
            \label{fig:MicroBooNEtheta}
        \end{minipage}
        \begin{minipage}{0.05\textwidth}
        \end{minipage}
        \begin{minipage}{0.45\textwidth}
            \centering
            \vspace{-5pt}
            \includegraphics[width=0.9\textwidth]{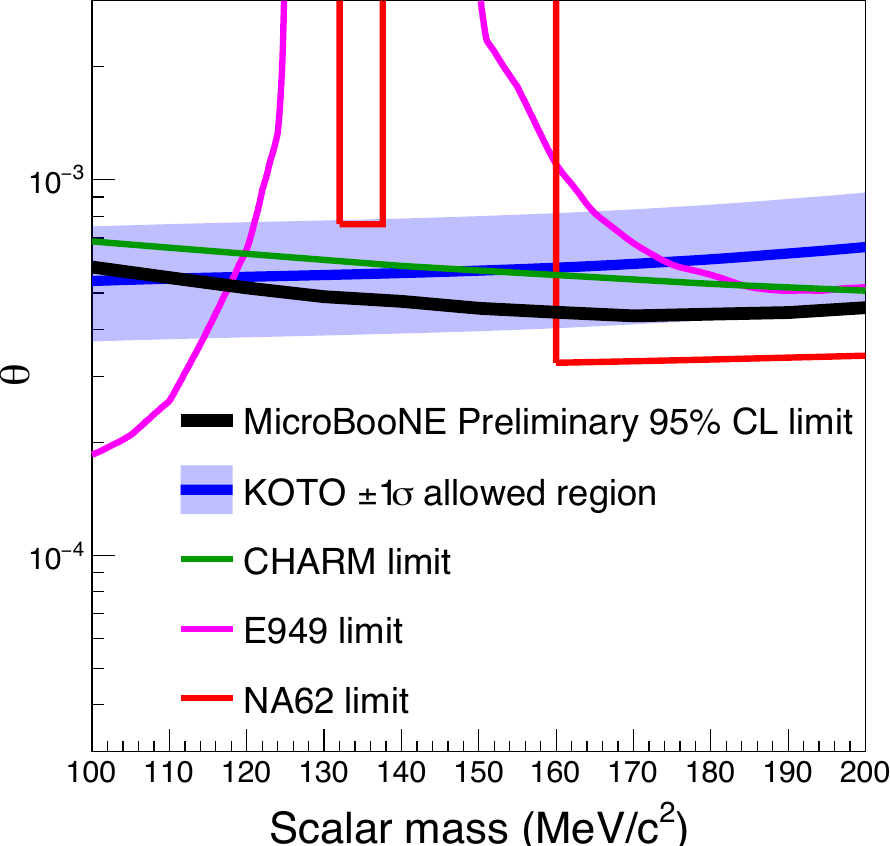}
            \label{fig:MicroBooNEthetaComp}
    \end{minipage}
    \caption{Upper limits at $95\%$ confidence level for the mixing parameter $\theta$ as a function of scalar mass $m_{S}$ obtained by MicroBooNE (left), and comparison to other results (right)~\cite{MicorBooNE_ScalarSearch}.}
    \label{fig:MicroBooNEresult}
\end{figure*}

\vskip 2mm
A search for $S\to e^+e^-$ decays has been performed using $1.93\times10^{20}$ POT delivered during Run 1 (2015--16) and Run 3 (2017--18).
A primary 120~GeV proton beam is delivered to a target to produce a secondary hadron beam which traverses a decay volume. Decays of $K^+$ and $\pi^+$ in this volume produce neutrinos which are used for the NO$v$A experiment. At the end of the decay volume the remaining hadrons are dumped into the NuMI hadron absorber in which a scalar $S$ may be produced and travel $\mathcal{O}(100\,\text{m})$ to the MicroBooNE detector. 
A search is performed for $S\to e^+e^-$ decays in the mass range $100$--$200\,\text{MeV}/c^{2}$ and for lifetimes $\mathcal{O}(1~\mu\text{s})$~\cite{MicorBooNE_ScalarSearch}. A signal candidate event has a distinctive `kink' signature of a pair of tracks, identified as $e^+e^-$, with their total momentum pointing back to the NuMI hadron absorber. The total expected background is $2.0\pm0.8$ events, and $5$ events are observed, consistent with the background-only hypothesis. Upper limits at $95\%$ confidence level  established on the mixing parameter $\theta$ as a function of $m_{S}$ are shown in figure~\ref{fig:MicroBooNEresult}. This result is complementary to the NA62 results discussed in Section~\ref{sec:NA62}.

\vskip 2mm
The MicroBooNE result is incompatible with the central $\theta$ values derived from the KOTO anomaly~\cite{Egana-Ugrinovic:2019wzj} over the range of scalar masses $110$--$200\,\text{MeV}/c^{2}$. It is noted that the values allowed by the KOTO anomaly are derived from KOTO results presented in 2019~\cite{KOTO_KAON19}, and may require reevaluation following the revised background estimates presented at ICHEP20~\cite{KOTO_ICHEP20}

\subsubsection{The Sea/Spin/DarkQuest Project at Fermilab}
\label{sec:QuestProject}
A project of fixed target beam dump experiments is ongoing at Fermilab, based on the existing SeaQuest spectrometer detector, and using a $120\,\text{GeV}$ proton beam from the Main Injector.

The SeaQuest experiment (E906, 2012--17) used an unpolarized fixed target to study study sea quark flavor asymmetry inside the nucleon ($\bar{d}$/$\bar{u}$). The detector, with existing SeaQuest components shown in black, is described in Ref.~\cite{Aidala:2017ofy}. 
In 2017 two dark photon scintillator strip trigger detectors~\cite{Liu:2017ryd} were installed and commissioning data was collected. 

The SpinQuest experiment (E1039, approved 2020--22), will use the SeaQuest spectrometer with upgraded DAQ systems, with primary objective of studying the sea-quarks’ Sivers transverse single spin asymmetry using a polarised target. However searches for dimuon final states arising from dark photons or dark scalars will also be performed with sensitivity to new phase-space expected with data taking ($1.4\times10^{18}$ POT planned) beginning in 2021.

Subsequently for the DarkQuest experiment (planned for 2023--25) an electromagnetic calorimeter (EMcal, re-purposed from a recently completed experiment) will be installed allowing detection of photons and electrons and allowing further searches for dark photons and $S\rightarrow e^{+}e^{-}$. With $10^{18}$--$10^{20}$ POT foreseen, a significant region of phase-space for both dark scalar and dark photons can be scrutinised~\cite{Berlin:2018pwi}~\cite{Tsai:2019mtm}. Comparable exclusion regions of phase-space for dark scalar searches are predicted for Spin/DarkQuest and NA62 in dump mode on a similar timescale (see figure~\ref{fig:DS_1}). 

A future LongQuest experiment is being considered~\cite{QuestLoI20} to increase capabilities for dark sector searches including for longer-lived particles. Further detector upgrades are proposed with additional EMcal, tracking and particle identification detectors.

\subsubsection{SHiP}
\label{sec:SHiP}
The Search for Hidden Particles (SHiP) experiment, to be constructed at a new beam dump facility at the CERN North Area, was proposed in 2015 with primary objective of searching for FIPs of mass below $\mathcal{O}(10\,\text{GeV})$~\cite{Anelli:2015pba}~\cite{SHiP_DesignRep19}. 

The experiment is designed to have background as close to zero as possible allowing for clear discovery of long-lived FIPs decaying to visible particles. 
In particular, the sensitivity for detection of long-lived scalars has been evaluated, suggesting an extensive new phase-space can be probed~\cite{SHiP_rep19}. 

\subsubsection{Summary and Conclusion}

The summary of the current landscape of searches for dark scalars is shown in figure~\ref{fig:DS_1}. 
No dark scalar has been observed so far but a number of recent results and results expected in the next few years will probe a significant new region of phase-space. The experiments at extracted beam lines considered here will make a significant contribution alongside others such as LHCb or Belle II.






\clearpage
\subsection{Results for the scalar portal}
\label{ssec:scalar-results}

\begin{figure*}[h]
\includegraphics[width=0.8\linewidth]{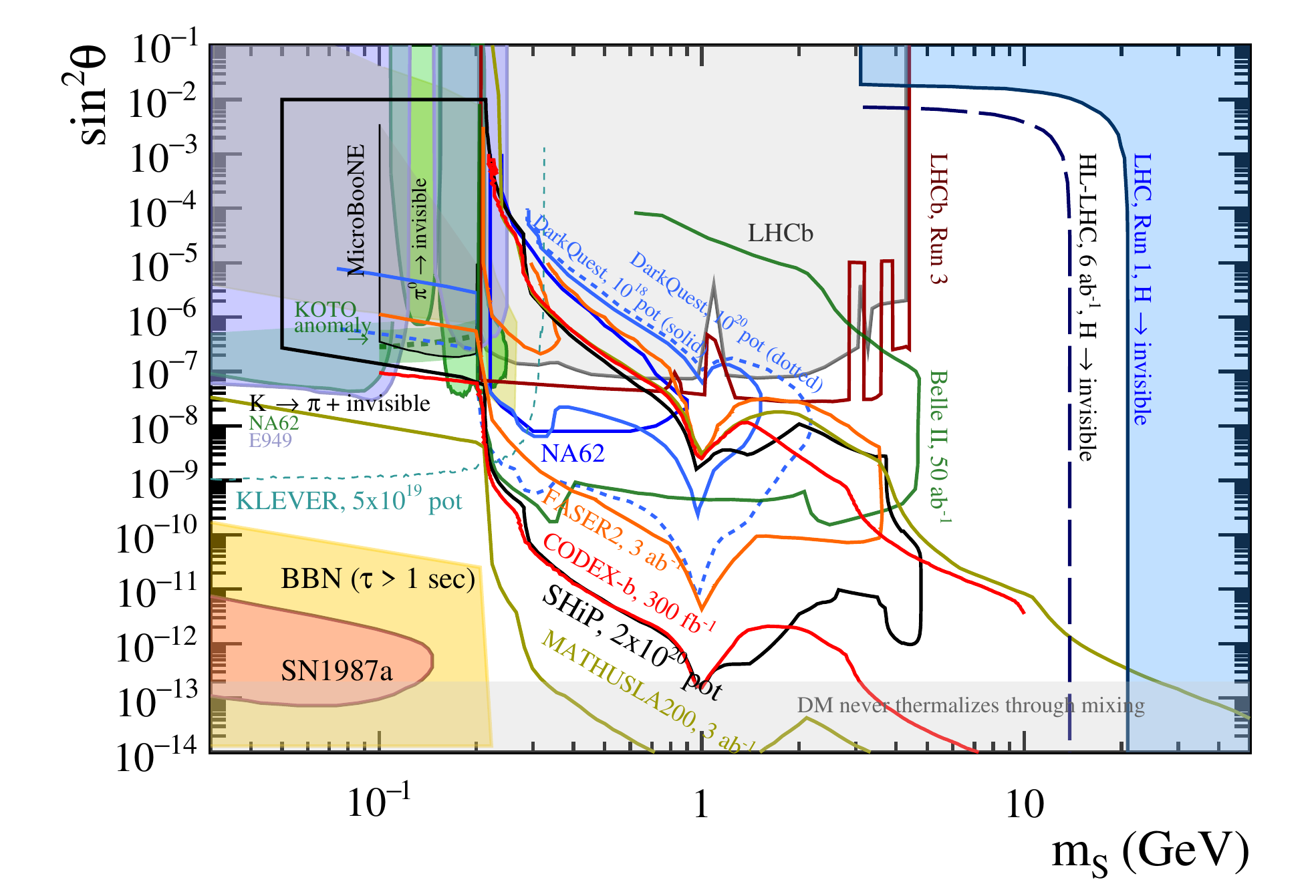}
\caption{ {\bf Dark scalar into visible final states}.
Shaded areas come from: reinterpretation~\cite{Winkler:2018qyg} of results from CHARM experiment~\cite{Bergsma:1985qz};
NA62~\cite{CortinaGil:2020fcx};
E949~\cite{Artamonov:2008qb,Dev:2019hho}; MicroBooNE~\cite{microboone} that excludes a light dark scalar as interpretation~\cite{Egana-Ugrinovic:2019wzj} of the KOTO anomaly; LHCb~\cite{Aaij:2016qsm,Aaij:2015tna} and Belle~\cite{Wei:2009zv}.
Coloured lines come from projections of existing/proposed experiments: 
NA62-dump~\cite{NA62:dump} and DarkQuest~\cite{Batell:2020vqn}, Belle II~\cite{Filimonova:2019tuy}, SHiP~\cite{Anelli:2015pba}, FASER2~\cite{Ariga:2018uku}, CODEX-b~\cite{Aielli:2019ivi}, MATHUSLA~\cite{Alpigiani:2020tva}, and KLEVER~\cite{Beacham:2019nyx}. Vertical lines come from the knowledge for the invisible Higgs width after Run~1 at the LHC and projections in the HL-LHC era (see~\cite{Strategy:2019vxc} and references therein). BBN and SN 1987A are from ~\cite{Fradette:2017sdd} and ~\cite{Dev:2020eam}, 
respectively. Figure revised from Ref.~\cite{Lanfranchi:2020crw}.}
\label{fig:DS_1}
\end{figure*}

The current status of experimental searches and projections for accelerator-based experiments for the minimal scalar portal model is shown in Figure~\ref{fig:DS_1}. 

The minimal scalar portal model operates with one extra singlet field $S$ and two types of couplings, $\mu$ and $\lambda$~\cite{OConnell:2006rsp},
\begin{equation}
\label{scalar}
{\cal L}_{\rm scalar} = {\cal L}_{\rm SM} + {\cal L}_{\rm DS} - (\mu S+ \lambda S^2)H^\dagger H \, .
\end{equation}
The dark sector Lagrangian may include the interaction with dark matter $\chi$, ${\cal L}_{\rm DS}= S\bar\chi \chi+...\,$. 
Most viable dark matter models in the sub-EW-scale range imply $2 m_\chi > m_S$ \cite{Krnjaic:2015mbs}.
At low energy, the Higgs field can be substituted for $H = (v + h)/\sqrt{2}$, where $v = 246$\,GeV
is the EW vacuum expectation value, and  $h$ is the field corresponding to the physical 125\,GeV Higgs boson.
The non-zero $\mu$ leads to the mixing of $h$ and $S$ states. In the limit of small mixing, it can be written as 
\begin{equation}
\theta = \frac{\mu v}{m_h^2- m_S^2} \, .
\end{equation}

The model assumed in Figure~\ref{fig:DS_1} is the established PBC benchmark $BC5$~\cite{Beacham:2019nyx} that we repeat here below for convenience.

\begin{itemize}
\item {\em BC5, Higgs-mixed scalar with large pair-production channel: }
  in this model, the parameter space is $\{\lambda, \theta, m_S\}$, and $\lambda$ is assumed to dominate
  the production via {\em e.g.} $h\to SS$, $B \to K^{(*)}SS$, $B^0 \to SS$ etc.
  In the sensitivity plots a value  of the branching fraction $BR({h \to SS})$
  close to $10^{-2}$ is assumed in order to be complementary to the LHC searches for the Higgs to invisible channels.
 \end{itemize}

\begin{figure*}[h]
\includegraphics[width=0.8\linewidth]{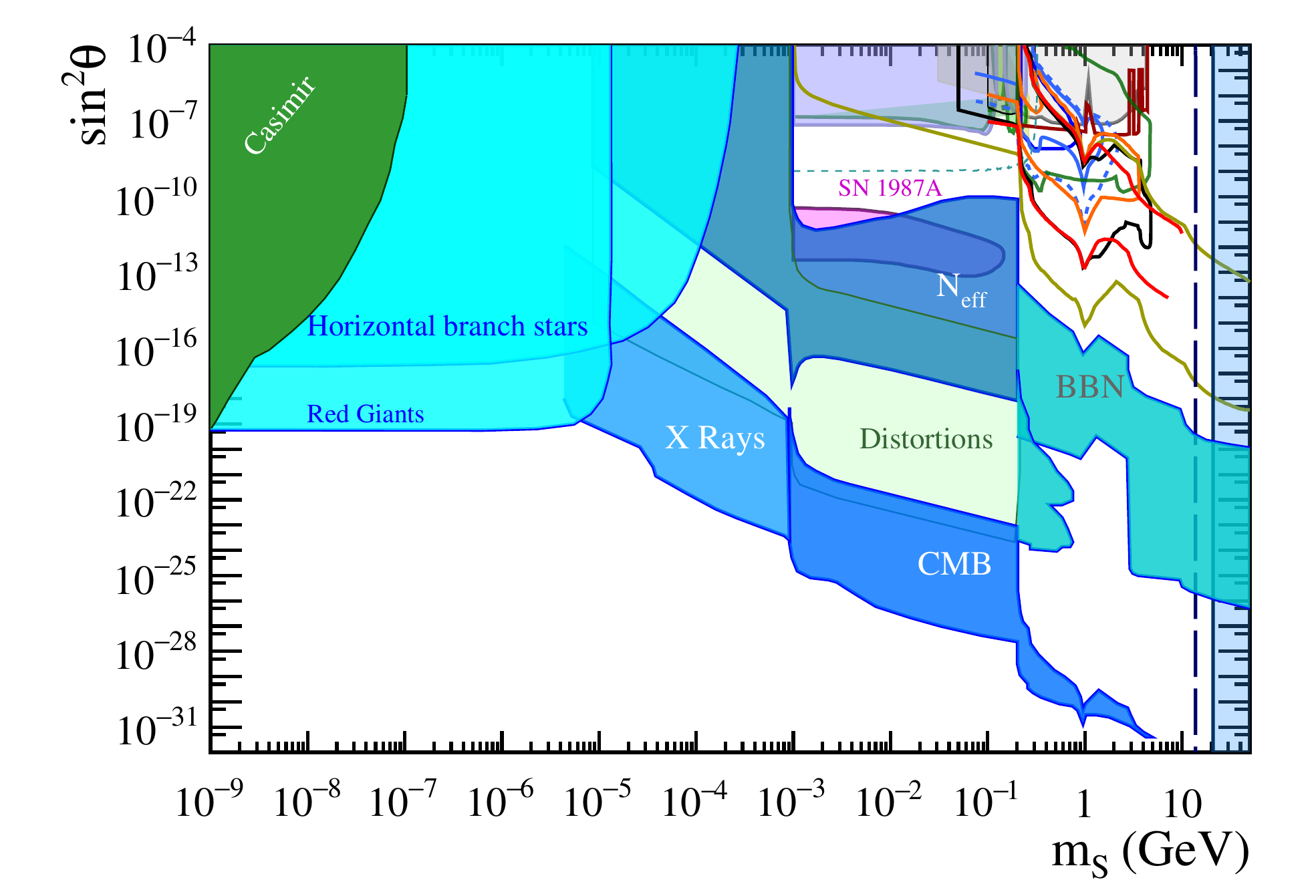}
\caption{An overview of the excluded parameter space for minimal  scalar portal, including the updated constraints to the diffuse X-ray background (X Rays), CMB anisotropies, spectral distortions, $N_{\rm eff}$ and BBN from Ref.~\cite{Fradette:2018hhl}.
  Constraints from new short-range forces~\cite{Kapner:2006si,Decca:2007jq,Geraci:2008hb,Sushkov:2011md} and stellar cooling~\cite{Hardy:2016kme} from other authors are also shown. We also display the projected SHiP sensitivity and an estimate of supernova (SN 1987A) constraints~\cite{Dev:2020eam}. 
  Figure revisited from Figure~\ref{Fig1} in Section~\ref{ssec:pospelov}}.
\label{fig:DS_2}
\end{figure*}

\vskip 2mm
In Figure~\ref{fig:DS_2}, the minimal scalar model is used to compute constraints to the diffuse X-ray background (XRay), CMB anisotropies, spectral distortions, $N_{\rm eff}$ and BBN from Ref.~\cite{Fradette:2018hhl}. 
Constraints from stellar cooling are also shown as in Ref.~\cite{Hardy:2016kme}.
The accelerator-based results are confined in the top-right corner.

\begin{figure*}[h]
\includegraphics[width=0.8\linewidth]{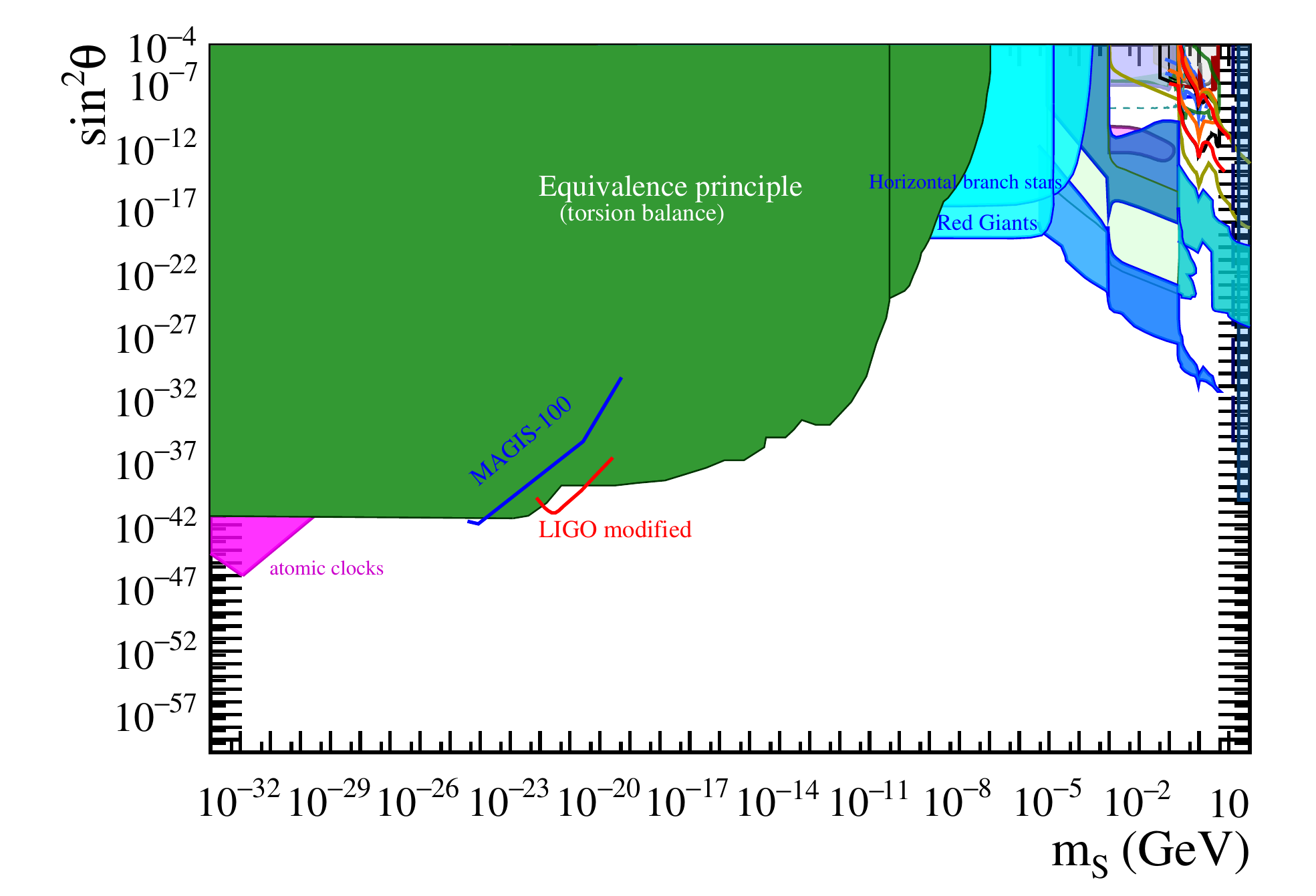}
\caption{
Bounds from tests of the equivalence principle with torsion balance experiments~\cite{Schlamminger:2007ht} and searches for dark matter with atomic clocks~\cite{Stadnik:2016zkf}, as well as the projected sensitivities of a modified LIGO experiment based on optical interferometry~\cite{Grote:2019uvn} and the MAGIS-100 experiment based on atom interferometry~\cite{Coleman:2018ozp}. 
Figure partially revised from Figure~\ref{Fig:Scalar_limits} in Section~\ref{ssec:stadnik}.} 
\label{fig:DS_3}
\end{figure*}

\vskip 2mm
Finally in Figure~\ref{fig:DS_3}, the mass scale is further expanded and 
bounds from tests of the equivalence principle with torsion balance experiments~\cite{Schlamminger:2007ht} and searches for dark matter with atomic clocks~\cite{Stadnik:2016zkf}, as well as the projected sensitivities of a modified LIGO experiment based on optical interferometry~\cite{Grote:2019uvn} and the MAGIS-100 experiment based on atom interferometry~\cite{Coleman:2018ozp} are shown for masses down to $10^{-24}$~eV. 

\vskip 2mm
In fact,
below the $\sim$~eV scale, spinless bosons form a coherently oscillating classical field  that can be searched for via non-gravitational interactions with SM fields (as explained in Section~\ref{ssec:stadnik}).
These non-gravitational interactions can be "scalar-type" interactions [see, e.g., Eq.~(\ref{linear_scalar_couplings})] if they are nominally associated with an even-parity spinless field. In this case, the singlet couples to SM particles through the mixing with the Higgs field. Depending on the mass and coupling of the scalar field $\phi$, the $\phi$-mediated attractive force can produce testable deviations from the $1/r^2$-gravitational force as well as  composition  dependence,  thus  violating the  equivalence  principle~\cite{Piazza:2010ye}.
It can also induce apparent oscillations of the electromagnetic fine-structure constant $\alpha$ and the fermion masses~\cite{Stadnik:2014tta}, which in turn  would induce oscillations of atomic transition frequencies and of lengths of solids that can be sought with a variety of precision techniques, including atomic clocks~\cite{VanTilburg:2015oza} and atom interferometry used for gravitational-wave detectors~\cite{Bertoldi:2019tck}.

\clearpage
\section{Feebly-interacting fermion particles}
\label{sec:fermion}

\subsection{Theoretical introduction to low scale seesaw models and their connection to leptogenesis}
\label{ssec:drewes}
{\it Author: Marco Drewes, <marco.drewes@uclouvain.be>} 

\renewcommand{\k}{\textbf{k}}
\renewcommand{\l}{\textbf{l}}
\renewcommand{\L}{l}
\newcommand{\Q}{q}
\renewcommand{\P}{p}
\newcommand{\K}{k}
\newcommand{\tm}{y}
\newcommand{\tp}{t}
\newcommand{\g}{h}
\newcommand{\G}{h_{(1,2)}}
\newcommand{\Dm}{\Delta^-}
\renewcommand{\Re}{\text{Re}}
\renewcommand{\Im}{{\text Im}}
\newcommand{\Dp}{\Delta^+}
\newcommand{\Pm}{\Pi^-}
\newcommand{\Pp}{\Pi^+}
\newcommand{\Wq}{\Omega_{\textbf{q}}(t)}
\newcommand{\wqi}{\omega_{\textbf{q}}(t_1)}
\newcommand{\q}{\textbf{q}}
\renewcommand{\k}{\textbf{k}}
\newcommand{\W}{\Omega}
\newcommand{\V}{\mathcal{V}}
\newcommand{\Mprime}{M_1}
\newcommand{\Vend}{\mathcal{V}_{\rm end}}
\newcommand{\fex}{\textit{e.g.}~}
\newcommand{\TI}{\tau_\text{int}}
\newcommand{\y}{\textbf{y}}
\newcommand{\T}{{\rm T}}
\newcommand{\B}{{\rm B}}
\newcommand{\sect}[1]{\emph{#1}.---}
\newcommand{\sectp}[1]{\emph{#1}.---}
\newcommand{\tphi}{\tilde{\phi}}
\newcommand{\tchi}{\tilde{\chi}}
\newcommand{\tildeg}{\tilde{g}}
\newcommand{\rephi}{A}
\newcommand{\imphi}{B}
\newcommand{\rechi}{C}
\newcommand{\imchi}{D}
\newcommand{\Rrad}{{\rm R}_{\rm rad}}
\newcommand{\zend}{z_{\rm end}}
\newcommand{\zreh}{z_{\rm re}}
\newcommand{\areh}{a_{\rm re}}
\newcommand{\aend}{a_{\rm end}}
\newcommand{\rhoend}{\rho_{\rm end}}
\newcommand{\rhoreh}{\rho_{\rm re}}
\newcommand{\rhotildegamma}{\tilde{\rho}_\gamma}
\newcommand{\wrehbar}{\bar{w}_{\rm re}}
\newcommand{\wreh}{w_{\rm re}}
\newcommand{\Nreh}{N_{\rm re}}
\newcommand{\X}{\mathcal{X}}
\newcommand{\MM}{{\rm M}}

\subsubsection{Neutrino masses as a key to New Physics}
\label{sssec:drewes-numasses-NP}

Neutrinos are very special particles in several regards. Within the minimal Standard Model of particle physics (SM), they are the only fermions that exist exclusively with left-handed chirality,
the only neutral fermions (that could potentially be Majorana fermions),
and the only massless fermions. The observed violation of lepton flavour numbers $L_a$ ($a=e,\mu,\tau$) in neutrino flavour oscillations, however, show that they do have tiny masses in the sub-eV range in reality. 

\vskip 2mm
The latter fact is, to date, the only firmly established piece of evidence for the existence of physics beyond the minimal SM that has been observed in the laboratory. 
Hence, studying neutrino masses and mixings may provide a key how the SM should be embedded in a more fundamental theory of nature.
In particular, questions of interest are the \emph{mass puzzle} and the \emph{flavour puzzle}, i.e., the questions why their masses $m_i$ are so much smaller than those of all other fermions, and why the neutrino mixing matrix $V_\nu$ looks so different from that of the quarks. 

\vskip 2mm
Another fundamental question is whether the neutrinos are Dirac or Majorana particles, i.e., whether the mass term is of the form 
\begin{eqnarray}\label{massterms}
\begin{tabular}{c c c}
$\overline{\nu_L} m_\nu\nu_R$ \; {\rm [Dirac]} 
& 

{\rm or} 
&
$\frac{1}{2}\overline{\nu_{L}}m_\nu \nu_{L}^{c}$ \; {\rm[Majorana]}
\end{tabular} 
\end{eqnarray}

\noindent
with 
$\nu_L^c=C\overline{\nu_L}^T$,  $C=i\gamma_2\gamma_0$.


\vskip 2mm
In a complete theory of neutrino masses the answers to the mass and flavour puzzles may be related. In absence of any knowledge about this underlying theory, different scenarios can be classified in terms of the scale $\Lambda$ of the new physics responsible for neutrino mass generation (where new particle appear), and in terms of symmetries in flavour space that explain the texture of $m_\nu$. 
Current neutrino oscillation data is consistent with the minimal hypothesis of three light neutrino mass eigenstates $\upnu_i$ with masses $m_i$ in the sub-eV range that are related to the weak interaction states via a unitary rotation $V_\nu=U_\nu$ in flavour space, with $U_\nu$ the unitary PMNS matrix. 
This suggests that the scale $\Lambda$ is much larger than the typical energy $E$ of neutrino oscillation experiments,
so that neutrino oscillations can be described in the framework of Effective Field Theory (EFT) in terms of operators $\mathcal{O}^{ [\mathrm{n}]}_\mathrm{i} = c^{[\mathrm{n}]}_\mathrm{i} \Lambda^{\mathrm{n}-4}$ of mass dimension $n>4$ that are suppressed by powers of $\Lambda^{n-4}$.
The  Wilson coefficients $c^{[\mathrm{n}]}_\mathrm{i}$ are tensors in flavour space. 
For example,  the only operator of dimension $\mathrm{n}=5$~\cite{Weinberg:1979sa},
\begin{equation}\label{weinbergoperator}
\frac{1}{2}\overline{\ell_{L}}\tilde{\Phi} c^{\rm [5]}\Lambda^{-1}\tilde{\Phi}^{T}\ell_{L}^{c} + h.c.,
\end{equation}
generates a Majorana mass term\footnote{
We note in passing that it is possible that neutrinos are Dirac particles (cf. e.g.~\cite{Broncano:2002rw} and~\cite{CentellesChulia:2018gwr,CentellesChulia:2018bkz}), and that leptogenesis is also feasible with Dirac neutrinos~\cite{Dick:1999je}. 
}
 \eqref{massterms}
with  $m_\nu=-v^2 c^{\rm [5]}\Lambda^{-1}$
via the Higgs mechanism, which can be seen be replacing $\Phi\rightarrow (0,v)^T$ in the unitary gauge, with $v = 174$ GeV.
Here $\ell_L$ is the SM lepton doublet and $\Phi$ the Higgs doublet, $\tilde{\Phi}=\epsilon\Phi^*$ 
with $\epsilon$ the antisymmetric SU(2) tensor. 

\begin{figure}
\centering
\includegraphics[width=0.42\textwidth]{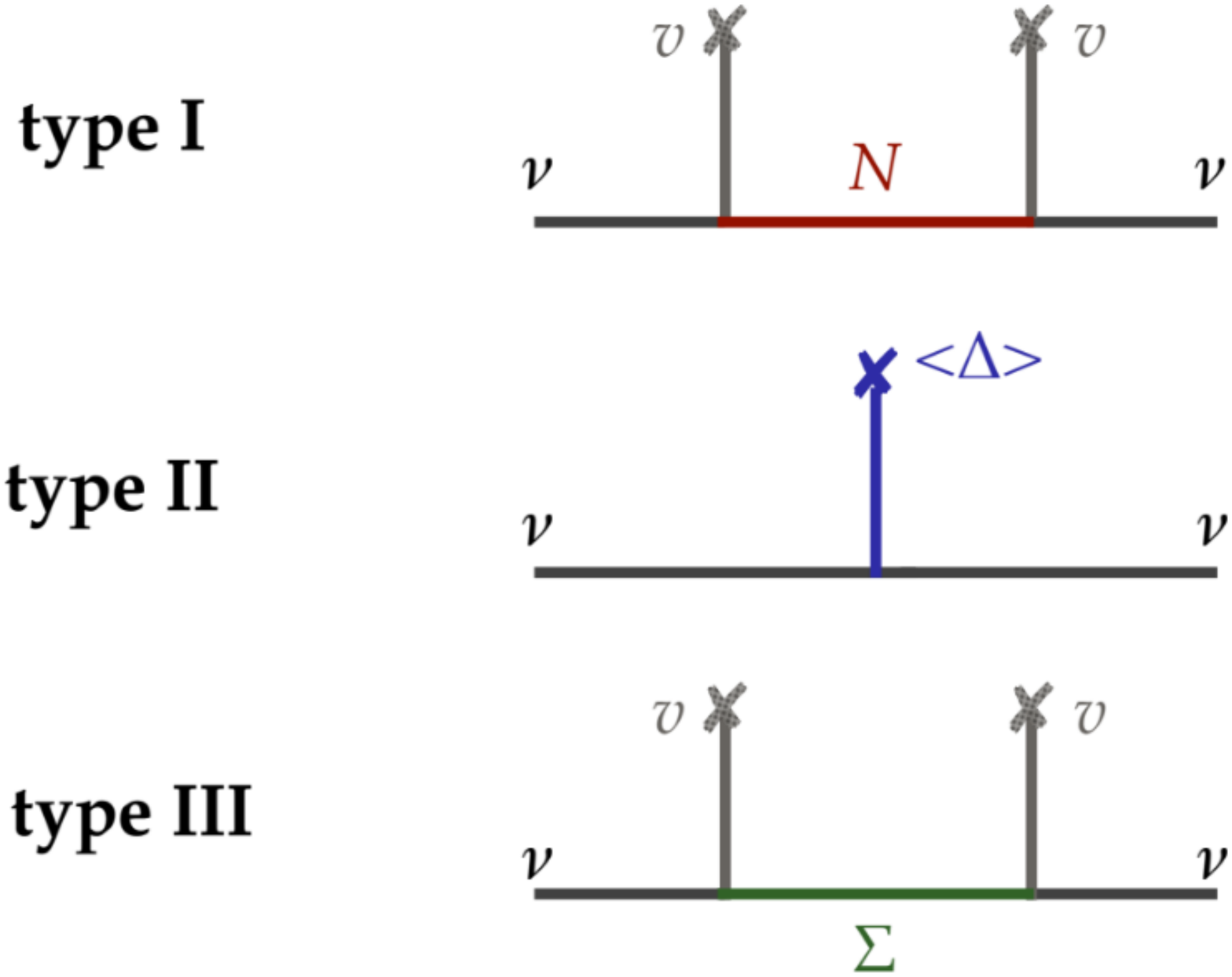}
\caption{\label{treelevelseesaws}Schematic  illustration of the three seesaw scenarios \eqref{seesaws}.}
\end{figure}

\vskip 2mm
The Majorana mass \eqref{massterms}
can be generated at tree level in three different ways~\cite{Ma:1998dn}, known as type-I~\cite{Minkowski:1977sc,Glashow:1979nm,GellMann:1980vs,Mohapatra:1979ia,Yanagida:1980xy,Schechter:1980gr} 
  type-II~\cite{Schechter:1980gr,Magg:1980ut,Cheng:1980qt,Lazarides:1980nt,Mohapatra:1980yp} and type-III~\cite{Foot:1988aq} seesaw (see ~Fig.~\ref{treelevelseesaws}),
  
  \begin{eqnarray}\label{seesaws}
  \begin{tabular}{ll}
  $\overline{\ell_L}Y_{\rm I}\nu_R\tilde{\Phi}$
  &
  {\rm [type \ I]} \\

  $\overline{\ell_L^c}Y_{\rm II}\mathrm{i}\sigma_2\Delta\ell_L$
  &
    {\rm [type \ II]} \\

$\overline{\ell_L}Y_{\rm III}\Sigma_L^c\tilde{\Phi}$
& 
    {\rm [type \ III]},
\end{tabular}
  \end{eqnarray}
  with $\nu_R$ a SM singlet fermion (``right handed neutrino''), 
  $\Delta$ a SU(2)$_L$ triplet scalar, and $\Sigma_L$ a fermionic SU(2)$_L$ triplet
  that couple to the SM via the coupling matrices $Y_{\rm I},  Y_{\rm II}, Y_{\rm III}$. After spontaneous symmetry breaking they generate contributions
  \begin{eqnarray}\label{LightNeutrinoMasses}
  \begin{tabular}{l}
  $m_\nu^{\rm I} = - v^2 Y_{\rm I} M_M^{-1} Y_{\rm I}^T$ \\
  $m_\nu^{\rm II} = - \sqrt{2} Y_{\rm II} v_\Delta$ \\
  $m_\nu^{\rm III} = - \frac{1}{2} v^2 Y_{\rm III} M_\Sigma^{-1} Y_{\rm III}^T$,
\end{tabular}
  \end{eqnarray}
with $M_M$ and $M_\Sigma$ Majorana masses for $\nu_R$ and $\Sigma_L$, respectively, and $v_\Delta$ the expectation value of $\Delta$. 
In the type I and type III cases it is straightforward to identify $\Lambda$ with the mass 
of the lightest relevant new fermion, while the Wilson coefficients are simply given by the couplings,
$\Lambda\sim (M_M)_{11}$ and $\Lambda\sim (M_\Sigma)_{11}$ 
in the mass basis, with $c^{\rm [5]}_{ab}=(Y_{\rm I})_{a1} (Y_{\rm I})_{b1}$ and 
$2c^{\rm [5]}_{ab}=(Y_{\rm III})_{a1} (Y_{\rm III})_{b1}$.
In the type II scenario, $\Lambda$ should be identified with the mass of the new scalar, $\Lambda\sim M_\Delta$,  with $c^{\rm [5]}_{ab}= \sqrt{2}(Y_{\rm II})_{ab} v_\Delta M_\Delta/v^2$, which for the simplest scalar potential is given by the coefficient in the term $\kappa\Phi^T \mathrm{i} \sigma_2\Delta^\dagger\Phi$ as $v_\Delta=\kappa v^2/(\sqrt{2}M_\Delta^2)$, i.e., $c^{\rm [5]}_{ab}=(Y_{\rm II})_{ab}\kappa/M_\Delta$. When expressing the dimensionful coefficient $\kappa \equiv\lambda_\kappa M_\Delta$ in units of $M_\Delta$, the Wilson coefficients in all three cases are simply given by dimensionless couplings $Y_{\rm I}, Y_{\rm II}, Y_{\rm III}, \lambda_\kappa$, while $\Lambda$ is given by the new particle masses $M_M, M_\Delta, M_\Sigma$. 

\vskip 2mm
Of course, any combination of these scenarios may coexist.
For instance, in the left-right symmetric model~\cite{Pati:1974yy,Mohapatra:1974gc,Senjanovic:1975rk}, a Majorana mass $M_M$ for the $\nu_R$ is generated through a type II mechanism in the right-handed sector, and then the $\nu_R$ generate the light neutrino masses trough the type I mechanism.  
Moreover, in addition to the three tree level mechanisms, 
the $\mathcal{O}^{[\mathrm{n}]}_\mathrm{i}$ may be generated due to quantum effects. 
A large number of radiative neutrino mass models is described in the literature, with some of the more prominent ones proposed in~\cite{Zee:1980ai,Witten:1979nr,Zee:1985id,Babu:1988ki,Ma:2006km}. Again, $\Lambda$ is given by the masses of new particles running in the loop, while the $c^{[\mathrm{n}]}_\mathrm{i}$ are typically given by a combination of coupling constants and loop factors. 

\vskip 2mm

A crucial point is that neutrino oscillation data is not sensitive to the magnitude of $\Lambda$, but only to the combinations $c^{[\mathrm{n}]}_\mathrm{i} \Lambda^{\mathrm{n}-4}$.
In principle $\Lambda\sim {\rm eV}$ is conceivable~\cite{deGouvea:2005er}.
The lack of observed deviations $\eta$ from unitarity in the light neutrino mixing matrix $V_\nu = \left(1 + \eta\right)U_\nu$ suggests that $\Lambda>$ MeV~\cite{Blennow:2016jkn}, while $\Lambda$ should be below the Planck scale $m_P$ if the underlying theory is required to preserve unitarity at the perturbative level~\cite{Maltoni:2000iq}, leaving a window of over 20 orders of magnitude.
Any stronger statements are entirely based on theoretical arguments.


\subsubsection{Why are the neutrino masses so small?}\label{sec:WhyMnuSmall}

The mass puzzle can then be explained by one or several of the following reasons:
a) $\Lambda$ is large, b) the entries of the matrices $c^{[\mathrm{n}]}_\mathrm{i}$ are small, c) there are cancellations between different terms in $m_\nu$.

\vskip 2mm
\noindent
\textbf{a) High scale seesaw mechanism}:  
Values of $\Lambda$ far above the electroweak scale parametrically suppress the $\mathcal{O}^{ [\mathrm{n}]}_\mathrm{i}$ and therefore the light neutrino masses by powers of $v/\Lambda$. 
The fact that larger $\Lambda$ lead to smaller $m_i$ has earned this idea the name \emph{seesaw mechanism} (``as $\Lambda$ goes up, the $m_i$ go down''). 

\vskip 2mm
\noindent
\textbf{b) Small numbers}: 
The $m_i$ can be made small for any value of $\Lambda$ if the entries of the matrices $c^{[\mathrm{n}]}_\mathrm{i}$ are small.
There are several ways how this may occur without tuning them ``by hand''.
 The probably most popular is the existence of an {\it approximate symmetry}. 
 For instance, in the type II seesaw, the total lepton number $L=\sum_a L_a$ is conserved in the limit $\kappa\to0$.\footnote{Strictly speaking it is $B-L$ that is conserved in the SM and broken here. Lepton number $L$ alone is already broken by anomalies in the SM~\cite{Adler:1969gk,Bell:1969ts}, but baryon number $B$ violating processes are highly suppressed in vacuum. In the early universe this makes a big difference, as $B+L$ violating processes occur frequently at temperatures above the electroweak scale~\cite{Kuzmin:1985mm}.\label{BminusLfootnote}} 
 In UV complete theories the $c^{[\mathrm{n}]}_\mathrm{i}$
 may be determined due to the spontaneous symmetry breaking of a flavour symmetry by one or several flavons~\cite{Froggatt:1978nt}. This may also help to address the flavour puzzle. 
 Small $c^{[\mathrm{n}]}_\mathrm{i}$ can also be justified if the $\mathcal{O}^{ [\mathrm{n}]}_\mathrm{i}$ are generated radiatively.
 In this case the smallness of the $m_i$ may be explained due to a combination of the suppression from the ``loop factors'' $(4\pi)^2$ and the couplings of the new particles in the loop.
More exotic explanations e.g. involve 
the gravitational anomaly~\cite{Dvali:2016uhn},
extra dimensions~\cite{Dienes:1998sb,ArkaniHamed:1998vp}, or string effects~\cite{Blumenhagen:2006xt,Antusch:2007jd}. 

\vskip 2mm
\noindent
\textbf{c) Protecting symmetry}: 
In principle the smallness of the light neutrino masses does not require the individual entries of the matrix $m_\nu$ to be small, 
as these entries are not directly constrained by neutrino oscillation experiments.
The physical neutrino mass squares $m_i^2$ are determined by the eigenvalues of the matrix $m_\nu^\dagger m_\nu$. 
If the $m_i$ are protected by a symmetry, then systematic cancellations can keep them much smaller than the individual entries of $m_\nu$, and no large $\Lambda$ or small $c^{[\mathrm{n}]}_\mathrm{i}$ are needed.
This is, for instance, the case if the type-I seesaw Lagrangian approximately respects a  $B-\bar{L}$ charge, where $\bar{L}$ is a generalisation of the SM lepton number $L$ under which the new particles may be charged. 
 Due to the strong suppression of baryon number $B$ violating processes in experiments the common jargon is to refer to these models as ``approximately lepton number conserving'', cf.~footnote~\ref{BminusLfootnote}.
Popular models that can incorporate this idea include the inverse~\cite{Mohapatra:1986aw,Mohapatra:1986bd,Bernabeu:1987gr} and linear~\cite{Akhmedov:1995ip,Akhmedov:1995vm} seesaws, scale invariant models~\cite{Khoze:2013oga} and the $\nu$MSM~\cite{Shaposhnikov:2006nn}.

\begin{figure*}[h]
\centering
\includegraphics[width=0.7\textwidth]{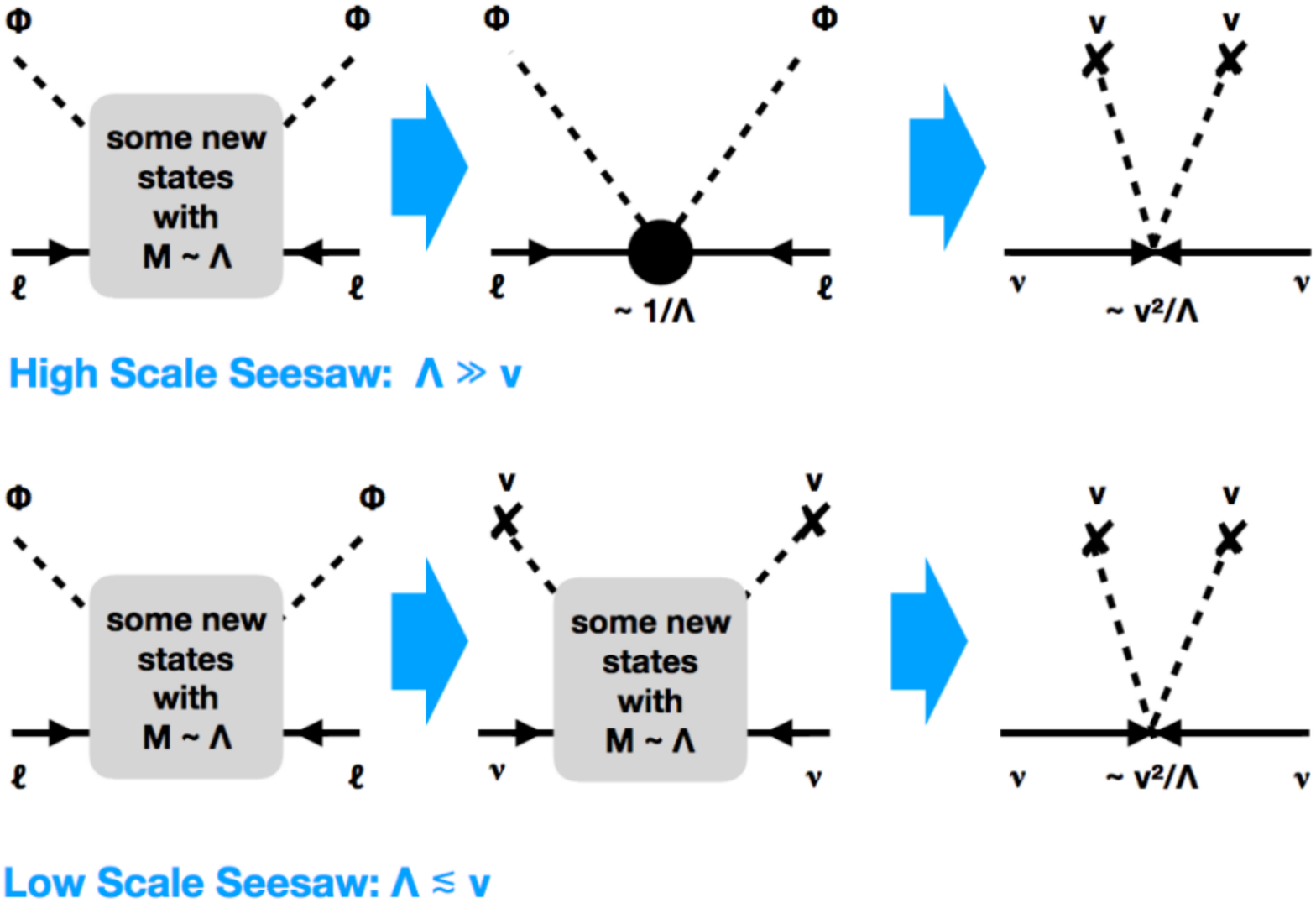}
\caption{\label{lowscaleseesaws}Schematic illustration of the generation of a Majorana mass tern \eqref{massterms} in low scale and high scale seesaws, figure from~\cite{Boyarsky:2018tvu}. The figure can e.g.~describe the type I and II seesaws in \eqref{seesaws}, where $m_\nu$ is generated due to the SM Higgs expectation value $v$.}
\end{figure*}

\vskip 3mm
To summarise, from an experimental viewpoint, there is no hint for why the neutrino masses are small. From a theory viewpoint, the suppression $v/\Lambda$ by a high seesaw scale $\Lambda$ is only one out of several possible explanations. In particular, symmetry-based arguments can provide alternative explanations in which the scale of new physics is low enough to be probed with existing experiments.

\subsubsection{What is the scale of the New Physics?}
\label{sssec:drewes-scale-NP}

We shall qualitatively distinguish between \emph{high scale seesaw} scenarios with $\Lambda \gg v$ and \emph{low scale seesaw} scenarios with $\Lambda \sim v$ or even $\Lambda < v$, cf.~figure~\ref{lowscaleseesaws}. In the former case the EFT description holds for all experiments that can be performed in the foreseeable future, in the latter case accelerator based experiments may resolve the microphysical origin of the $\mathcal{O}^{ [\mathrm{n}]}_\mathrm{i}$.
Since all possibilities are viable, one may wonder whether there are any theoretical reasons that favour specific choices of $\Lambda$. 

\vskip 2mm
High scale seesaws are theoretically appealing because they can explain the observed $m_i$ with dimensionless couplings 
of order unity and values of $\Lambda\sim 10^{14}$ GeV, not too far away from the scale of Grand Unification. Moreover, the simplest ``vanilla leptogenesis'' scenario suggests a rather high seesaw scale~\cite{Davidson:2002qv}.  
On the other hand, a high scale seesaw destabilises the Higgs mass~\cite{Vissani:1997ys,Clarke:2015gwa}
 unless it is embedded in an extended framework  in which these contributions are cancelled, such as electroweak scale supersymmetry.
Several arguments may be brought forward that make a low scale seesaw appealing.

\vskip 2mm
\emph{\bf Lamp post approach.} From an experimental viewpoint low scale seesaws are attractive simply because they are testable. 
The underlying philosophy may be described as agnosticism - since there is no clear hint what the origin of neutrino mass is, 
one may consider all theoretically consistent ideas as equally likely, and turn every stone to fully exploit the discovery potential of existing experimental facilities.  

\vskip 2mm
\emph{\bf Ockham's razor.
} 
Arguments involving different interpretations of Ockham's razor can be used to motivate both, low and high scale physics. When applied to the choice of gauge group, one may argue that a unified group such as $SO(10)$ is simpler than the somewhat arbitrary product group $SU(3)\times SU(2)\times U(1)$ of the SM. In such a ``top down'' approach Ockham's razor would favour high scale seesaws that appear in models that, in principle, can be derived from a small number of fundamental principles (such as supersymmetric grand unified theories that appear as the low energy limit of string theory).
Alternatively one may apply Ockham's razor to the number of free parameters, the number of new physical concepts, or the number of new particles. Such ``bottom up'' approaches tend to favour low scale seesaw models because 
high scale (e.g.~string-inspired) theories tend to predict many new particles, and involve a comparably large number of parameters in the scalar potentials responsible for the spontaneous symmetry breaking and the geometrical properties of compactified extra dimensions. 

\vskip 2mm
To give an explicit example, one may apply Ockham's razor to the number of physical scales in nature and speculate that $\Lambda$ could be connected to one of the three scales that we already know - the QCD scale, the Planck scale or the electroweak scale. This assumption seems theoretically well-motivated if nature is classically scale invariant, and all dimensionful scales are generated through quantum effects, as observed in QCD. In~\cite{Khoze:2013oga} a specific realisation of this idea has been worked out in which the electroweak scale and the seesaw scale are related because they are both created via a Coleman-Weinberg like mechanism in a hidden sector. 
Alternatively one may apply Ockham's razor to the number of new degrees of freedom that are needed to solve open problems in particle physics and cosmology, which, in addition to the neutrino masses, e.g.~include the Dark Matter and the matter-antimatter asymmetry of the universe. 
A well-known representative of this line of thought is the $\nu$MSM~\cite{Asaka:2005pn,Asaka:2005an}, which can explain all these problems with the addition of a right-handed partner $\nu_R$ for each SM neutrino, and no other new particles.

\vskip 2mm
\emph{\bf Technical naturalness.} 
Arguments related to ``naturalness'' have served as guidelines in the search for a more fundamental theory of nature for decades, see~\cite{Giudice:2008bi} and references therein. 
One of the more well-defined implementations of the naturalness idea is the so-called technical naturalness~\cite{tHooft:1979rat}, which practically states that large cancellations between different contributions to a physical quantity do not happen by accident, so that small quantities only occur when the symmetry of a physical system increases if the respective quantity were zero.
Specifically in the context of the seesaw mechanisms, one can argue that values of $\Lambda$ that are much smaller than other relevant scales in nature (such as the Planck scale) are natural only if $\Lambda\to 0$ increases the symmetry. 
This is indeed the case in seesaw models that generate Majorana masses for the neutrinos, which feature a global  $U(1)_{B-L}$ symmetry in the limit $M_M\to 0$ or $\kappa\to0$. Hence, a low seesaw scale can be considered natural. The fact that $U(1)_{B-L}$ is a global symmetry of the SM can be viewed as a hint that this reasoning is correct. 
An independent naturalness-related argument for a 
low scale Majorana mass is the fact that it does not contribute to the electroweak hierarchy problem~\cite{Vissani:1997ys,Clarke:2015gwa}.\footnote{
An intriguing possibility is that the Higgs potential is scale invariant at tree level, and the corrections from the seesaw are in fact the origin of electroweak symmetry breaking~\cite{Brivio:2017dfq}.
} 
This requirement points towards a seesaw scale below $10^7$ GeV in the type I scenario~\cite{Bambhaniya:2016rbb}
and values in the TeV range for the type II scenario~\cite{Dev:2017ouk}.

One may argue that the naturalness criterion should not only be applied to $\Lambda$, but also to the coefficients $c^{[\mathrm{n}]}_\mathrm{i}$. If one starts from the high scale seesaw (\emph{option a)} in section~\ref{sec:WhyMnuSmall}) and performs a ``naive scaling'' of \eqref{weinbergoperator} to $\Lambda \sim v$, one finds $c^{[\mathrm{n}]}_\mathrm{i} \sim 10^{-12}$, corresponding to dimensionless couplings $\sim10^{-7}$ in the tree level seesaws \eqref{seesaws}, which may be considered ``unnatural''.
However, two comments are in place. 
First, such small numbers may be natural if they can be related to an approximate symmetry in the underlying UV theory, as discussed in \emph{option b)} in section~\ref{sec:WhyMnuSmall}. Such symmetry may in any case be needed to explain the small electron Yukawa coupling. 
Second, the naive scaling argument ignores the flavour structure of \eqref{weinbergoperator}, and the possibility that the smallness of the $m_i$ may be due to an approximate symmetry (\emph{option c)}). 
To illustrate these arguments with a few numbers, values $\Lambda\sim $ TeV and dimensionless couplings 
$Y_{\rm I}, Y_{\rm II}, Y_{\rm III}, \lambda_\kappa \sim 10^{-5}$ (similar to that of the electron) are sufficient to generate a viable low scale type I or type III seesaw mechanism without the need of a protecting symmetry. 
In the presence of a protecting $B-\bar{L}$ symmetry, Yukawa couplings $Y_{\rm I}$ of order one can be consistent with the observed $m_i$ for almost any value of $\Lambda$.
Hence, a low value of $\Lambda$ is theoretically well-motivated from the viewpoint of naturalness, and it can be realised without resorting to tiny couplings constants if the smallness of the $m_i$ is the result of an approximate symmetry. 

\vskip 2mm
\emph{\bf No conspiracy theories.}
The electroweak hierarchy problem is based on the observation that the Higgs mass would be destabilised by radiative corrections due to particles with masses at some intermediate scale 
between $v$ and $m_P$.\footnote{We shall not consider radiative corrections from physics at the Planck scale itself here, but only from intermediate scales, because the framework of quantum field theory within which the hierarchy problem is formulated certainly beaks down and has to be replaced by a theory of quantum gravity at $m_P$.}
This argument implicitly assumes the existence of new physical scales between the electroweak and Planck scales; in absence of an intermediate scale there is no hierarchy problem~\cite{Bardeen:1995kv}.
The statement that the SM could be a valid EFT all the way up to the Planck scale~\cite{Ellis:2009tp,Bezrukov:2012sa,Degrassi:2012ry} is absolutely not trivial: For the observed top quark mass $m_t$, a larger Higgs mass $m_H$ would imply a Landau pole well below $m_P$, while a smaller $m_H$ would imply that the lifetime of the electroweak vacuum is considerably shorter than the age of the universe.

This remarkable conspiracy of numbers could be seen as a sign that there actually is no new scale between the electroweak scale and the Planck scale (or at least some very high energy scale). 
The rather accurate prediction of the Higgs mass ~\cite{Shaposhnikov:2009pv,Bezrukov:2012sa} that can be made based on the assumption that there is no intermediate scale which affects the running of couplings can be viewed as circumstantial evidence supporting this idea.
This would imply that the neutrino masses must either be explained by new physics at the Planck scale, or that the seesaw scale is low~\cite{Shaposhnikov:2007nj}.
The mechanism of neutrino mass generation may then be embedded in a comparably light hidden sector, which can also contain the solutions to other puzzles in particle physics and cosmology. 
From a theoretical viewpoint this possibility would, for instance, fit well into a broader context of classically scale invariant theories~\cite{Bardeen:1995kv,Khoze:2013oga}.
From an experimental viewpoint it would resolve the apparent contradiction between the need for new physics near the electroweak scale to resolve the electroweak hierarchy problem and the lower bounds on $\tilde{\Lambda}$ from flavour observables~\cite{Alpigiani:2017lpj},\footnote{
It has also been proposed that a low scale seesaw may resolve the long-standing flavour anomalies in B meson decays~\cite{Asadi:2018wea}.
} 
and it would explain the lack of new particle discoveries in conventional LHC searches as well as direct DM detection experiments. 



\subsubsection{Low scale seesaw models: Minimal and non-minimal models}
\label{sssec:drewes-MinimalNonMinimal}

The previous considerations have motivated a large number of theoretical and phenomenological investigations into low scale seesaw scenarios, many of which are embedded in extended hidden sectors. 
A partial overview can be found in~\cite{Alekhin:2015byh,Curtin:2018mvb,Beacham:2019nyx,Alimena:2019zri,Anamiati:2018cuq,SnowmassTestable,SnowmassNonMinimal}.

Phenomenological studies are often based on ``minimal'' scenarios in which the SM is only extended by 
$\nu_R$, $\Delta$ or $\Sigma_L$ needed to generate the terms \eqref{seesaws}, with no other new particles added. 
In general this assumption is questionable for two reasons. First, most UV completions of the SM that are based on ``top down'' considerations, such as grand unification or supersymmetry, contain additional new particles and interactions. 
For instance, left-right symmetric models do not only contain a combination of the type I and II seesaws, but also additional $SU(2)_R$ gauge interactions.
Second, even from a ``bottom up'' viewpoint the existence of additional new particles is well-motivated by cosmological problems, including DM, baryogenesis, and cosmic inflation. 

If we label the scale at which new particles other than those directly involved in the generation of neutrino masses 
 appear by $\tilde{\Lambda}$, one can qualitatively distinguish four scenarios,
 \begin{itemize}
\item[{\bf i)}] $\tilde{\Lambda}, \Lambda \gg {\rm TeV}$ corresponds the classic high scale seesaws, in which \eqref{weinbergoperator} is  only experimentally accessible operator.
\item[{\bf ii)}] $\tilde{\Lambda} \gg  {\rm TeV} > \Lambda$ corresponds to the pure low scale seesaws, which are entirely described by the renormalisable operators that can be constructed from SM fields and $\nu_R$, $\Delta$, $\Sigma_L$.
The type I scenario can be identified with the fermion portal to hidden sectors, while the type II and III scenarios do not quite fit into the framework of renormalisable hidden sector portal couplings used in~\cite{Beacham:2019nyx} because the $\Delta$ and $\Sigma_L$ are not gauge singlets. 
\item[{\bf iii)}] For   $\tilde{\Lambda} \gtrsim  {\rm TeV} > \Lambda$ the 
$\nu_R$, $\Delta$, or $\Sigma_L$
are the only new particles that are experimentally accessible, but new interactions described by non-renormalisable operators that are suppressed by powers of $1/\tilde{\Lambda}$ are phenomenologically relevant.
\item[{\bf iv)}] For $ {\rm TeV} > \tilde{\Lambda}, \Lambda$ there exists an extended hidden sector at collider-accessible energies, to which neutrinos open a portal. The EFT description in powers of $1/\tilde{\Lambda}$ cannot be applied, and phenomenological studies require to specify details of the hidden sector. 
\end{itemize}

 \subsubsection{An explicit example of low scale seesaw models: The pure type I seesaw}
 \label{sssec:drewes-seesaw-typeI}

In the following we shall illustrate the aforementioned arguments for the minimal implementation of the type I seesaw, as described by the most general renormalisable Lagrangian that can be constructed from SM fields and $n$ flavours of right-handed neutrinos, 
\begin{eqnarray}
    \mathcal L
  &= & \mathcal L_{\rm SM} + \mathrm{i} \overline{\nu_R}_i \slashed\partial \nu_{Ri}
  - Y_{ai} \overline{\ell_{L a}} \varepsilon \Phi^* \nu_{Ri}
  - Y_{ai}^* \overline{\nu_R}_i \Phi^T \varepsilon^\dagger \ell_{L a}
    \nonumber \\
  &-&  \frac{1}{2} \left( \overline{\nu_R^c}_i(M_M)_{ij}\nu_{Rj}
    + \overline{\nu_R}_i(M_M^\dagger)_{ij}\nu_{R}^c \right) 
\, , \label{MinimalSeesaw}
\end{eqnarray}
where we have simplified the notation to $Y\equiv Y_{\rm I}$. 
In general the model \eqref{MinimalSeesaw} should be viewed as an EFT with some cutoff scale $v < \tilde{\Lambda} < m_P$. 
We here consider the case ii) defined in section~\ref{sssec:drewes-MinimalNonMinimal}, i.e., the pure type I seesaw with $\tilde{\Lambda}\gg {\rm TeV}$. 
In spite of its simplicity, this model exhibits a rich phenomenology (see e.g.~~\cite{Drewes:2013gca}),
and it can accommodate  many features discussed in Sections~\ref{sssec:drewes-numasses-NP}-\ref{sssec:drewes-scale-NP}. 

\vskip 2mm
The most important parameters in  \eqref{MinimalSeesaw} are the number $n$ of right-handed neutrino flavours and the choice of eigenvalues of $M_M$, i.e., the seesaw scale(s). One $\nu_{R i}$ flavour is needed to generate each non-zero $m_i$ (``seesaw fair play rule''). 
Hence, the minimal choice needed to explain the observed two light neutrinos mass splittings is $n=2$, which automatically implies that the lightest neutrino is massless ($m_{\rm lightest} = 0$). 
Another popular choice is $n=3$, which permits $m_{\rm lightest} > 0$ and has a certain appeal in view of the fact that it provides a right-handed partner $\nu_R$ for each of the left-handed neutrinos $\nu_L$ in the SM.
If the UV completion of \eqref{MinimalSeesaw} contains a local $U(1)_{B-\bar{L}}$ gauge symmetry, the choice $n=3$ is needed for theoretical consistency. 
After electroweak symmetry breaking the model \eqref{MinimalSeesaw} contains three light Majorana neutrinos $\upnu_i$ with a mass matrix $m_\nu^{\rm I} $ in \eqref{LightNeutrinoMasses}
and $n$ heavy neutrinos with masses roughly 
given by the eigenvalues of $M_M$,
described by the Majorana spinors
$\upnu \simeq U_{\nu}^{\dagger} (\nu_{L} - \theta \nu_{R}^c) + \text{c.c.}$
and
$N \simeq U_N^\dagger ( \nu_{R} +  \theta^T\nu_{L}^{c}) + \text{c.c.},$
 where $U_N$ is the equivalent of $U_\nu$ in the sterile sector. 
Here we have defined the mixing matrix $\theta = v Y M_M^{-1}$ and neglected 
corrections $\mathcal{O}[\theta^2]$.

\vskip 2mm
For the sake of definiteness, let us consider the case $n=3$. 
It is instructive to parameterise the masses and couplings in \eqref{MinimalSeesaw} for $n=3$ as
 $Y_{a2}=Y_a(1+ \epsilon_a)$, 
  $Y_{a3}=\mathrm{i} Y_a(1 - \epsilon_a)$,
 $Y_{a1}=\epsilon'_{a}Y_a$ and $\bar{M}_2,\bar{M}_3 = \bar{M}\pm\mu$.
In the limit $\epsilon_a, \epsilon'_a, \mu \to 0$ the model exhibits an global $U(1)_{B-\bar{L}}$ symmetry~\cite{Shaposhnikov:2006nn,Kersten:2007vk,Moffat:2017feq}, and 
one can assign the $\bar{L}$ charges $\pm1$ to the spinors $\nu_{R {\rm s}}\equiv \frac{1}{\sqrt{2}}(\nu_{R 2} + i \nu_{R 3})$,
and
$\nu_{R {\rm w}}\equiv \frac{1}{\sqrt{2}}(\nu_{R 2} - i \nu_{R 3})$, respectively, and $0$ to
$\nu_{R 3}$. 
The spectrum of neutrinos 
before electroweak symmetry breaking consists 
of three massless Weyl fermions $\nu_{La}$,
a Dirac fermion 
$\psi=(\nu_{R {\rm s}} + \nu_{R {\rm w}}^c)$,
and a Majorana fermion $\nu_{R1}$ that does not couple to the SM. 
After electroweak symmetry breaking the light neutrinos remain massless ($m_\nu=0$), but mix with a heavy Dirac neutrino with mixings given by 
$Y_a v/\bar{M}$, while $\nu_{R1}$ remains decoupled. 
Switching on the $B-\bar{L}$ violating parameters $\epsilon_a, \epsilon'_a, \mu$ results in non-zero light neutrino masses, 
splits the heavy Dirac neutrino into two heavy Majorana neutrinos, and gives 
$\epsilon'_a$-suppressed interactions to $\nu_{R1}$.
The complete Lagrangian \eqref{MinimalSeesaw} in this parameterisation can be written as
\begin{eqnarray}\label{PseudoDiracL}
\mathcal{L}&=&\mathcal{L}_{\rm SM} + \overline{\psi}(\mathrm{i}\slashed{\partial} - \bar{M})\psi 
+\overline{\nu_{R 1}}\mathrm{i}\slashed{\partial}\nu_{R 1}
\nonumber \\
&&- Y_a^*\overline{\psi}\Phi^T\varepsilon^\dagger \ell_{L a} - Y_a \overline{\ell}_{L a} \varepsilon\Phi^* \psi\nonumber\\
&&-\epsilon_a^* Y_a^*\overline{\psi^c_N}\Phi^T\varepsilon^\dagger \ell_{L a} 
- \epsilon_a Y_a\overline{\ell}_{L a} \varepsilon\Phi^* \psi^c
\nonumber \\
&&  - \epsilon'_{a}Y_a\overline{\ell_{L a}}\varepsilon\Phi^* \nu_{R 1}
  - \epsilon_a^{' *}Y_a^*\overline{\nu_{R 1}}\Phi^T \varepsilon^\dagger \ell_{L a}\nonumber\\
  &&-\mu\bar{M} \frac{1}{2}\left(\overline{\psi^c}\psi + \overline{\psi}\psi^c\right) - \bar{M}_1 \overline{\nu_{R 1}^c}\nu_{R 1} \,.
\end{eqnarray}
Specific models that exhibit an approximate $U(1)_{B-\bar{L}}$ symmetry, such as the inverse seesaw or linear seesaw, 
give a prescription how to take the limit $\epsilon_a, \epsilon'_a, \mu\to 0$. 
As far as the $\nu_R$ are concerned they can all be described by \eqref{PseudoDiracL}.

\vskip 2mm
The above symmetry considerations do not fix the seesaw scale $\Lambda=\bar{M}$ or the size of the Yukawa couplings $Y_a$. 
Eigenvalues $\bar{M} \gg v$   reproduce the classic high scale seesaw.
However, 
since the neutrino masses are protected by the $U(1)_{B-\bar{L}}$ symmetry, there is no need for $\bar{M}$ to be large or for the $Y_a$ to be small. 
   Experimental constraints~\cite{Chrzaszcz:2019inj} permit $|Y_a| \sim 0.5$ for $\bar{M} = 1 \ {\rm TeV}$, or $|Y_a| \sim 10^{-4} - 10^{-3}$ for $\bar{M} $ at the QCD scale.
 Such comparably low values of the seesaw scale can be explained by the arguments outlined Sections~\ref{sssec:drewes-numasses-NP}-\ref{sssec:drewes-scale-NP},
such as 
scale invariance, 
technical naturalness, 
 or by demanding that other puzzles in particle physics or cosmology can be explained.
 Further, the symmetry arguments do not fix the value of $M_1$. 
 From a phenomenological ``bottom up'' viewpoint one may distinguish three scenarios.
The choice $M_1\gg \bar{M}$ practically reproduces the model with $n=2$ if the $\epsilon'_a$ are small, and introduces a high scale seesaw contribution in case they are not small.
For $M_1\simeq \bar{M}$ the Majorana mass exhibits an additional $SO(3)$ symmetry, sometimes referred to as \emph{mass communism},
which can lead to interesting leptogenesis scenarios~\cite{Abada:2018oly,Drewes:2021nqr},
and can lead to sizeable rates for lepton number violating processes at colliders even if $\epsilon_a, \epsilon'_a, \mu\ll 1$~\cite{Drewes:2019byd}, cf.~figures~\ref{experiments} and~\ref{leptogenesisn2}.
The choice $M_1 \ll \bar{M}$ corresponds to the $\nu$MSM.
A collection of ``top down'' considerations that can also address the flavour puzzle can be found in~\cite{King:2013eh,Feruglio:2019ktm,Xing:2019vks,Xing:2020ald} and references therein. 
 
 \vskip 2mm
 Quite remarkably, combining all these arguments leads to a self-consistent picture in which the model \eqref{PseudoDiracL} is a complete EFT up to the Planck scale that~\cite{Bezrukov:2012sa} can simultaneously explain the neutrino masses, baryogenesis and DM~\cite{Asaka:2005pn,Shaposhnikov:2008pf,Canetti:2012kh,Ghiglieri:2020ulj}
 for $\bar{M} \sim v$ and $M_1$ in the keV range. 
 
\subsubsection{Low scale Leptogenesis}
\label{sssec:drewes-lowscale-leptogenesis}
The observable universe contains almost no antimatter~\cite{Canetti:2012zc}, and the number density of photons exceeds that of baryons by almost 10 orders of magnitude~\cite{Zyla:2020zbs,Aghanim:2018eyx}. 
These two observations indicate that the baryonic matter in the present universe is the remnant of a small matter-antimatter asymmetry in the primordial plasma, known as baryon asymmetry of the universe (BAU), that survived the mutual annihilation of all particles and antiparticles after the temperature $T$ in the early universe dropped below the pair creation thresholds. 

\vskip 2mm
Explaining the BAU requires three ingredients, known as Sakharov conditions~\cite{Sakharov:1967dj}, 
I) violation of $B$,
II) violation of $C$ and $CP$,
and III) a deviation from thermal equilibrium.
In the SM $B+L$ is violated in the early universe via electroweak sphalerons \cite{Kuzmin:1985mm}, while the $C$ and $CP$ violation~\cite{Gavela:1993ts,Huet:1994jb,Gavela:1994dt} as well as the deviation from equilibrium~\cite{Gurtler:1997hr,Bochkarev:1987wf,Kajantie:1995kf,Laine:1998jb,Csikor:1998eu,Aoki:1999fi}  are too small and require physics beyond the SM. 
Leptogenesis relates the solution of these shortcomings to the mechanism of neutrino mass generation. 
Originally this idea was proposed in the type I scenario~\cite{Fukugita:1986hr}, but it can e.g.~also be realised with scalar~\cite{Ma:1998dx} or fermionic~\cite{Hambye:2003rt} triplets, cf.~\cite{Hambye:2012fh} for a review.
The observation of neutrino masses and the tentative hints for $CP$ violation in their oscillations have lead to a huge interest in leptogenesis scenarios~\cite{Bodeker:2020ghk}, the computational methods to describe them~\cite{Garbrecht:2018mrp}, and their phenomenology~\cite{Chun:2017spz}.

\begin{figure*}[h]
\centering
\includegraphics[width=0.7\textwidth]{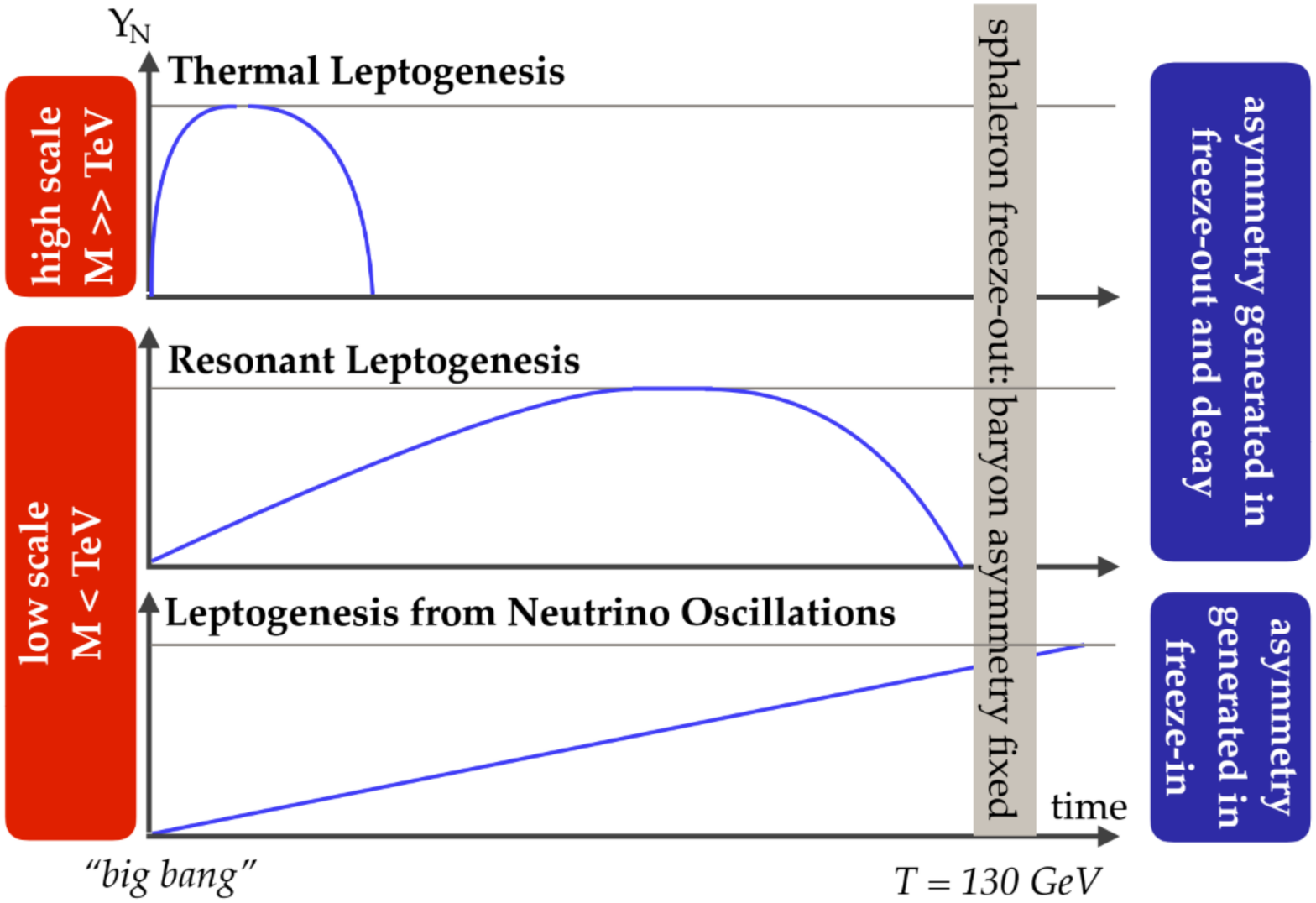}
\caption{
\label{fig:leptogenesi-freezein-freezeout}
Schematic illustration of the evolution of the $N_i$ abundance $Y_N$ in freeze-in and freeze-out leptogenesis scenarios. The BAU is fixed when electroweak sphalerons become inefficient at $T_{\rm sph}\simeq 130$ GeV.}
\end{figure*}

\vskip 2mm
Leptogenesis models can be classified according to the scale $\Lambda$ (low scale vs.~high scale), and according to the way how the Sakharov conditions are fulfilled (freeze-in vs.~freeze-out), cf.~Figure~\ref{fig:leptogenesi-freezein-freezeout}. 
Low scale leptogenesis
can be made possible in several ways  \cite{Hambye:2001eu,Chun:2017spz,Garbrecht:2018mrp} that may be realised in different models. 
The most relevant features can be illustrated in the pure type I model \eqref{MinimalSeesaw}.
The lower bound on $M_i$ \cite{Davidson:2002qv} is typically avoided by a resonant enhancement
of the contributions from the $N_i$-mode $\textbf{p}$ to the asymmetry generation by a factor
 $\sim \Gamma_{\rm p}/\Delta M$ \cite{Klaric:2021cpi},
with $\Gamma_{\rm p}$ is the thermal width of the $N_i$-mode $\textbf{p}$.
\footnote{The precise shape of the regulator
in the limit where $M_i^2 = M_j^2$ 
depends on the details  of the scenario under consideration \cite{Dev:2017wwc}.}
For non-relativistic momenta 
this requires quasi-degeneracies in the spectrum of $M_i$ \cite{Flanz:1994yx,Covi:1996wh},
 in the relativistic regime it is sufficient that $T\gg M_i$ \cite{Drewes:2012ma}.
\vskip 2mm

In all cases the second Sakharov condition is fulfilled by the 
$CP$ violating interactions of the heavy neutrinos, which generate lepton asymmetries $L_a$ that are then converted into a baryon asymmetry $B$ by electroweak sphaleron processes. For the sphalerons to act quicker than the rate of Hubble expansion, as required by condition I), this process must occur at temperatures above $T_{\rm sph}=131$ GeV~\cite{DOnofrio:2014rug}. 
The key difference lies in the way how the third Sakharov condition is realised. 
Assuming that the heavy neutrinos are not produced efficiently during reheating~\cite{Bezrukov:2008ut,Enomoto:2020lpf,Shaposhnikov:2020aen} and there are no pre-existing asymmetries~\cite{Domcke:2020quw}, the radiation dominated epoch starts at some temperature $T_R$ with a $CP$ symmetric plasma of SM particles and no heavy neutrinos $N_i$. 
In the minimal scenario $N_i$ are produced thermally from the plasma and can induce lepton asymmetries via their $CP$-violating interactions with SM particles.

\vskip 2mm
The details of the asymmetry generation can be quite complicated even in the simple model \eqref{MinimalSeesaw}, as one in principle has to track the $CP$-odd and $CP$-even deviations from equilibrium separately for each combination of heavy neutrinos flavours and helicities, and at the same time follow the various SM charges and spectator effects, cf.~\cite{Garbrecht:2018mrp} for a recent review. 
One can  categorise different scenarios according to the ordering of the temperatures at which the $N_i$ 
start performing coherent flavour oscillations ($T=T_{\rm osc}$),
enter thermal equilibrium ($T=T_{\rm in}$), 
become non-relativistic ($T=M_i$),
decouple from the plasma ($T=T_{\rm out}$), 
and decay ($T=T_{\rm dec}$),
as well as the weak sphaleron freeze-out temperature ($T=T_{\rm sph}$)
and the temperatures when the interaction rates for flavour dependent interactions exceed the Hubble rate ($T=T_a$):
For instance, vanilla leptogenesis corresponds to the scenario where all processes happen at very high temperatures, $T_R > T_{\rm in} > M > T_{\rm dec} > T_a \gg T_{\rm sph}$. 
In low scale scenarios the relevant temperatures tend to be lower because the seesaw relation \eqref{LightNeutrinoMasses} tends to imply smaller couplings for a lower seesaw scale (though this is not necessarily true in the presence of protecting symmetries, cf.~section~\ref{sec:WhyMnuSmall}), resulting in a slower approach to equilibrium. 
The qualitatively most important difference is whether the BAU is generated during the approach of the $N_i$ to equilibrium~\cite{Akhmedov:1998qx,Asaka:2005pn} (\emph{freeze-in scenario}, $T_{\rm dec}, T_{\rm in} < T_{\rm sph}$)
or in their decays (\emph{freeze-out scenario}, $T_{\rm dec} > T_{\rm sph}$)~\cite{Fukugita:1986hr}, cf.~figure~\ref{fig:leptogenesi-freezein-freezeout}.
Realising the freeze-out scenario with a sub-TeV seesaw scale requires a degenerate heavy neutrino mass spectrum to resonantly enhance the asymmetry~\cite{Covi:1996wh,Dev:2017wwc} (``resonant leptogenesis''~\cite{Pilaftsis:2003gt}),
while no degeneracy is needed in the freeze-in scenario~\cite{Drewes:2012ma}. 
Both of these mechanisms can be realised in low scale seesaw scenarios, and the mass ranges where they can operate overlap~\cite{Klaric:2020lov,Klaric:2021cpi,Drewes:2021nqr}, so that leptogenesis is feasible for any mass $M_i$ above a few tens of MeV \cite{Canetti:2012kh,extracted_beams:drewes_mev_lep}.
An attractive property of low scale leptogenesis is that it can be tested with laboratory experiments~\cite{Chun:2017spz}, cf.~Section~\ref{sssec:drewes-testing}.

\begin{figure*}
\centering
\includegraphics[width=0.49\textwidth]{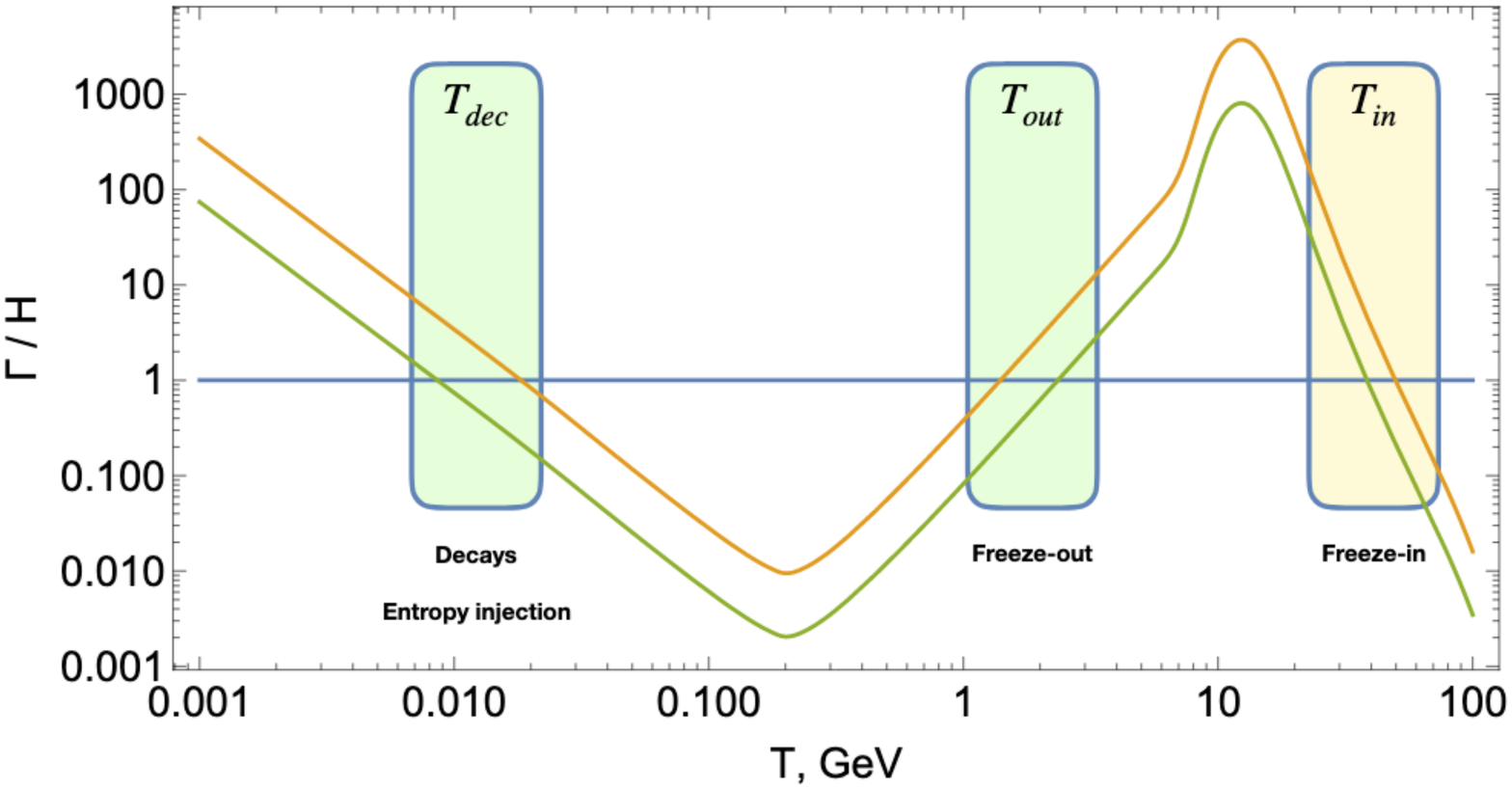}
\includegraphics[width=0.49\textwidth]{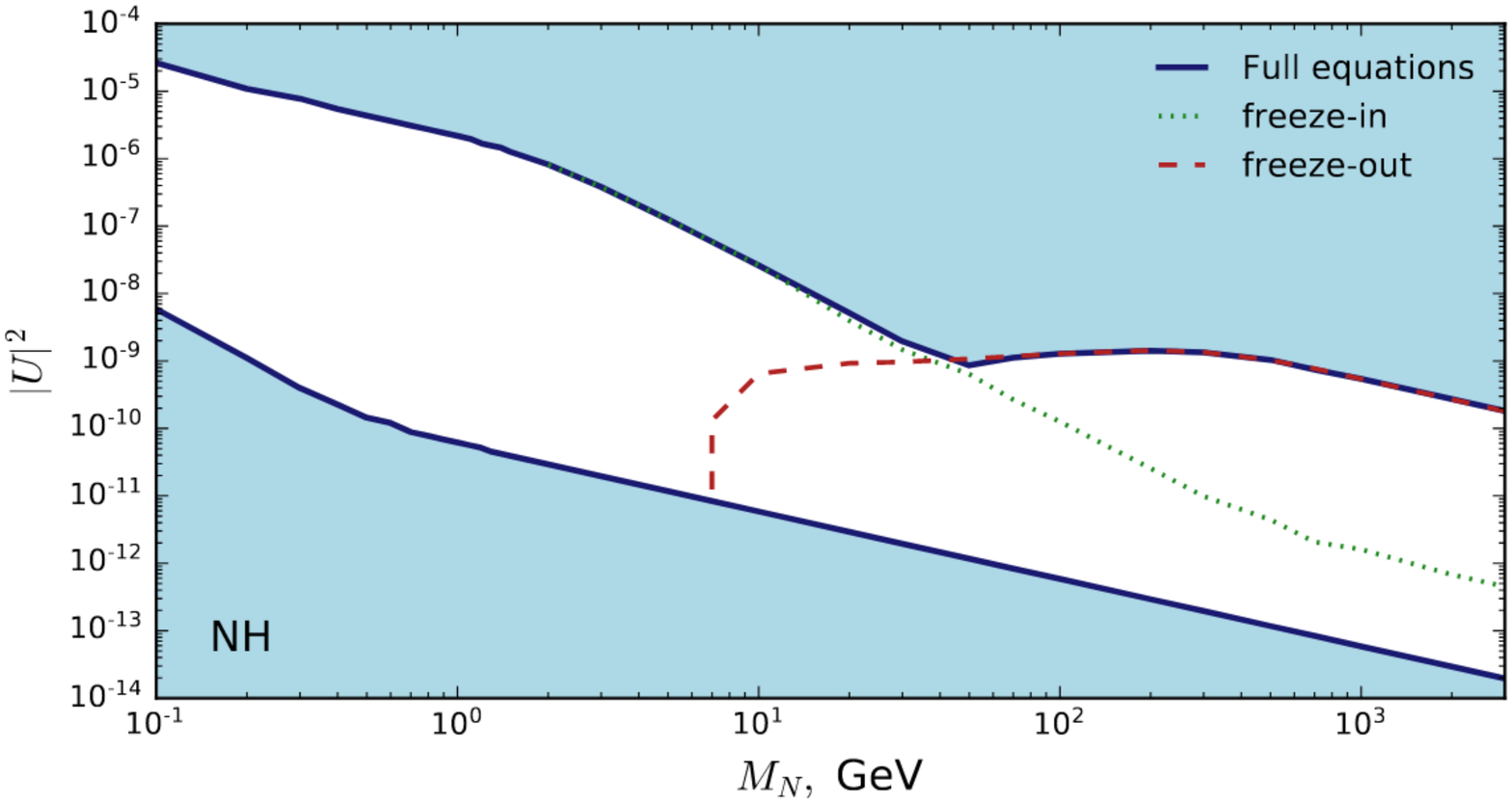}
\caption{\label{leptogenesisn2}
Left: Evolution of the heavy neutrino interaction rates and relevant temperatures in units of the Hubble rate in the $\nu$MSM~\cite{Eijima:2020shs}.
Right: Range of mass and mixing where the freeze-in and freeze-out mechanism can be operational for $n=2$~\cite{Klaric:2020lov}.
For $n=3$ the viable parameter range is much larger and extends to mixings up to the current experimental bounds~\cite{Abada:2018oly,Drewes:2021nqr}, cf.~Fig.~\ref{experiments}.
}
\end{figure*}

\subsubsection{Experimental tests of the low scale seesaw and leptogenesis}
\label{sssec:drewes-testing}

\begin{figure*}
\centering
\includegraphics[width=0.49\textwidth]{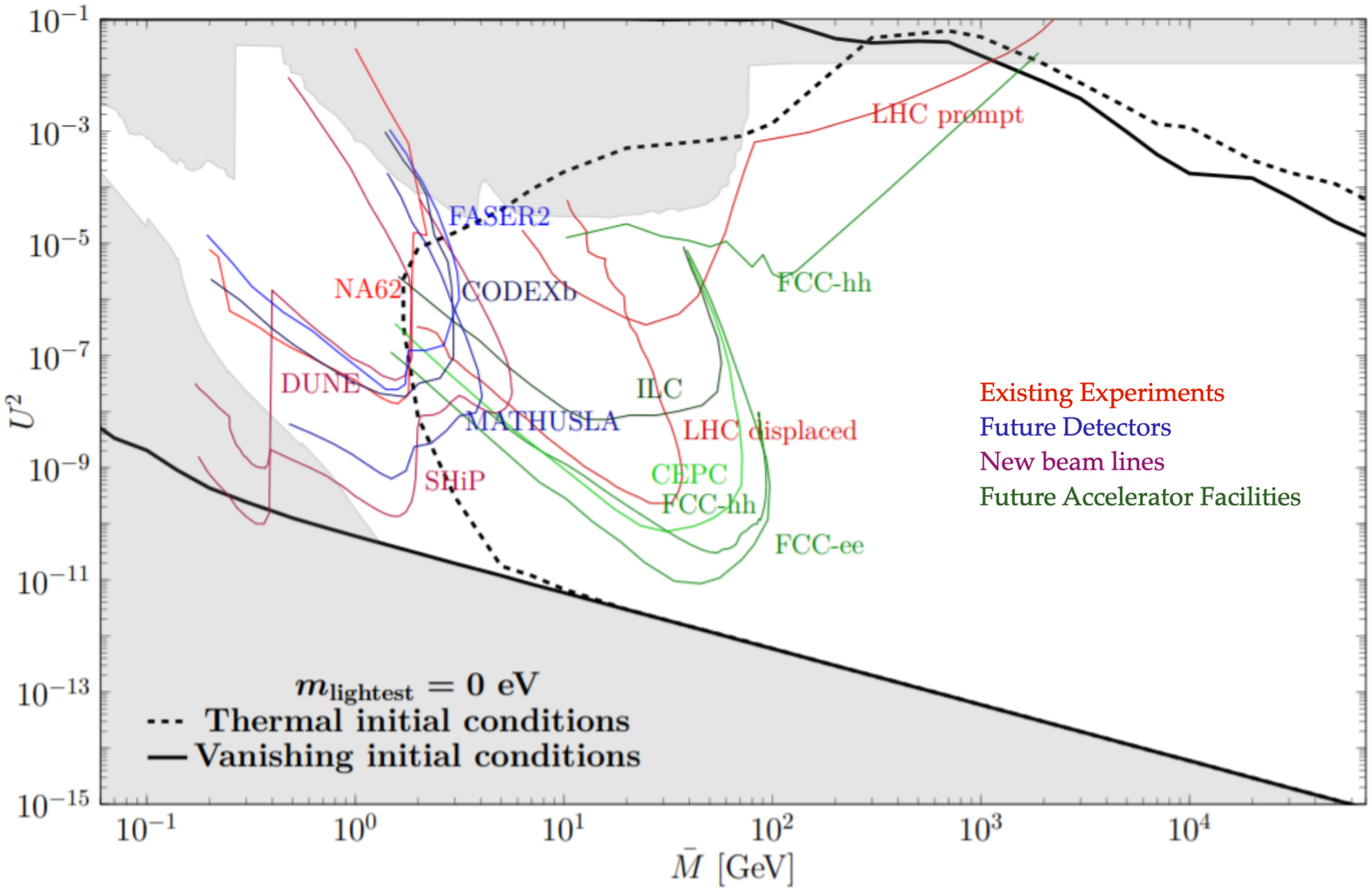}
\includegraphics[width=0.49\textwidth]{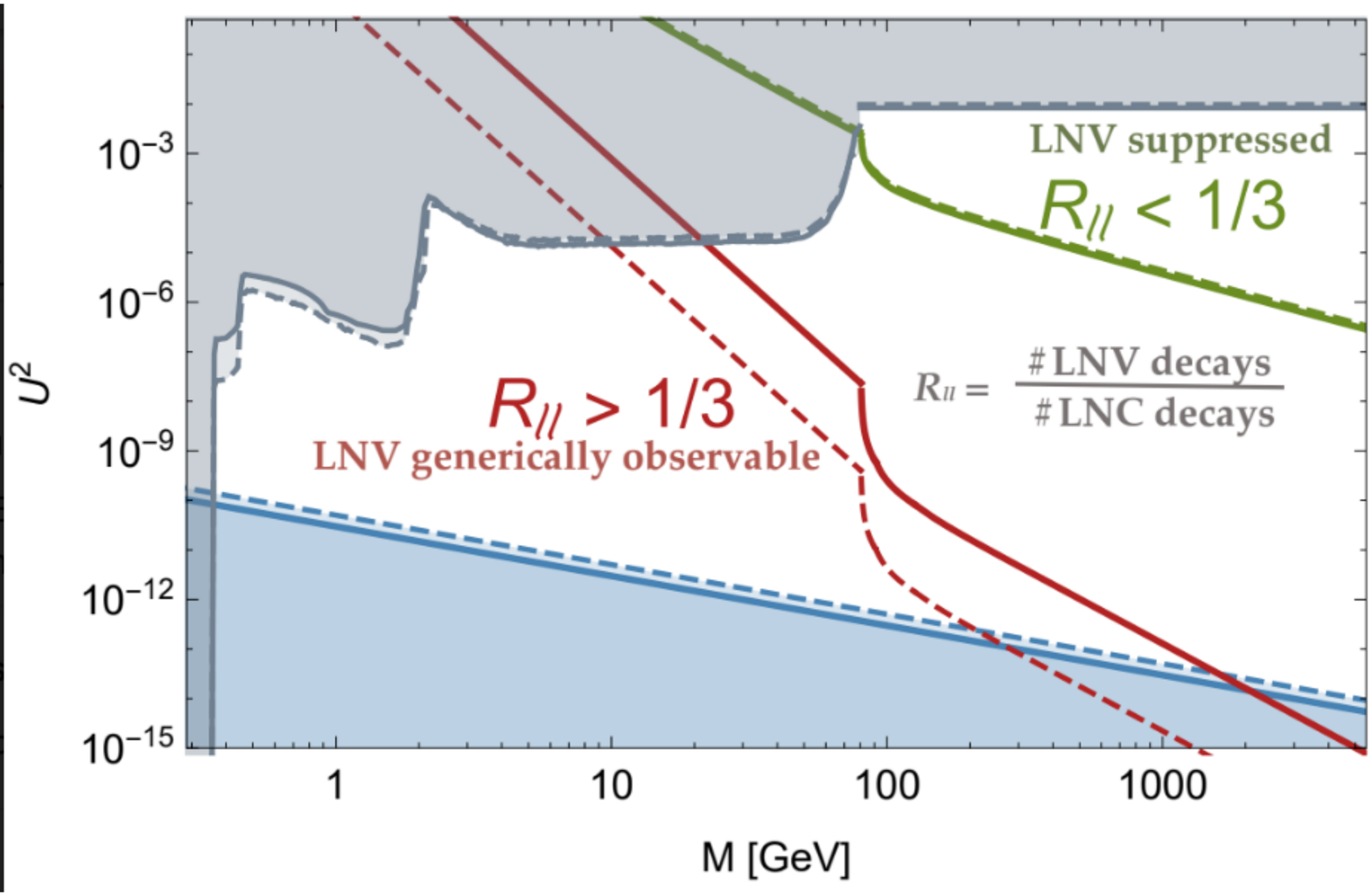}
\caption{\label{experiments}
Left: 
Parameter range in which leptogenesis with three $\nu_R$ is possible \cite{Drewes:2021nqr} (between the black lines), compared to the  
region ruled out in the global scan \cite{Chrzaszcz:2019inj} (The upper bound is dominated by the direct searches \cite{Abreu:1996pa,NA62:2020mcv,Artamonov:2014urb,Bergsma:1985is,Bernardi:1987ek,Ruchayskiy:2011aa,Sirunyan:2018mtv,Aad:2019kiz,Sirunyan:2018xiv,Antusch:2017hhu,Shuve:2016muy,Vaitaitis:1999wq})
complemented by the updated BBN bounds from~\cite{Sabti:2020yrt,Boyarsky:2020dzc}
(gray area) and the estimated sensitivities of the LHC main detectors (taken from \cite{Izaguirre:2015pga,Drewes:2019fou,Pascoli:2018heg}) and 
		NA62 \cite{Drewes:2018gkc} as well as the sensitivities of selected planned or proposed experiments (DUNE \cite{Ballett:2019bgd},
			FASER2 \cite{Ariga:2018uku},
			SHiP \cite{SHiP:2018xqw,Gorbunov:2020rjx}
			MATHUSLA \cite{Curtin:2018mvb},
		Codex-b \cite{Aielli:2019ivi}) as well as
		future lepton colliders \cite{Antusch:2017pkq} or proton colliders \cite{Antusch:2016ejd}.
Right: Regions of mass- and mixing where $L$-violating processes are expected to be observable in the $\nu$MSM~\cite{Drewes:2019byd}, quantified by the ratio $R_{ll}$ of $L$-violating to $L$-conserving event rates.}
\end{figure*}

\vskip 2mm
An attractive feature of low scale seesaw scenarios is that it is possible to discover the new particles experimentally \cite{Chun:2017spz}, 
including at existing colliders~\cite{Atre:2009rg,Deppisch:2015qwa,Cai:2017mow},
future colliders~\cite{CEPCStudyGroup:2018ghi,Abada:2019zxq,Antusch:2016ejd},
new detectors and non-collider experiments ~\cite{Beacham:2019nyx}\footnote{Specific discussions of the neutrino portal also can be found in the SHiP~\cite{Alekhin:2015byh} and MATHUSLA~\cite{Curtin:2018mvb} physics case papers.}
as well as indirect probes~\cite{Gorbunov:2014ypa,Drewes:2016jae,Antusch:2014woa,Lindner:2016bgg,Fernandez-Martinez:2016lgt,Chrzaszcz:2019inj}, cf.~figure~\ref{experiments}.
An interesting question in this context is to what degree experiments would be able to distinguish different underlying models.  
Supposed that one or several new particles are discovered experimentally, a key question will be whether or not these particles are responsible for the origin of neutrino mass. 

\begin{figure*}
\centering
\includegraphics[width=0.8\textwidth]{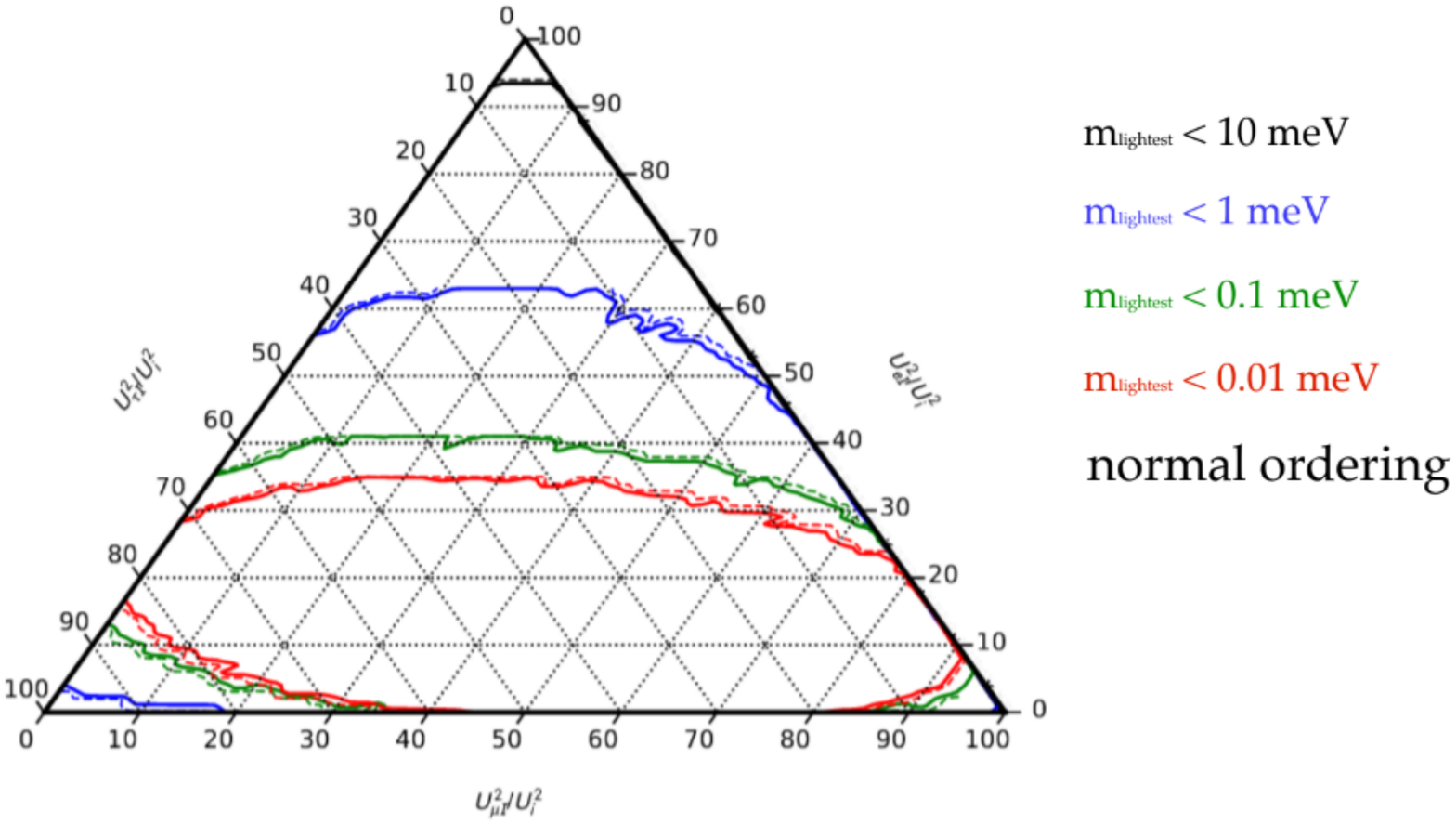}
\caption{\label{triangleplot}
Range of flavour mixing patterns $U_{ai}^2/U_i^2$ expected in the type I seesaw model with $n=3$ for different values of $m_{\rm lightest}$~\cite{Chrzaszcz:2019inj}.}
\end{figure*}

\vskip 2mm
The first question would be whether the properties of these particles are consistent with the hypothesis that they are responsible for neutrino mass generation. On one hand this requires to reproduce the flavour mixing pattern in the light neutrino mixing matrix $V_\nu$. This typically requires the determination of the new particles' couplings to individual SM generations. For instance, in the pure type I scenario \eqref{MinimalSeesaw}, 
this requirement can be turned into predictions for the ratios $U_{ai}^2/U_i^2$ (with 
$U_{ai}^2=|\Theta_{ai}|^2$,
$U_i^2=\sum_a U_{ai}^2$) 
\cite{Hernandez:2016kel,Drewes:2016jae,Caputo:2017pit,Drewes:2018gkc,Chrzaszcz:2019inj}, cf.~figures~\ref{triangleplot} and~\ref{fig:triangle}, and $CP$ properties~\cite{Cvetic:2015naa}.
However, this alone does not prove that the heavy neutrinos are the origin of neutrino mass: in the limit $\epsilon_a, \epsilon'_a, \mu = 0$ the model \eqref{PseudoDiracL} yields vanishing $m_i$, even if the mixings $\theta_{ai}$ are large. 
Observing $L$-violation in the decay of the $N_i$ would be a convincing way to test the Majorana nature of the $N_i$ and their relation to the $m_i$.\footnote{
$L$-violation can only be seen explicitly if the final state is full reconstructable, which is the case for charged current mediated $N_i$ decays into hadrons and a lepton. Indirect probes of $L$ violation include the angular distribution of $N_i$ decay products~\cite{Arbelaez:2017zqq,Balantekin:2018ukw} and using the flavour mixing pattern~\cite{Dib:2016wge}.
} 
In the minimal model \eqref{MinimalSeesaw} this observation is not trivial because the $L$-violating processes are parametrically suppressed by the smallness of $\epsilon_a, \epsilon'_a, \mu$~\cite{Kersten:2007vk,Moffat:2017feq}. However, 
it turns out that coherent oscillations of the $N_i$ within the detector can overcome this suppression~\cite{Boyanovsky:2014una,Cvetic:2015ura,Anamiati:2016uxp}. Requiring the absence of fine-tuning in the generation of light neutrino masses permits to identify the regions in the mass-mixing plane where the $L$-violation is expected to be observable \cite{Drewes:2019byd}, cf.~figure~\ref{experiments}.
This would provide strong evidence that the discovered particles are the origin of neutrino mass.
In the minimal type I seesaw with $n=2$ one can, if nature is kind enough to choose this scenario and picks experimentally accessible parameters, make even stronger statements~\cite{Hernandez:2016kel,Drewes:2016jae}. 
In this particular scenario, in principle all parameters in the Lagrangian \eqref{MinimalSeesaw} can be constrained experimentally. 
To see this it is convenient to consider the Casas-Ibarra parameterisation
$Y = \mathrm{i} U_\nu {\rm diag}(m_1,m_2,m_3)^{1/2}\mathcal{R}{\rm diag}(M_1,M_2)^{1/2}$, with $\mathcal{R}\mathcal{R}^T=1$. The unknown parameters in \eqref{MinimalSeesaw} are $M_1$, $M_2$, a complex mixing angle $\omega$ that parameterises $\mathcal{R}$ and the Dirac and Majorana phase in $U_\nu$.\footnote{
Note that $m_{\rm lightest}=0$ for $n=2$, so that the light neutrino masses $m_i$ are fixed once the mass ordering is determined in oscillation experiments.
} 
In principle all of these can be measured experimentally~\cite{Drewes:2016jae}: The $M_i$ can be measured kinematically, 
the Dirac phase can be measured in neutrino oscillation experiments, 
the Majorana phase can then be extracted from the $U_{ai}^2/U_i^2$~\cite{Drewes:2016jae,Caputo:2016ojx}, 
${\rm Im}\omega$ is given by the overall magnitude of $U_i^2$, 
${\rm Re}\omega$ can be extracted from $U_{a1}^2/U_{a2}^2$. 
In practice, the symmetry ($\epsilon_a, \mu \ll1$) makes it rather difficult to directly measure the mass splitting $|M_1-M_2|$ and ${\rm Re}\omega$. 
However, information about $|M_1-M_2|$ can in principle be extracted by resolving the $N_i$ oscillations in the detector~\cite{Cvetic:2015ura,Antusch:2017ebe,Cvetic:2018elt,Tastet:2019nqj}, cf.~figure~\ref{leptogenesisn2}, 
and neutrinoless double $\beta$-decay is sensitive to ${\rm Re}\omega$~\cite{Asaka:2013jfa,Hernandez:2016kel,Drewes:2016lqo}. 
Though the parameter region where the experimental sensitivity is sufficient to extract meaningful constraints on all parameters is quite limited, this  provides a proof-of-principle that neutrino mass models can be fully testable if one combines data from various experiments, emphasising the importance of the complementarity between the different experimental programs.
In addition, these measurements can also be used to test the hypothesis that the heavy neutrinos are responsible for the origin of matter through low scale leptogenesis~\cite{Hernandez:2016kel,Drewes:2016jae,Antusch:2017pkq}, cf.~figures~\ref{leptogenesisn2} and~\ref{fig:masterplot}.

\begin{figure*}
\centering
\includegraphics[width=0.49\textwidth]{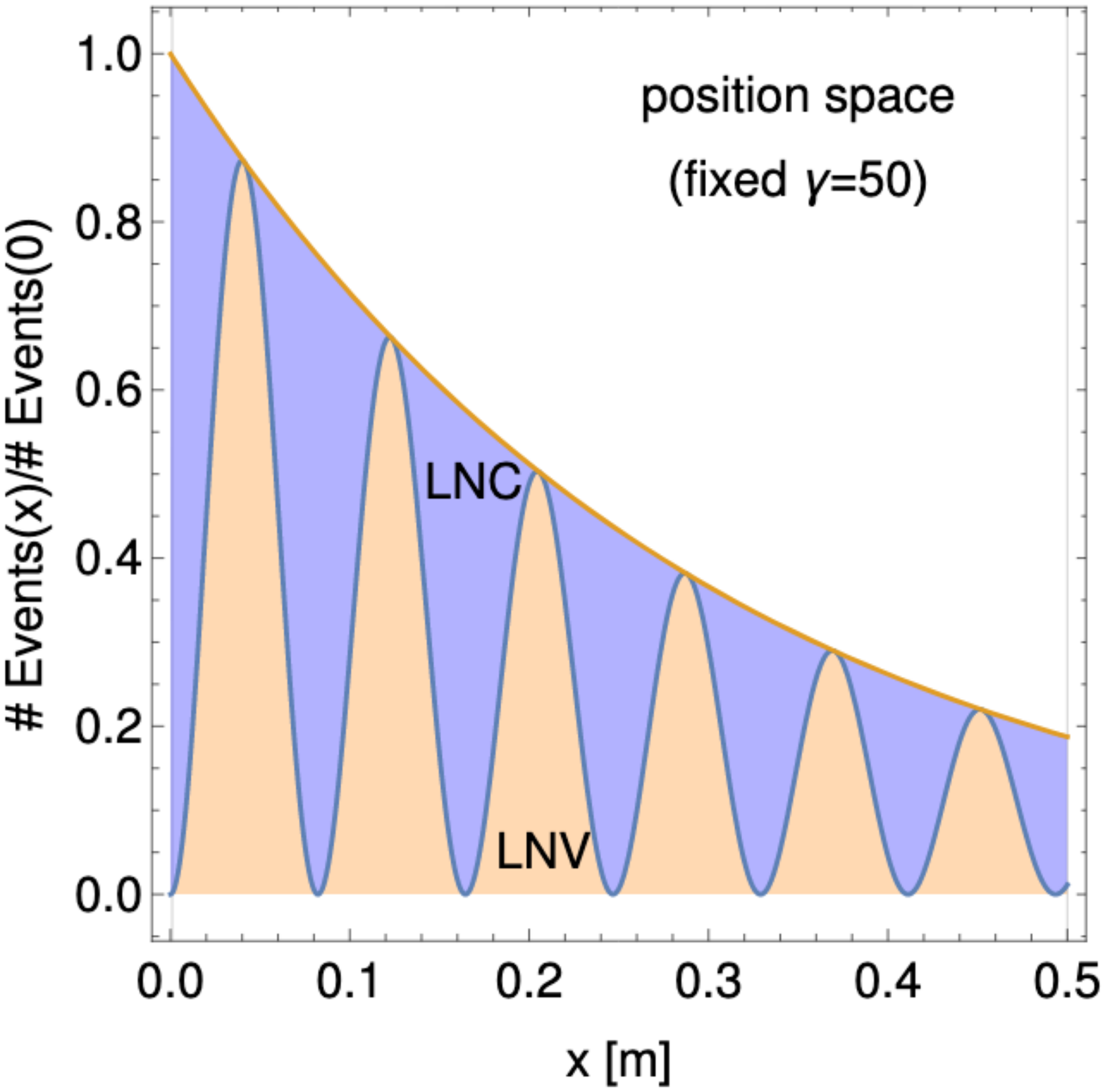}
\includegraphics[width=0.49\textwidth]{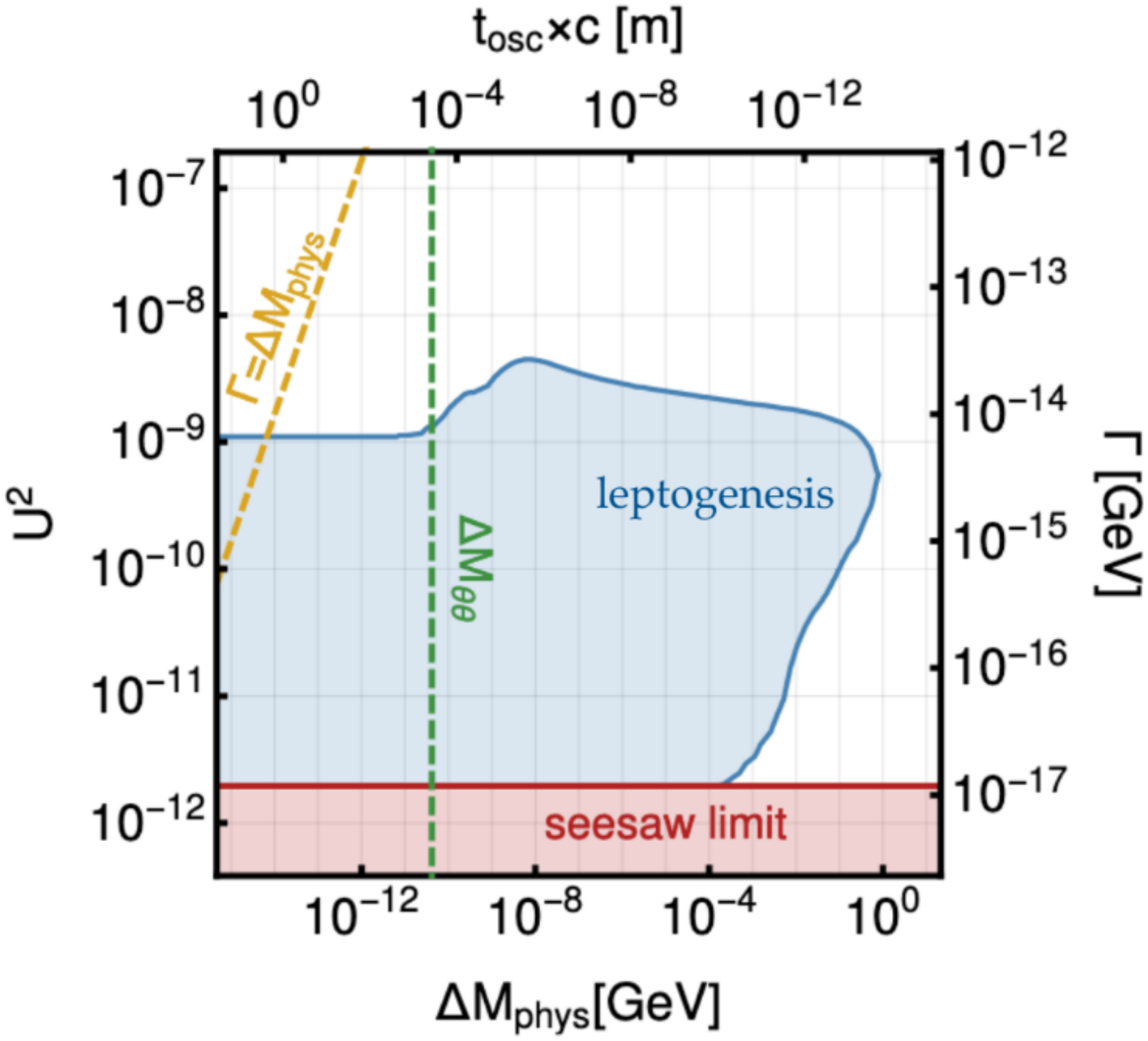}
\caption{\label{leptogenesisn2}
Left: Idealised picture of the ratio of $L$-violating (LNV) and $L$-conserving (LNC) heavy neutrino decays as a function of position. The ratio oscillates due to coherent $N_i$ flavour oscillations~\cite{Antusch:2017ebe}.
Right: Range of physical heave neutrino mass splitting $\Delta M_{\rm phys}$ and mixing $U^2=\sum_iU_i^2$ for which leptogenesis is feasible for $\bar{M}=30$ GeV (blue region), compared to the contribution $\Delta M_{\theta\theta}$ to $\Delta M_{\rm phys}$ from the Higgs mechanism~\cite{Antusch:2017pkq}.
}
\end{figure*}

\vskip 2mm
Non-minimal models tend to offer a tradeoff between discovery potential and testability, as they tend to open up new production channels, but the extended parameter space makes them less predictive. 
For instance, in left-right symmetric models, the exchange of right-handed $W$-bosons can considerably enhance the $N_i$ production cross section at colliders~\cite{Keung:1983uu} (cf.~\cite{Nemevsek:2018bbt} for a recent discussion), but the scalar potential contains a considerable number of parameters and has a complicated vacuum structure~\cite{Dev:2018foq}. Similarly, models with an extended dark sector can lead to a richer phenomenology (cf.~e.g.~\cite{Blennow:2019fhy,Ballett:2019cqp}),
but this also makes it more difficult to exploit the complementarity between different measurements~\cite{deGouvea:2015euy}.

\clearpage
\subsection{HNLs and their relation (or non-relation) to active neutrino physics}
\label{ssec:lopezpavon}
{\it Author: Jacobo Lopez-Pavon, <Jacobo.Lopez@ific.uv.es>} 
\subsubsection{Formalism}

The main goal of this contribution is to discuss the connection between Heavy Neutral Lepton (HNL) physics and the active neutrino sector. As we will see, this connection is always present whenever HNLs account for light neutrino masses and is stronger in the minimal models. 

In order to introduce HNLs in our theory, the Standard Model (SM) field content should be extended with the addition of these new particles (just fermion singlets) and, thus, the most general renormalizable Lagrangian compatible with the SM gauge symmetries that can be built is given by 
   \begin{eqnarray}
{\cal L} = {\cal L}_{SM}- \sum_{\alpha,i} \bar L^\alpha Y^{\alpha i} \tilde\Phi N^i_R - \sum_{i,j} \bar N^{ic}_R M^{ij} N_R^j+ h.c., \nonumber
\label{eq:lag}
\end{eqnarray}
where $Y$ is the $3\times n$ complex Yukawa matrix and $M$ a diagonal $n\times n$ real matrix, with $n$ the number of extra HNLs ($N_R^j$). If we do not add any extra symmetry, the introduction of HNLs implies that a New Physics (NP) scale, given by the mass $M$ of the new states, and Lepton number violation are also introduced via the Majorana mass term in the above Lagrangian. This opens the possibility of generating both the observed light neutrino masses and mixing (via the so called seesaw mechanism~\cite{Minkowski:1977sc,GellMann:1980vs,Yanagida:1979as,Mohapatra:1979ia}
) and the Baryon asymmetry of the Universe (via Leptogenesis~\cite{Fukugita:1986hr}). Indeed, introducing HNLs in our theory automatically generates a contribution to the light neutrino masses given by

\begin{equation}
m_\nu \simeq \frac{v^2}{2}YM^{-1}Y^T=\theta M \theta^T,\nonumber
\label{eq:mnu}
\end{equation}
where $\theta$ is the mixing among active neutrinos and HNLs. Assuming that there is no contribution from another NP sector, the generation of light neutrino masses and mixing imposes the following constraint 

\begin{equation}
m_\nu \simeq \theta M \theta^T=UmU^T,
\label{eq:mnu2}
\end{equation}
between the active (right hand side) and HNL (left hand side) sectors, where $U$ is the PMNS matrix. This can be translated into a constraint on the HNL mixing with the active neutrinos~\cite{Casas:2001sr}

\begin{equation}
\theta\simeq iU m^{1/2}R^\dagger M^{-1/2},
\label{eq:theta}
\end{equation}
where $R$ is a $3\times n$ complex orthogonal matrix associated to the HNL sector. Therefore, the mixing  $\theta$ depends not only on parameters associated to the NP sector but also on the active neutrino parameters contained in $U$ and $m$. In other words, the information about $U$ and $m$ that can be extracted from neutrino oscillation searches and experiments sensitive to the light neutrino mass scale, as neutrinoless double beta decay or cosmological observables like BBN and CMB, could partially determine the HNL mixing with the active neutrinos $\theta$. Notice that the HNL phenomenology is driven by $\theta$, since it essentially represents the coupling to the visible sector, and the HNL masses $M_j$.
\medskip 

\subsubsection{Interplay among cosmological observations, neutrino oscillations and neutrinoless double beta decay searches.}

\medskip 

It is well known that HNLs can have an impact in the early universe and thus constraints on their mass and mixing can be extracted from cosmological observables as BBN and CMB. Interestingly, the lightest active neutrino mass scale $m_{\rm{lightest}}$ plays a very important role in this context if the generic model with $n=3$ HNLs~\cite{Hernandez:2014fha}, where neutrino masses are induced via Eq.~(\ref{eq:mnu2}), is considered. Due to the connection to the active neutrino sector via Eq.~(\ref{eq:theta}), two states always have a large enough mixing $\theta$ that force them to thermalize, while the thermalization of the third state depends
on $m_{\rm{lightest}}$ since there is a lower theoretical bound on the mixing ultimately controlled by this parameter. Two general conclusions can be extracted~\cite{Hernandez:2014fha}: (i) a general bound excluding HNLs masses below $100$ MeV applies to the three states provided $m_{\rm{lightest}}\gtrsim 10^{-3}$~eV; (ii) if $m_{\rm{lightest}}\lesssim10^{-3}$~eV, the same bound still applies for two HNLs but the third state can still have a mass below $100$ MeV depending on the particular value of its corresponding mixing. 

\begin{figure}
\begin{center}
\includegraphics[width=\columnwidth]{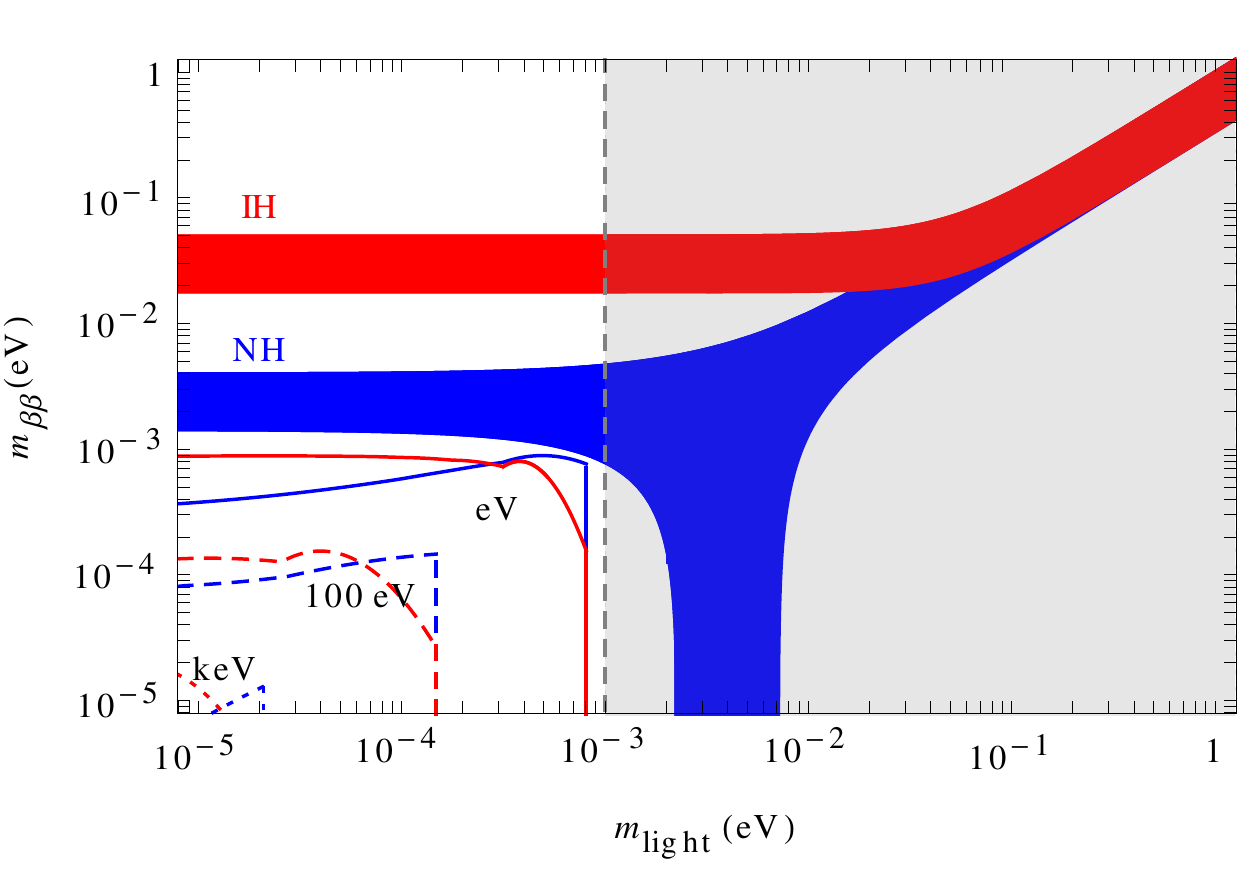}
\caption{\label{fig:nuless} 
 The read and blue regions are the allowed contributions from the active neutrinos dictated by neutrino oscillation data. The maximum contribution from HNLs with $M_1=1$~eV (solid), $100$~eV (dashed), $1$~KeV (dotted) for Normal Hierarchy (NH, blue) and Inverted Hierarchy (IH, red) and $M_{2,3}\gg100$ MeV is also shown. The shaded region is ruled out by the thermalization bound on the lightest neutrino mass $m_{lightest}\leq 10^{-3}$~eV. See~\cite{Hernandez:2014fha} for details.}
\end{center}
\end{figure}

\vskip 2mm
The information from cosmology is highly complementary to that coming from neutrinoless double beta decay ($0\nu\beta\beta$) and neutrino oscillation searches. In particular, HNLs can mediate $0\nu\beta\beta$ decay and give a sizeable contribution (analogous to the long range active neutrino one) for mass scales below the typical momentum exchange of the process ($\lesssim 100$ MeV). That region is precisely subject to strong constraints from cosmology as it is shown in Fig.~\ref{fig:nuless}. As it can be observed in the figure, the HNL contribution to the $0\nu\beta\beta$ rate (encoded in the effective mass $m_{\beta\beta}$) becomes subleading as compared with the active neutrino one, once the information from cosmology and neutrino oscillations is included. There is still room for a sizeable HNL contribution to this process for masses in the range $100\, \rm{MeV}\lesssim M_j \lesssim 10 \,\rm{GeV}$, and even for heavier scales if a fine tuned cancellation among the tree and one loop induced contributions to the light neutrino masses is invoked~\cite{Mitra:2011qr,LopezPavon:2012zg,Lopez-Pavon:2015cga,Bolton:2019pcu}. 
\medskip 

\subsubsection{Direct searches and flavor structure in minimal model}

\begin{figure}[t!]
\begin{center}
\includegraphics[width=\columnwidth]{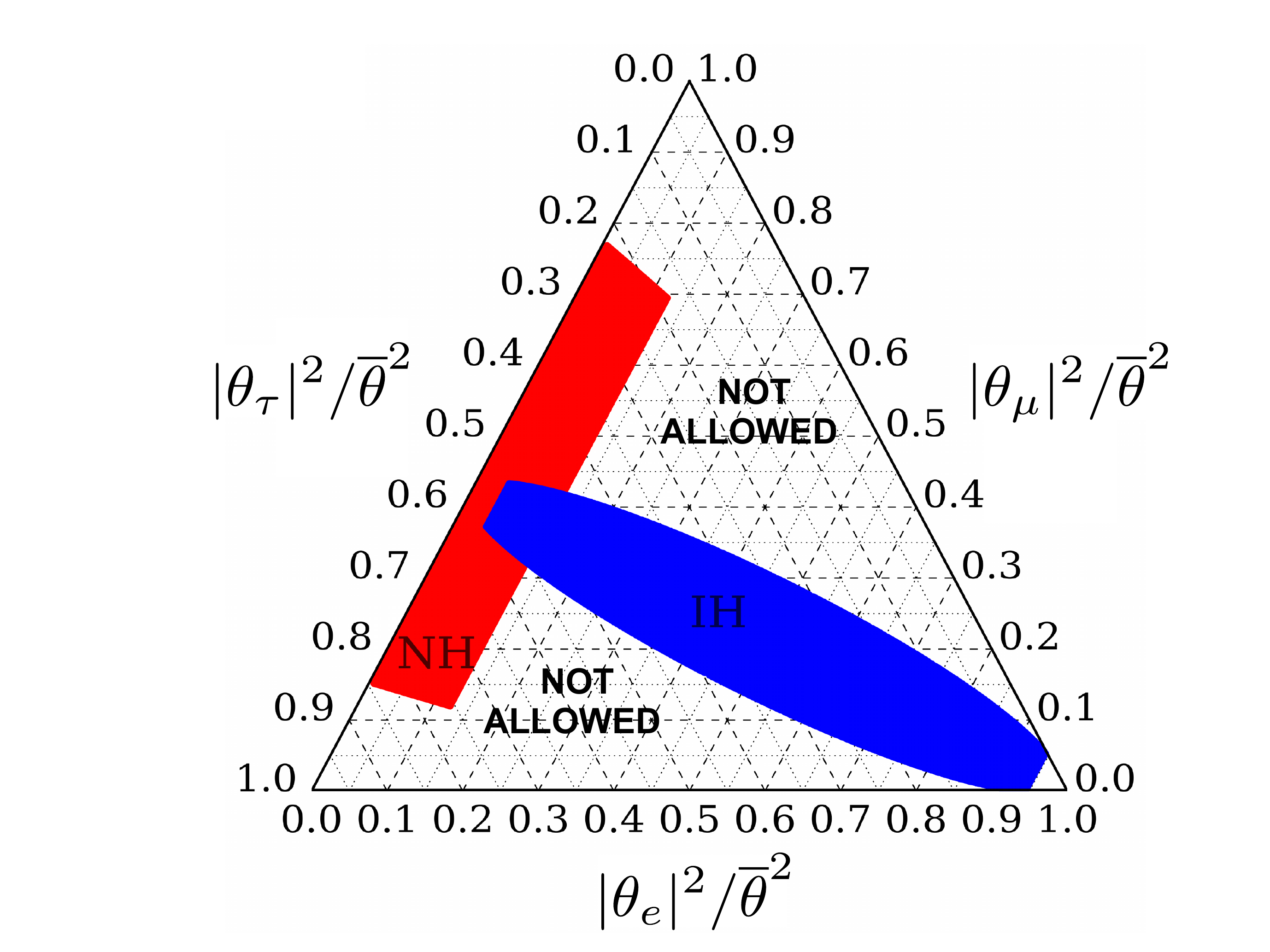}
\caption{\label{fig:triangle} Ternary diagram for the normalized flavour mixings $|\theta_{\alpha h}|^2/\bar{\theta}^2$
(in the large mixing regime)~\cite{Caputo:2017pit}  with $\bar{\theta}^2\equiv\sum_{\alpha} |\theta_{\alpha h}|^2$, fixing the known oscillation parameters to their best 
fit values~\cite{Esteban:2016qun} and varying the CP phases from $\left[0,2\pi\right]$ for NH (red) and IH (blue) in the minimal seesaw model.}
\end{center}
\end{figure}

In the minimal model with $n=2$ HNLs, the correlation between the NP and active neutrino sectors given by Eq.~(\ref{eq:theta}) becomes particularly strong, leading to a quite constrained flavour structure of the heavy mixing $\theta$, as 
shown in Fig.~\ref{fig:triangle}. This is realized for larger mixings than the naive seesaw scaling  
$|\theta_{\alpha h}|^2>\mathcal{O}\left(m_\nu/M_h\right)$, the region of the parameter space that can be directly probed by near future beam dump or collider experiments for HNL scales in the range $\mathcal{O}\left(0.1-100\,\rm{GeV}\right)$. This large mixing regime 
can be realized in a 
technically natural way for symmetry protected scenarios as the inverse or direct seesaws~\cite{Mohapatra:1986aw,Mohapatra:1986bd,Bernabeu:1987gr,Branco:1988ex,Malinsky:2005bi,Gavela:2009cd}. 
Given the constrained heavy mixing flavour structure, a measurement of the flavour ratios compatible with the coloured regions in
Fig~\ref{fig:triangle} would provide strong indication of the connection between light neutrino masses and the observed HNLs. Notice that neutrino oscillation experiments will surely provide a measurement of the light neutrino hierarchy (NH or IH), the octant and the Dirac CP phase, which will significantly improve the predictivity of the minimal model, considerably reducing the allowed regions in~Fig.~\ref{fig:triangle}.

Furthermore, the flavour ratios basically only depend on the 
phases of the PMNS matrix~\cite{Hernandez:2016kel,Caputo:2016ojx} and, therefore, a precise experimental determination would also provide an indirect
way to determine the CP phases in the leptonic sector~\cite{Caputo:2016ojx}. This is an extremely interesting feature of the minimal model since future facilities will open a new window for leptonic CP violation with sensitivity not only to the 
Dirac CP phase $\delta$, accessible in neutrino oscillation experiments, but also to the Majorana CP phase whose measurement is very challenging. Further details on the future prospects for the measurement of the PMNS CP phases in direct HNL searches can be found in~\cite{Caputo:2016ojx}.

\subsubsection{Impact of active neutrino measurements in Leptogenesis}

Leptonic CP violation is required in order to generate the observed Baryon asymmetry of the universe via Leptogenesis. This CP violation can come from the active neutrino sector or/and the HNL sector. Regardless the generic requirement that CP needs to be violated, the PMNS Dirac CP phase that can be measured in near future neutrino oscillation experiments becomes particularly relevant in two possible general scenarios: (i) Flavor Models in which the structure of the Yukawa couplings (and thus the $R$ matrix) is theoretically constrained by flavor symmetries (see for instance~\cite{Hagedorn:2017wjy,Merlo:2018rin}); (ii) Most minimal version of the low scale type-I seesaw, including just two HNLs, where the total number of CP phases is reduced to three (two PMNS phases and one associated to the HNL sector) and the lightest neutrino mass is zero. 

\begin{figure}[t!]
 \begin{center}
 \includegraphics[width=\columnwidth]{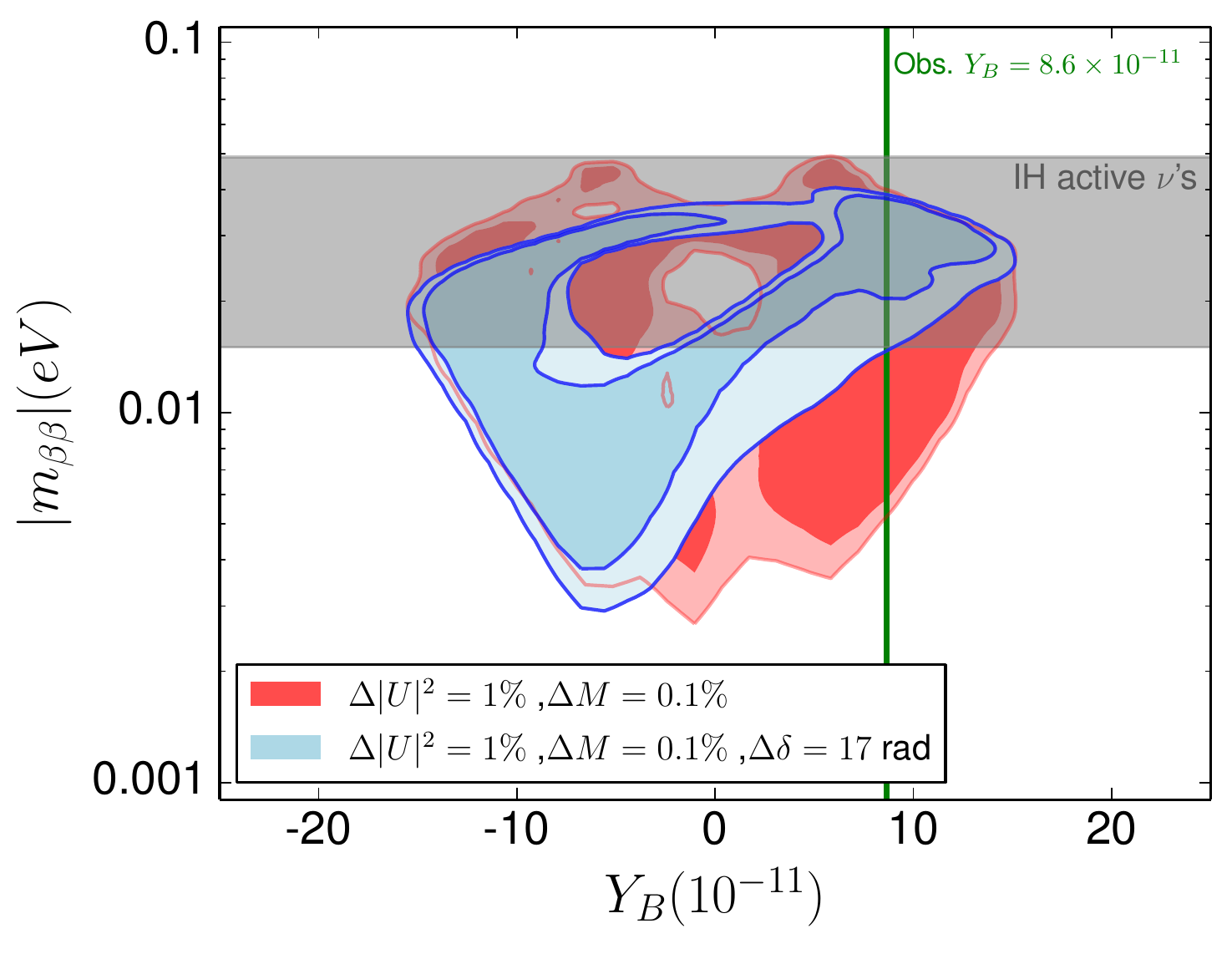} 
\caption{\label{fig:masterplot} Posterior probabilities in the $|m_{\beta\beta}|$ vs $Y_B$ plane from a putative measurement at SHiP (red), assuming
$0.1\%, 1\%$ uncertainty on the measurement of the HNL masses and mixing respectively, and the additional putative measurement of $\delta$ by T2HK or 
DUNE (blue). See~\cite{Hernandez:2016kel} for further details.}
\end{center}
\end{figure}

In order to briefly discuss how the combined information from different observables can potentially allow us to test low scale leptogenesis scenarios~\cite{Akhmedov:1998qx,Asaka:2005pn,Shaposhnikov:2008pf} here we will focus on scenario (ii). This is illustrated in Fig.~\ref{fig:masterplot}, where a
putative measurement at SHiP (red region) associated to the parameters of the model denoted by the star in Fig.~7 of ref.~\cite{Hernandez:2016kel} is assumed, showing the correlation
between the corresponding effective neutrinoless double beta decay mass $m_{\beta\beta}$ (which includes the contribution of both light and heavy states) 
and Baryon asymmetry $Y_B$ generated in the model. In the blue regions a putative measurement of $\delta$ from future neutrino
oscillation experiments is also included. Taking into account that future neutrinoless double beta decay experiments are expected to be sensitive to 
effective neutrino mass at the level of $10^{-2}$~eV, Fig.~\ref{fig:masterplot} opens two possible future scenarios for the case under consideration: (i) if no neutrinoless double beta decay
signal is observed, accommodating the observed baryon asymmetry in the minimal model would be experimentally disfavored; (ii) if a  
signal is instead found, the baryon asymmetry generated could be predicted up to a sign, being compatible with the observed value. 

\subsubsection{Deviations from minimal model predictions}

The results presented here regarding the constrained flavour structure shown in Fig.~\ref{fig:triangle}, the indirect observation of PMNS leptonic CP violation and the predictivity for the Baryon asymmetry generated via low scale Leptogenesis, rely to a very large
extent on the minimality of the model under consideration. 
An interesting question arising in this context is to what extent can these predictions be modified in the presence of additional NP. This 
question was addressed in~\cite{Caputo:2017pit} 
where the impact of generic heavier NP 
was studied using the effective field theory approach~\cite{Graesser:2007yj,delAguila:2008ir,Aparici:2009fh}. 
In~\cite{Caputo:2017pit} it was found that the deviations with respect to the flavor structure given in Fig.~\ref{fig:triangle} are driven by $m_{\rm{lightest}}$. Significant deviations can be expected for 
$m_{\rm{lightest}}\gtrsim \sqrt{\Delta m^2_{\rm{solar}}}$ ($m_{\rm{lightest}}\gtrsim \sqrt{\Delta m^2_{\rm{atm}}}$) for IH (NH). Measuring $m_{\rm{lightest}}$ above this value would surely imply that the minimal model predictions are no longer applicable. This has been confirmed in~\cite{Chrzaszcz:2019inj} for the particular case of the next to minimal model with $n=3$ HNLs. 

The extra NP could also come from low energies implying the existence of new light particles and interactions. In such a case, the light neutrino mass generation would still impose theoretical constraints but the minimal model predictions would no longer be valid. In particular, the new sector can have a very relevant impact in the early universe and leptogenesis~\cite{Caputo:2018zky}. 

\subsubsection{Conclusions}
In summary, from a model independent perspective, the HNL masses and mixing with different neutrino flavors can be considered as independent free parameters. However, the light neutrino mass generation imposes theoretical constraints introducing a link between the HNL and active neutrino sectors that could be considered as a sort of guidance regarding near future experimental searches. Complementarity among different observables is a key issue in this context.

\clearpage
\subsection{The Majoron and its connection to Majorana neutrinos}
\label{ssec:heeck}
{\it Author: Julian Heeck, <heeck@virginia.edu>} \\ 

As the Goldstone boson associated with lepton number, the Majoron is intimately connected and coupled to Majorana neutrinos. At loop level it receives couplings to all other Standard Model particles, including flavor-changing lepton and quark couplings. This makes rare decays a prime hunting ground for these axion-like particles that could ultimately help us to understand the seesaw mechanism.

\vskip 2mm
The Majoron $J$ is the Goldstone boson of lepton number~\cite{Chikashige:1980ui,Schechter:1981cv}; in the simplest realization, it resides in a singlet complex scalar $\sigma = (f+ \sigma^0 + i J)/\sqrt{2}$ that carries $L=-2$, $f$ being the lepton-number breaking scale. Further introducing three right-handed neutrinos~$N_R$, the Lagrangian reads
\begin{eqnarray}
	\mathcal{L} &=& \mathcal{L}_\text{SM}+ i\overline{N}_R \gamma^\mu\partial_\mu N_R + (\partial_\mu \sigma)^\dagger (\partial^\mu \sigma) \nonumber \\
	&-& V(\sigma)-\left(\overline{L} y N_R H +\tfrac12\overline{N}_R^c\lambda  N_R \sigma +\text{h.c.}\right) .
	\label{eq:lagrangian}
\end{eqnarray} 
Symmetry breaking then yields the famous seesaw neutrino mass matrix $M_\nu \simeq - M_D M_R^{-1} M_D^T$ with $M_D = y v/\sqrt{2}$ and $M_R = \lambda f/\sqrt{2} \gg M_D$.
In addition to explaining neutrino masses, the Majoron is itself an interesting particle and can be viewed as a well-motivated simple renormalizable realization of an axion-like particle. 

\vskip 2mm
The tree-level couplings of $J$ include the couplings $J \overline{\nu}_j i \gamma_5 \nu_j m_j/(2f)$ to the light neutrino mass eigenstates $\nu_j$, severely suppressed by the tiny neutrino masses. At one-loop level~\cite{Chikashige:1980ui,Pilaftsis:1993af,Garcia-Cely:2017oco}, the Majoron also obtains couplings to charged leptons $\ell$ and quarks~$q$, parametrized as $i J \bar{f}_1 (g_{Jf_1 f_2}^S+ g_{Jf_1 f_2}^P \gamma_5) f_2$ with coefficients
\begin{align}
	g_{J q q'}^P &\simeq \frac{m_q }{8\pi^2 v}\delta_{q q'} T^q_3 \, \text{tr} K\,, \\
	g_{J q q'}^S &= 0 \,,\\
	g_{J\ell{\ell'}}^P &\simeq \frac{m_\ell+m_{\ell'}}{16\pi^2 v} \left(\delta_{\ell \ell'} T^\ell_3\, \text{tr} K +  K_{\ell\ell'} \right), \\
	g_{J\ell{\ell'}}^S &\simeq \frac{m_{\ell'}-m_\ell}{16\pi^2 v}  K_{\ell\ell'} \,,
	\label{eq:fermion_couplings}
\end{align}
where $T^{d,\ell}_3= - T^u_3 = -1/2$ and  we introduced  the dimensionless Hermitian coupling matrix $K \equiv  M_D M_D^\dagger/(v f)$. The Majoron couplings to charged fermions are hence determined by the seesaw parameters $M_D M_D^\dagger$, which are \emph{independent} of the neutrino masses~\cite{Davidson:2001zk} and can in particular be much bigger than the naive one-generation expectation $M_\nu M_R$. Perturbativity sets an upper bound on $K$ of order $4\pi v/f$, and since $K$ is furthermore positive definite we have the inequalities $|K_{\ell\ell'}|\leq \text{tr} K/2$. The couplings to charged fermions are obviously crucial for phenomenology and in principle even offer a new avenue to reconstruct the seesaw parameters. Note in particular the off-diagonal lepton couplings, which will lead to lepton flavor violation, e.g.~$\ell\to\ell' J$~\cite{Pilaftsis:1993af,Garcia-Cely:2017oco}.
At two-loop level~\cite{Heeck:2019guh} $J$ further receives off-diagonal couplings to quarks, inducing meson decays such as $K\to\pi J$. If $J$ is a pseudo-Goldstone with mass $m_J$ it also inherits a coupling to two photons at the two-loop level, severely suppressed compared to other axion-like particles.

\begin{figure*}[t]
\center
	\includegraphics[width=0.7\textwidth]{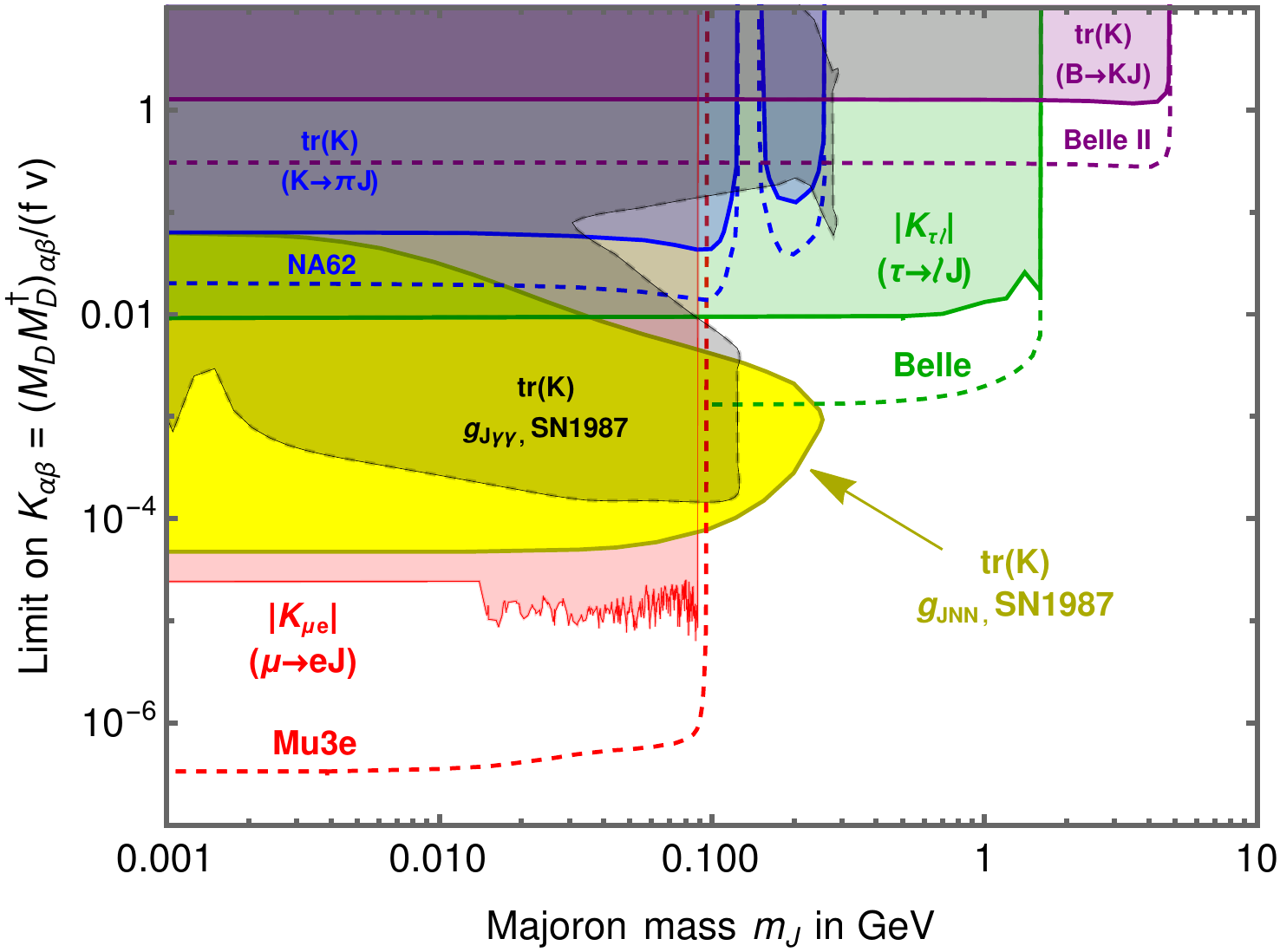}
	\caption{
		Upper limits on combinations of $K_{\alpha\beta}= (M_D M_D^\dagger)_{\alpha\beta}/(v f) $ for Majoron masses above MeV. The shaded regions exclude $|K_{\alpha\beta}|$ or $\text{tr} (K)$ by non-observed rare decays, the dashed lines show the potential future reach; see Ref.~\cite{Heeck:2019guh} for details. $K$ is further subject to the inequalities $|K_{\alpha\beta} | < \text{tr} (K)/2$.
}
	\label{fig:light_Majoron_constraints}
\end{figure*}

\vskip 2mm
Having established all relevant Majoron couplings we can study the phenomenology. The couplings to nucleons, photons, and electrons are subject to astrophysical bounds, for example from stellar cooling or SN1987.
The \emph{off-diagonal} Majoron couplings to leptons and quarks offer more compelling signatures in the form of rare decays $\ell\to\ell' J$ and $K^+\to\pi^+ J$. In this region of parameter space $J$ is long-lived and thus leaves the detector as missing energy, forcing us to look exclusively at the monoenergetic charged decay partner $\ell'$ or $\pi^+$ on top of the continuous Standard Model background.\footnote{Channels with more tagging potential, e.g.~$\ell \to \ell' J, J\to \text{visible}$, can improve sensitivity for other models~\cite{Heeck:2017xmg}.}
Current limits for the rare lepton decays translate into $|K_{\mu e}|\lesssim 10^{-5}$, $|K_{\tau \ell}|\lesssim \mathcal{O}(10^{-3})$, with great prospects for improvement at Mu3e and Belle~\cite{Heeck:2016xkh,Calibbi:2020jvd}, see Fig.~\ref{fig:light_Majoron_constraints}.
We stress that lepton flavor violation with Majorons depends on a different combination of seesaw parameters than the more commonly studied heavy-neutrino induced $\ell\to \ell' \gamma$ and are hence complementary~\cite{Heeck:2019guh}. 

\vskip 2mm
Despite the additional suppression by loops and CKM matrix elements, rare meson decays such as $K^+\to\pi^+ J$ also provide relevant constraints on the Majoron parameter space, see Fig.~\ref{fig:light_Majoron_constraints}. It is important to note that these probe different $K$ entries and should not be compared directly to the stronger constraints from rare lepton decays, but rather through the inequalities $|K_{\alpha\beta} | < \text{tr} (K)/2$.
Future improvements in NA62 ($K^+\to\pi^+ J$) and Belle II ($B\to K J$) are very promising as well~\cite{Gavela:2019wzg}.

\vskip 2mm
In summary, the singlet Majoron model inherits some nice properties from the seesaw Lagrangian, namely small Majorana neutrino masses and leptogenesis, while providing a new phenomenological handle. The loop-induced Majoron couplings to charged particles are precisely given by the seesaw parameters that are impossible to determine from the neutrino mass matrix, which could in principle allow us to reconstruct the seesaw with low-energy measurements. Rare decays in both the lepton and quark sector are crucial to explore the parameter space of this well-motivated axion-like particle.

\clearpage
\subsection{Search for HNLs at extracted beams}
\label{ssec:izmaylov}
{\it Author: Alexander Izmaylov, <izmaylov@inr.ru>} 
\subsubsection{Overview}
\label{sssec:overview}
Among different methods to probe a possible existence of Heavy Neutral Leptons (HNLs, heavy neutrinos), experiments at extracted beams play an important role. Observable signatures can arise in a range of current and upcoming experiments, in this summary we focus mainly on neutrino accelerator projects. In this case an intense proton beam hits a target to produce daughter particles which subsequent decays give rise to a neutrino beam. Decays of pion, kaon and heavier mesons can potentially also lead to a production  of HNLs that would then travel along the neutrino beam. Their further decays thus can be searched for with a detector,  usually a near detector located close $\mathcal{O}(100)$ m to a beam-producing target and originally serving to measure SM (Standard Model) neutrino flux and SM neutrino interactions with matter. Modified flux and kinematics of HNLs with respect to SM neutrinos can be used to separate signal from backgrounds, which in this case are mainly composed from SM neutrino interactions that can mimic HNLs signatures. Current neutrino projects are usually able to probe HNLs from light mesons $\mathcal{O}(100)$ MeV, although future experiments can potentially extend the analysis to larger masses. 

In this summary the experiments for rare meson decays studies are also included with the focus on NA62 project which leads the field in the kaon sector. In this case a modified kinematics of meson decays with respect to SM processes can be used to directly probe a potential HNL production. 

\subsubsection{Current neutrino experiments}

\begin{itemize}
    \item \textbf{T2K}: the Tokai-to-Kamioka~\cite{Abe:2011ks} is a long-baseline neutrino experiment located in Japan with the primary goal of measuring muon (anti-)neutrino oscillations using Super-Kamiokande as its far detector. The T2K neutrino beam is produced at the J-PARC center by colliding 30~GeV protons on a graphite target. The pions and kaons produced are focused and selected by charge with magnetic horns and subsequently decay in flight to neutrinos. Depending on the polarity of the current in the horns, the experiment can be run either in neutrino or anti-neutrino mode. Search for heavy neutrinos is performed by means of the near detector complex ND280. Located 280 metres from  the  proton  target, the complex is  composed  of  several sub-detectors with a 0.2~T magnet. Analysis is performed with the central tracker consisting of three time projection chambers (TPCs) and two scintillator-based fine-grained detectors (FGDs)  surrounded by an electromagnetic calorimeter (ECal).  T2K published results~\cite{Abe:2019kgx} on the search for heavy neutrinos from kaon parents ( $K^{\pm}\to \ell^{\pm}_{\alpha}N, \alpha = e,\mu$). All the possible decay modes $N\to\ell^{\pm}\pi^{\mp}$ and $N\to\ell^{\pm}\ell^{\mp}\overline{\nu}$ were considered. The neutral current decay modes $N \to e^+ e^- \overline{\nu}_{\tau}$ and $N \to \mu^+ \mu^- \overline{\nu}_{\tau}$ are directly sensitive to the mixing element $U_{\tau}^2$. Production and decay branching ratios were extracted from~\cite{Gorbunov:2007ak}. Effects related to heavy neutrino polarization~\cite{Levy:2018dns} and delayed arrival time (with respect to light neutrinos) are taken into account.
    
    \vskip 2mm
    The selection was developed to isolate the signal events from the background expected from SM neutrino interactions with matter. In order to significantly improve the signal to background ratio, which is inversely proportional to the density of the medium, only events occurring in the TPC gas volume are considered for this analysis. Events are preselected by identifying two tracks of opposite charge originating from a vertex in a TPC, energy loss in a TPC is used for particle identification, then a number of selection criteria is applied on the kinematics of the two tracks, ECal information is used in some cases. Upstream detector activity is also vetoed. Five signal channels are then identified: $\mu^{\pm}\pi^{\mp}$, $e^- \pi^+$, $e^+ \pi^-$, $e^+ e^-$, $\mu^+ \mu^-$. 
    
    \vskip 2mm
    One of the dominant background contributions is then neutrino-induced coherent pion production on argon nuclei in the TPC gas. Additional background sources include other types of neutrino interactions in the gas and interactions outside the gas, e.g. photon conversion.
    
    \vskip 2mm
    In this analysis, the production of heavy neutrinos from kaon decays in data taken from November 2010 to May 2017 are considered (2:1 of neutrino to anti-neutrino mode running). No events are observed in the defined signal regions, which is consistent with the background-only hypothesis. Several approaches are applied to obtain final limits on the $U_{\alpha}$ mixing elements. First, each heavy neutrino production/decay mode is considered independently and the corresponding analysis channel is used to put limits on the associated mixing elements. A “combined” approach is also applied, in which all the heavy neutrino production and decay modes and the ten different analysis channels (five for each beam mode) are considered simultaneously. The results are shown in Fig.~\ref{fig:T2K_HNL_results}. 
    
    \begin{figure*}[h]
            \centering
            \includegraphics[width=0.55\textwidth]{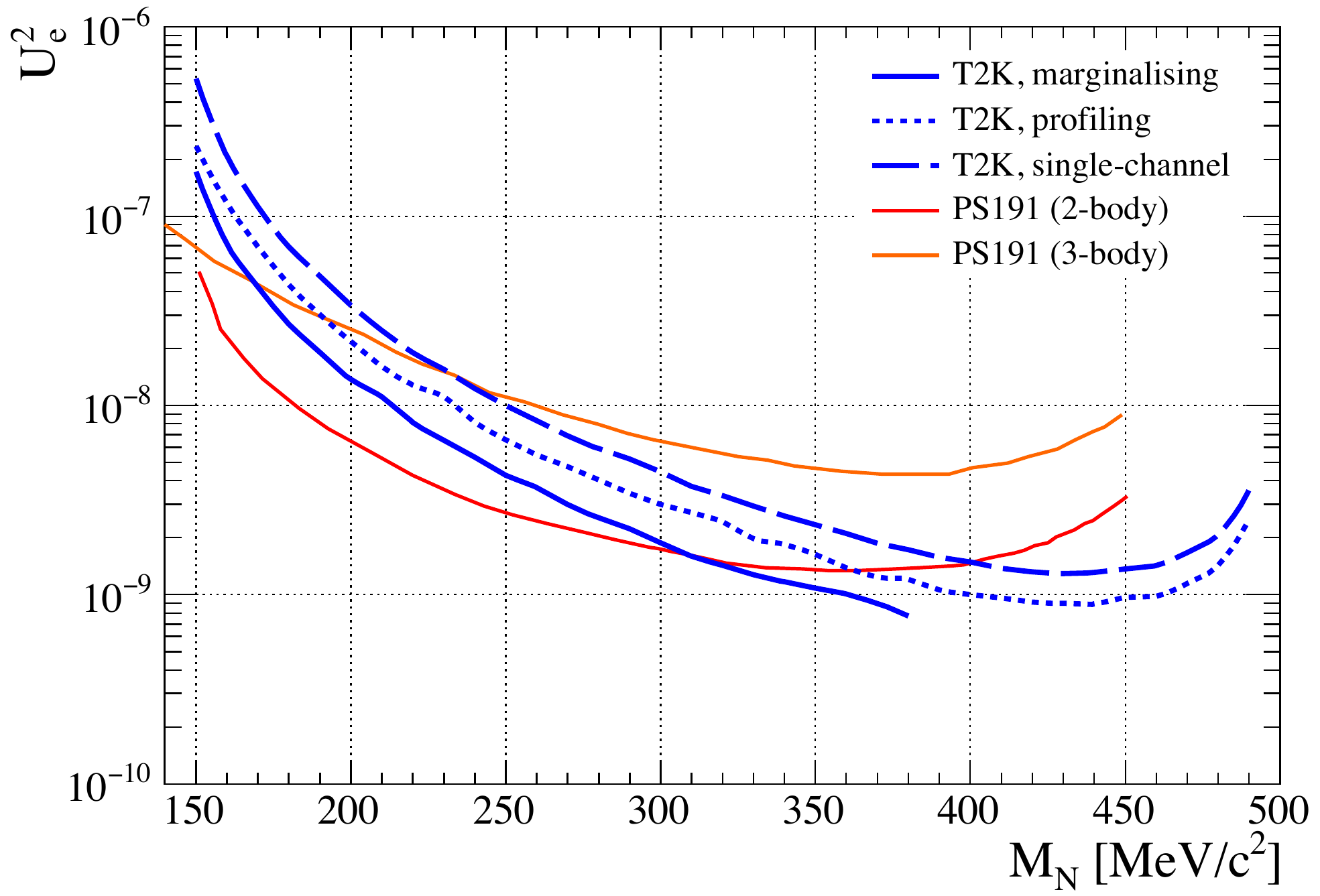}
            \includegraphics[width=0.55\textwidth]{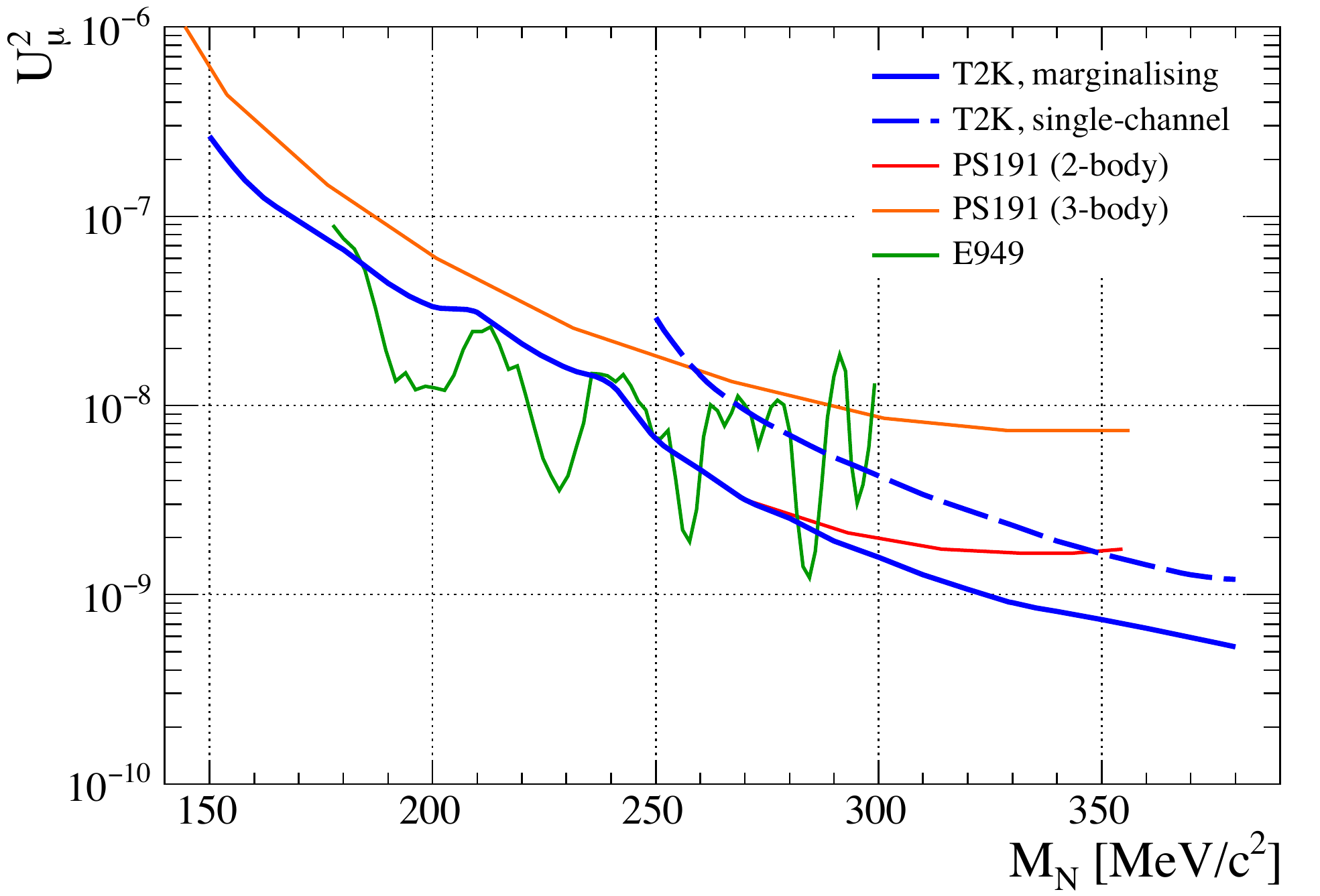}   
            \includegraphics[width=0.55\textwidth]{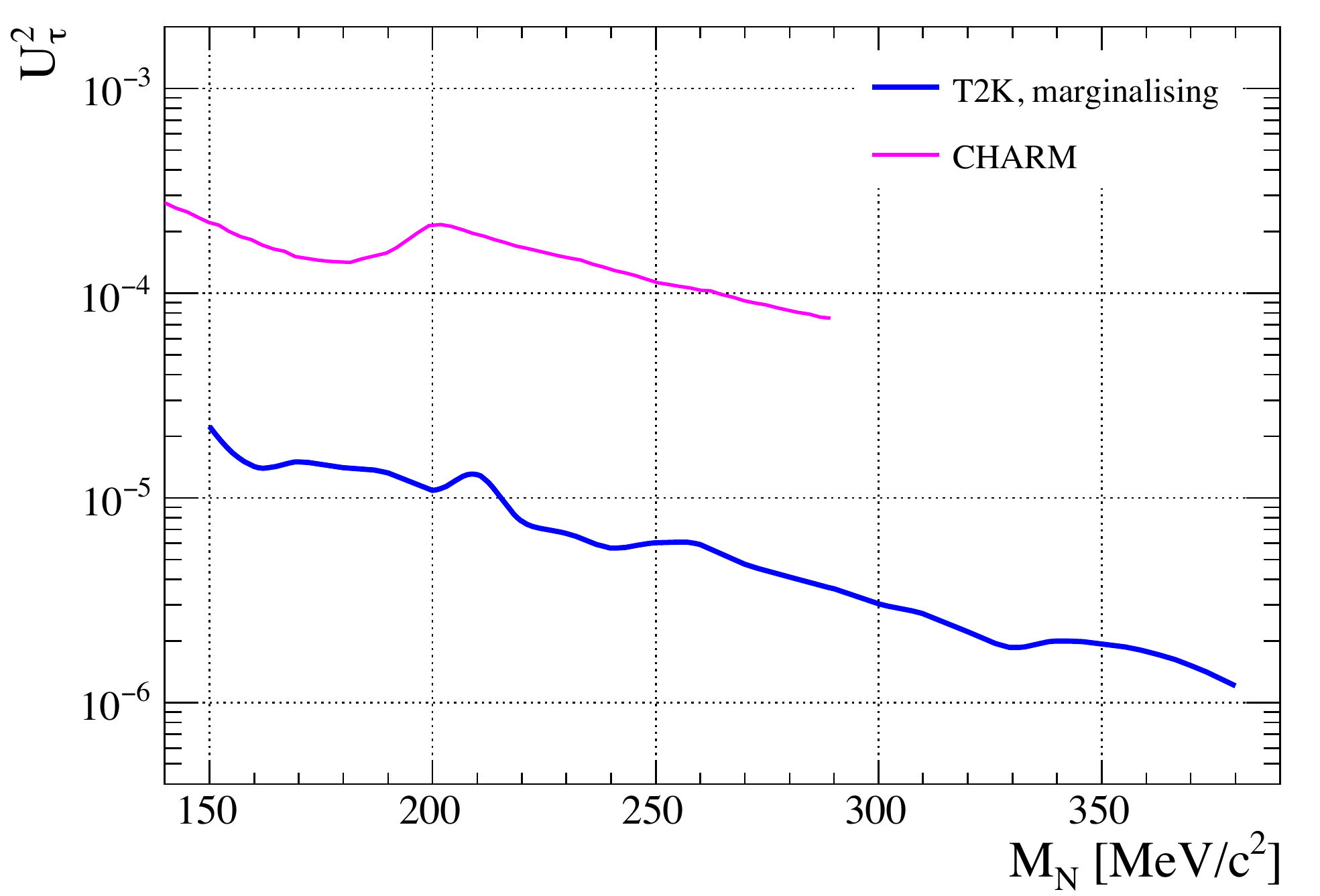}
            \caption{90\% upper limits on the mixing elements $U_e^2$ (top), $U_{\mu}^2$ (middle), $U_{\tau}^2$ (bottom) as a function of heavy neutrino mass, obtained with the combined approach. The blue dashed lines corresponds to the results of the single-channel approach . The blue solid lines are obtained after marginalisation over the two other mixing elements. In the top plot, the additional blue dotted line corresponds to the case where profiling is used ($U_{\mu}^2=U_{\tau}^2=0$). The limits are compared to the ones of other experiments: PS191~\cite{Bernardi:1985ny,Bernardi:1987ek},
            E949~\cite{Artamonov:2014urb},
            T2K~\cite{extracted_beams:T2K_HNL_results},
            CHARM~\cite{extracted_beams:CHARM}.} 
            \label{fig:T2K_HNL_results}
    \end{figure*}

    Results apply to any model with heavy neutrinos with masses between $140$\,MeV/c$^2$ and $493$\,MeV/c$^2$.  The limits are competitive with those of previous experiments such as PS191~\cite{Bernardi:1985ny,Bernardi:1987ek}, E949~\cite{Artamonov:2014urb} and CHARM~\cite{Orloff:2002de}, especially in the high-mass region (above $300$ MeV/c$^2$).
    
    \item \textbf{MicroBooNE}: the MicroBooNE detector~\cite{Acciarri:2016smi} started  data taking with the Booster Neutrino Beam (BNB, 8 GeV proton beam, Fermilab, USA) in 2015, making it the first fully operational detector of the three liquid-Argon time projection chambers comprising the Short-Baseline Neutrino Program (SBN)~\cite{Machado:2019oxb}. The SBN program aims to address the short-baseline anomalies observed by the MiniBooNE and LSND collaborations. In addition to the studies of the effects of light eV-scale sterile neutrinos, the energy range of the BNB makes possible to extend the sensitivity of the detectors to the production and decay of HNLs with masses of $\mathcal{O}(100)$ MeV.  So far the analysis only considers HNLs produced with $K^+\to\mu^+N$ mode and which decay into $\mu\pi$ pairs (no sensitivity to charge signs). 
    
    The detector is located 463~m downstream from the neutrino production target and due to their mass, some of the HNLs are expected to arrive late compared to the arrival of the BNB spill.  Data collected with a dedicated HNL trigger is used to suppress background from SM neutrino interactions.  The trigger was commissioned in 2017 and is used to search for late signatures occurring after the arrival of the SM neutrino beam spill. A two-track vertex is required inside the MicroBooNE TPC. Further pre-selection criteria is applied on the two tracks: fiducial volume, containment, kinematics. Finally a boosted decision tree (BDT) is trained on a set of kinematic variables to discriminate between signal candidates and backgrounds, a BDT is also validated with SM data. For background studies ``off-beam'' data sets are used which contain mainly cosmic rays and no SM neutrino interactions.
    
    \vskip 2mm
     Analysis based on data taken in 2017 and 2018 revealed no data excess in the signal region with high BDT scores. The corresponding upper limits at the 90$\%$ CL on the element $|U_{\mu 4}|^2$ of the extended PMNS mixing matrix in the range $|U_{\mu 4}|^2 < (6.6-0.9) \times  10^{-7}$ for Dirac HNLs and $|U_{\mu 4}|^2 < (4.7-0.7) \times  10^{-7}$ for Majorana HNLs, assuming HNL masses between 260 and 385 MeV and $|U_{e 4}|^2=|U_{\tau 4}|^2=0$, are obtained~\cite{Abratenko:2019kez}  and shown in Fig.~\ref{fig:MicroBooNE_HNL_results}. 
    \begin{figure*}[h]
        \centering
        \includegraphics[width=0.7\textwidth]{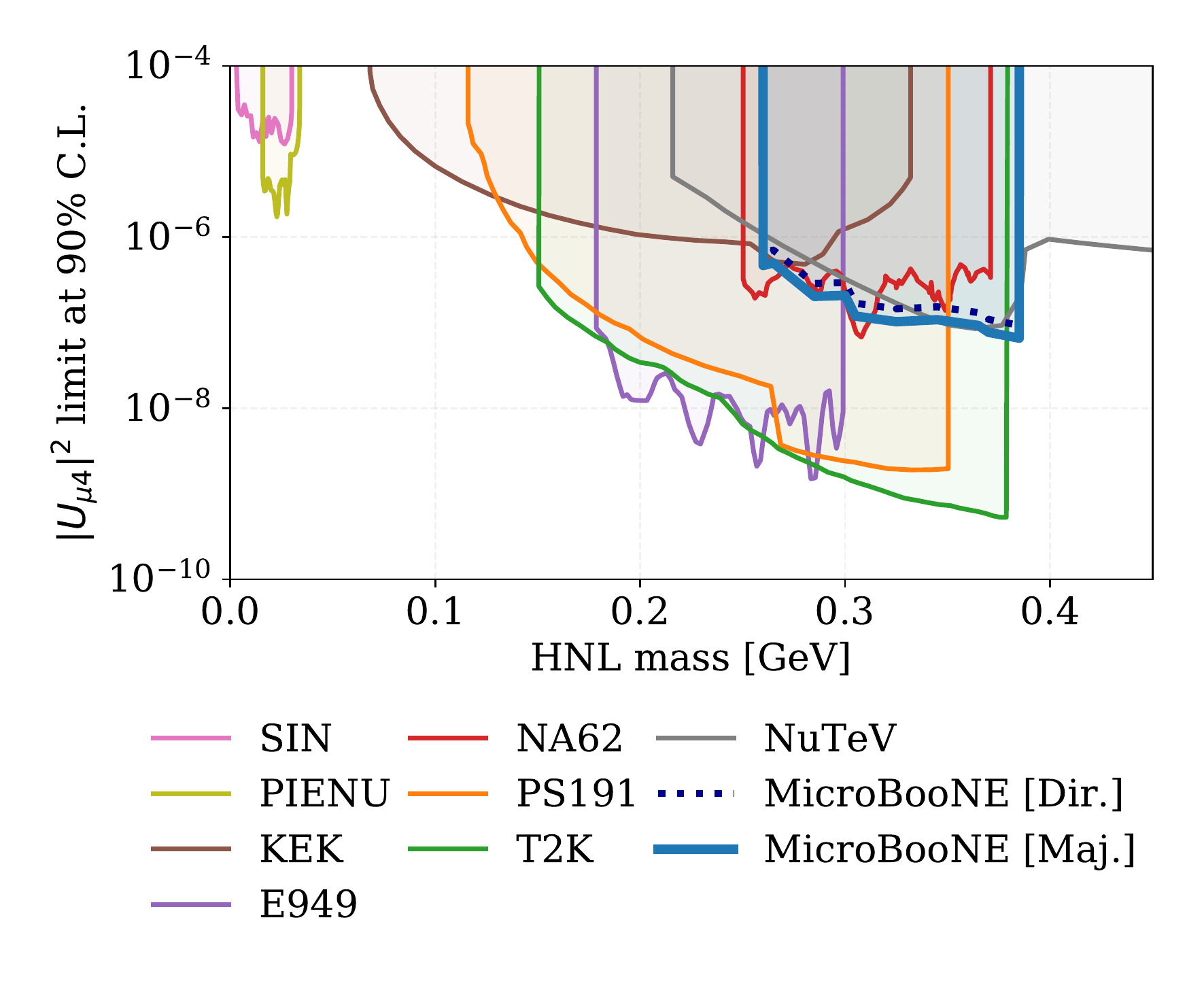}
        \caption{Limits on $|U_{\mu 4}|^2$ at the 90\% confidence level as a function of mass for Majorana and Dirac HNLs decaying into $\mu\pi$ pairs according to~\cite{extracted_beams:microboone_lake_louise_conf}. The limits are compared to the ones of other experiments:
        T2K~\cite{Abe:2019kgx},
        PS191~\cite{Bernardi:1985ny,Bernardi:1987ek}, E949~\cite{Artamonov:2014urb},
        NA62 (2015 data)~\cite{extracted_beams:NA62_2015},
        KEK~\cite{extracted_beams:KEK_results},
        PIENU~\cite{Aguilar-Arevalo:2017vlf},
        SIN~\cite{extracted_beams:SIN_results_1,extracted_beams:SIN_results_2},
        NUTEV~\cite{extracted_beams:nutev_results}.
        Note that for NA62 not the latest limits are used on the figure.}
        \label{fig:MicroBooNE_HNL_results}
    \end{figure*}
    
\end{itemize}
\subsubsection{Kaon decay experiments: NA62}

\begin{figure*}[h]
        \centering
        \includegraphics[width=0.7\textwidth]{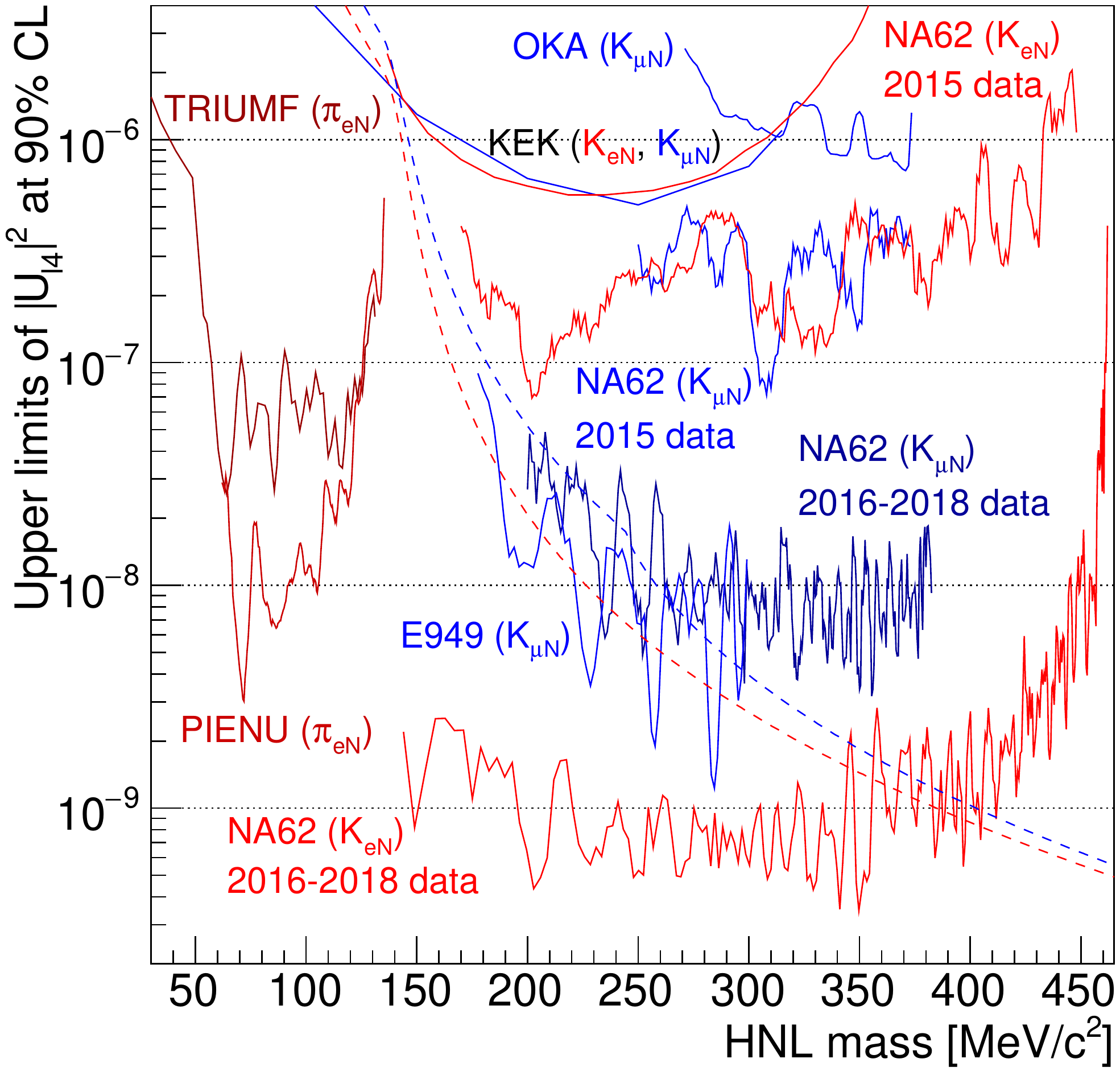}
        \caption{Upper limits at 90\% CL of $|U_{\ell 4}|^2$ obtained for each assumed HNL mass in production searches in $\pi/K\to\ell N$ decays: 
        TRIUMF~\cite{Britton:1992xv,Britton:1992pg},
        KEK~\cite{extracted_beams:KEK_results},
        NA62 (2015 data)~\cite{CortinaGil:2017mqf},
        NA62 (2007 data)~\cite{Lazzeroni:2017fza},
        NA62 ($K_{\mu N}$) (2016--2018)~\cite{CortinaGil:2021gga},
        NA62 ($K_{eN}$)
        (2016--2018)~\cite{NA62:2020mcv},
        PIENU~\cite{Aguilar-Arevalo:2017vlf},
        E949~\cite{Artamonov:2014urb},
        OKA~\cite{Sadovsky:2017qsr}.}
        \label{fig:NA62_HNL_results}
    \end{figure*}

The NA62 project (kaon factory)~\cite{extracted_beams:na62_apparatus} is the ultra-rare kaon decay experiment which aims to measure the decay $K^+\to\pi^+\nu\bar{\nu}$ at the CERN SPS to extract a 10\% measurement of the CKM parameter $|V_{td}|$ and perform a precise direct test of the SM. A 400 GeV/c proton beam from the SPS is used to perform the experiment. An unseparated secondary beam of $\pi^+$ (70\%), protons (23\%) and $K^+$ (6\%) is created by directing 400 GeV/c protons extracted from the CERN SPS onto a beryllium target. The central beam momentum is 75~GeV/c. Beam kaons are tagged by a differential Cherenkov counter and their kinematic information is measured  by a silicon pixel spectrometer consisting of three stations. The beam is delivered into a vacuum tank, which contains a 75 m long fiducial decay volume. Momenta of charged particles produced by kaon decays in the FV are measured by a magnetic spectrometer located downstream of the FV. A ring-imaging Cherenkov detector (RICH) is used for the identification of charged particles and time measurements, there are also a liquid krypton (LKr) electromagnetic calorimeter, an iron/scintillator sampling hadronic calorimeter and a muon detector which are used for particle identification and (LKr) for photon detection. The NA62 experiment carried out a search for HNL production using the 2016--2018 data in $K^+$ decays in the muon mode~\cite{CortinaGil:2021gga} and the positron mode~\cite{NA62:2020mcv}. The analysis is based on peak searches in the squared missing mass: $m_{miss}^2=(P_K - P_{\ell})^2$, where $P_K$ is a measured kaon 4-momentum, and $P_{\ell}$ is a lepton 4-momentum. No HNL signal was observed for either decay mode and  the corresponding upper limits of the decay branching fractions and the mixing parameters $|U_{\ell 4}|^2$ have been established (Fig.~\ref{fig:NA62_HNL_results}). 

\subsubsection{Future prospects}
\begin{itemize}
    \item \textbf{T2K-II and beyond}: the reported T2K results~\cite{Abe:2019kgx} are still statistically limited. The experiment got an approval to further collect data up to 2026, i.e. the beginning of the Hyper-Kamiokande era, so called T2K-II project. The project includes an upgrade of the near detector complex ND280-Upgrade~\cite{extracted_beams:ND280Upgrade} as well as an increase of the beam power. ND280Upgrade detector will have a part of the central $\pi^0$-detector replaced with a new highly granular scintillator detector (Super-FGD), two new TPCs (High-Angle TPC) and six TOF planes. Additional TPCs  will provide more active volume for HNLs search. Based on an improved detector performance and new data collected the HNL results are expected to further improve by a factor of 3-5. Additional data will also allow the background treatment to be improved by using more populated control regions. 
    
    By considering heavy neutrino production from pion decays, it would also be possible to extend the phase space down to a few MeV/c$^2$. When combined with a better understanding of the expected background, this may permit the low-mass heavy neutrino phase space ($10<M_N<493$\,MeV/c$^2$ and $U_{e,\mu}^2>10^{-11}-10^{-10}$) to be fully explored. Thus T2K can potentially improve current low-mass limits~\cite{extracted_beams:drewes_mev_lep} in the near future.
    \item \textbf{SBN program}: the SBN~\cite{extracted_beams:SBN_program} project is based on three detectors placed in the BNB at different (short) baseline distances: SBND  at 110 m from the target, MicroBooNE at 470 m and ICARUS at 600 m. All three detectors utilize LArTPC technology with powerful event reconstruction capabilities allowing for a significantly improved understanding of background processes compared to predecessor technologies. With this design, SBN has been shown to be able to extend the current bounds on light oscillating sterile neutrinos, thoroughly exploring the eV-scale sterile neutrino mass region, whilst also pursuing many other physics goals including a search for heavy neutral leptons~\cite{extracted_beams:SBN_hnl}.  It is expected that the SBN program can extend existing bounds on well constrained channels such as $N\to\nu\ell^+\ell^-$ and $N\to\ell^{\pm}\pi^{\mp}$ and due to the unique particle identification capabilities of LAr technology, also place bounds on often neglected channels such as $N\to\nu\gamma$ and $N\to\nu\pi^0$. Interesting to note is a phenomenological impact of improved event timing information at the SBN detectors: if the light-detection systems in SBND and ICARUS can achieve nanosecond timing resolution, the effect of finite sterile neutrino mass could be directly observable. The predicted 90\% CL upper limits for the combined SBN detectors are shown in Fig.~\ref{fig:SBN_HNL_limits}.
    
    \begin{figure*}[h]
        \centering
        \includegraphics[width=0.9\linewidth]{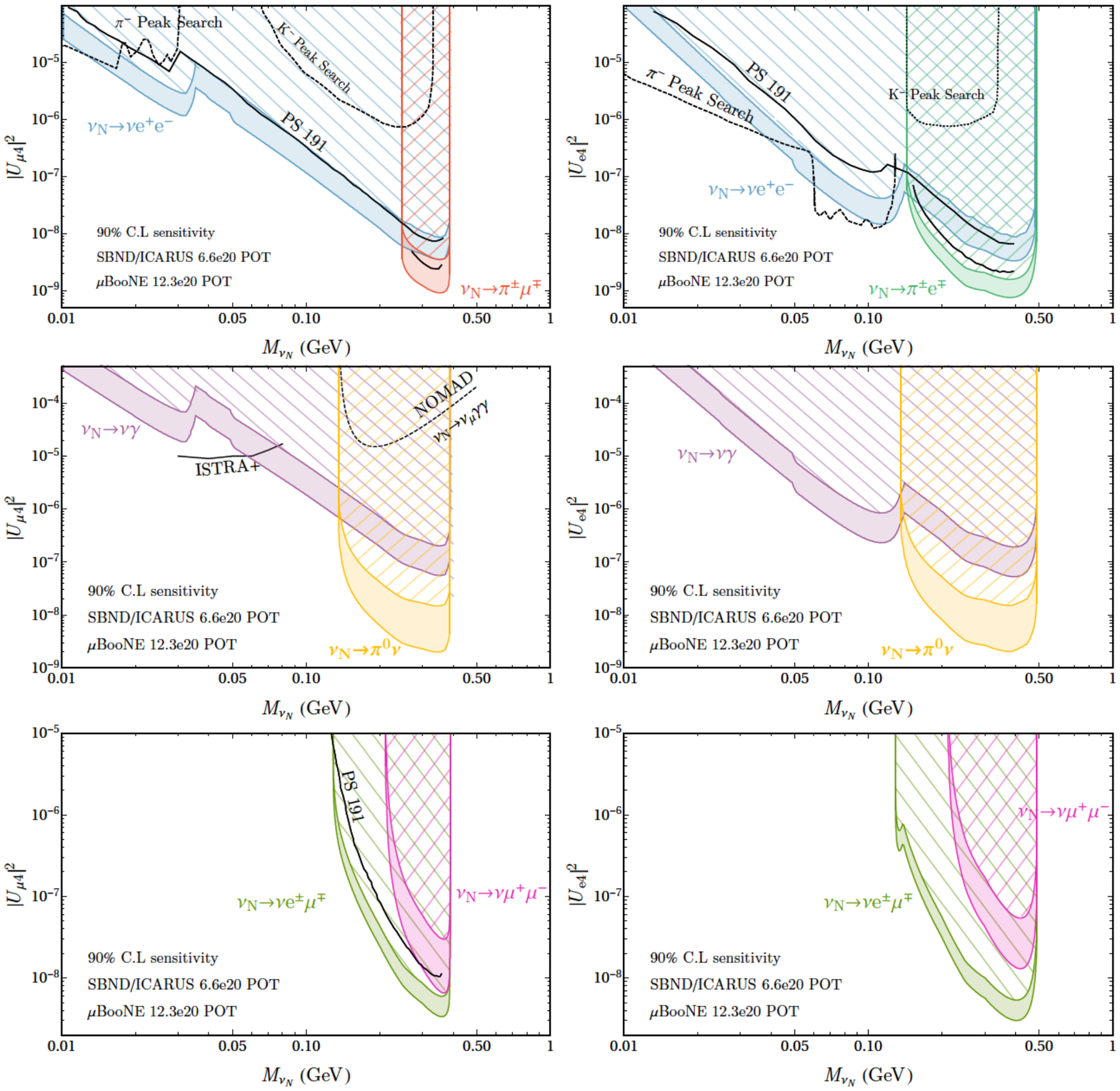}
        \caption{The predicted 90\% CL upper limit contours for the combined SBN detectors according to~\cite{extracted_beams:SBN_hnl}, which utilises the minimal SM extension model. Shown also in black solid lines are the bounds from PS-191~\cite{Bernardi:1985ny,Bernardi:1987ek} as well as bounds from lepton peak searches in pion and kaon decays~\cite{Britton:1992xv,Britton:1992pg} (dashed black lines). The photonic channels have little or no direct bounds, one can mention ISTRA+~\cite{extracted_beams:istra_limits}  bounding the radiative decay and re-interpreted  $N\to\nu\gamma\gamma$ bounds at NOMAD on $N\to\nu\pi^0$~\cite{extracted_beams:nomad_hnl_gamma_limits}.
        }
        \label{fig:SBN_HNL_limits}
    \end{figure*}
    
    As for the project status the MicroBooNE detector is already in operation, ICARUS was shipped to Fermilab in 2017  and is under commissioning, the SBND detector is expected to start data taking in near future.  
    
    As already mentioned the MicroBooNE detector is the first LArTPC to publish the results on a HNL search. It is planned to further extend the analysis with more data, more production and decay channels considered and as well an inclusion of the on-beam events.
    
     \item \textbf{DUNE}: the Deep Underground Neutrino Experiment (USA)~\cite{extracted_beams:dune_general} is a mega-scale future international neutrino project and will be a flagship experiment to achieve transforming discoveries with definitive determinations of neutrino properties. LArTPCs of tenths of kilotons scale will be used as far detectors and the near detector (ND) complex will be composed of a LArTPC together with a Multi-Purpose Detector (MPD, High-pressure Ar TPC and fine-grained tracker). The latter complex together with the  high  intensity  of  the  Long-Baseline Neutrino Facility (LBNF, 60-120 GeV tunable $\sim$MW scale)  beam  and the  production  of  charm  and  bottom  mesons  in  the beam  enables  DUNE  to  search  for  a  wide  variety  of light weight ``exotic'' particles by  looking  for topologies of rare event interactions and decays inside ND. The  main  background to this search comes from SM neutrino–nucleon scattering events in which the hadronic activity at the vertex is below threshold. The DUNE collaboration prepared a summary~\cite{extracted_beams:dube_bsm} of the BSM physics potential of the experiment which also includes a possible search for HNLs, in its turn studied in details in~\cite{extracted_beams:dune_ballet_hnl}.  The expected sensitivities are shown in Fig.~\ref{fig:DUNE_HNL_limits}. 
      \begin{figure*}[h]
        \centering
        \includegraphics[width=\textwidth]{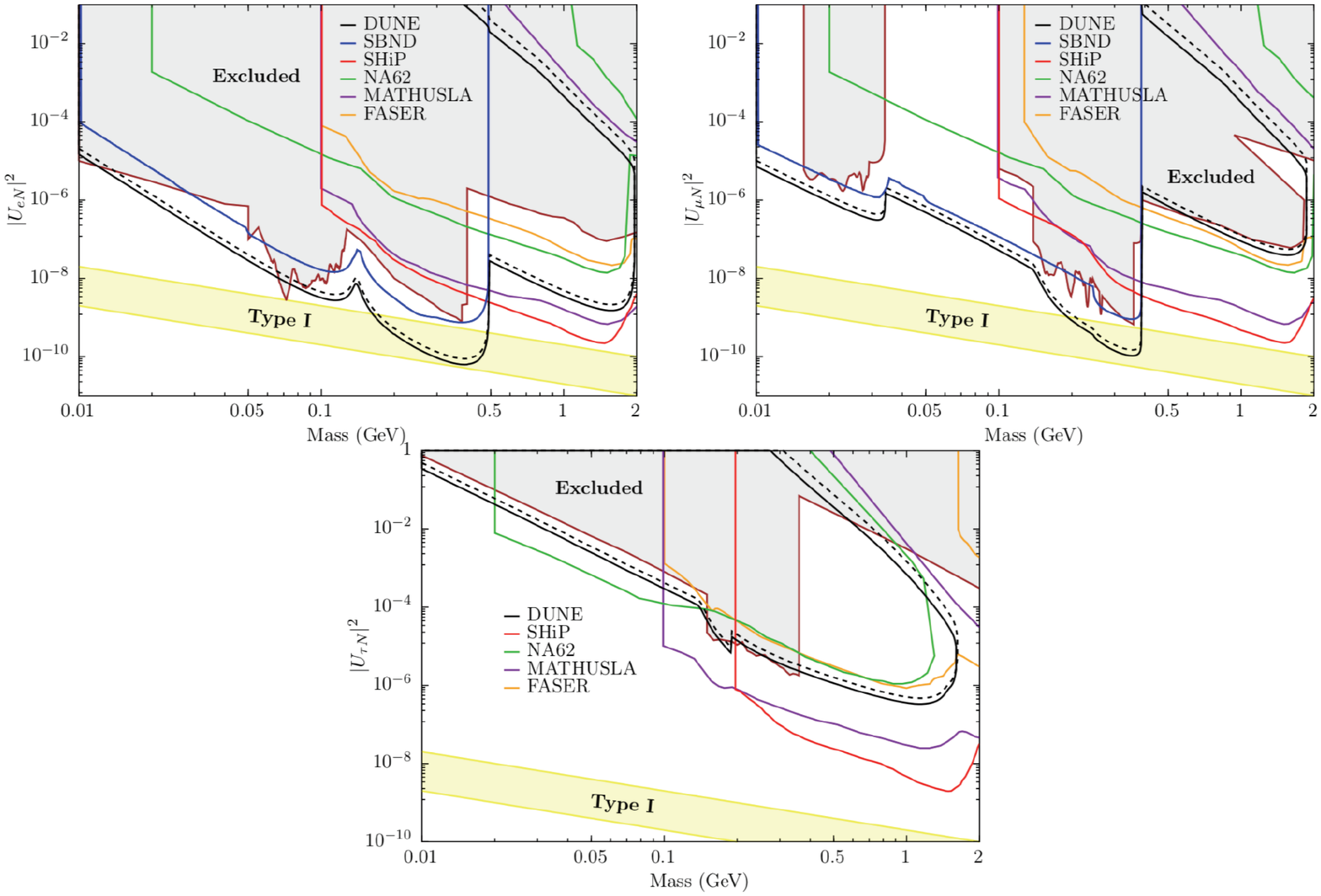}
        \caption{The predicted 90\% CL upper limit contours for the mixing parameters $|U_{eN}|^2$ (top left), $|U_{\mu N}|^2$ (top right) and $|U_{\tau N}|^2$ (bottom) for DUNE ND (black) according to~\cite{extracted_beams:dube_bsm},
        from~\cite{extracted_beams:dune_ballet_hnl}. The study is performed for Majorana neutrinos (solid) and Dirac neutrinos (dashed), and no background is assumed. A combination of $N\to$ $\nu ee$, $\nu\mu\mu$, $\nu e\mu$, $e\pi$, $\mu\pi$ and $\nu\pi^0$ channels is used. Already excluded (grey) region corresponds to the results of the following experiments: PS-191~\cite{Bernardi:1985ny,Bernardi:1987ek}, NuTEV~\cite{Vaitaitis:1999wq},  T2K~\cite{Abe:2019kgx},
        CHARM~\cite{Orloff:2002de},
        DELPHI~\cite{Abreu:1996pa},
        pion and kaon peak searches~\cite{Artamonov:2014urb,Aguilar-Arevalo:2017vlf,Britton:1992xv,Britton:1992pg}. The sensitivity for DUNE ND is compared to the predictions of future experiments, SBN~\cite{extracted_beams:SBN_hnl} (blue), SHiP~\cite{extracted_beams:ship_hnl_sensitivity} (red), NA62 (beam-dump mode)~\cite{Drewes:2018gkc} (green), MATHUSLA~\cite{Curtin:2018mvb} (purple), and the Phase II of FASER~\cite{Kling:2018wct} (dark yellow). For reference, a band corresponding to the contribution light neutrino masses between 20~meV and 200~meV in a single generation see-saw type I model is shown (yellow).
        }
        \label{fig:DUNE_HNL_limits}
    \end{figure*}
     \item \textbf{NA62++}: as mentioned earlier the new results of the experiment on $K^+\to \mu^+ N$ search with 2016-2017 data are preliminary, they will be published in 2021. Further improvement in sensitivity of the analyses can only be expected with future NA62 data. A unique opportunity for searches for MeV-GeV mass ``hidden-sector'', ``exotic'' candidates can be opened with the optimised NA62++ setup: beam-dump mode operation~\cite{}. In this mode the  beryllium  target  is  lifted  and  the  proton  beam  is  dumped  after closing the movable beam-defining collimator, further facility optimisations are foreseen as well. The idea is to boost meson production including heavy, charm/beauty ones by 400 GeV intense proton beam.  Long-lived mediators faintly coupled to the SM fields can be produced in the meson decays giving the experiment a possibility to gain high sensitivities in probing a variety of new-physics scenarios, including HNLs.  Test data was collected in dump-mode in 2016-2018 period which includes an implementation of the dedicated long-bandwidth triggers. A possibility to have physics runs in beam-dump mode in 2023 is considered: $\mathcal{O}(3)$ months of data taking is expected to be needed to achieve enough statistics for analyses. Sensitivity estimations for HNLs searches with NA62++ can be found in Fig.~\ref{fig:DUNE_HNL_limits}.
\end{itemize}


\clearpage
\subsection{Search for HNLs at LHCb, ATLAS, CMS: status and prospects}
\label{ssec:shschutska}
{\it Author: Shchutska Lesya, <lesya.shchutska@epfl.ch>}  
\subsubsection{Summary of the latest results}
The searches for heavy neutral leptons (HNL or N) at the LHC have a long tradition on one hand, but are currently undergoing  significant strategy changes on the other hand. These strategies can be globally classified by the three aspects of the HNL phenomenology: 

\begin{description}
\item[production mechanism:] in decays of W bosons (W$\to\ell$N) or b quarks (b$\to{\rm X}\ell$N, where X stands for an u or a c quark);
\item[decays:] N$\to$W$\ell$, N$\to$Z$\nu$, or N$\to$H$\nu$, where W, Z, and H can be either on-shell of off-shell particles, 
which in turn decay leading to detectable final states;
\item[lifetime:] from very short-lived (prompt decays) to long-lived N leading to macroscopic distances from production vertex (displaced decays) 
as HNL lifetime $\tau\propto\abs{V_{\ell{\rm N}}}^{-2}m_{\rm N}^{-5}$.
\end{description}

\begin{figure*}[h!]
\center
	\includegraphics[height=0.3\textwidth]{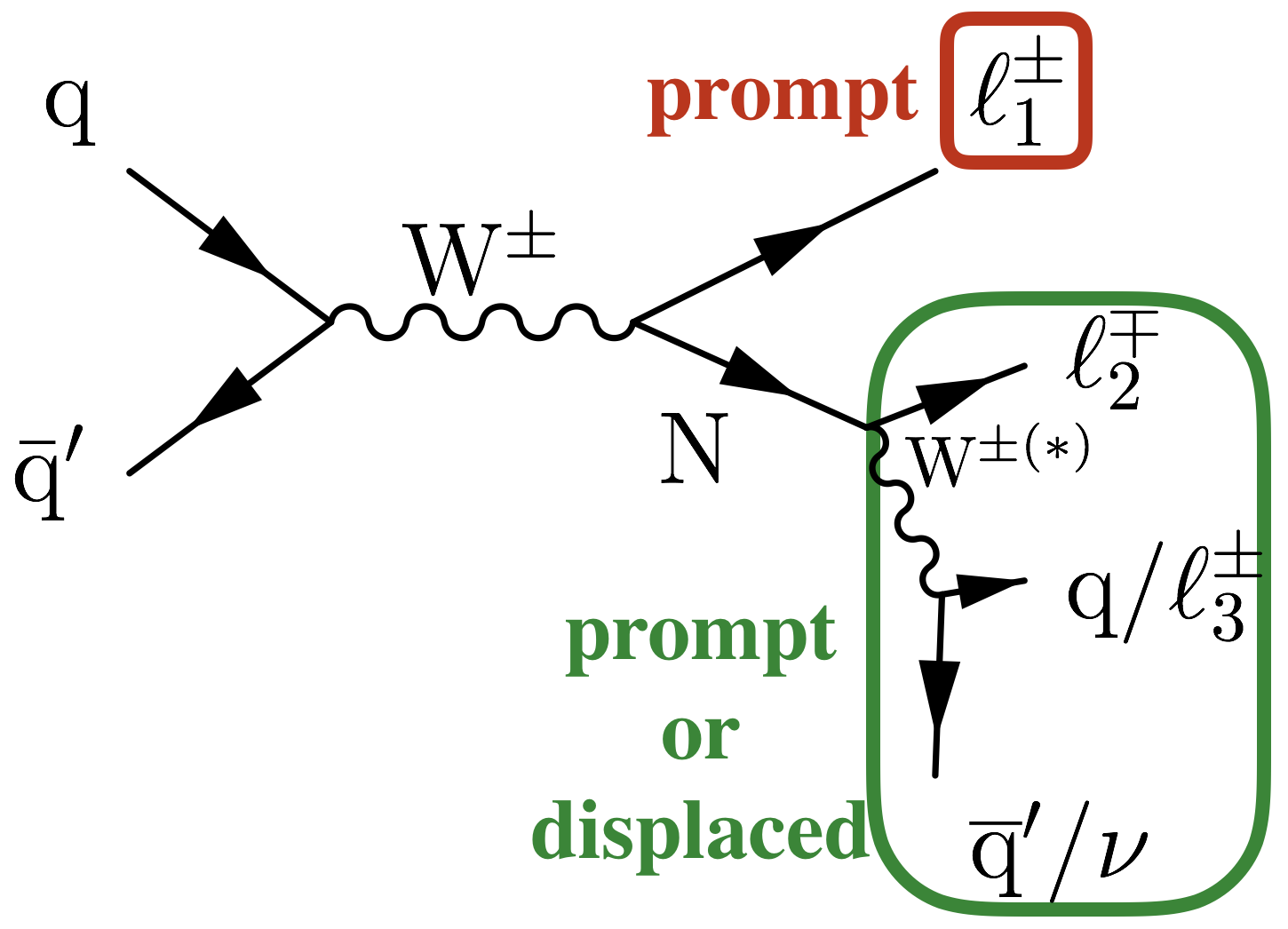} \hspace*{1.0cm}
	\includegraphics[height=0.4\textwidth]{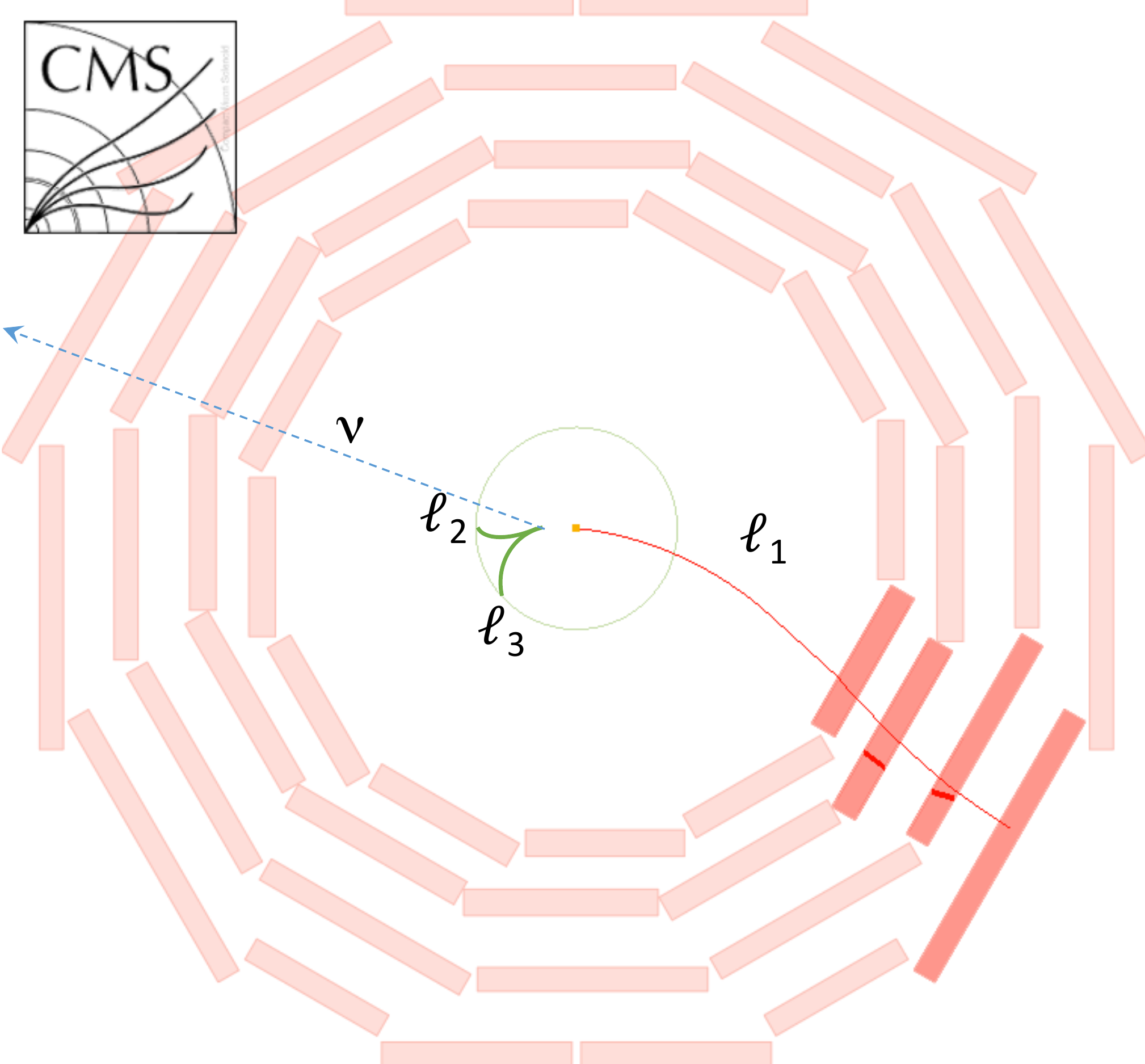}
	\caption{
	(left) A diagram showing an HNL production in a W boson decay, and its subsequent decay via a W$^{(*)}$ boson. 
	(right) Schematic event display of a long-lived HNL decay in the CMS detector.
}
	\label{fig:HNL_cartoon}
\end{figure*}

An example diagram showing the N production in the W boson decay, which subsequently decays to a W boson and a lepton, is shown in Fig.~\ref{fig:HNL_cartoon} (left). In case the HNL is sufficiently long-lived it could lead to a signature with a displaced vertex in the detector as shown in Fig.~\ref{fig:HNL_cartoon} (right). Presence of a prompt lepton allows for an efficient trigger strategy,
and a requirement for a displaced vertex formed with two leptons significantly suppresses SM background.

Promptly decaying HNLs produced in W boson decays have been explored both by ATLAS and CMS collaborations in Run 1 of the LHC, in the same-sign dilepton with jets signature ($\ell^\pm\ell^\pm$+jets). 
Promptly decaying HNLs coming from b quark decays were looked for by the LHCb collaboration
in Run 1 by concentrating on a search for a B$^\pm\to\pi^\mp\mu^\pm\mu^\pm$ decay~\cite{Aaij:2014aba}.

Since recently, all the collaborations started to explore non-traditional approaches for the HNL searches, 
and several new results have been already published, while more analyses are underway. 
In the following, the most recent results from the LHC collaborations which are using non-conventional approaches, are discussed.

\subsubsection{Searches with prompt N decays by CMS} 

Sterile neutrinos searches targeting Majorana neutrino signatures have traditionally been carried out with the same-sign dilepton signature. This final state appears with neutrino decaying to a lepton and two quarks, and its advantage is that it allows to fully
reconstruct a neutrino mass peak in the invariant mass spectrum of a lepton and two jets. However, such searches are sensitive to rather heavy neutrinos as e.g. jets transverse momenta have to be rather high (typically $p_\text{T} > 30$~GeV), and they are sensitive only to lepton-number-violating (LNV) neutrino decays. 

\vskip 2mm
Recently, a new approach using trilepton (e or $\mu$) final state has been introduced for heavy neutrino searches~\cite{Sirunyan:2018mtv}.
Though trilepton final state cannot provide a clear heavy neutrino mass peak due to the escaping $\nu$, it benefits from a lower standard model background, and allows to detect decay products of very light N as lower reconstruction thresholds for the lepton transverse momenta can go as low as 3--5 GeV. Finally, such signature is sensitive to both lepton-number-conserving (LNC) and LNV heavy neutrino decays.

\begin{figure*}[h!]
\centering
	\includegraphics[width=0.35\textwidth]{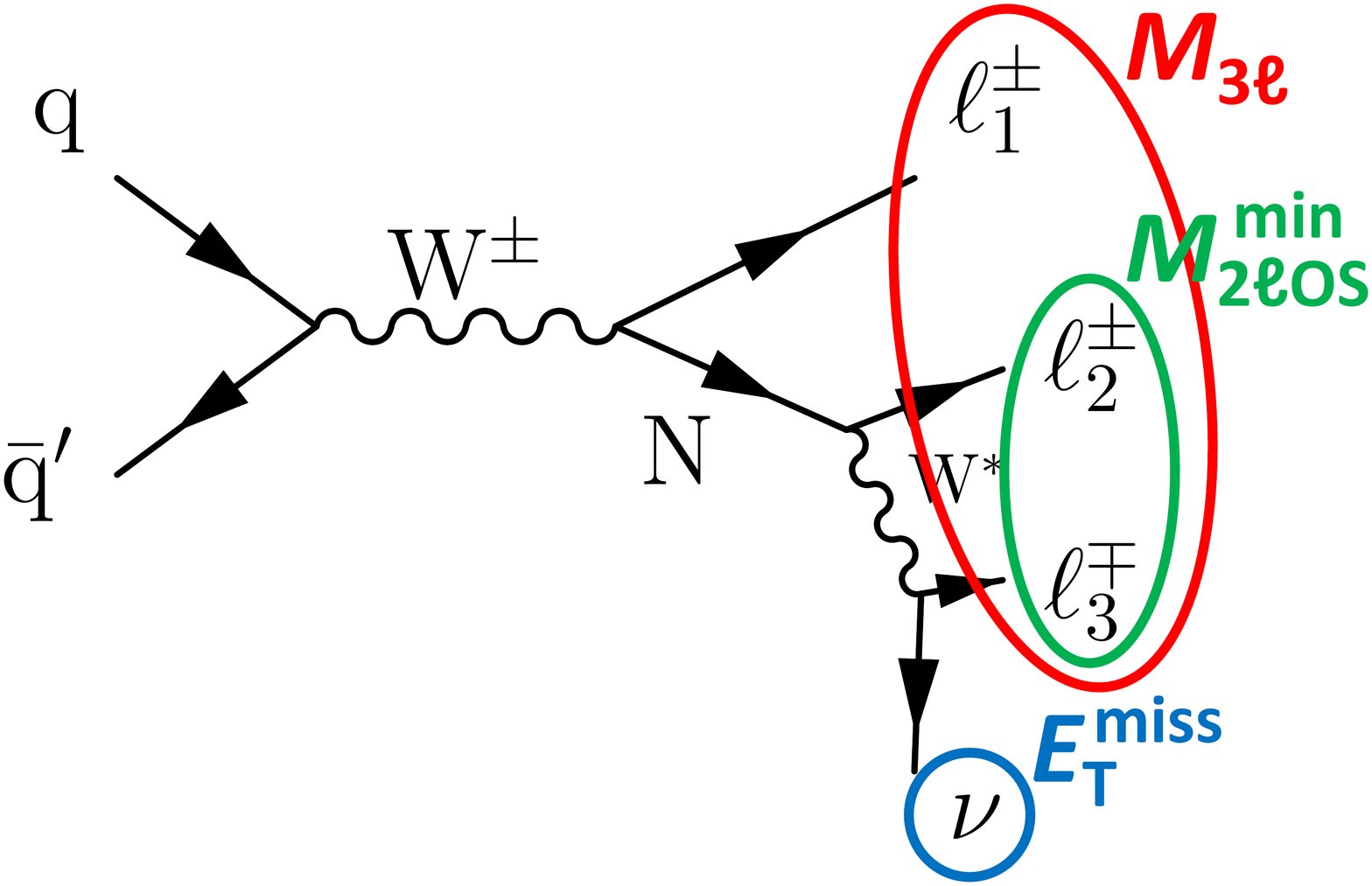} \\
	\includegraphics[width=0.32\textwidth]{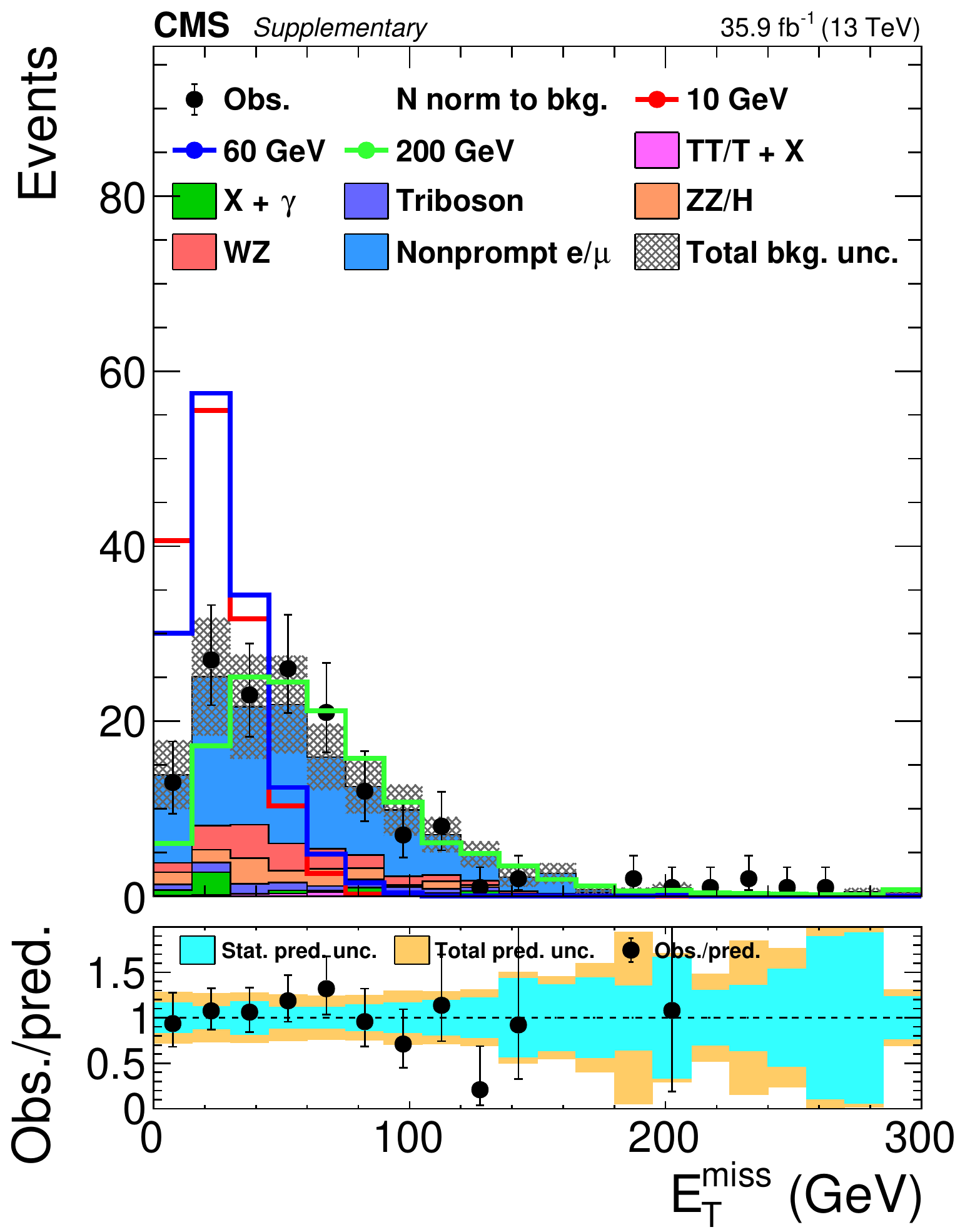} 
	\includegraphics[width=0.32\textwidth]{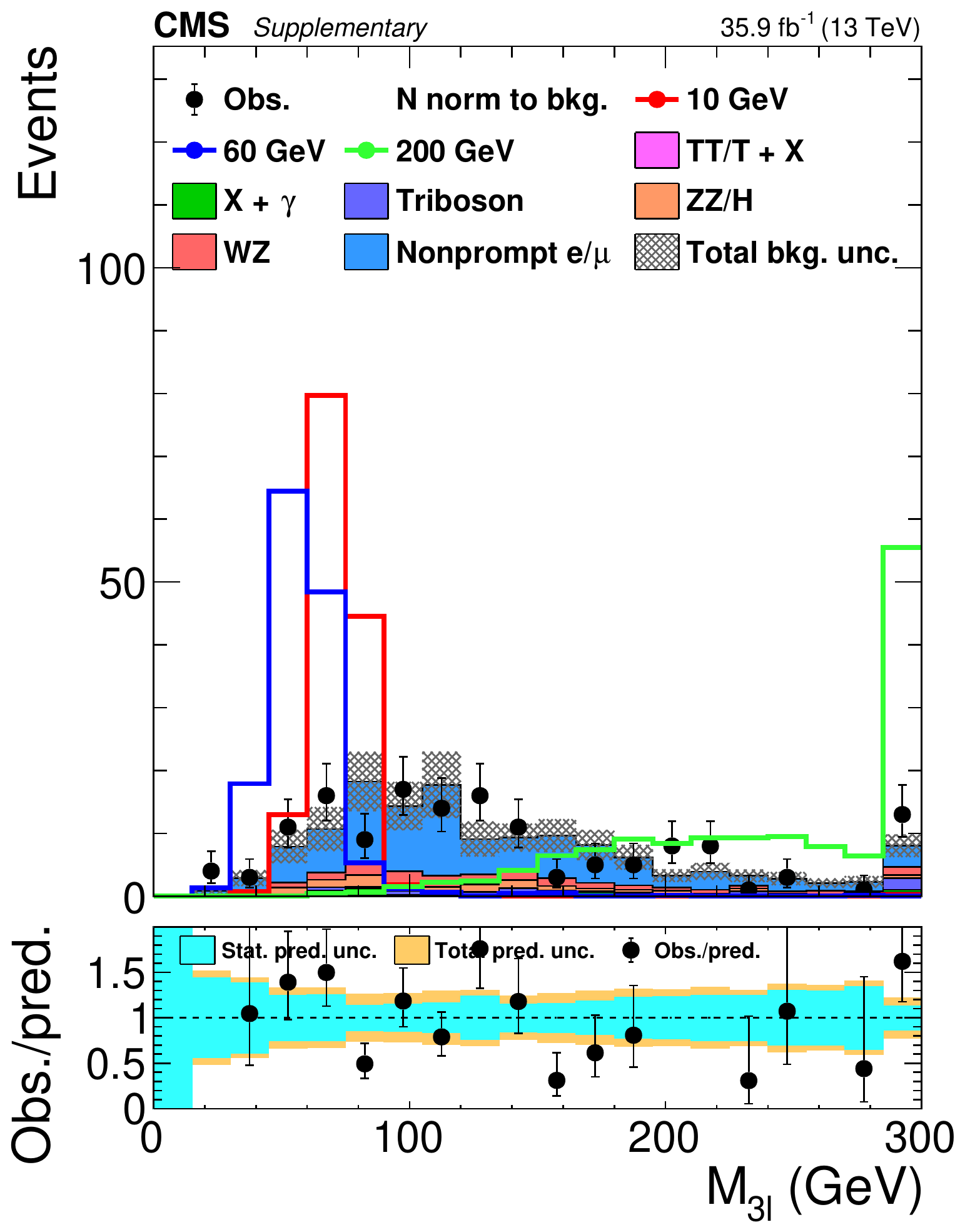} 
	\includegraphics[width=0.32\textwidth]{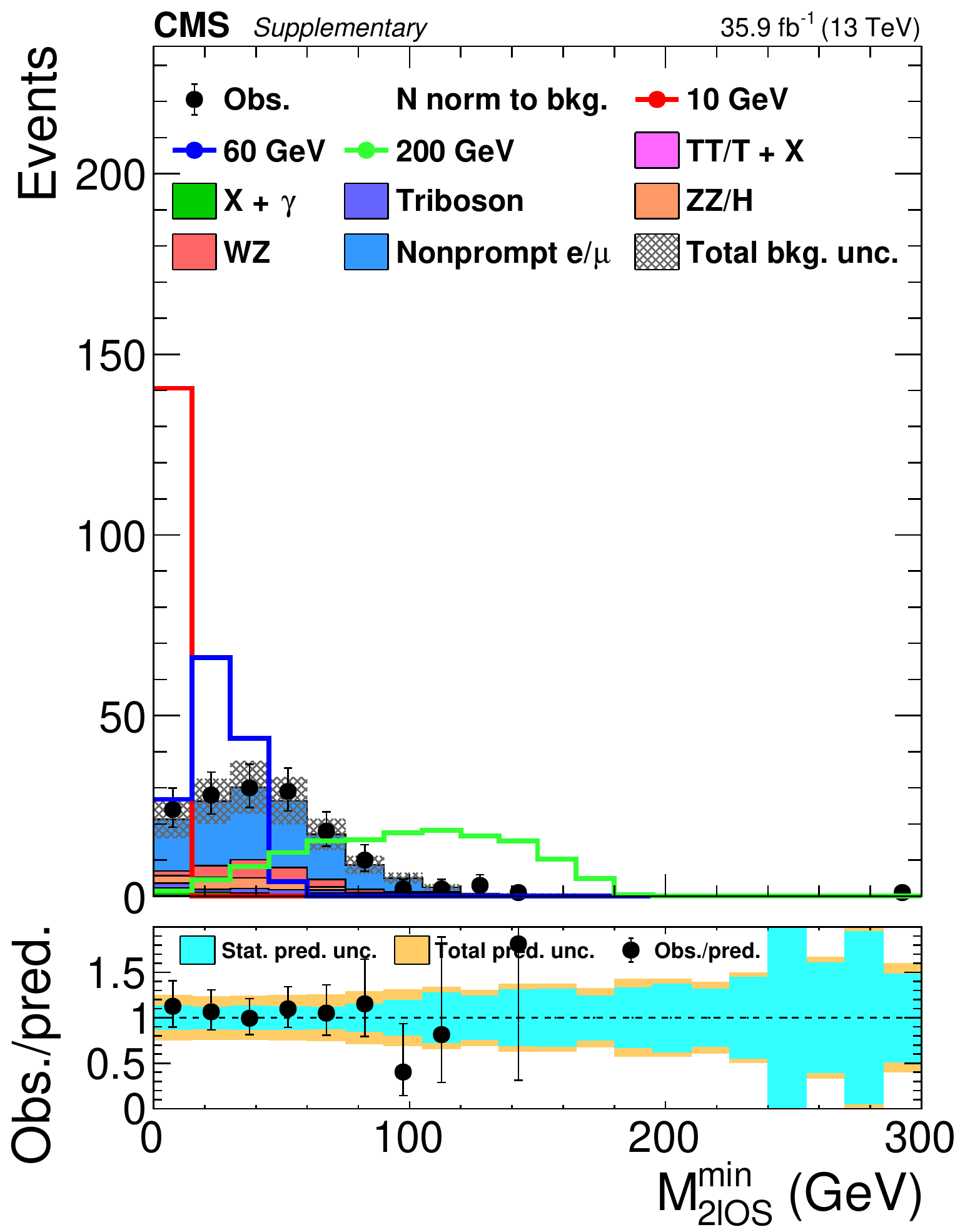} 
	\caption{ (top) A diagram for a targeted N production and decay processes which shows the meaning of designed kinematic 
	variables of the analysis in Ref.~\cite{Sirunyan:2018mtv}. Bottom panel shows the comparison of the estimated standard model
	background, heavy neutrino signal distribution for several N mass hypotheses, and observed data for three kinematics variables
	used in the analysis: 
	(left) missing transverse momentum $E_\text{T}^\text{miss}$, 
	(center) invariant mass of three charged leptons $M_{3\ell}$, 
	and (right) minimal invariant mass of two oppositely-charged leptons $M_{2\ell\text{OS}}^\text{min}$~\cite{Sirunyan:2018mtv}.	
}
	\label{fig:CMS_data}
\end{figure*}

\vskip 2mm
The trilepton search has been carried out with the integrated luminosity of 36 fb$^{-1}$ of the proton collision data recorded by the CMS detector in 2016. The search employed dedicated kinematic variables to discriminate between the heavy neutrino
signal and the SM background. Figure~\ref{fig:CMS_data} shows the distribution of the main search variables used in the analysis. 
Dedicated search regions are designed to target N with masses below or above a W boson mass.

\begin{figure*}[h!]
\centering
	\includegraphics[height=0.34\textwidth]{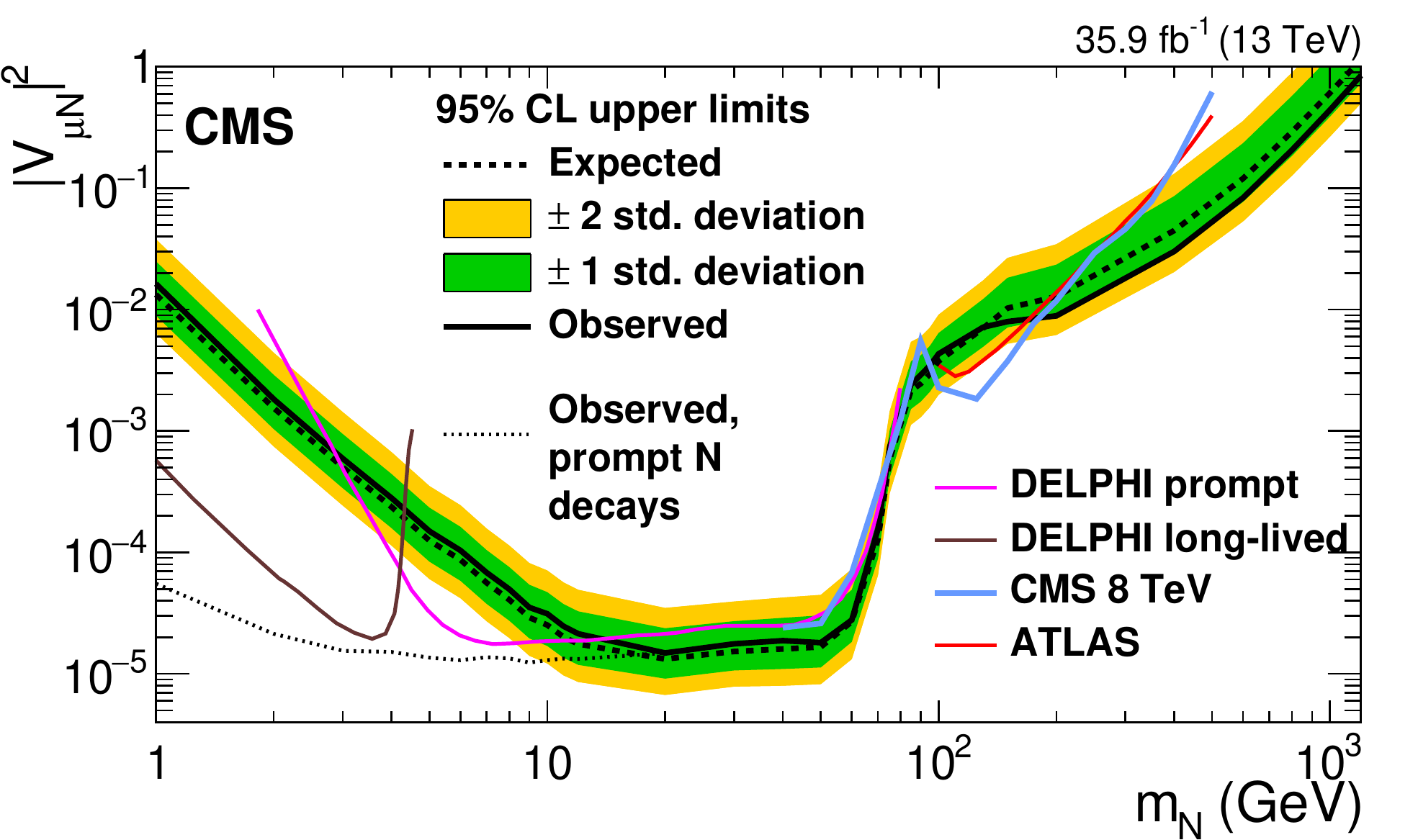} 
	\includegraphics[height=0.34\textwidth]{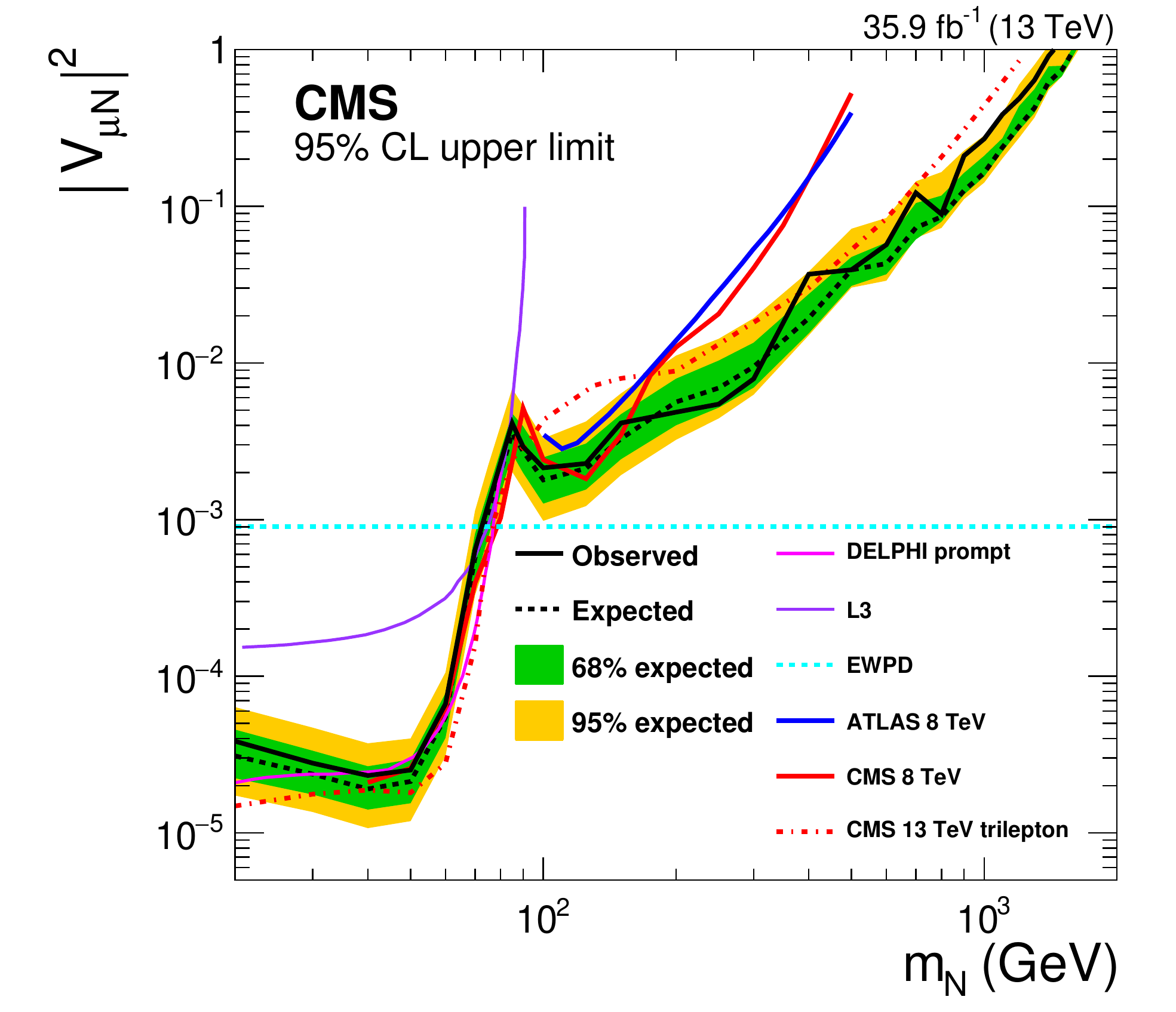}
	\caption{Upper limits on the mixing parameter between the muon neutrino 
	and a sterile Majorana neutrinos $\abs{V_{\mu\text{N}}}^2$
	at the 95\% CL obtained in the trilepton analysis~\cite{Sirunyan:2018mtv} (left) 
	and in the same-sign dilepton analysis~\cite{Sirunyan:2018xiv} (right).
}
	\label{fig:CMS_results}
\end{figure*}

The final interpretation is obtained for the LNV scenarios with a Majorana neutrino having an exclusive mixing with either an electron or a muon neutrino,
and the upper limits on $\abs{V_\text{eN}}^2$ and $\abs{V_{\mu\text{N}}}^2$ (Fig.~\ref{fig:CMS_results} left) are computed.
As leptons used in the analysis are required to originate from the primary vertex, the analysis is not sensitive to scenarios when heavy neutrino is long-lived. Therefore an acceptance correction is performed in the interpretation to take into account this effect.

The same dataset is used in the most recent same-sign dilepton with jets search, and the results are reported as the upper limits on $\abs{V_\text{eN}}^2$ and $\abs{V_{\mu\text{N}}}^2$ (Fig.~\ref{fig:CMS_results} right). 
In the high N mass region the same-sign dilepton analysis is more powerful due to larger signal acceptance,
while in the low N mass region the trilepton analysis is more sensitive due to the lower SM background to this final state.

\subsubsection{Search with prompt N decays by LHCb} 

A heavy neutrino search in dimuon with one jet final state has been carried out by the LHCb collaboration with the integrated luminosity of 3.0~fb$^{-1}$ of proton collision data at 7 and 8~TeV recorded by the LHCb detector between 2010 and 2012. 
For the first time, oppositely charge muon pairs are used in such a search. One jet signature corresponds to the cases whether two quarks from a heavy neutrino decay form one jet, or when one of the jets is outside of the detector acceptance.
Due to the ability of LHCb to work with low-$p_\text{T}$ jets ($>$~10 GeV in this analysis), probed heavy neutrinos can be rather light.
Inclusion of both oppositely-charged and same-charged muon pairs allows to probe both LNV and LNC heavy neutrino decays.

\begin{figure*}[h!]
\centering
	\includegraphics[height=0.3\textwidth]{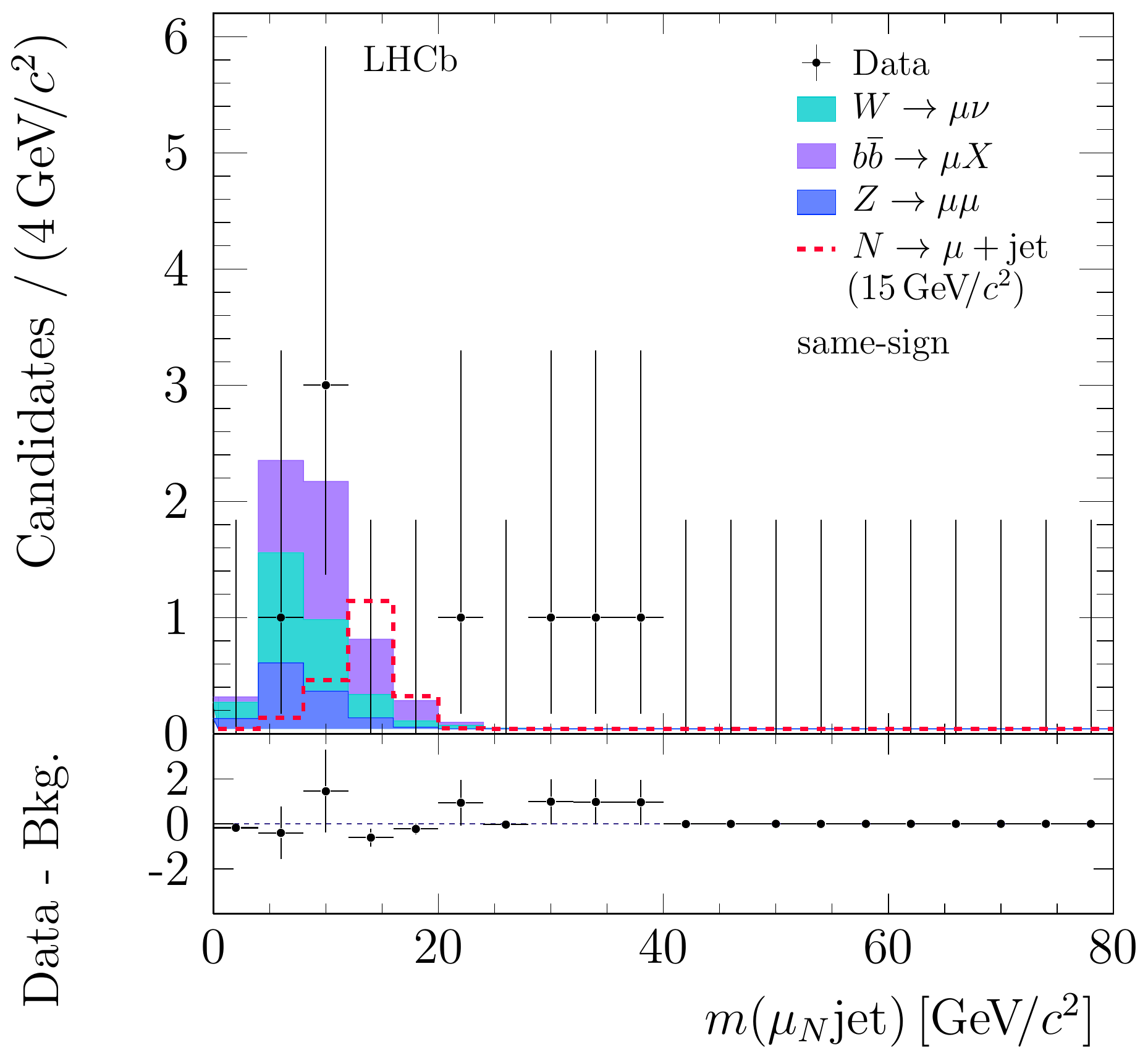} 
	\includegraphics[height=0.3\textwidth]{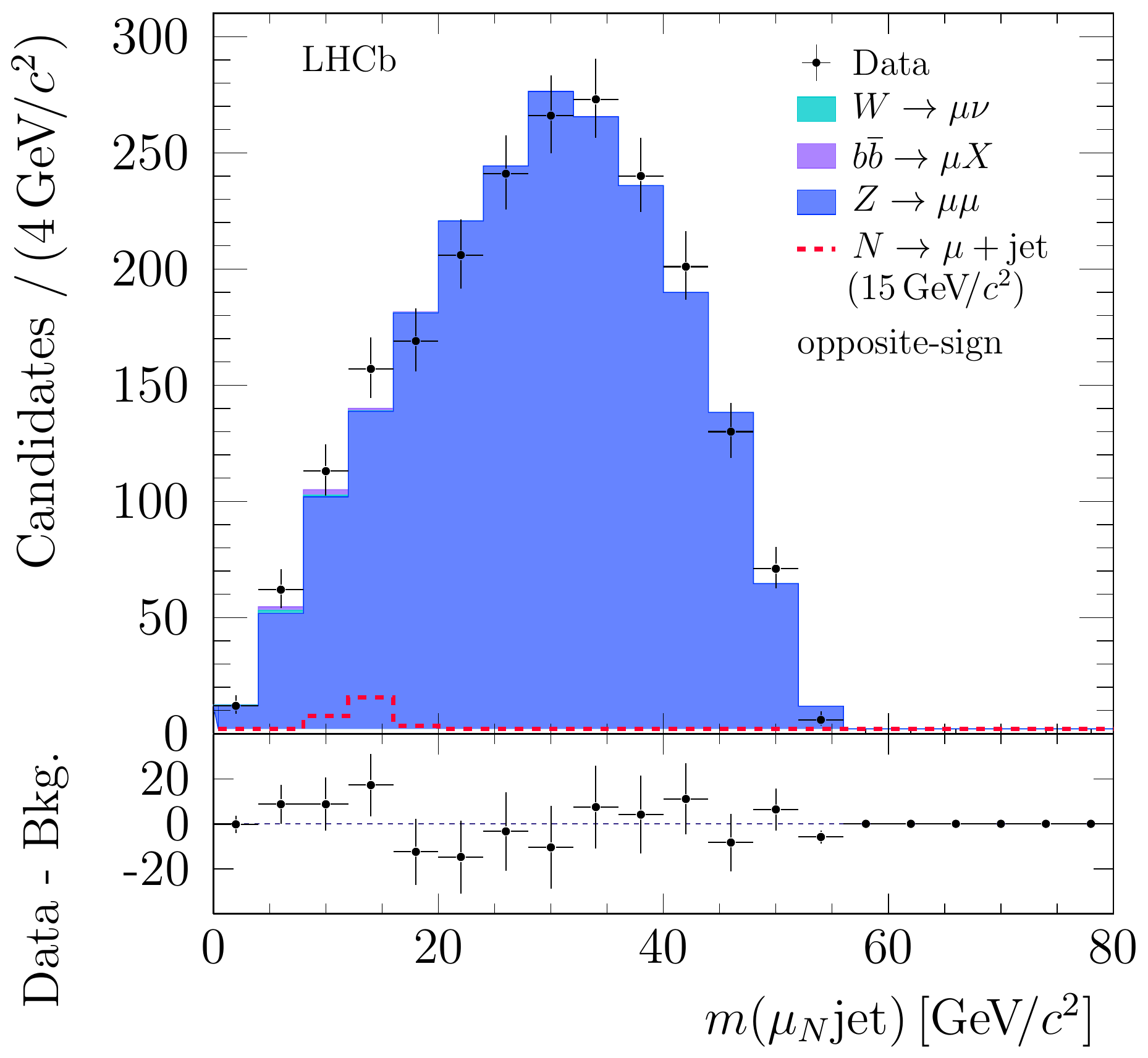}
	\includegraphics[height=0.3\textwidth]{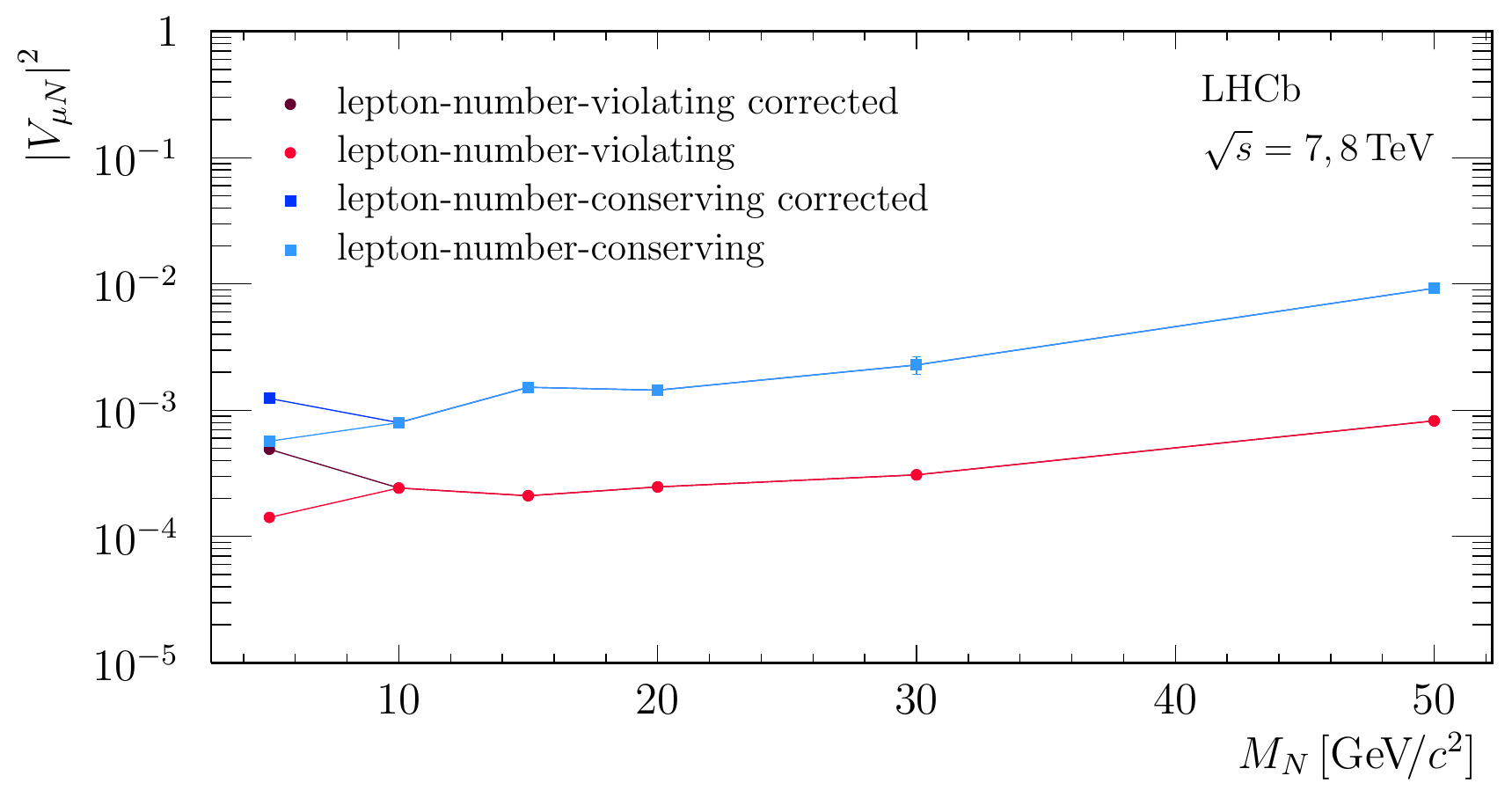}
	\caption{Top panel: invariant mass distributions for the heavy neutrino candidates as expected from the SM and as observed in data for 
	same-sign muon (left) and opposite-sign muon (right) final state.
	Bottom panel: upper limits on the mixing parameter between the muon neutrino 
	and a sterile Majorana neutrinos $\abs{V_{\mu\text{N}}}^2$
	at the 95\% CL for the LNC and LNV scenarios~\cite{Aaij:2020ovh}.
}
	\label{fig:LHCb_results}
\end{figure*}

\vskip 2mm
Observed neutrino candidates invariant mass distributions are reported in the top panel of Fig.~\ref{fig:LHCb_results}.
The upper limits on the $\abs{V_{\mu\text{N}}}^2$ mixing parameter in the two scenarios is shown in the bottom panel of Fig.~\ref{fig:LHCb_results}.
Similarly to the previously discussed CMS result, in the low N mass region the analysis sensitivity has to be corrected to take into account long-lived heavy neutrinos.

\subsubsection{First displaced HNL search at the LHC: ATLAS result} 

Though the LHC detectors were designed to study particles produced in the proton collisions at (or close to) the primary vertex, their large volume allows to probe rather long-lived particles, for which dedicated reconstruction algorithms have been developed. 
ATLAS and CMS detectors offer ``decay volume'' up to about 3~m (10~ns) in the transverse plane. The LHCb detector has longer baseline but it is not instrumented by the magnetic field after the tracking stations. So effectively the same 2-3~m can be used
as a ``decay volume'' for reconstructible long-lived particles. 

\vskip 2mm
Recently, for the first time since LEP result~\cite{extracted_beams:delphi_limits}, the ATLAS collaboration has published the search for long-lived heavy neutrinos in leptonic final states~\cite{Aad:2019kiz}. This search requires at least one prompt muon from W$\to\mu\text{N}$ 
with $p_\text{T}^\mu>28$~GeV. Though again, no clear N mass peak is visible due to the escaping $\nu$, a sufficiently displaced
dilepton ($\mu\mu$ or $\mu\text{e}$) vertex allows to clearly separate heavy neutrino signal from the SM background.
The vertex displacement can be between 4 and 300 mm from the primary vertex in the transverse plane. 
The analysis is sensitive to very light N as lepton $p_\text{T}$ thresholds are low ($>4-5$ GeV for $\mu$, and $>7$~GeV for e).
Both LNC and LNV signatures are taken into consideration.

\begin{figure*}[h!]
\centering
	\includegraphics[width=0.48\textwidth]{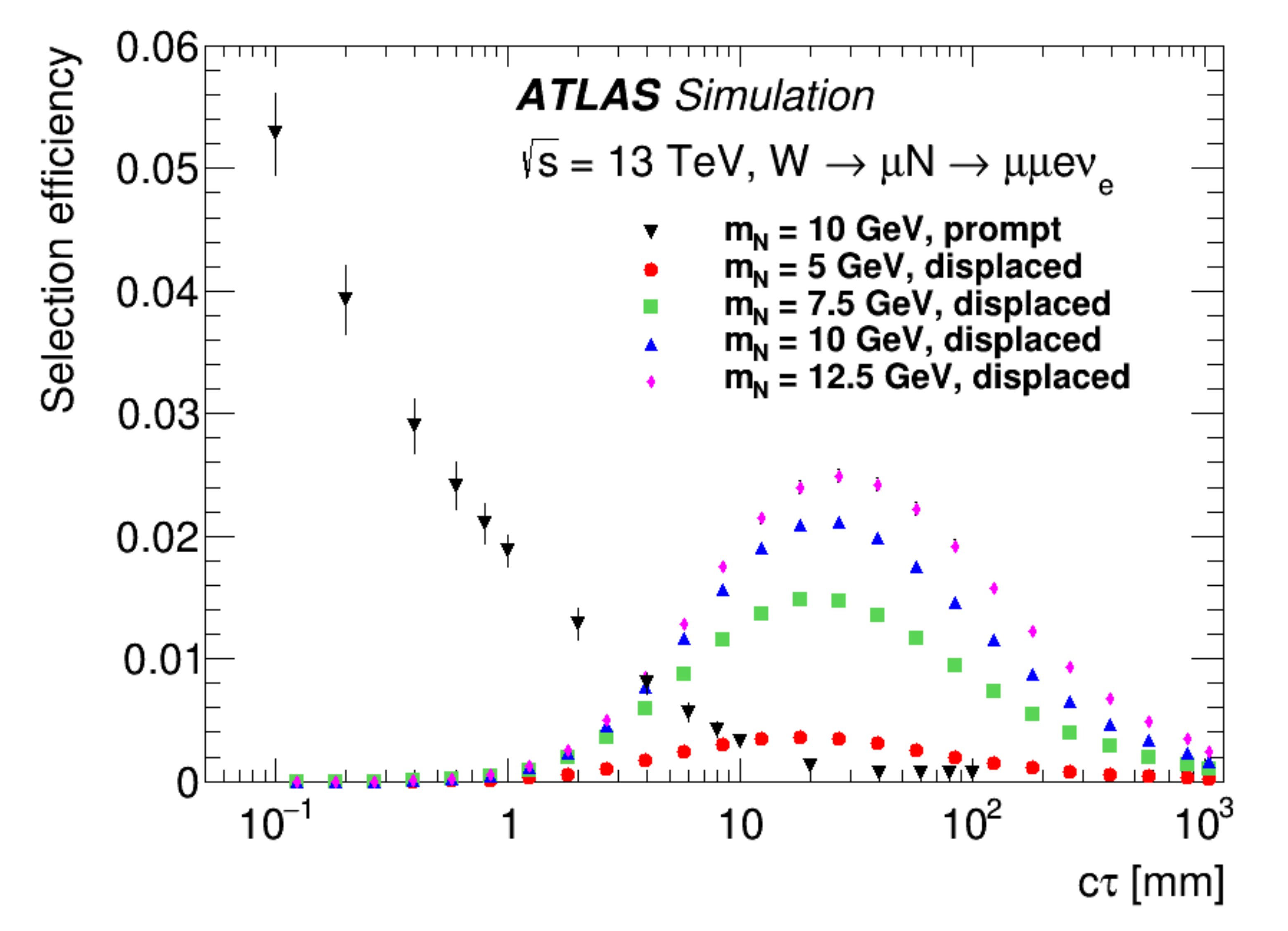} 
	\includegraphics[width=0.48\textwidth, height=6cm]{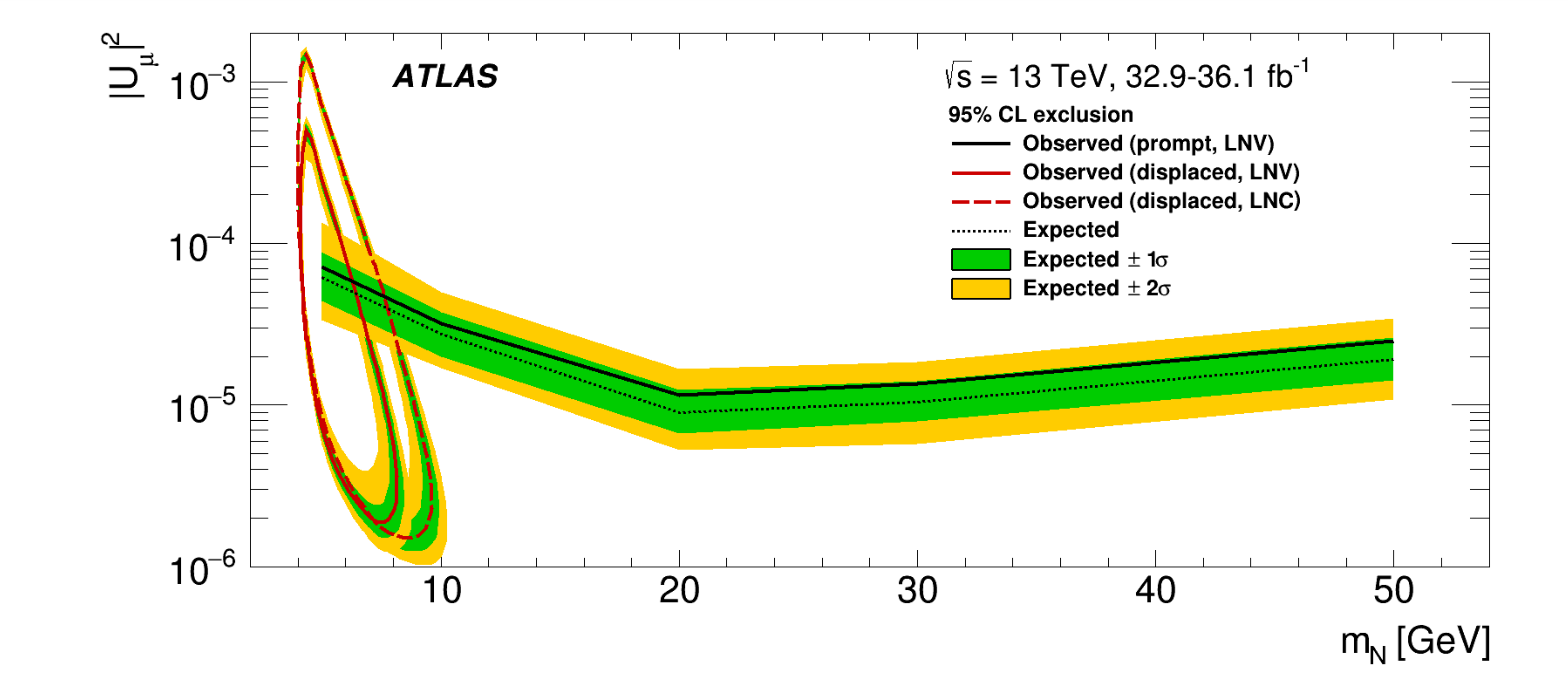}
	\caption{
	(left) Heavy neutrino selection efficiencies for the prompt search (10 GeV neutrinos), and for the displaced search (various neutrino masses). 
	(right) Upper limits on the mixing parameter between the muon neutrino 
	and a sterile Majorana neutrinos $\abs{V_{\mu\text{N}}}^2$
	at the 95\% CL for the LNC and LNV scenarios~\cite{Aad:2019kiz}.
}
	\label{fig:ATLAS_results}
\end{figure*}

\vskip 2mm
The reconstruction efficiency for the long-lived particles typically is lower than those decaying promptly as can be seen in Fig.~\ref{fig:ATLAS_results} (top). Nonetheless, achieved performance allows to capture heavy neutrinos up to $c\tau\approx1$~m. 
The interpretation for $\abs{V_{\mu\text{N}}}^2$ mixing parameter in the two scenarios is shown in Fig.~\ref{fig:ATLAS_results} (bottom).
The analysis probes N masses between 4.5 and 50 GeV. Below 4.5~GeV, the SM background is dominated by b-hadron 
decays and conversions, which lowers the analysis sensitivity; while for higher N masses the search is practically background-free 
thanks to the displaced signature. Prompt analysis targets only the LNV scenario due to the high SM background in the LNC case. 

\subsubsection{Phenomenological projections for the (HL-)LHC} 

While there are no projections for heavy neutrino searches at the (HL-)LHC from the LHC collaborations,
there are several phenomenology papers offering new ideas and their sensitivity estimates for already collected LHC data 
or for the future HL-LHC dataset~\cite{Boiarska:2019jcw,Drewes:2019fou,Bondarenko:2019tss}.

\begin{figure*}[ht!]
\centering
	\includegraphics[height=0.45\textwidth]{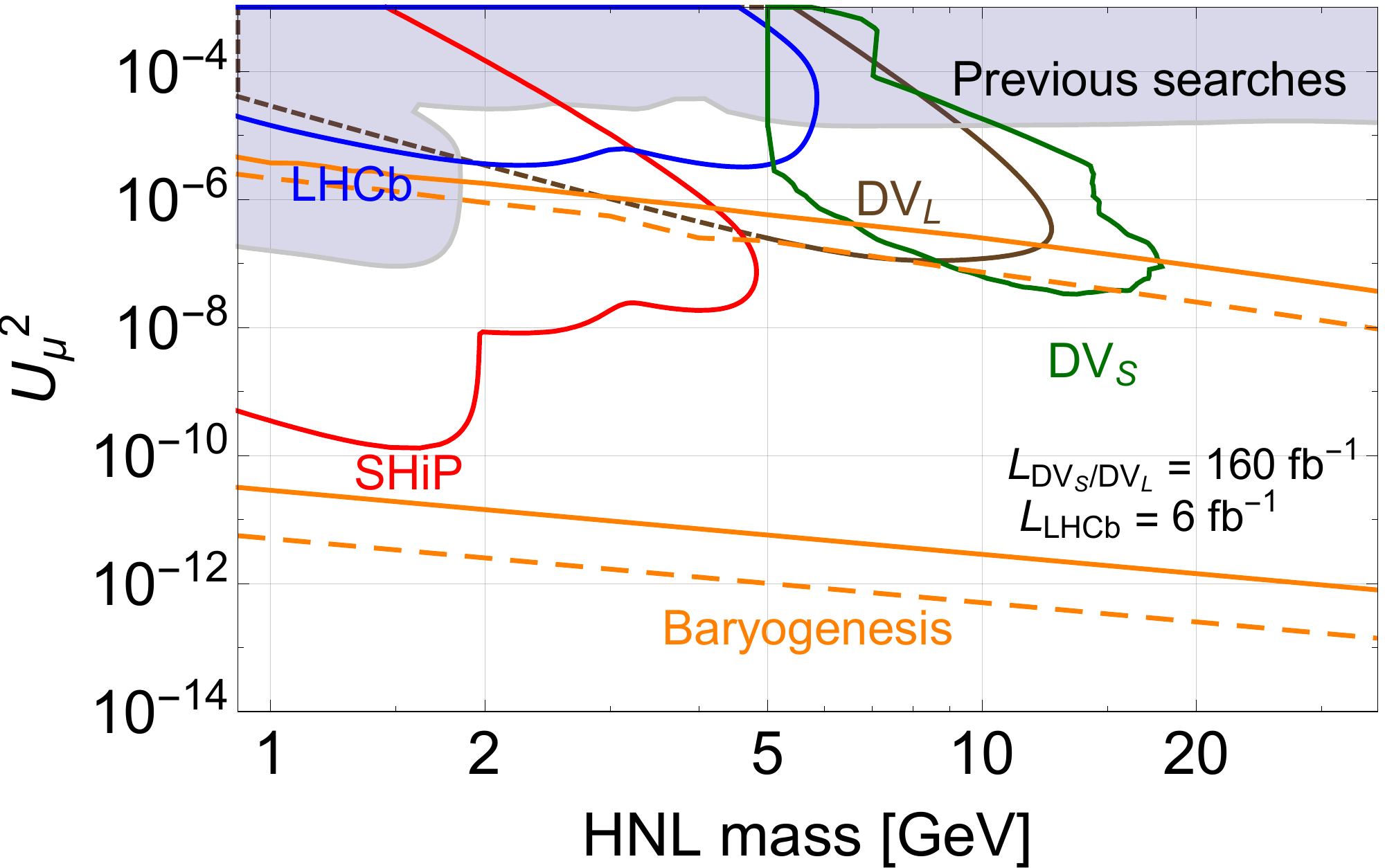} 
	\includegraphics[height=0.45\textwidth]{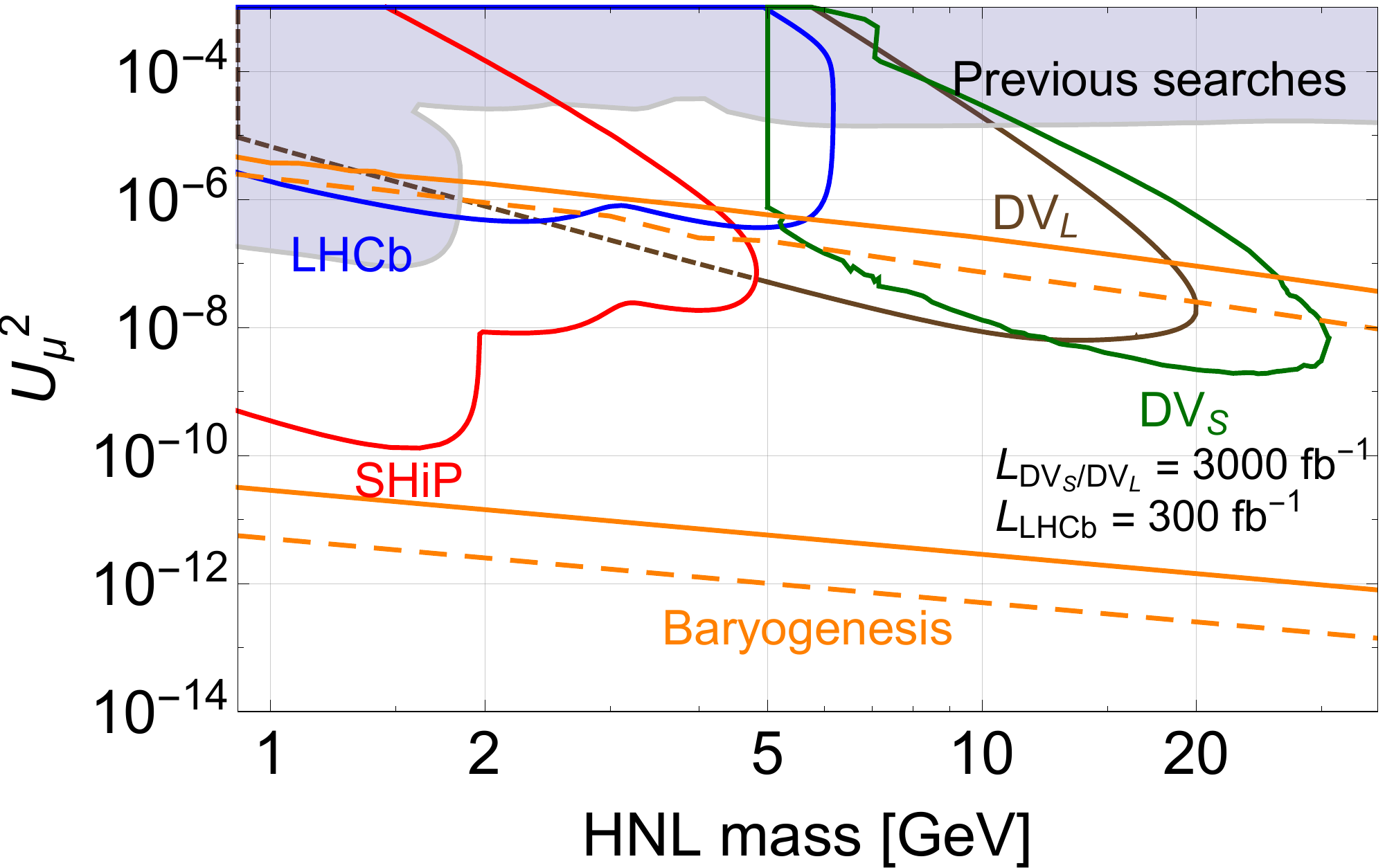}
	\caption{
	 Upper limits on the mixing parameter between the muon neutrino 
	and a sterile Majorana neutrinos $\abs{V_{\mu\text{N}}}^2$ which can be obtained in various new searches discussed 
	in the text with the existing LHC data (left), and with the HL-LHC data (right)~\cite{Boiarska:2019jcw}.
}
	\label{fig:future_results}
\end{figure*}

An example of such estimates from Ref.~\cite{Boiarska:2019jcw} is shown in Fig.~\ref{fig:future_results}.
In these projections, long-lived heavy neutrinos are considered. It is assumed that signal displacement completely
removes SM bkg. The signal acceptance is estimated with the generator-level MC information, applying reconstruction
and identification efficiencies reported by the LHC collaborations. Three types of searches are considered. 
A displaced vertex search in a tracker volume by the ATLAS or CMS is shown as DV$_S$;
a displaced vertex search with muon system at CMS is shown as DV$_L$.
Finally, LHCb reach is estimated for the inclusive heavy neutrino search in b-hadrons decays.
It is worth to highlight that the data available already now after the Run 2 of the LHC should allow
to probe HNL parameter space interesting from baryogenesis considerations. 

Future datasets and new analyses and ideas would significantly improve the reach in the sensitivity.

\clearpage
\subsection{Results and new benchmarks for heavy neutral lepton searches}
\label{ssec:hnls-results}
\subsubsection{Results}
\label{sssec:hnl-results}

\begin{figure*}[h]
\centering
\includegraphics[width=0.8\linewidth]{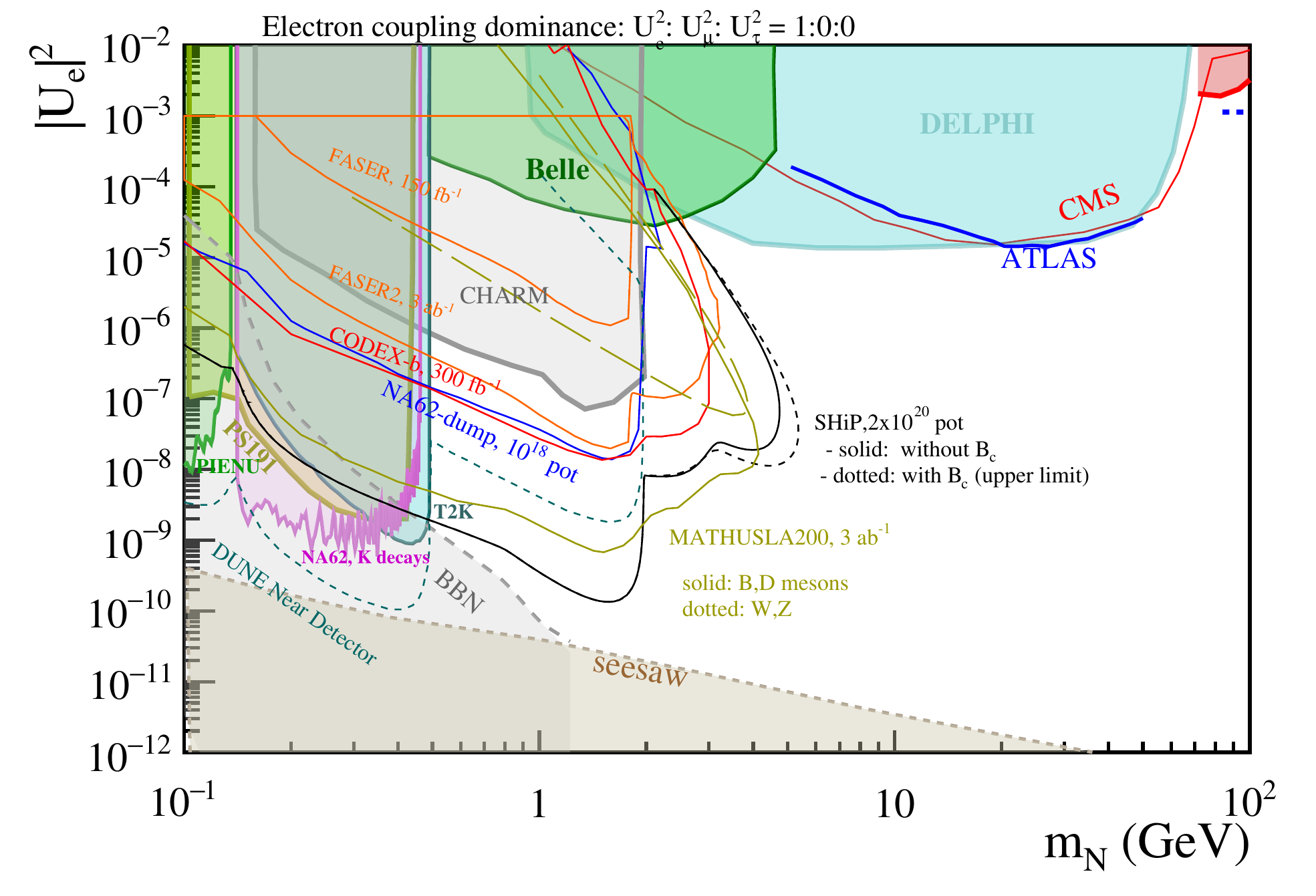}
\includegraphics[width=0.8\linewidth]{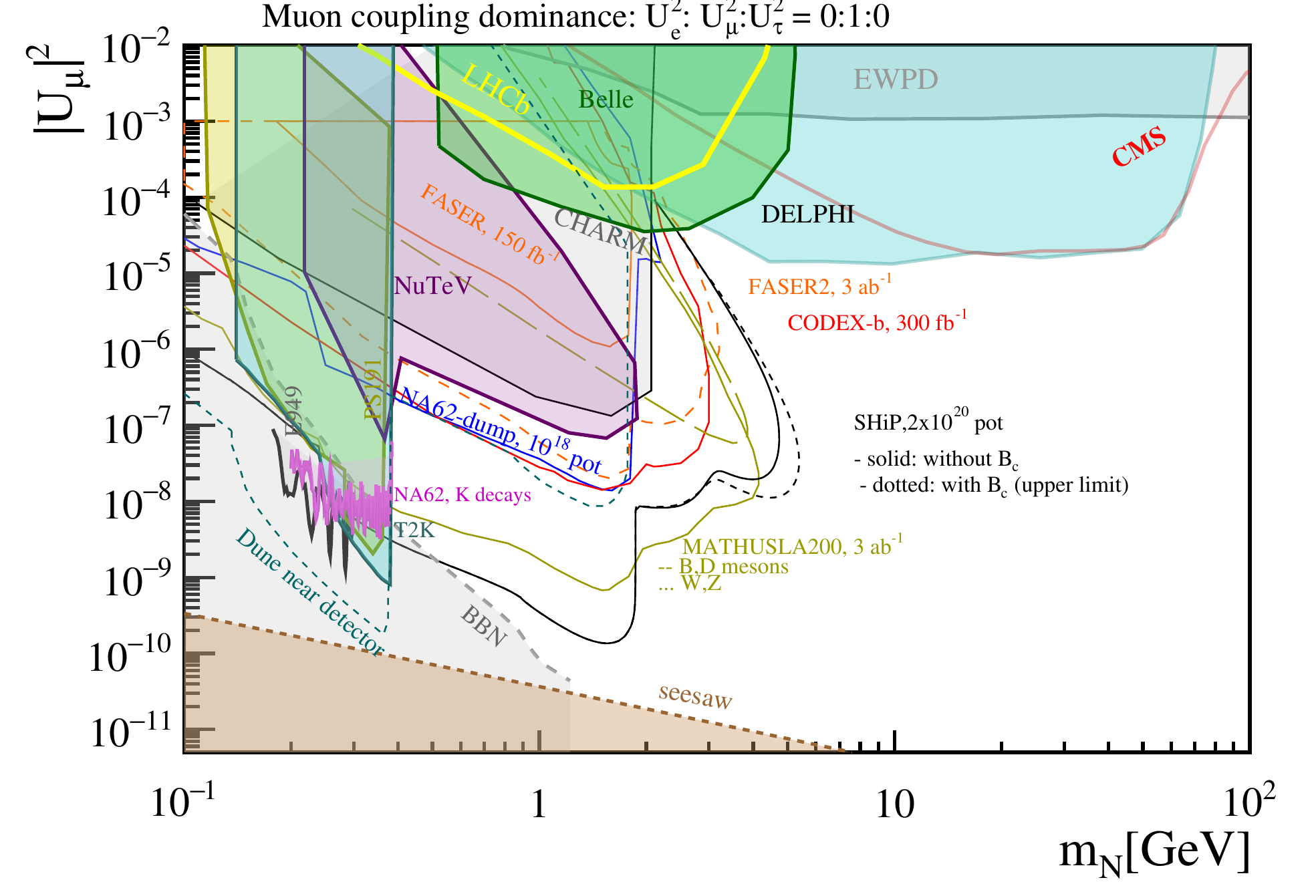}
\caption{Heavy Neutral Leptons with coupling to the first (top) and second (bottom) lepton generation. Filled areas are existing bounds from:
PS191~\cite{Bernardi:1987ek}, CHARM~\cite{Bergsma:1985qz}, PIENU~\cite{Aguilar-Arevalo:2017vlf},
NA62 ($K_{eN}$)~\cite{NA62:2020mcv},
NA62 ($K_{\mu N}$)~\cite{CortinaGil:2021gga},
T2K~\cite{extracted_beams:T2K_HNL_results},
Belle~\cite{Liventsev:2013zz}; DELPHI~\cite{Abreu:1996pa}, ATLAS~\cite{Aad:2019kiz} and CMS~\cite{Sirunyan:2018mtv}.
Coloured curves are projections from:  NA62-dump~\cite{NA62:dump,Beacham:2019nyx},
DarkQuest~\cite{Batell:2020vqn},
Belle-II~\cite{Dib:2019tuj},
FASER and FASER2~\cite{Ariga:2018uku}; SHiP~\cite{SHiP:2018xqw}, DUNE near detector~\cite{Ballett:2019bgd}, CODEX-b~\cite{Aielli:2019ivi}, and MATHUSLA200~\cite{Alpigiani:2020tva}.
The BBN bounds are from \cite{Sabti:2020yrt}. The seesaw bounds are computed under the hypothesis of two HNLs mixing with 
active neutrinos, and should be considered only indicative.
Figures revised from ref.~\cite{Lanfranchi:2020crw}.
\label{fig:HNL_e_mu}}
\end{figure*}

\begin{figure*}[h]
\centering
\includegraphics[width=0.8\linewidth]{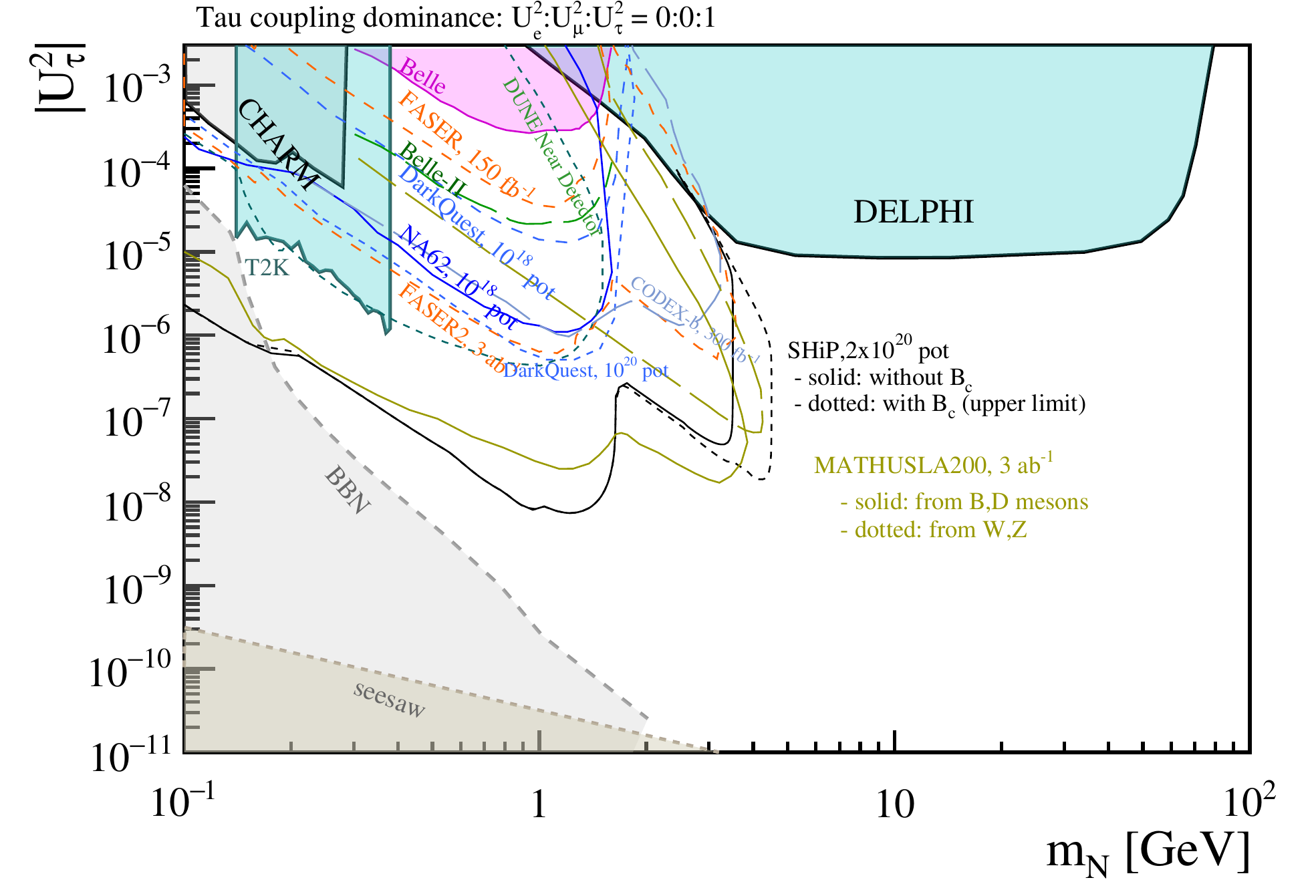}
\caption{Heavy Neutral Leptons with coupling to the third lepton generation. Filled areas are existing bounds from:
 CHARM~\cite{Bergsma:1985qz};  Belle~\cite{Liventsev:2013zz}; DELPHI~\cite{Abreu:1996pa},;
 T2K~\cite{extracted_beams:T2K_HNL_results}
Coloured curves are projections from:  NA62-dump~\cite{NA62:dump,Beacham:2019nyx},
DarkQuest~\cite{Batell:2020vqn},
Belle-II~\cite{Dib:2019tuj},
FASER and FASER2~\cite{Ariga:2018uku}; SHiP~\cite{SHiP:2018xqw},  CODEX-b~\cite{Aielli:2019ivi}, and MATHUSLA200~\cite{Alpigiani:2020tva}.
The BBN bounds are from \cite{Sabti:2020yrt}. The seesaw bounds are computed under the hypothesis of two HNLs mixing with 
active neutrinos, and should be considered only indicative.
Figures revised from ref.~\cite{Lanfranchi:2020crw}.}
\label{fig:HNL_tau}
\end{figure*}

Figures~\ref{fig:HNL_e_mu} and ~\ref{fig:HNL_tau} show the current status of searches and projected sensitivities for HNL coupled with only one lepton generation at the time.\footnote{The seesaw and BBN bounds depend stronger on the choice of model parameters than the direct search bounds (e.g.~on the number of HNL the flavours, relative size of the mixings to individual SM flavours, and the mass of the lightest SM neutrino), cf.~e.g.~\cite{Drewes:2019mhg,Boyarsky:2020dzc} for a discussion. As a result, the plots shown here are very conservative regarding the viable parameter space. The projection of the viable parameter region in the pure type I seesaw model with three heavy neutrinos on the mass-mixing plane is considerably larger and extends down to a few tens of MeV \cite{Domcke:2020ety}.}
These benchmarks correspond to the established PBC benchmarks
$BC6$, $BC7$ and $BC8$ as reported in~\cite{Beacham:2019nyx}
and repeated here for convenience.

\begin{itemize}

\item{\em BC6, Single HNL, electron dominance:}
  Assuming one Majorana HNL state $N$, and the predominant mixing with electron neutrinos, 
all production and decay can be determined as function of parameter space $(m_N, |U_e|^2 )$. 
\item{\em BC7, Single HNL, muon dominance:}  Assuming one Majorana HNL state $N$, and the predominant mixing with muon neutrinos, 
all production and decay can be determined as function of parameter space $(m_N, |U_\mu|^2 )$. 

\item{\em BC8, Single HNL, tau dominance:}  One Majorana HNL state with predominant mixing to tau neutrinos. Parameter space is 
$(m_N, |U_\tau|^2 )$.
 
\end{itemize}

\subsubsection{Final considerations and proposal of new benchmarks} 
\label{sssec:hnl-recommendations}

The aim of this Section is to harmonise future experimental results related to heavy neutrino searches at extracted beam lines and colliders with recent theoretical advances and results from neighbouring fields, as active neutrino physics, astroparticle, and cosmology.

\vskip 2mm
To this aim, a set of recommendations 
are outlined that might be used as guidelines for the next round of experimental results for minimal HNL models. 

\begin{enumerate}

\item Given the fast progress in model building, the experimental results should be presented also in a form that allows an easy re-interpretation of the same data set in different models. To this aim {\it we advocate the publication of the maps of experimental efficiencies and background yields in a 2D plane lifetime versus mass and, in general, of any information needed to recast the same data sets in new models following the fast theory progress.}

\item Historically the exclusion bounds are presented at the 90\% CL by experiments operating at extracted beam lines, and at the 95\% CL by experiments operating at colliders. {\it We suggest to use everywhere the 95\% CL convention for presenting exclusion bounds.}

\item {\it Minimal models - single-flavor dominance:}\\
Currently the exclusion bounds for HNL searches are presented in the single-flavor dominance approximation.

The choice of these benchmarks was historically driven by simplicity and allowed a fast and easy comparison across a large spectrum of experimental results. Moreover these benchmarks are important in case the HNLs are decoupled from the mechanism of mass generation for light neutrinos.

{\it We advocate to keep the hypothesis of single-flavor dominance as benchmarks for the experimental reaches also in the future.}

\item {\it Minimal models - new benchmarks}\\
If the HNLs are heavy neutrinos (hence they are related to the origin of neutrino masses and oscillations) the single-flavor dominance approximation cannot reflect reality, as at least two flavor couplings are necessary to explain the mixing and masses of light neutrinos.
If the HNLs are responsible of the active neutrino mass generation mechanism, at least two HNL-$\nu$ couplings should be active. The values of HNL-$\nu$ couplings must be compatible with the active neutrino mass and mixing parameters. The ranges of couplings where this condition is satisfied are shown in Figure~\ref{fig:triangle} (for two HNL case) and Figure~\ref{triangleplot} (for three HNL case).
{\it We consider the case of 2 HNLs (Figure~\ref{fig:triangle}) because this case is more conservative as far as the allowed couplings are concerned}. 

For the two HNL case, new benchmarks involve the choice of two set of ratios of couplings, one for normal hierarchy (NH)  and one for inverted hierarchy (IH) for the light neutrino masses. In addition to the constraints from active neutrino parameters, the new working points should also be compatible with the existing experimental bounds and constraints from BBN.

{\it We advocate the use of the following two new benchmarks for the next round of experimental results:}
\begin{itemize}
    \item[-] IH: $U^2_e : U^2_{\mu} : U^2_{\tau} = 1/3 : 1/3 : 1/3$;
    \item[-] NH: $U^2_e : U^2_{\mu} : U^2_{\tau} = 0.06 : 0.48 : 0.46$

\end{itemize}
These two working points are the central values of the blue (for IH) and red (for NH) regions of Figure~\ref{fig:triangle}, that are the regions  compatible with the active neutrino mixing parameters.
The BBN curves compatible with this new choice and a preliminary recast of the existing data up to 5 GeV  are shown in Ref.~\cite{Bondarenko:2021cpc}.

\end{enumerate}

\clearpage
\section{Feebly-interacting particles: Further Searches}
\label{sec:non-minimal}

\subsection{Search for FIPs at LHCb}
\label{ssec:cidvidal}
{\it Author: Xabier Cid Vidal, <xabier.cid.vidal@cern.ch>} 
\subsubsection{Introduction}

In this Section I present searches for FIPs done at the LHCb experiment at CERN \cite{detector_lhcb,perform_lhcb} 
combined with the introduction of some new ideas for future searches, to make the most of the LHCb potential. These futures ideas will involve not only the short-term, but also potential future upgrades of the detector \cite{lhcb_eoi}.

LHCb is a forward spectrometer whose special features are particularly useful for the detection of light FIPs. Examples of these features include an advanced vertexing, and excellent momentum resolution and particle identification (PID). Furthermore, the LHCb trigger is capable of triggering on very soft objects compared to other LHC experiments \cite{trigger_upgrade}. This will be reinforced after the current upgrade of the detector, when this will be readout at every LHC bunch-crossing, allowing a trigger fully-based on software, which is very convenient to detect FIPs. A very important aspect of this new trigger will be the use of GPUs \cite{gpu}, which will allow maintaining soft $p_{T}$ thresholds. 

\subsubsection{LHCb results}

In the last years LHCb has put in place a strong program to search for FIPs \cite{exotica}, exploring different signatures and final states that target a large number of theoretical models. Rather than reviewing all of these results, here I give examples of some relevant analyses, focusing in the most recent ones or those showing the capabilities of LHCb.

The first examples, which I will cover only briefly, include a search for decays of Higgs-like bosons to a $\mu \tau$ final state \cite{LHCb-PAPER-2018-030} and that of a long-lived particle (LLP) decaying to a pair of jets \cite{LHCb-PAPER-2016-065}. For the first, it provides limits in the Higgs mass range $40-100$ GeV/c$^ 2$, not looked into at other experiments, and involves four different reconstruction modes of the $\tau$ lepton. For the second, it is able to cover a region in the LLP mass-lifetime plane where other LHC experiments have no sensitivity. The future sensitivity of LHCb for this search is studied in Ref.~\cite{LHCb:2018hiv}.

\vskip 2mm
The second group of results I will review in these proceedings are more recent, and involve the search for FIPs in the $\mu^+\mu^-$ final state, where the FIP decay can be displaced from the primary vertex (PV) or occurring promptly. Depending on the FIP model addressed the experimental strategy becomes slightly different. This is explained in more detail in the next paragraphs.

\vskip 2mm
\noindent
One of the models for which this final state provides more sensitivity is that of Dark Photons (DP). Dark photons couple via kinetic mixing to the SM photons and provide a ``portal'' that connects a hidden dark sector to the SM. The magnitude of this coupling affects not only the production rate of dark photons, but also their lifetime. Different proposals exist to search for dark photons at LHCb \cite{Ilten:2015hya,Ilten:2016tkc}. Following these, two different searches have been performed at LHCb using the datasets taken in the 2016-2018 period \cite{LHCb-PAPER-2017-038,Aaij:2019bvg}. The main advantage of searching for dark photons in the $\mu\mu$ final state is that these self-normalize to the SM $\gamma^* \rightarrow \mu\mu$ process, so that no absolute efficiencies must be computed in general. Although no dark photon has been found, the limits achieved are world-best in different regions of the parameter space of these particles, in particular in the mass region between 0.2 and 10 GeV/c$^2$. 

\begin{figure*}
    \centering
    \includegraphics[width=0.8\textwidth]{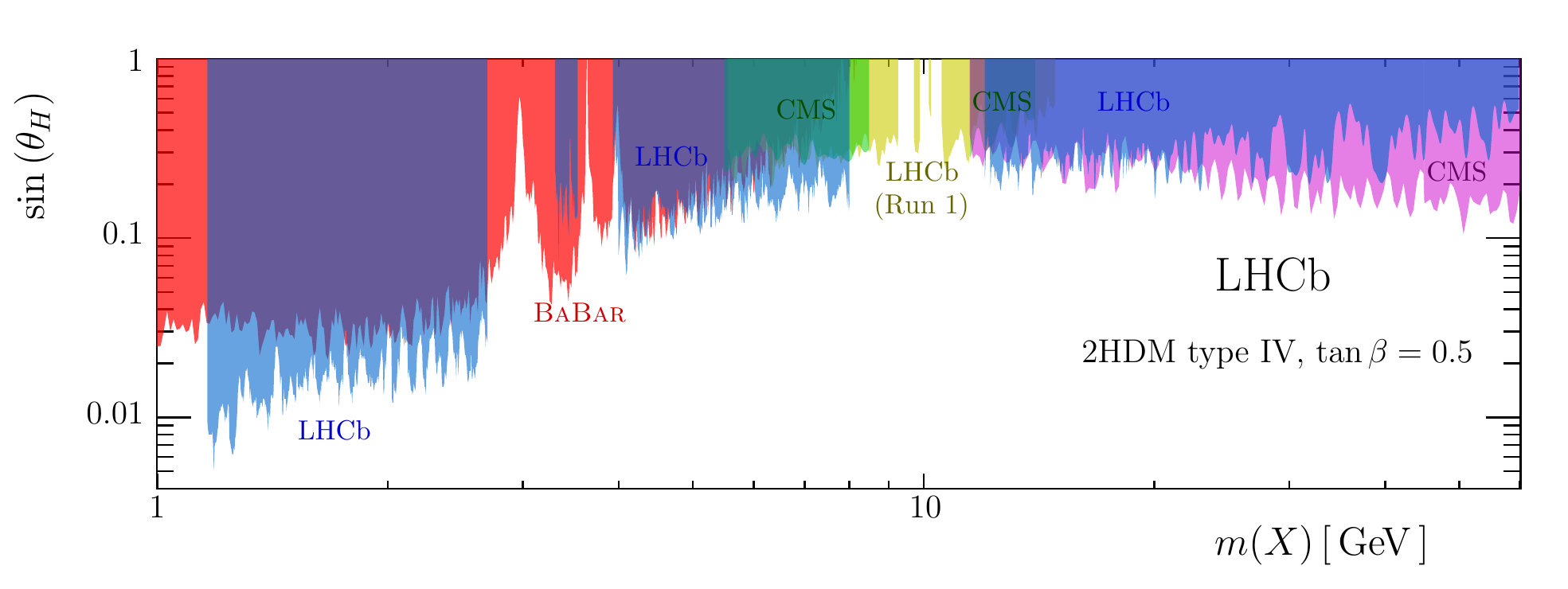}
    \caption{The FIP $(X)$ - Higgs mixing angle vs FIP mass space for a realization of the 2HDM model \cite{Haisch:2016hzu}. Figure taken from Ref.~\cite{Aaij:2020ikh}. The experimental exclusions include results from LHCb \cite{Aaij:2020ikh,LHCb-PAPER-2018-008}, CMS \cite{Chatrchyan:2012am,Sirunyan:2019wqq}, and BaBar \cite{Lees:2012iw}.}
    \label{fig:2hdm_lhcb}
\end{figure*}

\vskip 2mm
Apart from dark photons, the $\mu^+\mu^-$ final state provides sensitivity to several other BSM models. In that regard, an independent more general search has also been performed at LHCb \cite{Aaij:2020ikh}. In this case, absolute efficiencies are obtained from simulation, and model-independent results are provided in terms of different parameters of the FIP decaying to a pair of muons, such as mass, $p_T$ or natural width. The search includes both displaced or prompt decays of the FIP, and also the fact that these might not point back to the PV. Furthermore, the case in which the FIP is produced in association to $b$-jets is also covered. Once again, even if no excess has been found, the results achieved are world-best in several regions of the relevant parameter space. Apart from the model independent results, interpretations are provided in a couple of theoretical models of interest \cite{Haisch:2016hzu,Pierce:2017taw}. For the first, which involves a realization of the 2HDM scenario, the regions excluded by this result are showed in Fig.~\ref{fig:2hdm_lhcb}.

\subsubsection{Other proposals}

Beyond these experimental analyses, the potential of LHCb has been explored in other publications, that propose future ambitious searches for FIPs at LHCb. Some examples are shown below.

\begin{figure*}
    \centering
    \includegraphics[width=0.8\textwidth]{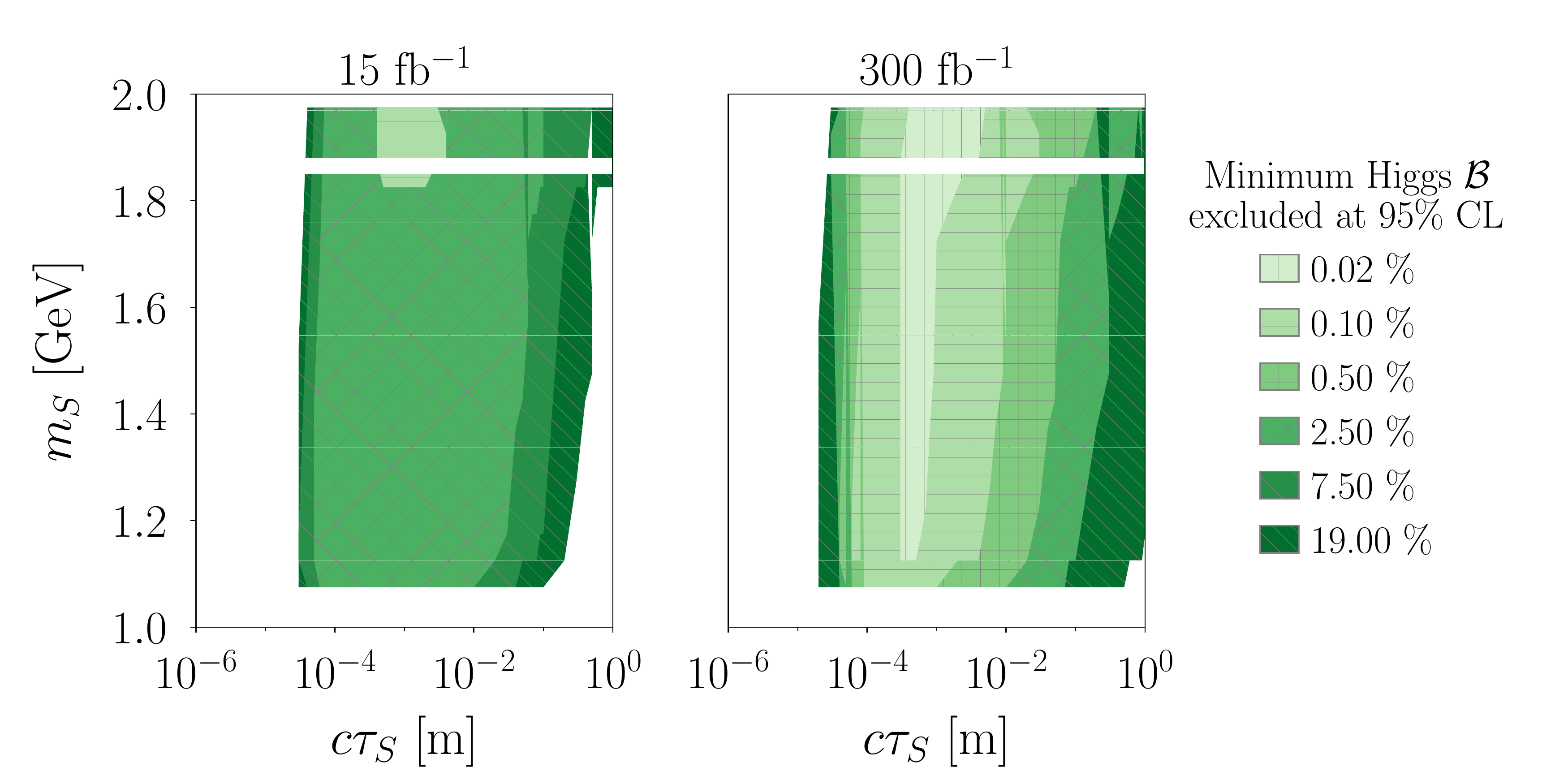}
    \caption{LHCb prospects for model independent limits on $\mathcal{B}(H\rightarrow SS)$, assuming $\mathcal{B}(S\rightarrow K^+K^-) = 100\%$ for different lifetimes and masses of the scalar $S$, which is the FIP in this case. Figure taken from Ref.~\cite{CidVidal:2019urm}. The integrated luminosities chosen as reference are those that LHCb is expected to collect by 2025 (15 fb$^{-1})$ or the end if its lifetime (300 fb$^{-1})$.}
    \label{fig:limits_exclusive}
\end{figure*}

LHCb has unique PID capabilities at the LHC. In particular, the presence of the RICH detectors \cite{Adinolfi:2012qfa} can be very useful to search for certain types of FIPs. An example of this is the capacity of RICH to distinguish kaons from pions. In this regard, if FIPs are light, their decay to charged kaons can have dominant branching fractions in certain models 
\cite{Winkler:2018qyg}. This is studied in more detail in Ref.~\cite{CidVidal:2019urm}, where the sensitivity of LHCb to reconstruct such FIPs, produced through exotic decays of the SM Higgs boson, is studied. The main results of this study are summarized in Fig.~\ref{fig:limits_exclusive}. Depending on the lifetime and mass of these FIPs, exotic branching fractions of the SM Higgs boson to this type of FIP could be excluded at the per-mille level with future LHCb data.

\vskip 2mm
The existence of $\mu^+\mu^-$ a bound state, know as True Muonium (TM), is well established in the SM, with its properties being just depending on QED \cite{Brodsky:2009gx}. Furthermore, BSM physics might affect these properties \cite{Smirnov:2020zwf}, which makes the study of this particle of great interest. The ortho-TM state is expected to be long-lived, decay predominantly to an $e^+e^-$ pair and be a fragile state, i.e., so that material interactions would produce the $\mu^+\mu^-$ separation. Taking this into account, a recent study \cite{CidVidal:2019qub} shows that LHCb could discover the TM with the dataset collected by $\sim2025$. For this, it would reconstruct TMs produced predominantly in $\eta\rightarrow \mu \mu \gamma$ decays.

\vskip 2mm
Another interesting example of the capabilities of LHCb in the area of FIPs arises from the study of light Axion-like particles (ALPs). ALPs are pseudo Nambu-Goldstone bosons, associated to spontaneously broken approximate symmetries, whose mass can be arbitrarily below the BSM scales currently probed at the LHC. Depending on the mass and final state to which ALPs decay, LHCb can provide unique potential for their discovery. More specifically, I will refer here to a certain type of ALPs that couple to gluons (allowing their production at the LHC) or photons (allowing their detection) \cite{Mariotti:2017vtv}. As showed in Ref.~\cite{CidVidal:2018blh}, if the mass of the ALP is in the range $3-10$ GeV/c$^2$, LHCb has the world-best sensitivity to discover such a particle. The main reason for this is the special trigger configuration at LHCb \cite{Benson:2019wvt}, which provides high efficiency to reconstruct relatively soft photon pairs. 

\vskip 2mm
The final case I will cover is the potential discovery of Dark Matter (DM) at LHCb. Indeed, a recently introduced mechanism would allow to explain the origin of DM and baryogenesis at once \cite{Elor:2018twp}, by proposing DM to be charged under the baryonic number and also to be produced in $b$ hadron decays ($b\rightarrow \rm{DM} + \rm{X^{SM}}$). In this type of models, the allowed branching fractions of the $b$ decays to DM would be large, at reach at LHCb. The expected signature would be a displaced and isolated SM resonance ($\rm{X^{SM}}$), with the decay point of the $b$ hadron determined thanks to the $\rm{X^{SM}}$ decay. Moreover, the production of the $b$ hadrons through heavier $b$ resonances has already been shown \cite{Aaij:2020mqb} to provide additional handles to discriminate against the SM background.

\subsubsection{Conclusions}
Although usually LHCb is simply considered a flavour factory, in the last years it has been shown that it could become the best tool to provide a FIP discovery, similar to other proposed experiments \cite{Beacham:2019nyx}, but the big advantage that it is already built and running. 

A plethora of searches has been performed in the area of FIPs, and several new ideas to enlarge these searches exist. The recent analyses using the $\mu\mu$ final state are the best example of the LHCb potential for FIP physics.

\clearpage
\section{Conclusions and Outlook}
\label{sec:conclusions}
Feebly-interacting particles (FIPs) may provide an answer to many open questions in particle physics,
astrophysics, and cosmology. These include the origin of the neutrino masses and oscillations, the baryon asymmetry of the universe, the dark matter, the strong CP problem, the cosmological inflation,
and the apparent fine tuning of the EW scale.

\vskip 2mm
The community behind these efforts is active and diversified: it includes particle physics theorists and astrophysicists with a wide range of interests, as well as particle physics experimentalists working on collider,
fixed-target, beam-dump, direct  DM detection experiments, as well as dedicated axion/ALP experiments.

\vskip 2mm
In the coming years a wealth of experimental results for FIPs are expected from all the major laboratories in the world.
These include accelerator based experiments (ATLAS, CMS, LHCb, NA64, NA62 at CERN; MEG-II and Mu3e at PSI; Belle II at KEK; BDX and HPS at JLab; MiniBooNE at FNAL; T2K ND280 and KOTO at J-Park) as well as dark matter direct detection experiments and dedicated
experiments searching for axions/ALPs in Europe and the US\@.
In addition to established experimental efforts, a multitude of new initiatives has recently emerged both at CERN (NA62 in dump mode, SHiP, MATHUSLA, FASER, CODEX-b) and elsewhere (LDXM at SLAC, long and short baseline neutrino near detectors at Fermilab, Beam Dump experiment at Mainz, and others) aiming at covering in the
coming decade uncharted regions of FIPs parameter space inaccessible to traditional experiments.

\vskip 2mm
No single experiment or experimental approach is sufficient alone to cover the large parameter space in terms of masses and
couplings that FIPs models can suggest: hence synergy and complementarity among a great variety of
experimental facilities are paramount, calling for a deep collaboration and cross-fertilisation across different communities.
For example, understanding the interplay between active neutrino mixing parameters derived from ongoing
and future neutrino experiments and the favoured ranges of HNL parameters would be important,
as would be more comprehensive comparisons of sub-GeV direct detection searches and accelerator based searches.
Likewise, the wealth of data coming from accelerator based experiments should be compared in greater detail
to existing and future astrophysics and cosmology driven sources of insight.

\vskip 2mm
Fundamental physics questions might be naturally intertwined and it is paramount to develop a common approach
within the FIPs paradigm. We believe that the multitude of experimental efforts today calls for a reassessment of the
theoretical framework for FIPs that emphasizes the most promising and relevant models going beyond the minimal portal framework.
Theoretical studies should continue to map out how the known portal operators are
embedded in complete models addressing the varied puzzles of the SM, and this will be important for
identifying additional promising sensitivity milestones around which to focus experimental efforts.

\vskip 2mm
The breadth of the open questions in particle physics and their deep interconnection, together with the failure so far of standard paradigms, requires today a diversified research programme with different experimental objectives and techniques, and with strong and focused theoretical involvement.
In this vein, the field of FIPs is casting a light on the sub-GeV frontier, and offers many opportunities for exciting and profound discoveries in the future.

\section{Acknowledgements}

We warmly thanks the neutrino platform secretariat at CERN for the support with the Workshop organization.
IGI acknowledges support from the European Research Council (ERC) under grant ERC-2017-AdG-788781 (IAXO+ project), as well as from the Spanish Agencia Estatal de Investigacion (AEI) under grant PID2019-108122GB-C31. FK  is  funded  by  the  Deutsche Forschungsgemeinschaft (DFG) through the Emmy Noether Grant No.\ KA 4662/1-1. The work of VMO is supported in part by the National Science Foundation CAREER Grant 1830832. 
The work of YVS was supported by the World Premier International Research Center Initiative (WPI), MEXT, Japan, and by the JSPS KAKENHI Grant Number JP20K14460. The research of SG is supported in part by the National Science Foundation CAREER grant PHY- 1915852.
The work of XCV is supported by MINECO (Spain) through the Ram\'{o}n y Cajal program RYC-2016-20073 and by XuntaGAL under the ED431F 2018/01 project. He has also received financial support from XuntaGAL (Centro singular de investigaci\'{o}n de Galicia accreditation 2019-2022), by European Union ERDF, and by the Mar\'{i}a de Maeztua Units of Excellence program MDM-2016-0692 and the Spanish Research State Agency.
The work of YS is supported by grants from  NSF-BSF, ISF  and the Azrieli foundation, YS is Taub fellow (supported by the Taub Family Foundation).
OR acknowledges support under the European Union's Horizon 2020 research and innovation programme (GA 694896) and from the Carlsberg foundation. JLP acknowledges support
from Generalitat Valenciana through the plan GenT program (CIDEGENT/2018/019) and from the Spanish MINECO under Grant FPA2017-85985-P.
JS is supported in part by DOE Grant DE-SC0017840. LS acknowledges support from the European Research Council under grant agreement 758316. 

\clearpage

\bibliographystyle{JHEP}
\bibliography{biblio/biblio.bib}

\end{document}